\documentclass[10pt, twoside, notitlepage]{book}

\usepackage[paper=a4paper,margin=2.5cm, bindingoffset=0.5cm]{geometry}
\usepackage[onehalfspacing]{setspace}
\usepackage{indentfirst}
\usepackage{braket}

\usepackage[utf8]{inputenc} 
\usepackage{appendix}

\usepackage{cancel}
\usepackage{slashed}
\usepackage{graphics}
\usepackage{bbold}

\usepackage{graphicx} % needed for inserting images in document
\usepackage{float} % improves interface for floating objects and tables
\usepackage{multirow} % allows to use multirows in tables
\usepackage{booktabs}
\usepackage{tabularray}
\usepackage[figurename=Fig.]{caption}
\usepackage[list=true, listformat=simple, labelformat=simple]{subcaption}

\graphicspath{ {images/} }

\usepackage{amsmath, amssymb,bm} % math text
\usepackage{amsfonts} % additional math fonts
\usepackage{nicefrac}
\usepackage{bbold}
\usepackage{slashed}
\usepackage{siunitx}

\DeclareMathOperator{\Tr}{Tr} 

\usepackage[backend=bibtex, style=phys,%
   articletitle=true,biblabel=brackets,%
   chaptertitle=false,pageranges=true,eprint=true]{biblatex}
    
\DeclareFieldFormat[thesis]{citetitle}{#1} %italic phdthesis
\DeclareFieldFormat[thesis]{title}{#1} %italic phdthesis

\usepackage[explicit]{titlesec}
\usepackage{xcolor}
\usepackage{lmodern}
\usepackage[colorlinks=false,citecolor=black,linkcolor=black,urlcolor=blue,bookmarksopen=true]{hyperref}
\usepackage{bookmark}

\DefTblrTemplate{contfoot-text}{default}{}
\DefTblrTemplate{conthead-text}{default}{}
\DefTblrTemplate{caption}{default}{}
\DefTblrTemplate{conthead}{default}{}
\DefTblrTemplate{capcont}{default}{}

\usepackage{epigraph}

\DeclareMathOperator{\diag}{diag}

\newcommand\MyBox[2]{
  \fbox{\lower0.75cm
    \vbox to 1.7cm{\vfil
      \hbox to 1.7cm{\hfil\parbox{1.4cm}{#1\\#2}\hfil}
      \vfil}%
  }%
}

\newcolumntype{K}[1]{>{\centering\arraybackslash}m{#1}}
\def\gsim{\raise0.3ex\hbox{$\;>$\kern-0.75em\raise-1.1ex\hbox{$\sim\;$}}}
\def\lsim{\raise0.3ex\hbox{$\;<$\kern-0.75em\raise-1.1ex\hbox{$\sim\;$}}}

\def\vev#1{\left\langle #1\right\rangle}
\def\SM{$\mathrm{SU(3)}_c \otimes \mathrm{SU(2)}_L \otimes \mathrm{U(1)}_Y$}
\def\EW{$\mathrm{SU}(2)_L \otimes \mathrm{U}(1)_Y$ }

\newcommand{\M}{\mathbf{M}}
\newcommand{\U}{\mathbf{U}}

\newcommand{\Y}{\mathbf{Y}}

\newcommand{\BR}{{\rm BR}}

\def\MR{\mathbf{M}_R}

\def\Ye{\mathbf{Y}_e}
\def\Y{\mathbf{Y}}

\def\YD{\mathbf{Y}_D}
\def\YR{\mathbf{Y}_R}
\def\YRp{\mathbf{Y}_R'}

\def\Mnu{\mathbf{M}_\nu}
\def\MD{\mathbf{M}_D}
\def\Mnuh{\widehat{\mathbf{M}}_\nu}

\def\Me{\mathbf{M}_e}

\def\Uh{\mathbf{U}_{\text{h}}}

\def\U{\mathbf{U}}

\newcommand {\dblack} {\color{black}}

\allowdisplaybreaks

\begin{document}

% Title page

%Draft

%\include{chapters/TitlePage_Draft}

%\thispagestyle{empty}

%Final

%--------------------------------------------------------------------
%				TITLE PAGE    
%--------------------------------------------------------------------

\newgeometry{left=1.5cm, right=1.5cm, top=1.5cm, bottom=1.5cm}
\begin{titlepage}
\begin{figure}[H]
	\raggedright
	\includegraphics[keepaspectratio, width=0.3\textwidth]{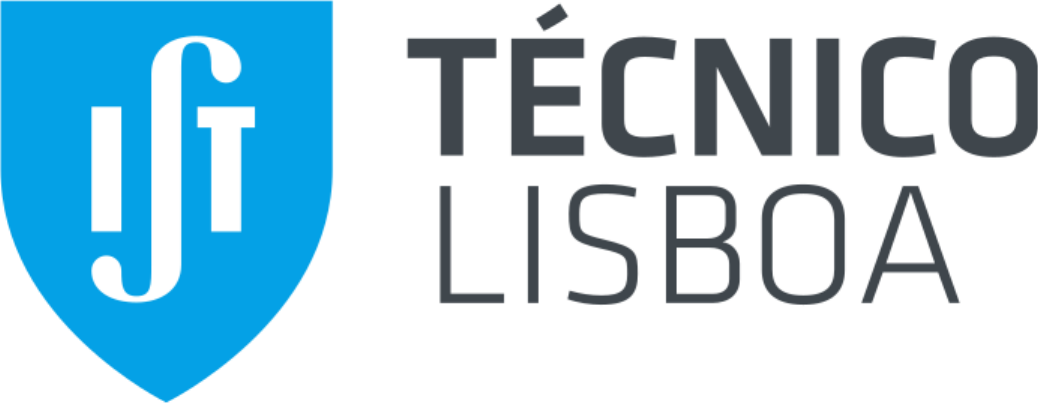}
\end{figure}

\begin{center}

	\LARGE\textbf{UNIVERSIDADE DE LISBOA \\ INSTITUTO SUPERIOR TÉCNICO}
	
	\vspace{2.5cm}
	
	\Large\textbf{From Dark Sectors to the \\ Axion-Neutrino Connection}
	
	\vspace{2cm}
	
	\Large\textbf{Henrique Pedro Fernandes de Noronha Brito Câmara}
	
	\vspace{2cm}
	
	\begin{tabular}{rl}
	\large \textbf{Supervisor:} & \large Doctor Filipe Rafael Joaquim \\
	\large \textbf{Co-supervisor:} & \large Doctor Ricardo Jorge González Felipe
	\end{tabular}
	
	\vspace{2cm}

    \large \textbf{Thesis approved in public session to obtain the PhD Degree in} \\
	\Large Physics
	
	\vspace{2cm}
	
	\large \textbf{Jury final classification:} Pass with Distinction and Honour 
	
	\vspace{3cm}

	\Large \textbf{2025}
	
\end{center}
\end{titlepage}
\restoregeometry

\thispagestyle{empty}

%--------------------------------------------------------------------
%				TITLE PAGE    
%--------------------------------------------------------------------

\newgeometry{left=2.5cm, right=2.5cm, top=1.5cm, bottom=1.5cm}
\begin{titlepage}
\begin{figure}[H]
	\raggedright
	\includegraphics[keepaspectratio, width=0.3\textwidth]{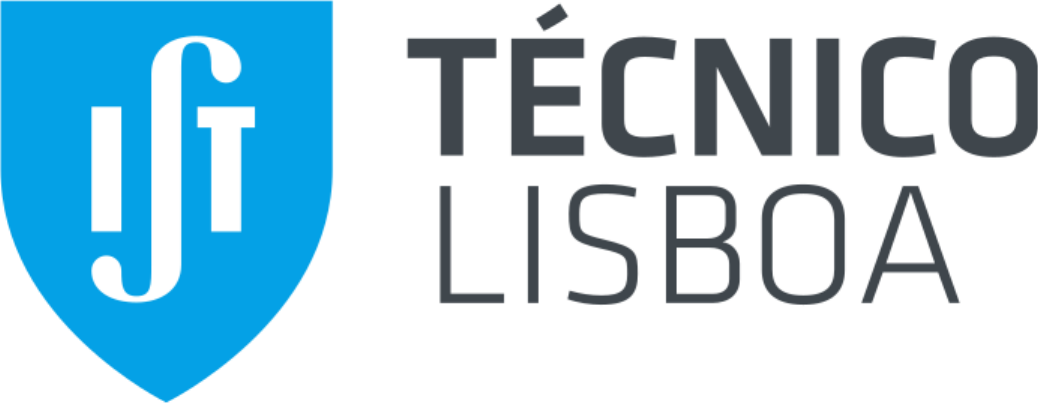}
\end{figure}

\begin{center}

	\LARGE\textbf{UNIVERSIDADE DE LISBOA \\ INSTITUTO SUPERIOR TÉCNICO}
	
	\vspace{1.5cm}
	
	\Large\textbf{From Dark Sectors to the \\ Axion-Neutrino Connection}
	
	\vspace{1cm}
	
	\Large\textbf{Henrique Pedro Fernandes de Noronha Brito Câmara}
	
	\vspace{1cm}
	
	\begin{tabular}{rl}
	\large \textbf{Supervisor:} & \large Doctor Filipe Rafael Joaquim \\
	\large \textbf{Co-supervisor:} & \large Doctor Ricardo Jorge González Felipe
	\end{tabular}
	
	\vspace{1cm}
	
	\large\textbf{Thesis approved in public session to obtain the PhD Degree in} \\
	\Large Physics
	
	\vspace{1cm}
	
	\large \textbf{Jury final classification:} Pass with Distinction and Honour
	
	\vspace{0.5cm}

	\large\textbf{Jury}
	
\end{center}

\begin{flushleft}
\normalsize
	\textbf{Chairperson:} 
	 Doctor João Paulo Ferreira da Silva, Instituto Superior Técnico, \\ \hspace{0.5cm} Universidade de Lisboa \\[0.25cm]
	\textbf{Members of the Committee:}  \\
	\hspace{0.5cm} Doctor António Joaquim Onofre de Abreu Ribeiro Gonçalves, Escola de Ciências, \\ \hspace{0.5cm} Universidade do Minho \\
	\hspace{0.5cm} Doctor Avelino Vicente Montesinos, Instituto de Física Corpuscular, \\ \hspace{0.5cm} CSIC -- Universitat de València, Spain \\
	\hspace{0.5cm} Doctor Filipe Rafael Joaquim, Instituto Superior Técnico, Universidade de Lisboa \\
	\hspace{0.5cm} Doctor Maria Margarida Nesbitt Rebelo da Silva, Instituto Superior Técnico, \\ \hspace{0.5cm} Universidade de Lisboa \\
\end{flushleft}

\begin{center}

	\vspace{0.5cm}

	\large\textbf{Funding Institution:} Fundação para a Ciência e a Tecnologia (FCT)
	
	\vspace{1cm}
	
	\Large \textbf{2025}
	
\end{center}
\end{titlepage}
\restoregeometry

\thispagestyle{empty}

% Pre-thesis
\pagestyle{plain}

\frontmatter % Roman numbering

%%%%%%%%%%%%%%%%%%%%%%%%%%%%%%%%%%%%%%%%%%%%%%%%%%%%%%%%%%%%%%%%%%%%%%%%

\null\vskip1cm%
\begin{center}
$\begin{aligned}
&\text{MAR PORTUGUÊS} \\
&\text{Ó mar salgado, quanto do teu sal} \\
&\text{São lágrimas de Portugal!} \\
&\text{Por te cruzarmos, quantas mães choraram,} \\
&\text{Quantos filhos em vão rezaram!} \\
&\text{Quantas noivas ficaram por casar} \\
&\text{Para que fosses nosso, ó mar!} \\
&\text{ } \\
&\text{Valeu a pena? Tudo vale a pena} \\
&\text{Se a alma não é pequena.} \\
&\text{Quem quer passar além do Bojador} \\
&\text{Tem que passar além da dor.} \\
&\text{Deus ao mar o perigo e o abismo deu,} \\
&\text{Mas nele é que espelhou o céu.} \\
\end{aligned}$
\end{center}
\begin{center}
  Fernando Pessoa, Mensagem (1934).
\end{center}

\null\vskip1cm%

\begin{flushleft}
Grande mer, toujours labourée, toujours vierge, ma religion avec la nuit! Elle nous lave et nous rassasie dans ses sillons stériles, elle nous libère et nous tient debout. À chaque vague, une promesse, toujours la même. Que dit la vague? Si je devais mourir, entouré de montagnes froides, renié par les miens, à bout de force enfin, la mer, au dernier moment, emplirait ma cellule, viendrait soutenir au-dessus de moi-même et m'aider à mourir sans haine.
\end{flushleft}
\begin{flushleft}
Albert Camus, L'Été (1954).
\end{flushleft}

\null\vskip1cm%

\begin{center}
$\begin{aligned}
&\text{- Il garde le mouvement des marées, les mouvements de la lumière.} \\
&\text{- Non.} \\
&\text{- Le mouvement des eaux. Le vent. Le sable.} \\
&\text{- Non.} \\
&\text{- Le sommeil ?} \\
&\text{- Non - elle hésite - rien. Le voyageur se tait.} \\
\end{aligned}$
\end{center}
\begin{center}
  Marguerite Duras, L'Amour (1971).
\end{center}

\vfill\newpage

%-------------------------------------------------------------------------------------
%				ACKNOWLEDGEMENTS    
%-------------------------------------------------------------------------------------

\chapter{Acknowledgements}

\vspace{-1.2cm}

First and foremost, I wish to express my deepest gratitude to my supervisor, Filipe Joaquim, whose guidance, encouragement, and mentorship have been a constant source of inspiration. The opportunities he provided have allowed me to grow not only as a researcher, physicist, and teacher, but also as a person. I am equally indebted to my co-supervisor, Ricardo Felipe, for his collaboration and the stimulating discussions we shared. I am profoundly grateful to both of them for the many conversations about physics and life, for the lessons they taught -- both scientific and personal -- for fostering a space where I could freely explore my ideas, and for their enduring support and friendship.

I gratefully acknowledge Fundação para a Ciência e a Tecnologia (FCT) for supporting my doctoral studies. I am also thankful to CFTP for providing an inspiring environment, and to Gui Rebelo and Sandra Oliveira for their dedication and support to the Centre and its members.

I acknowledge my international collaborators José Valle, Juan Antonio Aguilar Saavedra, Rahul Srivastava, and Newton Nath, thanks to whom I gained invaluable insights in physics and enriched the work presented in this thesis. I am also grateful for my visits to IFIC, Valencia, and IFT, Madrid, where I benefited from fruitful exchanges of ideas, guidance, and the opportunity to know José and Juan beyond the professional context. I would like to express my gratitude to Adam Smetana for hosting me during my multiple visits to IEAP, CTU in Prague.

I thank my fellow CFTP PhD student colleagues, and above all friends, with whom I collaborated: Débora Barreiros, Aditya Batra, João Seabra, and Miguel Bento. My experience at CFTP would not have been as enriching without their contribution to such an intellectually stimulating environment. I also thank Bernardo Gonçalves, Miguel Levy, Miguel Oliveira and Andreas Trautner for all the insightful discussions during lunch and at the quiosque. I am grateful to my CFTP MSc students José Rocha and Ricardo Amadeu, with whom I learned a great deal and developed a lasting friendship. Our collaboration with José, fueled by his relentless dedication, led to the publication of two papers.

À minha família, não há palavras que consigam expressar a minha gratidão. Obrigado, Mãe, Pai e Irmão, por acreditarem sempre em mim, por me apoiarem em todos os momentos e pela vossa dedicação incansável. Agradeço também à Avó Dilar, Avó Liv, Avô João, Tia Berta, Tia Polaca e Tio Paulo. Estiveram ao meu lado nos momentos mais importantes, oferecendo-me tudo o que precisei para perseguir os meus sonhos. Desde sempre, guiaram a minha curiosidade pelo mundo e alimentaram a minha paixão pela Física. Proporcionando carinho e apoio incondicional, ajudaram-me a crescer, a aprender e a tornar-me melhor a cada dia. Obrigado, de coração.

Estou também agradecido aos meus fiéis companheiros desde o MEFT, Milton e Ricardo, por todos os jantares, convívios e conversas que começaram entre relatórios de laboratório e continuam até hoje, tornando a experiência no IST numa jornada mais leve, divertida e inesquecível.

Mes remerciements s'étendent aussi à mes amis du lycée -- Antonio, Boloch, Henrique et Jean-Baptiste -- avec qui j’ai partagé tant de rires et de souvenirs, toujours présents dans les bons comme dans les mauvais moments, et avec qui j’ai construit une amitié sincère et durable pour la vie.

To my partner in life, Alžběta, my guiding star in this vast Universe, whose love fills every corner of my world with meaning and beauty. To your family -- Ladislav, Maria, Dany, and little Viktor -- for their kindness, warmth, and for always making me feel at home.

\pagebreak

\vspace*{\fill}
This work is supported by Fundação para a Ciência e a Tecnologia (FCT, Portugal) through the projects CFTP FCT Unit UIDB/00777/2020 and UIDP/00777/2020, CERN/FIS-PAR/0019/2021 and 2024.02004.CERN, which is partially funded through POCTI (FEDER), COMPETE, QREN and EU, and the PhD FCT grant 2021.06340.BD.

%----------------------------------------------------------------
%				RESUMO    
%----------------------------------------------------------------

\chapter{Resumo}

O Modelo Padrão (MP) da física de partículas fornece uma descrição bem-sucedida das partículas fundamentais e suas interações, mas não explica fenómenos como oscilações de neutrinos, matéria escura (ME) e a assimetria bariónica do Universo. Estas indicações claras de física para além do MP motivam extensões com novas partículas e simetrias. Questões teóricas como o \textit{puzzle} do sabor --- massas dos fermiões e padrões de mistura --- e o problema de carga e paridade (CP) forte também guiam modelos para além do MP.

Esta tese explora cenários para além do MP com dois princípios: construir paradigmas que abordem simultaneamente múltiplos problemas em aberto na (astro)física de partículas e cosmologia, enfatizando a sua testabilidade com análises fenomenológicas detalhadas. Esta abordagem revela conexões entre setores aparentemente desconexos, promovendo uma visão mais completa da física fundamental.

Investigamos modelos do setor escuro na origem das massas dos neutrinos, propondo um novo cenário: o \emph{seesaw} linear escuro, que liga a geração radiativa da massa dos neutrinos a candidatos de ME e prevê violação do sabor em leptões carregados. Nestes modelos, partículas do setor escuro podem constituir ME do tipo \emph{weakly interacting massive particle} (WIMP), testável em experiências de deteção direta. Estudamos violação espontânea de CP induzida por um singleto escalar, como fonte comum de efeitos de violação de CP de baixa e alta energia relevantes para leptogénese. Analisamos também um modelo de Nelson-Barr que resolve o problema CP forte, gera radiativamente a matriz CKM e fornece ME escalar tipo WIMP. Apresentamos modelos unificados de axiões onde um setor colorido gera radiativamente massas de neutrinos. Estes preveem acoplamentos distintivos do axião a fotões e fermiões, com cenários de ME em cosmologias pré- e pós-inflacionárias. Finalmente, exploramos simetrias Peccei-Quinn de sabor mínimas, ligando o \textit{puzzle} do sabor, massas de neutrinos e ME num cenário preditivo e testável.

\vfill
\textbf{Palavras-Chave:} Física para além do Modelo Padrão; Modelos de massa de neutrinos; Matéria escura; Violação de CP; Axiões.

%------------------------------------------------------------------
%				ABSTRACT    
%------------------------------------------------------------------

\chapter{Abstract}

The Standard Model (SM) of particle physics provides a successful description of fundamental particles and their interactions but fails to explain phenomena such as neutrino oscillations, dark matter (DM), and the baryon asymmetry of the Universe. These clear signs of physics beyond the SM (BSM) motivate extensions introducing new particles and symmetries. Theoretical questions like the flavor puzzle -- fermion masses and mixing -- and the strong charge-conjugation and parity (CP) problem further guide BSM frameworks.

This thesis explores BSM scenarios based on two guiding principles: constructing unified frameworks that address multiple open problems in (astro)particle physics and cosmology, and emphasizing their experimental testability through detailed phenomenological analyses. This approach uncovers deep connections between seemingly disconnected sectors, offering a more complete view of fundamental physics.

We investigate dark sector models for the origin of neutrino masses, proposing a novel setup: the dark linear seesaw, which radiatively links neutrino mass generation to viable DM candidates and predicts charged lepton flavor violation. In these scenarios, dark sector particles can constitute weakly interacting massive particle~(WIMP) DM, testable via direct detection experiments. We study spontaneous CP violation induced by a complex scalar singlet, acting as a common origin of low- and high-energy CP violating effects relevant for leptogenesis. We also analyze a Nelson-Barr model that solves the strong CP problem, generates a realistic CKM matrix radiatively, and yields scalar WIMP DM. Additionally, we present unified axion frameworks in which a colored sector radiatively generates neutrino masses. These models predict distinctive axion couplings to photons and fermions and accommodate axion DM in both pre- and post-inflationary cosmologies. Finally, we explore minimal flavored Peccei-Quinn symmetries that link the flavor puzzle, neutrino masses and DM within a predictive and testable BSM framework.

\vfill
\textbf{Keywords:} Physics beyond the Standard Model; Neutrino mass models; Dark Matter; CP Violation; Axions.

\tableofcontents

\listoftables
\addcontentsline{toc}{chapter}{\listtablename}

\listoffigures
\addcontentsline{toc}{chapter}{\listfigurename}

%------------------------------------------------------------------------
%			LIST OF ABBREVIATIONS    
%------------------------------------------------------------------------

\chapter{List of Abbreviations}
	\begin{center}
	\begin{longtblr}[caption={}, label=none, entry=none]{}
        \textbf{2HDM}    & Two-Higgs-doublet model \\ 
         \textbf{ALP} & Axion-like particle \\ 
         \textbf{BAO} & Baryon Acoustic Oscillations \\ 
        \textbf{BAU}      & Baryon asymmetry of the Universe \\ 
        \textbf{BBN} & Big Bang Nucleosynthesis \\ 
        \textbf{BBP}     &  Bento-Branco-Parada \\ 
        \textbf{BFB}     &  Bounded from below \\ 
        \textbf{BR}      & Branching ratio \\ 
        \textbf{BH}      & Black Hole \\ 
        \textbf{BSM}     & Beyond the standard model \\ 
        \textbf{BE}      & Boltzmann equation \\ 
        \textbf{BEH} & Brout-Englert-Higgs \\ 
        \textbf{$\chi$PT} & Chiral perturbation theory \\ 
        \textbf{C}      & Charge-conjugation \\ 
        \textbf{CC}      & Charged current \\ 
        \textbf{CDM}      & Cold Dark Matter \\ 
        \textbf{CE$\nu$NS} & Coherent Elastic Neutrino-Nucleus Scattering \\ 
        \textbf{CKM}     & Cabibbo-Kobayashi-Maskawa \\ 
        \textbf{CL}      & Confidence level \\ 
        \textbf{cLFV}    & Charged lepton flavor violation \\ 
        \textbf{CP}      & Charge-conjugation and parity \\ 
         \textbf{CMB} & Cosmic Microwave Background \\ 
        \textbf{DM}      & Dark matter \\ 
        \textbf{d.o.f} & degrees of freedom\\ 
        \textbf{DFSZ}    & Dine-Fischler-Srednicki-Zhitnitsky \\ 
        \textbf{DD}      & Direct detection \\ 
        \textbf{DIGA}    & Dilute instanton gas approximation \\ 
        \textbf{DW}      & Domain wall \\ 
        \textbf{EFT}     & Effective Field Theory \\ 
        \textbf{EM}      & Electromagnetic \\ 
        \textbf{EOM}     & Equations of motion \\ 
        \textbf{EW}      & Electroweak \\ 
        \textbf{EWSB}    & Electroweak symmetry breaking \\ 
        \textbf{FCNC}    & Flavor changing neutral current \\ 
        \textbf{FRW}     & Friedman-Robertson-Walker \\ 
        \textbf{GB}      & Goldstone boson \\ 
        \textbf{GUT}     & Grand Unified Theory \\ 
        \textbf{H.c.}    & Hermitian conjugate \\ 
        \textbf{HB}    & Horizontal Branch \\ 
        \textbf{ID}      & Indirect detection \\ 
        \textbf{IO}      & Inverted ordering \\ 
        \textbf{ISS}     & Inverse seesaw \\ 
        \textbf{KSVZ}    & Kim-Shifman-Vainshtein-Zakharov \\ 
        \textbf{LSS}     & Linear seesaw \\ 
         \textbf{LCPV}   & Leptonic CP violation \\ 
          \textbf{LEP} & Large Electron-Positron Collider \\ 
        \textbf{LH}      & Left-handed \\ 
        \textbf{LHC}     & Large hadron collider \\ 
        \textbf{LNV}     & Lepton number violation \\ 
        \textbf{LO}     & Leading order \\ 
        \textbf{NC}      & Neutral current \\ 
        \textbf{NLO}     & Next-to-leading order \\ 
        \textbf{nEDM}    & Neutron electric dipole moment \\ 
        \textbf{NB}      & Nelson-Barr \\ 
        \textbf{NO}      & Normal ordering \\ 
        \textbf{NS}      & Neutron star \\ 
        $0_\nu \beta \beta$ & Neutrinoless double beta decay \\ 
        \textbf{P}      & Parity \\ 
        \textbf{PMNS} & Pontecorvo-Maki-Nakagawa-Sakata \\ 
        \textbf{PQ}      & Peccei-Quinn \\ 
        \textbf{QCD} &  Quantum Chromodynamics \\ 
        \textbf{QED} &  Quantum Electrodynamics \\ 
        \textbf{QFT}     & Quantum field theory \\ 
        \textbf{RGB}      & Red Giant Branch \\ 
        \textbf{RH}      & Right-handed \\ 
        \textbf{RIS} & Real Intermediate States \\ 
        \textbf{SCPV}    & Spontaneous CP violation \\ 
        \textbf{SD} & Spin Dependent \\ 
 \textbf{SI} & Spin Independent \\ 
        \textbf{SM}      & Standard Model \\ 
        \textbf{SSB}     & Spontaneous symmetry breaking \\ 
        \textbf{SN} & Supernova \\ 
        \textbf{UV} & Ultraviolet \\ 
        \textbf{VEV} & Vacuum expectation value \\ 
        \textbf{VLQ}  & Vector-like quark \\ 
        \textbf{WD}   & White dwarfs \\ 
        \textbf{WDLF}  & WD Luminosity function \\ 
        \textbf{WIMP}  & Weakly Interacting Massive Particle

	\end{longtblr}
	\end{center}

% Main part
\mainmatter % Normal numbering

%------------ 
% Chapter 00    
%------------ 

%%%%%%%%%%%%%%%%%%%%%%%%%%%%%%%%%%%%%%%%%%%%%%%%%%%%%%%%%%%%%%%%%%%%%%%%%%%%%
\chapter*{Preface}
%%%%%%%%%%%%%%%%%%%%%%%%%%%%%%%%%%%%%%%%%%%%%%%%%%%%%%%%%%%%%%%%%%%%%%%%%%%%%

\addcontentsline{toc}{chapter}{Preface}

In this thesis we address some of the open problems in particle and astroparticle physics -- neutrino masses and mixing, dark matter~(DM), matter-antimatter asymmetry, strong charge-conjugation and parity~(CP) problem and flavor puzzle -- by establishing synergies among them in unified frameworks beyond the Standard Model~(BSM).

The work presented here is organized as follows. The first chapter has an introductory purpose. It starts with a brief overview of the building blocks of the Standard Model~(SM) of particle physics to motivate the need to go BSM by pointing out the key experimental evidences that require a more complete description of Nature. Namely, the discovery of neutrino flavor oscillations -- establishing that neutrinos are massive and leptons mix -- already provides unambiguous evidence for BSM physics. Moreover, cosmological and astrophysical observations reinforce this need: the measured abundance of cold dark matter (CDM) and the baryon asymmetry of the Universe (BAU) cannot be accounted for within the SM framework. In addition to these empirical indications, several theoretical considerations also suggest the existence of new physics. Among them are the flavor puzzle, reflecting the absence of a guiding principle in the SM capable of explaining the observed pattern of fermion masses and mixings, and the strong CP problem, which raises the question of why the strong sector of the SM appears to preserve CP symmetry despite the lack of any protective symmetry principle.

The second chapter starts with a summary of the current status of the neutrino oscillation observables: masses, mixing and leptonic CP violation~(LCPV), followed by a discussion on the experimental setups that aim at testing the Majorana character of neutrinos. Subsequently, are presented neutrino mass generation models focusing on the simplest tree-level and radiative ultraviolet~(UV) complete realizations of the Weinberg operator. Namely, at tree-level are briefly reviewed the seesaw mechanisms, as well as the low-scale variants -- inverse and linear seesaws. For radiative schemes the study is centered on the scotogenic model where a dark-sector, containing DM candidates, is at the origin of neutrino masses. The end of the chapter is based on our publication: 
\begin{itemize}
    \item A.~Batra, H.~B.~C\^amara and F.~R.~Joaquim, \emph{Dark linear seesaw mechanism}, Phys. Lett. B \textbf{843}, 138012 (2023)~\cite{Batra:2023bqj} ,
\end{itemize}
which introduces a novel dark-sector seeded neutrino mass generation mechanism. DM relic density is generated via thermal freeze-out in the standard weakly interacting massive particle~(WIMP) paradigm. Direct detection~(DD) DM constraints, as well as charged-lepton flavor violation~(cLFV) are also analyzed.

Chapter three is dedicated to spontaneous CP violation~(SCPV) stemming from the complex vacuum expectation value~(VEV) of a scalar singlet. In order for such scenario to have observable physical consequences additional fermionic content is required. This vacuum can be at the origin of various CP-violating effects. In fact, the first part of this chapter analyzes scalar-singlet assisted leptogenesis within the type-I seesaw framework, following our work:
\begin{itemize}
    \item D.~M.~Barreiros, H.~B.~C\^amara, R.~G.~Felipe and F.~R.~Joaquim, \emph{Scalar-singlet assisted leptogenesis with CP violation from the vacuum}, JHEP \textbf{01}, 010 (2023)~\cite{Barreiros:2022fpi},
\end{itemize}
where we show that the singlet VEV leads to non-trivial LCPV and sufficient CP violation to generate the observed BAU. The second part of the chapter introduces the minimal Nelson–Barr (NB) model, in which SCPV driven by a complex singlet addresses the strong CP problem. Subsequently, the NB mechanism is realized within a dark-sector framework, based on:
\begin{itemize}
    \item H.~B.~C\^amara, F.~R.~Joaquim and J.~W.~F.~Valle, \emph{Dark-sector seeded solution to the strong CP problem}, Phys. Rev. D \textbf{108} no.9, 095003 (2023)~\cite{Camara:2023hhn}.
\end{itemize}
In this scenario the strong CP phase is further protected at higher order in
perturbation theory and weak quark sector CP violation is generated radiatively. The implications for the NB quality problem are discussed, and the associated model phenomenology is analyzed with respect to flavor physics and WIMP DM.

Chapter four is a review on the axion solution to the strong CP problem with the objective of understanding the key theoretical properties of axions and how experiments search for these types of particles. The discussion begins by showing that axions are the pseudo-Goldstone bosons~(GB) of an associated Peccei–Quinn (PQ) symmetry. The main QCD axion models -- Kim–Shifman–Vainshtein–Zakharov (KSVZ) and Dine–Fischler–Srednicki–Zhitnitsky (DFSZ) -- are then presented. Axion DM is analyzed in both pre and post-inflationary cosmological scenarios. An overview of current experiments searching for axions and axion-like particles~(ALPs), as well as prospects for future searches, is also provided. The final part of the chapter contains a brief discussion of flavor-violating axions.

In the fifth chapter the axion-neutrino connection is explored, following the results of our papers: 
\begin{itemize}
    \item A.~Batra, H.~B.~C\^amara, F.~R.~Joaquim, R.~Srivastava and J.~W.~F.~Valle, \emph{Axion Paradigm with Color-Mediated Neutrino Masses}, Phys. Rev. Lett. \textbf{132} no.5, 051801 (2024)~\cite{Batra:2023erw},
    \item A.~Batra, H.~B.~C\^amara, F.~R.~Joaquim, N.~Nath, R.~Srivastava and J.~W.~F.~Valle, \emph{Axion framework with color-mediated Dirac neutrino masses}, Phys. Lett. B \textbf{868}, 139629 (2025)~\cite{Batra:2025gzy},
\end{itemize}
where in the context of the KSVZ framework, which features exotic quarks, we propose that these fermions together with scalar leptoquarks generate color-mediated Majorana and Dirac neutrino masses at the radiative level. The predictions for the axion-to-photon coupling, as well as its flavor-violating couplings to quarks, are studied.

The sixth chapter is based on:
\begin{itemize}
    \item J.~R.~Rocha, H.~B.~C\^amara, R.~G.~Felipe and F.~R.~Joaquim, \emph{Minimal U(1) two-Higgs-doublet models for quark and lepton flavor}, Phys. Rev. D \textbf{110} no.3, 035027 (2024)~\cite{Rocha:2024twm},
    
    \item J.~R.~Rocha, H.~B.~C\^amara and F.~R.~Joaquim, \emph{Flavored Peccei-Quinn symmetries in the minimal $\nu$DFSZ model}, Phys. Rev. D \textbf{112} no.7, 075053 (2025)~\cite{Rocha:2025ade},
\end{itemize}
with the discussion being centered on the second article above, which studies the DFSZ axion model extended with two right-handed neutrinos. In this $\nu$DFSZ model we systematically classify the minimal quark and lepton flavor patterns compatible with masses, mixing and CP violation data, realized by flavored PQ symmetries. Minimal flavored PQ symmetries provide a natural framework to suppress axion flavor-violating couplings to fermions.

Finally, the last chapter is a summary of the most relevant conclusions of this thesis.

As a last note, the research conducted during this PhD has also resulted in the following publications, which are not discussed in detail in this thesis:
\begin{itemize}
    \item D.~M.~Barreiros, H.~B.~Camara and F.~R.~Joaquim, \emph{Flavour and dark matter in a scoto/type-II seesaw model}, JHEP \textbf{08}, 030 (2022)~\cite{Barreiros:2022aqu},
    \item J.~A.~Aguilar-Saavedra, H.~B.~C\^amara, F.~R.~Joaquim and J.~F.~Seabra, \emph{Confronting the 95 GeV excesses within the U(1)'-extended next-to-minimal 2HDM}, Phys. Rev. D \textbf{108} no.7, 075020 (2023)~\cite{Aguilar-Saavedra:2023tql},
    \item M.~P.~Bento, H.~B.~C\^amara and J.~F.~Seabra, \emph{Unraveling particle dark matter with Physics-Informed Neural Networks}, Phys. Lett. B \textbf{868}, 139690 (2025)~\cite{Bento:2025agw}.
\end{itemize}
%

% %%%%%%%%%%%%%%%%%%%%%%%%%%%%%%%%%%%%%%%%%%%%%%%%%%%%%%%%%%%%%%%%%%%%%%%%%%%%%

%------------
% CHAPTER 01    
%------------

%%%%%%%%%%%%%%%%%%%%%%%%%%%%%%%%%%%%%%%%%%%%%%%%%%%%%%%%%%%%%%%%%%%%%%%%%%%%%
\chapter{The Standard Model and Beyond} 
\label{chpt:BSM}
%%%%%%%%%%%%%%%%%%%%%%%%%%%%%%%%%%%%%%%%%%%%%%%%%%%%%%%%%%%%%%%%%%%%%%%%%%%%%

The SM of particle physics is one of the greatest achievements in Physics, being a well established theory that describes fundamental particles and their interactions with extraordinary precision, in remarkable agreement with experiments. In this introductory chapter are briefly presented the SM particle content and symmetries, the electroweak~(EW) interactions, Higgs mechanism, fermion masses and mixing, as well as CP violation in the strong sector. Next, the need for physics BSM is motivated, by presenting some of the open problems in particle and astroparticle physics, starting with the experimental observations -- neutrino oscillation, DM, BAU -- that point towards the existence of a more complete picture of Nature. Also, from a theoretical perspective, the flavor puzzle and strong CP problem hint at possible avenues for extensions of the SM.

%%%%%%%%%%%%%%%%%%%%%%%%%%%%%%%%%%%%%%%%%%%%%%%%%%%%%%%%%%%%%%%%%%%%%%%%%%%%%
\section{Standard Model of Particle Physics}
\label{sec:SM}
%%%%%%%%%%%%%%%%%%%%%%%%%%%%%%%%%%%%%%%%%%%%%%%%%%%%%%%%%%%%%%%%%%%%%%%%%%%%%

During the second half of the twentieth century, the work of many physicists led to extraordinary ideas culminating in the SM of particle physics. The pioneering work on non-Abelian theories, by Yang and Mills in 1954~\cite{Yang:1954ek}, set the fundamental mathematical building blocks of the SM. In 1961, S.L. Glashow~\cite{Glashow:1961tr}, proposed the unification of the electromagnetic~(EM) and weak interactions. A few years later, in 1964-66, was formulated the famous Brout-Englert-Higgs~(BEH)~mechanism~\cite{Englert:1964et,Guralnik:1964eu,Higgs:1964pj,Higgs:1966ev}. It was shown that spontaneous symmetry breaking~(SSB) of a local gauge symmetry could generate massive vector bosons together with massless unphysical scalars, the so-called Nambu-Golstone bosons~\cite{Nambu:1961tp,Goldstone:1961eq,Goldstone:1962es}, ensuring gauge invariance. In 1967-68, S.~Weinberg and A.~Salam~\cite{Weinberg:1967tq,Salam:1968rm} implemented the Higgs mechanism in the non-Abelian $\text{SU}(2)_L\times\text{U}(1)_Y$ gauge theory of EW interactions. Moreover, the quark model, proposed independently by Gell-Mann and Zweig in 1964~\cite{GellMann:1964nj,Zweig:1964jf}, addressed some properties of hadrons and strong interactions. Finally, the renormalizability of the EW theory was proved by 't~Hooft and Veltman in 1971-72~\cite{tHooft:1971qjg,tHooft:1972tcz}.

%%%%%%%%%%%%%%%%%%%%%%%%%%%%%%%%%%%%%%%%%%%%%%%%%%%%%%%%%%%%%%%%%%%%%%%%%%%%%
\subsection{Particle content and electroweak interactions}
\label{sec:partLag}
%%%%%%%%%%%%%%%%%%%%%%%%%%%%%%%%%%%%%%%%%%%%%%%%%%%%%%%%%%%%%%%%%%%%%%%%%%%%%

The SM is a non-Abelian gauge quantum field theory (QFT) based on the symmetry group $G_{\text{SM}} = \text{SU}(3)_{c} \times \text{SU}(2)_L \times \text{U}(1)_Y$, described by the following Lagrangian locally invariant under $G_{\text{SM}}$,
\begin{equation}
\mathcal{L}_{\text{SM}} = \mathcal{L}_{\text{gauge}} + \mathcal{L}_{\text{fermions}} + \mathcal{L}_{\Phi} + \mathcal{L}_{\text{Yuk.}} + \mathcal{L}_{\overline{\theta}}\; ,
\label{eq:lagsm}
\end{equation}
and with field content summarized in Table~\ref{tab:SMpartcont}.

\begin{table}[t!]
\renewcommand*{\arraystretch}{1.5}
	\centering
	\begin{tabular}{| K{2cm} | K{2cm} | K{5.5cm} |}
		\hline
&Fields&$\text{SU}(3)_{c} \times \text{SU}(2)_L \times \text{U}(1)_Y$   \\
\hline
		\multirow{3}{*}{Gauge bosons} 
&$G_{\mu}$&($\mathbf{8},\mathbf{1},0$)   \\
&$W_{\mu}$&($\mathbf{1},\mathbf{3},0$)   \\
&$B_{\mu}$&($\mathbf{1},\mathbf{1},0$)   \\ 
\hline
\multirow{3}{*}{Quarks} 
&$q_L$&($\mathbf{3},\mathbf{2},1/3$)   \\
&$u_R$&($\mathbf{3},\mathbf{1},4/3$)    \\
&$d_R$&($\mathbf{3},\mathbf{1},-2/3$)  \\ \hline
		\multirow{2}{*}{Leptons} 
&$\ell_L$&($\mathbf{1},\mathbf{2}, {-1/2}$)   \\
&$e_R$&($\mathbf{1},\mathbf{1}, {-1}$)    \\
\hline
Scalars  &$\Phi$&($\mathbf{1},\mathbf{2}, 1$)  \\
\hline
\end{tabular}
\caption{SM field content and their corresponding transformation properties under $G_{\text{SM}}$.}
\label{tab:SMpartcont}
\end{table}

The $\text{SU}(3)_c$ gauge group corresponds to the strong interactions, where $c$ denotes color, and its associated QFT is Quantum Chromodynamics~(QCD). Invariance under this group requires the introduction of eight gluons $G_{\mu}^a$~($a=1,...,8$) one for each group generator. In Sec.~\ref{sec:strongCP}, the strong sector is described specifically in regards to CP violation encoded in $\mathcal{L}_{\overline{\theta}}$. The EW sector gauge group is $\text{SU}(2)_L \otimes \text{U}(1)_Y$, where $L$ stands for left-handedness and $Y$ is the hypercharge. Invariance under this group requires the introduction of four EW gauge bosons: three  $W_{\mu}^i$~($i=1,2,3$) for $\text{SU(2)}_L$ and one $B_{\mu}$ for $\text{U}(1)_Y$.

In contrast to the gauge bosons, the matter content -- number of fermions and scalars -- is not fixed by~$G_{\text{SM}}$, being determined empirically. In the SM, there are three generations of fermions, comprised of quarks and leptons, while there is only one scalar doublet in the theory that generates all the masses of the particles. All fields transform under the SM gauge group, namely quarks transform as triplets $\mathbf{3}$ under $\text{SU}(3)_c$, all left-handed~(LH) fermions $q_L$, $\ell_L$ and the Higgs doublet $\Phi$ are organized in doublets $\mathbf{2}$ of $\text{SU}(2)_L$, while right-handed~(RH) fermions $q_R$, $e_R$ are singlets $\mathbf{1}$ of $\text{SU}(2)_L$:
\begin{equation}
 q_{\alpha L} = \begin{pmatrix} u_{\alpha L}  \\ d_{\alpha L}\end{pmatrix} \; , \; \begin{aligned}
 &u_{\alpha R} = u_R, c_R, t_R\\
 &d_{\alpha R} = d_R, s_R, b_R\
 \end{aligned} \; ; \;  \ell_{\alpha L} = \begin{pmatrix} \nu_{\alpha L}  \\ e_{\alpha L}\end{pmatrix} \; , \; e_{\alpha R} = e_R, {\mu}_R, {\tau}_R \; ; \; \Phi = \begin{pmatrix} \phi^+ \\ \phi^0 \end{pmatrix}.
\label{eq:SMfields}
\end{equation}

Gauge invariance is ensured by replacing the ordinary derivative $\partial_{\mu}$ by the covariant derivative,
\begin{equation}
   D_{\mu} = \partial_{\mu} - i g^\prime \frac{Y}{2} B_{\mu} - i g \sum_{i=1}^3 \frac{\tau_i}{2} W_{\mu}^i - i g_s \sum_{a=1}^8 \frac{\lambda_a}{2}  G_{\mu}^a \; ,
\end{equation}
where $g^\prime$, $g$ and $g_s$ are the coupling constants associated to the $\text{U}(1)_{Y}$, $\text{SU}(2)_{L}$ and $\text{SU}(3)_{c}$ groups, respectively. The generators of these groups are respectively~$Y/2$, $T_i = \tau_i/2$ with $\tau_i$~($i=1,2,3$) being the Pauli matrices, and $\lambda_a/2$ with $\lambda_a$~($a=1,\cdots,8$) being the Gell-Mann matrices. Using the raising and lowering operators~$T_{\pm} = \left( T_1 \pm i T_2 \right)/ \sqrt{2}$ of~$\text{SU}(2)$ we define
\begin{equation}
W_{\mu}^{\pm} = \frac{1}{\sqrt{2}} \left(W_{\mu}^1 \mp i W_{\mu}^2 \right) \rightarrow T_1 W_{\mu}^1 + T_2 W_{\mu}^2 = T_{+} W_{\mu}^{+} + T_{-} W_{\mu}^{-} \; .
\end{equation}
Furthermore, $W_{\mu}^3$ and $B_{\mu}$ mix among themselves through the weak mixing angle $\theta_W$ given by,
\begin{equation}
\cos\theta_W = \frac{e}{g^\prime} \; , \; \sin\theta_W = \frac{e}{g} \; , 
\end{equation}
where $e$ is the electric charge of the positron. Then, $W_{\mu}^3$ and~$B_{\mu}$ are related to the physical bosons $Z_{\mu}$ and $A_{\mu}$ via the rotation
\begin{equation}
\begin{pmatrix} Z_{\mu} \\ A_{\mu} \end{pmatrix} = \begin{pmatrix} c_W & s_W\\ - s_W & c_W \end{pmatrix}  \begin{pmatrix} W_{\mu}^3 \\ B_{\mu} \end{pmatrix}  \rightarrow g^\prime \frac{Y}{2} B_{\mu} + g T_3 W_{\mu}^3 = - e Q A_{\mu} + \frac{g}{c_W} \left( T_3 - s_W^2 Q\right) Z_{\mu} \; ,
\end{equation}
where $c_W \equiv \cos\theta_W$ and  $s_W \equiv \sin\theta_W$ and $Q$ is the electric charge operator defined through the Gell-Nishijima formula as $Q = T_3 + Y/2$~\cite{Nakano:1953zz}.
Finally, the covariant derivative can be written as
\begin{equation}
   D_{\mu} = \partial_{\mu} + i e Q A_{\mu}  - \frac{ig}{c_W} \left(T_3 - s_W^2 Q \right) Z_{\mu} - ig \left( W_{\mu}^{+} T^+ + W_{\mu}^{-} T^{-} \right) - i g_s \sum_{a=1}^8 \frac{\lambda_a}{2}  G_{\mu}^a \; .
\end{equation}
In $\mathcal{L}_{\text{SM}}$~\eqref{eq:lagsm}, the Lagrangian encoding kinetic terms, triple and quartic self-interactions of the EW gauge bosons is written as
\begin{equation}
    \mathcal{L}_{\text{gauge}} = - \frac{1}{4} B^{\mu \nu} B_{\mu \nu} - \frac{1}{4} W^{\mu \nu}_{i} W_{\mu \nu}^{i} - \frac{1}{4} G_{\mu \nu}^a G^{a \mu \nu} \; ,
\label{eq:laggauge}
\end{equation}
with the field stress-tensors being
\begin{align}
B^{\mu \nu} &= \partial^{\mu} B^{\nu} - \partial^{\nu} B^{\mu} \; , \nonumber \\
W^{\mu \nu}_{i} &= \partial^{\mu} W^{\nu}_{i} - \partial^{\nu} W^{\mu}_{i} + g \; \varepsilon_{i j k} W^{\mu}_{j} W^{\nu}_{k} \; , \nonumber \\
G^{a}_{\mu \nu} &= \partial_\mu G_\nu^a - \partial_\nu G_\mu^a + g_s \; f_{a b c} G_{\mu}^b G_{\nu}^c \; ,
\label{eq:stresstensors}
\end{align}
where $\varepsilon_{i j k}$ is the rank-3 Levi-Civita tensor and $f_{abc}$ the SU(3) structure constants.

Next, the Lagrangian containing the kinetic terms of the fermions and their interactions with the gauge bosons is
\begin{equation}
    \mathcal{L}_{\text{fermions}} = \overline{q_{\alpha L}} \left(i \slashed{D}\right) q_{\alpha L} + \overline{u_{\alpha R}} \left(i \slashed{D}\right) u_{\alpha R} +\overline{d_{\alpha R}} \left(i \slashed{D}\right) d_{\alpha R} +\overline{\ell_{\alpha L}} \left(i \slashed{D}\right) \ell_{\alpha L} +\overline{e_{\alpha R}} \left(i \slashed{D}\right) e_{\alpha R} \; ,
\end{equation}
where $\slashed{D} = \gamma^{\mu} D_{\mu}$. The above expression contains the charged-current~(CC), EM and weak neutral-current~(NC) interactions, mediated by the $W^{\pm}$, $A$ (photon) and $Z$ boson, respectively. Namely, in the weak basis, the CC interactions are given by
\begin{equation}
    \mathcal{L}_{\text{CC}} = \frac{g}{\sqrt{2}} W_{\mu}^{-} \left( \overline{e_{\alpha}}  \gamma^{\mu} P_L \nu_{\alpha} + \overline{d_{\alpha}} \gamma^{\mu} P_L u_{\alpha} \right) + \text{H.c.} \; ,
\label{eq:CCweaksm}
\end{equation}
where $P_{L,R} = \left(1 \mp \gamma_5 \right)/2$ are the chiral projectors and $\text{H.c.}$ stands for Hermitian conjugate. The CC interactions are chiral since they involve only LH fields. Furthermore, the NC Lagrangian in the weak basis is contains the following terms:
\begin{align}
\mathcal{L}_{\text{NC}}^{\text{EM}} &= - e A_{\mu} \sum_{f} Q_f  \overline{\psi_{f}} \gamma^{\mu} \psi_f \; , \\
\mathcal{L}_{\text{NC}}^{\text{weak}} &= \frac{g}{c_W} Z_{\mu} \sum_{f} \overline{\psi_{f}} \gamma^{\mu} \left(g_L^f P_L + g_R^f P_R \right) \psi_{f} \; , \; g_L^f =  T_3^f - Q_f s_W^2 \; , \; g_R^f = - Q_f s_W^2 \; ,
\label{eq:weakNC}
\end{align}
where the sum in $f$ runs over all fermions $\psi_f$ with charge $Q_f$ and third component of weak isospin~$T_3^f$.

%%%%%%%%%%%%%%%%%%%%%%%%%%%%%%%%%%%%%%%%%%%%%%%%%%%%%%%%%%%%%%%%%%%%%%%%%%%%%
\subsection{Electroweak symmetry breaking}
\label{sec:EWSB}
%%%%%%%%%%%%%%%%%%%%%%%%%%%%%%%%%%%%%%%%%%%%%%%%%%%%%%%%%%%%%%%%%%%%%%%%%%%%%

The scalar sector of the SM is defined by the Lagrangian,
\begin{equation}
  \mathcal{L}_{\Phi} = (D_{\mu} \Phi)^{\dagger} (D^{\mu} \Phi) -  V\left( \Phi \right) \; , \;  V\left( \Phi \right) = m_{\Phi}^2 \left(\Phi^\dagger \Phi\right) + \lambda_{\Phi} (\Phi^{\dagger}\Phi)^2 \; ,
 \label{eq:lagphism}
\end{equation} 
where the scalar potential $V\left( \Phi \right)$ is guaranteed to be bounded from below for $\lambda_{\Phi}>0$.

The $\text{SU}(2)_L \otimes \text{U}(1)_Y$ symmetry is not manifest in Nature since at low energies only the subgroup $\text{U}(1)_{Q}$ of electromagnetism remains unbroken. The transition from the full EW gauge group to the EM one is realized through the mechanism of electroweak symmetry breaking~(EWSB). This takes place when a scalar field acquires a non-zero VEV. Such a field must be a scalar in order to preserve Lorentz invariance, and electrically neutral to ensure electric charge conservation. The SM adopts the the most economical and consistent realization of SSB achieved through the introduction of a single scalar doublet $\Phi$.

The minimum of $V\left( \Phi \right)$ for $ m_{\Phi}^2 < 0$ is given by
\begin{equation}
\left<\Phi^{\dagger} \Phi \right> = \frac{v^2}{2} \; , \; v^2 = - \frac{m_{\Phi}^2}{\lambda_{\Phi}} > 0 \; ,
\label{eq:smhiggspot}
\end{equation}
where the VEV $v$ is taken to be real without loss of generality. Before SSB, the theory consists of four massless gauge bosons, each with two transversal polarization degrees of freedom (d.o.f), and the scalar doublet with four d.o.f. giving a total of twelve d.o.f.. Upon EWSB, the neutral scalar field $\phi^0$ in Eq.~\eqref{eq:SMfields} acquires a non-zero VEV. Since only the charge operator $Q$ leaves the vacuum invariant, the scalar doublet will contain three unphysical GB, $G^{\pm}$~and~$G^0$~\cite{Goldstone:1961eq,Goldstone:1962es}, as well as one massive real scalar field -- the Higgs boson $h$. The unphysical fields $G^{\pm}$~and~$G^0$, will be, respectively, absorbed in the longitudinal components of the gauge bosons $W^{\pm}$ and $Z$, providing them with mass, while the photon~$A$ remains massless. This is known as the Higgs mechanism. Overall, the number of physical d.o.f. is maintained. 

The SM scalar doublet can be parameterized as
\begin{equation}
 \Phi = \frac{1}{\sqrt{2}} \begin{pmatrix} \sqrt{2} G^+ \\ v + h + i G^0 \end{pmatrix} \; , \; \left<\Phi \right> = \frac{1}{\sqrt{2}} \begin{pmatrix} 0 \\ v  \end{pmatrix} \; ,
\label{eq:parthiggs}
\end{equation}
with $v \simeq 246.2 \ \text{GeV}$~\cite{ParticleDataGroup:2024cfk}. The scalar potential in Eq.~\eqref{eq:lagphism} encodes the scalar self-interactions, the remaining kinetic term with the covariant derivative describes the interactions between the scalar d.o.f. and the gauge bosons which, after EWSB, yields the following mass terms:
\begin{align}
\mathcal{L}_{\Phi}^{\text{mass}} &= - \frac{m_{h}^2}{2} h^{2} + \frac{m_Z^2}{2} Z_{\mu} Z^{\mu} + m_{W^\pm}^2 W_{\mu}^{-} W^{+ \mu} \; , \nonumber \\
m_{h} &= \sqrt{-2 m_{\Phi}^2} \; , \;   m_Z = \frac{g v}{2 c_W} \; , \; m_{W^\pm} = \frac{g v}{2} \; ,
\label{eq:Higgsmassmechanism}
\end{align}
with the photon being massless. The gauge boson mass values are $m_Z = 91.1880 \pm 0.0020$ GeV and $m_{W^\pm} = 80.3692 \pm 0.0133$ GeV~\cite{ParticleDataGroup:2024cfk}. The CP-even Higgs boson $h$ discovered at the Large Hadron Collider~(LHC) by the ATLAS~\cite{ATLAS:2012yve} and CMS~\cite{CMS:2012qbp} collaborations in 2012 has mass $m_{h} = 125.20 \pm 0.11 \ \text{GeV}$~\cite{ParticleDataGroup:2024cfk}.

The relation between the $W^\pm$ and $Z$ boson masses defines the parameter
\begin{equation}
	\rho \equiv \frac{m_W^2}{m_Z^2\cos^2\theta_W} \; ,
	\label{eq:rhoSM}
\end{equation}
which is equal to unity at tree level in the SM. In BSM theories, featuring extended particle content, the tree-level $\rho$ parameter value can deviate from one. Namely, considering the case of an arbitrary number $n$ of scalar multiplets that acquire a non-zero VEV, one has~\cite{Lee:1972gj}
\begin{equation}
	\rho = \frac{\sum_{i=1}^n|v_i|^2\left[T^i\,\left(T^i+1\right)-\left(T_3^i\right)^2\right]}{2\sum_{i=1}^n|v_i|^2\left(T_3^i\right)^2} \; ,
\label{eq:rhoMultiHiggs}
\end{equation}
where $v_i$ is the VEV of the $i$-th multiplet, $T^i$ its isospin and $T_3^i$ its respective third component. Note that the SM prediction $\rho=1$ at tree-level is always verified regardless of the number of SU(2) scalar singlets and doublets added to the SM. In contrast, multiplets of higher isospin have their VEVs severely constrained by the experimental value $\rho = 1.00038\pm 0.00020$~\cite{ParticleDataGroup:2024cfk}.

%%%%%%%%%%%%%%%%%%%%%%%%%%%%%%%%%%%%%%%%%%%%%%%%%%%%%%%%%%%%%%%%%%%%%%%%%%%%%
\subsection{Fermion masses, mixing and CP violation}
\label{sec:fermionmassmix}
%%%%%%%%%%%%%%%%%%%%%%%%%%%%%%%%%%%%%%%%%%%%%%%%%%%%%%%%%%%%%%%%%%%%%%%%%%%%%

In the SM, a Dirac mass term generically written as
\begin{equation}
 - m_{\psi} \overline{\psi} \psi = - m_{\psi} \left(\overline{\psi_{L}} \psi_{R} + \overline{\psi_{R}} \psi_{L} \right) \; ,
\end{equation}
is not consistent with gauge principles. Namely, this term is not invariant under $\text{SU}(2)_L \otimes \text{U}(1)_Y$ since it is not a $\text{SU}(2)_L$ singlet and the LH and RH fermionic fields have different hypercharge values. However, it is invariant under $\text{U}(1)_{Q}$, which hints at the possibility that this term can be generated through SSB. In fact, the Yukawa interaction terms among the fermion fields and the Higgs doublet are given by
\begin{equation}
    -\mathcal{L}_{\text{Yuk.}} =  \overline{\ell_L} \mathbf{Y}_{e} \Phi e_R + \overline{q_L} \mathbf{Y}_{u} \tilde{\Phi} u_R +  \overline{q_L} \mathbf{Y}_{d} \Phi d_R + \text{H.c.} \; ,
\label{eq:lagyuksm}
\end{equation}
where $\tilde{\Phi} = i \tau_2 \Phi^*$ and $\mathbf{Y}_{e,u,d}$ are $3 \times 3$ arbitrary complex Yukawa coupling matrices. After EWSB, in the weak basis, the fermion mass terms are
\begin{equation}
   - \mathcal{L}_{\text{mass}} =  \overline{e_L}  \mathbf{M}_{e} e_R +  \overline{u_L} \mathbf{M}_{u} u_R +  \overline{d_L} \mathbf{M}_{d} d_R + \text{H.c.} \; , \; \mathbf{M}_{e,u,d} = \frac{v}{\sqrt{2}} \mathbf{Y}_{e,u,d} \; .
\end{equation}
Notice that it is not possible to construct a Yukawa term for neutrinos, since there are no RH neutrinos in the SM. Consequently, neutrinos are strictly massless in the SM. 

The fermion mass matrices are brought to the physical basis via the unitary transformations
\begin{align}
    d_{L,R} &\rightarrow \mathbf{U}_{L,R}^d \ d_{L,R} \Rightarrow \mathbf{U}_L^{d \dagger} \mathbf{M}_d \mathbf{U}_R^d = \mathbf{D}_d = \text{diag}(m_d,m_s,m_b) \; , 
    \nonumber \\
    u_{L,R} &\rightarrow \mathbf{U}_{L,R}^u \ u_{L,R} \Rightarrow \mathbf{U}_L^{u \dagger} \mathbf{M}_u \mathbf{U}_R^u = \mathbf{D}_u = \text{diag}(m_u,m_c,m_t) \; , \nonumber \\
    e_{L,R} &\rightarrow \mathbf{U}_{L,R}^e \ e_{L,R} \Rightarrow \mathbf{U}_L^{e \dagger} \mathbf{M}_e \mathbf{U}_R^e = \mathbf{D}_e = \text{diag}(m_e,m_\mu,m_\tau) \; ,
    \label{eq:massdiag}
\end{align}
where $m_{d,s,b}$, $m_{u,c,t}$ and $m_{e,\mu,\tau}$ denote the physical down-type, up-type quark and charged-lepton masses with their current values given in Table~\ref{tab:quarkdata}. The above unitary matrices are obtained by diagonalizing the Hermitian matrices $\mathbf{H}_{f}= \mathbf{M}_{f} \mathbf{M}_{f}^\dagger$ and $\mathbf{H}_{f}^\prime= \mathbf{M}_{f}^\dagger \mathbf{M}_{f}$ ($f = d,u,e$) as follows,
\begin{align}
    \mathbf{U}_L^{f \dagger} \mathbf{H}_{f} \mathbf{U}_L^{f} = \mathbf{U}_R^{f \dagger} \mathbf{H}_{f}^\prime \mathbf{U}_R^{f} = \mathbf{D}_{f}^2 \; .
    \label{eq:hermitianmassdiag}
\end{align}
These field rotations do not affect NC interactions, ensuring there are no flavor-changing neutral currents~(FCNC) at tree level in the SM. However, in the quark sector, CC interactions are described by a $3\times 3$ unitary mixing matrix $\mathbf{V}$, such that
\begin{equation}
    \mathcal{L}_{\text{CC}} \supset \frac{g}{\sqrt{2}} W_{\mu}^+ \overline{u}_L \gamma^{\mu} \mathbf{V} d_L + \text{H.c.} \; , \; \mathbf{V}=\mathbf{V}^{u\dagger}_L\mathbf{V}^{d}_L \; ,
\label{eq:VCKMdef}
\end{equation}
where $\mathbf{V}$ is the Cabibbo-Kobayashi-Maskawa (CKM) matrix~\cite{Cabibbo:1963yz,Kobayashi:1973fv}. In general, an $n \times n$ unitary matrix depends on $n^2$ parameters comprised of~$n(n-1)/2$~mixing angles and~$n(n+1)/2$~phases. Therefore, the CKM matrix contains \textit{à priori} nine parameters, namely three mixing angles and six phases. However, since there is the freedom to rephase the LH fields without altering the interactions terms, five unphysical phases can be removed. Overall, the CKM matrix is described by four physical parameters: three mixing angles~$\theta_{12}^q$,~$\theta_{13}^q$,~$\theta_{23}^q$ and a Dirac CP-violating phase~$\delta^q$. The standard parameterization of this matrix is given by~\cite{Chau:1984fp}
\begin{align}
\mathbf{V}  = \begin{pmatrix}  c_{12}^q c_{13}^q & s_{12}^q c_{13}^q & s_{13}^q e^{-i\delta^q}  \\
- s_{12}^q c_{23}^q - c_{12}^q s_{23}^q s_{13}^q e^{i\delta^q} & c_{12}^q c_{23}^q - s_{12}^q s_{23}^q s_{13}^q e^{i\delta^q} & s_{23}^q c_{13}^q  \\ 
 s_{12}^q s_{23}^q - c_{12}^q c_{23}^q s_{13}^q e^{i\delta^q} &- c_{12}^q s_{23}^q - s_{12}^q c_{23}^q s_{13}^q e^{i\delta^q} & c_{23}^q c_{13}^q \\ 
\end{pmatrix} \; , 
\label{eq:VCKMparam}
\end{align}
where $c_{i j}^q \equiv \cos \theta_{i j}^q$,~$s_{i j}^q \equiv \sin \theta_{i j}^q$ and we take without loss of generality $\theta_{i j}^q \in \left[0, \pi/2 \right]$~and~$\delta^q \in \left[0, 2\pi \right[$. The current CKM data is given in Table~\ref{tab:quarkdata}.
\begin{table}[t!]
\renewcommand*{\arraystretch}{1.5}
\centering
\begin{tabular}{|c c|}
        \hline 
        Parameter & Best fit $\pm 1 \sigma$ \\
        \hline
        $m_e (\times \; \text{keV})$ \; \; & $510.99895000\pm 0.00000015$ \\
        $m_\mu (\times \; \text{MeV})$ \; \;& $105.6583755\pm 0.0000023$   \\
        $m_\tau (\times \; \text{GeV})$ \; \;& $1.77686\pm 0.00012$ \\ \hline
        $m_d (\times \; \text{MeV})$ \; \; & $4.67^{+0.48}_{-0.17}$ \\
        $m_s (\times \; \text{MeV})$ \; \;& $93.4^{+8.6}_{-3.4}$  \\
        $m_b (\times \; \text{GeV})$ \; \;& $4.18^{+0.03}_{-0.02}$ \\
        $m_u (\times \; \text{MeV})$ \; \;& $2.16^{+0.49}_{-0.26}$ \\
        $m_c (\times \; \text{GeV})$ \; \;& $1.27 \pm 0.02$  \\
        $m_t (\times \; \text{GeV})$\; \; & $172.69 \pm 0.30$ \\ 
        $\theta_{12}^q (^\circ)$ \; \;& $13.04\pm0.05$ \\
        $\theta_{23}^q (^\circ)$ \; \;& $2.38\pm0.06$ \\
        $\theta_{13}^q (^\circ)$ \; \;& $0.201\pm0.011$ \\
        $\delta^q (^\circ)$ \; \;& $68.75\pm4.5$ \\
        \hline 
\end{tabular}
\caption{Current charged-lepton masses and quark data: masses, mixing angles and Dirac CP phase~\cite{ParticleDataGroup:2024cfk}.}
\label{tab:quarkdata}
\end{table}

The non-trivial flavor structure of the CKM matrix leads to the only known source of CP violation in Nature, first observed in the neutral kaon system $K^0-\overline{K^0}$ in 1964~\cite{Christenson:1964fg}. To understand how quark CC interactions violate CP, it is necessary to have a closer look on how parity~(P), charge-conjugation~(C) and CP transformations act on the fields with the aim to derive a basis-independent quantity to describe CP violation. Given the Lorentz spacetime vector $x=(x^0,\vec{x})$, a P transformation leads to a sign inversion in the space coordinates $x \rightarrow x_{\text{P}} = (x^0,-\vec{x})$. The way P acts on a fermionic field described by a spinor $\psi(x)$ is
\begin{equation}
    \psi(x) \xrightarrow{\text{P}} \xi_{\text{P}} \gamma^0 \psi(x_{\text{P}}) \; ,
\end{equation}
where $\xi_{\text{P}} = \pm, \pm i$. A P transformation flips a spinor's helicity, hence P acts on the LH component of a field by transforming it in its RH component. Furthermore, C transforms a particle into its antiparticle,
\begin{equation}
    \psi(x) \xrightarrow{\text{C}} \xi_{\text{C}} C \overline{\psi}^T(x) \; ,
\end{equation}
with $\xi_{\text{C}}$ being an overall phase factor and $C$ the charge conjugation operator that exhibits the properties $C^\dagger = C^{-1}$ and $C^T = - C$. Under C the LH chiral component of a spinor is transformed into the conjugate of the RH component. In the SM, since SU(2)$_L$ acts only on LH fermions, both C and P symmetries are explicitly and maximally broken. Combining both transformations, CP acts on a spinor as follows,
\begin{equation}
    \psi(x) \xrightarrow{\text{CP}} \xi_{\text{CP}} \gamma^0 C \overline{\psi}^T(x_{\text{P}}) = - \xi_{\text{CP}} C \psi^\ast(x_{\text{P}}) \; ,
\end{equation}
where $\xi_{\text{CP}}$ is an overall phase factor and the last equality is obtained by making use of $C \gamma^{0 T} C^\ast \gamma^{0 \dagger} = 1$. Hence CP transforms a field into its antiparticle and inverts its helicity. Therefore under CP a LH fermion is transformed into its charge conjugate. 

To translate the above for the quark fields one needs to define the CP transformation from the CP-preserving part of the SM Lagrangian. Namely, there is the freedom to make use of weak-basis transformations that leave the gauge interactions and kinetic terms invariant. In general for quark fields~\cite{Branco:1999fs}:
\begin{align} 
u_L(x) &\xrightarrow{\text{CP}} e^{i \xi_W} \mathbf{X}_L \gamma^0 C \overline{u_L}^T(x_{\text{P}}) \; , \nonumber \\
d_L(x) &\xrightarrow{\text{CP}} \mathbf{X}_L \gamma^0 C \overline{d_L}^T(x_{\text{P}}) \; , \nonumber \\
u_R(x) &\xrightarrow{\text{CP}} \mathbf{X}_R^u \gamma^0 C \overline{u_R}^T(x_{\text{P}})  \; , \nonumber \\
d_R(x) &\xrightarrow{\text{CP}} \mathbf{X}_R^d \gamma^0 C \overline{d_R}^T(x_{\text{P}})  \; ,
\label{eq:CPtransformations}
\end{align}
where $\mathbf{X}_L$ and $\mathbf{X}_R^{u,d}$ are $3 \times 3$ unitary matrices in flavor space and the $W^\pm$ bosons transform as
\begin{align}
    W_\mu^+ \xrightarrow{\text{CP}} - e^{i \xi_W} W^{\mu -} \; ,
\end{align}
such that the pure gauge kinetic interactions of Eq.~\eqref{eq:laggauge} are unchanged.

Consider first that the quark mass matrices are in the diagonal basis and that $\mathbf{X}_L$ and $\mathbf{X}_R^{u,d}$ are diagonal complex matrices. When the above CP transformations are applied to the quark CC interactions, the following CP-invariance condition in terms of the CKM matrix elements are obtained:
\begin{align}
    \mathbf{V}_{\alpha \beta} e^{i(\xi_\beta - \xi_\alpha + \xi_W)} = \mathbf{V}_{\alpha \beta}^\ast \; ,
\end{align}
which for a quartet translates into
\begin{align}
    Q_{\alpha \beta \gamma \delta} = \mathbf{V}_{\alpha \beta} \mathbf{V}_{\gamma \delta} \mathbf{V}_{\gamma \beta}^\ast \mathbf{V}_{\alpha \delta}^\ast = Q_{\alpha \beta \gamma \delta}^\ast \; ,
\end{align}
which is independent of the arbitrary CP phases $\xi_{\text{CP}}$. Thus if CP is conserved, all quartets of the CKM matrix need to be real. Taking the imaginary part of the quartets leads to the so-called Jarlskog invariant~\cite{Jarlskog:1985ht}:
\begin{align}
    J_{\text{CP}} \sum_{\sigma,\rho} \varepsilon_{\alpha \gamma \sigma} \varepsilon_{\beta \delta \rho} = \text{Im} \left[\mathbf{V}_{\alpha \beta} \mathbf{V}_{\gamma \delta} \mathbf{V}_{\gamma \beta}^\ast \mathbf{V}_{\alpha \delta}^\ast \right] \; ,
\end{align}
which is a basis-invariant way to characterize CP violation in the SM. Given the parameterization of the CKM matrix of Eq.~\eqref{eq:VCKMparam} one obtains
\begin{align}
    J_{\text{CP}} = \frac{1}{8} \sin (2 \theta_{12}^q) \sin (2 \theta_{23}^q) \sin (2 \theta_{13}^q) \cos \theta_{13}^q \sin \delta^q \; .
    \label{eq:jarlskog}
\end{align}
For CP to be conserved it is required that $J_{\text{CP}}=0$. However, the current value for the modulus of the Jarlskog invariant is $J_{\text{CP}}=\left(3.08^{+0.15}_{-0.13}\right) \times 10^{-5}$~\cite{ParticleDataGroup:2024cfk}. It is then clear that for three quark generations, the non-trivial mixing of the CKM matrix leads to CP violation in the weak sector of the SM~\cite{Branco:1999fs}.

If one applies the general CP transformations of Eq.~\eqref{eq:CPtransformations} to the quark Yukawa Lagrangian of Eq.~\eqref{eq:lagyuksm}, the CP-invariance condition reads
\begin{align}
    \mathbf{X}_L^\dagger \mathbf{M}_u \mathbf{X}_R^u = \mathbf{M}_u^\ast \; \wedge \; \mathbf{X}_L^\dagger \mathbf{M}_d \mathbf{X}_R^d = \mathbf{M}_d^\ast \; .
\end{align}
If CP is conserved it should be possible to find unitary matrices $\mathbf{X}_L$ and $\mathbf{X}_R^{u,d}$ that obey the above conditions. However, this is not possible since there are three quark generations. By making use of the Hermitian matrices $\mathbf{H}_{u,d}$, the above relations can be written in a weak-basis invariant way, resulting in~\cite{Bernabeu:1986fc}:
\begin{equation}
    \text{Tr} \left[\mathbf{H}_{u},\mathbf{H}_{d}\right]^n = 0 \; ,
\end{equation}
which is a non-trivial condition as long as $n \geq 3$. For $n=3$ and three fermion families the above equations is a necessary and sufficient condition for CP invariance~\cite{Bernabeu:1986fc}. Thus, applying the minimal condition $\text{Tr} \left[\mathbf{H}_{u},\mathbf{H}_{d}\right]^3 = 0$, to the SM, implies that in order to have CP violation all quark masses must be non-degenerate and $J_{\text{CP}} \neq 0$, which is observed experimentally -- see Table~\ref{tab:quarkdata}.

In the lepton sector there is the freedom to perform the neutrino field rotation $\nu_L \rightarrow \mathbf{U}_L^e \nu_L$ [see Eq.~\eqref{eq:massdiag}] such that no lepton mixing matrix appears in the leptonic CC interactions. Thus, there is no lepton mixing in the SM. Furthermore, there is conservation of lepton number for each flavor and consequently total lepton number is conserved in the SM, being an accidental symmetry of the theory.

%%%%%%%%%%%%%%%%%%%%%%%%%%%%%%%%%%%%%%%%%%%%%%%%%%%%%%%%%%%%%%%%%%%%%%%%%%%%%
\subsection{Strong sector and CP violation}
\label{sec:strongCP}
%%%%%%%%%%%%%%%%%%%%%%%%%%%%%%%%%%%%%%%%%%%%%%%%%%%%%%%%%%%%%%%%%%%%%%%%%%%%%

As seen in the previous section, CP symmetry is broken in the weak sector of the SM, specifically by the quark CC interactions. However, the strong sector Lagrangian 
\begin{equation}
    \mathcal{L}_{\text{QCD}} = - \frac{1}{4} G_{\mu \nu}^a G^{a \mu \nu} + \overline{u} (i \slashed{D} - \mathbf{M}_u) u + \overline{d} (i \slashed{D} - \mathbf{M}_d) d \; ,
    \label{eq:LQCD}
\end{equation}
leads to an additional term known as the QCD anomalous or topological term, that violates P and CP, given by [see Eq.~\eqref{eq:lagsm}]
\begin{align}
\mathcal{L}_{\overline{\theta}} = \overline{\theta} \frac{\alpha_s}{8 \pi} G^{a}_{\mu \nu} \widetilde{G}^{a \mu \nu} \; ,
\label{eq:topological}
\end{align}
where $\alpha_s = g_s^2 / (4 \pi)$ and $\widetilde{G}^{a \mu \nu} = \epsilon^{\mu \nu \rho \sigma} G^{a}_{\rho \sigma}/2$ is the dual of the gluonic field stress tensor given in Eq.~\eqref{eq:stresstensors}. The parameter $\overline{\theta}$ is the so-called strong CP phase. In order to understand the origin of this term, we will start by briefly reviewing some general aspects of symmetries, which play a crucial role in Physics, as well exemplified by the SM.

%%%%%%%%%%%%%%%%%%%%%%%%%%%%%%%%%%%%%%%%%%%%%%%%%%%%%%%%%%%%%%%%%%%%%%%%%%%%%
\subsubsection{From classical to anomalous symmetries}
%%%%%%%%%%%%%%%%%%%%%%%%%%%%%%%%%%%%%%%%%%%%%%%%%%%%%%%%%%%%%%%%%%%%%%%%%%%%%

Consider a classical field theory in which the fields undergo an infinitesimal transformation, \(\phi \rightarrow \phi + \delta \phi\), causing the Lagrangian \(\mathcal{L}(x) = \mathcal{L}[\phi(x), \partial_\mu \phi(x)]\) to change as 
\(\mathcal{L} \rightarrow \mathcal{L} + \delta \mathcal{L}\). Using the stationary principle of the action $S = \int d^4 x \mathcal{L}$, one can derive the classical Euler-Lagrange equations of motion~(EOM). This allows to identify the associated current $J^\mu$ as
\begin{equation}
    J^{\mu} = \frac{\partial \mathcal{L}}{\partial(\partial_\mu \phi)} \delta \phi\; , \; \partial_{\mu} J^\mu = \delta \mathcal{L} \; .
\end{equation}
Noether's theorem states that for every generator of a  global (continous) symmetry of the theory there is a corresponding conserved current $\partial_\mu J^\mu = 0$ and time independent charge $Q = \int d^3 x J^0$. This happens if a field transformation leaves the action invariant i.e., $\delta \mathcal{L} = 0$.

In QFT all the information is encoded in the generating functional $Z[J]$. The time-ordered correlator of an arbitrary operator $\mathcal{O}$ is related to the path integral as follows,
\begin{equation}
    Z[J] = \int \mathcal{D} \phi \, e^{i \int d^4 x ( \mathcal{L} +  J \phi) } \; , \; \left<\mathcal{O}\right> = \frac{1}{Z_0} \int \mathcal{D} \phi \, e^{i S} \mathcal{O} \; ,
    \label{eq:Zobs}
\end{equation}
where $J$ is a source for $\phi$, $Z_0 \equiv Z[0]$ and $\left<\mathcal{O} \right> \equiv \bra{\Omega} T \mathcal{O} \ket{\Omega}$, with $\Omega$ being the vacuum state. The fields~$\phi$ in the path integral are classical. The quantum character lies in the functional measure $\mathcal{D} \phi$. In fact, the integral sums over all possible paths in field space weighted by~$e^{i S}$, whereas in a classical theory only the path with least action is considered. Making an infinitesimal transformation of the fields and considering $\delta \mathcal{L} = 0$, leads to the Ward-Takahashi identity
\begin{equation}
i \partial_\mu \left<J^\mu_x \phi_{x_1} \cdots \phi_{x_n} \right> = \sum_{j=1}^{n} \left< \phi_{x_1} \cdots \delta^{(4)}_{x-x_j} \cdots \phi_{x_n} \right>.
\label{eq:current1}
\end{equation}
Thus, the Noether current is conserved apart from contact terms. Considering now the correlator involving the current and an arbitrary operator 
\(\mathcal{O}\), which is invariant under the transformation, the Schwinger–Dyson 
equations for current conservation are obtained as,
\begin{equation}
\partial_\mu \left<J^\mu_x \mathcal{O}_{x_1\cdots x_n} \right> = 0 \; .
\label{eq:current2}
\end{equation}
Note that, to derive both of the above equations, it is assumed that the field transformation leaves the path integral measure invariant. However, as seen shortly, this does not hold for all transformations. 

A symmetry can be classically conserved but not persist at the quantum level, which makes it anomalous. This has very strong implications. In fact, gauge invariant theories must be anomaly free, otherwise this would lead to inconsistencies. The prime example is the SM that respects the gauge anomaly cancellation equations. In this section, we focus on global symmetries which can be anomalous without leading to problems in a theory. However, other interesting consequences arise.

To derive the implications of the $\text{U}(1)_{\text{A}}$ axial anomaly we consider Quantum Electrodynamics~(QED), then generalize the results to QCD. QED describes the interactions between a fermion $\psi$ of mass $m$ and the photon $A_\mu$, with Lagrangian
\begin{equation}
    \mathcal{L}_{\text{QED}} = - \frac{1}{4} F_{\mu \nu} F^{\mu \nu} + \overline{\psi} (i \slashed{D} - m) \psi \; ,
\end{equation}
where $F_{\mu \nu} = \partial_\mu A_\nu - \partial_\nu A_\mu$ and $D_\mu = \partial_\mu + i e A_\mu$. In the massless limit, $\mathcal{L}_{\text{QED}}$ is invariant under
\begin{equation}
\text{U}(1)_{\text{V}}: \psi \rightarrow e^{i \alpha} \psi \; , \; \text{U}(1)_{\text{A}}: \psi \rightarrow e^{i \beta \gamma^5} \psi \; .
\end{equation}
These classical symmetries have associated currents,
\begin{align}
    \partial_\mu J^\mu_{\text{V}} = 0 \; , \;
    \partial_\mu J^\mu_{\text{A}} = 2 i m \overline{\psi} \gamma^5 \psi \; ,
    \label{eq:currentsEOM}
\end{align}
where $J^\mu_{\text{V}}=\overline{\psi} \gamma^\mu \psi$ and $J^\mu_{\text{A}}=\overline{\psi} \gamma^\mu \gamma^5 \psi$, are the vector and axial current, respectively. Hence, $\text{U}(1)_{\text{V}}$ is exactly conserved, while $\text{U}(1)_{\text{A}}$ is recovered for $m=0$. We will show that the~$\text{U}(1)_{\text{A}}$ symmetry is actually anomalous. 

%%%%%%%%%%%%%%%%%%%%%%%%%%%%%%%%%%%%%%%%%%%%%%%%%%%%%%%%%%%%%%%%%%%%%%%%%%%%%
\subsubsection{Axial Anomaly}
\label{sec:axial}
%%%%%%%%%%%%%%%%%%%%%%%%%%%%%%%%%%%%%%%%%%%%%%%%%%%%%%%%%%%%%%%%%%%%%%%%%%%%%

There are two ways to derive the $\text{U}(1)_{\text{A}}$ axial anomaly. The first one follows the historically determination of~$\Gamma(\pi^0 \rightarrow \gamma \gamma)$, that is the computation of triangle diagrams that contribute to the two-photon matrix element of the divergence of the axial vector current. Reaching a correct decay rate for the $\pi^0 \rightarrow \gamma \gamma$ and its relation with the axial anomaly was a significant challenge for physicists for quite some time~\cite{Bell:1969ts}. The subtleties in the computation of the triangle diagrams have to do with the properties of linearly divergent integrals~\footnote{For more details on the derivations in this section see Refs.~\cite{Peskin:1995ev,Schwartz:2014sze}. Namely, the very interesting historical development in obtaining the correct decay rate for $\pi^0 \rightarrow \gamma \gamma$ and, subtleties in its calculation, are very instructively discussed in Ref.~\cite{Schwartz:2014sze}.}. The second approach, adopted here, works directly with the path integral measure, emphasizing its key ideas and subtleties. Namely, Fujikawa~\cite{PhysRevLett.42.1195} showed that anomalies arise when there are symmetries of the action that are not symmetries of the path integral measure.

The goal is to understand how the functional measure for fermion fields is altered for the axial transformation $\psi(x) \rightarrow e^{i \beta(x) \gamma^5} \psi(x)$ and $\overline{\psi}(x) \rightarrow e^{-i \beta(x) \gamma^5} \overline{\psi}(x)$. Since the fields are Grassmann-valued, the measure transforms as $\mathcal{D} \overline{\psi} \mathcal{D} \psi \rightarrow \mathcal{J}^{-2} \mathcal{D} \overline{\psi} \mathcal{D} \psi$. The Jacobian is given by~\footnote{For a general linear transformation of the fields $\psi(x) \rightarrow \Lambda(x) \psi(x)$ and $\overline{\psi}(x) \rightarrow \Lambda^\dagger(x) \overline{\psi}(x)$, the Jacobian is $\mathcal{J} = \det(\Lambda) = e^{\Tr[\ln(\Lambda)]}$, where the trace sums over the eigenvalues of $\ln(\Lambda)$.},
\begin{equation}
    \mathcal{J} = \det\left[e^{i \beta(x) \gamma^5}\right] = \exp\left[i \int d^4 x \ \beta(x) \Tr \gamma^5 \right] \; .
\end{equation}
Since, $\Tr \gamma^5=0$, it seems that the Jacobian is $1$, and therefore $\mathcal{D} \overline{\psi} \mathcal{D} \psi$ is invariant. However, it is necessary to examine the measure more closely and define it properly.

The idea is to expand the fermion fields in a basis $\{\ket{x}\}$, of one-particle Hilbert space, which are eigenstates of $\slashed{D}$. Then, the trace in the Jacobian is evaluated as
\begin{equation}
   \ln(\mathcal{J}) = i \int d^4 x \Tr[ \bra{x} \beta(x) \gamma^5 \ket{x}] \; .
\end{equation}
To regulate the divergent integral above, a Gaussian exponential is used $e^{(i \slashed{D})^2/M^2}$, with $M$ being a cutoff scale:
\begin{equation}
\Tr[ \bra{x} \beta \gamma^5 \ket{x} ] = \lim_{M \to \infty} \Tr[ \bra{x} \beta \gamma^5 e^{\frac{(i \slashed{D})^2}{M^2}} \ket{x} ] \; .
\end{equation}
By knowing that $(i \slashed{D})^2 = - D^2 + (e/2) \sigma^{\mu \nu} F_{\mu \nu}$, with $\sigma^{\mu \nu} = (i/2) [\gamma^{\mu}, \gamma^{\nu}]$, one can expand the above exponential in powers of~$\alpha_e = e^2/(4 \pi)$, keeping the lowest-order non-vanishing contribution. Given that the trace of $\gamma^5$ alone, with $\gamma^{\mu} \gamma^{\nu}$ or an odd number of Dirac matrices, vanishes, the first non-zero contribution will be $\propto \alpha_e (i \slashed{D})^4/M^4$ i.e., the exponential needs to be expanded up to second order. Using~$\Tr[\gamma^5 \gamma^\mu \gamma^\nu \gamma^\rho \gamma^\sigma] = - 4 i \epsilon^{\mu \nu \rho \sigma}$, leads to
\begin{equation}
\Tr[ \bra{x} \beta \gamma^5 \ket{x} ] = i 4 \pi \alpha_e F \widetilde{F} \lim_{M \to \infty} \frac{\bra{x} e^{-\frac{\partial^2}{M^2}} \ket{x}}{M^4} \; ,
\end{equation}
where $\widetilde{F}^{\mu \nu} = \epsilon^{\mu \nu \rho \sigma} F_{\rho \sigma}/2$. At one-loop order $\bra{x} e^{-\frac{\partial^2}{M^2}} \ket{x} = i M^4/(16 \pi^2)$, hence,
\begin{equation}
 \mathcal{J}^{-2} = \exp\left[i \int d^4 x \beta \frac{\alpha_e}{2 \pi} F_{\mu \nu} \widetilde{F}^{\mu \nu} \right] \; .
\label{eq:shift}
\end{equation}
Thus, the axial transformation of the fields alters the path integral measure, adding the above term to the Lagrangian.

The EOM for current conservation at the quantum level will include the mass contribution of Eq.~\eqref{eq:currentsEOM}, as well as, the one-loop anomalous term of Eq.~\eqref{eq:shift}. The result reads~\footnote{The result is presented in abbreviated notation, it should be understood as relations between correlators as in Eqs.~\eqref{eq:current1} and~\eqref{eq:current2}.}:
\begin{equation}
    \partial_\mu J^\mu_{\text{A}} = 2 i m \ \overline{\psi} \gamma^5 \psi - \frac{\alpha_e}{2 \pi} F_{\mu \nu} \widetilde{F}^{\mu \nu} \; ,
    \label{eq:fullJA}
\end{equation}
the last term being the $\text{U}(1)_{\text{A}}$ anomalous contribution. One may wonder if the axial current will receive further higher-order loop corrections. The answer is no. The Adler-Bardeen theorem~\cite{PhysRev.177.2426,PhysRev.182.1517} states that the chiral anomaly is one-loop exact with all higher-order corrections vanishing.

We are now set to generalize the above result for QCD, which is the theory based on the Lagrangian~\eqref{eq:LQCD}. Performing the unitary rotations of the LH and RH quark fields, such that $\mathbf{M}_{u,d}$ become diagonal and real, leads to the CKM quark mixing matrix as shown in Sec.~\ref{sec:fermionmassmix}. Since the rotations are chiral -- distinct between LH and RH fields -- these are anomalous. Hence, an additional CP-violating term appears in $\mathcal{L}_{\text{QCD}}$ similar to the axial anomaly of Eq.~\eqref{eq:shift}, namely,
\begin{align}
\mathcal{L}_{\theta_F} = \theta_F \frac{\alpha_s}{8 \pi} G^{a}_{\mu \nu} \widetilde{G}^{a \mu \nu} \; , \; \theta_F = \arg[\det(\mathbf{M}_u)] + \arg[\det(\mathbf{M}_d)] \; .
\label{eq:thetaF}
\end{align}

The above equation can be written as a total derivative $G_{\mu \nu}^a \widetilde{G}^{a \mu \nu} = \partial_{\mu} J^\mu_{\text{CS}}$, with the Chern-Simmons current~\cite{Chern:1974ft},
\begin{equation}
     J^\mu_{\text{CS}} = 2 \epsilon^{\mu \nu \rho \sigma} \Tr \left(A_\nu \partial_\rho A_\sigma - \frac{i 2}{3} A_\nu A_\rho A_\sigma\right) \; .
    \label{eq:JCS}
\end{equation}
Being a total derivative, this extra CP-violating term has no impact in perturbation theory. Thus, it may seem that it has no physical meaning. However, QCD is strongly coupled at low energies, its vacuum properties cannot be fully understood through perturbative approaches. We will show that the above result stems from non-perturbative effects, namely instantons, allowed by the QCD-vacuum structure.

%%%%%%%%%%%%%%%%%%%%%%%%%%%%%%%%%%%%%%%%%%%%%%%%%%%%%%%%%%%%%%%%%%%%%%%%%%%%%
\subsubsection{Instantons and QCD vacuum}
\label{sec:instantons}
%%%%%%%%%%%%%%%%%%%%%%%%%%%%%%%%%%%%%%%%%%%%%%%%%%%%%%%%%%%%%%%%%%%%%%%%%%%%%

Consider the Euclidean action of a pure Yang--Mills (YM) theory in four spatial dimensions~\footnote{This section follows the discussions of Refs.~\cite{Coleman:1985rnk,Srednicki:2007qs}.},  
\begin{equation}
 S_{\text{E}} = \int d^4x F_{\mu \nu} F^{\mu \nu} \Rightarrow D_{\mu} F^{\mu \nu} = 0 \; .
\end{equation}
The latter relation corresponds to the usual classical EOM for the gauge fields. A trivial solution is given by~$A_\mu = 0$, up to a gauge transformation $U$, namely 
\begin{equation}
 A_\mu(x) = \frac{i}{g} U(x) \, \partial_\mu U^\dagger(x) \, ,
\end{equation} 
where $g$ denotes the coupling constant of the gauge group $G$. These are the so called pure-gauge configurations. We are interested in finding non-trivial solutions that lead to a non-zero finite action. It can be shown that, if these solutions exist, they can only be constructed in four dimensional Euclidean YM theories~\cite{Derrick:1964ww}. 

The action $S_{\text{E}}$ corresponds to a volume integral in four-dimensional space, which can be recast as a surface integral over a three-sphere $S^3_\infty$ at infinite radius~$r = |x| \to \infty$, via Gauss’ theorem. Finiteness of the action requires the integrand to vanish sufficiently rapidly as $r \to \infty$. Consequently, the gauge potential must decay as $A_\mu \sim 1/r^{1+\epsilon}$, with~$\epsilon$ an infinitesimal quantity. However, its general form is defined apart from a gauge transformation, 
\begin{equation}
 A_\mu(x) \sim \frac{1}{r^{1+\epsilon}} \;+\; \frac{i}{g} U(x) \partial_\mu U^\dagger(x) \, .
\end{equation} 
In the asymptotic region, the latter expression approaches a pure-gauge configuration. These solutions therefore define a mapping of the three-sphere at infinity into the gauge group: $U(x): S^3_{\infty} \rightarrow G$. Such mappings cannot, in general, be continuously deformed into the trivial configuration, since they fall into distinct homotopy classes. The disconnected homotopy classes of maps $U(x): S^3_{\infty} \rightarrow G$, are classified by the third homotopy group $\pi_3(G)$. In the case of interest, one has $\pi_3[\text{SU}(n)] = \mathbb{Z}$ for $n \geq 2$, with $\mathbb{Z}$ denoting the set of integers. Hence, non-trivial configurations are gauge equivalent to zero (pure gauge solutions), but cannot be removed away due to topological obstruction.

The winding number $\omega$ is the integer that characterizes each topologically disconnected solution. For the specific maps of interest i.e., $U(x): S^3_{\infty} \rightarrow G$:
\begin{align}
    \omega & = \frac{1}{24 \pi^2} \int_{S^3} d S_{\mu} \epsilon^{\mu \nu \rho \sigma} \Tr[(U \partial_\nu U^\dagger) (U \partial_\rho U^\dagger) (U \partial_\sigma U^\dagger)] = \frac{ i g^3}{24 \pi^2} \int_{S^3} d S_{\mu} \epsilon^{\mu \nu \rho \sigma} \Tr[A_\nu A_\rho A_\sigma] \; .
\end{align}
Notice that $\omega$ only depends on the boundary condition of the sphere at infinity i.e., on the gauge field values. Also, gauge transformations relate a given map to another one, within the same homotopy class, hence with same $\omega$, which is, therefore, a gauge-invariant quantity. Writing the above as a volume integral, it can be shown that
\begin{equation}
    \omega = \frac{g^2}{32 \pi^2} \int d^4 x \partial_\mu J_{\text{CS}}^\mu = \frac{g^2}{32 \pi^2} \int d^4 x F_{\mu \nu} \widetilde{F}^{\mu \nu} \; ,
    \label{eq:winding}
\end{equation}
with $J_{\text{CS}}^\mu$ being a Chern-Simmons current [see Eq.~\eqref{eq:JCS}]. Solving the classical EOM of a YM theory in Euclidean space yields finite-action, non-trivial solutions, called instantons. These give rise to a non-zero integral over the anomalous term [see Eq.~\eqref{eq:thetaF}], which corresponds to a topological charge, the winding number. To establish the connection to QCD, it is necessary to examine the vacuum structure of this theory.

A given classical vacuum $\ket{n}$, labeled by its winding number $n$, has vanishing energy density, corresponding to trivial solutions. As mentioned, pure gauge configurations cannot relate two classical vacua with distinct winding number. However, quantum effects can alter this picture. In fact, the Euclidean instanton configurations can be seen as paths in Minkowski spacetime, involving non-zero energy density, relating a classical vacuum to another. Instantons evolve over time, mediating tunneling between the $\ket{n}$ states with a probability $\propto e^{-S_{\text{E}}} \sim e^{-1/g^2}$. This exponential behavior exhibits the characteristic inverse dependence on the coupling constant, which is typical of non-perturbative effects in QFT. Furthermore, the winding number difference $n-m$, between two states $\ket{n}$ and $\ket{m}$, corresponds to a gauge-invariant quantity, namely, the winding number $\omega = n-m$ of the Euclidean instanton. The degeneracy among classical vacua is lifted thanks to these tunneling effects.

Consequently, the true QCD-vacuum, labeled here as $\ket{\theta}$, will be the infinite coherent superposition of all classical states~$\ket{n}$, namely
    \begin{equation}
    \ket{\theta} = \sum_{n=- \infty}^{+ \infty} e^{-i n \theta} \ket{n} \; .    
    \end{equation}
Imposing gauge invariance on $\ket{\theta}$ amounts at obtaining the same state apart from a phase. Note that different values of $\theta$ do not label different states of a same theory, but correspond to distinct theories, leading to different physics. Overall, the parameter $\theta$ alone characterizes the vacuum of a theory. The physical impact of the QCD-vacuum on any observable, described generically by an operator~$\mathcal{O}$, is determined by evaluating the VEV:
\begin{align}
\bra{\theta} \mathcal{O} \ket{\theta} & = \sum_{l} e^{i \theta l} \bra{l} \mathcal{O} \ket{0} \; ,
\label{eq:theta0}
\end{align}
where $l$ is the winding number difference between vacua. In Mikowski spacetime, $l$ is the same as the Euclidean instanton, i.e., $l = \omega$ [see Eq.~\eqref{eq:winding}]. Hence, by writing the correlator in terms of the path integral as in Eq.\eqref{eq:Zobs}, summing over all states $l$, and defining the measure as $\mathcal{D} A = \sum_l \mathcal{D} A_l$, one obtains
\begin{align}
\bra{\theta} \mathcal{O} \ket{\theta} & = \frac{1}{Z_0} \int \mathcal{D} A \ \exp\left[i \int d^4 x \left(\mathcal{L}_{\text{QCD}} + \theta \frac{\alpha_s}{8 \pi} G_{\mu \nu}^a \widetilde{G}^{a \mu \nu} \right) \right] \; .
\label{eq:theta}
\end{align}
The physical impact, of a given~$\ket{\theta}$, on any observable, amounts at adding to the Lagrangian a topological term, with the same form as the one in Eq.~\eqref{eq:thetaF}, but with different coupling parameter. Thus, this CP-violating term in QCD cannot be neglected (see Sec.~\ref{sec:strongCPnedmsol}).

In summary, the anomalous chiral rotations of the quark fields lead to the $\theta_F$-term of Eq.~\eqref{eq:thetaF}, while non-perturbative instantons generate the same  term with a parameter $\theta$ presented in Eq.~\eqref{eq:theta}. Hence, the strong CP phase in Eq.~\eqref{eq:topological} is given by
\begin{align}
\overline{\theta} = \theta + \theta_F = \theta + \arg[\det(\mathbf{M}_u)] + \arg[\det(\mathbf{M}_d)]\; ,
\label{eq:thetafully}
\end{align}
being the basis invariant quantity characterizing CP violation in the strong sector of the SM.

%%%%%%%%%%%%%%%%%%%%%%%%%%%%%%%%%%%%%%%%%%%%%%%%%%%%%%%%%%%%%%%%%%%%%%%%%%%%%
\subsection{Accidental symmetries and sphalerons}
\label{sec:sphalerons}
%%%%%%%%%%%%%%%%%%%%%%%%%%%%%%%%%%%%%%%%%%%%%%%%%%%%%%%%%%%%%%%%%%%%%%%%%%%%%

The SM Lagrangian exhibits several global accidental symmetries. Among these, U$(1)_B$ and $\text{U}(1)_{L_e} \times \text{U}(1)_{L_\mu} \times \text{U}(1)_{L_\tau}$, associated respectively with baryon number \( B \) and individual lepton flavor numbers \( L_\alpha \), are conserved at the classical level. Consequently, total lepton number \( L \equiv L_e + L_\mu + L_\tau \) is also conserved at tree-level. Thus, the classical Lagrangian is invariant under global baryon and lepton number transformations given by
\begin{equation}
    \text{U}(1)_B \ \text{:} \ \mathcal{Q} \rightarrow e^{i\alpha_B} \mathcal{Q} \; , \; \text{U}(1)_L \ \text{:} \ \mathcal{L} \rightarrow  e^{i\alpha_L} \mathcal{L} \; ,
\end{equation}
where we define $\mathcal{Q} = \{q_L, u_R, d_R\}$ and $\mathcal{L} = \{\ell_L, e_R\}$. Note that quarks (antiquarks) have \( B = 1/3 \) (\( B = -1/3 \)) and \( L = 0 \), while leptons (antileptons) have \( B = 0 \) and \( L = 1 \) (\( L = -1 \)). The corresponding Noether currents are
\begin{equation}
J_B^\mu = \frac{1}{N_c} \sum_{\alpha} \overline{\mathcal{Q}_\alpha} \gamma^\mu \mathcal{Q}_\alpha \; , \; J_L^\mu = \sum_{\alpha} \overline{\mathcal{L}_\alpha} \gamma^\mu \mathcal{L}_\alpha \; ,
\label{eq:BandLcurrents}
\end{equation}
where $N_c = 3$ is the number of quark colors and we sum over generations. At tree level, $J_{B,L}^\mu = 0$ in the SM. The conservation of baryon number forbids processes such as proton decay and neutron--antineutron oscillations, both of which remain unobserved and are tightly constrained by experimental data~\cite{Super-Kamiokande:2011idx,SNO:2018ydj}. However, as we will discuss in Sec.~\ref{sec:neutrino}, the observation of neutrino flavor oscillations requires that neutrinos have mass and there is lepton mixing implying that flavor lepton number is violated. Furthermore, from a theoretical perspective, the explicit violation of total lepton number opens up the possibility for Majorana neutrino mass generation which will be studied in Chapter~\ref{chpt:neutrinodarksectors}.

As seen in Sec.~\ref{sec:strongCP}, a symmetry can be classically conserved but broken by the quantum character of the theory, thus making it anomalous. Due to the chiral anomaly of the SM fermion currents under the EW gauge group, the currents of Eq.~\eqref{eq:BandLcurrents} are written in terms of anomalous divergences, in a similar way as for the QCD $\overline{\theta}$ term. Namely,
\begin{align}
\partial_\mu J^\mu_{B} =  \partial_\mu J^\mu_{L} =  \frac{3}{8 \pi^2} \left(g^2 W^i_{\mu\nu} \widetilde{W}^{i \mu\nu} + g^{\prime 2} B_{\mu\nu} \widetilde{B}^{\mu\nu} \right) \; ,
\end{align}
where $\widetilde{W}^{i \mu \nu} = \epsilon^{\mu \nu \rho \sigma} W^{i}_{\rho \sigma}/2$ and $\widetilde{B}^{\mu \nu} = \epsilon^{\mu \nu \rho \sigma} B_{\rho \sigma}/2$ are the duals of their respective field stress tensors given in Eq.~\eqref{eq:stresstensors}. From this, it is clear that $B$, $L$ and $B + L$ are violated. However, $B-L$ is exactly preserved in the SM. 

We also showed, in Sec.~\ref{sec:strongCP}, that non-Abelian gauge theories have a non-trivial vacuum structure resulting in non-perturbative effects -- instantons. For $\text{U}(1)_Y$ there are no non-trivial vacuum configurations since $\pi_3[\text{U}(1)] = 0$. However, SU$(2)_L$ has infinitely many degenerate vacua, being distinguished by an integer value topological invariant known as the Chern-Simmons number $N_{\text{CS}}$~\cite{Chern:1974ft}. Transitions between vacua differing by $\Delta N_{\text{CS}} = \pm 1$ induce a change in $B + L$ as
\begin{equation}
\Delta(B + L) = 6 \, \Delta N_{\text{CS}} \; , 
\end{equation}
where a minimal violation of $\Delta(B + L) = \pm 6$, corresponds to processes that violate $B$ and $L$ each by three units. The explicit twelve fermion interaction, induced by \( \Delta N_{\text{CS}} = \pm 1 \) transitions, involves nine (three different generations with three different color states) LH quarks (or RH antiquarks) and three (three different generations with the same isospin) LH leptons (or RH antileptons):
\begin{equation}
\text{vacuum} \Rightarrow \varepsilon^{ij} \varepsilon^{kl} \varepsilon^{a b c} \sum_{\text{generations}} q_{L}^{a i} q_{L}^{b j} q_{L}^{c k} \ell_{L}^l \; ,
\end{equation}
where, as usual, $i, j, k, l$ and $a, b, c$ are SU$(2)_L$ and SU$(3)_c$ indices, respectively. Transitions between vacua are mediated by instantons at zero temperature and by sphalerons at finite temperature. At zero temperature, the tunneling rate of such $(B+L)$-violating processes is exponentially suppressed as~\cite{tHooft:1976rip}
\begin{equation}
\Gamma^{\text{inst}}_{\cancel{B+L}} \sim e^{- 4 \pi/\alpha_W} \sim \mathcal{O}(10^{-165}) \; ,
\end{equation}
where $\alpha_W = g^2/(4 \pi)$. However, at high temperatures above the EW scale, i.e. $T > T_{\text{EW}} \sim 100 \, \text{GeV}$, thermal fluctuations allow sphaleron transitions to overcome the energy barrier with height $E_{\text{sph}} \sim \text{TeV}$, leading to unsuppressed $(B + L)$-violating interactions with rates of order~\cite{Arnold:1987mh,Arnold:1987zg}
\begin{equation}
\Gamma^{\text{sph}}_{\cancel{B+L}} \sim 250 \alpha_W^5 T \; .
\label{eq:sphalerons}
\end{equation}

Overall, while the SM forbids perturbative $B$ and $L$ violating processes, and exactly preserves $B-L$, it allows for non-perturbative $B + L$ violation through EW sphalerons. As we will discuss in Sec.~\ref{sec:BAU}, sphalerons provide a fundamental ingredient in mechanisms that account for the observed matter-antimatter asymmetry of the Universe.

%%%%%%%%%%%%%%%%%%%%%%%%%%%%%%%%%%%%%%%%%%%%%%%%%%%%%%%%%%%%%%%%%%%%%%%%%%%%%
\section{Avenues Beyond the Standard Model}
\label{sec:BSMavenues}
%%%%%%%%%%%%%%%%%%%%%%%%%%%%%%%%%%%%%%%%%%%%%%%%%%%%%%%%%%%%%%%%%%%%%%%%%%%%%

The discovery of the last missing piece of the SM, the Higgs boson, at the LHC was announced in 2012 by the CERN ATLAS and CMS collaborations~\cite{ATLAS:2015yey}. This established the SM as a very successful theory in describing Nature, with its predictions being in extraordinary agreement with experimental results. Since then no other fundamental particle has been discovered. However, apart from colliders, other facilities found clear evidences that point towards a more complete picture of Nature. The purpose of this section is to review such evidences.

%%%%%%%%%%%%%%%%%%%%%%%%%%%%%%%%%%%%%%%%%%%%%%%%%%%%%%%%%%%%%%%%%%%%%%%%%%%%%
\subsection{Neutrino masses, mixing and CP violation}
\label{sec:neutrino}
%%%%%%%%%%%%%%%%%%%%%%%%%%%%%%%%%%%%%%%%%%%%%%%%%%%%%%%%%%%%%%%%%%%%%%%%%%%%%

As seen previously, RH neutrinos are absent in the SM. Therefore, we begin our discussion of massive neutrinos by adding three RH neutrino fields $\nu_R$, singlets under the SM gauge group~\footnote{Note that, adding an arbitrary number of gauge singlets to the SM particle content does not affect the anomaly-cancellation constraints.}. A Yukawa term for neutrinos, analogous to the one of the up-type quarks of Eq.~\eqref{eq:lagyuksm}, is written as
\begin{equation}
    \mathcal{L}_{\text{Yuk.}} \supset -  \overline{\ell_L} \mathbf{Y}_{\nu} \tilde{\Phi} \nu_R  + \text{H.c.} \xrightarrow{\text{EWSB}}  -  \overline{\nu_L} \mathbf{M}_{\nu} \nu_R  + \text{H.c.} \; , \; \mathbf{M}_{\nu} = \frac{v}{\sqrt{2}} \mathbf{Y}_{\nu} \; ,
\end{equation}
where $\mathbf{M}_{\nu}$ is the $3 \times 3$ complex neutrino mass matrix of Dirac type, which can be bidiagonalized as in Eq.~\eqref{eq:massdiag}. The unitary rotations,
\begin{equation} 
\nu_{L,R} \rightarrow \mathbf{U}^{\nu}_{L,R} \nu_{L,R} \Rightarrow  \mathbf{U}^{\nu\dagger}_L \mathbf{M}_{\nu} \mathbf{U}^{\nu}_R = \text{diag}\left(m_1, m_2, m_3\right) \; ,
\label{eq:rotDiracnu}
\end{equation}
with $m_{1,2,3}$, the real and positive neutrino masses, relate the flavor neutrinos $\nu_{\alpha}$ ($\alpha = e,\mu,\tau$) and mass eigenstates $\nu_{i}$ ($i=1,2,3$). The matrices $\mathbf{U}^{\nu}_{L,R}$ are obtained by diagonalizing the Hermitian matrices $\mathbf{H}_{\nu}= \mathbf{M}_{\nu} \mathbf{M}_{\nu}^\dagger$ and $\mathbf{H}_{\nu}^\prime= \mathbf{M}_{\nu}^\dagger \mathbf{M}_{\nu}$, respectively. Note that, there will be mixing in the leptonic CC interactions analogous to the quarks. Thus, total lepton number remains conserved, but flavor lepton number is now violated.

Although the actual neutrino mass values have not been measured yet, we quote a couple of recent bounds. The first is the cosmological constraint $\sum m_{\nu} < 0.12 \ \text{eV} \ (95 \% \ \text{CL})$ on the sum of neutrino masses provided by the Planck collaboration~(2018)~\cite{Planck:2018vyg} and the second one is the upper neutrino mass limit of $0.8~\text{eV} \ (90 \% \ \text{CL})$ obtained by the KATRIN collaboration~(2022)~\cite{KATRIN:2021uub}. Since neutrino masses are extremely tiny (around $10^{6}$ times smaller than the mass of the lightest charged fermion, the electron) in order to generate small Dirac neutrino masses one has to assume unnaturally small Yukawa couplings $\mathbf{Y}_{\nu}$ of the order of $10^{-12}$. According to 't Hooft's naturalness criterium which states that~\cite{tHooft:1979rat}
\begin{align}
&\emph{``at any energy scale $\mu$, a physical parameter or set of physical parameters $\alpha_{i}(\mu)$ is allowed to be} \nonumber \\ 
&\emph{very small if the replacement $\alpha_{i}(\mu)=0$ would increase the symmetry of the system",} \nonumber
\end{align}
a Dirac mass term for neutrinos is unnatural. Indeed, taking the limit $\mathbf{Y}_{\nu} \rightarrow 0$, the theory does not exhibit a new symmetry. Nonetheless, there is an alternative way to describe the nature of neutrinos that allows for naturally small couplings.

%%%%%%%%%%%%%%%%%%%%%%%%%%%%%%%%%%%%%%%%%%%%%%%%%%%%%%%%%%%%%%%%%%%%%%%%%%%%%
\subsubsection{Majorana fermions}
\label{sec:Majorana}
%%%%%%%%%%%%%%%%%%%%%%%%%%%%%%%%%%%%%%%%%%%%%%%%%%%%%%%%%%%%%%%%%%%%%%%%%%%%%

As first proposed in 1937 by E.~Majorana~\cite{Majorana:1937vz} neutrinos can be their own antiparticle, i.e., they can be Majorana particles. Generically, given a Majorana fermion $\psi$ with mass $m$ one can write a Majorana mass term of the form,
\begin{equation}
- m \overline{\psi} \psi = - \frac{m}{2} \overline{\psi} \psi^c + \text{H.c.} = - \frac{m}{2} \overline{\psi} C \overline{\psi}^T + \frac{m^*}{2} \psi^T C^{-1} \psi \; ,
\end{equation}
where $\psi^c = C \overline{\psi}^T$ and the bilinear $\overline{\psi} \psi^c$ is Lorentz invariant. Compared to a Dirac particle, the chiral components of a Majorana field are not independent, $\psi_R = \psi_L^c$, requiring only half of the d.o.f. to be fully described. Furthermore, if the fields $\psi$ carry a charge $q$ under a $\text{U}(1)$ symmetry, $\psi \rightarrow e^{i q \phi} \psi$, then the Majorana mass term breaks this symmetry by $2 q$.

Consider the following Majorana mass terms for $\nu_L$ and $\nu_R$,
\begin{equation}
    -\frac{1}{2} \overline{\nu_L^c} \mathbf{M}_L \nu_L -\frac{1}{2} \overline{\nu_R} \mathbf{M}_R \nu_R^c + \text{H.c.} \; .
\label{eq:Majoranamassgene}
\end{equation}
Due to the anticommuting character of fermionic fields and the antisymmetric property of~$C$, Majorana mass matrices are symmetric. In addition, these terms violate total lepton number by two units. Consequently, the smallness of a Majorana mass for active neutrinos is natural according to 't Hooft, since by taking $\mathbf{M}_{L} \rightarrow 0$, total lepton number conservation is recovered. Notice that $\overline{\nu_R} \nu_R^c$ is invariant under $\text{SU}(2)_L \times \text{U}(1)_Y$ since RH neutrinos are gauge singlets. In contrast, $\overline{\nu_L^c} \nu_L$ belongs to a triplet of $\text{SU}(2)_L$ and has $Y=-2$. As we will see in Chapter~\ref{chpt:neutrinodarksectors}, to generate at tree level a singlet out of this term the SM has to be extended with a scalar triplet with $Y=+2$ (see Sec.~\ref{sec:TypeII}).

%%%%%%%%%%%%%%%%%%%%%%%%%%%%%%%%%%%%%%%%%%%%%%%%%%%%%%%%%%%%%%%%%%%%%%%%%%%%%
\subsubsection{Effective neutrino masses}
\label{sec:Weinberg}
%%%%%%%%%%%%%%%%%%%%%%%%%%%%%%%%%%%%%%%%%%%%%%%%%%%%%%%%%%%%%%%%%%%%%%%%%%%%%

A Majorana mass term for the LH neutrinos $\nu_L$ as the one presented in Eq.~\eqref{eq:Majoranamassgene} is not allowed by the SM symmetries. The origin of such term can be explained considering the SM as a low-energy effective theory resulting from a more complete theory at a high-energy scale~$\Lambda \gg v$. Any given UV completion of the SM is composed of additional heavy fields, denoted here by~$N_i$, where $i$ runs over the number of extra particles with mass scale of the order of~$\Lambda$. Using the path integral formalism these heavy states can be integrated out leading to an effective Lagrangian~$\mathcal{L}_{\text{eff}}$. The effective action $S_{\text{eff}}$ is defined as~\cite{Broncano:2002rw,Gavela:2009cd,Abada:2007ux}
\begin{equation}
e^{i S_{\text{eff}}} = \exp\left( i \int d^4 x \mathcal{L}_{\text{eff}}(x) \right) \equiv \int \mathcal{D} N \mathcal{D} \overline{N} e^{i S} = e^{i S_{\text{SM}}}  \int \mathcal{D} N \mathcal{D} \overline{N} e^{i S_{N} (N)} \; ,
\label{eq:effS}
\end{equation}
where $\mathcal{D} N$ is the integral measure and $S$ is the full action, which is separated in the terms involving the SM fields $S_{\text{SM}}$ and the terms involving the heavy fields $S_{N}$. Expanding the action $S_{N}(N)$ around its stationary point given by the minimum energy configuration $N_0$, leads to
\begin{equation}
e^{i S_{\text{SM}}}  \int \mathcal{D} N \mathcal{D} \overline{N} e^{i S_{N} (N)} =  e^{i S_{\text{SM}}}  \int \mathcal{D} N \mathcal{D} \overline{N} e^{i \left[S_{N} (N_0) + \delta S_{N} (N_0) + \delta^2 S_{N} (N_0) + \cdots \right]} \simeq e^{i \left[S_{\text{SM}} + S_{N} (N_0)\right]} \; ,
\end{equation}
where $\delta S_{N} (N_0)$ vanishes due to the stationary condition and, in the last expression, higher-order terms are neglected. The effective action is
\begin{equation}
S_{\text{eff}} \simeq \int d^4 x \left[ \mathcal{L}_{\text{SM}} + \mathcal{L}_{N} (N_0) \right]  = S_{\text{SM}} + S_{N} (N_0) \; .
\end{equation}
The stationary fields are obtained by solving the classical Euler-Lagrange EOM,
\begin{equation}
 \frac{\delta S}{\delta \overline{N_i}}\bigg\vert_{\overline{N_i}=\overline{N_{0 i}}} = 0 \; , \; \frac{\delta S}{\delta N_i}\bigg\vert_{N_i = N_{0 i}} = 0 \; .
\label{eq:EOMEL}
\end{equation}
Inserting the solutions in $S_N(N_0)$ leads to an effective Lagrangian valid at scales much lower than~$\Lambda$,
\begin{equation}
    \mathcal{L}_{\text{eff}} = \mathcal{L}_{\text{SM}} + \mathcal{L}_{N}(N_0) = \mathcal{L}_{\text{SM}} + \mathcal{L}_{d=5} + \mathcal{L}_{d=6} + \cdots \; ,
\label{eq:effL}
\end{equation}
where the non-renormalizable dimension $d=n > 4$ operators $\mathcal{L}_{d=n}$ are invariant under $G_{\text{SM}}$ and suppressed by $\Lambda^{4-n}$.

The lowest dimension operator of such kind is the single gauge-invariant dimension-five Weinberg operator~\cite{Weinberg:1979sa}, which leads, after EWSB, to a Majorana mass term for $\nu_L$ as the one given in Eq.~\eqref{eq:Majoranamassgene}:
\begin{equation}
\mathcal{L}_{d=5} = \frac{c_{\alpha \beta}^{d=5}}{2} \left(\overline{\ell^c_{{\alpha}_L}} \tilde{\Phi}^{*} \right) \left(\tilde{\Phi}^{\dagger} \ell_{{\beta}_L}\right) + \text{H.c.} \xrightarrow{\text{EWSB}} -\frac{1}{2} \overline{\nu_L^c} \mathbf{M}_{\nu} \nu_L + \text{H.c.} \; , \; \mathbf{M}_{\nu} = - \frac{v^2}{2} c_{\alpha \beta}^{d=5} \; ,
\label{eq:weinbergop}
\end{equation}
where $c^{d=5}$ is suppressed by $\Lambda^{-1}$ and $\mathbf{M}_{\nu}$ is the effective light-neutrino mass matrix. Note that a Majorana neutrino mass term is the lowest-order effect of high-energy physics. Thus, neutrinos open a new window to physics BSM. Higher-order operators are crucial to explore other low-energy effects, e.g. dimension-six operators encode non-unitarity of the lepton mixing matrix and cLFV.

%%%%%%%%%%%%%%%%%%%%%%%%%%%%%%%%%%%%%%%%%%%%%%%%%%%%%%%%%%%%%%%%%%%%%%%%%%%%%
\subsubsection{Lepton sector observables}
\label{sec:neutrinoobservables}
%%%%%%%%%%%%%%%%%%%%%%%%%%%%%%%%%%%%%%%%%%%%%%%%%%%%%%%%%%%%%%%%%%%%%%%%%%%%%

Consider that $\nu_L$ is a Majorana field, with the mass term given in Eq.~\eqref{eq:Majoranamassgene}. Besides the unitary rotations of the charged-lepton fields of Eq.~\eqref{eq:massdiag}, one also performs the following unitary rotations,
\begin{equation}
\nu_{{\alpha}_L} \rightarrow (\mathbf{U}_{\nu})_{\alpha j} \nu_{{j}_L} \; \Rightarrow \; \mathbf{U}_{\nu}^{T} \mathbf{M}_{L} \mathbf{U}_{\nu} = \mathbf{D}_\nu = \text{diag}(m_1, m_2, m_3) \; ,
\label{eq:leptonmixing}
\end{equation}
with the unitary rotation obtained by diagonalizing the Hermitian matrix $\mathbf{H}_{\nu}^\prime= \mathbf{M}_{L}^\dagger \mathbf{M}_{L}$ and the mass eigenstate Majorana neutrinos $\nu_j$ have real and positive masses $m_{1,2,3}$. The field rotation will affect the CC interactions leading to the $3 \times 3$ unitary Pontecorvo-Maki-Nakagawa-Sakata~(PMNS) lepton mixing matrix $\mathbf{U}$~\cite{Pontecorvo:1957cp,Maki:1962mu},
\begin{equation}
    \mathcal{L}_{\text{CC}} \supset \frac{g}{\sqrt{2}} \overline{e}_L \gamma^{\mu} \mathbf{U}\, \nu_L W_{\mu}+ \text{H.c.} \; , \; \mathbf{U} = \mathbf{U}_{L}^{e \dagger} \mathbf{U}_{\nu} \; ,
    \label{eq:CCleptoU}
\end{equation}
containing a total of six parameters: three mixing angles $\theta_{12}^\ell$, $\theta_{23}^\ell$, and $\theta_{13}^\ell$, and three CP-violating phases: one Dirac-type $\delta^\ell$ and two Majorana-type $\alpha_{21}$ and $\alpha_{31}$. The Dirac neutrino ($\alpha_{21} = \alpha_{31} = 0$) case is similar to what happens for quarks. We can parameterize $\mathbf{U}$ as~\cite{Schechter:1980gr,Rodejohann:2011vc}
\begin{equation}
 \mathbf{U} = \begin{pmatrix}
c_{12}^\ell c_{13}^\ell&s_{12}^\ell c_{13}^\ell&s_{13}^\ell\\
-s_{12}^\ell c_{23}^\ell-c_{12}^\ell s_{23}^\ell s_{13}^\ell e^{i\delta^\ell}&c_{12}^\ell c_{23}^\ell-s_{12}^\ell s_{23}^\ell s_{13}^\ell e^{i\delta^\ell}&s_{23}^\ell c_{13}^\ell e^{i\delta^\ell}\\
s_{12}^\ell s_{23}^\ell-c_{12}^\ell c_{23}^\ell s_{13}^\ell e^{i\delta^\ell}&-c_{12}^\ell s_{23}^\ell-s_{12}^\ell c_{23}^\ell s_{13}^\ell e^{i\delta^\ell}&c_{23}^\ell c_{13}^\ell e^{i\delta^\ell}
\end{pmatrix}
\begin{pmatrix}  1 & 0 & 0 \\
0 &  e^{i\frac{\alpha_{21}}{2}} & 0 \\ 
0 & 0 & e^{i\frac{\alpha_{31}}{2}} \\ 
\end{pmatrix} \; ,
\label{eq:PMNSparam}
\end{equation}
where $c_{i j}^\ell \equiv \cos \theta_{i j}^\ell$,~$s_{i j}^\ell \equiv \sin \theta_{i j}^\ell$ and we take without loss of generality $\theta_{i j}^\ell \in \left[0, \pi/2 \right]$, $\delta^\ell \in~\left[0, 2\pi \right[$, and $\alpha_{21,31} \in~\left[0, 2\pi \right[$.

Neutrinos can undergo a quantum mechanical phenomenon known as neutrino oscillations~\cite{Pontecorvo:1957cp,Pontecorvo:1957qd}, with its discovery being awarded the 2015 Physics Nobel prize~\cite{McDonald:2016ixn,Kajita:2016cak}. The vacuum oscillation probability for a neutrino of flavor $\alpha$, created at a source with energy $E$ and traveling a distance $L$, to be detected as flavor $\beta$ is given by the well-known expression~\cite{Giunti}
\begin{equation}
P\left(\nu_{\alpha} \rightarrow \nu_{\beta} \right) = \sum_{j,k} \mathbf{U}^{*}_{\alpha j} \mathbf{U}_{\beta j} \mathbf{U}_{\alpha k} \mathbf{U}^{*}_{\beta k} \exp \left( -i \frac{\Delta m_{j k}^2 L}{2 E} \right) \; , \; \Delta m_{j k}^2 = m_j^2 - m_k^2 \; .
\end{equation}
Neutrino oscillations are lepton number conserving and therefore the transition probability above does not involve the Majorana phases. Hence, oscillation experiments cannot determine whether neutrinos are of Dirac or Majorana type. Moreover, these experiments are only sensitive to the neutrino mass-squared differences $\Delta m_{j k}^2$, providing no information on the absolute neutrino mass scale. Nevertheless, information on the mixing angles and the Dirac CP-violating phase is obtained from these facilities. Current oscillation experiments analyze a multitude of neutrino sources: solar, atmospheric, accelerator and reactor neutrinos. Thus, a common notation for the observables is $\Delta m_{21}^2 \equiv \Delta m_{\text{sol}}^2$, $\theta_{12}$ (solar), $\Delta m_{31}^2 \equiv \Delta m_{\text{atm}}^2$, $\theta_{23}$ (atmospheric) and $\theta_{13}$ is obtained from data analysis of nuclear reactor and accelerator experiments. 

\begin{table}[t!]
\renewcommand*{\arraystretch}{1.5}
\centering
\begin{tabular}{|c c|}  
        \hline 
        Parameter  & Best Fit $\pm 1 \sigma$ \\ \hline
        $\Delta m_{21}^2 \left(\times 10^{-5} \ \text{eV}^2\right)$ \; \; & $7.50^{+0.22}_{-0.20}$  \\
        $\left|\Delta m_{31}^2\right| \left(\times 10^{-3}  \ \text{eV}^2\right) [\text{NO}]$ \; \; & $2.55^{+0.02}_{-0.03}$  \\
        $\left|\Delta m_{31}^2\right| \left(\times 10^{-3} \ \text{eV}^2\right) [\text{IO}]$ \; \; & $2.45^{+0.02}_{-0.03}$ \\
        $\theta_{12}^\ell (^\circ)$ \; \; & $34.3\pm1.0$ \\
        $\theta_{23}^\ell (^\circ) [\text{NO}]$ \; \; & $49.26\pm0.79$ \\
        $\theta_{23}^\ell (^\circ) [\text{IO}]$ \; \; & $49.46^{+0.60}_{-0.97}$ \\
        $\theta_{13}^\ell (^\circ) [\text{NO}]$ \; \; & $8.53^{+0.13}_{-0.12}$ \\
        $\theta_{13}^\ell (^\circ) [\text{IO}]$ \; \; & $8.58^{+0.12}_{-0.14}$ \\
        $\delta^\ell  (^\circ) [\text{NO}]$ \; \; & $194^{+24}_{-22}$ \\
        $\delta^\ell  (^\circ) [\text{IO}]$ \; \; & $284^{+26}_{-28}$ \\
        \hline 
\end{tabular}
\caption{Current neutrino data: mass-squared differences, mixing angles and Dirac CP phase, obtained from the global fit of neutrino oscillation data of Ref.~\cite{deSalas:2020pgw} (see also Refs.~\cite{Esteban:2024eli} and~\cite{Capozzi:2025wyn}).}
\label{tab:leptondata}
\end{table}

Global data fits of oscillation phenomena in a three-neutrino mixing scheme~\cite{deSalas:2020pgw,Esteban:2024eli,Capozzi:2025wyn} allow to obtain the most up to date values for the neutrino observables. In this thesis, we take as reference the data obtained from the Valencia group~\cite{deSalas:2020pgw}. The results are presented in Table~\ref{tab:leptondata} which, among others, include the data from the solar neutrino experiments Sudbury Neutrino Observatory~(SNO)~\cite{Aharmim:2011vm} and  KamLAND~\cite{Gando:2010aa}; the atmospheric neutrino experiments Super-Kamiokande~\cite{Abe:2017aap} and IceCube DeepCore~\cite{Aartsen:2017nmd,Aartsen:2019tjl}; the reactor experiments RENO~\cite{Bak:2018ydk} and Daya Bay~\cite{Adey:2018zwh}; the long-baseline accelerator experiments NO$\nu$A~\cite{Acero:2019ksn}, T2K~\cite{Abe:2019ffx,Abe:2019vii}, MINOS~\cite{Adamson:2014vgd} and K2K~\cite{Ahn:2006zza}.

Notice from Table~\ref{tab:leptondata} that $\Delta m_{21}^2 \ll \Delta m_{31}^2$ and the sign of $\Delta m_{31}^2$ is still unknown, leaving the possibility for two neutrino mass orderings: normal~ordering~(NO)~where~$m_1< m_2 \ll m_3$, and inverted~ordering~(IO)~where $m_3 \ll m_1 < m_2$. For both cases the neutrino mass-squared differences in terms of the lightest neutrino states can be written as,
\begin{align}
\text{NO :}& \ m_2 = \sqrt{m_1^2 + \Delta m_{21}^2 }\; , \; m_3 = \sqrt{m_1^2 + \left| \Delta m_{31}^2 \right|} \; , \label{eq:NOIOmass1} \\ 
\text{IO :} &\ m_1 = \sqrt{m_3^2 +  \left| \Delta m_{31}^2 \right|} \; , \; m_2 = \sqrt{m_3^2 + \Delta m_{21}^2 +  \left| \Delta m_{31}^2 \right|} \; .
\label{eq:NOIOmass2}
\end{align}
At this point it is also worthwhile to remark that, since only information on $\Delta m_{jk}^2$ is obtained experimentally, there is the possibility for a massless neutrino state. Note that, for the BSM scenarios presented in this thesis, we will identify the minimal neutrino mass generation setups which contain a lightest neutrino state that is massless, so that $m_1 = 0$ for NO and~$m_3 = 0$~ for IO.

As  mentioned  before, the most stringent bound on the absolute neutrino mass scale is the cosmological constraint $\sum m_{\nu} < 0.12 \ \text{eV}$ ($0.54 \ \text{eV}$) at $95 \% \ \text{CL}$, on the sum of neutrino masses, provided by the Planck collaboration, if one considers TT,TE+lowE+lensing+BAO (TT+lowE) data~\cite{Akrami:2018vks}. If one neutrino is massless, one has $\sum m_{\nu} \simeq 0.059 \ \text{eV} $ for~NO and  $\sum m_{\nu} \simeq 0.099 \ \text{eV}$ for IO, both satisfying the cosmological bound. However, this is an indirect limit since it is model dependent. The most important bounds come from the direct measurements of the $\beta$-decay endpoint of tritium $^{3}\text{H} \rightarrow {^{3}\text{He}} + e^- + \overline{\nu_e}$. This provides an upper neutrino mass limit of $0.8~\text{eV} \ (90 \% \ \text{CL})$ obtained by the KATRIN collaboration~\cite{Aker:2019uuj}. The KATRIN experiment projected sensitivity is to improve this bound down to $0.2$ eV.

Alongside the outstanding issues of the absolute neutrino mass scale and ordering, CP violation in the lepton sector has yet to be firmly established. Currently, while some tension between NO$\nu$A and T2K results exist regarding the Dirac CP phase for a NO, maximal CP violation seems to be preferred for an inverted spectrum (for a recent review see Ref.~\cite{Rahaman:2022rfp}). In the next decades, long-baseline neutrino experiments like DUNE~\cite{DUNE:2016hlj} and Hyper-Kamiokande~\cite{Hyper-Kamiokande:2018ofw} will provide much more information on LCPV~\cite{Branco:2011zb}.

\begin{table}[!t]
\renewcommand*{\arraystretch}{1.5}
    \begin{minipage}{.5\linewidth}
      \centering
      \vspace{-1.2cm}
\begin{tabular}{|c c|} 
\hline
\multicolumn{2}{|c|}{Present limit}\\
        \hline 
Experiment & $m_{\beta \beta}$ (meV) \\ \hline
GERDA~\cite{Agostini:2020xta} & \hspace{+0.3cm} $79 - 180$ \\
CUORE~\cite{Adams:2019jhp} & \hspace{+0.3cm} $73 - 350$ \\
EXO-200~\cite{Anton:2019wmi} & \hspace{+0.3cm} $93 - 286$ \\
KamLAND-Zen 800~\cite{KamLAND-Zen:2022tow} & \hspace{+0.3cm} $36 - 156$ \\
		\hline
	\end{tabular}
	 \end{minipage}%
    \begin{minipage}{.5\linewidth}
      \centering
      %\vspace{-2.5cm}
\begin{tabular}{|c c|} 
\hline
\multicolumn{2}{|c|}{Future sensitivity}\\
\hline
Experiment & $m_{\beta \beta}$ (meV) \\
\hline
AMORE II~\cite{Lee:2020rjh} & \hspace{+0.3cm} $15 - 30$ \\
CUPID~\cite{Wang:2015raa} & \hspace{+0.3cm} $10 - 15$ \\
LEGEND~\cite{Abgrall:2017syy} & \hspace{+0.3cm} $15 - 50$ \\
SNO+~I~\cite{Andringa:2015tza} & \hspace{+0.3cm} $41 - 99$ \\
KamLAND2-Zen~\cite{KamLAND-Zen:2016pfg} & \hspace{+0.3cm} $25 - 70$ \\
nEXO~\cite{Albert:2017hjq} & \hspace{+0.3cm} $8 - 18$ \\
PandaX-III~\cite{Chen:2016qcd} & \hspace{+0.3cm} $20 - 50$  \\ \hline
	\end{tabular}
    \end{minipage}%
\caption{Present limits and future sensitivities for the effective Majorana mass~$m_{\beta \beta}$ of several~$0_\nu \beta \beta$ experiments.}
\label{tab:dataNDBD}
\end{table}
\begin{figure}[!t]
    \centering
      \includegraphics[scale=0.85]{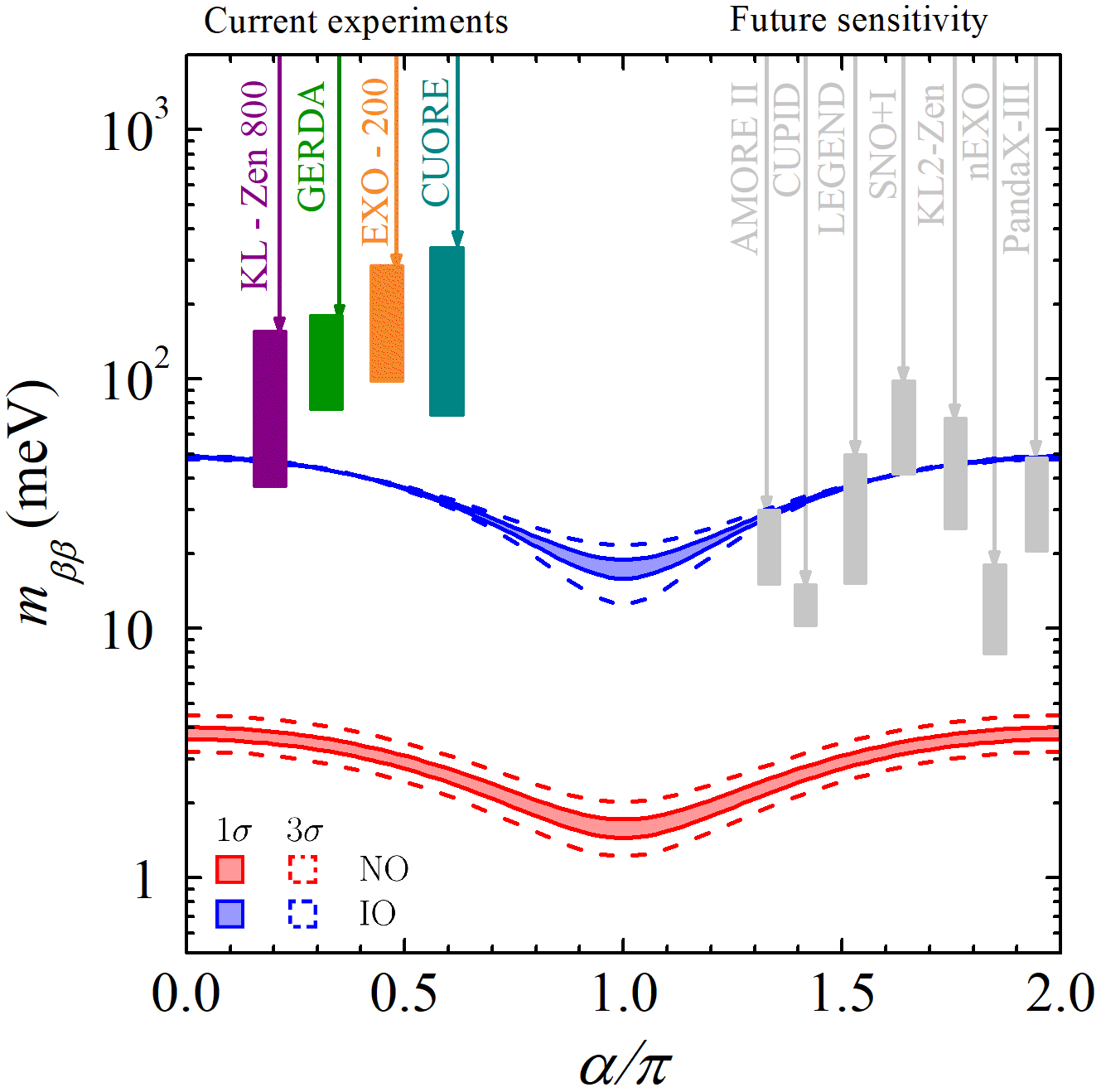}
    \caption{Allowed $m_{\beta \beta}$ regions as a function of $\alpha$ for the case where the lightest neutrino is massless obtained using Eqs.~\eqref{eq:NOIOmbb1min} and~\eqref{eq:NOIOmbb2min} and varying low-energy neutrino observables within their allowed 1$\sigma$ and 3$\sigma$ intervals by neutrino oscillation data presented in Table~\ref{tab:leptondata} (see also Ref.~\cite{deSalas:2020pgw}). Results for NO (IO) are shown in red (blue). Current most constraining $0_\nu \beta \beta$ experimental upper-bound ranges are represented by vertical colored bars while future sensitivities are indicated in gray -- see Table~\ref{tab:dataNDBD}.}
    \label{fig:mbb_alpha}
\end{figure}

Elucidating the intrinsic nature of neutrinos remains a formidable task, with no experimental information yet available on the Majorana phases. Investigations of lepton-number violating~(LNV) processes, particularly neutrinoless double beta decay ($0_\nu \beta \beta$), offer a pathway to determine the fundamental nature of neutrinos (see Refs.~\cite{Bilenky:2014uka,DellOro:2016tmg,Dolinski:2019nrj} for recent reviews on this subject). Namely, a positive observation of $0_\nu \beta \beta$ would not only prove the Majorana nature of neutrinos by the black-box theorem~\cite{Schechter:1981bd}, but could potentially determine Majorana CP-violation~\cite{Branco:2002ie}. The amplitude for this rare process is proportional to the effective Majorana mass parameter~\cite{Giunti},
\begin{equation}
m_{\beta \beta} = \left| \sum_{j=1}^{3} \mathbf{U}_{e j} m_j \right| = \left| \left(m_1 c_{12}^2 + m_2 s_{12}^2 e^{2 i \alpha_{21}} \right) c_{13}^2 + m_3 s_{13}^2 e^{2 i \alpha_{31}} \right| \; ,
\label{eq:mbbformula}
\end{equation}
which can be written in terms of the lightest neutrino mass for both spectrum orderings as
\begin{align}
\text{NO :}& \ m_{\beta \beta} = \left| \left(m_1 c_{12}^2 + \sqrt{m_1^2 + \Delta m_{21}^2 } \ s_{12}^2 e^{2 i \alpha_{21}} \right) c_{13}^2 + \sqrt{m_1^2 + \left| \Delta m_{31}^2 \right|} \ s_{13}^2 e^{2 i \alpha_{31}} \right| \; , \label{eq:NOIOmbb1} \\ 
\text{IO :}& \ m_{\beta \beta} = \left| \left(\sqrt{m_3^2 + \left| \Delta m_{31}^2 \right|} \ c_{12}^2  + \sqrt{m_3^2 + \Delta m_{21}^2 +  \left| \Delta m_{31}^2 \right|} \ s_{12}^2 e^{2 i \alpha_{21}} \right) c_{13}^2 + m_3 s_{13}^2 e^{ 2 i \alpha_{31}} \right| \; .
\label{eq:NOIOmbb2}
\end{align}
Table~\ref{tab:dataNDBD} displays the present upper limits on~$m_{\beta\beta}$ reported by GERDA~\cite{Agostini:2020xta}, CUORE~\cite{Adams:2019jhp}, EXO-200~\cite{Anton:2019wmi} and KamLAND-Zen 800~\cite{KamLAND-Zen:2022tow}. Also are shown the future $m_{\beta\beta}$ sensitivities projected by the upcoming experiments~AMORE~II~\cite{Lee:2020rjh}, CUPID~\cite{Wang:2015raa}, LEGEND~\cite{Abgrall:2017syy}, SNO+~I~\cite{Andringa:2015tza}, KamLAND2-Zen~\cite{KamLAND-Zen:2016pfg}, nEXO~\cite{Albert:2017hjq} and PandaX-III~\cite{Chen:2016qcd}. In the case of a massless neutrino, one of the phases is unphysical and the sole physical Majorana phase is $\alpha=\alpha_{31}-\alpha_{21}$ for NO or $\alpha=\alpha_{21}$ for IO. This minimal scenario offers promising testability at future experiments looking for the $0_\nu \beta \beta$ process. The Majorana mass parameter $m_{\beta \beta}$, in this case, can be written, for both mass orderings, as~\cite{Joaquim:2003pn}:
\begin{align}
\text{NO :}& \ m_{\beta \beta} = \left| \sqrt{\Delta m_{21}^2 } \ s_{12}^2 c_{13}^2 e^{-i \alpha}  + \sqrt{\left| \Delta m_{31}^2 \right|} \ s_{13}^2 \right| \; , \label{eq:NOIOmbb1min} \\ 
\text{IO :}& \ m_{\beta \beta} = \left|\sqrt{\left| \Delta m_{31}^2 \right|} \ c_{12}^2  c_{13}^2 + \sqrt{\Delta m_{21}^2 +  \left| \Delta m_{31}^2 \right|} \ s_{12}^2 c_{13}^2 e^{-i \alpha}\right| \; ,
\label{eq:NOIOmbb2min}
\end{align}
where it is clear that the resulting allowed regions, by the oscillation data, for $m_{\beta \beta}$ will correlate with the Majorana phase~$\alpha$, as shown in Fig.~\ref{fig:mbb_alpha}. Results for NO and IO are presented in red and blue, respectively, with the colored (dashed) regions obtained by varying the neutrino observables within their allowed 1$\sigma$ (3$\sigma$) intervals (see Table~\ref{tab:leptondata} and Ref.~\cite{deSalas:2020pgw}). We notice that in this minimal scenario there is no cancellation in the $0_\nu \beta \beta$ amplitude even for NO (red) neutrino masses~\cite{Barreiros:2018bju}. Namely, the effective Majorana mass parameter~\cite{Barreiros:2018bju} has a lower bound for NO neutrino masses $m_{\beta \beta} \sim 1.5$ meV for $\alpha \sim \pi$. Most interestingly for IO (blue) the predictions $m_{\beta \beta} \in [15,50]$ meV, fall within the projected sensitivities of upcoming experiments (gray bars).

Overall, if neutrinos are Majorana particles, the lepton sector contains a total of twelve parameters: three charged lepton masses, three light neutrino masses, three mixing angles and three phases. If one neutrino is massless, the number of parameters is reduced to ten. Any extension of the SM that aim at studying neutrinos must take into account the experimental observables presented in this section. 

As a last note we wish to point out that besides all of the aforementioned open questions in the neutrino sector -- absolute mass scale, ordering, LCPV, Majorana or Dirac -- certain quantitative questions arise when the lepton sector is compared to quarks. For example, the CKM matrix exhibits small inter-generational mixing being close to the identity matrix, while mixing in the lepton sector is larger among different generations. These intriguing mixing patterns are part of the so-called flavor puzzle. Remarkably, the flavor sector of the SM, including neutrino masses from an effective perspective, remains the most unconstrained sector of the theory. Thus, from a theoretical viewpoint such questions can serve as a guide in building extensions of the SM, where addressing the peculiar fermion mass and mixing patterns can provide useful insight into other open problems in particle  and astroparticle physics. Namely, we will address the flavor puzzle in Sec.~\ref{sec:lepto}, connected to neutrino mass generation and matter-antimatter asymmetry while in Chapter~\ref{chpt:flavoraxion}, we address this puzzle in a unified model for neutrino masses, DM and a solution to the strong CP problem.

%%%%%%%%%%%%%%%%%%%%%%%%%%%%%%%%%%%%%%%%%%%%%%%%%%%%%%%%%%%%%%%%%%%%%%%%%%%%%
\subsection{Dark matter}
\label{sec:DM}
%%%%%%%%%%%%%%%%%%%%%%%%%%%%%%%%%%%%%%%%%%%%%%%%%%%%%%%%%%%%%%%%%%%%%%%%%%%%%

One of the most striking experimental evidences for new physics is the existence of DM, established through indirect astrophysical measurements and cosmological observations (for a recent review see Ref.~\cite{Cirelli:2024ssz}). Zwicky, in 1933, was the first to point out the existence of some kind of DM~\cite{Zwicky:1933gu}. By analyzing the velocity dispersion of galaxies in the Coma Cluster, Zwicky inferred a total mass far exceeding the amount derived from luminous matter, suggesting the presence of a DM halo. Decades later, observations of galaxy rotation curves~\cite{Rubin:1970zza} revealed that the rotational velocities of stars in the outer regions of spiral galaxies remain constant with radius, contrary to expectations if only visible matter were present. These flat rotation curves provided strong, independent evidence for a dominant, unseen mass component. Gravitational lensing studies have further substantiated the presence of DM by revealing a discrepancy between the gravitational mass inferred from light deflection and the mass attributable to luminous matter~\cite{Bartelmann:1999yn}. X-ray observations of hot intracluster gas in galaxy clusters also suggest the need for a substantial dark component to explain the observed hydrostatic equilibrium of the gas~\cite{Sarazin:1988fb}. A particularly compelling case is the Bullet Cluster, where the spatial offset between the X-ray emitting baryonic matter and the gravitational potential derived from weak lensing maps strongly supports the collisionless nature of DM~\cite{Clowe:2006eq}. In addition, the anisotropies in the Cosmic Microwave Background (CMB), as measured by experiments such as WMAP and Planck, provide precise constraints on the DM relic abundance, consistent with a cold, non-baryonic component~\cite{Planck:2018vyg}. First detected in 1964 by Penzias and Wilson~\cite{Penzias:1965wn}, the CMB consists of relic photons that decoupled from baryonic matter approximately 380,000 years after the Big Bang. These photons, having traversed the cosmos freely, preserve detailed information about the last scattering surface. The CMB exhibits an almost perfectly isotropic black-body spectrum with a temperature $T \sim 2.726$ K~\cite{Fixsen_2009}. However, there are small anisotropies in the CMB temperature spectrum of the order $\delta T / T \sim 10^{-5}$, arising from acoustic oscillations in the primordial plasma. These are the result of the interplay between radiation pressure from photons and the gravitational potential wells created by baryonic matter and DM, ultimately giving rise to the large-scale structure observed in the present Universe~\cite{Hu:1997mn}. The evidence for DM is further reinforced by a range of complementary cosmological probes, including large-scale structure surveys~\cite{BOSS:2013kxl,Beutler:2011hx,Blake:2011en,Padmanabhan:2012hf,Percival:2009xn,Parkinson:2012vd}, baryon acoustic oscillations~(BAO)~\cite{Eisenstein:2005su}, Type Ia supernovae observations~\cite{Riess:1998cb}, weak gravitational lensing~\cite{Kilbinger:2014cea}, and precise measurements of the Hubble constant~\cite{Freedman:2019jwv}, among others.

The fit of $\Lambda$CDM, which describes the evolution of our Universe, its matter content, structure formation and accelerated expansion, based on the Big-Bang theory, to the CMB power spectrum allows to determine the CDM relic abundance. Overall, about $27\%$ of the total energy in the Universe is DM, being about $5$ times more abundant than ordinary baryonic matter. Precisely, the current values for the baryon and CDM abundances obtained from the Planck combined TT,TE,EE+lowE+lensing satellite data are~\cite{Planck:2018vyg}
\begin{align}
\Omega_{B} h^2  & = 0.02237 \pm 0.00015 \; ,
   \label{eq:Oh2BaryonPlanck} \\
   \Omega_{\text{CDM}} h^2  & = 0.1200 \pm 0.0012 \; ,
   \label{eq:Oh2PlanckOG}
\end{align}
at $68\%$ CL where $h = H_0 / (100$ km$/$s$/$Mpc) is the reduced Hubble constant with $H_0=67.36\pm0.54 \ \text{km} \ \text{s}^{-1} \ \text{Mpc}^{-1}$. 

From the aforementioned astrophysical and cosmological evidence, it remains the question about the nature of DM. Albeit there is the possibility of explaining some DM observations by modifying the laws of gravity~\cite{Clifton:2011jh}, in this thesis we consider DM to be a particle. From the data, a viable particle DM candidate has the following characteristics:
\begin{itemize}
    \item Its abundance needs to satisfy the DM relic density value of Eq.~\eqref{eq:Oh2PlanckOG};
    \item Cannot interact strongly, i.e., is non-baryonic;
    \item Electrically neutral~\cite{McDermott:2010pa};
    \item Weakly interacting with ordinary matter;
    \item Stable at the cosmological scale~\cite{Audren:2014bca,Mambrini:2015sia,Baring:2015sza,Slatyer:2016qyl,Jin:2020hmc,Acciari:2018sjn};
    \item Is mostly non-relativistic, constituting CDM, such that large-scale structure formation in the early Universe complies with observation~\cite{Peebles:1982ff};
    \item The amount of DM self-interactions is constrained by cluster collision data~\cite{Clowe:2006eq}.
\end{itemize}
It is clear that in the SM there is no particle with these properties. Consequently, we must explore BSM extensions featuring viable DM candidates. The paradigmatic WIMP scenario~\cite{Kolb:1990vq} stands out among the numerous particle DM proposals, which we briefly review in what follows.

%%%%%%%%%%%%%%%%%%%%%%%%%%%%%%%%%%%%%%%%%%%%%%%%%%%%%%%%%%%%%%%%%%%%%%%%%%%%%
\subsubsection{WIMP paradigm}
\label{sec:WIMPDM}
%%%%%%%%%%%%%%%%%%%%%%%%%%%%%%%%%%%%%%%%%%%%%%%%%%%%%%%%%%%%%%%%%%%%%%%%%%%%%

Consider a case where the DM candidate is initially in thermal equilibrium with the SM particle bath (visible sector), with no initial asymmetry between particles and anti-particles and with no sizable self-interactions, that undergoes thermal freeze-out. Its relic abundance will be produced via annihilation and inverse annihilation processes $\text{DM} \ \text{DM} \leftrightarrow \text{SM} \ \text{SM}$. This DM candidate is commonly referred in the literature as WIMP. The evolution of the WIMP DM number density is governed by the following Boltzmann equation~(BE)~\cite{Kolb:1990vq} (general aspects related to BEs are reviewed in Appendix~\ref{chpt:genBEs}):
\begin{equation}
    \frac{d Y}{d z} = - \frac{1}{z^2} \frac{\langle \sigma v \rangle S(m)}{H(m)} \left( Y^2 - Y_{\text{eq}}^2\right) \; , \; Y(0)=Y_{\text{eq}}(0) \; ,
    \label{eq:freezeout1}
\end{equation}
written in terms of the particle number density per comoving volume $Y \equiv n / S$, also known as yield, and $z \equiv m/T$, with $m$ being the mass of the DM candidate and $T$ the temperature. In the above equation, the particle interaction processes $\text{DM} \ \text{DM} \leftrightarrow \text{SM} \ \text{SM}$ are characterized by the total thermally average cross section $\langle \sigma v \rangle$. The Hubble parameter and entropy density are given in Eqs.~\eqref{eq:Hubble} and~\eqref{eq:ournotationDM}, respectively. The equilibrium value for the DM particle number density per comoving volume $Y_{\text{eq}}$ is given by [see Eq.~\eqref{eq:ournotationDM}]:
\begin{equation}
    Y_{\text{eq}}(z) = \frac{45}{4 \pi^4} \frac{g}{g_{\ast S}} z^2 \mathcal{K}_2(z) \xrightarrow[z \gg 3]{} \frac{45}{\sqrt{32 \pi^7}} \frac{g}{g_{\ast S}} z^{3/2} e^{-z} \; ,
    \label{eq:Yeq}
\end{equation}
where $g$ is the DM candidate d.o.f., e.g. for a real scalar $g=1$. The above expression shows the asymptotic limit $z \gg 3$ of $Y_{\text{eq}}$, useful for the case under consideration of cold relics, i.e. non-relativistic species.

    \begin{figure*}[t!]
        \centering
        \includegraphics[scale=0.7]{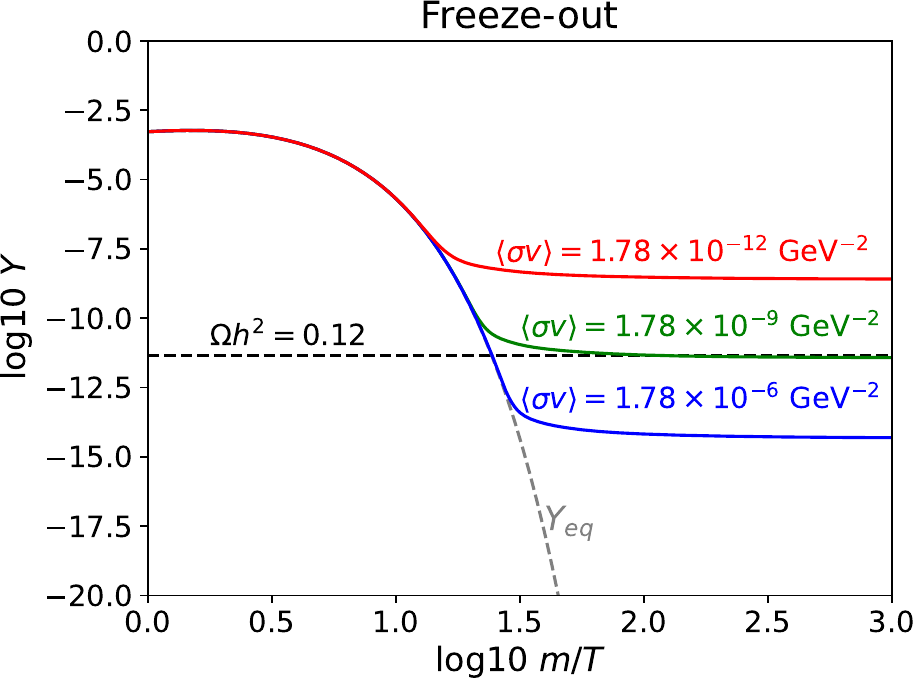}
        \caption{Evolution of a WIMP DM candidate yield $Y$ in terms of~$m/T$ (see text for details). We set $m=100$ GeV and present three distinct $\langle \sigma v\rangle$ cases via colored contours. The dashed black horizontal line indicates the value of $Y$ leading to $\Omega h^2 = 0.12$ [see Eq.~\eqref{eq:Oh2PlanckOG}] and the equilibrium yield $Y_{\text{eq}}$ [see Eq.~\eqref{eq:Yeq}] is shown via a dashed gray contour.}
    \label{fig:conventional}
    \end{figure*}
The solution of the BE~\eqref{eq:freezeout1} provides the DM particle candidate yield $Y(z)$, from which by taking the $z \rightarrow \infty$ limit the relic density value at present time is obtained,
\begin{equation}
    \Omega h^2 = \frac{S_0}{\rho_{\text{crit}}^0/h^2} m Y(z \rightarrow \infty) \; ,
    \label{eq:relic}
\end{equation}
with $S_0 = 2.89 \times 10^3$ cm$^{-3}$ being the entropy at present time, $\rho_{\text{crit}}^0 = 1.05 h^2 \times 10^{-5}$ GeV$/$cm$^{3}$ the present critical energy density~\cite{ParticleDataGroup:2024cfk}.

Analytically, one can obtain an approximate solution of Eq.~\eqref{eq:freezeout1} given by~\cite{Kolb:1990vq}
\begin{equation}
  \Omega h^2 \simeq \frac{S_0}{\rho_{\text{crit}}^0/h^2} \sqrt{\frac{45}{\pi}} \frac{g_\ast^{1/2}}{g_{\ast S}} \frac{z_f}{M_{\text{Pl}}\langle \sigma v \rangle} \simeq 0.12 \left( \frac{z_f}{25} \right) \left(\frac{1.78 \times 10^{-9} \ \text{GeV}^{-2}}{\langle \sigma v \rangle}\right) \; ,
\end{equation}
where $z_f \equiv m/T_f$ characterizes the time at which $Y$ ceases to track the equilibrium $Y_{\text{eq}}$. In Fig.~\ref{fig:conventional}, we present the numerical results for the WIMP DM yield $Y$, respectively, in terms of $m/T$, where we take $m=100$ GeV. We show three distinct solutions for values of $\langle \sigma v \rangle$, with the green contour matching the observed values for the CDM abundance obtained by Planck [see Eq.~\eqref{eq:Oh2PlanckOG}], which corroborate the analytical approximation. Freeze-out DM is initially in thermal equilibrium with the visible bath, however once its interactions $\text{DM} \ \text{DM} \leftrightarrow \text{SM} \ \text{SM}$ can no longer compete with the expansion of the Universe, the particle yield freezes to a constant value. We notice from the figure the inverse dependence of the relic abundance for WIMPs on the thermally average cross section, as derived in the analytical approximation. Furthermore, the value of $\langle \sigma v \rangle$ resulting in $\Omega h^2 = 0.12$ is around $\mathcal{O}(10^{-9})$ GeV$^{-2}$ or equivalently $\mathcal{O}(1)$ pb, a typical weak scale scattering cross section value. Considering the cross section to be $\sigma \sim g_{\text{DM}}^4/m_{\text{DM}}^2$, leads to couplings of the strength of the weak coupling for masses around the EW scale, i.e. between $100$ GeV and $10$ TeV, in order to reproduce the observed CDM abundance. This so-called ``WIMP miracle" makes it a testable paradigm at various experimental facilities which we briefly describe in what follows.

%%%%%%%%%%%%%%%%%%%%%%%%%%%%%%%%%%%%%%%%%%%%%%%%%%%%%%%%%%%%%%%%%%%%%%%%%%%%%
\subsubsection{Experimental searches for WIMPs}
\label{sec:DDexperiment}
%%%%%%%%%%%%%%%%%%%%%%%%%%%%%%%%%%%%%%%%%%%%%%%%%%%%%%%%%%%%%%%%%%%%%%%%%%%%%

The WIMP paradigm is being tested at three main categories of experimental setups:
\begin{itemize}

    \begin{figure}[!t]
    \centering
    \includegraphics[scale=0.55]{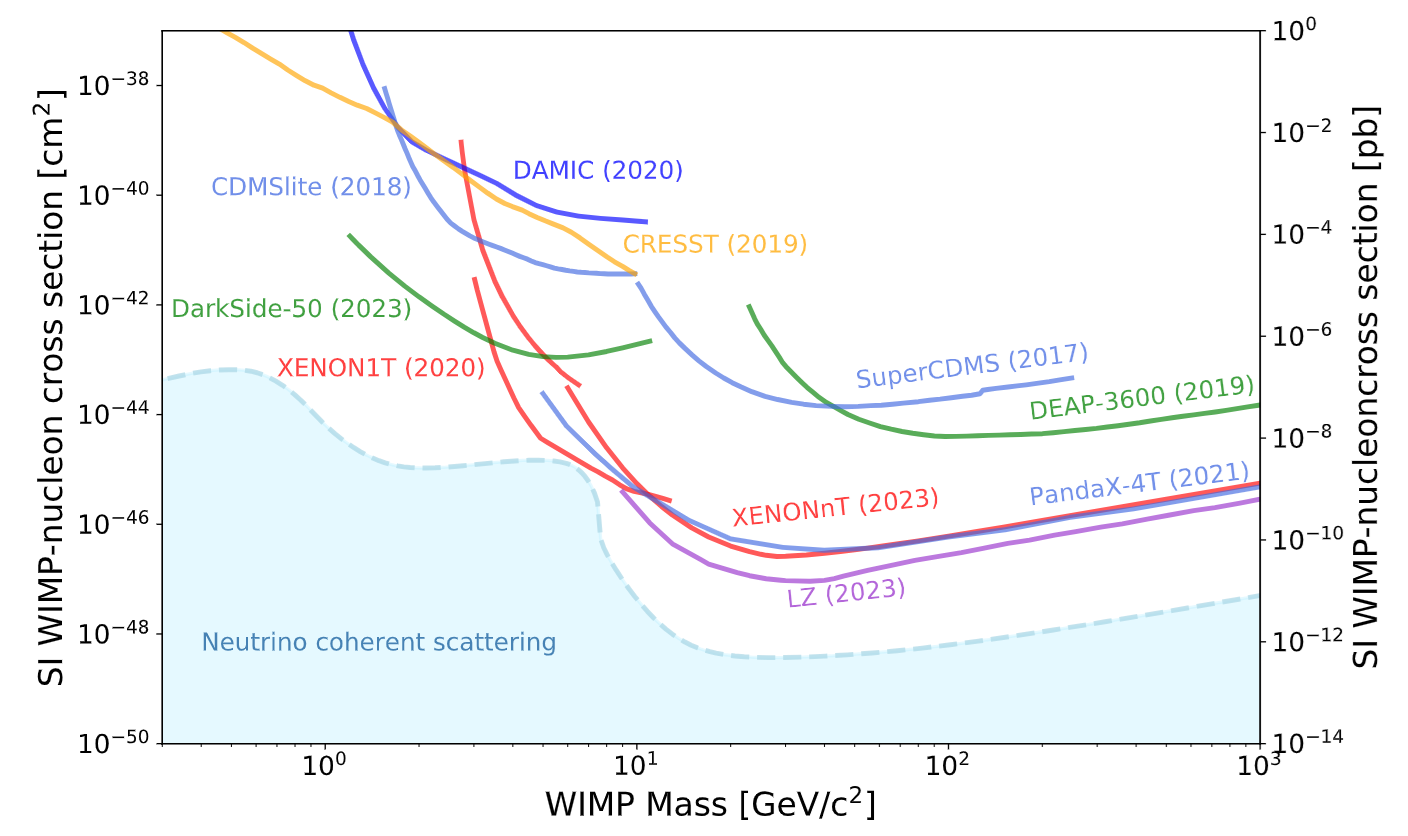}
    \caption{Landscape of current limits from various DD experiments on the SI WIMP-nucleon elastic scattering cross section versus the WIMP DM mass -- taken from Ref.~\cite{ParticleDataGroup:2024cfk}. The blue shaded region is the expected background set by CE$\nu$NS also know as ``neutrino floor"~\cite{Billard:2013qya}.}
    \label{fig:DMDDDsummary}
    \end{figure}
    \item \textbf{Direct detection (DD)} experiments aim to observe the scattering between DM and SM particles -- $\text{DM} + \text{SM} \rightarrow \text{DM} + \text{SM}$ -- via the measurement of nuclear recoil energy when WIMPs scatter off atomic nuclei in detectors on Earth (a very complete review on the subject can be found in Ref.~\cite{Billard:2021uyg}). These WIMPs are assumed to originate from the Milky Way’s DM halo and interact with detectors due to the Earth's motion through the galaxy~\cite{lewin1996review}. The detection rate depends on the local WIMP flux, mass and interaction cross section~\cite{jungman1996supersymmetric}. WIMP-nucleon cross sections are categorized as either spin-independent~(SI) or spin-dependent~(SD), depending on the nature of the coupling to nuclei. SI interactions benefit from coherent scattering, which enhances sensitivity for heavier elements like Xe and Ge~\cite{goodman1985detectability}. The SD interaction limits are weaker than SI due to the absence of coherent enhancement.
    
    No experiment has yet observed a definitive WIMP signal. Instead, upper limits on cross-sections as a function of WIMP mass have been set. In Fig.~\ref{fig:DMDDDsummary} we show the current limits from various DD experiments on the SI WIMP-nucleon elastic scattering cross section in terms of the WIMP DM mass~\cite{ParticleDataGroup:2024cfk}. As shown in the figure, the strongest constraints for masses between 10 and 1000 GeV stem from  XENON1T~\cite{XENON:2019rxp}, PandaX-4T~\cite{PandaX-4T:2022zrn} and LUX-ZEPLIN~(LZ)~\cite{LZ:2022lsv}. Note that, in 2024, PandaX-4T~\cite{PandaX:2024muv} and LZ~\cite{LZ:2024zvo} further improved their bounds with LZ currently holding the most stringent constraint. Future detectors like DarkSide-20k~\cite{DarkSide-20k:2017zyg}, ARGO~\cite{TGADM2018} and DARWIN \cite{DARWIN:2016hyl} are under development to further increase sensitivity. For lower masses $\lsim 10$  GeV, the best limits come from CRESST~\cite{CRESST:2015txj}, CDMSLite~\cite{CRESST:2015txj}, DarkSide-50~\cite{CRESST:2015txj}, and XENON1T~\cite{CRESST:2015txj}. Next-generation cryogenic experiments like SuperCDMS~\cite{SuperCDMS:2018gro}, SNOLAB~\cite{SuperCDMS:2016wui} and DarkSide-LowMass~\cite{CRESST:2015txj} aim to further probe this regime. Eventually, these searches will reach the ``neutrino floor", a background caused by coherent elastic neutrino-nucleus scattering~(CE$\nu$NS) from solar, atmospheric, and supernova~(SN) neutrinos~\cite{Billard:2013qya}. Although challenging, this is not a fundamental limit and may be overcome with improved detectors and analysis, and is recently often referred to as ``neutrino fog"~\cite{OHare:2021utq}. 
    
    \item \textbf{Indirect detection (ID)} aims at analyzing products resulting from DM annihilation into SM particles -- $\text{DM} + \text{DM} \rightarrow \text{SM} + \text{SM}$. WIMPs can be gravitationally captured by massive astrophysical bodies such as the Sun, Earth, or Galactic Center through elastic scattering with nucleons. This accumulation increases the annihilation probability and results in the emission of secondary particles such as $\gamma$-rays, neutrinos, and charged cosmic rays in the form of electrons, positrons, protons and antiprotons. The flux of these annihilation products provides a means to constrain the annihilation cross section $\langle \sigma v \rangle$.
    
    The Fermi Large Area Telescope~(Fermi-LAT) has placed stringent limits on $\langle \sigma v \rangle$ for masses up to~$\sim$1 TeV through analyses of $\gamma$-ray emissions from DM-dominated dwarf spheroidal galaxies~\cite{FermiLAT:2014sfa}. At higher masses, the strongest constraints come from the H.E.S.S. experiment, which has observed $\gamma$-rays from the Galactic Center and dwarf spheroidal galaxies, setting upper bounds on $\langle \sigma v \rangle$ down to $\sim 10^{-24}~\text{cm}^3/\text{s}$ depending on the annihilation channel and DM profile assumptions~\cite{HESS:2014rky}. These searches are crucial in probing the parameter space of heavy WIMP scenarios. In the future the Cherenkov Telescope Array~(CTA) is expected to enhance the sensitivity to DM annihilation signals over a broad mass range $\sim$10 GeV to $\sim 300$ TeV.

     \item \textbf{Collider searches} look for WIMPs in production channels -- $\text{SM} + \text{SM} \rightarrow \text{DM} + \text{DM}$. Since WIMPs are stable, neutral and non-baryonic, a positive signature would give rise to large missing transverse energy $\slashed{E}_T$. In fact, at the LHC, ATLAS and CMS are looking to identify such events, focusing on final states where DM is produced in association with a visible SM particle, resulting in so-called mono-\(X\) signatures, where \(X\) is a jet, photon, EW boson, heavy quark, or Higgs boson. These processes are generically denoted as \(pp \rightarrow \slashed{E}_T + X\)~\cite{Abercrombie:2015wmb}. ATLAS and CMS have set stringent bounds on DM-SM interaction cross sections using both effective field theory~(EFT) and simplified models, see e.g. Refs~\cite{Aaboud:2017phn, Sirunyan:2017jix}. 
     
     Additional constraints are derived from EW precision data. In particular, LEP measurements of the \(Z\) and \(W\)-boson decay widths exclude new light invisible particles~\cite{ALEPH:2005ab}. Furthermore LHC searches for invisible decays of the Higgs boson provide strong limits on Higgs-portal scenarios~\cite{ATLAS:2020cjb, CMS:2018yfx}. Ultimately, collider searches play a crucial role by probing regions of parameter space that are complementary to those explored by DD and ID experiments~\cite{Boveia:2018yeb}.

\end{itemize}
In this thesis we study WIMP DM in the context of radiative neutrino mass generation and the strong CP problem in Secs.~\ref{sec:darkLSS} and~\ref{sec:darkNB}, respectively. Thus, we will make use of the aforementioned DD bounds, precision and Higgs data, to constrain the parameter space of the models presented in those sections. Furthermore, in Chapter~\ref{chpt:axions} we will study another paradigmatic DM candidate: the axion.

%%%%%%%%%%%%%%%%%%%%%%%%%%%%%%%%%%%%%%%%%%%%%%%%%%%%%%%%%%%%%%%%%%%%%%%%%%%%%
\subsection{Matter-antimatter asymmetry}
\label{sec:BAU}
%%%%%%%%%%%%%%%%%%%%%%%%%%%%%%%%%%%%%%%%%%%%%%%%%%%%%%%%%%%%%%%%%%%%%%%%%%%%%

The observable Universe is overwhelmingly composed of matter. One of the most compelling evidence supporting this matter-antimatter asymmetry is derived from cosmic ray observations. Indeed, the flux of cosmic ray protons exceeds that of antiprotons by a factor of approximately $\sim 10^4$~\cite{Adriani:2008zq}. Furthermore, the non-detection of antinuclei in cosmic rays serves as a critical observational constraint, reinforcing the hypothesis that the Universe lacks significant regions of primordial antimatter~\cite{Beach:2001ub,Fuke:2005it,Abe:2012tz}. Additionally, the existence of large-scale antimatter domains is further disfavored by precise measurements of the CMB and the cosmic $\gamma$-ray background. Annihilation at the boundaries between matter and hypothetical antimatter regions would produce observable distortions in the CMB and contribute excess flux to the $\gamma$-ray background. However, current data suggests no such anomalies~\cite{Cohen:1997ac}. Analysis of the CMB radiation provides a robust method for probing matter-antimatter asymmetry, which can be quantified through the baryon-to-photon ratio,
\begin{align}
    \eta_B \equiv \frac{n_B - n_{\bar{B}}}{n_\gamma} \; ,
\end{align}
with $n_B$, $n_{\bar{B}}$ and $n_\gamma$ being, respectively, the number densities of baryons, antibaryons and photons. The present value for $\eta_B$ can be obtained from the $\Lambda$CDM fit to the power spectrum of the CMB anisotropies. From the combined Planck TT,TE,EE+ lowE+lensing data~\cite{Planck:2018vyg}:
\begin{align}
    \eta_B^0 = (6.12 \pm 0.04) \times 10^{-10} \; ,
    \label{eq:etab0}
\end{align}
at 68\% CL. Another fundamental approach to quantify the BAU involves comparing the observed abundances of light elements -- Deuterium, Helium-$3$ and $4$, and Lithium-$7$ -- which are highly sensitive to $\eta_B$, with the theoretical predictions of Big Bang Nucleosynthesis~(BBN)~\cite{Fields:2019pfx}. The BBN prediction yields $\eta_B^0 = (6.143 \pm 0.190) \times 10^{-10}$ at the $1\sigma$ level, which is in remarkable agreement with the CMB result, further validating the standard cosmological model.

From a theoretical perspective the mechanism responsible for generating the matter-antimatter asymmetry is still unknown. The observed BAU must have arisen dynamically from an initially symmetric state. This assertion is grounded in two main arguments. Most notably, cosmological inflation -- an epoch of exponential expansion shortly after the Big Bang (\(10^{-33}\) to \(10^{-32}\) s) -- would have erased any preexisting asymmetry, yielding a matter-antimatter symmetric Universe~\cite{Guth:1980zm,Linde:1981mu,Albrecht:1982wi}. Moreover, even if a small initial asymmetry had survived inflation, it would require extreme fine-tuning (a quark-antiquark ratio of \(6{,}000{,}001{:}6{,}000{,}000\)) to match the observed $\eta_B^0$ value~\cite{Kolb:1990vq}. Thus, a dynamical mechanism -- baryogenesis -- must have operated at the post-inflationary epoch to generate the baryon asymmetry from an initial symmetric state. In 1967, Sakharov showed that three conditions need to be fulfilled in order to successfully generate a baryonic asymmetry from a symmetric state~\cite{Sakharov:1967dj}. Namely, baryogenesis requires:
\begin{itemize}

    \item \textbf{i) $B$ violation:} If all interactions are $B$ conserving these cannot alter the initial symmetric state and therefore no baryon asymmetry can be generated;
    
    \item \textbf{ii) C and CP violation:} If $B$-violating interactions occur at the same rate as their C and CP-conjugates then no net asymmetry would be generated;
    
    \item \textbf{iii) Departure from thermal equilibrium:} Equilibrium conditions enforce an overall balance in the interaction rates, which would otherwise erase any generated asymmetry. To better understand this argument consider the process \( X \to Y + Z \), where $X$ and $Y$ have $B=0$ while $Z$ carries baryon number $B\neq0$. In thermal equilibrium, the forward and reverse rates are equal, \( \Gamma(X \to Y + Z) = \Gamma(Y + Z \to X) \), resulting in no net asymmetry. However, as the Universe cools and \( m_X \gg T \), the inverse reaction becomes Boltzmann-suppressed, \( \Gamma(Y + Z \to X) \sim e^{-m_X/T} \). This leads to a departure from equilibrium when the interactions cannot keep up with the expansion rate of the Universe, i.e. \( \Gamma \lsim H(T) \) -- the Hubble parameter is given by Eq.~\eqref{eq:Hubble}. Under these conditions, with C and CP violation present, the out-of-equilibrium decay $\Gamma(X \to Y + Z)$, dynamically generates a net baryon asymmetry.
    
\end{itemize}
The SM, despite satisfying all the Sakharov conditions, cannot successfully generate the observed BAU. As discussed in Sec.~\ref{sec:sphalerons}, although $B$ is conserved at the perturbative level, it is violated non-perturbatively via EW sphaleron transitions induced by chiral anomalies in the SU$(2)_L$ sector~\cite{Klinkhamer:1984di,Kuzmin:1985mm}. These processes violate $B+L$ while conserving $B-L$, and remain unsuppressed at temperatures above the EW scale ($T_{\text{EW}} \gtrsim 100$~GeV). Furthermore, as seen in Sec.~\ref{sec:fermionmassmix}, the SM weak interactions maximally violate C and CP violation arises in quark CC interactions due to the complex character of the CKM matrix. The resulting generated baryon asymmetry at $T_{\text{EW}}$ can be estimated as~\cite{Gavela:1994ts,Huet:1994jb}:
\begin{equation}
    \eta_B \lsim J_{\text{CP}} \frac{(m_t^2 - m_c^2) (m_t^2 - m_u^2) (m_c^2 - m_u^2) (m_b^2 - m_s^2) (m_b^2 - m_d^2) (m_s^2 - m_d^2) }{T_{\text{EW}}^{12}} \sim 10^{-20} \; ,
    \label{eq:etaBSMCPV}
\end{equation}
written in terms of the Jarlskog invariant $J_{\text{CP}}$ of Eq.~\eqref{eq:jarlskog} and products of squared quark mass differences (see Table~\ref{tab:quarkdata}). It is clear that the CP-violating effects present in the SM are too small to generate the observed value of $\eta_B$ given in Eq.~\eqref{eq:etab0}. Moreover, the EW phase transition in the SM is a smooth crossover for the physical Higgs mass $m_h \simeq 125$~GeV, and not strongly first-order, as required to achieve sufficient departure from thermal equilibrium~\cite{Kajantie:1996mn,Dine:1992wr}. In a first-order phase transition, expanding bubbles of broken phase create CP-asymmetric environments near their walls, which seed a chiral asymmetry later converted into baryon number via sphaleron processes~\cite{Cohen:1993nk,Shaposhnikov:1986jp}. However, for the asymmetry to be preserved, sphaleron processes must be suppressed within the bubbles, requiring $v(T)/T \gtrsim 1$~\cite{Shaposhnikov:1986jp}, with $T$ being the temperature at which bubble nucleation occurs, and the Higgs VEV in the broken phase is $v(T)$. This condition cannot be fulfilled for $m_h \gtrsim 80$~GeV~\cite{Kajantie:1996mn}. Hence any generated asymmetry is washed out. Consequently, the SM alone cannot explain the observed BAU, pointing towards the necessity of physics BSM.

A plethora of baryogenesis mechanisms have been proposed in extensions of the SM (for reviews see Refs.~\cite{Buchmuller:2005eh,Buchmuller:2004nz,Davidson:2008bu,Fong:2012buy,Hambye:2012fh,Bodeker:2020ghk}). In this thesis, in Sec.~\ref{sec:lepto}, we will explore baryogenesis via type-I seesaw leptogenesis~\cite{Fukugita:1986hr} in a scenario where CP is spontaneously broken by the VEV of a complex scalar singlet.

%%%%%%%%%%%%%%%%%%%%%%%%%%%%%%%%%%%%%%%%%%%%%%%%%%%%%%%%%%%%%%%%%%%%%%%%%%%%%
\subsection{Strong CP problem}
\label{sec:strongCPnedmsol}
%%%%%%%%%%%%%%%%%%%%%%%%%%%%%%%%%%%%%%%%%%%%%%%%%%%%%%%%%%%%%%%%%%%%%%%%%%%%%

As discussed in Sec.~\ref{sec:strongCP}, the QCD sector of the SM contains a P and CP-violating interaction $G \tilde{G}$, dubbed as the anomalous or topological QCD term, with parameter $\overline{\theta}$. The strong CP phase has physical repercussions on observables sensitive to CP-violating effects. Notably, $\overline{\theta}$ contributes to the neutron electric dipole moment (nEDM) $d_n$~\cite{Pospelov:2005pr}. The theoretical calculation for this observable has been made through various techniques such as chiral perturbation theory~\cite{Baluni:1978rf,Crewther:1979pi,Pich:1991fq}, QCD sum-rules~\cite{Pospelov:1999mv}, holography~\cite{Bartolini:2016jxq} and lattice QCD~\cite{Abramczyk:2017oxr,Dragos:2019oxn}. To illustrate how $d_n$ depends on $\overline{\theta}$, consider the chiral Lagrangian where the nucleons i.e., neutron and proton, form an isospin doublet $\Psi$ and couple to the pion $\pi$ through the terms:
\begin{equation}
\mathcal{L}_{\pi n n} = \pi^a \overline{\Psi} \left( i \gamma^5 g_{\pi n n} + \bar{g}_{\pi n n} \right) \tau_a \Psi \; , 
\end{equation}
with $\tau_a$ being the Pauli matrices, $g_{\pi n n}$ the Yukawa coupling representing the strong interaction among nucleons and $\bar{g}_{\pi n n}$ is proportional to $\overline{\theta}$ since it describes a CP-violating term~\cite{Crewther:1979pi}. The pion loop contribution to the nEDM $d_n$ yields
\begin{equation}
    d_n = \frac{m_n}{4 \pi^2} g_{\pi n n} \bar{g}_{\pi n n} \ln\left(\frac{m_n}{m_\pi}\right) \simeq (5 \times 10^{-16} \  e \cdot \text{cm}) \ \overline{\theta} \; ,
    \label{eq:nEDM}
\end{equation}
where $m_n$ and $m_\pi$ are the neutron and pion mass, respectively. Current experimental searches did not observe the nEDM, setting a very stringent bound of $|d_n| < 3 \times 10^{-26} \; e \cdot \text{cm}$~\cite{PhysRevLett.97.131801,Pendlebury:2015lrz}, implying
\begin{equation}
    \left|\overline{\theta}\right| \lsim 10^{-10} \; .
    \label{eq:thetabound}
\end{equation}
Future searches are expected to improve the sensitivity on this observable by up to two orders of magnitude~\cite{Filippone:2018vxf}. Hence, QCD does not seem to violate CP. The strong CP problem is the lack of theoretical explanation for the smallness of $\overline{\theta}$.

\subsubsection{Solutions to the strong CP problem}

There are three possible solutions to the strong CP problem, which we briefly enumerate:
\begin{itemize}

    \item \textbf{Massless up-quark:} This is the simplest solution~\cite{tHooft:1976rip}. For~$m_u = 0$ we obtain~$\theta_F = 0$ [see Eq.~\eqref{eq:thetaF}]. The chiral rotation~$\text{U}(1)_{\text{A}}^u : u \rightarrow e^{i \beta_u \gamma^5}$ of the up-quark field is classically conserved since $u$ is massless. However, this symmetry is explicitly broken at the quantum level, altering the path integral measure, such that $ \theta \rightarrow \theta + 2 \beta_u$  [see Eq.~\eqref{eq:shift}]. Thus, $\overline{\theta}$ is removed by setting $\theta = - 2 \beta_u$.

    This proposal does not require additional particle content. Although it was shown that an up-quark mass could be generated through higher-order contributions from the chiral Lagrangian~\cite{PhysRevLett.56.2004}, the value does not match the current measurement $m_u = 2.16^{+0.49}_{-0.26}$ MeV~\cite{Zyla:2020zbs} (see Table~\ref{tab:quarkdata}). Furthermore, lattice calculations rule out a massless quark~\cite{Nelson:2001bhq}. Consequently, this solution is phenomenologically non-viable.
    
    \item \textbf{Nelson-Barr mechanism:} Assuming that P or CP are symmetries of an UV theory sets $\overline{\theta}=0$. This class of solutions are commonly referred as Nelson and Barr~(NB) models~\cite{Nelson:1983zb,Barr:1984qx,Barr:1984fh}. Usually, these frameworks require some parameter tuning and/or involved model building. This is due to the fact that these symmetries need to be subsequently spontaneously broken. Namely, P breaking is needed to account for the SM chiral structure, while CP breaking is required in order to generate weak CP violation encoded in the CKM quark mixing matrix (see Sec.~\ref{sec:fermionmassmix}). Furthermore, the parameter $\overline{\theta}$ will receive non-zero loop corrections at a certain order in perturbation theory known as threshold corrections, becoming therefore a calculable parameter. Requiring that the strong CP phase respects Eq.~\eqref{eq:thetabound}, leads to some tuning of the UV theory couplings and particle masses.

    In Chapter~\ref{chpt:SCPV}, we will study how SCPV from the complex VEV of a scalar singlet can address the strong CP problem in calculable $\overline{\theta}$ scenarios. We will start, in Chapter~\ref{sec:darkNB}, by reviewing the minimal NB realization known as the Bento-Branco-Parada~(BBP) model~\cite{Bento:1991ez}, and subsequently connect this idea to a dark sector.

    \item \textbf{Peccei-Quinn mechanism:} An extensively explored and conceptually elegant approach involves the Peccei–Quinn~(PQ) mechanism~\cite{Peccei:1977hh,Peccei:1977ur}, which introduces a global, QCD-anomalous $\text{U}(1)_{\text{PQ}}$ symmetry. Its spontaneous breaking gives rise to a pseudo-GB -- the axion~\cite{Weinberg:1977ma,Wilczek:1977pj} -- which acquires mass through non-perturbative QCD effects. The ground state of the axion potential is such that it dynamically drives $\bar{\theta} \to 0$, thereby solving the strong CP problem.

    Chapter~\ref{chpt:axions} is entirely dedicated to the axion solution to the strong CP problem, where are discussed the properties of axions, paradigmatic UV complete models and how experimental facilities search for this particle. In Chapters~\ref{chpt:axionneutrino} and~\ref{chpt:flavoraxion}, we will connect axions to neutrinos building unified frameworks that simultaneously address multiple BSM problems.
    
\end{itemize}
%

%------------
% CHAPTER 02    
%------------

%%%%%%%%%%%%%%%%%%%%%%%%%%%%%%%%%%%%%%%%%%%%%%%%%%%%%%%%%%%%%%%%%%%%%%%%%%%%%
\chapter{Majorana neutrino mass generation and dark sectors} 
\label{chpt:neutrinodarksectors}
%%%%%%%%%%%%%%%%%%%%%%%%%%%%%%%%%%%%%%%%%%%%%%%%%%%%%%%%%%%%%%%%%%%%%%%%%%%%%

The discovery of neutrino flavor oscillations requires the existence of neutrino masses and leptonic mixing, thus providing evidence for physics BSM. Determining the mechanism behind neutrino mass generation is one of the key challenges in particle physics, particularly as oscillation parameters are being measured with increasing precision. As shown in Sec.~\ref{sec:neutrino}, despite recent advancements in neutrino experiments, several questions remain unanswered: Are neutrinos Dirac or Majorana particles? What is the absolute scale of neutrino mass? What is the ordering of the neutrino mass spectrum? Is there LCPV? From a theory viewpoint, looking at the SM as an effective theory the Weinberg operator points at Majorana neutrino mass generation, which is the object of study in this chapter. Among the plethora of UV complete realizations, at tree-level, the most popular ones rely on the seesaw mechanism which we discuss in its type-I, II and III incarnations. Furthermore, low-scale -- inverse and linear -- variants, rely on small LNV parameters to trigger neutrino mass generation, offering more testability than the canonical seesaws. Interestingly, the connection between a dark sector, containing viable DM candidates, and neutrino masses has been explored in radiative realizations of the effective Weinberg operator. Among all the possible one-loop models standout the scotogenic scenario. Having motivated these popular Majorana neutrino mass models the end of the chapter is dedicated to our work of Ref.~\cite{Batra:2023bqj} where a novel dark-sector seeded neutrino mass mechanism is presented: dark linear seesaw.

%%%%%%%%%%%%%%%%%%%%%%%%%%%%%%%%%%%%%%%%%%%%%%%%%%%%%%%%%%%%%%%%%%%%%%%%%%%%%
\section{Seesaw mechanisms}
\label{sec:seesaw}
%%%%%%%%%%%%%%%%%%%%%%%%%%%%%%%%%%%%%%%%%%%%%%%%%%%%%%%%%%%%%%%%%%%%%%%%%%%%%

\begin{figure}[t!]
        \centering
        \includegraphics[scale=0.7]{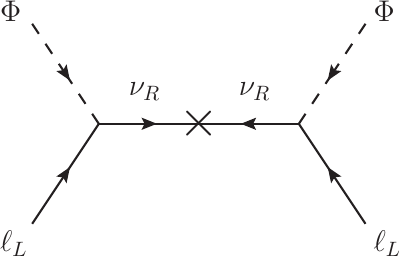} \hspace{+0.1cm} \includegraphics[scale=0.7]{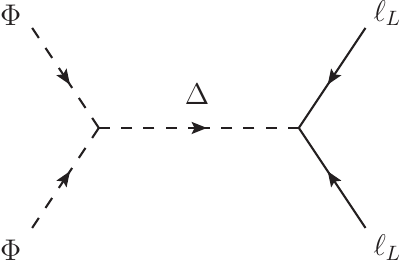} \hspace{+0.1cm} \includegraphics[scale=0.7]{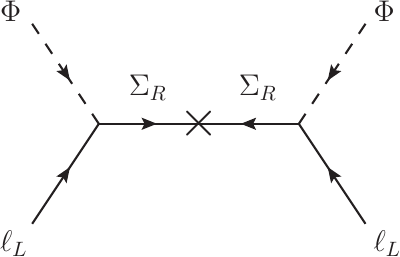}
        \caption{From left to right: diagramatic representation of the type I, II and III seesaw interactions, mediated by a singlet fermion $\nu_R$, scalar triplet $\Delta$ and fermionic triplet $\Sigma_{R}$, respectively.}
    \label{fig:seesaw}
    \end{figure}

The idea of the seesaw mechanism offers a natural and elegant way of explaining the origin and smallness of neutrino masses. These scenarios are UV complete realizations of the Weinberg operator that generate light neutrino masses with a tree-level exchange of a heavy field with mass scale $\Lambda \gg v$ [see Eq.~\eqref{eq:weinbergop}]. The smallness of neutrino masses is explained by the heaviness of the extra fields. The nature of these neutrino mass mediators depends on the type of renormalizable interactions added to the SM Lagrangian. Namely, the possible non-zero invariant vertices that can be constructed with the SU(2)$_L$ doublets $\ell_L$ and $\Phi$ in the Weinberg operator are:
\begin{align}
    (\ell_L \Phi)_{\mathbf{1}} (\ell_L \Phi)_{\mathbf{1}} \; , \; (\ell_L \ell_L)_{\mathbf{3}} (\Phi \Phi)_{\mathbf{3}} \; , \; (\ell_L \Phi)_{\mathbf{3}} (\ell_L \Phi)_{\mathbf{3}} \; , \;
\end{align}
where the subscript indicates the SU(2) representation of the group product of doublets inside the parenthesis. Lorentz and gauge invariance under $G_{\text{SM}}$ require that the field mediating the interactions above be a fermion singlet $\nu_R \sim (\mathbf{1}, \mathbf{1}, 0)$, scalar triplet $\Delta \sim (\mathbf{1}, \mathbf{3}, 2)$ and fermionic triplet $\Sigma_R \sim (\mathbf{1}, \mathbf{3}, 0)$, respectively. These models correspond to the canonical type-I, II and III seesaw mechanisms, diagrammatically depicted in Fig.~\ref{fig:seesaw}.

%
%%%%%%%%%%%%%%%%%%%%%%%%%%%%%%%%%%%%%%%%%%%%%%%%%%%%%%%%%%%%%%%%%%%%%%%%%%%%%
\subsection{Type-I seesaw}
\label{sec:TypeI}
%%%%%%%%%%%%%%%%%%%%%%%%%%%%%%%%%%%%%%%%%%%%%%%%%%%%%%%%%%%%%%%%%%%%%%%%%%%%%

The simplest and most common framework to generate small neutrino masses and lepton mixing is the type-I seesaw mechanism~\cite{Minkowski:1977sc,Yanagida:1979as,GellMann:1980vs,Glashow:1979nm,Mohapatra:1979ia}. In this case, consider $n_R$ RH neutrinos $\nu_R$ are added to the SM leading to the most general Lagrangian invariant under $G_{\text{SM}}$ given by
\begin{equation}
    \mathcal{L}_{\text{I}} = \mathcal{L}_{\text{SM}} + i \overline{\nu_R} \slashed{\partial} \nu_R - \left( \overline{\ell_L} \tilde{\Phi} \mathbf{Y}_{D}^* \nu_R + \frac{1}{2} \overline{\nu_R^c} \mathbf{M}_{R} \nu_R + \text{H.c.} \right) \; ,
\label{eq:TypeILag}
\end{equation}
where the $n_R \times n_R$ mass matrix $\mathbf{M}_R$ is of Majorana type and the complex $3 \times n_R$ matrix $\mathbf{Y}_D$ is of Dirac~type.
After EWSB the neutrino mass Lagrangian reads
\begin{equation}
    -\mathcal{L}_{\text{I}} \supset \overline{\nu_L} \mathbf{M}_{D}^{*} \nu_R + \frac{1}{2} \overline{\nu_R^c} \mathbf{M}_{R} \nu_R + \text{H.c.} \; , \; \mathbf{M}_{D} = \frac{v \mathbf{Y}_{D}}{\sqrt{2}} \; .
\end{equation}
Organizing the neutrino fields in the vector $N_{L}=\left(\nu_{L}, \nu^c_{R} \right)^{T}$ of dimension $n_f = 3+n_R$, with $\nu_{L}=\left(\nu_{e L}, \nu_{\mu L}, \nu_{\tau L} \right)^T,\, \nu_{R}=\left(\nu_{R 1}, ... \ , \nu_{R n_R} \right)^T$, leads to the simplified form
\begin{equation}
-\mathcal{L}_{\text{I}} \supset \frac12\,\overline{N_{L}^c}\,\bm{\mathcal{M}}_\nu \,N_{L}+\text{H.c.} \; , \; \bm{\mathcal{M}}_\nu= \begin{pmatrix}
0 & \mathbf{M}_D  \\
\mathbf{M}_D^{T} & \mathbf{M}_{R}
\end{pmatrix} \; ,
\label{eq:TypeILagmass}
\end{equation}
where $\bm{\mathcal{M}}_\nu$ is the full $n_f \times n_f$ Majorana-type mass matrix being diagonalized by the  $n_f\times n_f$ unitary transformation,
\begin{align}
N_L = \bm{\mathcal{U}}_\nu \, (\nu_1, \dots \ , \nu_{n_f})^T_L \; ,
\label{eq:fullUnu}
\end{align}
relating the weak-basis states $N_{L}$ to the mass eigenstates~$(\nu_1,...,\nu_{n_f})^T$,
such that
\begin{align}
\bm{\mathcal{U}}_\nu^{T} \bm{\mathcal{M}}_\nu \ \bm{\mathcal{U}}_\nu = \bm{\mathcal{D}}_{\nu} = \text{diag}\,(m_1, \dots \ , m_{n_f})\,,
\label{diagnutr}
\end{align}
with the unitary rotation being obtained by diagonalizing the Hermitian matrix $\bm{\mathcal{H}}_\nu^\prime= \bm{\mathcal{M}}_\nu^\dagger \bm{\mathcal{M}}_\nu$ and  $m_{1,\dots,n_f}$ denote the $n_f$ (real and positive) neutrino masses. Notice that, in general, the light-active (heavy-sterile) neutrino masses are labeled as $m_{1,2,3}$ ($m_{4,...,n_f}$).

The effective light neutrino mass matrix can be derived via the diagonalization and the heavy-state integration methods.
\subsubsection{Diagonalization method}

Assuming the mass hierarchy $M_D \ll M_{R}$ for the mass matrix entries, the full neutrino mass matrix can be block-diagonalized through the unitary matrix $\bm{\mathcal{W}}_\nu$, such that~\cite{Schechter:1980gr,Grimus:2000vj}
\begin{equation}
\bm{\mathcal{W}}_\nu = \begin{pmatrix} 
 \sqrt{\mathbf{1}-\mathbf{\Theta}_\nu \mathbf{\Theta}_\nu^{\dagger}} &  \mathbf{\Theta}_\nu \\
 - \mathbf{\Theta}_\nu^{\dagger} & \sqrt{\mathbf{1}-\mathbf{\Theta}_\nu^{\dagger} \mathbf{\Theta}_\nu}
\end{pmatrix} \Rightarrow
\bm{\mathcal{W}}_\nu^T \bm{\mathcal{M}}_\nu \ \bm{\mathcal{W}}_\nu =
\begin{pmatrix} \mathbf{M}_{\nu} & 0 \\
 0 & \mathbf{M}_{\text{h}}
\end{pmatrix} \; ,
\label{eq:UB}
\end{equation}
where $\mathbf{\Theta}_\nu$ is a $3 \times n_R$ complex matrix, $\mathbf{M}_\nu$ is the $3 \times 3$ effective light neutrino mass matrix and $\mathbf{M}_{\text{h}}$ is the $n_R \times n_R$ heavy neutrino mass matrix. The following system of matrix equations is given at first order in $\mathbf{\Theta}_\nu \sim \mathcal{O}\left(M_R^{-1}\right)$:
\begin{align}
\mathbf{M}_\nu & \simeq - \mathbf{\Theta}_\nu^* \mathbf{M}_{R}  \mathbf{\Theta}_\nu^{\dagger} - \mathbf{\Theta}_\nu^* \mathbf{M}_{D}^T + \mathbf{M}_{D} \mathbf{\Theta}_\nu^{\dagger} \; , \nonumber \\
0 & \simeq  \mathbf{M}_{D}
- \mathbf{\Theta}_\nu^* \mathbf{M}_{R} - \mathbf{\Theta}_\nu^* \mathbf{M}_{D}^T  \mathbf{\Theta}_\nu \; , \nonumber \\
0 & \simeq \mathbf{M}_{D}^T - \mathbf{M}_{R} \mathbf{\Theta}_\nu^{\dagger} - \mathbf{\Theta}_\nu^T \mathbf{M}_{D} \mathbf{\Theta}_\nu^{\dagger} \; , \nonumber \\ \mathbf{M}_{\text{h}} & \simeq  \mathbf{M}_{R} + \mathbf{\Theta}_\nu^T \mathbf{M}_{D} + \mathbf{M}_{D}^T \mathbf{\Theta}_\nu \; , 
\label{eq:TypeIF}
\end{align}
with the expansion parameter being
\begin{equation}
\mathbf{\Theta}_\nu \simeq \mathbf{M}_{D}^* (\mathbf{M}_{R}^*)^{-1} \; .
\label{eq:thetanuI}
\end{equation}
From the above the light and heavy neutrino mass matrices are obtained as,
\begin{equation}
\mathbf{M}_{\nu} = - \mathbf{\Theta}_\nu^* \mathbf{M}_{R}  \mathbf{\Theta}_\nu^{\dagger} = - \mathbf{M}_D \mathbf{M}_{R}^{-1} \mathbf{M}_D^T \; , \; \mathbf{M}_{\text{h}} = \mathbf{M}_{R} \; , 
\label{eq:TypeIMeff}
\end{equation}
at first order in the seesaw approximation. Notice that the effective light neutrino mass formula is inversely proportional to the mass of the heavy neutrinos. Therefore, for natural Dirac Yukawa couplings of the order of unity, a mass scale~$M_{R} \sim 10^{14}~\text{GeV}$ is required to explain the smallness of neutrino masses $m_{\nu} \sim 0.1~\text{eV}$.

The resulting mass matrices in Eq.~\eqref{eq:TypeIMeff} are diagonalized through the following unitary matrices,
\begin{align}
&\nu_{L}  \rightarrow \mathbf{U}_{\nu}\, \nu_{L} \Rightarrow \mathbf{U}_{\nu}^{ T}\, \mathbf{M}_{\nu}\, \mathbf{U}_{\nu}  = \mathbf{D}_{\nu} = \text{diag}\left(m_1, \dots , m_{3}\right) \; , \label{eq:TypeIMeffdiag} \\
&\nu_{R}^c  \rightarrow \mathbf{U}_{\text{h}}\, \nu_{R}^c \Rightarrow \mathbf{U}_{\text{h}}^{ T}\, \mathbf{M}_{\text{h}}\, \mathbf{U}_{\text{h}}  = \mathbf{D}_{\text{h}} = \text{diag}\left(M_1, \dots , M_{n_R}\right) \; ,
\label{eq:TypeIMheavy}
\end{align}
where $m_{i}$ ($i=1, \cdots, 3$) are the real and positive light neutrino masses, while $M_j$ ($j=1,\cdots,n_R$) are the heavy neutrino masses, in the seesaw approximation. Therefore, the full unitary matrix $ \bm{\mathcal{U}}_\nu$ of Eq.~\eqref{eq:fullUnu} is expressed in the seesaw approximation by
\begin{equation}
\bm{\mathcal{U}}_\nu = \begin{pmatrix} 
 \sqrt{\mathbf{1}-\mathbf{\Theta}_\nu \mathbf{\Theta}_\nu^{\dagger}} &  \mathbf{\Theta}_\nu \\
 - \mathbf{\Theta}_\nu^{\dagger} & \sqrt{\mathbf{1}-\mathbf{\Theta}_\nu^{\dagger} \mathbf{\Theta}_\nu}
\end{pmatrix}
\begin{pmatrix} 
\mathbf{U}_{\nu} &  0 \\
 0 & \mathbf{U}_{\text{h}}
\end{pmatrix} \; .
\label{eq:UparamI}
\end{equation}

The rectangular $3 \times n_f$ matrix $(\mathbf{W}_\nu)_{\alpha j}\equiv (\bm{\mathcal{U}}_\nu)_{\alpha j}$ ($\alpha=e,\mu,\tau$, $j=1,\dots,n_f$), can be decomposed in the form
\begin{equation}
\mathbf{W}_\nu = \left(\sqrt{\mathbb{1}-\mathbf{\Theta}_\nu \mathbf{\Theta}_\nu^{\dagger}} \,\U_\nu\,,\, \mathbf{\Theta}_\nu \U_\text{h}\right) \; .
\label{eq:WnucLFV}
\end{equation}
From the above it is clear that, due to the additional sterile neutrino states $\nu_R$, the lepton mixing matrix is no longer unitary~\cite{FernandezMartinez:2007ms}:
\begin{align}
&\mathbf{U}^\prime = \mathbf{U}_{L}^{e \dagger} \ \sqrt{\mathbf{1}-\mathbf{\Theta}_\nu \mathbf{\Theta}_\nu^{\dagger}} \ \mathbf{U}_{\nu} = \left(\mathbb{1} - \boldsymbol{\eta} \right) \mathbf{U} \; , \label{eq:unitpmnsI}
\end{align}
where $\mathbf{U}$ is the unitary lepton mixing matrix given in Eq.~\eqref{eq:CCleptoU} and the Hermitian matrix $\boldsymbol{\eta}$ encodes deviations from unitarity. In the seesaw approximation, expanding the square root in Eq.~\eqref{eq:unitpmnsI}, up to second order in $\mathbf{\Theta}_\nu$, i.e., up to $\mathcal{O}\left(M_R^{-2}\right)$, one obtains
\begin{equation}
\bm{\eta} = \frac{1}{2} \mathbf{U}_{L}^{e \dagger} \mathbf{\Theta}_\nu \mathbf{\Theta}_\nu^{\dagger} \mathbf{U}_{L}^{e} \simeq \frac{1}{2} \mathbf{U}_{L}^{e \dagger} \mathbf{M}_D^* (\mathbf{M}_R^{*})^{-1}  \mathbf{M}_R^{-1} \mathbf{M}_D^{T} \mathbf{U}_{L}^{e} \; .
\label{eq:epsilondevuniI}
\end{equation}
Deviations from unitarity of the lepton mixing matrix affect multiple EW precision observables and flavor processes, which are able to experimentally constrain the entries of $\bm{\eta}$ with current experiments setting upper bounds of the order of $10^{-5} - 10^{-3}$~\cite{Antusch:2006vwa,Fernandez-Martinez:2016lgt}. In the type-I seesaw, which features heavy mediators with masses around the Grand Unified Theory~(GUT) scale, deviations from unitarity are negligible.

Furthermore, since the mass spectrum is composed of light and heavy neutrinos, it becomes clear from Eq.~\eqref{eq:UparamI} that, in the basis where $\mathbf{M}_{e}$ is diagonal, the heavy-light mixing will be given by the $3 \times n_R$ matrix,
\begin{equation} \mathbf{U}_{L}^{e \dagger} \mathbf{\Theta}_\nu \mathbf{U}_{\text{h}} \simeq \mathbf{U}_{L}^{e \dagger} \mathbf{M}_D^{*} (\mathbf{M}_R^{*})^{-1} \mathbf{U}_{\text{h}} \; ,
\label{eq:HeavyLightI}
\end{equation}
at first order in $\mathbf{\Theta}_\nu$.

\subsubsection{Heavy-state integration method}
Using the approach described in~Eqs.~\eqref{eq:effS}–\eqref{eq:effL}, the heavy states can be integrated out to yield the effective neutrino mass matrix in Eq.~\eqref{eq:TypeIMeff}. Without loss of generality, $\mathbf{M}_R$ is assumed to be real and diagonal. The heavy fields are defined as $N = \nu_R + \nu_R^c$, and~\eqref{eq:TypeILag} is rewritten as follows:
\begin{equation}
\mathcal{L}_{\rm I} = \mathcal{L}_{\text{SM}} + \frac{1}{2} \left[\overline{N} \left(i \slashed{\partial} - \mathbf{M}_R \right) N - \left( \overline{N}  \mathbf{Y}_{D}^T  \tilde{\Phi}^\dagger \ell_L +  \overline{N} \mathbf{Y}_{D}^\dagger \tilde{\Phi}^T \ell_L^c  + \text{H.c.} \right) \right] \; .
\label{eq:LagIHM}
\end{equation}
From Eq~\eqref{eq:EOMEL} one obtains the EOM for the stationary fields $N_0$,
\begin{align}
\left(i \slashed{\partial} - \mathbf{M}_R \right) N_0 =   \left(  \mathbf{Y}_{D}^T  \tilde{\Phi}^\dagger \ell_L +  \mathbf{Y}_{D}^\dagger \tilde{\Phi}^T \ell_L^c  \right) \; .
\end{align}
Reinserting the solution back into~\eqref{eq:LagIHM} leads to
\begin{equation}
\mathcal{L}_{\rm I} = \mathcal{L}_{\text{SM}} + \mathcal{L}_{N}\left(N_0 \right) = \mathcal{L}_{\text{SM}} - \frac{1}{2} \left(\overline{\ell_L} \tilde{\Phi} \mathbf{Y}_{D}^* + \overline{\ell_L^c} \tilde{\Phi}^* \mathbf{Y}_{D} \right) N_0 \; .
\end{equation}
The higher-order effective operators are obtained by expanding the propagator of the heavy fields:
\begin{equation}
\left(i \slashed{\partial} - \mathbf{M}_R \right)^{-1} = - \mathbf{M}_R^{-1} - i \slashed{\partial} \ \mathbf{M}_R^{-2} + \cdots \; ,
\label{eq:propexp}   
\end{equation}
where the first-order term yields the dimension-five operator and the second-order one will provide dimension-six operators. For our purposes it is enough to limit the analysis up to the dimension-six operators. 

Using the expansion above, the coefficient of the Weinberg operator will produce the effective neutrino mass matrix as in Eq.~\eqref{eq:TypeIMeff}. Namely
\begin{equation}
c^{d=5} = \mathbf{Y}_D \mathbf{M}_R^{-1} \mathbf{Y}_D^T \xrightarrow{\text{EWSB}} \mathbf{M}_{\nu} = - \frac{v^2}{2} c^{d=5} = - \mathbf{M}_D \mathbf{M}_{R}^{-1} \mathbf{M}_D^T \; .
\end{equation}

For this particular UV completion -- type-I seesaw -- there is a single dimension-six operator obtained by using the expansion in Eq.~\eqref{eq:propexp}:
\begin{equation}
\mathcal{L}_{d=6} = c_{\alpha \beta}^{d=6}  \left(\overline{\ell_{\alpha L}} \tilde{\Phi} \right) i \slashed{\partial}  \left( \tilde{\Phi}^{\dagger} \ell_{\beta L} \right) \; , \; c^{d=6} = \mathbf{Y}_D^* \mathbf{M}_R^{-2} \mathbf{Y}_D^{T} \; .
\end{equation}
After EWSB, the dimension-six operator essentially modifies the LH neutrinos kinetic term where $c^{d=6}$ encodes deviations from unitarity as in Eq.~\eqref{eq:epsilondevuniI},
\begin{equation}
\mathcal{L}_{d=6} = i \frac{v^2}{2} c_{\alpha \beta}^{d=6} \overline{\nu_{\alpha L}}  \slashed{\partial} \nu_{\beta L} \rightarrow \bm{\eta} = \frac{v^2}{4} \mathbf{U}_{L}^{e \dagger} c^{d=6} \mathbf{U}_{L}^{e} \; .
\label{eq:TypeIL6}
\end{equation}
An important feature of the type-I seesaw is the fact that both $c^{d=5}$ and $c^{d=6}$ depend quadratically on the Dirac-type mass, and that $c^{d=5}$ is suppressed by $\mathcal{O}\left(M_R^{-1}\right)$, while $c^{d=6}$ is further suppressed by $\mathcal{O}\left(M_R^{-2}\right)$.

%%%%%%%%%%%%%%%%%%%%%%%%%%%%%%%%%%%%%%%%%%%%%%%%%%%%%%%%%%%%%%%%%%%%%%%%%%%%%
\subsection{Type-II seesaw}
\label{sec:TypeII}
%%%%%%%%%%%%%%%%%%%%%%%%%%%%%%%%%%%%%%%%%%%%%%%%%%%%%%%%%%%%%%%%%%%%%%%%%%%%%

The type-II seesaw mechanism~\cite{Konetschny:1977bn,Schechter:1980gr,Cheng:1980qt,Lazarides:1980nt,Mohapatra:1980yp,Schechter:1981bd} extends the SM particle content with a scalar triplet $\Vec{\Delta} = \left(\Delta_1 ,\Delta_2 ,\Delta_3\right)^T$ with $Y=+2$. The triplet is in the adjoint representation of $\text{SU}(2)_L$ with generators
\begin{equation}
T_1 = \begin{pmatrix} 0& 0&0 \\
0& 0 & -i\\
0&i & 0 \end{pmatrix} \; , \; T_2 = \begin{pmatrix} 0& 0&i \\
0& 0 & 0\\
-i&0 & 0 \end{pmatrix} \; , \; T_3 = \begin{pmatrix} 0& -i&0 \\
i& 0 & 0\\
0&0 & 0 \end{pmatrix} \; .
\end{equation}
To work with the charge eigenfields, the charge operator \(Q\) must be diagonalized by 
choosing a basis in which the generator \(T_3\) is diagonal. Namely, via the similarity transformation~$T_a^\prime = \mathbf{K} T_a  \mathbf{K}^\dagger$ ($a=1,2,3$), the matrix $\mathbf{K}$ relates the flavor fields to the charge eigenstate ones as
\begin{equation}
\mathbf{K} = \frac{1}{\sqrt{2}} \begin{pmatrix}
 - 1& i &0 \\
0 & 0& \sqrt{2}\\
 1& i& 0
\end{pmatrix} \rightarrow \mathbf{K} \Vec{\Delta} = \begin{pmatrix}
 \frac{- \Delta_1 + i \Delta_2}{\sqrt{2}}\\
 \Delta_3 \\
 \frac{\Delta_1 + i \Delta_2}{\sqrt{2}}
\end{pmatrix} = \begin{pmatrix}
 - \Delta^{++}\\
 \Delta^{+}\\
 \Delta^0
\end{pmatrix} \; ,
\end{equation}
where the charge of the field combinations above was determined by applying the diagonal charge operator to $\mathbf{K} \Vec{\Delta}$. Hence, the fields above can be written in the matrix representation
\begin{equation}
\Delta \equiv i \tau_2 \left(\Vec{\tau} \cdot \Vec{\Delta} \right)  = \begin{pmatrix}
 \Delta_1 + i \Delta_2 & - \Delta_3\\
- \Delta_3 & -\Delta_1 + i \Delta_2
\end{pmatrix} = \begin{pmatrix}
 \sqrt{2} \Delta^{0} &  -\Delta^{+}\\
 -\Delta^{+} & - \sqrt{2} \Delta^{++}
\end{pmatrix}
\label{eq:2times2repforD} \; .
\end{equation}
The type-II seesaw Lagrangian is given by
\begin{equation}
\mathcal{L}_{\text{II}} = \mathcal{L}_{\text{SM}} - \left[ \overline{\ell_L^c} \mathbf{Y}_{\Delta} \Delta \ell_L + \text{H.c.} \right] + \frac{1}{2} \text{Tr}\left[ \left(D_{\mu} \Delta \right)^\dagger \left(D^{\mu} \Delta \right) \right]-  V(\Phi, \Delta) \; ,
\label{eq:TypeIILag}
\end{equation}    
where $\mathbf{Y}_{\Delta}$ is a $3 \times 3$ symmetric Yukawa coupling matrix and the scalar potential is written as
\begin{align}
V(\Phi, \Delta) &= \frac{1}{2} m_{\Delta}^2 \text{Tr}\left(\Delta^\dagger \Delta\right) + \left(\mu_{\Delta} \tilde{\Phi}^T \Delta \Phi + \text{H.c.} \right) \nonumber \\
&+ \lambda_{\Delta 1} \text{Tr}\left(\Delta^\dagger \Delta \right)^2 + \lambda_{\Delta 2} \text{Tr}\left(\Delta^\dagger \Delta \Delta^\dagger \Delta \right) + \lambda_{\Delta 3} \left(\Phi^\dagger \Phi\right) \text{Tr}\left(\Delta^\dagger \Delta \right) + \lambda_{\Delta 4} \left(\Phi^\dagger \Delta^\dagger \Delta \Phi\right) \; .
\end{align}

In this model, neutrino mass generation is triggered when $\Delta$ acquires a very small VEV~$v_{\Delta}$, induced after EWSB by the Higgs doublet VEV $v$, via the $\mu_\Delta$ parameter. The vacuum configuration is the following:
\begin{equation}
  \left<\Delta \right>  = \begin{pmatrix}
 \sqrt{2} \left<\Delta^{0}\right> &  -\left<\Delta^{+}\right>\\
 -\left<\Delta^{+}\right> & - \sqrt{2} \left<\Delta^{++}\right>
\end{pmatrix} =  \begin{pmatrix}
 v_{\Delta} & 0\\
 0 & 0
\end{pmatrix} \; .
\label{eq:TypeIIvev}
\end{equation}
Minimizing the scalar potential yields
\begin{equation}
v_{\Delta} \simeq - \mu_{\Delta}^{*} \frac{v^2}{m_{\Delta}^2 + \lambda_{\Delta 3} v^2} \simeq - \mu_{\Delta}^{*} \frac{v^2}{m_{\Delta}^2} \; ,
\label{eq:TypeIIvevvalue}
\end{equation}
where the approximate expression is valid for $\lambda_{\Delta 3} v^2 \ll m_{\Delta}^2$. From the above equation, one obtains the effective light neutrino mass matrix,
\begin{equation}
\mathbf{M}_{\nu} = 2 v_{\Delta} \mathbf{Y}_{\Delta} \simeq - 2 \mu_{\Delta}^{*} \frac{v^2}{m_{\Delta}^2} \mathbf{Y}_{\Delta} \; .
\end{equation}
Note an important distinction with the type-I seesaw. In the type-II case the effective light neutrino mass matrix is inversely proportional to $m_{\Delta}^2$. Considering heavy scalars $m_{\Delta} \gg \mu_{\Delta}$ and $\mu_\Delta/v \ll 1$ naturally small in the 't Hooft sense~\cite{tHooft:1979rat}, the VEV hierarchy $v_{\Delta}/v \ll 1$, is naturally obtained. Note also that since $\Delta$ is a SU(2) triplet it will modify the $\rho$ parameter at tree-level -- see Eq.~\eqref{eq:rhoMultiHiggs} -- with current experimental constraints setting the upper bound on the VEV $v_{\Delta} \lsim 2.6$ GeV~\cite{ATLAS:2018ceg,CMS:2017fhs}. Hence, for natural Yukawa couplings of the order of unity and a coupling $\mu_{\Delta} \sim 1~\text{eV}$, the masses of the additional scalars need to be~$m_{\Delta} \sim \text{TeV}$, to explain the small neutrino masses $m_\nu \sim 0.1$ eV. Therefore, the type-II seesaw has a low-scale implementation enabling feasible experimental detection of direct new physics signals -- for a recent comprehensive phenomenological study of the type-II seesaw see Ref.~\cite{Mandal:2022zmy}. 

%%%%%%%%%%%%%%%%%%%%%%%%%%%%%%%%%%%%%%%%%%%%%%%%%%%%%%%%%%%%%%%%%%%%%%%%%%%%%
\subsection{Type-III seesaw}
%%%%%%%%%%%%%%%%%%%%%%%%%%%%%%%%%%%%%%%%%%%%%%%%%%%%%%%%%%%%%%%%%%%%%%%%%%%%%

The type-III seesaw~\cite{Foot:1988aq} consists of adding to the SM $n_{\Sigma}$ fermionic triplets $\Vec{\Sigma}_R^i = \left(\Sigma_{1}^i, \Sigma_{2}^i, \Sigma_{3}^i\right)^T$ ($i= 1, \dots, n_{\Sigma}$), with $Y=0$. As outlined in the previous section, these fields can be written in an SU(2) matrix representation analogous to Eq.~\eqref{eq:2times2repforD}. Namely,
\begin{equation}
i \tau_2 \left(\Vec{\tau} \cdot \Vec{\Sigma}_R \right)  =  \begin{pmatrix}
\Sigma_1 + i \Sigma_2 & - \Sigma_3\\
- \Sigma_3 & -\Sigma_1 + i \Sigma_2
\end{pmatrix}  = 
\begin{pmatrix}
\sqrt{2} \Sigma^{-} & - \Sigma^0\\
- \Sigma^0 & - \sqrt{2} \Sigma^{+}
\end{pmatrix} \; .
\end{equation}
The Lagrangian is given by
\begin{equation}
\mathcal{L}_{\text{III}} = \mathcal{L}^{\text{SM}} + \overline{\Vec{\Sigma}_R} \left(i \slashed{D} \right) \Vec{\Sigma}_R- \left[ \overline{\Vec{\Sigma}_R} \cdot \left(\mathbf{Y}_{\Sigma}^T \tilde{\Phi}^{\dagger} \Vec{\tau} \ell_L \right) + \frac{1}{2} \overline{\Vec{\Sigma}_R^c} \mathbf{M}_{\Sigma}^* \Vec{\Sigma}_R + \text{H.c.} \right] \; ,
\end{equation}  
where the $n_{\Sigma} \times n_{\Sigma}$ mass matrix $\mathbf{M}_{\Sigma}$ is of Majorana type and the complex $3 \times n_{\Sigma}$ matrix $\mathbf{Y}_{\Sigma}$ is of Dirac~type. After EWSB, the neutrino mass Lagrangian reads as
\begin{equation}
-\mathcal{L}_{\text{III}} \supset \overline{\nu_L} \mathbf{M}_{D}^{*} \Sigma_R^0 + \frac{1}{2} \overline{\Sigma_R^{0 c}} \mathbf{M}_{\Sigma}^* \Sigma_R^0 + \text{H.c.} \; , \; \mathbf{M}_D= \frac{v}{\sqrt{2}} \mathbf{Y}_{\Sigma} \; .
\end{equation}
Organizing the neutrino fields in the vector $N_{L}=\left(\nu_{L}, \Sigma^{0 c}_{R} \right)^{T}$ of dimension $n_f = 3+n_{\Sigma}$, with $\nu_{L}=\left(\nu_{e L}, \nu_{\mu L}, \nu_{\tau L} \right)^T,\, \Sigma^{0}_{R}=\left(\Sigma^{0}_{ R 1}, ... \ , \Sigma^{0}_{ R n_{\Sigma}} \right)^T$, leads to the simplified form
\begin{equation}
-\mathcal{L}_{\text{III}} \supset \frac12\,\overline{N_{L}^c}\,\bm{\mathcal{M}}_\nu \,N_{L}+\text{H.c.} \; , \; \bm{\mathcal{M}}_\nu = \begin{pmatrix}
0 & \mathbf{M}_D  \\
\mathbf{M}_D^{T} & \mathbf{M}_{\Sigma}
\end{pmatrix} \; .
\label{eq:TypeIIILagmass}
\end{equation}
Note that the full neutrino mass matrix has the same form as the one for the type-I seesaw case in Eq.~\eqref{eq:TypeILagmass}. Thus, following a similar block-diagonalization procedure, in the seesaw approximation $M_D \ll M_{\Sigma}$, yields
\begin{equation}
\mathbf{M}_{\nu} = - \mathbf{M}_D \mathbf{M}_{\Sigma}^{-1} \mathbf{M}_D^T \; .
\end{equation}
In both the type-I and type-III seesaw mechanisms, the neutrino masses are inversely proportional to the mass of the particles added to the SM. For natural values of the Yukawa couplings, it is required a mass scale $M_{\Sigma}$ near the GUT scale to be able to explain the tiny light neutrino masses $m_\nu \sim 0.1$ eV. However, unlike the type-I seesaw mechanism, the gauge interactions of the fermion triplets enable distinctive collider signatures. The mixing of the heavy charged triplet components with SM charged leptons gives rise to tree-level contributions to rare cLFV processes such as $\mu \to 3e$ and $\mu - e$ conversion in nuclei -- for recent analyses of the triplet fermion seesaw phenomenology see Refs.~\cite{Biggio:2019eeo,Das:2020uer,Ashanujjaman:2021jhi}.

%%%%%%%%%%%%%%%%%%%%%%%%%%%%%%%%%%%%%%%%%%%%%%%%%%%%%%%%%%%%%%%%%%%%%%%%%%%%%
\section{Low-scale seesaw schemes}
\label{sec:lowseesaw}
%%%%%%%%%%%%%%%%%%%%%%%%%%%%%%%%%%%%%%%%%%%%%%%%%%%%%%%%%%%%%%%%%%%%%%%%%%%%%

As seen in the previous sections, the canonical type-I and III seesaw scenarios require very heavy particles or unnaturally tiny Dirac Yukawa couplings in order to generate small neutrino masses and lepton mixing. In contrast, the type-II seesaw allows for a low-scale implementation where the masses of the additional particles are testable at experiments such as the LHC. This section briefly reviews the paradigmatic low-scale seesaw variants in their inverse and linear realizations. These mechanisms can be implemented by extending the SM particle content with $n_R$ RH neutrinos $\nu_{R}$ and $n_S$ sterile fermion singlets $S_R$, with the fields $\nu_L$, $\nu_R$ and $S_R$ carrying lepton-number $+1$, $+1$ and $-1$, respectively. The key idea, common to both mechanisms, is neutrino mass suppression triggered by small LNV parameters. The lightness of neutrinos stems from an approximate lepton-number symmetry which is restored when those parameters are set to zero. Thus, these provide a natural framework for neutrino-mass generation in the 't Hooft sense~\cite{tHooft:1979rat}.

%%%%%%%%%%%%%%%%%%%%%%%%%%%%%%%%%%%%%%%%%%%%%%%%%%%%%%%%%%%%%%%%%%%%%%%%%%%%%
\subsection{Inverse seesaw mechanism}
\label{sec:ISS}
%%%%%%%%%%%%%%%%%%%%%%%%%%%%%%%%%%%%%%%%%%%%%%%%%%%%%%%%%%%%%%%%%%%%%%%%%%%%%

The inverse seesaw~(ISS)~\cite{Mohapatra:1986aw,Mohapatra:1986bd,GonzalezGarcia:1988rw}, denoted as ISS$(n_R,n_S)$, stems from the following neutrino mass Lagrangian:
\begin{equation}
\begin{split}
-\mathcal{L}_{\text{ISS}} & = \overline{\nu_L}\, \mathbf{M}_{D}^\ast \nu_R+ \overline{\nu_R^c}\, \mathbf{M}_{R}  S_R + \frac{1}{2} \overline{S_R^c} \ \mathbf{m}_{S} S_R + \text{H.c.} \; ,
\end{split}
\label{eq:LmassISS}
\end{equation}
where $\nu_{L}=\left(\nu_{e L}, \nu_{\mu L}, \nu_{\tau L} \right)^T$, $\nu_{R}=\left(\nu_{R1}, ... \ , \nu_{Rn_R} \right)^T$,\, $S_R=\left(S_{R1}, ...\ , S_{Rn_S} \right)^T$. In the above equation, $\mathbf{M}_{D}$ is a $3\times n_R$ Dirac-type mass matrix, $\mathbf{M}_{R}$ is a $n_R \times n_S$ matrix, and $\mathbf{m}_{S}$ is a LNV $n_S \times n_S$ Majorana mass matrix. The latter can be naturally small in the 't Hooft~\cite{tHooft:1979rat} sense, since lepton number conservation is restored in the limit where $\mathbf{m}_{S}$ vanishes. Note that the minimal ISS scenario compatible with neutrino data is ISS$(2,2)$.

Defining $N_{L}=\left(\nu_{L}, \nu^c_{R}, S^c_{R} \right)^{T}$ of dimension $n_f = 3+n_R+n_S$, $\mathcal{L}_{\text{ISS}}$ is written in the compact form
\begin{equation}
\begin{split}
-\mathcal{L}_{\text{ISS}} &=\frac12\,\overline{N_{L}^c}\,\bm{\mathcal{M}}_\nu\,N_{L}+\text{H.c.}\;\;,\;\; 
\bm{\mathcal{M}}_\nu= \begin{pmatrix}
0 & \mathbf{M}_D & 0 \\
\mathbf{M}_D^{T} & 0 &\mathbf{M}_{R}\\
0 & \mathbf{M}^{T}_{R} & \mathbf{m}_S
\end{pmatrix}\,,
\label{eq:bigmISS}
\end{split}
\end{equation}
where $\bm{\mathcal{M}}_\nu$ is the full $n_f \times n_f$ neutrino mass matrix.

In the ISS approximation limit where $m_S, M_D \ll M_{R}$, for the matrix entries, the effective light neutrino mass matrix can be derived following the diagonalization method outlined in Sec.~\ref{sec:TypeI} for the type-I seesaw case. However, here $\mathbf{M}_{\text{h}} \simeq \mathbf{M}_R^\prime$ is a $(n_R+n_S) \times (n_R+n_S)$ matrix given by
\begin{equation} 
\mathbf{M}_{R}^\prime = \begin{pmatrix} 
0 & \mathbf{M}_R  \\
\mathbf{M}_R^{T} & \mathbf{m}_S
\end{pmatrix} \; ,
\end{equation}
which is diagonalized by a $(n_R+n_S) \times (n_R+n_S)$ unitary matrix $\mathbf{U}_{\text{h}}$, yielding $n_R + n_S$ heavy neutrinos. Its inverse is
\begin{equation} 
\mathbf{M}_{R}^{\prime -1} = \begin{pmatrix} 
- \left( \mathbf{M}_{R}\, \mathbf{m}_S^{-1} \mathbf{M}_{R}^{T} \right)^{-1} & (\mathbf{M}_R^{T})^{-1}  \\
(\mathbf{M}_R)^{-1} & 0
\end{pmatrix} \; ,
\end{equation}
with $\mathbf{M}_R$ and $\mathbf{m}_S$ invertible. Furthermore, by defining the $3 \times (n_R + n_S)$ matrix
\begin{equation}
\mathbf{M}_D^\prime = \left(\mathbf{M}_D \; , 0\right) \; ,
\end{equation}
the expansion parameter $\mathbf{\Theta}_\nu$ of Eq.~\eqref{eq:thetanuI} is here, at first order in the seesaw approximation, given by
\begin{equation}
\mathbf{\Theta}_\nu \simeq \mathbf{M}_{D}^{\prime \ast} (\mathbf{M}_{R}^{\prime *})^{-1} \simeq \left(0,\ \mathbf{M}_D^\ast (\mathbf{M}_R^{\dagger})^{-1}\right) \; .
\label{eq:FISS}
\end{equation}
This leads to the $3\times 3$ effective light-neutrino mass matrix
\begin{equation}
\mathbf{M}_{\nu} = - \mathbf{\Theta}_\nu^* \mathbf{M}_{R}^\prime  \mathbf{\Theta}_\nu^{\dagger}  = - \mathbf{M}_D \,\left( \mathbf{M}_{R}\, \mathbf{m}_S^{-1} \mathbf{M}_{R}^{T} \right)^{-1}\mathbf{M}_D^{T} \; ,
\label{eq:invss}
\end{equation}
being diagonalized through a unitary rotation of the active neutrino fields as in Eq.~\eqref{eq:TypeIMeffdiag}. Furthermore, deviations from unitarity of the lepton mixing matrix are given by [see Eqs.~\eqref{eq:unitpmnsI} and~\eqref{eq:epsilondevuniI}]
\begin{equation}
\bm{\eta} = \frac{1}{2} \mathbf{U}_{L}^{e \dagger} \mathbf{\Theta}_\nu \mathbf{\Theta}_\nu^{\dagger} \mathbf{U}_{L}^{e} \simeq \dfrac{1}{2} \mathbf{U}_L^{e \dagger}\mathbf{M}_D^\ast (\mathbf{M}_R^{\dagger})^{-1}  \mathbf{M}_R^{-1} \mathbf{M}_D^{T}\mathbf{U}_L^e \; ,
\label{eq:etaiss}
\end{equation}
while active-sterile neutrino mixing is described, at first order in $\mathbf{\Theta}_\nu$, as [see Eq.~\eqref{eq:HeavyLightI}]
\begin{equation}
\mathbf{U}_{L}^{e \dagger} \mathbf{\Theta}_\nu \mathbf{U}_{\text{h}} \simeq \mathbf{U}_{L}^{e \dagger} \left(0,\ \mathbf{M}_D^\ast (\mathbf{M}_R^{\dagger})^{-1}\right) \mathbf{U}_{\text{h}} \; .
\label{eq:VLW}
\end{equation}
The expressions above are very similar to ones for the type-I seesaw, the relevant differences lying in the order of magnitude of the mass scale $M_R$ due to the presence of the small LNV parameter~$m_S$. Namely, by comparing the effective light neutrino mass matrix formula of Eq.~\eqref{eq:invss} with the one for the type-I case of Eq.~\eqref{eq:TypeIMeff}, it is clear that in the latter case the smallness of neutrino masses is explained through the sole suppression $M_R^{-1}$ whereas in the ISS case, thanks to the addition of a second species of sterile fermions, there are two suppressing factors $M_R^{-1} M_D$ and $m_S$. Hence, for natural Dirac Yukawa couplings, $M_D \sim v/\sqrt{2}$, and having a small LNV parameter $m_S \sim \text{eV}$, one only needs $M_R \sim \text{TeV}$, in order to explain the light neutrino mass scale $m_\nu \sim 0.1$ eV. As a result, the mixing between the (active) light neutrinos and the new (sterile) states can be sizable for sterile neutrino masses lying not far from the EW scale. The presence of new neutral fermions interacting with SM leptons and gauge bosons motivates phenomenological studies beyond the SM, making the ISS a perfect theoretical framework to guide new physics probes. In particular, experimental searches for cLFV processes like $\mu\rightarrow e \gamma$~\cite{TheMEG:2016wtm,Baldini:2018nnn}, $\mu\rightarrow e e e$~\cite{Bellgardt:1987du} and $\mu-e$ conversion in heavy nuclei~\cite{Dohmen:1993mp,Honecker:1996zf,Bertl:2006up,Kuno:2013mha,Natori:2014yba} have been studied in the ISS framework~\cite{Deppisch:2005zm,Abada:2014vea,Arganda:2014dta,Abada:2015oba,Abada:2014cca,DeRomeri:2015ipa,Arganda:2015ija,Abada:2018nio,Camara:2020efq} with the purpose of understanding at which extent our current knowledge on those processes is able to constrain the ISS parameter space. Depending on their masses and mixing with the SM degrees of freedom, sterile neutrinos may also lead to interesting signals potentially observable at the LHC, as well as at other experiments sensitive to new physics effects induced by the presence of those particles~\cite{delAguila:2008cj,Das:2012ze,Deppisch:2015qwa,Antusch:2016ejd,Caputo:2017pit,Bhardwaj:2018lma,Bolton:2019pcu}.

%%%%%%%%%%%%%%%%%%%%%%%%%%%%%%%%%%%%%%%%%%%%%%%%%%%%%%%%%%%%%%%%%%%%%%%%%%%%%
\subsection{Linear seesaw mechanism}
\label{sec:LSS}
%%%%%%%%%%%%%%%%%%%%%%%%%%%%%%%%%%%%%%%%%%%%%%%%%%%%%%%%%%%%%%%%%%%%%%%%%%%%%

The linear seesaw~(LSS)~\cite{Akhmedov:1995ip,Akhmedov:1995vm,Malinsky:2005bi}, denoted as LSS$(n_R,n_S)$, stems from the following neutrino mass Lagrangian:
\begin{equation}
\begin{split}
-\mathcal{L}_{\text{LSS}} & = \overline{\nu_L}\, \mathbf{M}_{D} \nu_R+ \overline{\nu_R^c}\, \mathbf{M}_{R}  S_R + \overline{\nu_L} \ \mathbf{m}_{S} S_R + \text{H.c.} \; ,
\end{split}
\label{eq:LmassLSS}
\end{equation}
where $\mathbf{m}_{S}$ is a LNV $3 \times n_S$ mass matrix, which can be naturally small in the 't Hooft~\cite{tHooft:1979rat} sense. Note that, the minimal LSS scenario compatible with neutrino data is LSS$(1,1)$. The above Lagrangian can be written in compact form [see Eq.~\eqref{eq:bigmISS}:
\begin{equation}
\begin{split}
-\mathcal{L}_{\text{ISS}} &=\frac12\,\overline{N_{L}^c}\,\bm{\mathcal{M}}_\nu\,N_{L}+\text{H.c.}\;\;,\;\; 
\bm{\mathcal{M}}_\nu= \begin{pmatrix}
0 & \mathbf{M}_D & \mathbf{m}_S \\
\mathbf{M}_D^{T} & 0 &\mathbf{M}_{R}\\
\mathbf{m}_S^T & \mathbf{M}^{T}_{R} & 0
\end{pmatrix}\, .
\label{eq:bigmLSS}
\end{split}
\end{equation}

In the LSS approximation limit where $m_S, M_D \ll M_{R}$, for the matrix entries, the effective light neutrino mass matrix can be derived following the diagonalization method outlined in Sec.~\ref{sec:TypeI} for the type-I seesaw case. However, here $\mathbf{M}_{\text{h}} \simeq \mathbf{M}_R^\prime$ is a $(n_R+n_S) \times (n_R+n_S)$ matrix, diagonalized by a $(n_R+n_S) \times (n_R+n_S)$ unitary matrix $\mathbf{U}_{\text{h}}$, yielding $n_R + n_S$ heavy neutrinos. The heavy neutrino mass matrix is defined as
\begin{equation} 
\mathbf{M}_{R}^{\prime} = \begin{pmatrix} 
0 & \mathbf{M}_R  \\
\mathbf{M}_R^{T} & 0
\end{pmatrix} \; ,
\end{equation}
with $\mathbf{M}_R$ invertible. Furthermore, by defining the $3 \times (n_R + n_S)$ matrix,
\begin{equation}
\mathbf{M}_D^\prime = \left(\mathbf{M}_D \; , \mathbf{m}_S\right) \; ,
\end{equation}
the expansion parameter $\mathbf{\Theta}_\nu$ of Eq.~\eqref{eq:thetanuI} is here, at first order in the seesaw approximation, given by
\begin{equation}
\mathbf{\Theta}_\nu \simeq \mathbf{M}_{D}^{\prime \ast} (\mathbf{M}_{R}^{\prime *})^{-1} \simeq \left(\mathbf{m}_S^\ast (\mathbf{M}_R^{\ast})^{-1} ,\ \mathbf{M}_D^\ast (\mathbf{M}_R^{\dagger})^{-1}\right) \; .
\label{eq:FLSS}
\end{equation}
This leads to the $3\times 3$ effective light-neutrino mass matrix
\begin{equation}
\mathbf{M}_{\nu} = - \mathbf{\Theta}_\nu^* \mathbf{M}_{R}^\prime  \mathbf{\Theta}_\nu^{\dagger}  = - \mathbf{M}_D \mathbf{M}_R^{-1} \mathbf{m}_S^T - \mathbf{m}_S \mathbf{M}_R^{-1} \mathbf{M}_D^T \; ,
\label{eq:MnuLSS}
\end{equation}
being diagonalized through a unitary rotation of the active neutrino fields as in Eq.~\eqref{eq:TypeIMeffdiag}. Furthermore, deviations from unitarity of the lepton mixing matrix are given by [see Eqs.~\eqref{eq:unitpmnsI} and~\eqref{eq:epsilondevuniI}]
\begin{equation}
\bm{\eta} = \frac{1}{2} \mathbf{U}_{L}^{e \dagger} \mathbf{\Theta}_\nu \mathbf{\Theta}_\nu^{\dagger} \mathbf{U}_{L}^{e} \simeq \dfrac{1}{2} \mathbf{U}_L^{e \dagger}\mathbf{M}_D^\ast (\mathbf{M}_R^{\dagger})^{-1}  \mathbf{M}_R^{-1} \mathbf{M}_D^{T}\mathbf{U}_L^e \; ,
\label{eq:etaLss}
\end{equation}
while active-sterile neutrino mixing is described, at first order in $\mathbf{\Theta}_\nu$, as [see Eq.~\eqref{eq:HeavyLightI}]
\begin{equation}
\mathbf{U}_{L}^{e \dagger} \mathbf{\Theta}_\nu \mathbf{U}_{\text{h}} \simeq \mathbf{U}_{L}^{e \dagger} \left(\mathbf{m}_S^\ast (\mathbf{M}_R^{\ast})^{-1} ,\ \mathbf{M}_D^\ast (\mathbf{M}_R^{\dagger})^{-1}\right)  \mathbf{U}_{\text{h}} \; .
\label{eq:VLW2}
\end{equation}
As for the ISS, the LSS is a low-scale seesaw mechanism. Taking natural Yukawa couplings of order unity and $m_S \sim \mathcal{O}(1\,{\rm eV})$, leads to light neutrino masses $\sim \mathcal{O}(0.1\,{\rm eV})$ for a heavy neutrino mass scale at $M_R \sim \mathcal{O}(1\,{\rm TeV})$. This has important phenomenological implications in regards to cLFV, as well as interesting collider signals involving the heavy neutral leptons, i.e. the heavy pseudo-Dirac sterile neutrino states -- see Refs.~\cite{Batra:2023ssq,Batra:2023mds} for a detailed study of the phenomenology of the simplest LSS. We will discuss the cLFV phenomenology of a radiative LSS implementation in Sec.~\ref{sec:darkLSS}.

%%%%%%%%%%%%%%%%%%%%%%%%%%%%%%%%%%%%%%%%%%%%%%%%%%%%%%%%%%%%%%%%%%%%%%%%%%%%%
\section{Scotogenic neutrino masses}
\label{sec:scotogenic}
%%%%%%%%%%%%%%%%%%%%%%%%%%%%%%%%%%%%%%%%%%%%%%%%%%%%%%%%%%%%%%%%%%%%%%%%%%%%%

%
    \begin{figure}[t!]
        \centering
        \includegraphics[scale=0.9]{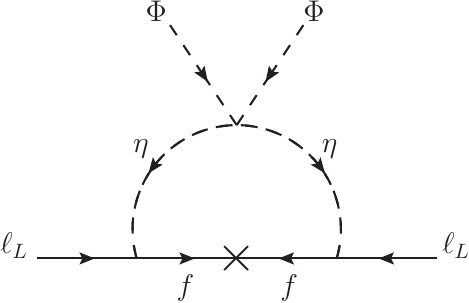} \hspace{+0.1cm} \includegraphics[scale=0.9]{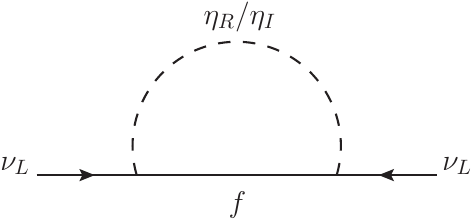}
        \caption{One-loop scotogenic neutrino mass diagram in the weak-basis (mass-basis) on the left (right).}
    \label{fig:scotogenic}
    \end{figure}
In radiative models, small neutrino masses are induced at the quantum level -- for a review on the subject see Refs.~\cite{Cai:2017jrq,Avila:2025qsc}. Interestingly, the particles in the loops may also be viable DM candidates, part of a so-called dark sector, stabilized by some symmetry as a simple $\mathcal{Z}_2$. This is the principle behind the scotogenic mechanism~\cite{Tao:1996vb,Ma:2006km} that combines neutrino mass generation and DM in a framework where DM acts as the ``seed'' of neutrino masses. 

In the canonical scotogenic model are added to the SM three sterile singlet fermions $f$ and an inert scalar doublet $\eta$. These particles are odd under a $\mathcal{Z}_2$ symmetry, forbidding tree-level neutrino mass generation, as well as preventing the lightest dark-sector particle to decay. Given these new fields and symmetry, the Lagrangian is
\begin{equation}
    \mathcal{L}_{\text{scoto}} = \mathcal{L}_{\text{SM}} + i \overline{f} \slashed{\partial} f - \left( \overline{\ell_L} \tilde{\eta} \mathbf{Y}_{f}^* f + \frac{1}{2} \overline{f} \mathbf{M}_{f} f^c + \text{H.c.} \right) + (D^\mu \eta)^\dagger (D_\mu \eta) - V(\Phi, \eta) \; ,
\label{eq:TypescotogenicLag}
\end{equation}
where $\mathbf{Y}_{f}$ is $3 \times 3$ complex Yukawa matrix, $\mathbf{M}_{f}$ is a $3 \times 3$ symmetric bare mass matrix, and $\tilde{\eta}= i \tau_2 \eta^\ast$. The SM scalar potential is extended by
    \begin{align}
    V(\Phi, \eta) &= m_{\eta}^2 \left(\eta^\dagger \eta\right) + \frac{\lambda_{\eta}}{2} \left(\eta^\dagger \eta\right)^2 + \lambda_{3} \left(\Phi^\dagger \Phi\right) \left(\eta^\dagger \eta\right) + \lambda_{4} \left(\Phi^\dagger \eta \right) \left(\eta^\dagger \Phi\right) + \left[\frac{\lambda_5}{2} \left(\Phi^\dagger \eta \right) + \text{H.c.} \right]\; ,
    \label{eq:Vscotogenic}
    \end{align}
where $\lambda_5$ can be made real without loss of generality. The doublet $\eta$ is parameterized as
\begin{align}
\eta =\begin{pmatrix}
\eta^{+} \\
\eta^0
\end{pmatrix}= \frac{1}{\sqrt{2}}  \begin{pmatrix}
 \sqrt{2} \eta^{+} \\
\eta_{\text{R}} + i \eta_{\text{I}}
\end{pmatrix} \; .
\label{eq:scalarsetascotogenic}
\end{align}
As usual, the Higgs doublet acquires a non-zero VEV where $\left< \phi^0 \right> = v/\sqrt{2}$. However, taking $m_\eta^2>0$, the doublet $\eta$ can be inert, i.e. $\left< \eta^0 \right> = 0$. This ensures that after EWSB the $\mathcal{Z}_2$ symmetry remains unbroken and stabilizes DM. The dark-scalar sector contains a charged $\eta^\pm$ and two neutral $\eta_{R,I}$ states, with masses
\begin{align}
m_{\eta^\pm}^2 &= m_\eta^2 + \frac{1}{2} v^2 \lambda_3 \; , \nonumber \\
m_{\eta_R}^2 &= m_\eta^2 + \frac{1}{2} v^2 \left(\lambda_3 + \lambda_4 + \lambda_5 \right) \; , \nonumber \\
m_{\eta_I}^2 &= m_\eta^2 + \frac{1}{2} v^2 \left(\lambda_3 + \lambda_4 - \lambda_5 \right) \; , 
\label{eq:scotogenicscalarmasses}
\end{align}
with the mass of the physical non-dark scalar being the one of the SM Higgs boson [see Eq.~\eqref{eq:Higgsmassmechanism}], since the dark $\mathcal{Z}_2$ forbids any $\Phi-\eta$ mixing.

Neutrino masses will be generated at the one-loop level via the diagrams shown in Fig.~\ref{fig:scotogenic}, in the weak (mass) basis, on the left (right). Considering, without loss of generality, that $\mathbf{M}_f$ is diagonal and real, the effective light neutrino masses are given by
\begin{align}
\left(\mathbf{M}_\nu\right)_{\alpha \beta} = \sum_{i=1}^3 \mathcal{F}_{\text{scoto}} \left(M_{f_i} , m_{\eta_R} , m_{\eta_I} \right) M_{f_i} \left(\mathbf{Y}_f\right)_{\alpha i} \left(\mathbf{Y}_f\right)_{\beta i} \; .
\label{eq:Mnuscotogenic}
\end{align}
with the loop factors being
\begin{align}
\mathcal{F}_{\text{scoto}} \left(M_{f_i} , m_{\eta_R} , m_{\eta_I} \right) = & \; \frac{1}{32 \pi^2} \left[\frac{m_{\eta_R}^2}{M_{f_i}^2 -m_{\eta_R}^2} \ln \left( \frac{M_{f_i}^2}{m_{\eta_R}^2} \right) - \frac{m_{\eta_I}^2}{M_{f_i}^2 - m_{\eta_I}^2} \ln \left( \frac{M_{f_i}^2}{m_{\eta_I}^2} \right) \right] \; .
\label{eq:Fscoto}
\end{align}
From the weak-basis neutrino mass generation diagram, shown on the left of Fig.~\ref{fig:scotogenic}, it is clear that in the limit $\lambda_5 \rightarrow 0$ neutrinos are massless. The smallness of neutrino masses is controlled by the mass splitting between $\eta_R$ and $\eta_I$. In fact, $\lambda_5$ is naturally small in the 't Hooft sense~\cite{tHooft:1979rat}, since in the limit $\lambda_5 \to 0$ the Lagrangian acquires an enhanced symmetry: the scalar potential reduces to a U(1)-symmetric two-Higgs-doublet model~(2HDM). Defining the quantity $m_0^2 = (m^2_{\eta_{R}} + m^2_{\eta_{I}})/2$, and in the limit of small $\lambda_5$, i.e. $m^2_{\eta_{R}} - m^2_{\eta_{I}} = v^2 \lambda_5 \ll m_0^2$, one can write Eq.~\eqref{eq:Mnuscotogenic}, up to first order in $v^2 \lambda_5$, as
\begin{align}
\left(\mathbf{M}_\nu\right)_{\alpha \beta} \simeq \sum_{i=1}^3 \frac{\lambda_5}{16 \pi^2} \frac{v^2\left(\mathbf{Y}_f\right)_{\alpha i} \left(\mathbf{Y}_f\right)_{\beta i}}{2 M_{f_i}} \left[ \frac{M_{f_i}^2}{m_0^2-M_{f_i}^2} + \frac{M_{f_i}^4}{(m_0^2-M_{f_i}^2)^2} \ln \left(\frac{M_{f_i}^2}{m_0^2}\right)\right] \; .
\label{eq:Mnuscotogeniclimit}
\end{align}
Compared to the canonical type-I seesaw, scotogenic neutrino masses are suppressed at least by an additional factor $\lambda_5/(16 \pi^2)$ [see Eq.~\eqref{eq:TypeIMeff}]. Considering dark fermion masses $M_{f_i} \sim 10$ TeV and a scalar mass scale $m_0 \sim 1$ TeV, the light neutrino mass scale of 0.1~eV can be accommodated with a naturally small $\lambda_5 \sim 10^{-5}$, while maintaining sizable Yukawa couplings $Y_f \sim 0.1$. Hence, in contrast to the type-I seesaw scenario, this radiative neutrino mass scheme offers interesting prospects for testability at experimental facilities.

Besides providing a natural explanation for the smallness of neutrino masses, this model also accommodates a viable DM candidate, stabilized by an exact $\mathcal{Z}_2$ symmetry that remains unbroken after EWSB. Depending on the spectrum, the DM particle may be either the lightest $\mathcal{Z}_2$-odd neutral fermion or the lightest neutral scalar. In both scenarios, assuming a WIMP framework, annihilation and coannihilation processes involving the DM candidate contribute to the thermally averaged cross section, allowing for a relic abundance consistent with observational data. Detailed analyses of the DM phenomenology for fermionic Majorana and scalar candidates can be found in Refs.~\cite{Karan:2023adm,Ahriche:2017iar,Hagedorn:2018spx,Karan:2023adm} and~\cite{Diaz:2015pyv,Borah:2017dfn,Avila:2021mwg}, respectively. Additional probes of the scotogenic model arise from the low-energy effects of the new $\mathcal{Z}_2$-odd particles. Notably, interactions with SM leptons can induce cLFV~\cite{Toma:2013zsa,Vicente:2014wga}, while couplings to EW gauge bosons give rise to distinctive collider signatures~\cite{Baumholzer:2019twf,Lozano:2025tst}. As seen in the upcoming Sec.~\ref{sec:darkLSS}, the combined constraints from DM relic density, cLFV limits, and collider searches are essential for determining the viable parameter space of dark-seeded neutrino mass models. 

As a last note, the interplay between the scotogenic and seesaw paradigm leads to the so-called "scoto-seesaw" models~\cite{Rojas:2018wym,Barreiros:2020gxu,Mandal:2021yph,Barreiros:2022aqu,Ganguly:2023jml,Ganguly:2022qxj,VanDong:2023xbd,Leite:2023gzl,Kumar:2023moh,VanDong:2024lry,Kumar:2025cte,}. This was initially motivated by the fact that in the minimal scenario the coexistence of both mechanisms allows for the generation of the atmospheric and solar neutrino mass scale from a distinct tree-level and radiative origin, respectively~\cite{Rojas:2018wym}. Interestingly, this framework allows for a natural implementation of SCPV~\cite{Barreiros:2020gxu}, which we have explored in the context where the seesaw is of type-II~\cite{Barreiros:2022aqu}.

%%%%%%%%%%%%%%%%%%%%%%%%%%%%%%%%%%%%%%%%%%%%%%%%%%%%%%%%%%%%%%%%%%%%%%%%%%%%%
\section{Dark linear seesaw mechanism}
\label{sec:darkLSS}
%%%%%%%%%%%%%%%%%%%%%%%%%%%%%%%%%%%%%%%%%%%%%%%%%%%%%%%%%%%%%%%%%%%%%%%%%%%%%

Following closely our work in Ref.~\cite{Batra:2023bqj}, in this section we explore a simple model where DM ``seeds'' lepton-number breaking in the LSS context with a minimal field and symmetry content. Namely, assuming a (softly-broken) lepton-number symmetry U$(1)_L$, we add three neutral fermion singlets with $L=\pm 1$. DM stability is ensured by a $\mathcal{Z}_2$ discrete symmetry under which one of the fermions is odd. The soft U$(1)_L$ breaking stems from a cubic interaction among the SM Higgs doublet and two dark scalars (a doublet and a real singlet). Hence, we consider the possibility that LNV in the neutrino sector is seeded at the quantum level by dark fields, allowing for a direct connection to the DM problem. After presenting our model -- dark linear seesaw -- we will study its phenomenological implications for cLFV and DM.

%%%%%%%%%%%%%%%%%%%%%%%%%%%%%%%%%%%%%%%%%%%%%%%%%%%%%%%%%%%%%%%%%%%%%%%%%%%%%
\subsection{Model}
%%%%%%%%%%%%%%%%%%%%%%%%%%%%%%%%%%%%%%%%%%%%%%%%%%%%%%%%%%%%%%%%%%%%%%%%%%%%%

%
\begin{table}[t!]
\renewcommand*{\arraystretch}{1.5}
	\centering
	\begin{tabular}{| K{1.5cm} | K{1cm} | K{2.5cm} | K{1.0cm} | K{1.0cm} | }
		\hline
&Fields&\EW&  U$(1)_L$ &  $\mathcal{Z}_2$  \\
		\hline
		\multirow{4}{*}{Fermions} 
&$\ell_L$&($\mathbf{2}, {-1}$)& $1$   &   $+$  \\
&$e_R$&($\mathbf{1}, {2}$)& {$1$}  &   $+$   \\
&$\nu_R$&($\mathbf{1}, {0}$)& {$1$}  &  $+$   \\
&$S_R$&($\mathbf{1}, {0}$)& {$-1$} &    $+$\\
&$f_{L,R}$&($\mathbf{1}, {0}$)& {$-1$}   & {$-$} \\
		\hline
\multirow{3}{*}{Scalars}  &$\Phi$&($\mathbf{2}, {1}$)& {$0$} &  $+$   \\
&$\eta$&($\mathbf{2}, {1}$)& {$-2$}  &  $-$   \\
&$\chi$&($\mathbf{1},0$)& {$0$} &  $-$   \\		
\hline
	\end{tabular}
	\caption{Field content and transformation properties under \EW and U$(1)_L \otimes \mathcal{Z}_2$.}
	\label{tab:model} 
\end{table}
As shown in Sec.~\ref{sec:LSS}, the LSS mechanism is implemented by adding sterile singlet fermions $\nu_R$ and $S_R$ to the SM, requiring in its minimal realization one of each to explain the two observed neutrino mass squared differences. In this framework, lepton number is violated by the $m_S\overline{\nu_L}S_R$ term. Neutrino masses are then proportional to $m_S$, which can be naturally small in the 't Hooft sense~\cite{tHooft:1979rat} and, hence, be the origin of neutrino mass suppression. Due to the existence of that small LNV parameter, low-scale neutrino mass generation can be envisaged, in contrast with the type-I seesaw which requires extremely heavy mediators (see Sec.~\ref{sec:TypeI}).

Our model is based on the symmetry and field content shown in Table~\ref{tab:model}, from which it is apparent that the global lepton number symmetry U$(1)_L$ forbids $m_S $ at tree level. To generate it radiatively, we introduce a Dirac vector-like singlet fermion $f_{L,R}$, as well as a scalar doublet $\eta$ and real singlet $\chi$. These three fields (the dark sector) are odd under a discrete $\mathcal{Z}_2$ symmetry. The most general fermion mass and Yukawa Lagrangian allowed by the SM gauge and U$(1)_L \otimes \mathcal{Z}_2$ symmetries is
\begin{align}
    -\mathcal{L} &\supset \mathbf{Y}_e \overline{\ell_L} \Phi e_R + \mathbf{Y}_D \overline{\ell_L} \tilde{\Phi} \nu_R + \mathbf{Y}_f \overline{\ell_L} \tilde{\eta} f_R \nonumber \\
    & + Y_S \overline{f_L} S_R \chi + Y_{R} \overline{f_R^c} \nu_R \chi + M_B \overline{\nu_R^c} S_R + M_f \overline{f_L} f_R + \text{H.c.} \; ,
    \label{eq:LYukDLSS}
\end{align}
where the Yukawa couplings $\mathbf{Y}_e$, $\mathbf{Y}_{D,f}$ and $Y_{R,S}$ are $3 \times 3$ matrices, $3 \times 1$ vectors and numbers, respectively. The mass $M_B$ is the typical heavy-neutrino mass scale, while $M_f$ is the bare mass of the dark fermion. Without loss of generality, we consider that the charged leptons are already in the physical basis and focus on the neutrino sector only.

The scalar content of our model contains, besides the usual Higgs doublet $\Phi$, a doublet $\eta$, which we define as in Eq.~\eqref{eq:scalarsetascotogenic}, as well as, a real scalar singlet $\chi$. The Higgs doublet acquires a non-zero VEV as usual, while the doublet $\eta$ and singlet $\chi$ are inert, i.e. $\left< \eta^0 \right> = \left< \chi \right> = 0$. The full scalar potential allowed by the symmetries of our model is
    \begin{align}
    V(\Phi,\eta,\chi) &=  V(\Phi) + m_{\eta}^2 \left(\eta^\dagger \eta\right) + m_\chi^2 \chi^2  + \lambda_{3} \left(\Phi^\dagger \Phi\right) \left(\eta^\dagger \eta\right) + \lambda_{4} \left(\Phi^\dagger \eta \right) \left(\eta^\dagger \Phi\right) \nonumber \\
    & + \frac{\lambda_{\eta}}{2} \left(\eta^\dagger \eta\right)^2 + \frac{\lambda_{\chi}}{2} \chi^4 + \lambda_{\Phi \chi} \left(\Phi^\dagger \Phi\right) \chi^2 +  \lambda_{\eta \chi} \left(\eta^\dagger \eta\right) \chi^2 + V_{\text{soft}} \; ,
    \label{eq:VpotentialfullDLSS}
    \end{align}
    where the origin of LNV stems from
    \begin{align}
    V_{\text{soft}} = \kappa \left(\eta^\dagger \Phi \right)\chi + \text{H.c.} \; ,
    \label{eq:VpotentialDLSSsoft}
    \end{align}
    which breaks U$(1)_L$ softly by two units, and $\kappa$ can be made real. Note that this is the only possible soft LNV term that can be written in the scalar potential. This is due to the restriction imposed by the $\mathcal{Z}_2$ symmetry which forbids any other soft LNV term that might lead to the decay of dark sector particles. The dark-scalar sector contains a charged $\eta^\pm$ and three neutral $\zeta_k$ ($k=1, 2, 3$) states. The dark-charged scalar mass expression is the same as in Eq.~\eqref{eq:scotogenicscalarmasses}. The mass matrix for the dark-neutral scalar states in the $\left(\eta_\text{R}, \chi, \eta_{\text{I}} \right)$ basis reads
\begin{align}
    \mathcal{M}^2_{\eta \chi}= \begin{pmatrix}
   m_\eta^2 + \frac{v^2}{2} \left(\lambda_3+\lambda_4\right) 
    & v \kappa & 0 \\
    \cdot&2 m_\chi^2 + \lambda_{\Phi\chi} v^2 &  0\\
    \cdot& \cdot &   m_\eta^2 + \frac{v^2}{2} \left(\lambda_3+\lambda_4\right)
    \end{pmatrix} \; ,
    \label{eq:darkscalarmassDLSS}
\end{align}
where `$\cdot$' reflects the symmetric nature of the matrix. The neutral components of the $\eta$ and $\chi$ are related to the mass eigenstates~$\zeta_{i}$ through the $3 \times 3$ orthogonal matrix $\mathbf{V}$:
\begin{equation}
(\eta_{\text{R}},
\chi,
\eta_{\text{I}})^T
 = \mathbf{V} 
(\zeta_1, \zeta_2, \zeta_3)^T \; , \; \sqrt{2} \eta^0 = \sum_{k=1}^{3} \left(\mathbf{V}_{1 k} + i \mathbf{V}_{3 k} \right) \zeta_k \; , \; \chi = \sum_{k=1}^{3} \mathbf{V}_{2 k} \zeta_k \; .
\label{eq:mixneutralscotoDMDLSS}
\end{equation}
The $\zeta_{i}$ masses are
\begin{align}
m_{\zeta_1}^2 = m_\eta^2 + \frac{v^2}{2} \left(\lambda_3+\lambda_4\right) \; , \; m_{\zeta_2}^2 = m_{\zeta_0}^2 - \frac{1}{4} \sqrt{\Lambda} \; , \; m_{\zeta_3}^2 = m_{\zeta_0}^2 + \frac{1}{4} \sqrt{\Lambda} \; ,
\end{align}
with
\begin{align}
m_{\zeta_0}^2 &= \frac{1}{2} \left(m_\eta^2 +2 m_\chi^2 \right) + \frac{v^2}{4} \left(\lambda_3+\lambda_4+ 2 \lambda_{\Phi \chi} \right) \; , \nonumber \\
\Lambda &= \left[2 \left(m_\eta^2 - 2 m_\chi^2 \right) + v^2 \left(\lambda_3+\lambda_4-2\lambda_{\Phi\chi} \right) \right]^2 + 16 v^2 \kappa^2 \; .
\end{align}
    \begin{figure}[t!]
        \centering
        \includegraphics[scale=0.9]{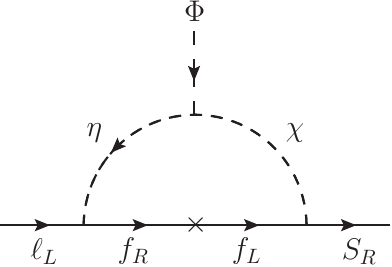} \hspace{+1cm} \includegraphics[scale=0.9]{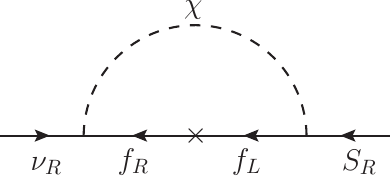}
        \caption{Lowest order one-loop diagrams for neutrino mass generation: dark-seeded LSS (left) and additional correction to the bare heavy neutrino mass (right) (see text for details).}
    \label{fig:neutrinoDLSS}
    \end{figure}
After EWSB, the full tree-level neutrino mass matrix $\bm{\mathcal{M}}_\nu^0$ defined in the $(\nu_L, \nu_R^c,S_R^c)$ basis is
\begin{align}
\bm{\mathcal{M}}_\nu^0 = \begin{pmatrix}
 0 & \mathbf{M}_D & 0    \\
 \mathbf{M}_D^{T} & 0 & M_B    \\
 0 & M_B & 0    \\
\end{pmatrix} \; , \; \mathbf{M}_D = \frac{v \mathbf{Y}_D}{\sqrt{2}} \; .
\label{eq:treemassmatrix}
\end{align}
Obviously, active neutrinos are massless at tree level due to lepton number conservation. However, the explicit lepton number breaking of Eq.~\eqref{eq:VpotentialDLSSsoft} triggers radiative neutrino mass generation with the $\mathcal{Z}_2$ odd fields $\eta$, $\chi$ and $f_{L,R}$ running in the loops, as shown in Fig.~\ref{fig:neutrinoDLSS}. Namely, the left diagram is responsible for inducing a nonzero $m_S$, whose smallness comes from the naturally-small LNV parameter $\kappa$ and dark-loop suppression. Furthermore, the diagram on the right corresponds to a dark radiative correction $\Delta M_B$ to the bare mass term $M_B$. Note that the dark fermion $f_{L,R}$ being a Dirac particle prevents the existence of any higher order corrections to Majorana mass terms for $\nu_L$, $\nu_R$ or $S_R$ (the diagonal entries in the neutrino mass matrix). Taking all this into account, we obtain
\begin{align}
\bm{\mathcal{M}}_\nu = \bm{\mathcal{M}}_\nu^0 + \Delta \bm{\mathcal{M}}_\nu = \begin{pmatrix}
 0 & \mathbf{M}_D & \mathbf{m}_S    \\
 \mathbf{M}_D^T & 0 & M    \\
 \mathbf{m}_S^T & M & 0    \\
\end{pmatrix} \; , \; M = M_B + \Delta M_B \; ,
\end{align}
where $\mathbf{m}_S$ (a $3\times 1$ vector) is the equivalent of the generic tree-level $\mathbf{m}_S$ in the regular LSS. The one-loop mass terms computed in the scalar mass basis can be written as
\begin{align}
 \mathbf{m}_S = \mathcal{F}_{S}(M_f,m_{\zeta_k}) M_f \; Y_S \mathbf{Y}_f \;, \; 
 \Delta M_B  = \mathcal{F}_{B}(M_f,m_{\zeta_k}) M_f \; Y_{R} Y_S \; ,
\label{eq:loops}
\end{align}
where the loop factors are
\begin{align}
\mathcal{F}_{S} \left(M_f , m_{\zeta_k}\right) = & \; \frac{1}{16 \sqrt{2} \pi^2} \sum_{k=1}^{3} \left(\mathbf{V}_{1 k} - i \mathbf{V}_{3 k} \right) \mathbf{V}_{2 k} \frac{m_{\zeta_k}^2}{M_f^2 - m_{\zeta_k}^2} \ln \left( \frac{M_f^2}{m_{\zeta_k}^2} \right) \; ,
\nonumber \\
\mathcal{F}_{B} \left(M_f , m_{\zeta_k}\right) = & \; \frac{1}{16 \pi^2} \sum_{k=1}^{3} \mathbf{V}_{2 k}^2 \frac{m_{\zeta_k}^2}{M_f^2 - m_{\zeta_k}^2} \ln \left( \frac{M_f^2}{m_{\zeta_k}^2} \right) \; .
\label{eq:Floop}
\end{align}
which depend on $M_f$, the dark neutral-scalar masses $m_{\zeta_k}$ ($k=1, 2, 3$) and the mixing matrix of the neutral components of $\eta$ and $\chi$ [see Eq.~\eqref{eq:mixneutralscotoDMDLSS}]. 

In the LSS approximation, i.e. for $m_S \ll M_D \ll M$, following the diagonalization procedure for $\bm{\mathcal{M}}_\nu $ outlined in Sec.~\ref{sec:TypeI}, we obtain the effective light neutrino mass matrix $\mathbf{M}_\nu$ shown in Eq.~\eqref{eq:MnuLSS}. Since we consider the minimal setup containing a single $\nu_R-S_R$ pair, the model presented here generates via a dark-sector the LSS(1,1) scenario. In this case $\mathbf{M}_\nu$ has rank $2$. Therefore, this scenario provides a massless neutrino at one-loop level, while with two $\nu_R-S_R$ all three light neutrinos would be massive. Minimal neutrino mass generation leads to $m_1=0$ for NO and $m_3=0$ for IO, with interesting testability prospects at experiments looking for the $0_\nu \beta \beta$ decay process -- see discussion in Sec.~\ref{sec:neutrinoobservables}. The heavy neutrinos stemming from $\nu_R$ and $S_R$ form a pseudo-Dirac pair, with masses $m_{4,5}$ approximately equal to $M$ apart from terms proportional to the small parameters $m_S$. Note that, as mentioned before in Sec.~\ref{sec:LSS}, the LSS can be accommodated as a low-scale mechanism. In fact, since the smallness of $m_S$ is provided by the LNV parameter $\kappa$ and loop suppression, we can have $m_S \sim \mathcal{O}(1\,{\rm eV})$, which leads to light neutrino masses $\sim \mathcal{O}(0.1\,{\rm eV})$ for a heavy neutrino mass scale at $M \sim \mathcal{O}(1\,{\rm TeV})$. This will have important implications in regards to cLFV~(see Sec.~\ref{sec:cLFV}).

%%%%%%%%%%%%%%%%%%%%%%%%%%%%%%%%%%%%%%%%%%%%%%%%%%%%%%%%%%%%%%%%%%%%%%%%%%%%%
\subsection{Phenomenological analysis: procedure and constraints}
\label{sec:pheno}
%%%%%%%%%%%%%%%%%%%%%%%%%%%%%%%%%%%%%%%%%%%%%%%%%%%%%%%%%%%%%%%%%%%%%%%%%%%%%
%
\begin{table}[!t]
\renewcommand*{\arraystretch}{1.5}
\centering
\begin{tabular}{|K{1.3cm}|K{2cm}|K{3cm}|}   
\hline
Sector & Parameters & Scan range \\
\hline
\multirow{3}{*}{Fermion} & $M_f$ & $[10 , 10^4]$ (GeV)  \\
& $M_B$ & $[10 , 10^5]$ (GeV) \\
& $Y_S, Y_{R}$ & $\left[10^{-3} , 1\right]$ \\
\hline
\multirow{3}{*}{Scalar} & $m_{\eta}^2 , m_{\chi}^2$ &  $[10^2 , 10^8]$ (GeV$^2$) \\
& $\kappa$ &  $[10^{-8}, 10^{2}]$ (GeV)  \\
& $\lambda_{3},|\lambda_{4}|,\lambda_{\Phi \chi}$ & $[10^{-8} , 1]$ \\
\hline
\end{tabular}
\caption{Input parameters of our model and corresponding ranges used in our numerical scan. }
\label{tab:ScanDLSS}
\end{table}

In the upcoming sections, we turn our attention to the cLFV and DM phenomenological implications of our model. We perform a numerical scan with scalar and fermion input parameters [see Eqs.~\eqref{eq:LYukDLSS} and~\eqref{eq:VpotentialfullDLSS}, respectively] as shown in Table~\ref{tab:ScanDLSS}. Since we are interested in a low-scale seesaw scenario, $M_{f,B}$ and $m_{\eta,\chi}^2$ are chosen to be at the GeV-TeV, requiring sizable Yukawa couplings $\mathbf{Y}_{f,D}$. Neutrino mass suppression stems from the loop generated $\mathbf{m}_S$, controlled by the naturally small soft-breaking $\kappa$ parameter taken to be in the $[10^{-8},1]$ range. For simplicity we assume that $Y_{R,S}$ are real and vary them within natural values $[10^{-3},1]$. As for the quartic scalar parameters, shown in the Table, we consider a wide range of values needed for the scalar DM analysis in Sec.~\ref{sec:DMDLSS}. Furthermore, we set the dark-scalar self-couplings to $\lambda_{\eta} = \lambda_{\chi} = \lambda_{\eta \chi} = 0.5$ since they do not modify our analysis.

The scalar sector masses and mixing are computed via Eqs.~\eqref{eq:scotogenicscalarmasses}, \eqref{eq:darkscalarmassDLSS} and~\eqref{eq:mixneutralscotoDMDLSS}. In order for the scalar potential of Eq.~\eqref{eq:VpotentialfullDLSS} to be bounded from below~(BFB), the quartic parameters must obey the following conditions:
    \begin{align}
    &\lambda_{\Phi},\lambda_{\eta},\lambda_{\chi} > 0\,, \nonumber \\
    &\lambda_{3} + \sqrt{\lambda_\Phi \lambda_\eta} > 0\;,\;\lambda_{3} + \lambda_{4} + \sqrt{\lambda_\Phi \lambda_\eta} > 0 \; , \; \lambda_{\Phi \chi} + \sqrt{\lambda_\Phi \lambda_{\chi}} > 0 \; , \; \lambda_{\eta \chi} + \sqrt{\lambda_\eta \lambda_{\chi}} > 0 \; , \nonumber \\
    &\lambda_{3}\lambda_{\chi} - \lambda_{\Phi \chi} \lambda_{\eta \chi} + \sqrt{(\lambda_\phi \lambda_{\chi} - \lambda_{\Phi \chi}^2) (\lambda_\eta \lambda_{\chi} - \lambda_{\eta \chi}^2)} > 0\;, \nonumber \\ &(\lambda_{3} + \lambda_4)\lambda_{\chi} - \lambda_{\Phi \chi} \lambda_{\eta \chi} + \sqrt{(\lambda_\phi \lambda_{\chi} - \lambda_{\Phi \chi}^2) (\lambda_\eta \lambda_{\chi} - \lambda_{\eta \chi}^2) } > 0\;. \label{eq:bounded} 
    \end{align}

    A convenient way to obtain Yukawa coupling vectors $\mathbf{Y}_{D,f}$ of Eq.~\eqref{eq:LYukDLSS}, that lead to compatibility with neutrino data, is via the Casas-Ibarra parameterization~\cite{Casas:2001sr}. We start by writing $\mathbf{M}_\nu$ of Eq.~\eqref{eq:MnuLSS} as follows,
\begin{equation}
  \mathbf{M}_{\nu} = - \frac{v^2}{2} \mathbf{Y} \; \mathbf{M}^{-1} \mathbf{Y}^T \; ,\; \mathbf{Y} = \left(\mathbf{Y}_{D} \; \mathbf{Y}_{f}\right) \; , \; \mathbf{M} = \begin{pmatrix}
      0 & |\tilde{M}| e^{i \phi}\\
      |\tilde{M}| e^{i \phi} & 0
  \end{pmatrix} \; , 
\end{equation}
where we define the $3\times 2$ Yukawa matrix $\mathbf{Y}$, the $2\times 2$ mass matrix $\mathbf{M}$ and
\begin{align}
    \tilde{M} = v \frac{M_B+\mathcal{F}_{B} M_f \; Y_{R} Y_S}{M_f} \frac{1}{\sqrt{2} Y_S \mathcal{F}_S} \; , \; \phi = \arg \tilde{M} \; ,
\end{align}
with the masses $M_B,M_f$ and Yukawa parameters $Y_R,Y_{S}$ and dark-loop functions $\mathcal{F}$ given in Eqs.~\eqref{eq:LYukDLSS} and~\eqref{eq:Floop}, respectively. The matrix $\mathbf{M}$ can be diagonalized as
\begin{align}
\U_M^{\dagger}\, \M\, \U_M^\ast &= \mathbf{D}_{M} = \text{diag}\left(|\tilde{M}| , |\tilde{M}| \right) \; , \;
\U_M = \frac{e^{i \phi}}{\sqrt{2}} \begin{pmatrix}
    i & 1 \\
    -i & 1
\end{pmatrix} \; . 
\label{eq:Mheavy}
\end{align}
From the above we find the following parameterization for $\mathbf{Y}$:
\begin{align} 
    \mathbf{Y} &= i \frac{\sqrt{2}}{v} \mathbf{U}_{\nu}^\ast\, \mathbf{D}_\nu^{1/2}\, \mathbf{R}\, \mathbf{D}_M^{1/2} \mathbf{U}_M^T \; ,
\end{align}
where $\mathbf{D}_\nu$, $\mathbf{U}_{\nu}$, $\mathbf{D}_M$ and $\mathbf{U}_M$ are given by Eqs.~\eqref{eq:leptonmixing} and~\eqref{eq:Mheavy}, respectively. Furthermore, $\mathbf{R}$ is an orthogonal $3\times 2$ matrix
with $z$ being a complex parameter, given by
\begin{align}
    \mathbf{R}_\text{NO} &= \begin{pmatrix}
        0 & 0 \\
        \cos (z) & - \sin (z) \\
        \pm \sin (z) & \pm \cos (z)
    \end{pmatrix} \; , \; 
   \mathbf{R}_\text{IO} = \begin{pmatrix}
        \cos (z) & - \sin (z) \\
        \pm \sin (z) & \pm \cos (z) \\
         0 & 0
    \end{pmatrix} \; . 
\end{align}
In our numerical analysis we take a real angle $z \in [0,2\pi]$, and $\alpha \in [0,2\pi]$ for the Majorana phase. Note that, we focus on the currently favored NO neutrino masses and vary the values for the neutrino oscillation observables (neutrino mass squared-differences, mixing angles and Dirac CP phase) within their $3 \sigma$ ranges using the results obtained by the global data fit of Ref.~\cite{deSalas:2020pgw}, which are shown in Table~\ref{tab:leptondata} of Sec.~\ref{sec:neutrino}.

We apply the following collider constraints in our numerical scan: 
\begin{itemize}
    
    \item \textit{$Z$-boson decay:} The precise LEP-I measurements on the $Z$-boson decay width leads to the lower bounds on the dark-neutral scalar masses, $m_{\zeta_i} > m_Z/2 = 45.6$ GeV and $m_{\zeta_i} + m_{\zeta_j} > m_Z = 91.2$ GeV ($i , j = 1, 2, 3$)~\cite{Cao:2007rm,Gustafsson:2007pc}. The latter ensures that the decays $Z \rightarrow \zeta_i \zeta_k$ are kinematically forbidden. Also, reinterpreting LEP-II results for chargino searches in the context of singly-charged scalar production, leads to the conservative bound on the dark-charged scalar mass $m_{\eta^\pm} > 70$ GeV~\cite{Pierce:2007ut}.

     \item \textit{Higgs invisible decay}: The odd-neutral scalars $\zeta_i$ ($i= 1, 2, 3$) contribute to the Higgs invisible decay width through the channels $h \rightarrow \zeta_i \zeta_j$ ($i,j= 1, 2, 3$), which open up for masses $m_{\zeta_i} + m_{\zeta_j} < m_{h}$. The decay width and BR are
    \begin{align}
    &\Gamma(h \rightarrow \text{inv}) = \frac{1}{2} \sum_{i, j} \Gamma(h \rightarrow \zeta_i \zeta_j) \; , \nonumber \\  
    & \BR(h \rightarrow \text{inv}) = \frac{\Gamma(h \rightarrow \text{inv})}{\Gamma_{h}^{\text{tot}}}  = \frac{\Gamma(h \rightarrow \text{inv})}{\Gamma_{h}^{\text{SM}} + \Gamma(h \rightarrow \text{inv})} \; ,
    \label{eq:BoundBRh1inv_1}
    \end{align}
    where $\Gamma_{h}^{\text{SM}} = 3.2^{+2.8}_{-2.2} \; \text{MeV}$ is the SM Higgs decay width~\cite{ParticleDataGroup:2024cfk}. The BR above is constrained by the LHC Higgs data, being the current bound~\cite{ParticleDataGroup:2024cfk}
    \begin{equation}
         \BR(h \rightarrow \text{inv}) \leq 0.19 \; .
         \label{eq:BoundBRh1inv}
    \end{equation}

    \item \textit{Higgs diphoton decay}: The odd-charged scalar $\eta^\pm$ will contribute at loop level to the diphoton Higgs decay $h \rightarrow \gamma \gamma$ We define the $h \rightarrow \gamma \gamma$ signal strength as
    \begin{equation}
    R_{\gamma \gamma} = \frac{\BR(h \rightarrow \gamma \gamma)}{\BR_{\text{SM}}(h \rightarrow \gamma \gamma)} \; ,
    \end{equation}
    where we use the following SM value for $\BR(h \rightarrow \gamma \gamma)$ and constraint on $R_{\gamma \gamma}$~\cite{ATLAS:2018hxb},
    \begin{equation}
        \BR_{\text{SM}}(h \rightarrow \gamma \gamma) = 2.27 \times 10^{-3} \; , \; R_{\gamma \gamma} = 0.99^{+0.15}_{-0.14} \; .
    \label{eq:BoundRgg}
    \end{equation}
    
\end{itemize}
%

%%%%%%%%%%%%%%%%%%%%%%%%%%%%%%%%%%%%%%%%%%%%%%%%%%%%%%%%%%%%%%%%%%%%%%%%%%%%%
\subsection{Charged lepton flavor violation}
\label{sec:cLFV}
%%%%%%%%%%%%%%%%%%%%%%%%%%%%%%%%%%%%%%%%%%%%%%%%%%%%%%%%%%%%%%%%%%%%%%%%%%%%%

%
    \begin{figure}[t!]
        \centering
        \includegraphics[scale=0.75]{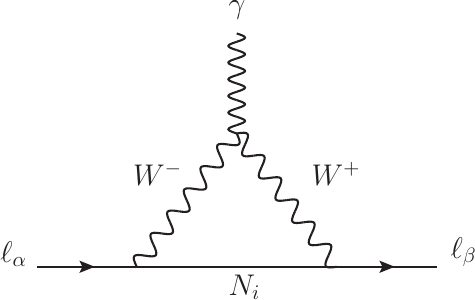} \hspace{+1cm} \includegraphics[scale=0.75]{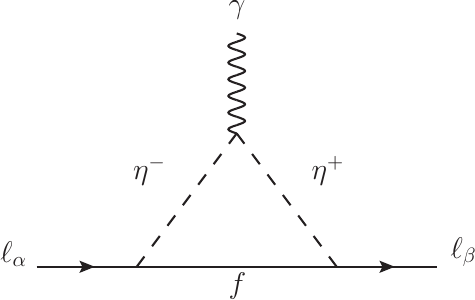}
        \caption{The $\gamma$-penguin diagrams that contribute to $\ell_{\alpha} \rightarrow \ell_{\beta} \gamma$: on the left (right) low-scale seesaw (dark-sector) contribution.}
    \label{fig:cLFVdiags}
    \end{figure}
Searches for lepton flavor violating processes like $\ell_{\alpha} \rightarrow \ell_{\beta} \gamma$, $\ell_{\alpha} \rightarrow 3 \ell_{\beta}$ and $\mu - e$ conversion in nuclei are important probes of physics beyond the SM. These become even more relevant in low-scale seesaw scenarios since heavy-light neutrino mixing can be sizable and induce non-negligible contributions to cLFV~\cite{Minkowski:1977sc,Marciano:1977wx,Cheng:1980tp,Lim:1981kv,Langacker:1988up,Ilakovac:1994kj,Alonso:2012ji,Abada:2015oba,Abada:2018nio,Hernandez-Tome:2019lkb,Camara:2020efq}. In our framework, there are additional loop contributions mediated by the dark-sector particles $\eta^\pm$ and $f$~\cite{Toma:2013zsa,Vicente:2014wga,Ahriche:2016cio,Hagedorn:2018spx,Barreiros:2022aqu}. Here, we will focus on radiative $\ell_{\alpha} \rightarrow \ell_{\beta} \gamma$ decays generated via the one-loop diagrams shown in Fig.~\ref{fig:cLFVdiags}. Namely, the left diagram corresponds to the contribution from heavy neutrino states via the $W^\pm$-boson loop, while the right diagram is the dark-sector contribution. The BR for this process is given by
\begin{align}
\frac{\BR(\ell_{\alpha} \rightarrow \ell_{\beta} \gamma)}{\BR\left(\ell_{\alpha} \rightarrow \ell_{\beta} \nu_{\alpha} \overline{\nu_{\beta}} \right)} & = \frac{3 \alpha_{e}}{2 \pi} \left| G_{W^\pm}^{\alpha \beta}+ G_{\eta^{\pm}}^{\alpha \beta}\right|^2 \nonumber \\
& = \frac{3 \alpha_{e}}{2 \pi} \left|\sum_{i=1}^5 (\mathbf{W}_\nu^\ast)_{\alpha i} (\mathbf{W}_\nu)_{\beta i} G_{W^\pm}\left(\frac{m_{i}^2}{M_W^2}\right)+\frac{v^2}{4 m_{\zeta^\pm}^2} \mathbf{Y}_{f}^{\beta} \mathbf{Y}_{f}^{\alpha \ast}  G_{\eta^{\pm}}\left(\frac{M_f^2}{m_{\zeta^{+}}^2}\right)\right|^2 \; ,
\label{eq:cLFV}
\end{align}
where $\alpha_{e} = e^2/(4\pi)$, $\BR\left(\mu \rightarrow e \nu_{\mu} \overline{\nu_{e}}\right) \simeq 1.0$, $\BR\left(\tau \rightarrow e \nu_{\tau} \overline{\nu_{e}}\right) \simeq 0.18$ and $\BR\left(\tau \rightarrow \mu \nu_{\tau} \overline{\nu_{\mu}}\right) \simeq 0.17$~\cite{ParticleDataGroup:2024cfk}. The mixing matrix $\mathbf{W}_\nu$ is given in Eq.~\eqref{eq:WnucLFV} with the LSS parameter $\mathbf{\Theta}_\nu$ of Eq.~\eqref{eq:FLSS}. Furthermore, $m_{i}$ are the neutrino masses, the dark charged-scalar mass $m_{\eta^{\pm}}$ is given by Eq.~\eqref{eq:scotogenicscalarmasses}, while $\mathbf{Y}_f$ and $M_f$ have been defined in Eq.~\eqref{eq:LYukDLSS}. The loop functions $G_{W^\pm}(x)$~\cite{Cheng:1980tp,Lim:1981kv,Langacker:1988up,Ilakovac:1994kj} and $G_{\eta^{\pm}}(x)$~\cite{Toma:2013zsa,Vicente:2014wga} are
\begin{align}
G_{W^\pm}(x) &= \frac{x (1-5x-2x^2)}{4 (1-x)^3} - \frac{3 x^3 \ln x }{2 (1-x)^4} \; ,  \nonumber \\ G_{\eta^{\pm}}(x) &= \frac{1 - 5 x - 2 x^2}{6 (1-x)^3} - \frac{x^2 \ln x}{(1-x)^4} \; .
\end{align}
\begin{figure}[!t]
   \centering
   \includegraphics[scale=0.33]{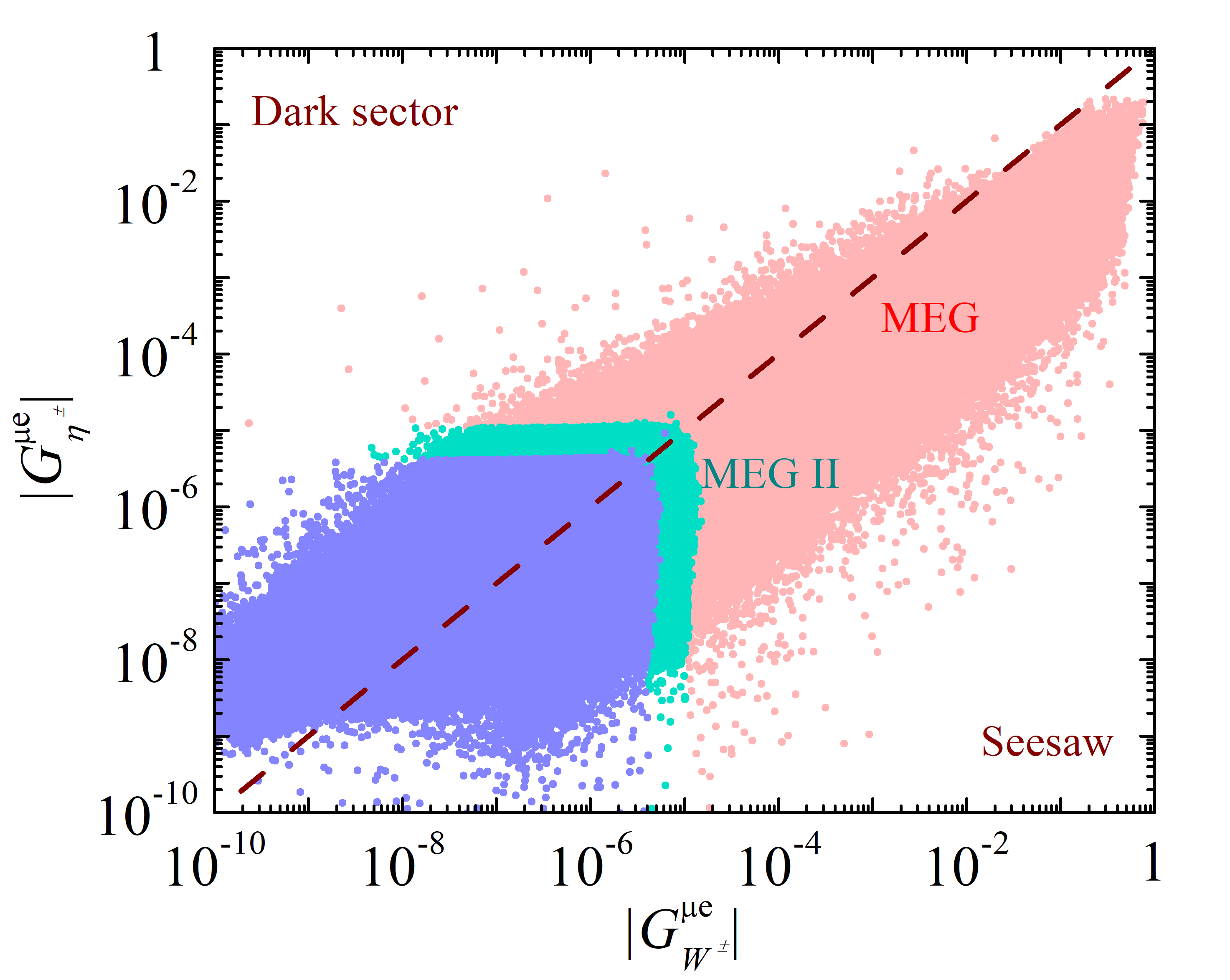} \\ 
  \vspace{+0.3cm} \includegraphics[scale=0.33]{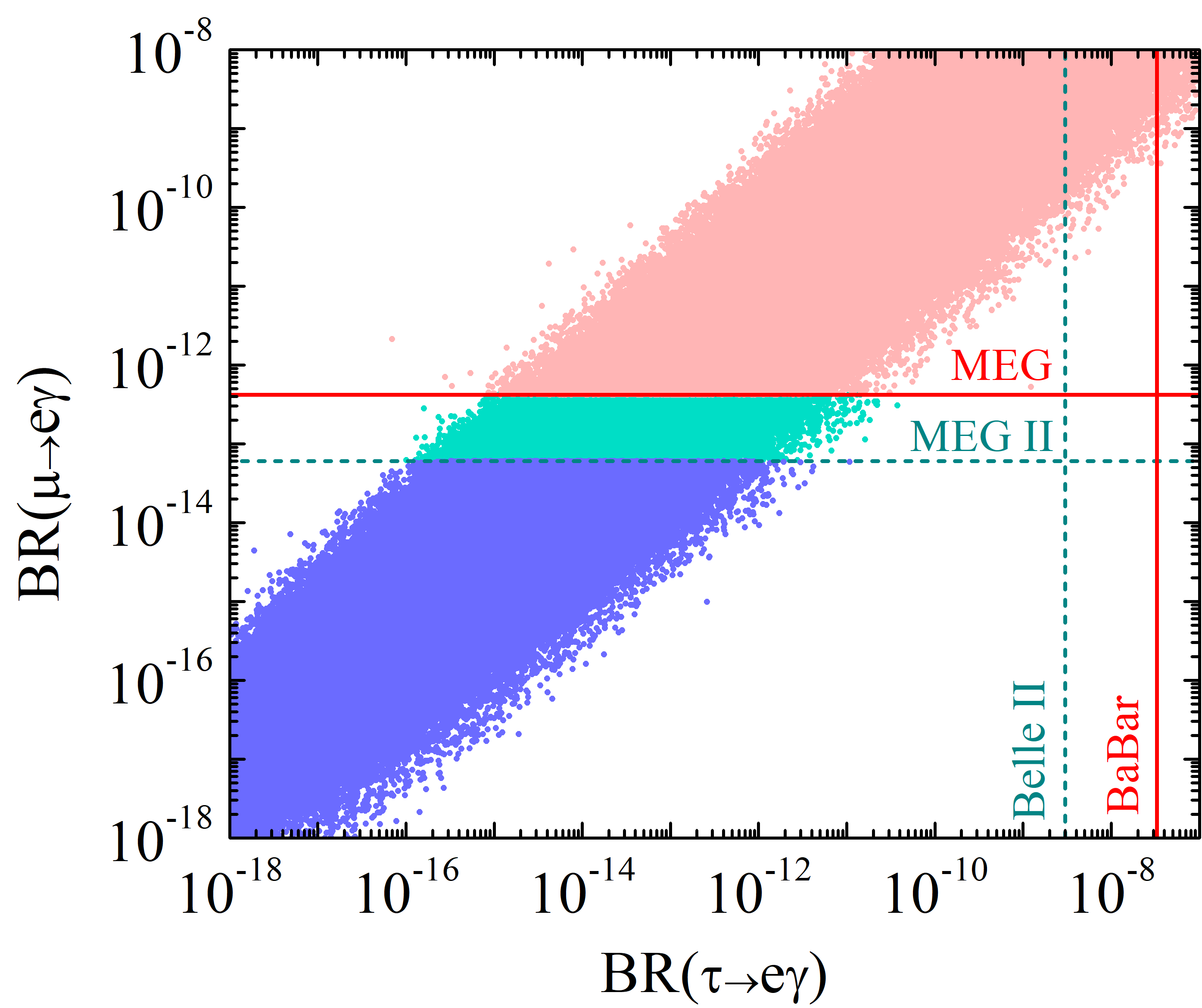} \includegraphics[scale=0.33]{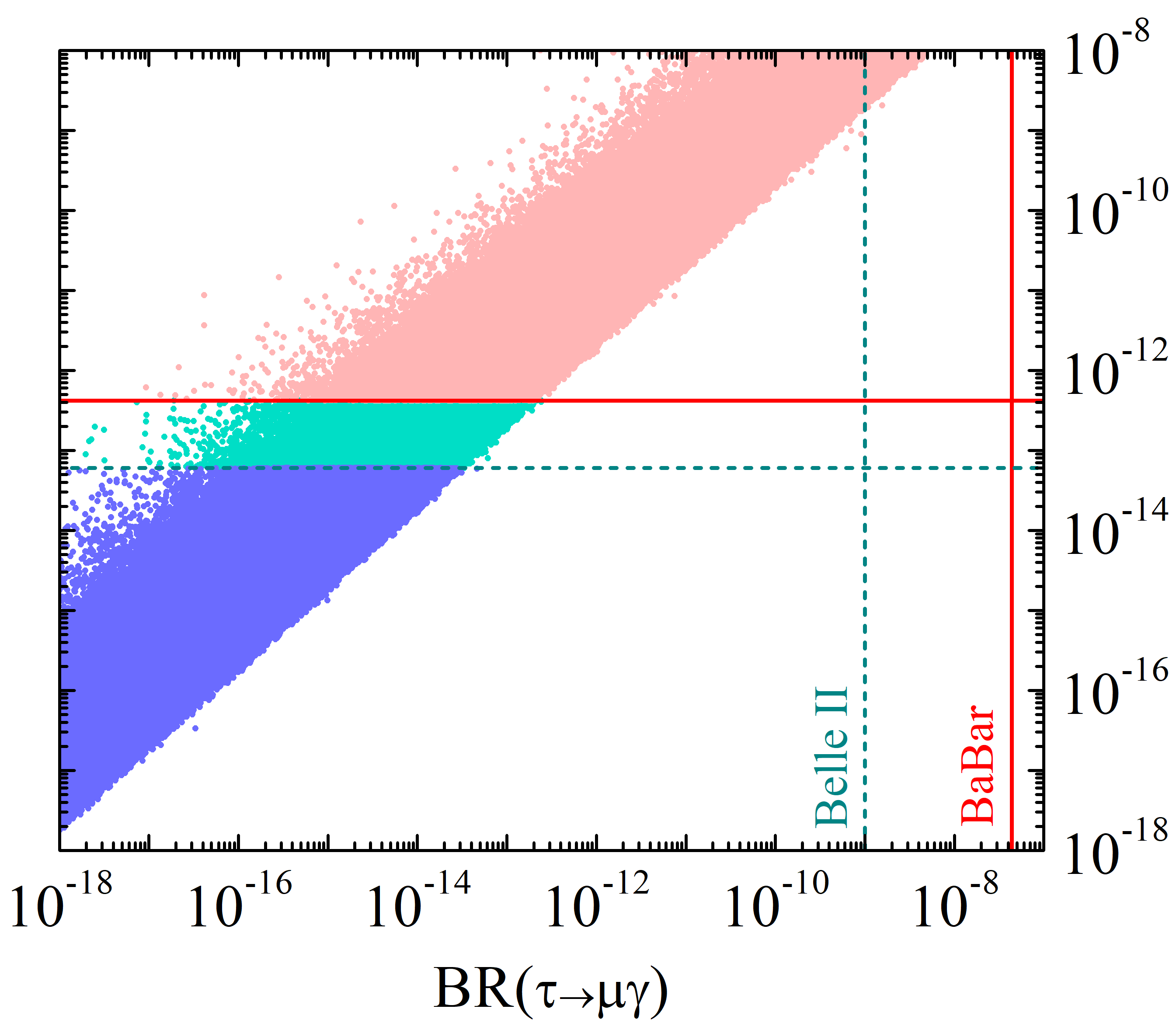}
  \caption{Top: $|G^{\mu e}_{\eta^\pm}|$ versus $|G^{\mu e}_{W^\pm}|$ [see Eq.~\eqref{eq:cLFV} and discussion therein]. The red points are excluded by MEG~\cite{MEG:2016leq} while the green ones can be probed by MEG II~\cite{MEGII:2018kmf}. Along the brown-dashed contour $|G^{\mu e}_{\eta^\pm}|=|G^{\mu e}_{W^\pm}|$ separating the dark-sector from the seesaw cLFV dominance regime. Bottom: $\BR(\mu \rightarrow e \gamma)$ versus $\BR(\tau \rightarrow e \gamma)$ (left) and $\BR(\tau \rightarrow \mu \gamma)$ (right). Current cLFV bounds and projected sensitivities are marked by a red-solid and green-dashed lines, respectively.}
  \label{fig:cLFV}
\end{figure}
Our numerical results are presented in Fig.~\ref{fig:cLFV}. In the upper plot, we show $|G^{\mu e}_{\eta^\pm}|$ versus $|G^{\mu e}_{W^\pm}|$ [see Eq.~\eqref{eq:cLFV}]. Along the brown-dashed contour $|G^{\mu e}_{\eta^\pm}| = |G^{\mu e}_{W^\pm}|$, above (below) which the dark (seesaw) contribution dominates, i.e. $|G^{\mu e}_{\eta^\pm}| > |G^{\mu e}_{W^\pm}|$ ($|G^{\mu e}_{\eta^\pm}| < |G^{\mu e}_{W^\pm}|$). The red points are excluded by the current MEG bound BR$(\mu \rightarrow e \gamma) < 4.2 \times 10^{-13}$~\cite{MEG:2016leq}, while the green ones are within the sensitivity reach of MEG II~\cite{MEGII:2018kmf}. Cancellations between $G^{\mu e}_{\eta^\pm}$ and $G^{\mu e}_{W^\pm}$ along the brown-dashed line allow for small BRs with large $|G^{\mu e}_{\eta^\pm}|$ and $|G^{\mu e}_{W^\pm}|$. In the lower plots, we present $\BR(\mu \rightarrow e \gamma)$ versus $\BR(\tau \rightarrow e \gamma)$ (left) and $\BR(\tau \rightarrow \mu \gamma)$ (right). It is apparent from our results that once the current MEG constraint is applied (horizontal red-solid line), the constraints on the $\tau$ radiative processes stemming from BaBar~\cite{BaBar:2009hkt} (current - vertical red-solid line) are automatically satisfied, leading to decay rates beyond the sensitivity reach of Belle II~\cite{Belle-II:2018jsg} (future - vertical green-dashed line).

%%%%%%%%%%%%%%%%%%%%%%%%%%%%%%%%%%%%%%%%%%%%%%%%%%%%%%%%%%%%%%%%%%%%%%%%%%%%%
\subsection{Dark matter}
\label{sec:DMDLSS}
%%%%%%%%%%%%%%%%%%%%%%%%%%%%%%%%%%%%%%%%%%%%%%%%%%%%%%%%%%%%%%%%%%%%%%%%%%%%%

%
\begin{figure}[t!]
    \centering
    \includegraphics[scale=0.7]{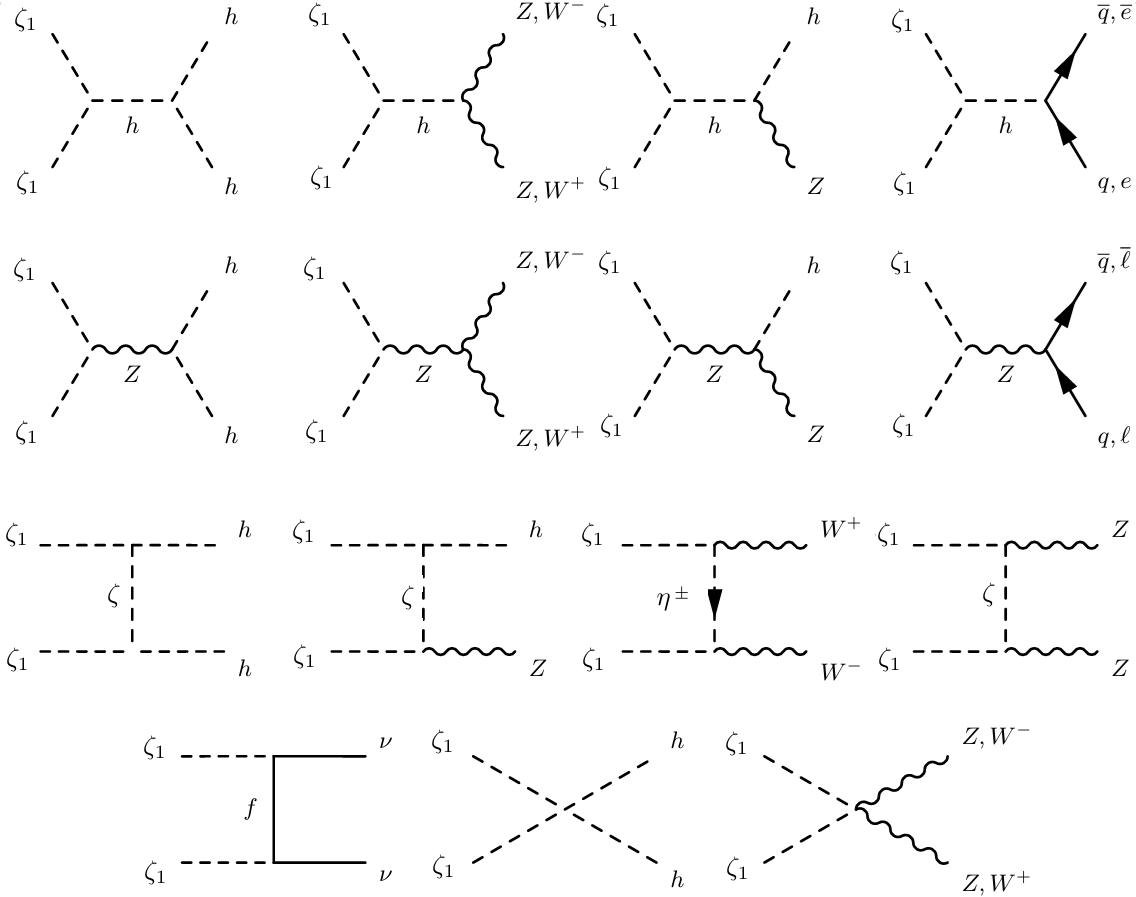}
    \caption{Annihilation diagrams for scalar DM. These coincide with the ones contributing to $\zeta_1$ and $\zeta_{k}$ ($k=2,3$) co-annihilation, after making the replacement of one initial $\zeta_1$ state by $\zeta_k$.}
    \label{fig:DiagAnnScalarDM}
\end{figure}
\begin{figure}[t!]
    \centering
    \includegraphics[scale=0.7]{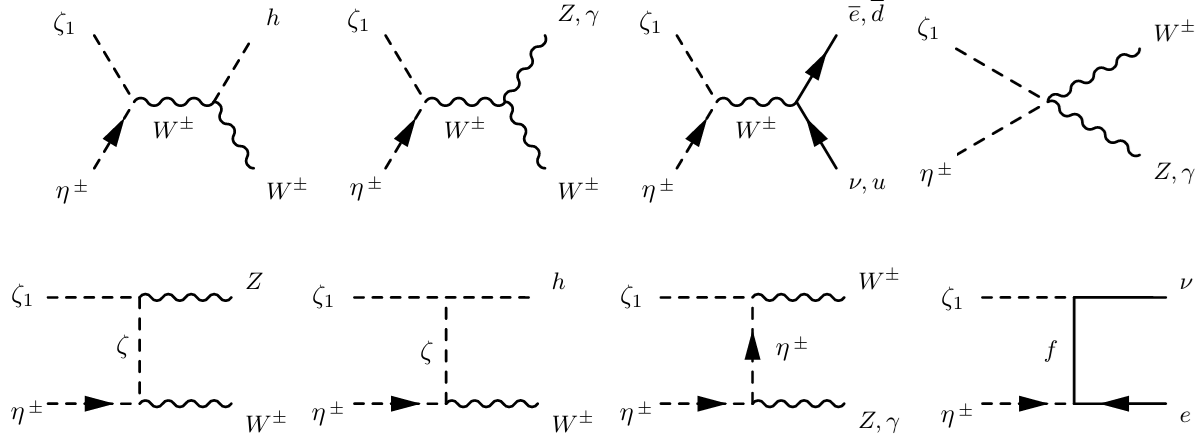}
    \caption{Diagrams contributing to co-annihilation of $\zeta_1$ with $\eta^\pm$.}
    \label{fig:DiagCoAnn1}
\end{figure}
\begin{figure}[t!]
    \centering
    \includegraphics[scale=0.7]{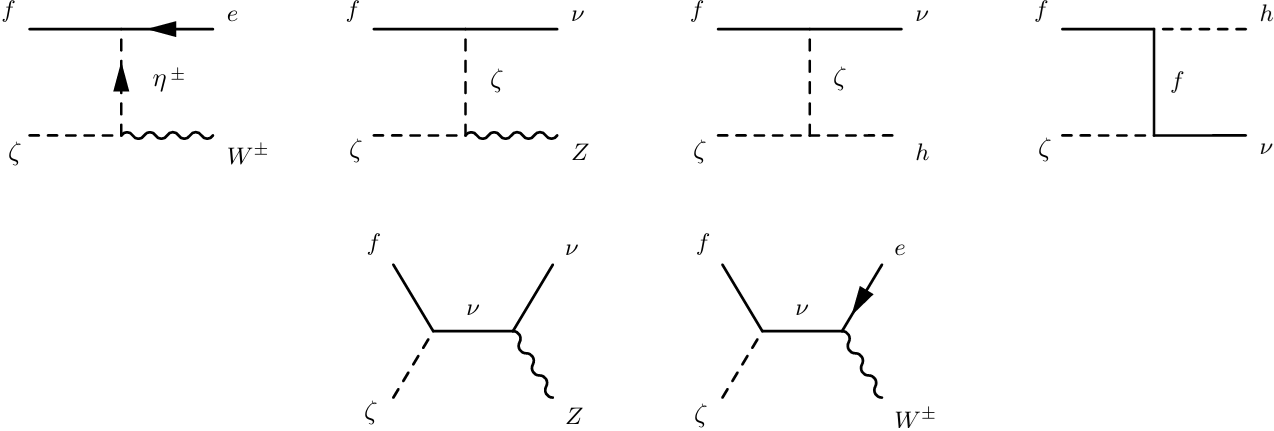}
    \caption{Diagrams contributing to co-annihilation of $f$ with $\zeta$.}
    \label{fig:DiagCoAnn3}
\end{figure}

The presence of the $\mathcal{Z}_2$ symmetry ensures the stability of the lightest $\mathcal{Z}_2$ odd particle. Thus, the fermion $f$ and the scalar $\zeta_1$ (which is a mixed state of the inert scalar doublet $\eta$ and singlet $\chi$) can be viable WIMP DM candidates (see Sec.~\ref{sec:WIMPDM}). Here, we focus on the scalar DM phenomenology, i.e. when the lightest dark particle is $\zeta_1$. In Fig.~\ref{fig:DiagAnnScalarDM} we show the $\zeta_1-\zeta_1$ annihilation channels into $\mathcal{Z}_2$ even particles and coannihilations $\zeta_1- \zeta_k$ ($k= 2, 3$), while in Figs.~\ref{fig:DiagCoAnn1} and~\ref{fig:DiagCoAnn3} are shown respectively the $\zeta_1-\zeta^{\pm}$ and $\zeta-f$ coannihilations diagrams. All these diagrams contribute to the thermaly-averaged cross section $\left<\sigma(\text{DM} \ \text{DM} \rightarrow \text{nonDM} \ \text{nonDM}) v\right>$, where ``nonDM" refers to SM and non-SM even particles, i.e. all particles that are not dark. Overall, the DM relic density will be produced via thermal freeze-out exhibiting the cross-section dependence $\Omega h^2 \propto \left<\sigma v\right>^{-1}$ -- see discussion in Sec.~\ref{sec:WIMPDM}. In our analysis we will use the $3 \sigma$ range for the CDM relic density obtained by the Planck satellite data~\cite{Planck:2018vyg},
\begin{equation}
   0.1126 \leq \Omega_{\text{CDM}} h^2 \leq 0.1246 \; .
   \label{eq:Oh2Planck}
\end{equation}

To check whether $\zeta_1$ is a viable DM candidate, we implement our model in \texttt{SARAH}~\cite{Staub:2013tta} and use \texttt{SPheno}~\cite{Porod:2003um} to obtain mass matrices, vertices and tadpole equations. The relic density $\Omega h^2$ and WIMP-nucleon SI elastic scattering cross-section $\sigma^{\text{SI}}$ are computed at tree-level by the \texttt{micrOMEGAs}~\cite{Belanger:2014vza} package.

\begin{figure}[!t]
    \centering
    \includegraphics[scale=0.45]{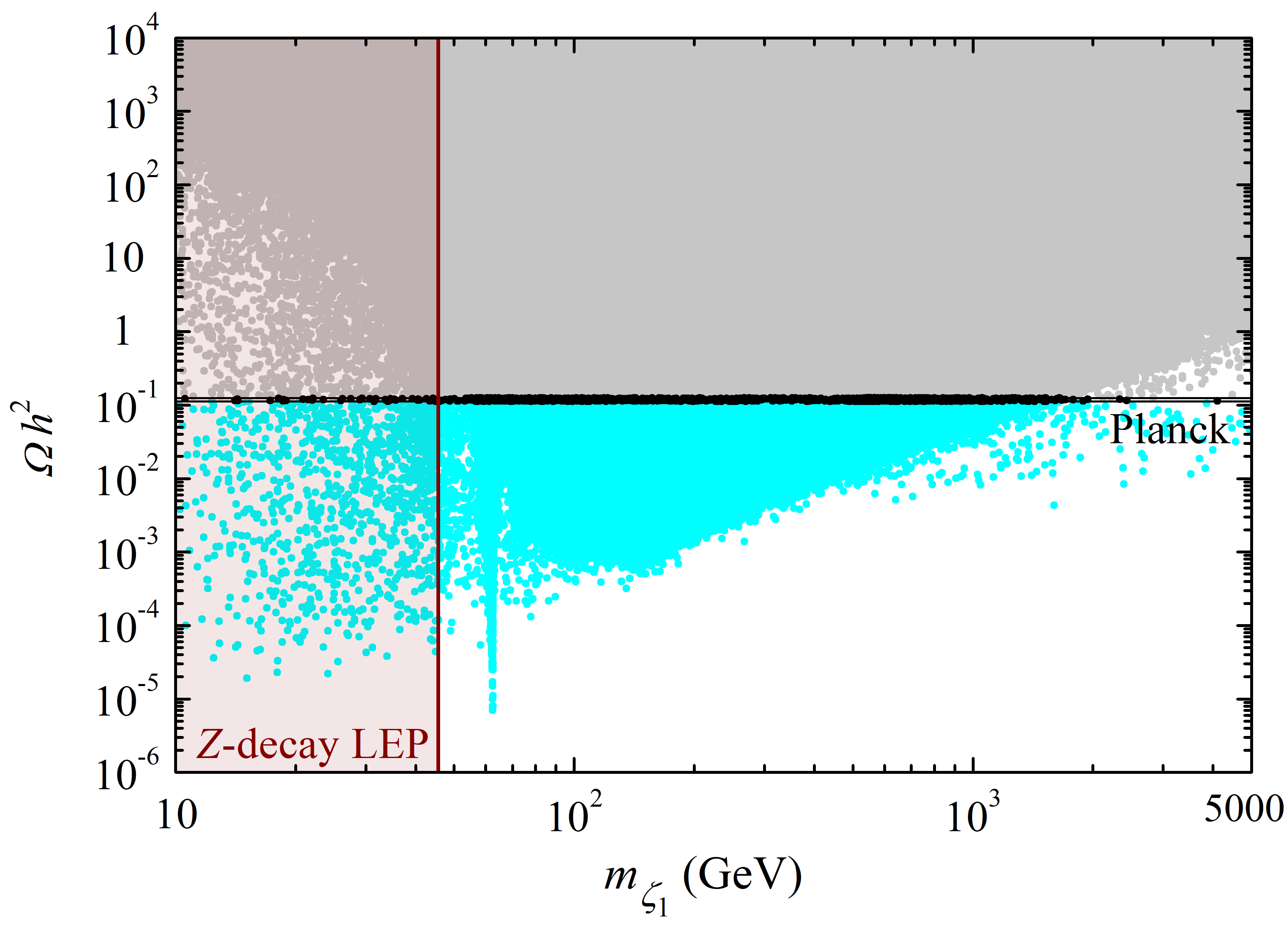}
    \caption{Relic density $\Omega h^2$ versus the scalar DM mass $m_{\zeta_1}$. Black points lie within the $3 \sigma$ range of the measured relic abundance obtained by Planck~\cite{Planck:2018vyg} [see Eq.~\eqref{eq:Oh2Planck}]. The grey (cyan) points depict overabundant (underabundant) DM. The red-shaded region is excluded by LEP constraints.}
    \label{fig:DMrelicDLSS}
\end{figure}
The results regarding DM abundance are presented in Fig.~\ref{fig:DMrelicDLSS}, where we show $\Omega h^2$ as a function of the scalar DM mass $m_{\zeta_1}$. The overabundant DM points shown in gray are excluded, while underabundance shown in cyan is still viable. The black points are within the 3$\sigma$ interval for the DM relic density obtained by Planck~\cite{Planck:2018vyg} [see Eq.~\eqref{eq:Oh2Planck}]. We remark that there is a dip at $m_{\zeta_1} \sim m_h/2$ due to DM annihilation mediated by the SM Higgs which becomes very efficient when it is on-shell (see Fig.~\ref{fig:DiagAnnScalarDM}). Furthermore, for masses $m_{\zeta_1} \gsim m_W, m_Z$, the annihilation processes into a pair of gauge bosons opens up leading to another dip in $\Omega h^2$. The most interesting feature of the model stems from the mixing between the neutral components of the dark scalars $\eta$ and $\chi$. Namely, when DM is dominated by the doublet $\eta$, the relic density behavior is similar to that in the inert/scotogenic model (see Sec.~\ref{sec:scotogenic}) as shown in the lower thick band in the figure~\cite{Belyaev:2016lok,Mandal:2021yph,Avila:2019hhv}. However, the presence of the singlet $\chi$ via mixing with the doublet $\eta$ causes the existence of several points above this curve and consequently we have correct relic density points for a large mass interval $45 \ \text{GeV} \lsim m_{\zeta_1} \lsim 2 \ \text{TeV}$.

    \begin{figure}[t!]
        \centering
        \includegraphics[scale=0.6]{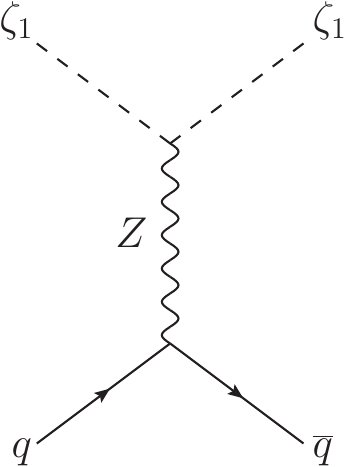} \hspace{+1cm} \includegraphics[scale=0.6]{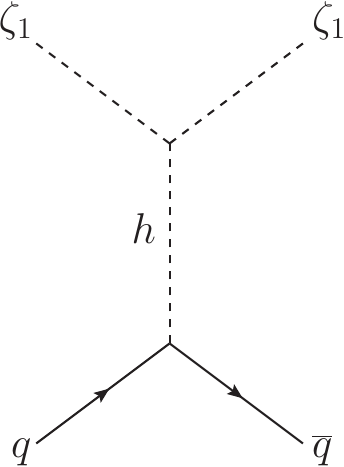}
        \caption{Tree-level diagrams contributing to WIMP-nucleon SI elastic cross-section for scalar DM.}
    \label{fig:DDdiags}
    \end{figure}
\begin{figure}[!t]
    \centering
    \includegraphics[scale=0.45]{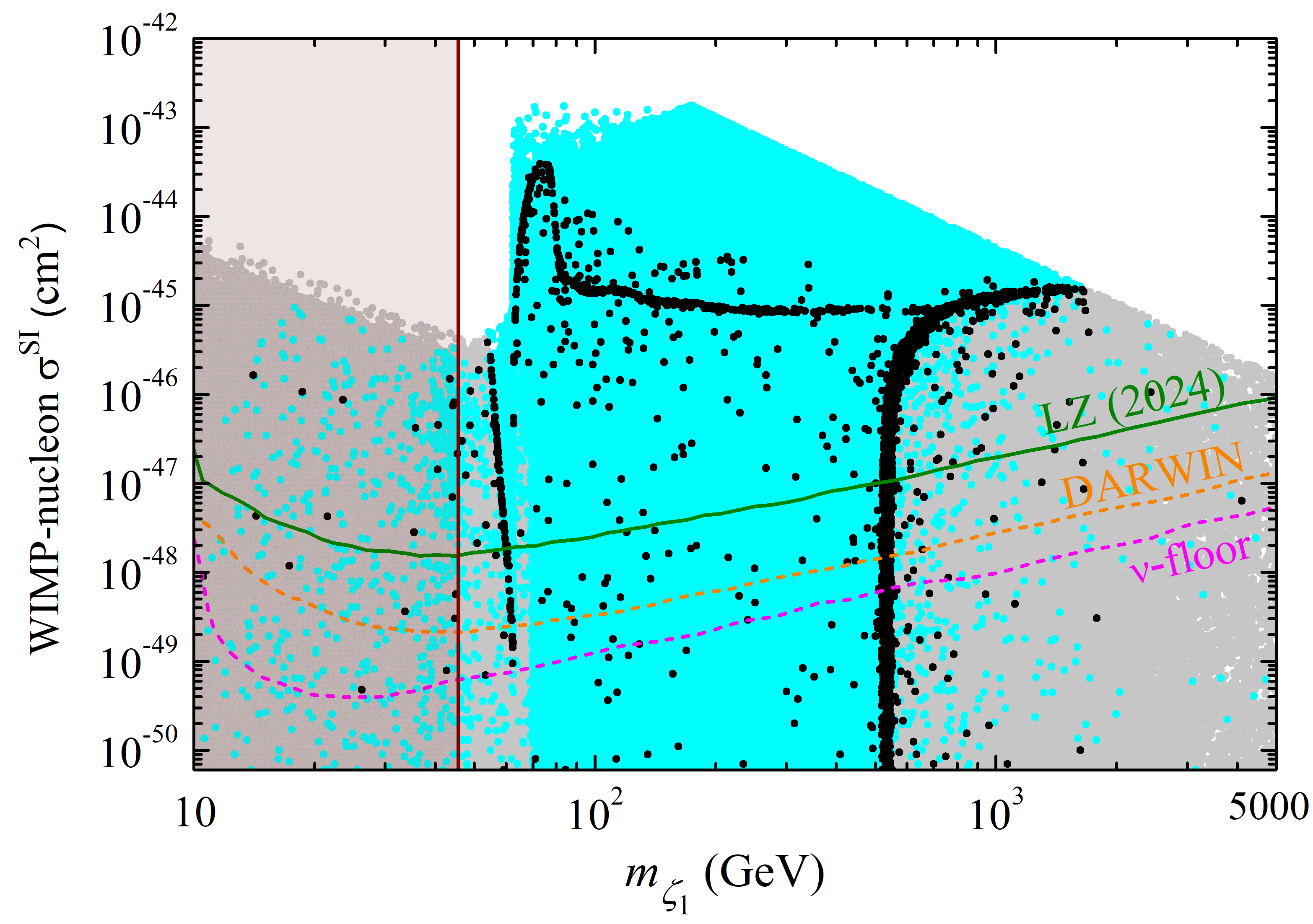}
    \caption{WIMP-nucleon SI elastic scattering cross-section $\sigma^{\text{SI}}$ versus the scalar DM mass $m_{\zeta_1}$. The solid green line indicates the current bound from the DD experiment LZ~\cite{LZ:2022lsv,LZ:2024zvo}. The orange-dashed contour indicates the projected sensitivity for DARWIN~\cite{DARWIN:2016hyl}. The pink-dashed line is the ``neutrino floor"~\cite{Billard:2013qya}. The remaining elements in the figure follow the color code of Fig.~\ref{fig:DMrelicDLSS}.}
    \label{fig:DMDDDLSS}
\end{figure}
In Fig.~\ref{fig:DMDDDLSS}, we show the WIMP-nucleon $\sigma^{\text{SI}}$ as a function of $m_{\zeta_1}$. Note that $\zeta_1$ will contribute to $\sigma^{\text{SI}}$ via the tree-level diagrams mediated by the SM Higgs boson $h$ and the $Z$-boson, shown in Fig.~\ref{fig:DDdiags}, being the dominant contribution the one mediated by $h$. Also, there are several points between $45 \ \text{GeV} \lsim m_{\zeta_1} \lsim 2 \ \text{TeV}$ with the correct relic density that satisfy the current most stringent DD bound imposed by the LZ experiment~\cite{LZ:2022lsv} (solid green line) (for details on DD experiments see discussion in Sec.~\ref{sec:DM}). Most of the points that evade the bound lie around $m_{\zeta_1} \sim 500$ GeV as in the canonical scotogenic model. Furthermore, a large part of the parameter space will be probed by future experiments such as DARWIN~\cite{DARWIN:2016hyl} (orange-dashed contour), down to the ``neutrino floor" (pink-dashed contour) coming from CE$\nu$NS~\cite{Billard:2013qya}.

The alternative fermion DM case, i.e. where the lightest dark particle is $f$, would be very similar to what happens in scotogenic-type models. Namely, for $M_f \gsim 45$ GeV the correct value for the relic density [see Eq.~\eqref{eq:Oh2Planck}] can be achieved as long as coannihilation channels between $f$ and $\zeta,\zeta^\pm$ are taken into account (these are important if the relative dark fermion-scalar mass difference is below $\sim 10 \%$~\cite{Mahanta:2019gfe,Ahriche:2016cio,Hagedorn:2018spx,Barreiros:2022aqu}).

%%%%%%%%%%%%%%%%%%%%%%%%%%%%%%%%%%%%%%%%%%%%%%%%%%%%%%%%%%%%%%%%%%%%%%%%%%%%%
\subsection{Key ideas and outlook}
\label{sec:concl}
%%%%%%%%%%%%%%%%%%%%%%%%%%%%%%%%%%%%%%%%%%%%%%%%%%%%%%%%%%%%%%%%%%%%%%%%%%%%%

We have presented a minimal model where neutrino masses stem from the LSS mechanism radiatively ``seeded'' by a dark sector. Namely, the required small LNV parameter arises at one-loop mediated by viable DM candidates. We have considered the simplest setup containing only one $\nu_R-S_R$ heavy neutrino pair, as well as a single dark vector-like fermion $f_{L,R}$. This is sufficient to explain the two neutrino mass splittings, predicting a massless light neutrino. Although the model may accommodate either fermionic or scalar DM, we focused on the latter possibility with DM being a mixed state of SU$(2)_L$ singlet and doublet scalars. Due to this feature, DM with the right relic abundance can be obtained for a wide mass range of $\sim [45~{\rm GeV},2~{\rm TeV}]$. This scenario can also be directly probed at future DD experiments such as LZ, XENONnT and DARWIN. Being a low-scale seesaw scheme, the dark-seeded LSS leads to sizable cLFV signatures making the seesaw paradigm testable at future cLFV facilities like MEG II and Belle II, with distinct contributions from the dark sector.

Overall, our setup is the simplest framework in which a dark-sector induces the small LNV parameter generating non-zero neutrino masses via the LSS mechanism, without contributing to any other mass term which could generate light neutrino masses through another mechanism. This feature is due to the Dirac character of $f_{L,R}$. Other approaches have been followed in the literature, for instance with a scalar triplet~\cite{CarcamoHernandez:2023atk}, in the context of gauged $B-L$ symmetry~\cite{Wang:2015saa,Das:2017ski} or in the $331$ model~\cite{CarcamoHernandez:2021tlv}, inevitably requiring more involved particle content and symmetries such as multiple copies of sterile singlet fermions, vector-like charged leptons for anomaly cancellation or non-minimal scalar sectors. These examples lead to distinct phenomenological signatures in regards to DM and cLFV, involving annihilation of DM candidates to new vector bosons. Lastly, we wish to point out that the idea of connecting DM with LNV has also been applied in the context of other low-scale seesaw schemes such as the ISS~\cite{Ma:2009gu,Bazzocchi:2009kc,Baldes:2013eva,Mandal:2019oth,Abada:2021yot} (see Sec.~\ref{sec:ISS}).

%------------
% CHAPTER 03    
%------------

%%%%%%%%%%%%%%%%%%%%%%%%%%%%%%%%%%%%%%%%%%%%%%%%%%%%%%%%%%%%%%%%%%%%%%%%%%%%%
\chapter{Spontaneous CP violation with a scalar-singlet} 
\label{chpt:SCPV}
%%%%%%%%%%%%%%%%%%%%%%%%%%%%%%%%%%%%%%%%%%%%%%%%%%%%%%%%%%%%%%%%%%%%%%%%%%%%%

In 1974, T.D. Lee~\cite{Lee:1974jb} proposed the idea of SCPV as a means of justifying the existence of CP violation at a time when only two generations of quarks were known. In models where the full Lagrangian is invariant under a CP symmetry, the origin of CP violation can be found in the complex VEV of BSM scalar fields. This extension is necessary because the VEV of the SM Higgs doublet is CP invariant~\cite{Branco:1999fs}. For example, by introducing new scalar singlets~\cite{Bento:1991ez,Branco:2003rt}, doublets~\cite{Branco:1983tn,Branco:1985aq}, triplets~\cite{Ferreira:2021bdj}, or more elaborate multiplets, SCPV can arise. SCPV is then communicated, under certain conditions, to the fermion sector through couplings between the fermion and these CP-breaking scalar fields. In this chapter we study models featuring SCPV stemming from the complex VEV of a scalar singlet.

Consider that the SM scalar sector is extended by a complex scalar singlet~$\sigma$, odd under a $\mathcal{Z}_2$ symmetry. Imposing the trivial CP transformations $\Phi \rightarrow \Phi^\ast$ and $\sigma \rightarrow \sigma^\ast$, the most general scalar potential is written as:
\begin{align}
V(\Phi, \sigma) &= V(\Phi) + m_\sigma^2 |\sigma|^2 + m_\sigma^{\prime 2} (\sigma^2 + \sigma^{\ast 2}) \nonumber \\
&+ \frac{\lambda_\sigma}{2} |\sigma|^4 + \lambda_\sigma^{\prime} (\sigma^4 + \sigma^{\ast 4}) + \lambda_\sigma^{\prime \prime} |\sigma|^2 (\sigma^2 + \sigma^{\ast 2}) \nonumber \\
&+ \lambda_{\Phi\sigma} (\Phi^\dagger \Phi) |\sigma|^2 + \lambda_{\Phi\sigma}^{\prime} (\Phi^\dagger \Phi)(\sigma^2 + \sigma^{\ast 2}) \; ,
\label{eq:VSCPVsinglet}
\end{align}
where all couplings are real due to the CP symmetry. The complex singlet field is parameterized as:
\begin{equation}
\sigma = \frac{1}{\sqrt{2}} \left(v_\sigma e^{i\varphi} + \sigma_{\text{R}} + i \sigma_{\text{I}}\right) \; , \; \langle \sigma \rangle = \frac{v_\sigma e^{i\varphi}}{\sqrt{2}} \; ,
\label{eq:VSCPVsingletvev}
\end{equation}
with $v_\sigma$ and $\varphi$ denoting the modulus and phase of the VEV of $\sigma$, respectively.

To determine whether the vacuum spontaneously breaks CP, one must minimize the above scalar potential. The minimization conditions yield:
\begin{align}
m_\Phi^2 + \frac{\lambda}{2} v^2 + \frac{\lambda_{\Phi\sigma}}{2} v_\sigma^2 + \lambda_{\Phi\sigma}^{\prime} v_\sigma^2 \cos(2\varphi) &= 0 \; ;
\label{eq:VSCPVsingletmin1} \\
m_\sigma^2 + \frac{\lambda_\sigma}{2} v_\sigma^2 + \frac{\lambda_{\Phi\sigma}}{2} v^2 + \left(2 m_\sigma^{\prime 2} + 2 \lambda_\sigma^{\prime \prime} v_\sigma^2 + \lambda_{\Phi\sigma}^{\prime} v^2\right)\cos(2\varphi) + 2 \lambda_\sigma^{\prime} v_\sigma^2 \cos(4\varphi) &= 0 \; ;
\label{eq:VSCPVsingletmin2} \\
-\sin(2\varphi) \left[m_\sigma^{\prime 2} + \frac{\lambda_\sigma^{\prime \prime}}{2} v_\sigma^2 + \frac{\lambda_{\Phi\sigma}^{\prime}}{2} v^2 + 2 \lambda_\sigma^{\prime} v_\sigma^2 \cos(2\varphi)\right] &= 0 \; .
\label{eq:VSCPVsingletmin3}
\end{align}
Solving the above equations for non-zero VEVs $v, v_\sigma \neq 0$, one finds the following possible solutions:
\begin{align}
\text{(i) :} \; m_\Phi^2 & = -\frac{\lambda_\Phi}{2} v^2 - \frac{v_\sigma^2}{2} (\lambda_{\Phi\sigma} + 2 \lambda_{\Phi\sigma}^{\prime}) \; , \nonumber \\ 
m_\sigma^2 &= - 2 m_\sigma^{\prime 2} - \frac{v_\sigma^2}{2} (\lambda_\sigma + 4 \lambda_\sigma^{\prime} + 4 \lambda_\sigma^{\prime \prime}) - \frac{v^2}{2} (\lambda_{\Phi \sigma} + 2 \lambda_{\Phi\sigma}^{\prime}) \; , \nonumber \\
\varphi &= k\pi \; , \; k \in \mathbb{Z} \; ;
\label{eq:VSCPVsingletsol1}
\end{align}
\begin{align}
\text{(ii) :} \; m_\Phi^2 &= -\frac{\lambda_\Phi}{2} v^2 - \frac{v_\sigma^2}{2} (\lambda_{\Phi\sigma} - 2 \lambda_{\Phi\sigma}^{\prime}) \; , \nonumber \\
m_\sigma^2 &= 2 m_\sigma^{\prime 2} - \frac{v_\sigma^2}{2} (\lambda_\sigma + 4 \lambda_\sigma^{\prime} - 4 \lambda_\sigma^{\prime \prime}) - \frac{v^2}{2} (\lambda_{\Phi \sigma} - 2 \lambda_{\Phi\sigma}^{\prime}) \; , \nonumber \\ \varphi &= \frac{\pi}{2} + k \pi \; , \; k \in \mathbb{Z} \; ;
\label{eq:VSCPVsingletsol2} 
\end{align}
\begin{align}
\text{(iii) :} \; m_\Phi^2 &= \frac{4 m_\sigma^{\prime 2} \lambda_{\Phi\sigma}^{\prime} + 2 v_\sigma^2( \lambda_\sigma^{\prime \prime} \lambda_{\Phi\sigma}^{\prime} - 2 \lambda_\sigma^{\prime} \lambda_{\Phi\sigma}) + 2 v^2(\lambda_{\Phi\sigma}^{\prime 2} - 2 \lambda_\Phi \lambda_\sigma^{\prime})}{8 \lambda_\sigma^{\prime}} \; , \nonumber \\
m_\sigma^2 &= \frac{8 m_\sigma^{\prime 2} \lambda_\sigma^{\prime \prime} + v_\sigma^2(4\lambda_\sigma^{\prime \prime 2} + 32 \lambda_\sigma^{\prime 2} - 8 \lambda_\sigma \lambda_\sigma^{\prime}) + 4 v^2(\lambda_\sigma^{\prime \prime} \lambda_{\Phi\sigma}^{\prime} - 2 \lambda_\sigma^{\prime} \lambda_{\Phi\sigma})}{16 \lambda_\sigma^{\prime}} \; , \nonumber \\
\cos(2\varphi) &= - \frac{4 m_\sigma^{\prime 2} + 2 u^2 \lambda_\sigma^{\prime \prime} + 2 v^2 \lambda_{\Phi\sigma}^{\prime}}{8 v_\sigma^2 \lambda_\sigma^{\prime}} \; .
\label{eq:VSCPVsingletsol3}
\end{align}
It can be shown that in cases (i) and (ii) the vacuum does not violate CP, since one can always define a CP transformation that leaves the vacuum invariant~\cite{Branco:1999fs}. Therefore, the only viable solution that leads to SCPV is (iii), which corresponds to global minimum of the potential if the following condition is satisfied:
\begin{equation}
\frac{16 m_\sigma^{\prime 4} - \left[2 v_\sigma^2 (4 \lambda_\sigma^\prime \pm \lambda_\sigma^{\prime \prime}) \pm 2 v^2 \lambda_{\Phi\sigma}^\prime\right]^2}{64 \lambda_\sigma^\prime} > 0 \; .
\label{eq:VSCPVsingletminimum}
\end{equation}
If the couplings $m_\sigma^{\prime 2}, \lambda_\sigma^{\prime \prime}, \lambda_{\Phi\sigma}^\prime$ are set to zero, one obtains:
\begin{equation}
\varphi = \frac{\pi}{4} + \frac{k\pi}{2}, \quad k \in \mathbb{Z} \; .
\label{eq:thetaexactOG}
\end{equation}
Although the vacuum in this case has a non-trivial phase, one needs to verify that it is successfully transmitted to the fermion sector. Since $\sigma$ is a singlet it is required to extend the fermion sector in order to communicate vacuum CP violation for example to generate a complex CKM matrix or in scenarios featuring neutrino mass generation to generate non-trivial LCPV effects. In fact, in Sec.~\ref{sec:lepto}, we will explore how SCPV can serve as the origin of both low-energy CP violation and the CP asymmetries required for successful leptogenesis in a complex singlet extension of the type-I seesaw model. While in Sec.~\ref{sec:darkNB} we will study how SCPV stemming from a complex scalar singlet can address the strong CP problem connected to a dark-sector.

%%%%%%%%%%%%%%%%%%%%%%%%%%%%%%%%%%%%%%%%%%%%%%%%%%%%%%%%%%%%%%%%%%%%%%%%%%%%%
\section{Scalar-singlet assisted type-I seesaw leptogenesis}
\label{sec:lepto}
%%%%%%%%%%%%%%%%%%%%%%%%%%%%%%%%%%%%%%%%%%%%%%%%%%%%%%%%%%%%%%%%%%%%%%%%%%%%%

The SM cannot account for the observed matter-antimmater asymmetry of the Universe -- see discussion in Sec.~\ref{sec:BAU}. It turns out that the amount of CP violation in the SM is too small [see Eq.~\eqref{eq:etaBSMCPV}] to successfully generate the observed BAU via EW baryogenesis~\cite{Gavela:1993ts,Gavela:1994ds}. This motivates the study of SM extensions with new sources of explicit or spontaneous CP violation, and new mechanisms to generate the BAU. In fact, within the type-I seesaw paradigm (see Sec.~\ref{sec:TypeI}), the excess of matter over antimatter can be explained through the leptogenesis mechanism~\cite{Fukugita:1986hr} (for reviews on leptogenesis see, e.g., Refs.~\cite{Buchmuller:2004nz,Davidson:2008bu,Fong:2012buy,Hambye:2012fh} and for detailed analyses in the context of the minimal type-I seesaw, see Refs.~\cite{Frampton:2002qc,Ibarra:2003up,GonzalezFelipe:2003fi,Joaquim:2005zv,Branco:2005jr,Abada:2006ea,Harigaya:2012bw,Zhang:2015tea,Siyeon:2016wro,Rink:2016vvl,Geib:2017bsw,Achelashvili:2017nqp,Shimizu:2017fgu,Shimizu:2017vwi,Covi:1996wh,Antusch:2011nz,Barreiros:2018ndn,Barreiros:2020mnr}). This is realized via the out-of-equilibrium LNV decays of heavy neutrinos in the early Universe, generating a lepton asymmetry which is subsequently converted into a baryon asymmetry by non-perturbative sphalerons~\cite{Kuzmin:1985mm} (see Sec.~\ref{sec:sphalerons}). 

In contrast to SM fermions, bare RH Majorana neutrino mass terms are invariant under the SM gauge group. Still, one can envisage scenarios where heavy neutrino masses are generated dynamically by adding scalar fields coupled to the RH neutrinos. After acquiring a non-zero VEV, heavy masses could be generated by those VEVs as, e.g., in Majoron models~\cite{Chikashige:1980qk,Chikashige:1980ui,Gelmini:1980re}. The existence of new scalar degrees of freedom coupled to heavy Majorana neutrinos induce new contributions to the CP asymmetries relevant for the generation of a lepton asymmetry in leptogenesis scenarios~\cite{Pilaftsis:2008qt,AristizabalSierra:2014uzi,LeDall:2014too,Alanne:2017sip,Alanne:2018brf,Abe:2021mfy}.

Following our work in Ref.~\cite{Barreiros:2022fpi}, we consider the case in which leptogenesis is assisted by complex scalar singlets. The latter may acquire complex VEVs which, being the sole source of CP violation, provide a common origin for CP-violating effects at low and high energies~\cite{Branco:2003rt}. We investigate the possibility that the spontaneous breaking of CP occurs at a scale above the leptogenesis scale, in such a way that the complex singlet VEVs give rise to complex RH neutrino mass terms. In the fermion and scalar mass eigenstate basis, this leads to non-trivial CP-violating scalar-heavy neutrino interactions which, in turn, trigger new contributions to the CP asymmetries in heavy neutrino decays. On the other hand, the evolution of particle number densities controlled by BEs is also affected by the presence of those new interactions. The whole setup is illustrated with a concrete model based on a SM extension with 2RH neutrinos, two Higgs doublets and a complex scalar singlet, supplemented with a discrete $\mathcal{Z}_8$ flavor symmetry. 

%%%%%%%%%%%%%%%%%%%%%%%%%%%%%%%%%%%%%%%%%%%%%%%%%%%%%%%%%%%%%%%%%%%%%%%%%%%%%
\subsection{Type-I seesaw and high-energy SCPV}
\label{sec:framework}
%%%%%%%%%%%%%%%%%%%%%%%%%%%%%%%%%%%%%%%%%%%%%%%%%%%%%%%%%%%%%%%%%%%%%%%%%%%%%

In the class of type-I seesaw models we are interested in, there are essentially two ways of breaking CP: i) explicitly by considering gauge-invariant complex Yukawa couplings and/or bare mass terms and ii) spontaneously through the complex VEVs of some scalar fields. In the latter case, the spin-0 complex fields can be singlets, doublets, triplets, or even more complicated multiplets, if the underlying symmetry groups are larger than the SM one. We will consider SM extensions with $n_R$ RH neutrinos $\nu_R$, $n_H$ scalar doublets $\Phi_a$ ($a=1, \cdots, n_H$) and $n_S$ complex scalar singlets $S_k$ ($k=1, \cdots, n_S$). Within this general setup, the Yukawa and mass terms allowed by the gauge symmetry are
\begin{align}
- \mathcal{L}_{\text{Yuk.}} = \overline{\ell_L}{\Ye^a} \Phi_a e_R+\overline{\ell_L} \YD^{a \ast} \tilde{\Phi}_a\nu_R+\dfrac{1}{2}\overline{\nu_R}\left(\MR^0 + \YR^k S_k+ \YRp^k S^*_k \right)\nu_R^c+\text{H.c.} \; ,
\label{eq:LYuk}
\end{align}
where $\Ye^a$, $\YD^{a \ast}$, $\YR^k$ and $\YRp^k$ are, respectively,  $3\times 3$, $3\times n_R$ and $n_R \times n_R$ complex Yukawa matrices, being the latter two symmetric. Bare RH neutrino masses are denoted by the $n_R \times n_R$ symmetric matrix $\MR^0$. The doublet and singlet  scalar fields $\Phi_a$ and $S_k$ are defined in the usual form:
\begin{align}
\Phi_a =\begin{pmatrix}
\phi^{+}_a \\
\phi^0_a
\end{pmatrix}= \frac{1}{\sqrt{2}}  \begin{pmatrix}
 \sqrt{2} \phi^{+}_a \\
 v_a e^{i \varphi_a} + \phi^0_{\text{R} a} + i \phi^0_{\text{I} a}
\end{pmatrix} \; , \;  S_k = \frac{1}{\sqrt{2}}\left( u_k\, e^{i \theta_k} + S_{\text{R} k} + i S_{\text{I} k}\right) \, ,
\label{eq:scalardef}
\end{align}
with $\tilde{\Phi}_a=i\tau_2 \Phi^\ast_a$ and the doublet VEVs are normalized as $v^2 = \sum_a |v_a|^2$. The $S_k\,\overline{\nu_R}\,\nu_R^c$ Yukawa interactions give rise to mass terms for the $\nu_R$'s if the scalar singlets develop non-zero VEVs $u_k\neq0$. After EWSB, the full fermion-mass Lagrangian is
\begin{equation}
\begin{split}
-\mathcal{L}_{\text{mass}} & = \overline{e_{L}}\, \Me\, e_{R} + \overline{\nu_L}\, \MD^\ast\, \nu_R + \dfrac{1}{2}\overline{\nu_R}\, \MR\,  \nu_R^c + \text{H.c.} \; ,
\end{split}
\end{equation}
where $\mathbf{M}_{e}$, $\mathbf{M}_{D}$ and $\mathbf{M}_{R}$ are the charged-lepton, Dirac neutrino and RH neutrino mass matrices, respectively given by:
\begin{align}
   \Me = \frac{v_a}{\sqrt{2}} e^{i \varphi_a} \Ye^a \; , \; \mathbf{M}_{D}^\ast = \frac{v_a}{\sqrt{2}} e^{- i \varphi_a} \YD^{a \ast} \; , \; \MR = \MR^0 + \frac{u_k}{\sqrt{2}} \left(\YR^k e^{i \theta_k} + \YRp^k e^{- i \theta_k}\right) \, ,
\label{eq:massyuk}
\end{align}
where sums over repeated indices are implicit. Defining the $n_f = 3+n_R$ component vector $N_{L}=\left(\nu_{L}, \nu^c_{R} \right)^{T}$ in flavor space, we can write $\mathcal{L}_{\text{mass}} $ as
\begin{align}
-\mathcal{L}_{\text{mass}} =\overline{e_{L}}\, \Me\, e_{R}+\frac{1}{2} \overline{N_{L}^c} \bm{\bm{\mathcal{M}}}_\nu N_{L} + \text{H.c.} \; , \; 
\bm{\bm{\mathcal{M}}}_\nu = 
\begin{pmatrix} 0 & \MD \\
\MD^{T} & \MR
\end{pmatrix} \, .
\label{eq:bigm}
\end{align}
The charged-lepton mass matrix is bidiagonalized through the unitary transformations $\mathbf{U}_{L,R}^e$ defined in Eq.~\eqref{eq:massdiag}. Furthermore, in the seesaw approximation limit, when $M_D \ll M_{R}$, the neutrino mass matrix $\bm{\bm{\mathcal{M}}}_\nu$ can be block-diagonalized following the procedure outlined in Sec.~\ref{sec:TypeI} yielding the well-known $3\times 3$ effective light neutrino mass matrix $\mathbf{M}_{\nu}$ of Eq.~\eqref{eq:TypeIMeff}. The effective matrix is then diagonalized by the unitary rotation $\mathbf{U}_\nu$ of the light-neutrino fields given in Eq.~\eqref{eq:TypeIMeffdiag}. Finally, the mass matrix~$\MR$ can be diagonalized through a unitary rotation $\mathbf{U}_{\text{h}}$ of the heavy neutrino fields given in Eq.~\eqref{eq:TypeIMheavy}, yielding $n_R$ heavy neutrinos $N_j$ with real and positive masses $M_j$.

As for the scalar potential, the new trilinear and quartic terms are
\begin{align}
V(\Phi,S) & \supset \mu_{a b, i} (\Phi_a^\dagger \Phi_b) S_i + \lambda_{a b, i k} (\Phi_a^\dagger \Phi_b) S_i S_k^\ast + \lambda_{a b, i k}^\prime (\Phi_a^\dagger \Phi_b) S_i S_k \nonumber \\  
&+ \mu_{i j k} S_i^\ast S_j S_k + \mu^\prime_{i j k} S_i S_j S_k \nonumber \\ 
& + \lambda_{i j k l} S_i^\ast S_j^\ast S_k S_l + \lambda_{i j k l}^\prime S_i^\ast S_j S_k S_l + \lambda_{i j k l}^{\prime \prime} S_i S_j S_k S_l + \text{H.c.}\; ,
\label{eq:cubicscalar}
\end{align} 
where the mass (dimensionless) $\mu$ ($\lambda$) are complex parameters. The scalar interactions stemming from \eqref{eq:cubicscalar} are of special interest since they will induce new contributions to the CP asymmetries in $N_i$ decays, and to the scattering processes entering the BEs (see Sec.~\ref{sec:leptogenesis})~\footnote{More details on the full scalar potential will be given for the specific case of two Higgs doublets and one scalar singlet in Sec.~\ref{sec:model}. For simplicity, hereafter we  neglect the quartic terms since they will not play a relevant role in our analysis.}.

We consider the scenario in which the singlets acquire complex VEVs at energies well above the EW scale $v$. At temperatures $T \gg v$, the Higgs doublets are VEVless, i.e. the EW symmetry is still unbroken. By imposing CP conservation at the Lagrangian level, all couplings in the Yukawa and scalar sectors are real. Consequently, under certain conditions, CP may be broken if the singlet fields $S_k$ develop complex VEVs, meaning that the sole source of CP violation in our framework comes from SCPV occurring at very high energies. This CP violation can be transmitted in a nontrivial way to the neutrino sector through the heavy neutrino-scalar portal $\overline{\nu_R}\, \nu_R^c\, S$, provided that the VEV phases appearing in the mass matrix $\M_R$, defined in Eq.~\eqref{eq:massyuk}, are not removable by field redefinitions. The link with low-energy LCPV effects is established when the Higgs doublets acquire non-zero VEVs and the EW symmetry is spontaneously broken. At this stage, the charged-lepton $\Me$ and Dirac-type neutrino $\MD$ mass matrices are generated, giving masses to the SM leptons and to the light neutrinos via the type-I seesaw mechanism. Since the complex matrix $\MR$ enters the expression for the effective light neutrino mass matrix $\mathbf{M}_{\nu}$ in Eq.~\eqref{eq:TypeIMeff}, non-trivial Dirac and Majorana phases may appear in the lepton mixing matrix. As we will show in the next section, besides explaining LCPV, vacuum CP violation can also lead to non-vanishing CP asymmetries in leptogenesis.

At the leptogenesis scale, the scalar degrees of freedom contained in the singlets will be massive with a typical mass of order $u \gg v$. In fact, there will be $2 n_S$ scalar mass eigenstates $h_i$ with mass matrix $\bm{\bm{\mathcal{M}}}_S$, being the mixing with the weak states given by
\begin{equation}
(S_{\text{R} 1}, \cdots, S_{\text{R} n_S}, S_{\text{I} 1}, \cdots, S_{\text{I} n_S})^T= \mathbf{V} (h_1, \cdots, h_{2 n_S})^T\;,
\label{eq:mixS}
\end{equation}
where $\mathbf{V}$ is a $2 n_S \times 2 n_S$ orthogonal matrix, such that
\begin{align}
\mathbf{V}^T \bm{\bm{\mathcal{M}}}_S^2 \mathbf{V} = \text{diag}(m_{h_1}^2, \cdots, m_{h_{2 n_S}}^2) \; .
\label{eq:mixingscalar}
\end{align}
As usually done in leptogenesis calculations, we will consider that, before EWSB, the heavy Majorana neutrinos $N_i$ and scalars $h_k$ are much heavier than the scalars stemming from the Higgs doublets $\Phi_a$, i.e. $m_{\Phi_a}\ll M_i, m_{h_k}$. For this reason, we neglect $m_{\Phi_a}$ in our calculations. Note also that the SM fermions are massless in the symmetric phase.

For computational purposes, we define some of the Yukawa and scalar couplings in the flavor-diagonal basis of charged-lepton Yukawa couplings, and on the mass basis of heavy neutrinos and scalar fields. Namely, for the $\overline{\ell_L} N \Phi_a$, $N N h_k$, $(\Phi_a^\dagger \Phi_b) h_k$ and $h_i h_j h_k$ couplings we now have\footnote{Notice that here we are already putting the quartic couplings of Eq.~\eqref{eq:cubicscalar} to zero. }
\begin{align}
\mathbf{Y}^{a\ast}\overline{\ell_L} N \Phi_a: \mathbf{Y}^{a\ast} & = \mathbf{U}_L^{e \dagger} \YD^{a \ast} \Uh \; , \; \mathbf{H}^a = \mathbf{Y}^{a \dagger} \mathbf{Y}^{a} \; , \label{eq:YHdef} \\
 \mathbf{\Delta}^k NNh_k: \mathbf{\Delta}^k &= \frac{1}{2 \sqrt{2}} \sum_{j=1}^{n_S} \Uh^\dagger \left[ \mathbf{V}_{j k} (\YR^j + \YRp^j) + i \mathbf{V}_{j+n_S k} (\YR^j - \YRp^j) \right] \Uh^\ast \; , \label{eq:Deltadef} \\
 \tilde{\mu}_{a b, k}(\Phi_a^\dagger \Phi_b) h_k: \tilde{\mu}_{a b, k} &= \sqrt{2} \sum_{j=1}^{n_S} \left[\text{Re}\left(\mu_{a b, j}\right) \mathbf{V}_{j k} - \text{Im}\left(\mu_{a b, j}\right) \mathbf{V}_{j+n_S k}\right] \; ,
\label{eq:muCPdef}
\\
 \tilde{\mu}_{i j k} h_i h_j h_k: \tilde{\mu}_{i j k} &= \frac{1}{\sqrt{2}} \sum_{l,p,q=1} \big[  \text{Re}\left(\mu_{lpq} + \mu_{lpq}^\prime \right)  \mathbf{V}_{l i} \left( \mathbf{V}_{p j} \mathbf{V}_{q k} - \mathbf{V}_{p+n_S j} \mathbf{V}_{q+n_S k} \right)\nonumber \\ 
 & + \text{Re}\left(\mu_{lpq} - \mu_{lpq}^\prime \right) \mathbf{V}_{l+n_S i} \left( \mathbf{V}_{p+n_S j} \mathbf{V}_{q k} + \mathbf{V}_{p j} \mathbf{V}_{q+n_S k} \right) \nonumber \\ 
 & - \text{Im}\left(\mu_{lpq} + \mu_{lpq}^\prime \right) \mathbf{V}_{l i} \left( \mathbf{V}_{p+n_S j} \mathbf{V}_{q k} + \mathbf{V}_{p j} \mathbf{V}_{q+n_S k} \right) \nonumber \\ 
 & + \text{Im}\left(\mu_{lpq} - \mu_{lpq}^\prime \right) \mathbf{V}_{l+n_S i} \left( \mathbf{V}_{p j} \mathbf{V}_{q k} - \mathbf{V}_{p+n_S j} \mathbf{V}_{q+n_S k} \right) \big]\,,
\label{eq:muhkCPdef}
\end{align}
where the field rotations were performed in Eqs.~\eqref{eq:LYuk} and \eqref{eq:cubicscalar} using the unitary matrices $\mathbf{U}_L^e$, $\Uh$ and $\mathbf{V}$ given in Eqs.~\eqref{eq:massdiag}, \eqref{eq:TypeIMheavy} and \eqref{eq:mixingscalar}, respectively.

%%%%%%%%%%%%%%%%%%%%%%%%%%%%%%%%%%%%%%%%%%%%%%%%%%%%%%%%%%%%%%%%%%%%%%%%%%%%%
\subsection{Vanilla CP-asymmetry contributions}
\label{sec:leptogenesis}
%%%%%%%%%%%%%%%%%%%%%%%%%%%%%%%%%%%%%%%%%%%%%%%%%%%%%%%%%%%%%%%%%%%%%%%%%%%%%

In the type-I seesaw framework, leptogenesis proceeds via the out-of-equilibrium decays of the heavy neutrinos $N_i$ in the early Universe. The resulting lepton asymmetry is then partially converted into a baryon asymmetry through the $(B+L)$-violating sphaleron transitions, leading to~\cite{Antusch:2011nz}
\begin{align}
    \eta_B = a_{\text{sph}} \frac{N_{B-L}^f}{N_{\gamma}^{\text{rec}}} \simeq 9.40 \times 10^{-3} N_{B-L}^f \; ,
    \label{eq:etaB}
\end{align}
where, for $3$ fermion generations, $a_{\text{sph}} = (24 + 4 n_H)/(66 + 13 n_H)$ is the sphaleron conversion factor~\cite{Khlebnikov:1988sr,Harvey:1990qw}, $N_{B-L}^f$ is the final asymmetry calculated in a comoving volume and $N_{\gamma}^{\text{rec}} \simeq 37.01$ is the number of photons in the same comoving volume at the recombination temperature. The approximation in \eqref{eq:etaB} corresponds to $n_H = 2$ which will be the case under study in Sec.~\ref{sec:model}.

\begin{figure}[t!]
    \centering
     \begin{subfigure}[b]{0.4\textwidth}
         \centering
         \includegraphics[scale=0.9,trim={5cm 23.5cm 12cm 1.0cm},clip]{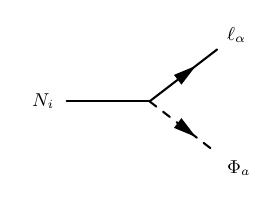}
         \caption{}
         \label{fig:treetypeI}
     \end{subfigure}
     \hspace{1cm}
     \begin{subfigure}[b]{0.4\textwidth}
         \centering
         \includegraphics[scale=0.9,trim={11.0cm 23.5cm 2cm 1.0cm},clip]{Figures/LeptoSCPV/diagrams_CPasymmetry.pdf}
         \caption{}
         \label{fig:vertextypeI}
     \end{subfigure}\\
     \begin{subfigure}[b]{0.9\textwidth}
         \centering
         \includegraphics[scale=0.9,trim={4.0cm 20.3cm 3.0cm 4.0cm},clip]{Figures/LeptoSCPV/diagrams_CPasymmetry.pdf}
         \caption{}
         \label{fig:wavetypeI}
     \end{subfigure} \\
    \caption{Usual diagrams contributing to the CP asymmetry in $N_i \rightarrow \ell_\alpha \Phi_a$ through interference. (a) Tree-level contribution $ \sim \mathcal{O}(Y^2)$. (b) Vertex contribution $\sim \mathcal{O}(Y^4)$. (c) Wavefunction contributions $\sim \mathcal{O}(Y^4)$.}
    \label{fig:usualcontribution_CPasym}
\end{figure}

A key ingredient in the generation of the BAU within the (standard) leptogenesis framework are the CP asymmetries $\varepsilon_{i \alpha}^a$ produced in the heavy neutrino decays $N_i \rightarrow \Phi_a \ell_\alpha$, where $\ell_\alpha$ represents one lepton of flavor $\alpha= e, \mu, \tau$ and $\Phi_a$ denotes a scalar-field component of the doublet with $a=1,...,n_H$. The CP asymmetries $\varepsilon_{i \alpha}^a$ are defined as follows~\cite{Covi:1996wh}
\begin{align}
    \varepsilon_{i \alpha}^a = \dfrac{\Gamma(N_i \rightarrow \Phi_a \ell_\alpha) - \Gamma(N_i \rightarrow \Phi_a^\dagger\, \overline{\ell}_\alpha)}{\sum\limits_{\beta=e,\mu, \tau} \sum\limits_{b=1}^{n_H} \left[ \Gamma(N_i \rightarrow \Phi_b \ell_\beta) + \Gamma(N_i \rightarrow \Phi_b^\dagger\, \overline{\ell}_\beta)\right]} \; .
\end{align}
Note that $\Gamma(N_i \rightarrow \Phi_a \ell_\alpha)$ and $\Gamma(N_i \rightarrow \Phi_a^\dagger\, \overline{\ell}_\alpha)$ are the heavy neutrino decays into leptons and antileptons of flavor $\alpha$, respectively. Summing $\varepsilon_{i \alpha}^a$ for all $\alpha$ and $a$, the total (unflavored) CP asymmetry $\varepsilon_i$ in the $N_i$ decays is obtained,
\begin{equation}
    \varepsilon_i = \sum_{\alpha=e,\mu,\tau} \sum_{a=1}^{n_H} \varepsilon_{i \alpha}^a \; .
    \label{eq:cptotunflavored}
\end{equation}
In the usual type-I leptogenesis scenario, the CP asymmetries are generated through the interference between the tree level and one-loop contributions of diagrams \subref{fig:treetypeI}-\subref{fig:wavetypeI} in Fig.~\ref{fig:usualcontribution_CPasym}. At tree level the $N_i \rightarrow \Phi_a \ell_\alpha $ decay width is given by
\begin{align}
\Gamma(N_i \rightarrow \Phi_a \ell_\alpha) = \Gamma(N_i \rightarrow \Phi_a^\dagger\, \overline{\ell}_\alpha) = M_i \frac{\mathbf{H}^a_{i i}}{16 \pi} \; ,
\label{eq:Decaytree}
\end{align}
where $\mathbf{H}^a$ has been defined in Eq.~\eqref{eq:YHdef}. The leading-order non-vanishing contributions to the CP asymmetries arising from the aforementioned interference is~\cite{Covi:1996wh,Branco:2011zb}

\begin{align}
    \varepsilon_{i \alpha}^a(\text{type-I}) & = - \frac{1}{8 \pi \sum\limits_{a=1}^{n_H} \mathbf{H}^a_{i i}}  \sum_{b=1}^{n_H} \bigg\{\sum_{j=1}^{n_R} \sum_{\beta=e, \mu, \tau} \text{Im}\left[\mathbf{Y}^{a \ast}_{\alpha i} \mathbf{Y}^{b \ast}_{\beta i} \mathbf{Y}^{b}_{\alpha j} \mathbf{Y}^{a}_{\beta j}\right] f(r_{j i}) \nonumber \\ &+\sum_{j\neq i=1}^{n_R} \text{Im}\left[\mathbf{Y}^{a \ast}_{\alpha i} \mathbf{H}^b_{i j} \mathbf{Y}^{a}_{\alpha j} \right] g(r_{j i}) +\sum_{j\neq i=1}^{n_R} \text{Im}\left[ \mathbf{Y}^{a \ast}_{\alpha i} \mathbf{H}^b_{j i} \mathbf{Y}^{a}_{\alpha j} \right] g^\prime(r_{j i}) \bigg\} \; ,
    \label{eq:CPasymtypeI}
\end{align}
where $r_{j i} = M_j^2/M_i^2$. The loop functions $f(x)$ and $g(x)$ correspond to the vertex correction [diagram \subref{fig:vertextypeI} in Fig.~\ref{fig:usualcontribution_CPasym}] and $g^\prime(x)$ to the self-energy one [diagrams \subref{fig:wavetypeI} in Fig.~\ref{fig:usualcontribution_CPasym}]. These functions are given by
\begin{align}
f(x) = \sqrt{x} \left[1-(1+x) \log\left(1 + \frac{1}{x}\right) \right] \; , \; g(x) = \sqrt{x} g^\prime(x)= \frac{\sqrt{x}}{1-x} \, .
\label{eq:loopI}
\end{align}

Due to the presence of the scalars $h_k$, which couple to a pair of RH neutrino fields and to $\Phi^\dag_a \Phi_b$ [see Eqs.~\eqref{eq:Deltadef} and~\eqref{eq:muCPdef}, respectively], there will be additional contributions to the one-loop $N_i \rightarrow \Phi_a \ell_\alpha$ decay diagrams, as depicted in Fig.~\ref{fig:newcontribution_CPasym}. Moreover, the interference of new $N_i \rightarrow \ell_\alpha h_k \Phi_a$ three-body decay diagrams, presented in Fig.~\ref{fig:newcontributiontree_CPasym}, must also be taken into account. These new contributions to the CP asymmetry were computed in Ref.~\cite{LeDall:2014too} for a single Higgs doublet and a real scalar singlet. We present the results for the general case of $n_R$ RH neutrino fields, $n_H$ Higgs doublets and $n_S$ complex scalar singlets. For the diagrammatic computations we have used the standard Majorana Feynman rules~\cite{Gluza:1991wj,Denner:1992vza,Denner:1992me}. 

\subsection{New singlet-induced contributions to the CP asymmetry}
\label{sec:asymmetry}

\begin{figure}[t!]
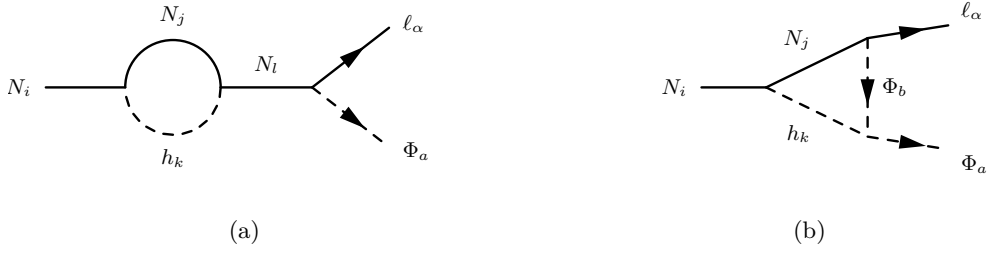

    \centering
     \begin{subfigure}[b]{0.4\textwidth}
         \centering
         \includegraphics[scale=0.9,trim={11.5cm 8.5cm 2.5cm 16cm},clip]{Figures/LeptoSCPV/diagrams_CPasymmetry.pdf}
         \caption{}
         \label{fig:newwave}
     \end{subfigure}
     \hspace{1cm}
      \begin{subfigure}[b]{0.4\textwidth}
         \centering
         \includegraphics[scale=0.9,trim={4.5cm 8.5cm 11.5cm 16cm},clip]{Figures/LeptoSCPV/diagrams_CPasymmetry.pdf}
         \caption{}
         \label{fig:newvertex}
     \end{subfigure}\\
    \caption{New one-loop diagrams contributing to the CP asymmetry in $N_i \rightarrow \ell_\alpha \Phi_a$ decays via the interference with the corresponding tree-level diagram (a) of Fig.~\ref{fig:usualcontribution_CPasym}. (a) Wavefunction contribution $\sim \mathcal{O}(Y^2 \Delta^2)$. (b) Vertex contribution $\sim \mathcal{O}(Y^2\Delta \mu)$.}
    \label{fig:newcontribution_CPasym}
\end{figure}

\begin{figure}[t!]
    \centering
    \includegraphics[scale=0.9,trim={4.5cm 12cm 2.5cm 12cm},clip]{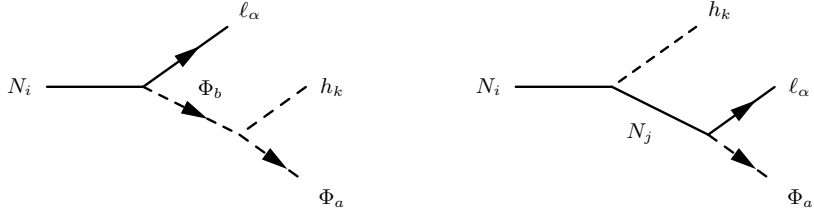}
    \caption{Interfering $N_i \rightarrow \ell_\alpha h_k \Phi_a$ tree-level 3-body decays. Left [right] diagram is of order~$\sim \mathcal{O}(Y \mu)$ [$\mathcal{O}(Y \Delta)$] in the couplings.  }
    \label{fig:newcontributiontree_CPasym}
\end{figure}

We start with the new $h_k$-mediated wavefunction contribution to the CP asymmetry $\varepsilon_{i \alpha}^a$ in the $N_i \rightarrow \Phi_a \ell_\alpha$ decays, for which the relevant diagrams are labeled as~\subref{fig:newwave} in Fig.~\ref{fig:newcontribution_CPasym}. Denoting generically the couplings in Eqs.~\eqref{eq:YHdef}-\eqref{eq:muCPdef} by $Y$, $\Delta$ and $\mu$, it is straightforward to see that these diagrams scale as $Y^2\Delta^2$. Their interference with the corresponding tree-level ones leads to the following wavefunction contribution to the CP asymmetry $\varepsilon_{i \alpha}^a$ at one-loop level:
\begin{align}
    \varepsilon_{i \alpha}^a(\text{wave})& = \frac{1}{8 \pi \sum\limits_{a=1}^{n_H}  \mathbf{H}^a_{i i}}\sum_{j,l \neq i = 1}^{n_R} \sum_{k=1}^{2 n_S} (1+\delta_{j l}) \bigg\{\text{Im}\left[ \mathbf{Y}_{\alpha l}^{a \ast} \mathbf{\Delta}_{l j}^k \mathbf{\Delta}_{j i}^{k \ast} \mathbf{Y}_{\alpha i}^{a} \right] \mathcal{F}_{\text{w}, L L}^{i j k l} \nonumber \\ 
    & + \text{Im}\left[\mathbf{Y}_{\alpha l}^{a \ast} \mathbf{\Delta}_{l j}^{k \ast} \mathbf{\Delta}_{j i}^{k \ast} \mathbf{Y}_{\alpha i}^{a} \right] \mathcal{F}_{\text{w}, L R}^{i j k l} 
    + \text{Im}\left[ \mathbf{Y}_{\alpha l}^{a \ast} \mathbf{\Delta}_{l j}^k \mathbf{\Delta}_{j i}^k \mathbf{Y}_{\alpha i}^{a} \right] \mathcal{F}_{\text{w}, R L}^{i j k l} \nonumber \\
    & + \text{Im}\left[\mathbf{Y}_{\alpha l}^{a \ast} \mathbf{\Delta}_{l j}^{k \ast} \mathbf{\Delta}_{j i}^k \mathbf{Y}_{\alpha i}^{a} \right] \mathcal{F}_{\text{w}, R R}^{i j k l}\bigg\} \; ,
    \label{eq:CPasymwave}
\end{align}
being the loop functions given by,
\begin{align}
\mathcal{F}_{\text{w}, L L}^{i j k l} &= \frac{\sqrt{\rho_{i j k}} \sqrt{\rho_{i j k} + 4 r_{j i}}}{2 (1-r_{l i})}  \, , \;  \mathcal{F}_{\text{w}, L R}^{i j k l} =  \frac{\sqrt{\rho_{i j k}} \sqrt{r_{j i}} \sqrt{r_{l i}}}{1-r_{l i}}\, , \nonumber \\
\mathcal{F}_{\text{w}, R L}^{i j k l} &=  \frac{\sqrt{\rho_{i j k}} \sqrt{r_{j i}}}{1-r_{l i}} \, , \; \mathcal{F}_{\text{w}, R R}^{i j k l} = \frac{\sqrt{\rho_{i j k}} \sqrt{r_{l i}} \sqrt{\rho_{i j k} + 4 r_{j i}}}{2 (1-r_{l i})} \, ,
\label{eq:loopwave}
\end{align}
where $\sigma_{k i} = m_{h_k}^2/M_i^2$ and $\rho_{i j k} = (1-r_{j i}-\sigma_{k i})^2 - 4 r_{ji} \sigma_{ki}$. Note that, in order for the CP-asymmetries induced by the $h_k$ scalars to be non-zero one must guarantee that the kinematic constraint $ M_i > M_j + m_{h_k} $, i.e.~$  \sqrt{r_{j i}} + \sqrt{\sigma_{k i}} <1 $, is verified. In fact, the one-loop and the tree-level diagrams depicted in Figs.~\ref{fig:newcontribution_CPasym} and~\ref{fig:newcontributiontree_CPasym}, respectively, only lead to a non-vanishing imaginary part for the CP-asymmetry if the latter kinematic constraint is fulfilled. Hence, the decay $N_i \rightarrow N_j h_k$ needs to be allowed.

The $h_k \Phi^\dag_a \Phi_b$ couplings of Eq.~\eqref{eq:muCPdef} will generate new one-loop vertex contributions to~$\varepsilon_{i \alpha}^a$ stemming from diagrams~\subref{fig:newvertex} of Fig.~\ref{fig:newcontribution_CPasym}, which are of order $\mathcal{O}(Y^2\Delta \mu)$. For this case we obtain
\begin{align}
    \varepsilon_{i \alpha}^a(\text{vertex}) &= \frac{1}{8 \pi M_i \sum\limits_{a=1}^{n_H}  \mathbf{H}^a_{i i}}\sum_{j\neq i = 1}^{n_R} \sum_{k=1}^{2 n_S} \sum_{b=1}^{n_H} \bigg\{\text{Im}\left[ \mathbf{Y}_{\alpha i}^{a } \mathbf{Y}_{\alpha j}^{b\ast} \mathbf{\Delta}_{i j}^k \tilde{\mu}_{a b, k} \right] \mathcal{F}_{\text{v}, L L}^{i j k} \nonumber \\
    &+ \text{Im}\left[ \mathbf{Y}_{\alpha i}^{a } \mathbf{Y}_{\alpha j}^{b\ast} \mathbf{\Delta}_{i j}^{k \ast} \tilde{\mu}_{a b, k} \right] \mathcal{F}_{\text{v}, R L}^{i j k} \bigg\} \; ,
    \label{eq:CPasymvertex}
\end{align}
where the loop functions read
\begin{align}
\mathcal{F}_{\text{v}, L L}^{i j k} &= - \sqrt{\rho_{i j k}} + r_{j i} \log \left[ \frac{\sqrt{\rho_{i j k} + 4 r_{j i} \sigma_{k i}} - \sqrt{\rho_{i j k}}}{\sqrt{\rho_{i j k} + 4 r_{j i} \sigma_{k i}} + \sqrt{\rho_{i j k}}} \right] \, , \nonumber \\
\mathcal{F}_{\text{v}, R L}^{i j k} &= \sqrt{r_{j i}} \log \left[ \frac{\sqrt{\rho_{i j k} + 4 r_{j i} \sigma_{k i}} - \sqrt{\rho_{i j k}}}{\sqrt{\rho_{i j k} + 4 r_{j i} \sigma_{k i}} + \sqrt{\rho_{i j k}}} \right] \, .
\label{eq:loopvertex}
\end{align}

The interference between the 3-body decay diagrams of Fig.~\ref{fig:newcontributiontree_CPasym} gives corrections to $\varepsilon_\text{CP}$ of the same order of the vertex contribution, i.e. of order $\mathcal{O}(Y^2 \Delta \mu)$. For this case the CP asymmetry is computed as
\begin{equation}
    \varepsilon_{i \alpha}^a(\text{3-body decay}) \simeq \frac{\sum\limits_{k=1}^{2n_S} \left[\Gamma(N_i \rightarrow \Phi_a \ell_\alpha h_k) - \Gamma(N_i \rightarrow \Phi_a^\dagger \overline{\ell}_\alpha h_k)\right]}{\sum\limits_{\beta=e,\mu,\tau} \sum\limits_{b=1}^{n_H} \left[ \Gamma(N_i \rightarrow \Phi_b \ell_\beta) + \Gamma(N_i \rightarrow \Phi_b^\dagger \overline{\ell}_\beta)\right]} \; ,
\end{equation}
where we neglected the three-body decay rate in the denominator since it is subdominant compared to the two-body decay, due to its reduced phase space. We obtain
\begin{align}
    \varepsilon_{i \alpha}^a(\text{3-body decay}) &=\frac{1}{8 \pi M_i \sum\limits_{a=1}^{n_H} \mathbf{H}^a_{i i}}\sum_{j\neq i = 1}^{n_R} \sum_{k=1}^{2 n_S} \sum_{b=1}^{n_H}\bigg\{\text{Im}\left[ \mathbf{Y}_{\alpha i}^{b \ast} \mathbf{Y}_{\alpha j}^a \mathbf{\Delta}_{i j}^k \tilde{\mu}_{a b, k} \right] \mathcal{F}_{\text{3BD}, L L}^{i j k} \nonumber \\
    &+ \text{Im}\left[ \mathbf{Y}_{\alpha i}^{b \ast} \mathbf{Y}_{\alpha j}^a \mathbf{\Delta}_{i j}^{k \ast} \tilde{\mu}_{a b, k} \right] \mathcal{F}_{\text{3BD}, R L}^{i j k} \bigg\} \; ,
    \label{eq:CPasym3body}
\end{align}
where,
\begin{align}
\mathcal{F}_{\text{3BD}, L L}^{i j k} &=  - \sqrt{\rho_{i j k}r_{j i}} + \sqrt{r_{j i}}\log \left[ \frac{\sqrt{\rho_{i j k} + 4 r_{j i} \sigma_{k i}} + 2\sigma _{ki}+ \sqrt{\rho_{i j k}}}{\sqrt{\rho_{i j k} + 4 r_{j i} \sigma_{k i}} + 2\sigma _{ki} - \sqrt{\rho_{i j k}}} \right] \, , \nonumber \\
\mathcal{F}_{\text{3BD}, R L}^{i j k} &= r_{j i} \log \left[ \frac{\sqrt{\rho_{i j k} + 4 r_{j i} \sigma_{k i}} + 2\sigma _{ki}+ \sqrt{\rho_{i j k}}}{\sqrt{\rho_{i j k} + 4 r_{j i} \sigma_{k i}} + 2\sigma _{ki} - \sqrt{\rho_{i j k}}} \right] \, .
\label{eq:loop3body}
\end{align}

Combining the different contributions shown above leads to
\begin{equation}
   \varepsilon_{i \alpha}^a = \varepsilon_{i \alpha}^a(\text{type-I}) + \varepsilon_{i \alpha}^a(\text{wave}) + \varepsilon_{i \alpha}^a(\text{vertex}) + \varepsilon_{i \alpha}^a(\text{3-body decay}) \; .
   \label{eq:fullCP}
\end{equation}
These results are consistent with the ones obtained in Ref.~\cite{LeDall:2014too} in the limit of a single Higgs doublet and one real scalar singlet, apart from the 3-body decay contribution in Eq.~\eqref{eq:CPasym3body} where the `$\text{Im}[ \cdots]$' coefficients in front of the $LL$ and $RL$ functions are exchanged.

As mentioned before, we are interested in imposing CP at the Lagrangian level and breaking it spontaneously through non-zero complex VEVs acquired by the singlets $S_k$ above the leptogenesis scale. Within this scenario, a few remarks can be made regarding the link between vacuum CP and the CP asymmetry:
\begin{itemize}

    \item By imposing CP conservation, the Yukawa couplings $\YD$, $\YR$, $\YR^\prime$, the bare mass term $\MR^0$ of Eq.~\eqref{eq:LYuk} and the scalar potential potential parameters are real. Consequently, the unitary matrix $\mathbf{U}_L^e$ of Eq.~\eqref{eq:massdiag} is orthogonal and $\tilde{\mu}_{a b, k}$ of Eq.~\eqref{eq:muCPdef} is real. Hence, only the matrix entries of $\mathbf{Y}^a$ and $\mathbf{\Delta}^k$, given by Eqs.~\eqref{eq:YHdef} and~\eqref{eq:Deltadef}, respectively, can be complex. If this is the case, the `$\text{Im}[\cdots]$' coefficients in the CP asymmetries do not vanish in general. More specifically, the rotation $\Uh$ provides the only connection to the CP violation encoded in the scalar complex VEVs, as explained in Sec.~\ref{sec:framework}.

    \item Focusing on the case where $\Uh$ is complex, i.e. when SCPV is successfully communicated to the heavy neutrino sector, $\mathbf{Y}^a$ and $\mathbf{\Delta}^k$ are a priori general complex matrices. Hence, one expects that the different contributions to the total $\varepsilon_{\text{CP}}$ of Eq.~\eqref{eq:fullCP} are non-zero. Interestingly, since the type-I CP-asymmetry contribution of Eq.~\eqref{eq:CPasymtypeI} depends on the product of four $\mathbf{Y}^a$ matrix elements, one can derive conditions such that this contribution vanishes.

    \item In the scenario where $\varepsilon(\text{type-I})=0$, one would generate the high-energy CP violation needed to explain the BAU entirely through the new singlet-assisted diagrams. To achieve this, one needs to guarantee that $\mathbf{Y}^{a}$ have some special properties. Namely, if the elements of a given row of the $\mathbf{Y}^a$ matrix have the same phase, i.e. $\arg[\mathbf{Y}^{a}_{\alpha i}] = \arg[\mathbf{Y}^{a}_{\alpha j}]$ for $i \neq j$ and for all $j$, the Hermitian matrix $\mathbf{H}^a$ is necessarily real and therefore the second and third `$\text{Im}[\cdots]$' terms in Eq.~\eqref{eq:CPasymtypeI} for $\varepsilon^a_{i \alpha}(\text{type-I})$ vanish. In such case, summing over all lepton flavors leads to a vanishing unflavored CP asymmetry, i.e. $\varepsilon^a_{i}(\text{type-I})=0$. In order for the flavored type-I CP asymmetry to vanish one needs to verify that the elements of a given row of the $\mathbf{Y}^a$ and $\mathbf{Y}^b$ matrices have the same phase, i.e. $\arg[\mathbf{Y}^{a}_{\alpha i}] = \arg[\mathbf{Y}^{b}_{\alpha j}]$ for $i \neq j$ and for all $j$ and $b$. Such specific properties of the Yukawa matrices $\mathbf{Y}^{a}$ may result, e.g., from a flavor symmetry.

\end{itemize}
In Sec.~\ref{sec:model} we will present a very simple flavor model in which the new CP asymmetries induced by the scalar states $h_k$ provide the only link between the vacuum CP phases and the BAU.

\subsection{Unflavored Boltzmann equations}
\label{sec:BEs}

\begin{figure}[t!]
    \centering
     \begin{subfigure}[b]{0.45\textwidth}
         \centering
         \includegraphics[scale=0.9]{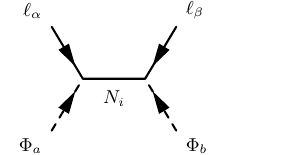}
         \caption{ }
         \label{fig:RIS1}
     \end{subfigure}
     \hspace{+1cm}
      \begin{subfigure}[b]{0.45\textwidth}
         \centering
         \includegraphics[scale=0.9]{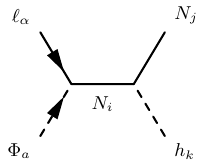}
         \caption{}
         \label{fig:RIS2}
     \end{subfigure}\\
     \vspace{+0.5cm}
     \begin{subfigure}[b]{0.45\textwidth}
         \centering
         \includegraphics[scale=0.9]{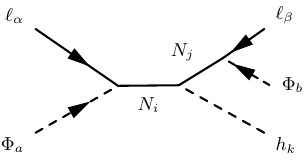}
         \caption{}
         \label{fig:RIS3}
     \end{subfigure}
     \hspace{+1cm}
      \begin{subfigure}[b]{0.45\textwidth}
         \centering
         \includegraphics[scale=0.9]{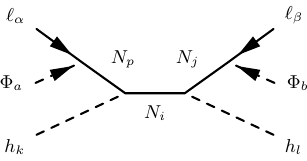}
        \caption{}
         \label{fig:RIS4}
     \end{subfigure}
    \caption{Scattering diagrams containing {\em real intermediate states}~(RIS). These must be subtracted in order to avoid inconsistent source terms in the BEs (see text for details). (a) Usual $\Delta L = 2$ scattering $\ell_\alpha \Phi_a \leftrightarrow \overline{\ell}_\beta \Phi_b^\dagger$ $s$-channel contribution $\sim \mathcal{O}(Y^4)$. (b) New $\Delta L = 1$ scattering $\ell_\alpha \Phi_a \leftrightarrow N_1 h_k$ $s$-channel contribution $\sim \mathcal{O}(Y^2 \Delta^2)$. (c) New $\Delta L = 2$ scattering $\ell_\alpha \Phi_a \leftrightarrow \overline{\ell}_\beta \Phi_b^\dagger h_k$ RIS contribution $\sim \mathcal{O}(Y^4 \Delta^2)$. (d) New $\Delta L = 2$ scattering $\ell_\alpha \Phi_a h_k \leftrightarrow \overline{\ell}_\beta \Phi_b^\dagger h_l$ RIS contribution $\sim \mathcal{O}(Y^4 \Delta^4)$.}
    \label{fig:diagsRIS}
\end{figure}

The lepton (and baryon) asymmetry produced through leptogenesis can be computed by solving the BEs that describe the out-of-equilibrium dynamics of the various processes involving the heavy Majorana neutrinos $N_i$. In this section, we derive the system of (classical) BEs relevant for the type of models we are interested in (general aspects related to BEs are reviewed in Appendix~\ref{chpt:genBEs}). Before presenting our results, the following comments are in order:  
\begin{itemize}
    \item We will restrict our analysis to the temperature regime $T > 10^{12}$~GeV, i.e. the unflavored scenario where the CP and lepton asymmetries are summed over all flavors (for reviews on flavor effects in leptogenesis see Refs.~\cite{Abada:2006fw,Nardi:2006fx,Abada:2006ea,Blanchet:2006be,Dev:2017trv}). For the sake of generality, we will present the BEs for an arbitrary number of RH neutrinos $N_{i}$ ($i=1, \cdots, n_R$), where the masses are ordered as $M_1 < M_2 < ... <M_{n_R}$, being $N_1$ the lightest heavy neutrino. We do not consider thermal corrections to the masses and to the CP asymmetries (for an analysis on this subject the reader is addressed to Ref.~\cite{Giudice:2003jh}).
    \item In the standard type-I seesaw leptogenesis, the CP asymmetries needed to generate the BAU are of order~$\mathcal{O}(Y^4)$ in the Yukawa couplings (see Fig.~\ref{fig:usualcontribution_CPasym}). Hence, in order to obtain BEs respecting the Sakharov conditions~\cite{Sakharov:1967dj}, one needs to take into account $\Delta L = 2$ scattering processes up to order $\mathcal{O}(Y^4)$. The corresponding diagrams exhibit what is known as {\em real intermediate states}~(RIS) that must be subtracted in order to obtain consistent BEs~\cite{Kolb:1979qa,Buchmuller:1997yu,Buchmuller:2000nq}.\footnote{Without such a procedure a $(B-L)$-asymmetry could be generated via source terms in thermal equilibrium, which is forbidden by the CPT symmetry~\cite{Dimopoulos:1978kv,Kolb:1979qa,Dolgov:1981hv}.} This is the case of the $s$-channel $N_i$ mediated $\Delta L = 2$ scattering  $\ell_\alpha \Phi_a \leftrightarrow \overline{\ell}_\beta \Phi_b^\dagger$ shown in diagram \subref{fig:RIS1} of Fig.~\ref{fig:diagsRIS}. 
    It has a RIS, since the mediating neutrino state can be produced on-shell, i.e. $\ell_\alpha \Phi_a \leftrightarrow N_i \leftrightarrow \overline{\ell}_\beta \Phi_b^\dagger$. Hence, this intermediate decay must be subtracted since it is already accounted for in the BEs.
    \item
    The presence of additional scalar mass-eigenstates coupling to heavy Majorana neutrinos opens up a new decay channel if $M_i>M_j + m_{h_k}$, namely $N_i \rightarrow N_j h_k$. The total decay widths for $N_i$ are
           \begin{align}
           \Gamma_i &= \sum_{a=1}^{n_H} \sum_{\alpha= e, \mu,\tau} \left[\Gamma(N_i \rightarrow \Phi_a \ell_\alpha) + \Gamma(N_i \rightarrow \Phi_a^\dagger \overline{\ell}_\alpha) \right]+ \sum_{j=1}^{n_R} \sum_{k=1}^{2 n_S} \Gamma(N_i \rightarrow N_j h_k) \; ,
           \label{eq:neutrinototal}
           \end{align}
          where the expressions for the decay rates are given in Eqs.~\eqref{eq:Decaytree} and~\eqref{eq:newdecayG}. Due to these extra interactions, compared to standard type-I seesaw framework, one needs to consider $\Delta L = 2$ scattering processes up to order $\mathcal{O}(Y^4 \Delta^4)$ and subtract the appropriate RIS to achieve consistent BEs. In fact, as shown in Fig.~\ref{fig:diagsRIS}, diagram~\subref{fig:RIS2} for the $\Delta L = 1$ scattering $\ell_\alpha \Phi_a \leftrightarrow N_j h_k$ process has a RIS corresponding to its $N_i$ ($i \neq j$) mediated $s$-channel. The $\Delta L = 2$ two-to-three body scattering $\ell_\alpha \Phi_a \leftrightarrow \overline{\ell}_\beta \Phi_b^\dagger h_k$ in Fig.~\ref{fig:diagsRIS}\subref{fig:RIS3} and three-to-three scattering $\ell_\alpha \Phi_a h_k \leftrightarrow \overline{\ell}_\beta \Phi_b^\dagger h_l$ in Fig.~\ref{fig:diagsRIS}\subref{fig:RIS4}, both contain RIS diagrams. By systematically subtracting these RIS contributions one reaches a set of coupled BEs meeting all the Sakharov conditions with no inconsistencies. The procedure outlined here was performed in detail in Ref.~\cite{LeDall:2014too} for the 2RH neutrino case ($n_R=2$), and we also followed the approach of Ref.~\cite{Giudice:2003jh}. Furthermore, there are additional $\Delta L = 1$ scatterings which contain $t$-channel RIS (see details in Appendix~\ref{chpt:scatterings}).

    \item For our purposes, we will neglect the three-body decay reaction densities in the BEs since these are subdominant when compared to usual two-body decays. Furthermore, the two-to-three and three-to-three scatterings are also subdominant when compared to the two-to-two scattering processes, due to their reduced phase space. Also, we will not consider the $\Delta L=2$ scattering contributions in the BEs.
    
\end{itemize}
Taking into account the general framework outlined above, the set of BEs for leptogenesis is given by~(for details see Appendix~\ref{chpt:genBEs})
\begin{align}
    \frac{d N_{N_i}}{d z} & = -(D_i + S_i) (N_{N_i} - N_{N_i}^{\text{eq}}) - \sum_{j=1}^{n_R} S_{i j} (N_{N_i} N_{N_j} - N_{N_i}^{\text{eq}} N_{N_j}^{\text{eq}}) \label{eq:BEsNi}  \\
    & + \sum_{j=1}^{n_R} \left[ \left(\frac{N_{N_i}^{\text{eq}}}{N_{N_j}^{\text{eq}}} D_{i j} + D_{j i}\right) (N_{N_j} - N_{N_j}^{\text{eq}}) - \left(D_{i j} + \frac{N_{N_j}^{\text{eq}}}{N_{N_i}^{\text{eq}}} D_{j i}\right) (N_{N_i} - N_{N_i}^{\text{eq}}) \right] \; , \nonumber \\
    \frac{d N_{B-L}}{d z} & = - \sum_{i=1}^{n_R} \varepsilon_i D_i (N_{N_i} - N_{N_i}^{\text{eq}}) - W N_{B-L} \;,\;\;\;\;N_{N_i}^{\text{eq}} = \frac{3}{8} z_i^2 \mathcal{K}_2(z_i) \,. \label{eq:BEsBL}
\end{align}
In the above equations, $z_i = M_i/T$ and we use the notation $z\equiv z_1$. The quantity $N_{N_i}$ ($N_{N_j}^{\text{eq}}$) is the (equilibrium) $N_i$ number density. The temperature-dependent quantities $D_i(z)$, $D_{j i}(z)$, $S_i(z)$, $S_{i j}(z)$ and $W(z)$ are, respectively, the decay, the scattering and the washout terms. In particular, the coefficient $D_{i j}$ vanishes if $N_i \rightarrow N_j h_k$ is not kinematically allowed, i.e. for~$i \leq j$. Note that $\varepsilon_i$ is the $N_i$ unflavored CP asymmetry computed via Eqs.~\eqref{eq:cptotunflavored} and~\eqref{eq:fullCP}. The above system is solved in order to compute~$N_{B-L}$ and determine the BAU using Eq.~\eqref{eq:etaB}. We will take as initial conditions $N_{N_i}(z=0) = N_{N_i}^{\text{eq}}(z=0) = 3/4$ and $N_{B-L}(z=0) = 0$. The latter corresponds to the case of a Universe with no initial $(B-L)$-asymmetry. As explained above, by including all necessary diagrams shown in Fig.~\ref{fig:diagsRIS}, we obtain BEs equations meeting all Sakharov conditions required to successfully generate an asymmetry from an initially symmetric state. Namely, the lepton-asymmetry production in the term~$D_i$, the CP violation in $\varepsilon_i$, and the departure from thermal equilibrium of $N_i$ through the term~$N_{N_i}-N_{N_i}^{\text{eq}}$. If any of these terms vanishes, a $(B-L)$-asymmetry cannot be generated. 

We now turn our attention to the specific formulae for the BE coefficients.
\begin{itemize}

    \item \textbf{Decays:}
    
    The expressions for the decay parameters $D_i(z)$ and $D_{i j}(z)$ are [see Eq.~\eqref{eq:Decaygen}]
    \begin{align}
    D_i(z) = K_i r_{i 1} z \frac{\mathcal{K}_1(z_i)}{\mathcal{K}_2(z_i)} \; , \; D_{i j}(z) =
        K_{i j} r_{i 1} z \frac{\mathcal{K}_1(z_j)}{\mathcal{K}_2(z_j)} \; ,
    \label{eq:decayparam}
    \end{align}
    with
    \begin{align}
    K_i = \sum_{\alpha=e, \mu, \tau} \sum_{a=1}^{n_H} \frac{\Gamma(N_i \rightarrow \Phi_a \ell_\alpha) + \Gamma(N_i \rightarrow \Phi_a^\dagger \overline{\ell}_\alpha)}{H(T=M_i)} \; , \; K_{i j} = \sum_{k=1}^{2 n_S} \frac{\Gamma(N_i \rightarrow N_j h_k)}{H(T=M_i)} \; , 
    \end{align}
    where the Hubble parameter is given by Eq.~\eqref{eq:Hubble} and the expression for the LNV decay entering $K_i$ is the one of Eq.~\eqref{eq:Decaytree}. Note that the LNV inverse decay must be taken into account while computing the washout parameter $W(z)$. Furthermore, the decay rate for the tree-level $N_i \rightarrow N_j h_k$ process is
    \begin{align}
    \Gamma(N_i \rightarrow N_j h_k) = \frac{M_i \sqrt{\rho_{j i k}}}{16 \pi} \left\{ (1 + r_{j i} - \sigma_{k i}) \left|\mathbf{\Delta}_{i j}^k\right|^2 + 2 \sqrt{r_{j i}} \text{Re}\left[ (\mathbf{\Delta}^k_{i j})^2 \right]\right\}\; ,
    \label{eq:newdecayG}
    \end{align}
    being kinematically allowed only for $M_i > M_j + m_{h_k}$. For the 2RH neutrino case, the result obtained in Ref.~\cite{LeDall:2014too} is recovered taking $i=2$ and $j=1$.

\begin{figure}[t!]
  \centering
     \begin{subfigure}[b]{0.9\textwidth}
        \centering
         \includegraphics[scale=0.95,trim={4cm 23.5cm 2.5cm 1cm},clip]{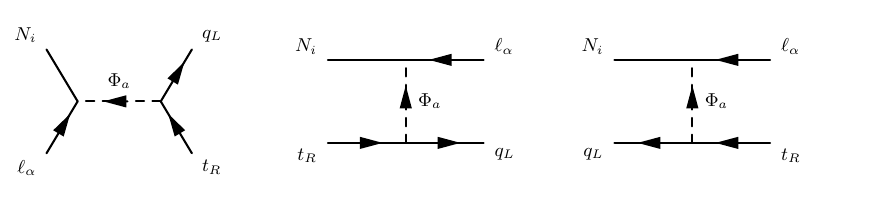}
         \caption{Usual type-I seesaw contributions: $N_i\ell_\alpha\leftrightarrow q_L t_R$, $N_it_R\leftrightarrow \ell_\alpha q_L $ and $N_iq_L\leftrightarrow \ell_\alpha t_R $.}
        \label{fig:DeltaLeq1_usual}
    \end{subfigure}\\
    \begin{subfigure}[b]{0.9\textwidth}
        \centering
       \includegraphics[scale=0.95,trim={4cm 20.1cm 2.5cm 4cm},clip]{Figures/LeptoSCPV/diagramsDeltaLeq1.pdf}
        \caption{$N_i\ell_\alpha\leftrightarrow\Phi_a h_k$}
         \label{fig:DeltaLeq1_Nltophih}
    \end{subfigure}
    \\
   \begin{subfigure}[b]{0.9\textwidth}
        \centering
        \includegraphics[scale=0.95,trim={4cm 16.9cm 2.5cm 7.5cm},clip]{Figures/LeptoSCPV/diagramsDeltaLeq1.pdf}
      \caption{$N_ih_k\leftrightarrow\ell_\alpha \Phi_a$}
        \label{fig:DeltaLeq1_Nhtophil}     
\end{subfigure}\\
    \begin{subfigure}[b]{0.9\textwidth}
\centering
       \includegraphics[scale=0.95,trim={4cm 13.9cm 2.5cm 11cm},clip]{Figures/LeptoSCPV/diagramsDeltaLeq1.pdf}
        \caption{$N_i\Phi_a\leftrightarrow\ell_\alpha h_k$}
         \label{fig:DeltaLeq1_Nphitohl}
     \end{subfigure}
    \caption{Scattering contributions with $\Delta L=1$.}
    \label{fig:DeltaLeq1_scattering}
\end{figure}

\item \textbf{Scatterings:}  
 
The Feynman diagrams for all $\Delta L=1$ two-body scattering processes included in our analysis are shown in Fig.~\ref{fig:DeltaLeq1_scattering}, with the corresponding reduced cross sections given in Appendix~\ref{chpt:scatterings}. Diagrams \subref{fig:DeltaLeq1_usual} are the usual ones of standard type-I seesaw leptogenesis, namely the $s$-channel $N_i\ell_\alpha\leftrightarrow q_L t_R$ and $t$-channel $N_it_R\leftrightarrow \ell_\alpha q_L$ and $N_iq_L\leftrightarrow \ell_\alpha t_R $~\cite{Buchmuller:2004nz}. Due to the presence of the new scalar particles $h_k$, new $\Delta L=1$ scatterings are allowed: \subref{fig:DeltaLeq1_Nltophih} $N_i\ell_\alpha\leftrightarrow\Phi_a h_k$, \subref{fig:DeltaLeq1_Nhtophil} $N_ih_k\leftrightarrow\ell_\alpha \Phi_a$ and \subref{fig:DeltaLeq1_Nphitohl} $N_i\Phi_a\leftrightarrow\ell_\alpha h_k$, with \subref{fig:DeltaLeq1_Nltophih} and \subref{fig:DeltaLeq1_Nhtophil} having an $s-$ and $t$-channel diagram, while \subref{fig:DeltaLeq1_Nphitohl} only occurs via $t$-channel mediation. As mentioned before, the $s$-channel $N_i\ell_\alpha\leftrightarrow\Phi_a h_k$ diagram features a RIS which must be subtracted to obtain a consistent set of BEs. Similarly, the $t$-channel Higgs mediated diagrams for $N_ih_k\leftrightarrow\ell_\alpha \Phi_a$ and $N_i\Phi_a\leftrightarrow\ell_\alpha h_k$ also contain RIS -- see Appendix~\ref{chpt:scatterings} for details on the subtraction procedure. In comparison to Ref.~\cite{LeDall:2014too}, where only the dominant contributions to these processes were included, here we consider and compute all the tree-level contributions for the aforementioned processes. The scattering parameters $S_i$ entering the BEs in Eqs.~\eqref{eq:BEsNi} and~\eqref{eq:BEsBL} are given by [see Eqs.~\eqref{eq:equi} and~\eqref{eq:Scatteringgen}]
    \begin{align}
        S_i = \frac{\gamma_i^\text{eq}}{n_{N_i}^\text{eq} H(z) z} \; , \; n_{N_i}^\text{eq}=\frac{M_i^2 T}{2 \pi^2} \mathcal{K}_2(z_i) \; ,
    \end{align}
    where $H(z) \equiv H(T= M_1/z)$. The reaction density is defined as
    \begin{align}
        \gamma_i^\text{eq} &= \sum_{\alpha=e, \mu, \tau} \left[\gamma^\text{eq}(N_i \ell_{\alpha} \leftrightarrow q_L t_R) + \gamma^\text{eq}(N_i t_R  \leftrightarrow \ell_{\alpha} q_L) + \gamma^\text{eq}(N_i q_L \leftrightarrow \ell_{\alpha} t_R) \right]  \\
        & + \sum_{\alpha=e, \mu, \tau} \sum_{a=1}^{n_H} \sum_{k=1}^{2 n_S} \left[\gamma^\text{eq}(N_i \ell_{\alpha} \leftrightarrow \Phi_a h_k) + \gamma^\text{eq}(N_i h_k \leftrightarrow \ell_{\alpha} \Phi_a) + \gamma^\text{eq}(N_i \Phi_a \leftrightarrow \ell_{\alpha} h_k) \right] \; , \nonumber
    \end{align}
    which is computed through Eq.~\eqref{eq:reacscattering} using the reduced cross sections presented in Appendix~\ref{chpt:scatterings} and taking into consideration all necessary RIS subtractions. It is worth stressing that the above LNV scattering processes will contribute to the washout coefficient~$W(z)$.

\begin{figure}[t!]
    \centering
     \begin{subfigure}[b]{0.9\textwidth}
         \centering
         \includegraphics[scale=0.95,trim={4cm 23.5cm 2.5cm 1cm},clip]{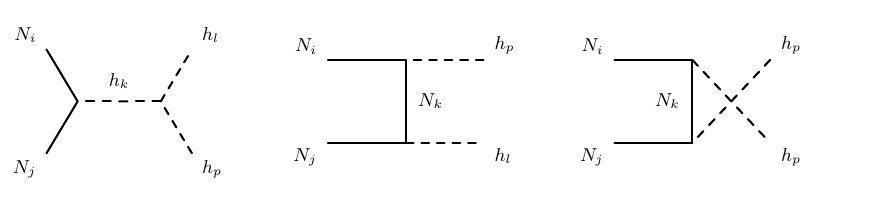}
        \caption{$N_i N_j \leftrightarrow h_p h_l$.}
         \label{fig:DeltaNeq2_NNtohh}
     \end{subfigure}\\
     \begin{subfigure}[b]{0.9\textwidth}
         \centering
        \includegraphics[scale=0.95,trim={4cm 20.5cm 2.5cm 4cm},clip]{Figures/LeptoSCPV/diagramsDeltaNeq2.pdf}
        \caption{$N_i N_j \leftrightarrow \Phi_a \Phi_b$}
         \label{fig:DeltaNeq2_NNtohhphiphi}
    \end{subfigure}
    \caption{Neutrino annihilation scattering contributions with $\Delta N=2$.}
    \label{fig:DeltaNeq2_scattering}
\end{figure}

In Fig.~\ref{fig:DeltaNeq2_scattering} we present the novel heavy-neutrino annihilation diagrams which, not being LNV, are dubbed as $\Delta N = 2$ processes.\footnote{Such~$\Delta N = 2$ interactions appear, e.g. in the context of supersymmetric SO(10) unification~\cite{Plumacher:1996kc,Plumacher:1997ru,Plumacher:1998ex}.} Their existence is due to the interactions between the scalar singlets and heavy Majorana neutrinos, as well as to the triple scalar potential couplings involving $h_k$ -- see eqs~\eqref{eq:LYuk} and~\eqref{eq:muCPdef}, respectively. The process $N_i N_j \leftrightarrow h_p h_l$ shown in diagram \subref{fig:DeltaNeq2_NNtohh} was considered in Ref.~\cite{AristizabalSierra:2014uzi} in the limit of a single heavy neutrino and one Yukawa coupling. The $s$-channel mediated diagram \subref{fig:DeltaNeq2_NNtohhphiphi} $N_i N_j \leftrightarrow \Phi_a \Phi_b$ was already computed in Refs.~\cite{LeDall:2014too,Alanne:2017sip,Alanne:2018brf} for the case of one Higgs doublet, while the $t$-channel, although present in standard type-I seesaw leptogenesis, is usually neglected since it does not influence the washout term. In fact, none of these scattering contributions enter the expression for $W(z)$. We present the general and complete scattering formulas for the aforementioned processes (see Appendix~\ref{chpt:scatterings}) and consider them in our numerical analysis of Sec.~\ref{sec:BAUmodel}. The $S_{ij}$ coefficients accounting for the $\Delta N = 2$ scatterings are then given by [see Eqs.~\eqref{eq:equi} and~\eqref{eq:Scatteringgen}]
    \begin{align}
        S_{i j} &= \frac{\gamma^\text{eq}_{i j}}{n_{N_i}^\text{eq} n_{N_j}^\text{eq}H(z) z} \; , \nonumber \\
        \gamma^\text{eq}_{i j} &= \sum_{a,b=1}^{n_H}  \gamma^\text{eq}(N_i N_j \leftrightarrow \Phi_a \Phi_b) + \sum_{p,l=1}^{2 n_S} \gamma^\text{eq}(N_i N_j \leftrightarrow h_p h_l)\; ,
    \end{align}
which are computed using Eq.~\eqref{eq:reacscattering} and the cross sections obtained in Appendix~\ref{chpt:scatterings}.
    
\item \textbf{Washout:}
    
The washout term only accounts for LNV processes, since these alter the ($B-L$) asymmetry, such as inverse decays $\ell_\alpha \Phi_a \rightarrow N_i$ and $\Delta L=1,2$ scatterings. Hence, the additional $N_j \rightarrow N_i h_k$ decays or $\Delta N =2$ scattering processes do not contribute to~$W(z)$. As mentioned before, we will not consider $\Delta L=2$ scatterings. Thus,
    \begin{align}
    W(z) &= W_{\rm ID}(z) + W_{\Delta L=1}(z) \; ,
    \end{align}
    with [see Eq.~\eqref{eq:InvDecaygen}],
    \begin{align}
    W_{\rm ID}(z) &= \sum_{i=1}^{n_R} \frac{1}{2} \frac{N_{N_i}^{\text{eq}}}{N_{\ell}^{\text{eq}}} D_i = \frac{1}{4} \sum_{i=1}^{n_R} K_i \sqrt{r_{i 1}} z_i^3 \mathcal{K}_1(z_i)  \; , \\
    W_{\Delta L=1}(z) &= \sum_{i=1}^{n_R} \frac{N_{N_i}^{\text{eq}}}{N_{\ell}^{\text{eq}}} \left(S_i^\prime + \frac{N_{N_i}}{N_{N_i}^{\text{eq}}} S_i^{\prime \prime} \right) \; ,
    \end{align}
    where $N_{\ell}^{\text{eq}}=3/4$ and 
    \begin{align}
        S_i^\prime = \frac{\gamma_i^{\text{eq} \; \prime}}{n_{N_i}^\text{eq} H(z) z} \; , \; \gamma_i^{\text{eq} \; \prime} &= \sum_{\alpha=e, \mu, \tau} \left[ \gamma^\text{eq}(N_i t_R  \leftrightarrow \ell_{\alpha} q_L) + \gamma^\text{eq}(N_i q_L \leftrightarrow \ell_{\alpha} t_R) \right] \nonumber \\ 
        & + \sum_{\alpha=e, \mu, \tau} \sum_{a=1}^{n_H} \sum_{k=1}^{2 n_S} \left[\gamma^\text{eq}(N_i h_k \leftrightarrow \ell_{\alpha} \Phi_a) + \gamma^\text{eq}(N_i \Phi_a \leftrightarrow \ell_{\alpha} h_k) \right] \; , \\
        S_i^{\prime \prime} = \frac{\gamma_i^{\text{eq} \; \prime \prime}}{n_{N_i}^\text{eq} H(z) z} \; , \; \gamma_i^{\text{eq} \; \prime \prime} &= \sum_{\alpha=e, \mu, \tau} \gamma^\text{eq}(N_i \ell_{\alpha} \leftrightarrow q_L t_R) \nonumber \\
        & + \sum_{\alpha=e, \mu, \tau} \sum_{a=1}^{n_H} \sum_{k=1}^{2 n_S} \gamma^\text{eq}(N_i \ell_{\alpha} \leftrightarrow \Phi_a h_k)\; ,
    \end{align}
    where $S_i^\prime$ ($S_i^{\prime \prime})$ contains all $\Delta L = 1$ scattering processes where the charged-lepton $\ell_\alpha$ appears in the final (initial) state.
    
\end{itemize}

Up to now we have set our theoretical framework and presented the expressions for the CP asymmetries and the unflavored BEs needed for the computation of the BAU. In the next section, we will present a simple model where leptogenesis is entirely due to the new scalar interactions fed by a single CP-violating phase of that scalar VEV. It turns out that (Dirac and Majorana) low-energy CP violation in the neutrino sector is also induced.

%%%%%%%%%%%%%%%%%%%%%%%%%%%%%%%%%%%%%%%%%%%%%%%%%%%%%%%%%%%%%%%%%%%%%%%%%%%%%
\subsection{A simple model for leptogenesis with high-energy SCPV}
\label{sec:model}
%%%%%%%%%%%%%%%%%%%%%%%%%%%%%%%%%%%%%%%%%%%%%%%%%%%%%%%%%%%%%%%%%%%%%%%%%%%%%

%
\begin{table}[t!]
\renewcommand{\arraystretch}{1.5}
	\centering
	\begin{tabular}{| K{1.8cm} | K{1.5cm} | K{3.0cm} |  K{1.5cm} | K{1.5cm} | K{1.5cm} | }
		\hline 
&Fields&\EW&  $\mathcal{Z}_{8}^{e}$ &  $\mathcal{Z}_{8}^{\mu}$  &  $\mathcal{Z}_{8}^{\tau}$ \\
		\hline 
		\multirow{8}{*}{Fermions} 
&$\ell_{e L} $&($\mathbf{2}, {-1/2}$)   & {$\omega^5$} & {$\omega^7$}  & {$\omega^6$}    \\
&$\ell_{\mu L}$&($\mathbf{2}, {-1/2}$)  & {$\omega^7$} & {$\omega^5$}  & {$\omega^5$}     \\
&$\ell_{\tau L}$&($\mathbf{2}, {-1/2}$) & {$\omega^6$} & {$\omega^6$}  & {$\omega^7$}     \\
&$e_R$&($\mathbf{1}, {-1}$)     & {$\omega^4$} & {$\omega^7$}  & {$\omega^6$}     \\
&$\mu_R$&($\mathbf{1}, {-1}$)   & {$\omega^7$} & {$\omega^4$}  & {$\omega^4$}    \\
&$\tau_R$&($\mathbf{1}, {-1}$)  & {$\omega^6$} & {$\omega^6$}  & {$\omega^7$}    \\
&$\nu_{R_1}$&($\mathbf{1}, {0}$)& {$\omega^6$} & {$\omega^6$}& {$\omega^6$}    \\
&$\nu_{R_2}$&($\mathbf{1}, {0}$)& {$1$}    & {$1$}    & {$1$}   \\
		\hline 
		\multirow{3}{*}{Scalars}
&$\Phi_1$&($\mathbf{2}, {1/2}$)&\multicolumn{3}{c|}{{$1$}}\\
&$\Phi_2$&($\mathbf{2}, {1/2}$)&\multicolumn{3}{c|}{{$\omega$}}\\
&$S$&($\mathbf{1}, {0}$)&\multicolumn{3}{c|}{{\ $\omega^2$}}\\
\hline
	\end{tabular}
	\caption{Field content of the model and corresponding transformation properties under the \EW gauge group. For the $\mathcal{Z}_8$ symmetry we have $\omega^k = e^{ik\pi/4}$.}
	\label{tab:part&sym} 
\end{table}

We now focus on a simple realization of the general framework described in the previous sections, i.e. a SM extension with two RH neutrinos $\nu_{R_{1,2}}$ ($n_R=2$) and one complex scalar singlet~$S$ ($n_S=1$). With this setup, SCPV can be achieved at high energies through the complex VEV of the scalar singlet $S$, provided we guarantee that the $S^4$ term in the scalar potential is present~\cite{Branco:1999fs}. As also noted in Refs.~\cite{Correia:2019vbn,Camara:2020efq}, at least two-Higgs doublets are necessary to implement Abelian flavor symmetries in this type of scenarios. Thus, we will add an extra scalar doublet to the field content of our model such that $n_H=2$. With $n_R=2$, $n_S=1$ and $n_H=2$, the most restrictive Abelian symmetry which can be implemented is a~$\mathcal{Z}_{8}$ which, as we will see shortly, will lead to testable low-energy predictions. 

The $\mathcal{Z}_{8}$ symmetry is suitable for two reasons: i) it allows for the $S^4$ term in the scalar potential needed for SCPV, provided $S$ transforms as $S \rightarrow \omega^2 S$ [$\omega=\text{exp}(i \pi / 4)$] and ii) with the aforementioned minimal particle content the $\mathcal{Z}_8$ is the lowest-order $\mathcal{Z}_n$ symmetry containing a sufficient number of charges to obtain non-trivial flavor predictions. The particle content of the model, together with the $\mathcal{Z}_{8}$ charge assignments, is summarized in Table~\ref{tab:part&sym}. The three cases $\mathcal{Z}_{8}^{e}$, $\mathcal{Z}_{8}^{\mu}$ and $\mathcal{Z}_{8}^{\tau}$ differ from each other by the $\mathcal{Z}_{8}$ charges of the SM lepton fields. As will be clear in the next sections, this minimal model illustrates the main idea: the single VEV phase $\theta$ of $S$ provides a common source for CP violation required to generate the BAU and low-energy leptonic CP violation.

%%%%%%%%%%%%%%%%%%%%%%%%%%%%%%%%%%%%%%%%%%%%%%%%%%%%%%%%%%%%%%%%%%%%%%%%%%%%%
\subsubsection{Compatibility with low-energy neutrino data}
\label{sec:neutrinodata}
%%%%%%%%%%%%%%%%%%%%%%%%%%%%%%%%%%%%%%%%%%%%%%%%%%%%%%%%%%%%%%%%%%%%%%%%%%%%%

With $n_{R,H}=2$ and $n_S=1$, the relevant couplings and mass parameters of Eq.~\eqref{eq:LYuk} are $\Ye^{1,2}$, $\YD^{1,2}$, $\YR^\prime$ and $\MR^0$ which are all real, since we impose CP symmetry at the Lagrangian level. As already mentioned, the three $\mathcal{Z}_8$ lepton charge assignments given in Table~\ref{tab:part&sym} will lead to different Yukawa couplings $\Ye^{1,2}$ and $\YD^{1,2}$. For the specific $\mathcal{Z}_8^{\mu}$ case, we have
\begin{align}
    \Ye^1=\begin{pmatrix}
    y_1&0&0\\
    0&0&0\\
    0&0&y_4
    \end{pmatrix},\; 
    &\Ye^2=\begin{pmatrix}
    0&0&y_2\\
    0&y_3&0\\
    0&0&0
    \end{pmatrix}, \;
    \YD^1=\begin{pmatrix}
    0 & 0\\
    0 & 0\\
    y_{D_3} & 0
    \end{pmatrix},\;
   \YD^2=\begin{pmatrix}
    0 & y_{D_1}\\
    y_{D_2} & 0 \\
    0 & 0
    \end{pmatrix}, \nonumber \\[0.2cm]
    &\MR^0=\begin{pmatrix}
    0&0\\
    .&m_R
    \end{pmatrix},\; 
    \YR'=\begin{pmatrix}
    0 & y_{R_S}\\
    .&0
    \end{pmatrix},\;
    \label{eq:yukawastructures}
\end{align}
where all parameters are real and the dots reflect the symmetric nature of the Majorana matrices. The corresponding matrices for $\mathcal{Z}_8^{e}$ and $\mathcal{Z}_8^{\tau}$ are obtained from the above by performing the permutations $\ell_L,e_R \rightarrow \mathbf{P}_{12}\, \ell_L,e_R$ and $\ell_L,e_R \rightarrow \mathbf{P}_{23}\, \ell_L,e_R$, respectively, being
\begin{equation}
\mathbf{P}_{12} = \begin{pmatrix} 
0 & 1 & 0 \\ 
1 & 0 & 0 \\
0 & 0 & 1
\end{pmatrix} \; , \; \mathbf{P}_{23} = \begin{pmatrix} 
1 & 0 & 0 \\ 
0 & 0 & 1 \\
0 & 1 & 0 
\end{pmatrix} \; .
\label{eq:P12P23}
\end{equation}
Notice that, due to the~$\mathcal{Z}_8$ symmetry, the term $\overline{\nu_R} \nu_R^c S$ is absent from the Lagrangian. However, since the $\overline{\nu_R} \nu_R^cS^\ast$ coupling $y_{R_S}$ is $\mathcal{Z}_{8}$ invariant, LCPV can, in principle, be successfully transmitted to the neutrino sector as long as~$S$ acquires a complex VEV. In fact, the charge assignments guarantee that $\MR^0$ and $\YR^\prime$ are not proportional to each other implying that $\Uh$ will be complex. Consequently, as we will see shortly, LCPV probed in neutrino oscillation experiments originates dynamically from the vacuum and so does the BAU. 

We consider the following VEV assignments for the neutral components of the two scalar doublets $\phi^0_{1,2}$ and for the complex singlet $S$:
\begin{align}
    \langle \phi^0_{1} \rangle = \frac{v_1}{\sqrt{2}} \; , \;
    \langle \phi^0_{2} \rangle = \frac{v_2}{\sqrt{2}} \; , \; 
    \langle S \rangle=\frac{u e^{i\theta}}{\sqrt{2}} \; ,
    \label{eq:vevs}
\end{align}
where $v_{1,2}$, $u$ and $\theta$ are real. The charged-lepton, Dirac neutrino and RH neutrino mass matrices, needed to compute low-energy neutrino masses and mixing, are given by:
\begin{align}
   \Me =\begin{pmatrix}
    a_1&0&a_2\\
    0&a_3&0\\
    0&0&a_4
    \end{pmatrix} \; , \;
   \MD=\begin{pmatrix}
    0 & m_{D_1}\\
    m_{D_2} & 0 \\
    m_{D_3} & 0
    \end{pmatrix} \; , \;
    \MR=\begin{pmatrix}
    0& m_{R_S} e^{-i \theta}\\
    .&m_R
    \end{pmatrix} \; ,
    \label{eq:masstructures}
\end{align}
where
\begin{equation}
\begin{aligned}
a_{1,4} = \frac{v_1 y_{1,4}}{\sqrt{2}} \; , \; a_{2,3} = \frac{v_2 y_{2,3}}{\sqrt{2}} \; , \; m_{D_{1,2}} = \frac{v_2 y_{D_{1,2}}}{\sqrt{2}} \; , \; m_{D_{3}} = \frac{v_1 y_{D_{3}}}{\sqrt{2}} \; , \; m_{R_S} = \frac{u y_{R_S}}{\sqrt{2}} \,.
\label{eq:pdef}
\end{aligned}
\end{equation}
The specific form of $\Me$ indicates that $a_3$ is directly the mass of one physical charged lepton, which we call $\ell_2$, and is decoupled from the other two $\ell_{1,3}$. Three out of the four $a_i$ can be written in terms of the charged-lepton masses $m_{\ell_{1,2,3}}$ and a single $a$ parameter, considered here to be $a_4$. Thus, we have
\begin{equation}
a_1^2 = \frac{m^2_{\ell_1} m^2_{\ell_3}}{a_4^2} \; , \; a_2^2 = \frac{(a_4^2-m^2_{\ell_1}) (m^2_{\ell_3}-a_4^2)}{a_4^2} \; , \; a_3^2 = m^2_{\ell_2} \; , \; m_{\ell_1}^2<a_4^2<m_{\ell_3}^2 \; ,
\end{equation}
The unitary matrix $\mathbf{U}_{L}^e$ that diagonalizes the Hermitian matrix $\mathbf{H}_{e} = \mathbf{M}_{e} \mathbf{M}_{e}^{\dagger}$ is given by
\begin{align}
\mathbf{H}_{e} = \begin{pmatrix} 
a_1^2 + a_2^2 & 0 & a_2 a_4\\ 
0 & a_3^2  & 0 \\
a_2 a_4 & 0 & a_4^2    
\end{pmatrix}\;,\; \mathbf{V}_{L} = \begin{pmatrix} 
c_L & 0 & s_L\\ 
0 & 1 & 0 \\
- s_L & 0 & c_L
\end{pmatrix} \; ,
\label{eq:VLmodel}
\end{align}
with $c_{L}\equiv \cos \theta_{L}$, $s_{L}\equiv \sin \theta_{L}$ and
\begin{equation}
\tan \left(2 \theta_{L}\right) = \dfrac{2 \sqrt{(a_4^2-m_{\ell_1}^2)(m_{\ell_3}^2-a_4^2)}}{(m_{\ell_1}^2+m_{\ell_3}^2)-2 a_4^2} \; .
\label{eq:tLexp}
\end{equation}
For each symmetry charge assignment $\mathcal{Z}_8^{e, \mu, \tau}$, one must consider the three possible choices for $\ell_2$, i.e. $\ell_2= e, \mu, \tau$. For the $\mathcal{Z}_8^{\mu}$ case, $\mathbf{U}_L^e$ in Eq.~\eqref{eq:VLmodel}  corresponds to $\ell_2 = \mu$. For electron (tau) decoupled i.e., $\ell_2 = e$ ($\ell_2 = \tau$), $\mathbf{U}_L^e$ is replaced by $\mathbf{P}_{12}\mathbf{U}_L^e$ ($\mathbf{P}_{23}\mathbf{U}_L^e$), with $\mathbf{P}_{12}$ $(\mathbf{P}_{23})$ given in Eq.~\eqref{eq:P12P23} and $\theta_L$ determined by Eq.~\eqref{eq:tLexp}.

Taking into account Eqs.~\eqref{eq:TypeIMeff} and~\eqref{eq:masstructures}, the effective neutrino mass matrix in the flavor basis reads
\begin{align}
\Mnu = 
e^{2 i \theta}\begin{pmatrix} 
0 & y\, e^{-i \theta}  & y \sqrt{\dfrac{z}{x}}\, e^{-i \theta}  \\
. &  x &  \sqrt{x z} \\
. & . &  z
\end{pmatrix}\;,\;
x=\frac{m_R m_{D_2}^2}{m_{R_S}^2}\;,\;
y=-\frac{m_{D_1}m_{D_2}}{m_{R_S}}\;,\;
z=\frac{m_R m_{D_3}^2}{m_{R_S}^2}\;.
%\label{eq:xyz}
\label{eq:Mnuxyz}
\end{align}
Performing the rotation to the charged-lepton mass basis with the unitary matrix $\mathbf{V}_{L}$ given in Eq.~\eqref{eq:VLmodel}, we obtain for $\mathcal{Z}_8^{\mu}$:
\begin{align}
&\mathbf{U}_L^{e T} \Mnu \mathbf{U}_L^e = \nonumber \\
& \begin{pmatrix} 
 - e^{-i \theta} y \sqrt{\dfrac{z}{x}} \sin(2 \theta_L) + z s_L^2\quad\quad & e^{-i \theta} y c_L-\sqrt{x z} s_L \quad\quad& e^{-i \theta} y \sqrt{\dfrac{z}{x}} \cos(2 \theta_L) - \dfrac{z}{2} \sin(2 \theta_L) \\
. & x &  e^{-i \theta} y s_L + \sqrt{x z} c_L\\
. & . & e^{-i \theta} y \sqrt{\dfrac{z}{x}} \sin(2 \theta_L) + z c_L^2
\end{pmatrix} \; ,
\label{eq:mnuclmassbasis}
\end{align}
while for $\mathcal{Z}_8^e$ ($\mathcal{Z}_8^\tau$) permutation  $\mathbf{P}_{12}$ ($\mathbf{P}_{23}$) in Eq.~\eqref{eq:P12P23} must be applied both on the left and right. We remark that the $\mathcal{Z}_8$ flavor symmetry reduces the total number of free effective parameters to five. Namely, $\theta_L$ (or equivalently $a_4$) from $\Me$ and $(x,y,z,\theta)$ from $\Mnu$. Notice that the singlet VEV phase $\theta$ cannot be removed via field redefinitions and, consequently, it may successfully lead to low-energy CP violation in the neutrino sector. Furthermore, since there are only two RH neutrinos, one of the light neutrinos is massless. This is readily seen by computing the eigenvalues of the matrices in Eqs.~\eqref{eq:Mnuxyz} and~\eqref{eq:mnuclmassbasis}.

In order to test the compatibility of our model with neutrino oscillation data, the mass matrix $\Mnu$ of Eq.~\eqref{eq:mnuclmassbasis} for the $\mathcal{Z}_8^{e,\mu,\tau}$ cases must be matched with that defined through low-energy parameters, namely
\begin{align}
    \Mnuh=\U^\ast\text{diag}(m_1,m_2,m_3)\U^\dagger,
\end{align}
where $\U$ is the lepton mixing matrix given in  Eq.~\eqref{eq:CCleptoU} and parameterized as in Eq.~\eqref{eq:PMNSparam}. We make use of the neutrino data given in Table~\ref{tab:leptondata} of Sec.~\ref{sec:neutrinoobservables}, obtained from the global fit of neutrino oscillation parameters Ref.~\cite{deSalas:2020pgw} (see also Refs.~\cite{Esteban:2024eli} and~\cite{Capozzi:2025wyn}). Since we are in the minimal type-I seesaw, the lightest neutrino is massless ($m_1=0$ for NO, and $m_3=0$ for IO) and only on Majorana phase $\alpha$ is physically relevant.

We tested the viability of cases $\mathcal{Z}_8^{e,\mu,\tau}$, for both NO and IO, using a standard $\chi^2$-analysis, by minimizing the function
\begin{align}
    \chi^2(x,y,z,\theta,\theta_L)=\sum_i\dfrac{[\mathcal{P}_i(x,y,z,\theta,\theta_L)-\mathcal{O}_i]^2}{\sigma_i^2},
\end{align}
with respect to the neutrino observables $\Delta m^2_{ij}$, $\theta_{ij}^\ell$ and $\delta^\ell$. In the above, $\mathcal{P}_i$ corresponds to the output value for the observable $i$ obtained by varying the input parameters $x$, $y$, $z$, $\theta$ and $\theta_L$, while $\mathcal{O}_i$ ($\sigma_i$) denotes the correspondent best-fit value ($1\sigma$ experimental uncertainty), indicated in Table~\ref{tab:leptondata}.

\begin{table}[!t]
\renewcommand{\arraystretch}{1.5}
\centering
\setlength{\tabcolsep}{10pt}
\begin{tabular}{|c|c|c|c|c|c|c|c|c|}  
\hline
\multirow{2}{*}{Case}  & $\theta_{12}^\ell$ & $\theta_{13}^\ell$ & $\theta_{23}^\ell$ & \multirow{2}{*}{$\delta^\ell/\pi$} & \multirow{2}{*}{$\alpha/\pi$} & $m_{\beta\beta}$&  $m_\beta$  & $\sum_i m_i$\\
&$(^\circ)$&$(^\circ)$&$(^\circ)$&&&(meV)&(meV)&(meV)\\
\hline
$\mathcal{Z}_8^\mu$  (IO) &$35.48$&8.60&$49.62$&$1.88$&$0.92$&$16.6$&$49.2$&$99.7$\\
\hline
\end{tabular}
\caption{$\theta_{12}^\ell$, $\theta_{13}^\ell$, $\theta_{23}^\ell$, $\delta^\ell$, $\alpha$, $m_{\beta\beta}$, $m_{\beta}$ and $\sum_i m_i$ best-fit values for the only case compatible with neutrino oscillation data at the $1\sigma$ level: $\mathcal{Z}_8^\mu$ muon-decoupled with IO neutrino masses. All the remaining cases, $\mathcal{Z}_8^e$, $\mathcal{Z}_8^\tau$ and $\mathcal{Z}_8^{\mu}$ for NO, and considering all possible decoupled charged-lepton states, are not compatible with data.}
\label{tab:compatibility}
\end{table}
\begin{table}[!t]
\renewcommand{\arraystretch}{1.5}
\centering
\setlength{\tabcolsep}{10pt}
\begin{tabular}{|c|c|c|c|}  
\hline
Case  & $\theta/\pi$ & $\theta_L/\pi$&$(x, \ y, \ z)$ (meV)\\
\hline
$\mathcal{Z}_8^\mu$  (IO) &$1.89$&$7.29\times10^{-2}$&$(0.325,\ 32.8, \ 0.426)$\\
\hline
\end{tabular}
\caption{Values of the $\theta$, $\theta_L$, $x$, $y$ and $z$ input parameters in Eq.~\eqref{eq:Mnuxyz} which lead to the best-fit case shown in Table~\ref{tab:compatibility}.}
\label{tab:compatibilityoutput}
\end{table}
In Table~\ref{tab:compatibility}, we show the $\theta_{12}^\ell$, $\theta_{13}^\ell$, $\theta_{23}^\ell$ and $\delta^\ell$ values for the case that best fits the data, i.e $\mathcal{Z}_8^\mu$ with muon decoupled and IO (the $\Delta m_{21,31}$ are on their best-fit values). We also indicate the predictions for the Majorana phase $\alpha$, effective masses $m_{\beta\beta}$ relevant for $0_\nu \beta \beta$ [see Eq.~\eqref{eq:NOIOmbb2min}] and $m_\beta$ ($\beta$-decay), as well as the sum of neutrino masses~$\sum_i m_i$. The remaining cases, $\mathcal{Z}_8^\mu$ for NO and $\mathcal{Z}_8^{e,\tau}$, are compatible with data at more than $3\sigma$. From these results, one can see that the model prefers $\theta_{23}^\ell$ in the second octant, and that $m_{\beta\beta}$, $m_\beta$ and $\sum_i m_i$ are well below the current most stringent limits from KamLAND-Zen 800 -- $m_{\beta\beta}< (36-156)$~meV ($90\%$ CL)~\cite{KamLAND-Zen:2022tow} (see Table~\ref{tab:dataNDBD}), KATRIN -- $m_{\beta}<0.8$~eV ($90\%$ CL)~\cite{KATRIN:2021uub}, and Planck -- $\sum_i m_i<(0.12-0.54)$~eV ($95\%$ CL)~\cite{Planck:2018vyg}, respectively. The corresponding values of the input parameters $(\theta,\theta_L,x,y,z)$ in Eq.~\eqref{eq:mnuclmassbasis} are given in Table~\ref{tab:compatibilityoutput} and will be used in the leptogenesis analysis of Sec.~\ref{sec:BAUmodel}. Note that the VEV phase $\theta$ is close to $340^\circ$ and the Dirac CP phase is predicted to be $\delta^\ell \simeq \theta$, which is $1.5\sigma$ away from the experimental best-fit for $\delta^\ell$. This relation between $\theta$ and $\delta^\ell$ stems from the underlying $\mathcal{Z}_8$ symmetry of our model. It is indeed remarkable that, with such a simple setup, a one-to-one correspondence between $\delta^\ell$ (low-energy Dirac CP phase) and $\theta$ (vacuum CP phase) can be established, connecting two completely different sectors of the model.

%%%%%%%%%%%%%%%%%%%%%%%%%%%%%%%%%%%%%%%%%%%%%%%%%%%%%%%%%%%%%%%%%%%%%%%%%%%%%
\subsubsection{BAU generation from vacuum CP violation}
\label{sec:BAUmodel}
%%%%%%%%%%%%%%%%%%%%%%%%%%%%%%%%%%%%%%%%%%%%%%%%%%%%%%%%%%%%%%%%%%%%%%%%%%%%%

In the model under consideration, CP is spontaneously broken at high energies by the complex VEV of the singlet $S$. The $\mathcal{Z}_8$-invariant scalar potential is\footnote{We added the soft-breaking term $\propto (\Phi_1^\dagger \Phi_2)$ to avoid a massless Goldstone boson which would stem from the Higgs doublets neutral degrees of freedom after EWSB.}
\begin{align}
V(\Phi_1,\Phi_2,S) &= m^2_{1} (\Phi_1^\dagger \Phi_1 ) + m^2_{2} (\Phi_2^\dagger \Phi_2 ) + m^2_{1 2} \left[ (\Phi_1^\dagger \Phi_2 ) + (\Phi_2^\dagger \Phi_1 ) \right] \nonumber \\
&+ \frac{\lambda_{1}}{2} (\Phi_1^\dagger \Phi_1 )^2 + \frac{\lambda_{2}}{2} (\Phi_2^\dagger \Phi_2 )^2 + \lambda_{3} (\Phi_1^\dagger \Phi_1 ) (\Phi_2^\dagger \Phi_2 ) + \lambda_{4} (\Phi_1^\dagger \Phi_2 ) (\Phi_2^\dagger \Phi_1) \nonumber \\
& + \lambda_{1 S} (\Phi_1^\dagger \Phi_1 ) |S|^2 + \lambda_{2 S} (\Phi_2^\dagger \Phi_2 ) |S|^2 \nonumber \\
&+ m^2_{S} |S|^2 + m^{\prime \; 2}_{S} \left(S^2 + {S^\ast}^2\right)  + \frac{\lambda_S}{2} |S|^4 + \lambda_S^\prime \left(S^4 + {S^\ast}^4 \right) \; ,
\label{eq:potentialmodel}
\end{align}
where all parameters are real, since CP invariance is imposed. We consider that at high-energies or, in other words, at temperatures much higher than the EW scale, only $S$ has non-zero VEV. In fact, within the unflavored leptogenesis scenario considered here, the heavy Majorana neutrino masses are such that $M_{1,2}\sim u \gtrsim 10^{12}$~GeV. Hence, the scalar $S$ is naturally decoupled from $\Phi_{1,2}$. For the above potential, we then obtain three non-trivial scalar potential minimization conditions:
\begin{align}
\text{(i) :} & \;  m_{S}^2 = - \frac{1}{2} \left[u^2 \left(\lambda_{S} + 4 \lambda_{S}^{\prime}\right) + 4 m_{S}^{\prime} \right] \; , \; \theta = k \pi \; , k \in \mathbb{Z} \; ; \\
\text{(ii) :} & \; m_{S}^2 = - \frac{1}{2} \left[u^2 \left(\lambda_{S} + 4 \lambda_{S}^{\prime}\right) - 4 m_{S}^{\prime} \right] \; , \; \theta = \frac{\pi}{2} + k \pi \; , k \in \mathbb{Z} \; ; \\
\text{(iii) :} & \; m_{S}^2 = - \frac{u^2}{2} \left(\lambda_{S} - 4 \lambda_{S}^{\prime} \right) \; , \; \cos (2 \theta) = - \frac{m^{\prime\,2}_S}{2 u^2 \lambda_{S}^{\prime}} \; .
\label{eq:SCPVsol}
\end{align}
As mentioned in the beginning of this chapter, in spite of (ii) leading to $\theta = \pi/2$, it can be shown that in this case the vacuum does not violate CP~\cite{Branco:1999fs}. Therefore, the only viable solution to implement SCPV is (iii). In the exact $\mathcal{Z}_8$-symmetric limit, i.e. if the soft-breaking parameter $m^{\prime\,2}_S$ vanishes, we have $\theta= \pi/4+k\pi/2$, still leading to SCPV [see Eq.~\eqref{eq:thetaexactOG}]. However, such value of $\theta$ is not compatible with neutrino data~\footnote{For this case, the solution which fits the data the best corresponds to $\theta=7\pi/4$, leading to best-fit value(s) of $\theta_{12}^\ell$ ($\theta_{13}^\ell$ and $\theta_{23}^\ell$) which is (are) approximately at $3\sigma$ ($1\sigma$) distance of its (their) experimental best fit.} (see Table~\ref{tab:compatibilityoutput}). Therefore, a non-vanishing $m^{\prime\,2}_S$ is required, which also helps avoiding the domain wall~(DW) problem arising from the spontaneous breaking of the $\mathcal{Z}_8$ discrete symmetry. The SCPV solution (\ref{eq:SCPVsol}) corresponds to the global minimum of the potential if $ \left(m_{S}^{\prime \; 4} - 4 u^4 \lambda_{S}^{\prime \; 2} \right) / (4 \lambda_{S}^{\prime}) > 0$. Also, in order for the scalar potential to be BFB we require:
\begin{align}
&\lambda_1, \lambda_2, \lambda_S > 0 \; , \nonumber \\
& \lambda_3 + \sqrt{\lambda_1 \lambda_2} > 0 \; , \; \lambda_3 + \lambda_4 + \sqrt{\lambda_1 \lambda_2} > 0 \; , \; \lambda_S - 4 |\lambda_S^\prime| > 0 \; , \nonumber \\
& \lambda_{1 S} + \sqrt{\lambda_1 \lambda_S} > 0 \; , \; \lambda_{2 S} + \sqrt{\lambda_2 \lambda_S} > 0 \; , \nonumber \\ &\lambda_{1 S} + \sqrt{\lambda_1 \left(\lambda_S - 4 |\lambda_S^\prime|\right)} > 0 \; , \; \lambda_{2 S} + \sqrt{\lambda_2 \left(\lambda_S - 4 |\lambda_S^\prime|\right)} > 0 \; .
\label{eq:bfb}
\end{align}

In Sec.~\ref{sec:leptogenesis}, we presented the expressions for the CP asymmetries in the fermion and $S$ scalar mass-eigenstate basis. For the model under discussion with $n_S=1$, $\bm{\bm{\mathcal{M}}}_S^2$ and $\mathbf{V}$ in Eqs.~\eqref{eq:mixS} and~\eqref{eq:mixingscalar} are $2\times2$ matrices. Namely, in the $(S_R,S_I)$ basis we have,
\begin{align}
\bm{\bm{\mathcal{M}}}_S^2 = u^2 \begin{pmatrix}
(\lambda_S + 4 \lambda_S^\prime ) \cos^2 \theta & (\lambda_S - 12 \lambda_S^\prime ) \cos \theta \sin \theta\\
(\lambda_S - 12 \lambda_S^\prime ) \cos \theta \sin \theta & (\lambda_S + 4 \lambda_S^\prime ) \sin^2 \theta
\end{pmatrix} \; ,
\end{align}
leading to the $h_{1,2}$ scalar masses
\begin{align}
m_{h_{1,2}}^2 = \frac{u^2}{2} \left(\lambda_S + 4 \lambda_S^\prime \mp \sqrt{\lambda_S^2 - 8 \lambda_S \lambda_S^\prime + 80 \lambda_S^{\prime 2} + 16 (\lambda_S - 4 \lambda_S^\prime ) \lambda_S^\prime \cos( 4 \theta) } \right)\,.
\label{eq:massesh1h2}
\end{align}

These states will be responsible for new contributions to the CP asymmetry, as seen in Sec.~\ref{sec:asymmetry}. In turn, scalar mixing is encoded by
\begin{align}
\mathbf{V} = \begin{pmatrix}
\cos \theta_S & \sin \theta_S\\
 -\sin \theta_S & \cos \theta_S
\end{pmatrix} \; , \; \tan(2 \theta_S) = - \frac{(\lambda_S - 12 \lambda_S^\prime) \tan (2 \theta)}{\lambda_S + 4 \lambda_S^\prime} \; .
\label{eq:thetaS1}
\end{align}
Inverting Eq.~\eqref{eq:massesh1h2}, one can write $\lambda_S$ and $\lambda_S'$ in terms of $m_{h_{1,2}}$ to find
\begin{align}
\tan(2 \theta_S) &= \mp \frac{\sqrt{1+r_h^4 - 6 r_h^2 - \left(1+r_h^2\right)^2 \cos(4 \theta)}}{\sqrt{2} \left(1 + r_h^2\right)\cos(2 \theta)} \;,\;r_h\equiv \dfrac{m_{h_1}}{m_{h_2}}\,,
\label{eq:thetaS2}
\end{align}
where the minus (plus) sign in $\mp$ is valid when $\sin(2\theta)$ is positive (negative). By requiring $\theta_S$ to be real, we get the following condition for $r_h$:
\begin{align}
r_h<{\rm min}\left\{|\tan\theta|,|\cot\theta|\right\}\,,
\label{eq:mh1mh2ratiolimit}
\end{align}
which, using the best-fit value of $\theta$ indicated in Table~\ref{tab:compatibilityoutput} for $\mathcal{Z}_8^\mu$, leads to $m_{h_1}/m_{h_2}\lesssim 0.37$.
Once more, we see that requiring compatibility with neutrino data imposes constraints on the scalar sector of the model, which is an interesting and uncommon feature.

The rotation to the heavy-neutrino mass-eigenstate basis is obtained by diagonalizing $\M_R$ in Eq.~\eqref{eq:masstructures} with the unitary matrix $\Uh$ defined as
\begin{align}
\Uh = \begin{pmatrix}
e^{-i \theta} \cos \theta_R & e^{-i \theta} \sin \theta_R\\
 -\sin \theta_R & \cos \theta_R
\end{pmatrix} \; , \; \tan(2 \theta_R) = 2 \frac{\sqrt{M_1 M_2}}{M_2-M_1} \; ,
\label{eq:mixMR}
\end{align}
being the masses of the heavy Majorana neutrinos $N_{1,2}$ given by:
\begin{align}
M_{1,2}^2 = \frac{1}{2} \left(m_R^2 + u^2 y_{R_S}^2 \mp m_R \sqrt{m_R^2 + 2 u^2 y_{R_S}^2} \right) \;.
\end{align}

\begin{figure}[t!]
    \centering
   \hspace*{-1.5cm}\includegraphics[scale=0.52]{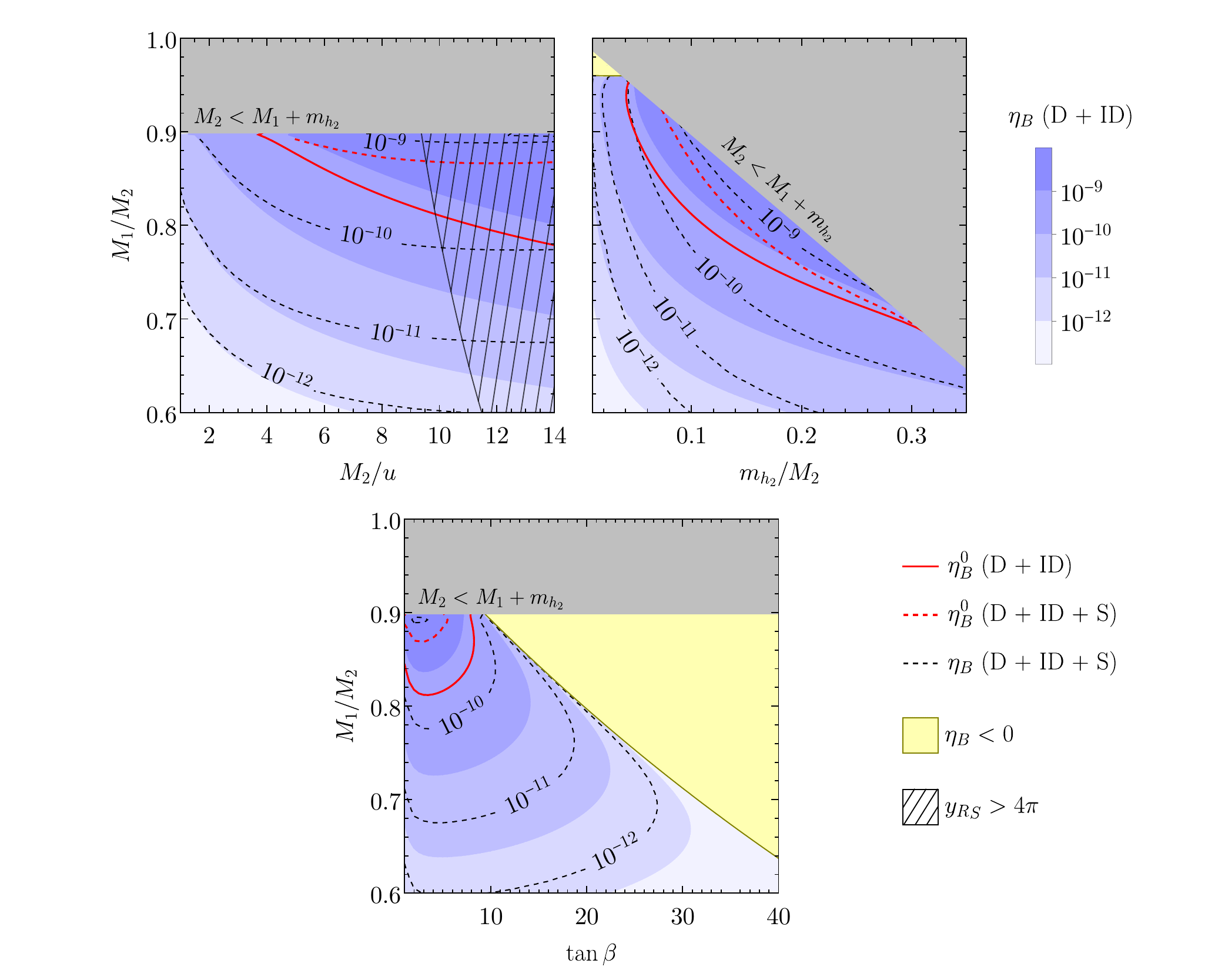}
    \caption{Top left: $\eta_B$ in the plane ($M_2/u,M_1/M_2$) with $m_{h_2}/M_2=0.1$ and $t_\beta=3$. Top right: $\eta_B$ in the plane ($m_{h_2}/M_2,M_1/M_2$) with $y_{R_S}=4\pi$ and $t_\beta=3$. Bottom: $\eta_B$ in the plane ($t_\beta,M_1/M_2$) with $y_{R_S}=4\pi$ and $m_{h_2}/M_2=0.1$. In all plots, $x$, $y$, $z$, $\theta_L$ and $\theta$ are fixed to their best-fit value for case $\mathcal{Z}_8^\mu$ (see Table~\ref{tab:compatibilityoutput}), $m_{h_1}/m_{h_2}=0.37$ and $u=10^{12}$~GeV. In the grey regions one has $M_2<M_1+m_{h_2}$, while in the hatched area $y_{R_S}>4\pi$. In the yellow region $\eta_\text{B}<0$. The black dashed contours and the blue regions correspond to $\eta_B$ values computed with and without $\Delta L=1$ and $\Delta N=2$ scattering terms in the BEs, respectively. The red dashed (solid) line indicates the observed baryon asymmetry [see Eq.~\eqref{eq:etab0}] when the same terms are (not) considered.}
    \label{fig:etaBwDeltaL=1}
\end{figure}

Finally, in the charged-lepton and heavy-neutrino  mass basis, one can use Eqs.~\eqref{eq:YHdef}, \eqref{eq:yukawastructures}, \eqref{eq:pdef}, \eqref{eq:VLmodel}, \eqref{eq:Mnuxyz}, and \eqref{eq:mixMR} to write the two Dirac neutrino Yukawa coupling matrices  $\Y^{1,2}$ as
\begin{gather}
    \Y^1=\dfrac{\sqrt{2zM_1}}{v_1\sqrt{1-r_{12}}}\begin{pmatrix}
        s_Le^{i\theta}\quad\quad&\sqrt[4]{r_{12}}s_Le^{i\theta}\\[0.5cm]
        0\quad\quad&0\\[0.5cm]
        -c_Le^{i\theta}\quad\quad&-\sqrt[4]{r_{12}}c_Le^{i\theta}
    \end{pmatrix}, \label{eq:Y1}\\[0.5cm]
    \Y^2=\dfrac{\sqrt{2xM_1}}{v_2\sqrt{1-r_{12}}}\begin{pmatrix}
        -\dfrac{y}{x}\left(1-\sqrt{r_{12}}\right)c_L\quad\quad&\dfrac{y}{x}\dfrac{1-\sqrt{r_{12}}}{\sqrt[4]{r_{12}}}c_L\\[0.5cm]
        - e^{i\theta}\quad\quad&-\sqrt[4]{r_{12}}e^{i\theta} \\[0.5cm]
      -\dfrac{y}{x}\left(1-\sqrt{r_{12}}\right)s_L\quad\quad&\dfrac{y}{x}\dfrac{1-\sqrt{r_{12}}}{\sqrt[4]{r_{12}}}s_L
    \end{pmatrix} \; , \; r_{12}=M_1^2/M_2^2 \; ,
    \label{eq:Y2}
\end{gather}
for case $\mathcal{Z}_8^\mu$. Furthermore, since $\YR=\mathbb{0}$, and matrices $\YR'$ and $\Uh$ are given, respectively, in Eqs.~\eqref{eq:yukawastructures} and~\eqref{eq:mixMR}, the couplings $ \mathbf{\Delta}^k $ in Eq.~\eqref{eq:Deltadef} are
\begin{align}
   \; \mathbf{\Delta}^1=\dfrac{M_2}{2 u }\dfrac{\sqrt[4]{r_{12}}}{1+\sqrt{r_{12}}}\begin{pmatrix}
    -2\sqrt[4]{r_{12}}&1-\sqrt{r_{12}}\\
    \cdot&2\sqrt[4]{r_{12}}
    \end{pmatrix}e^{i(\theta_S+\theta)},\quad \mathbf{\Delta}^2=-i\mathbf{\Delta}^1 \; .
    \label{eq:Deltasimp}
\end{align}
Having defined all interactions relevant for the CP asymmetries in $N_{1,2}$ decays, it is worth commenting on some of their properties before proceeding to a detailed numerical analysis of the parameter space:
\begin{itemize}

    \item Plugging $\Y^1$ and $\Y^2$ in the expression for $\varepsilon_{i \alpha}^a(\text{type-I})$ of Eq.~\eqref{eq:CPasymtypeI}, one sees that  $\varepsilon_{i \alpha}^a(\text{type-I}) = 0$. In fact, the products $\mathbf{Y}^{a \ast}_{\alpha i} \mathbf{Y}^{b \ast}_{\beta i} \mathbf{Y}^{b}_{\alpha j} \mathbf{Y}^{a}_{\beta j}$, $\mathbf{Y}^{a \ast}_{\alpha i} \mathbf{H}^b_{i j} \mathbf{Y}^{a}_{\alpha j}$ and $\mathbf{Y}^{a \ast}_{\alpha i} \mathbf{H}^b_{j i} \mathbf{Y}^{a}_{\alpha j}$ are real for $i=1,2\neq j$ and $a,b=1,2$. This happens because the matrix entries in each row of $\Y^1$ and $\Y^2$ have the same phase. Thus, we conclude that in our $\mathcal{Z}_8$ model the usual type-I seesaw diagrams of Fig.~\ref{fig:usualcontribution_CPasym} do not contribute to the CP asymmetry. Consequently, the $(B-L)$-asymmetry will be exclusively generated through the new $h_k$-induced interactions, namely the scalar-fermion portal $N N h_k$ and scalar triple coupling $\Phi^\dagger \Phi h_k$.
    
    \item As for $\varepsilon_{i \alpha}^a(\text{wave})$ of Eq.~\eqref{eq:CPasymwave}, using the above expressions for $\mathbf{Y}^{1,2}$ and $\mathbf{\Delta}^{1,2}$, we notice that the products $\mathbf{Y}_{\alpha l}^a \mathbf{\Delta}_{l j}^k \mathbf{\Delta}_{j i}^{k \ast} \mathbf{Y}_{\alpha i}^{a \ast}$ and $\mathbf{Y}_{\alpha l}^a \mathbf{\Delta}_{l j}^{k \ast} \mathbf{\Delta}_{j i}^k \mathbf{Y}_{\alpha i}^{a \ast}$ are real, for any combination of indices $i$, $j$ and $l$. Consequently, the $LL$ and $RR$ contributions to~$\varepsilon_{i \alpha}^a(\text{wave})$ vanish.

    \item Trilinear scalar terms $(\Phi_a^\dagger \Phi_b) S$ as those of Eq.~\eqref{eq:cubicscalar} are forbidden by the $\mathcal{Z}_8$ symmetry -- see Table~\ref{tab:part&sym}. Hence, the effective coupling $\tilde{\mu}_{a b,k}$ defined in the mass-eigenstate basis as shown in Eq.~\eqref{eq:muCPdef}, will be generated via the $\lambda_{1 S,2S} (\Phi_{1,2}^\dagger \Phi_{1,2} ) |S|^2$ quartics, once $S$ aquires a VEV at the leptogenesis scale. Consequently, in our model only $\tilde{\mu}_{a a,k}$ for $a,k =1,2$ will be non-zero, being proportional to $\lambda_{1 S}u$ and $\lambda_{2 S}u$. The $\tilde{\mu}_{ijk}$ couplings of Eq.~\eqref{eq:muhkCPdef} will originate from $\lambda_{S} |S|^4$ and $\lambda_{S}^\prime (S^4 + S^{\ast 4})$, being proportional to $ \lambda_{S} u $ and $\lambda_{S}^\prime u $. Moreover, the $\varepsilon_{i \alpha}^a(\text{vertex})$ and $\varepsilon_{i \alpha}^a(\text{3- body decay})$ CP-asymmetries in Eqs.~\eqref{eq:CPasymvertex} and~\eqref{eq:CPasym3body} are suppressed by $\tilde{\mu}_{a b, k}/M_i \propto u \lambda/M_i$ if $M_i \lsim u$. This has been  checked numerically and holds even for $\lambda \sim \mathcal{O}(1)$. For example, taking $u/M_i \sim 0.1$ (see Fig.~\ref{fig:etaBwDeltaL=1}) and $\lambda = 0.01$, we have $|\varepsilon_{i \alpha}^a(\text{wave})| \sim 10^3 \times |\varepsilon_{i \alpha}^a(\text{vertex}) + \varepsilon_{i \alpha}^a(\text{3-body decay})|$.
    
\end{itemize}
In conclusion, the only relevant contribution to the CP asymmetry in the $\mathcal{Z}_8$ model comes from the new wave diagrams~\subref{fig:newwave} of Fig.~\ref{fig:newcontribution_CPasym}, which are possible due to the presence of $S$. Thus, in this simple model, $S$ is responsible for SCPV and for the CP asymmetries in the heavy neutrino decays. Hence, in the present framework, leptogenesis is assisted by $S$ with CP violation coming from its VEV. 

By replacing Eqs.~\eqref{eq:Y1},~\eqref{eq:Y2} and~\eqref{eq:Deltasimp} in the general expression for $\varepsilon_{i\alpha}^a(\text{wave})$ in Eq.~\eqref{eq:CPasymwave} with $n_H=2$ and $n_S=1$ we have
\begin{align}
   \varepsilon_\text{CP}\equiv\varepsilon_2&= - \dfrac{1}{8\pi}\dfrac{M_2^2}{ u^2}\dfrac{r_{12}\left[1-\sqrt{r_{12}}\right]\left[(1-\sqrt{r_{12}})^2y^2-x^2\sqrt{r_{12}}-xzt_\beta^2\sqrt{r_{12}}\right]}{\left[1+\sqrt{r_{12}}\right]^2\left[r_{12}x^2+(\sqrt{r_{12}}-1)^2y^2+xzt_\beta^2r_{12}\right]}\sin[2(\theta_S+\theta)]\nonumber\\[0.2cm]
    &\times\left[\mathcal{F}_{\text{w}, L R}^{2 1 1 1}-\mathcal{F}_{\text{w}, L R}^{2 1 2 1} -\mathcal{F}_{\text{w}, R L}^{2 1 1 1}+\mathcal{F}_{\text{w}, R L}^{2 1 2 1}\right] \; ,
    \label{eq:CPasymmetrymodel}
\end{align}
after summing over flavor $\alpha=e,\mu,\tau$ and number of Higgs doublets~$a=1,2$ [see Eq.~\eqref{eq:cptotunflavored}]. Here, the doublet VEVs are related as follows,
\begin{equation}
t_\beta \equiv \tan\beta = \frac{v_2}{v_1} \; ,\;
v^2 = v_1^2 + v_2^2=246\,{\rm GeV}\,.
\label{eq:tanbeta}
\end{equation}
Let us highlight some features of the obtained CP asymmetry:
\begin{itemize}

    \item The CP asymmetry generated by the decay of $N_1$ is identically zero, i.e. $\varepsilon_1=0$, due to the kinematic constraint $M_2>M_1+m_{h_k}$ imposed by the decay $N_2\to N_1+h_k$.
    
    \item A direct connection between high and low-energy CP violation is established through the explicit dependence of $\varepsilon_\text{CP}$ on the CP violating phase $\theta$ [recall that $\theta_S$ depends as well on $\theta$ as shown in Eqs.~\eqref{eq:thetaS1} and~\eqref{eq:thetaS2}]. As long as $r_{12}\neq 0,1$ and $m_{h_1}/m_{h_2}\neq0,1$, a non-trivial low-energy CP violating phase is required to have non-zero CP asymmetry. This is allowed by neutrino oscillation data (see Table~\ref{tab:compatibility}).
    
    \item A negative CP asymmetry, required such as $\eta_B>0$ [see Eqs.~\eqref{eq:etaB} and \eqref{eq:BEsBL}], is obtained for
\begin{align}
\begin{cases}
    0<r_{12}<{r_{12}}_\text{lim}\\
    \sin[2(\theta_S+\theta)]<0
\end{cases}\quad
\text{ or }\quad\quad
\begin{cases}
    {r_{12}}_\text{lim}<r_{12}<1\\
    \sin[2(\theta_S+\theta)]>0
\end{cases},
\label{eq:negcpasym}
\end{align}
where the limiting $\sqrt{r_{12}}$ value reads
\begin{align}
    \sqrt{{r_{12}}_\text{lim}}=\dfrac{x^2+2y^2+t_\beta^2 x z-\sqrt{(x^2+t_\beta^2 xz)(x^2+4y^2+t_\beta^2 x z)}}{2y^2} \; .
    \label{eq:r12lim}
\end{align}
In the scenario where $r_{12}=0,1$ one gets $\varepsilon_\text{CP}=0$. 

\item The $\varepsilon_\text{CP}$ dependence on the scalar masses $m_{h_1}$ and $m_{h_2}$ can be analyzed from the $LR$ and $RL$ wave loop functions in Eq.~\eqref{eq:loopwave} and the $\theta_S$ expression in Eq.~\eqref{eq:thetaS2}. When $m_{h_1}/m_{h_2}\rightarrow 0$, $\theta_S \rightarrow - \theta$, and $\varepsilon_\text{CP} \rightarrow 0$. Also, if $m_{h_1}/m_{h_2}\rightarrow1$ then $\mathcal{F}_{\text{w}, L R}^{2 1 1 1}=\mathcal{F}_{\text{w}, L R}^{2 1 2 1}$ and  $\mathcal{F}_{\text{w}, R L}^{2 1 1 1}=\mathcal{F}_{\text{w}, R L}^{2 1 2 1}$ and $\varepsilon_\text{CP}$ vanishes as well. For $0<m_{h_1}/m_{h_2}<|\tan\theta|$ [see Eq.~\eqref{eq:mh1mh2ratiolimit}], the CP asymmetry grows with $m_{h_1}/m_{h_2}$ and, thus, one should consider $m_{h_1}/m_{h_2}=|\tan\theta|$ in order to maximize $\varepsilon_\text{CP}$. Furthermore, $|\varepsilon_\text{CP}|$ is enhanced for high $\sigma_{k2}$, $k=1,2$. This can be seen by looking at the $\rho_{ijk}$ dependence on $\sigma_{ki}$ -- see Eq.~\eqref{eq:loopwave}. Moreover, $\varepsilon_\text{CP}$ scales with $M_2^2/u^2$.

\end{itemize}

To obtain BAU predictions in the present model, we numerically solve the set of BEs given in~\eqref{eq:BEsNi} and \eqref{eq:BEsBL} for $n_R=2$ and $M_2>M_1$. Namely, 
\begin{align}
    \frac{d N_{N_1}}{d z} & = -(D_1 + \frac{N_{N_2}^{\text{eq}}}{N_{N_1}^{\text{eq}}} D_{2 1} + S_1) (N_{N_1} - N_{N_1}^{\text{eq}}) + D_{2 1} (N_{N_2} - N_{N_2}^{\text{eq}}) \nonumber \\
    & - S_{1 1} [(N_{N_1})^2 - (N_{N_1}^{\text{eq}})^2] - S_{1 2} (N_{N_1} N_{N_2} - N_{N_1}^{\text{eq}} N_{N_2}^{\text{eq}}) \; , \label{eq:BEsN1model} \\
    \frac{d N_{N_2}}{d z} & = - (D_2 + D_{2 1} + S_2)(N_{N_2} - N_{N_2}^{\text{eq}}) + \frac{N_{N_2}^{\text{eq}}}{N_{N_1}^{\text{eq}}} D_{2 1} (N_{N_1} - N_{N_1}^{\text{eq}}) \nonumber \\
    & - S_{2 2} [(N_{N_2})^2 - (N_{N_2}^{\text{eq}})^2] - S_{1 2} (N_{N_1} N_{N_2} - N_{N_1}^{\text{eq}} N_{N_2}^{\text{eq}}) \; , \label{eq:BEsN2model} \\
    \frac{d N_{B-L}}{d z} & = - \varepsilon_2 D_2 (N_{N_2} - N_{N_2}^{\text{eq}}) - W N_{B-L} \; ,
    \label{eq:BEsBLmodel}
\end{align}
which agree with those obtained in Refs.~\cite{LeDall:2014too,Alanne:2017sip,Alanne:2018brf}. We will take the values of  $x$, $y$, $z$, $\theta_L$ and $\theta$ fixed to their best-fit values (see Table~\ref{tab:compatibilityoutput}) for case $\mathcal{Z}_8^\mu$. Since in our model $\varepsilon_1=0$, $N_{B-L}$ will be exclusively generated by the decay of~$N_2$ into leptons and scalar doublets, when interactions involving $N_2$ go out of equilibrium. The lepton asymmetry is then washed out by the $L$-violating inverse decays and scatterings involving $N_1$ and $N_2$. The resulting baryon-to-photon ratio $\eta_B$ depends on the strength of these washout processes controlled by the interaction couplings and mass ratios of the participating states. After solving the BEs, we take $N_{B-L}(z\rightarrow\infty)$ and use Eq.~\eqref{eq:etaB} to compute~$\eta_B$.

In Fig.~\ref{fig:etaBwDeltaL=1}, we show our baryon-to-photon ratio $\eta_B$ results in the planes $(M_2/u,M_1/M_2\equiv\sqrt{r_{12}})$, $(m_{h_2}/M_2\equiv\sqrt{\sigma_{22}}$, $M_1/M_2=\sqrt{r_{12}})$ and $(t_\beta,M_1/M_2=\sqrt{r_{12}})$ on the upper-left, upper-right and bottom figures, respectively. In the upper-left (upper-right) [bottom] plot we considered $m_{h_2}/M_2=0.1$ and $t_\beta=3$ ($y_{R_S}=4\pi$ and $t_\beta=3$) [$m_{h_2}/M_2=0.1$ and $y_{R_S}=4\pi$]. In all cases we set $v=246$~GeV [see Eq.~\eqref{eq:tanbeta} for relation with $t_\beta$], $m_{h_1}/m_{h_2}=0.37$ and $u=10^{12}$~GeV (unflavored scenario). The parameter space regions where the kinematic constraint $M_2>M_1+m_{h_2}$ is not met are depicted in gray. Also, the yellow regions in the upper-right and bottom plots indicate parameter space where $\eta_B<0$ (corresponding to $\sqrt{r_{12}}\gsim 0.94$ for $t_\beta=3$ in this benchmark). Additionally, we imposed the perturbativity constraints $|\lambda_i| \leq 4 \pi$ and $|y_{D_i}|,|y_{R_S}| \leq 4 \pi$, and required the scalar potential to be bounded from below [see Eq.~\eqref{eq:bfb}]. For the chosen benchmark, among these conditions only $y_{R_S}$ perturbativity fails, which is indicated by the black hatched area in the upper-left plot. For the remaining figures $y_{R_S}=4\pi$ is imposed to maximise the ratio $M_2/u$, which in turn maximises $\eta_\text{B}$. The blue contour regions show the $\eta_B$ results obtained when considering only the contribution of decays and inverse decays to the final asymmetry. The black dashed contours correspond to $\eta_B$ values computed considering $\Delta L=1$ and $\Delta N=2$ scattering terms in the BEs. The red dashed (solid) line indicates the contour for $\eta_B=\eta_B^0$ -- see Eq.~\eqref{eq:etab0} -- when scattering effects are (not) included in the~BEs.

From the results shown in that figure we conclude the following:
\begin{itemize}
\item From the upper-left plot one can see that, for the considered benchmark with $t_\beta=3$ and $m_{h_2}/M_2=0.1$, the experimental value of the BAU $\eta_B^0$ is obtained for $M_1\gtrsim 0.81\, M_2$ ($M_1\gtrsim 0.86\, M_2$) and $M_2\gtrsim 3.5\, u$ ($M_2\gtrsim 5.2\, u$), when decays and inverse decays (decays, inverse decays and scatterings) are included in the BEs. In the upper-right plot, for which $y_{R_S}=4\pi$, $\eta_B=\eta_B^0$ for $M_1\gtrsim 0.69\, M_2$ ($M_1\gtrsim 0.71 \,M_2$) and $m_{h_2}\gtrsim 4\times 10^{-2} M_2$ ($m_{h_2}\gtrsim 7\times 10^{-2} M_2$). Choosing $m_{h_2}/M_2=0.1$ and keeping $y_{R_S}=4\pi$ (bottom plot), the observed value for the baryon-to-photon ratio lies in the region $M_1\gtrsim 0.81\, M_2$ ($M_1\gtrsim 0.86\, M_2$) and $t_\beta\lesssim 8.4$ ($t_\beta\lesssim 5.6$). The fact that $\eta_\text{B}^0$ can only be recovered in a small portion of the parameter space, even though $\varepsilon_\text{CP} \sim [10^{-5},10^{-2}]$, is due to the very strong washout regime (with $K_i\gsim 2\times 10^2$). Thus, only for high $M_1/M_2$, $M_2/u$, and $m_{h_2}/M_2$, where $\varepsilon_\text{CP}$ is maximised, ($B-L$)-asymmetry is sufficiently high. As seen in the upper-right plot of Fig.~\ref{fig:etaBwDeltaL=1}, the lower limit on~$M_1/M_2$ decreases when the ratio $m_{h_2}/M_2$ increases. Moreover, by looking at the yellow region in the bottom plot we notice that as $t_\beta$ increases the available parameter space where the baryon-to-photon ratio is positive shrinks. In fact, since the SCPV phase is predicted to be $\theta\simeq 1.89\pi$ we have $\sin[2(\theta_S+\theta)]<0$ ($\sim-0.54$ for $m_{h_1}/m_{h_2}=|\tan\theta|\simeq0.37$), which allows for $\eta_B>0$ when $0<r_{12}<{r_{12}}_\text{lim}$ [see Eq.~\eqref{eq:negcpasym}]. For low values of $t_\beta$, ${r_{12}}_\text{lim}\simeq 1$ and, consequently, the available parameter space for successful leptogenesis is larger.

\item Some of the scattering processes presented in Sec.~\ref{sec:BEs} (for which the reduced cross section are given in Appendix~\ref{chpt:scatterings}) have negligible effect on the evolution of $N_i$ number density and ($B-L$)-asymmetry. In fact, we numerically checked that in our model the $\Delta N=2$ neutrino annihilation scatterings (see Fig.~\ref{fig:DeltaNeq2_scattering}) are subdominant compared to the $\Delta L=1$ scatterings (see Fig.~\ref{fig:DeltaLeq1_scattering}). As mentioned before, in our scenario the effective triple scalar coupling $\tilde{\mu}_{a b, k} \propto u \lambda$ is naturally suppressed compared to the heavy neutrino masses~$M_{1,2}$. Hence, among the different contributions to $\Delta L=1$ scatterings, the ones involving this triple scalar interaction mediated by $\Phi_b$ will be subdominant compared to the corresponding $N$-mediated one (this has been verified numerically). Thus, among $\Delta L=1$ scatterings, the dominant ones are: the usual type-I seesaw diagrams of Fig.~\ref{fig:DeltaLeq1_usual}, the heavy neutrino mediated $t$-channel $N_i h_k \leftrightarrow \Phi_a \ell_\alpha$ and $N_i \Phi_a \leftrightarrow \ell_\alpha  h_k$ processes shown in Figs.~\ref{fig:DeltaLeq1_Nltophih} and \ref{fig:DeltaLeq1_Nphitohl}, respectively, and the $s$-channel contribution to $N_i  \ell_\alpha \leftrightarrow \Phi_a h_k$ in Fig.~\ref{fig:DeltaLeq1_Nhtophil}.

\item From the upper-left plot in Fig.~\ref{fig:etaBwDeltaL=1}, we distinguish two different scenarios: for $M_2/u\lsim 3$ scatterings are negligible while for $M_2/u\gsim3$ they significantly lower the value of $\eta_B$. In the former case, the dominant scattering effects stem from the regular type-I seesaw contributions of Fig.~\ref{fig:DeltaLeq1_usual}, which scale as $\mathcal{O}(Y^2Y_t^2)$. For low values of $M_2/u$, we verify that $\mathcal{O}(\Delta)\ll\mathcal{O}(Y_t)$ and, thus, the new scattering contributions of order $\mathcal{O}(Y^2\Delta^2)$ are subdominant. However, due to compatibility with neutrino data, we are in a strong washout regime with $K_i\gsim 2\times 10^2$, and the impact of the usual top-quark scatterings is negligible (as already remarked in Refs.~\cite{Buchmuller:2004nz,Hahn-Woernle:2009jyb}). Instead, for $M_2/u\gsim3$, the scatterings with the new scalars $h_k$ are significantly enhanced [note the $\mathbf{\Delta}_{1,2}$ dependence on this ratio in Eq.~\eqref{eq:Deltasimp}], becoming out of equilibrium much later in the early Universe. This leads to a stronger washout of the $(B-L)$-asymmetry and, consequently, to a lower $\eta_\text{B}$. In this region of the parameter space, we checked that the dominant scattering contribution corresponds to the $N$-mediated $s$-channel process $N h \leftrightarrow  \ell \Phi$. The dominance of this process over the usual $\Delta L=1$ type-I seesaw contributions is explained by the fact that $\mathcal{O}(Y_t)\ll \mathcal{O}(\Delta)$. Moreover, the new $t$-channel contributions $N_i \ell_\alpha \leftrightarrow \Phi_a h_k$ and $N_i \Phi_a \leftrightarrow \ell_\alpha h_k$ mediated by $N$ are naturally subdominant when compared to the (also $N$-mediated) $s$-channel process $N_i h_k \leftrightarrow \ell_\alpha \Phi_a$, due to the logarithmic dependence on the mediator mass.

\item From Eq.~\eqref{eq:CPasymmetrymodel}, it is clear that $\varepsilon_\text{CP}$ strongly depends on the SCPV phase $\theta$ through the factor $\sin[2(\theta+\theta_S)]$. By varying $\theta$ in the allowed 3$\sigma$ range for neutrino oscillation data we get $\theta\sim [1.89,1.96]\,\pi$, being the best-fit point $\theta=1.89 \pi$ (see Table~\ref{tab:compatibilityoutput}). Increasing $\theta$ up to its largest allowed value, leads to a lower $|\varepsilon_\text{CP}|$, and consequently, to a lower BAU. Hence, using the best-fit value for the vacuum CP phase in our analysis maximizes $\eta_\text{B}$.

\end{itemize}

To finalize our discussion, a few comments are in order regarding cLFV. At low-energies, i.e. at the EW scale, the two Higgs doublets $\Phi_{1,2}$ will develop non-zero VEVs $v_{1,2}$, as indicated in Eq.~\eqref{eq:vevs}. The angle $\beta$ [see Eq.~\eqref{eq:tanbeta}] diagonalizes the charged scalar mass matrix leading to the $W$-type Goldstone boson and a charged Higgs $H^{\pm}$. Furthermore, $\beta$ defines the Higgs basis where $\Phi_{1}$ matches the SM Higgs doublet~\cite{Branco:2011iw}. Note that, since $u \gg v_{1,2}$, the singlet~$S$ is decoupled from the doublets $\Phi_{1,2}$. So, we work in the alignment limit where we set $\beta-\alpha = \pi/2$, being $\alpha$ is the angle which which rotates the neutral doublet degrees of freedom to their mass basis [see Eq.~\eqref{eq:scalardef}]. The new scalars will mediate the cLFV decays $\ell_{\alpha}^{-} \rightarrow \ell_{\beta}^{-} \ell_{\gamma}^{+} \ell_{\delta}^{-}$ and $\ell_{\alpha} \rightarrow \ell_{\beta} \gamma$ at tree and one-loop level, respectively. For the $\mathcal{Z}_8^\mu$ case discussed above, all contributions to the $\mu \rightarrow 3 e$, $\mu \rightarrow e \gamma$ and $\mu-e$ conversion in nuclei, vanish. This is due to the presence of zeros imposed by the flavor symmetry on the charged lepton Yukawa matrices~$\Ye^{1,2}$ of Eq.~\eqref{eq:yukawastructures} and, consequently, on the mass matrix $\Me$ shown in Eq.~\eqref{eq:masstructures} -- see Refs.~\cite{Correia:2019vbn,Camara:2020efq,Rocha:2024twm,Rocha:2025ade} and Chapter~\ref{chpt:flavoraxion}. For this $\mu$-decoupled case, the neutral scalars only contribute to~$\tau \rightarrow 3 e$ and $\tau \rightarrow e \gamma$, whose current bounds are orders of magnitude above the stringent muon cLFV ones. Naturally, these contributions are suppressed for large scalar masses or if they are quasi-degenerate. Lastly, the $W^{\pm}$ and  $H^{\pm}$ will also contribute radiatively to the aforementioned cLFV processes, due to their interactions with Majorana neutrinos. For low-scale seesaw scenarios, as discussed in Sec.~\ref{sec:lowseesaw}, where heavy neutrinos can have masses of the order~$\mathcal{O}(1\,{\rm TeV})$, these contributions are testable at current and future indirect cLFV experiments, as studied e.g. in Ref.~\cite{Camara:2020efq} for the minimal ISS. However, for the canonical type-I scenario we analyzed here, the heavy neutrino masses are around $ 10^{12}$ GeV and, consequently, the contributions to cLFV from charged bosons are naturally suppressed.

%%%%%%%%%%%%%%%%%%%%%%%%%%%%%%%%%%%%%%%%%%%%%%%%%%%%%%%%%%%%%%%%%%%%%%%%%%%%%
\subsection{Key ideas and outlook}
\label{sec:conclLeptoSCPV}
%%%%%%%%%%%%%%%%%%%%%%%%%%%%%%%%%%%%%%%%%%%%%%%%%%%%%%%%%%%%%%%%%%%%%%%%%%%%%

In this section we explored thermal leptogenesis within the canonical type-I seesaw model extended with complex scalar singlets. Provided that CP invariance is imposed at the Lagrangian level, the complex VEVs of scalar singlets will be the unique source of both Dirac and Majorana CP violation, at the EW scale, and high-energy CP violation at the leptogenesis scale. These scalars unlock novel radiative corrections to the CP asymmetry generated when the heavy neutrinos decay into leptons, and provide new tree-level CP-violating three-body decay processes. We generalized the CP asymmetry calculation for an arbitrary number of RH neutrinos, complex scalar singlets and Higgs doublets, as shown in Eq.~\eqref{eq:fullCP}. Furthermore, we studied the unflavored BEs taking into account decays, $\Delta L=1$ and $\Delta N =2$ scatterings for an arbitrary number of RH neutrinos. The new complex scalar singlets participate in additional tree-level decays and scattering processes besides the ones considered in vanilla type-I seesaw leptogenesis. In order to compute the final values of the baryon-to-photon ratio, one has to solve the BEs presented in  Eqs.~\eqref{eq:BEsNi} and~\eqref{eq:BEsBL}. The reduced cross-sections for all included processes in the BEs are collected in Appendix~\ref{chpt:scatterings}.

To illustrate how SCPV can simultaneously lead to non-trivial low- and high-energy CP-violation, we studied a simple model where the SM is extended with one complex singlet, two RH neutrinos and a new scalar doublet. The parameters in the Lagrangian are further constrained by a $\mathcal{Z}_8$ flavor symmetry. This corresponds to the minimal particle content charged under the simplest discrete symmetry that allows for the possibility of SCPV and compatibility with neutrino oscillation data. The $\mathcal{Z}_8$ leads to constraints in the effective neutrino mass matrix which we tested against data. We concluded that out of the three possible charge assignments, $\mathcal{Z}_8^e$, $\mathcal{Z}_8^\mu$ and $\mathcal{Z}_8^\tau$, and considering all possible decoupled charged-lepton states ($e$, $\mu$ or $\tau$), the best case is the $\mathcal{Z}_8^{\mu}$ with muon decoupled and IO neutrino masses (see Tables~\ref{tab:compatibility} and~\ref{tab:compatibilityoutput}), which requires a vacuum singlet phase $\theta\sim 1.89\pi$.

Due to the constrained structure of the Yukawa couplings $\mathbf{Y}^{1,2}$ and the couplings of heavy neutrinos to the new scalar singlets, $\mathbf{\Delta}^{1,2}$, the CP asymmetries in the $N_2$ decays stem from the interference between the tree-level and the one-loop self-energy diagrams mediated by the $S$ singlet (the usual type-I seesaw diagrams are forbidden and the new 3-body decay and vertex contributions negligible). The expression for the total CP asymmetry in this model is presented in Eq.~\eqref{eq:CPasymmetrymodel}, where the link between low and high-energy CP violation is explicit. To compute the value of the BAU, we solved numerically the unflavored BEs for low-energy parameters that best fit neutrino data, including scattering processes. We concluded that for case $\mathcal{Z}_8^\mu$ successful leptogenesis is achieved when $M_2 \gsim 8\, u$, for $u=10^{12}$ GeV, as shown in the upper-left plot of Fig.~\ref{fig:etaBwDeltaL=1}. The exact $M_2/u$ lower bound increases slightly when scatterings are included, thanks to the new washout sources. Furthermore, we showed that scatterings are only relevant for $M_2\gsim 3\,u$, lowering significantly $\eta_\text{B}$. The values of $M_1/M_2$ above which the observed BAU is recovered strongly depend on $m_{h_2}/M_2$ and $t_\beta=v_2/v_1$, as shown in the upper-right and bottom plots of Fig.~\ref{fig:etaBwDeltaL=1}. Thus, we conclude that case $\mathcal{Z}_8^\mu$ recovers the observed BAU for part of the parameter space mainly depending on the ratios $M_1/M_2$, $m_{h_2}/M_2$ and $M_2/u$.

The general setup of scalar-singlet assisted leptogenesis with SCPV discussed here can be applied in a straightforward way to the type-III seesaw framework, as well as to the canonical scotogenic model. The Majorana mass term provides the link between SCPV induced by the scalar singlet VEVs and LCPV. Regarding leptogenesis, the expressions for the CP-asymmetry and BEs are essentially the same as the ones obtained here. The generalization for the type-II seesaw case can also be done, but the CP-asymmetries, BEs and portal linking SCPV and the neutrino sector are distinct.

%%%%%%%%%%%%%%%%%%%%%%%%%%%%%%%%%%%%%%%%%%%%%%%%%%%%%%%%%%%%%%%%%%%%%%%%%%%%%
\section{Dark-sector seeded solution to the strong CP problem}
\label{sec:darkNB}
%%%%%%%%%%%%%%%%%%%%%%%%%%%%%%%%%%%%%%%%%%%%%%%%%%%%%%%%%%%%%%%%%%%%%%%%%%%%%

The simplest extension to the SM, satisfying the Nelson and Barr criteria~\cite{Nelson:1983zb,Barr:1984qx,Barr:1984fh}, is the BBP model~\cite{Bento:1991ez}, where are added to the SM a single down-type vector-like-quark~(VLQ) $B_{R,L}$ (for a recent review on iso-singlet VLQs see Ref.~\cite{Alves:2023ufm}) and complex scalar singlet $\sigma$, both odd under a discrete $\mathcal{Z}_2$ symmetry. The idea is to impose CP at the Lagrangian level, making all parameters real and forbidding the $\theta$-term -- see Sec.~\ref{sec:strongCPnedmsol}. The weak sector CP violation is generated spontaneously. In fact, the scalar potential is given by Eq.~\eqref{eq:VSCPVsinglet}, where the terms $\sigma^2$, $\sigma^2 (\Phi^\dagger \Phi)$, $|\sigma|^2 \sigma^2$ and $\sigma^4$ allow for SCPV through the singlet VEV $\left< \sigma \right> = v_\sigma e^{i \varphi}/\sqrt{2}$ as shown in Eq.~\eqref{eq:VSCPVsingletsol3}. The Yukawa couplings involving the new fields read
\begin{align}
- \mathcal{L}_{\text{Yuk}} &\supset \overline{B}_L \left( \mathbf{Y}_{Bd} \; \sigma + \mathbf{Y}_{Bd}^\prime \; \sigma^\ast \right) d_{R} + m_B \; \overline{B_L} B_{R} + \text{H.c.} \; ,
\end{align}
where $\Y_{Bd}^{(\prime)}$ are $1 \times 3$ vectors and $m_B$ is a bare mass term. After the Higgs and singlet acquire VEVs, the down-type quark mass Lagrangian is given by,
    \begin{equation}
    \mathcal{L}_{\text{mass}} \supset \left(\overline{d}_L , \overline{B}_L \right) \bm{\mathcal{M}}_{d} \begin{pmatrix} d_R \\ B_R \end{pmatrix} \; , \; \bm{\mathcal{M}}_{d} = \begin{pmatrix} \mathbf{M}_{d}  & 0 \\ \mathbf{M}_{B d} & m_B \end{pmatrix} \; ,
    \end{equation}
    where $\mathbf{M}_{B d} = v_\sigma \left(\mathbf{Y}_{B d} e^{i \varphi} + \mathbf{Y}_{B d}^\prime e^{- i \varphi}\right) / \sqrt{2}$. Since $\bm{\mathcal{M}}_d$ contains an off-diagonal zero entry $\theta_F = 0$ [see Eq.~\eqref{eq:thetaF}]. Consequently, $\overline{\theta}$ vanishes at tree-level. 

    Block-diagonalizing the $\bm{\mathcal{M}}_{d}$ matrix leads to a heavy quark mass and light down-quark mass matrix. The latter being:
    \begin{align}
    \mathbf{M}_{\text{light}}^2 \simeq \mathbf{M}_d \mathbf{M}_d^T - \frac{\mathbf{M}_d \mathbf{M}_{Bd}^\dagger \mathbf{M}_{Bd} \mathbf{M}_d^T}{m_B^2 + |\mathbf{M}_{Bd}|^2} \; , \label{eq:MlightBBP} 
    \end{align}
    with the only complex parameter $e^{i \varphi}$ being in $\mathbf{M}_{Bd}$. The expression above, shows that SCPV can be successfully communicated to the quark sector, generating the weak CP-violating phase of the CKM matrix. However, one needs to guarantee $|\mathbf{M}_{Bd}|/m_B \gsim 1$ such that $\delta^q$ is not suppressed by~$\sim |\mathbf{M}_{Bd}|^2/m_B^2$. Also, to satisfy the collider bound on the down-type VLQ mass of Eq.~\eqref{eq:collider} (see analysis in Sec.~\ref{sec:CKMpheno}) we need $|\mathbf{M}_{Bd}| \gsim m_B \sim \mathcal{O}(1)$ TeV, implying for $\mathbf{Y}_{B d}^{(\prime)} \sim \mathcal{O}(1)$ that $v_\sigma \gsim \mathcal{O}(1)$~TeV.

The SCPV solution to the strong CP problem, leads to a calculable~$\overline{\theta}$. In fact, the down-type quark mass matrix will receive corrections beyond tree-level that can induce $\overline{\theta}$ which, ultimately, needs to satisfy the nEDM constraint $\left|\overline{\theta}\right| \lsim 10^{-10}$ of Eq.~\eqref{eq:thetabound}. We define the corrections to the full down-type quark mass matrix as,
\begin{equation} 
  \Delta \bm{\mathcal{M}}_d =
  \begin{pmatrix} \Delta\mathbf{M}_d  & \Delta \mathbf{M}_{d B} \\ \Delta \mathbf{M}_{Bd} & \Delta m_B \end{pmatrix} \; .
\label{eq:Lmassloopcorrection}
\end{equation}
Considering the limit $\Delta \bm{\mathcal{M}}_d \ll \bm{\mathcal{M}}_d$, the threshold corrections to $\overline{\theta}$ are written as,
\begin{equation}
\Delta \overline{\theta} = \text{Im} \left\{ \text{Tr} \left[\bm{\mathcal{M}}_d^{-1} \Delta \bm{\mathcal{M}}_d\right] \right\} \; ,
\end{equation}
from which we can derive,
\begin{align}
&\Delta \overline{\theta}|_{\Delta \mathbf{M}_{d}} =  \text{Im} \left\{ \text{Tr} \left[\mathbf{M}_{d}^{-1} \Delta \mathbf{M}_{d}\right] \right\} \; , \; \Delta \overline{\theta}|_{\Delta \mathbf{M}_{d B}} = - \text{Im} \left\{ \text{Tr} \left[ m_B^{-1} \mathbf{M}_{Bd} \mathbf{M}_d^{-1} \Delta \mathbf{M}_{d B} \right] \right\}\; , \nonumber \\
&\Delta \overline{\theta}|_{\Delta m_B} = \text{Im} \left[ m_B^{-1} \Delta m_B \right] \; .
\end{align}
    \begin{figure}[t!]
    \centering
    \includegraphics[scale=0.75]{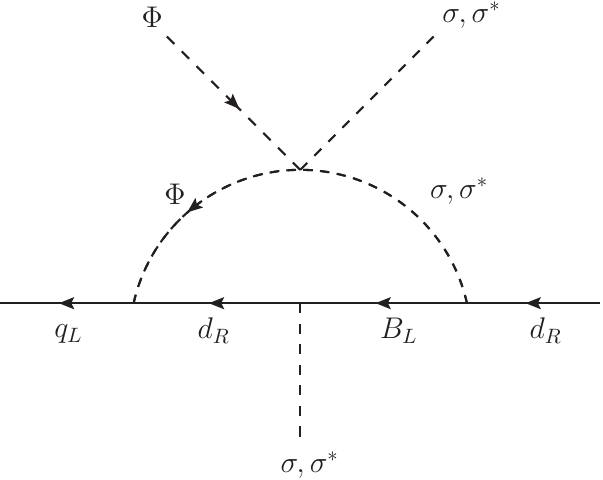} \\
    \vspace{+0.3cm}
    \includegraphics[scale=0.755]{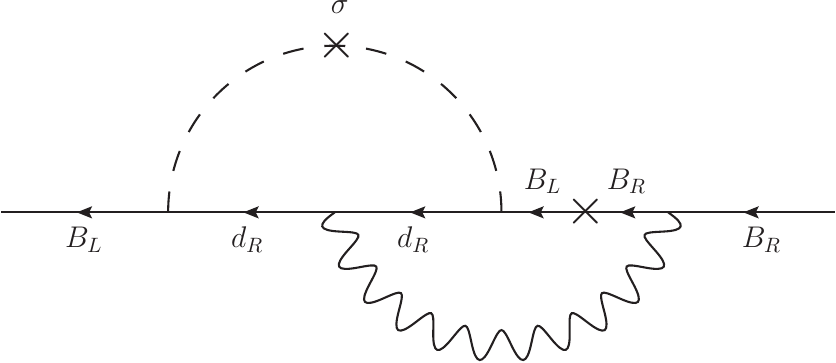}
    \caption{Examples of threshold corrections to $\overline{\theta}$, stemming from one and two loop corrections to $\mathbf{M}_d$ and $m_B$, respectively shown by the top and bottom diagram (see text for details).}
    \label{fig:thresholdBBP}
    \end{figure}
Although numerous diagrams contribute to the various one-loop self-energies that lead to mass corrections to the down-quark mass matrix, most yield manifestly real contributions and therefore do not affect the QCD $\overline{\theta}$ parameter~\cite{Bento:1991ez,Dine:2015jga,Asadi:2022vys}. The exceptions are the classes of diagrams shown in Fig.~\ref{fig:thresholdBBP}. Two-loop diagrams mediated by $\gamma/Z$ EW gauge bosons, as the one shown on the bottom of the figure, are know as "dead duck" diagrams~\cite{Nelson:1983zb,Dine:2015jga}. The one-loop diagram shown in the figure leads to complex corrections to $\mathbf{M}_d$ inducing contributions to $\bar{\theta}$, estimated as~\cite{Dine:2015jga}
    \begin{align}
     \Delta \overline{\theta}|_{\Delta \mathbf{M}_{d}}&\sim \frac{1}{16 \pi^2} \; \lambda_{\Phi \sigma}\, |\mathbf{Y}_{Bd}^{(\prime)}|^2 \,\frac{v_\sigma^2}{\tilde{m}_\sigma^2} \,,
     \end{align}
    where $\tilde{m}_\sigma^2 \sim \tilde{\lambda}_\sigma v_\sigma^2$ denotes a heavy scalar mass with $\tilde{\lambda}_\sigma$ being a scalar quartic coupling. In order for the above to respect the nEDM constraint $\left|\overline{\theta}\right| \lsim 10^{-10}$, we need to tune the Yukawa couplings $\mathbf{Y}_{Bd}^{(\prime)}$ and/or $\lambda_{\phi \sigma}$ to be small. Furthermore, as noted in Ref.~\cite{Bento:1991ez}, introducing $\sigma$ to break CP spontaneously will shift the Higgs mass term at tree-level once $\sigma$ acquires a non-zero VEV due to the $\lambda_{\Phi \sigma}$ quartic coupling [see Eq.~\eqref{eq:VSCPVsingletsol3}]. Hence, the quartic coupling  $\lambda_{\Phi \sigma}$ needs to be such that $\lambda_{\Phi \sigma}\lsim v^2/v_\sigma^2$ for $v$ to remain at the EW scale. The larger the scale at which CP breaks, the smaller $\lambda_{\Phi \sigma}$ needs to be. As in Ref.~\cite{Bento:1991ez}, we assume that whatever physics solves the Higgs hierarchy problem also provides a naturally small $\Phi-\sigma$ coupling. A natural suppression for $\lambda_{\Phi \sigma}$ can be obtained in, e.g. supersymmetric NB models~\cite{Dine:1993qm,Barr:1996wx,Dine:2015jga}. Additionally, for the two-loop diagram shown in the figure the complex contribution to $\Delta m_B$, leads to:
    \begin{align}
     \Delta \overline{\theta}|_{\Delta m_B}&\sim \frac{g^2}{(16 \pi^2)^2} \; \tilde{\lambda}_\sigma \, |\mathbf{Y}_{Bd}^{(\prime)}|^2 \,\frac{v_\sigma^2}{\tilde{m}_\sigma^2} \,.
     \end{align}
     Once again, in order for the above to respect the nEDM constraint, we need to tune the Yukawa and/or scalar quartic couplings to be small.

Beyond radiative corrections, NB models are also subject to a potential “quality problem”~\cite{Choi:1992xp,Dine:2015jga,Asadi:2022vys,Murgui:2025scx}. In the BBP model, $\Delta \mathbf{M}_{d B}$ and $\Delta m_B$ receive contributions from dim-5 operators of the type:
\begin{equation}
\mathcal{L} \supset \frac{y_\Lambda}{\Lambda} \overline{
B_L} B_R \sigma^2 + \frac{\mathbf{Y}_\Lambda}{\Lambda} \overline{
q_L}\Phi B_R \sigma + \text{H.c} \; ,
\label{eq:dim5}
\end{equation}
respectively, where $\mathbf{Y}_\Lambda$ and $y_\Lambda$ are dimensionless couplings and $\Lambda$ denotes the cutoff of the effective theory. The estimates of the contributions of the above operators on $\overline{\theta}$ are, 
\begin{align}
 \Delta \overline{\theta}|_{\Delta m_{B}} \sim y_\Lambda \frac{v_\sigma }{m_B} \left( \frac{v_\sigma}{\Lambda} \right) \; , \; \Delta \overline{\theta}|_{\Delta \mathbf{M}_{d B}} \sim \frac{y_\Lambda}{y_d} \frac{|\mathbf{M}_{Bd}|}{m_B}   \left( \frac{v_\sigma}{\Lambda} \right)  \; .
\end{align}
From the left estimate, assuming $y_\Lambda \sim \mathcal{O}(1)$ and $m_B \sim v_\sigma$, satisfying $\Delta \overline{\theta} \lesssim 10^{-10}$ enforces $v_\sigma \lesssim 10^8~\text{GeV}$, for a cutoff $\Lambda$ at the Planck scale. Furthermore, in the right estimate if we consider $y_d \sim 10^{-5}$, being the typical SM down quark Yukawa coupling, and since a nonzero CKM phase requires $|\mathbf{M}_{Bd}|/m_B \gsim 1$, the upper bound on the SCPV scale is even more drastic, $v_\sigma \lsim 10^3$ GeV. Evidently, the magnitude of this latter contribution depends on the flavor structure of the Yukawa couplings which can alleviate this naive estimate. Nonetheless, it is clear that the SCPV scale is bounded $v_\sigma \lsim 10^{3} - 10^{8}$~GeV for a cutoff $\Lambda$ at the Planck scale~\cite{Dine:2015jga}. This hierarchy, between $v_\sigma$ the SCPV scale and $\Lambda$ the cutoff scale of a more complete UV theory, required to control dangerous contributions to the strong CP phase, is the essence of the "NB quality problem".

Inspired by the BBP model, in what follows we propose a novel implementation of the NB mechanism, in which CP violation in the CKM matrix is generated at the quantum level via one-loop corrections mediated by a dark sector, odd under a $\mathcal{Z}_8$ symmetry. After SCPV induced by the complex VEV of a scalar singlet $\sigma$, the dark particles remain odd under a stabilizing $\mathcal{Z}_2$ symmetry. We show that the lightest of these particles (a scalar) is a viable WIMP DM particle -- see Sec.~\ref{sec:DM}. The crucial aspect of our dark-mediated solution to the strong CP problem is that threshold corrections to $\overline{\theta}$ arise only at the two-loop level. At the same time, we argue that the ``NB quality" problem is alleviated. The upcoming discussion follows closely our work of Ref.~\cite{Camara:2023hhn}.

%%%%%%%%%%%%%%%%%%%%%%%%%%%%%%%%%%%%%%%%%%%%%%%%%%%%%%%%%%%%%%%%%%%%%%%%%%%%%
\subsection{The model at tree level}
\label{sec:modelDLSS}
%%%%%%%%%%%%%%%%%%%%%%%%%%%%%%%%%%%%%%%%%%%%%%%%%%%%%%%%%%%%%%%%%%%%%%%%%%%%%

%
\begin{table}[t!]
\renewcommand{\arraystretch}{1.5}
	\centering
	\begin{tabular}{|K{1.5cm} | K{1.5cm} | K{4.5cm} |  K{3cm} |}
		\hline 
 &Fields&\SM&  $\mathcal{Z}_8 \to \mathcal{Z}_2$ \\
		\hline
		\multirow{6}{*}{Fermions}
&$q_L$&($\mathbf{3},\mathbf{2}, {1/6}$)& {$\omega^2$}   $\to$  $+$  \\
&$u_R$&($\mathbf{3},\mathbf{1}, {2/3}$)& {$\omega^2$}   $\to$  $+$  \\
&$d_{R}$&($\mathbf{3},\mathbf{1}, {-1/3}$)& {$\omega^2$}   $\to$  $+$  \\
&$B_{L,R}$&($\mathbf{3},\mathbf{1}, {-1/3}$)& {$\omega^6$}   $\to$  $+$  \\
&$D_{1 L,1R}$&($\mathbf{3},\mathbf{1}, {-1/3}$)& {$\omega^7$}   $\to$  $-$  \\
&$D_{2 L,2R}$&($\mathbf{3},\mathbf{1}, {-1/3}$)& {$\omega^3$}   $\to$  $-$  \\
		\hline
		\multirow{4}{*}{Scalars}
&$\Phi$&($\mathbf{1},\mathbf{2}, {1/2}$)&{$1$}    $\to$ $+$ \\	
&$\sigma$&($\mathbf{1},\mathbf{1}, {0}$)&{$\omega^2$}  $\to$  $+$ \\
&$\chi$&($\mathbf{1},\mathbf{1}, {0}$)&{$\omega^3$}    $\to$ $-$ \\	
&$\xi$&($\mathbf{1},\mathbf{1}, {0}$)&{$\omega$}  $\to$  $-$ \\
\hline
	\end{tabular}
	\caption{Field content and their transformation properties under the SM gauge and $\mathcal{Z}_8$ symmetries, where $\omega^k = e^{i\pi k/4}$, and under the remnant $\mathcal{Z}_2$ after spontaneous $\mathcal{Z}_8$ breaking.}
\label{tab:modelDSNB}
\end{table}
To implement our DM mediated NB solution to the strong CP problem, we extend the SM with three down-type VLQs $B_{L,R}$ and $D_{i L,iR}$ ($i=1,2$). Three complex scalar singlets $\sigma$, $\chi$ and $\xi$ are also introduced, besides the SM Higgs doublet $\Phi$. A $\mathcal{Z}_8$ symmetry, under which the several fields transform as indicated in Table~\ref{tab:modelDSNB}, is imposed to implement SCPV. As discussed later, this $\mathcal{Z}_8$ will be broken down to a dark $\mathcal{Z}_2$ by the VEV of $\sigma$ (the $\mathcal{Z}_2$ parities for each field are indicated in the last column of the table). Besides the usual SM quark Yukawa interactions of Eq.~\eqref{eq:lagyuksm}, in our model the SM quarks will couple to the new fields as:
\begin{align}
- \mathcal{L}_{\text{Yuk}} &\supset \Y_\xi  \overline{D_{2 L}} d_R \xi + \Y_\chi \;  \overline{D_{1 L}} d_R \chi^\ast+ \text{H.c.} \;,
\label{eq:LYukSMDSNB}
\end{align}
where $\Y_{\chi,\xi}$ are $1 \times 3$ vectors. The Yukawa couplings involving only new fields read
\begin{align}
- \mathcal{L}_{\text{Yuk}} &\supset  y_\chi \;  \overline{B_L} D_{2 R} \chi + y_\xi \;  \overline{B_L} D_{1 R} \xi^\ast + y_\chi^\prime \;  \overline{D_{2 L}} B_R \chi^\ast + y_\xi^\prime \;  \overline{D_{1 L}} B_R \xi + \text{H.c.} \;,
\label{eq:LYukchi}
\end{align}
where $y_{\chi,\xi}^{(\prime)}$ are numbers, and bare VLQ mass terms are
\begin{align}
- \mathcal{L}_{\text{mass}} &= m_B \; \overline{B_L} B_R + m_{D_{1,2}} \; \overline{D_{1,2 L}} D_{1,2 R}  + \text{H.c.} \,.
\label{eq:Lbare}
\end{align}
Notice that Eqs.~\eqref{eq:LYukSMDSNB}-\eqref{eq:Lbare} contain all gauge-invariant Yukawa and mass terms which respect the $\mathcal{Z}_8$ symmetry. CP invariance of the Lagrangian implies that all coupling and mass parameters are real.

The scalar content of our model contains, besides the usual Higgs doublet $\Phi$, three singlets $\sigma$, $\chi$ and $\xi$ defined as:
\begin{align}
\Phi &=\begin{pmatrix}
\phi^{+} \\
\phi^0
\end{pmatrix}= \frac{1}{\sqrt{2}}  \begin{pmatrix}
 \sqrt{2} \phi^{+} \\
 v + \phi^0_{\text{R}} + i \phi^0_{\text{I}}
\end{pmatrix} \; ; \nonumber \\ 
\sigma & = \frac{1}{\sqrt{2}}\left( v_\sigma\, e^{i \varphi} + \sigma_{\text{R}} + i \sigma_{\text{I}}\right) \; ; 
\chi = \frac{\chi_{\text{R}} + i \chi_{\text{I}}}{\sqrt{2}} \; ; \;  \xi = \frac{\xi_{\text{R}} + i \xi_{\text{I}}}{\sqrt{2}} \; .
\end{align}
The vacuum configuration is, 
\begin{align}
  \vev{ \phi^0} = \frac{v}{\sqrt{2}} \; , \; \vev{ \sigma} = \frac{v_\sigma\, e^{i \varphi}}{\sqrt{2}} \; , \; \vev{ \chi} = \vev{ \xi} = 0 \; ,
  \label{eq:vacuum}
\end{align}
where the dark-scalars $\chi$ and $\xi$ carry no VEV. SCPV is triggered by the complex VEV of $\sigma$. Imposing CP invariance at the Lagrangian level, the most general scalar potential allowed by the symmetries of our model is written as,
\begin{align}
V(\Phi,\sigma,\chi,\xi) &=  V(\Phi) + m_{\sigma}^2 |\sigma|^2 + m_{\sigma}^{\prime \; 2} \left(\sigma^2 + \sigma^{\ast 2} \right) +m_\chi^2 |\chi|^2 + m_\xi^2 |\xi|^2 \nonumber \\
&+\mu_{\chi} \left(\sigma \chi^2 + \sigma^\ast \chi^{\ast 2} \right) +\mu_{\xi} \left(\sigma \xi^{\ast 2} + \sigma^\ast \xi^2\right) +\mu_{\chi \xi} \left(\sigma \chi^\ast \xi + \sigma^{\ast} \chi \xi^\ast \right) \nonumber \\
& + \frac{\lambda_{\sigma}}{2} |\sigma|^4 + \lambda_{\sigma}^\prime \left(\sigma^4 + \sigma^{\ast 4} \right)  + \frac{\lambda_{\chi}}{2} |\chi|^4  + \frac{\lambda_{\xi}}{2} |\xi|^4 + \lambda_{\chi \xi} |\chi|^2 |\xi|^2  \nonumber \\
&+ \lambda_{\Phi \sigma} \left(\Phi^\dagger \Phi\right) |\sigma|^2 + \lambda_{\Phi \chi} \left(\Phi^\dagger \Phi\right) |\chi|^2 + \lambda_{\Phi \xi} \left(\Phi^\dagger \Phi\right) |\xi|^2 + \lambda_{\sigma \chi} |\sigma|^2 |\chi|^2 + \lambda_{\sigma \xi} |\sigma|^2 |\xi|^2 \nonumber \\
&+\lambda_{\sigma \chi \xi} \left(\sigma^2 \chi \xi + \sigma^{\ast 2} \chi^\ast \xi^\ast \right) + \lambda_{\sigma \chi \xi}^\prime \left(\sigma^{2} \chi^\ast \xi^\ast + \sigma^{\ast 2} \chi \xi \right) \; ,
\label{eq:Vpotential}
\end{align}
where all parameters are real. Note that, the spontaneous breaking of an exact discrete symmetry could lead to cosmological DW problems~\footnote{This might not be an issue if our mechanism is embedded in a more general framework providing a solution to that problem~(see e.g. Ref.~\cite{McNamara:2022lrw}).}. We, thus, consider a scenario in which the $\mathcal{Z}_8$ is softly broken by the bilinear term $m_\sigma^{2}(\sigma^2+\sigma^{*2})$, fixing the DW problem. The minimization of the scalar potential leads to three distinct solutions,
\begin{align}
 \; m_{\Phi}^2 & = - \frac{1}{2} \left[ v^2 \lambda_{\Phi} + v_\sigma^2 \lambda_{\Phi \sigma} \right] \; , \label{eq:mincondmphi}\\
\text{(i) :} \;  m_{\sigma}^2 & = - \frac{1}{2} \left[v^2 \lambda_{\Phi \sigma} + v_\sigma^2 \left(\lambda_{\sigma} + 4 \lambda_{\sigma}^{\prime}\right) + 4 m^{\prime \; 2}_\sigma \right]\; , \; \varphi = k \pi \; , k \in \mathbb{Z} \; ; \\
\text{(ii) :} \; m_{\sigma}^2 & = - \frac{1}{2} \left[v^2 \lambda_{\Phi \sigma} + v_\sigma^2 \left(\lambda_{\sigma} + 4 \lambda_{\sigma}^{\prime}\right) - 4 m^{\prime \; 2}_\sigma\right]\; , \; \varphi = \frac{\pi}{2} + k \pi \; , k \in \mathbb{Z} \; ; \\
\text{(iii) :} \; m_{\sigma}^2 & = - \frac{1}{2} \left[ v^2 \lambda_{\Phi \sigma} + v_\sigma^2 \left(\lambda_{\sigma} - 4 \lambda_{\sigma}^{\prime} \right)\right] \; , \; m^{\prime \; 2}_\sigma \neq 0 \Rightarrow \cos (2 \varphi) = - \frac{m^{\prime\,2}_\sigma}{2 v_\sigma^2 \lambda_{\sigma}^{\prime}} \; .
\end{align}
As mentioned before, the only viable solution to implement SCPV is (iii) which corresponds to the global minimum of the potential if~$ \left(m_{\sigma}^{\prime \; 4} - 4 v_\sigma^4 \lambda_{\sigma}^{\prime \; 2} \right) / (4 \lambda_{\sigma}^{\prime}) > 0$. Note that, in the limit of exact $\mathcal{Z}_8$ invariance, the only phase-sensitive term in the scalar potential is $\lambda_\sigma(\sigma^4+\sigma^{*4})$. Minimization leads to $\varphi=\pi/4+k\pi/2\,(k \in \mathbb{Z})$ [see Eq.~\eqref{eq:thetaexactOG}]. However, this possibility does not violate CP, since a generalized CP transformation can be defined such that the vacuum remains invariant. 

The scalar content of our model is decomposed into two classes, we can have $\mathcal{Z}_2$-even or $\mathcal{Z}_2$-odd dark-scalars. Among the even scalars, there will be the usual Goldstone bosons $G^0$ and $G^{\pm}$, as well as three massive neutral scalars. The mass matrix of the non-dark neutral states, in the $\left(\phi^0_\text{R}, \sigma_\text{R}, \sigma_\text{I} \right)$ basis, is given by
\begin{align}
    \bm{\mathcal{M}}^2_{\phi\sigma}=\begin{pmatrix}
    v^2\,\lambda_\Phi&v_\sigma\,v\,\lambda_{\Phi \sigma} \cos \varphi&v_\sigma\,v\,\lambda_{\Phi \sigma} \sin \varphi\\
   \cdot&v_\sigma^2\,\left(\lambda_\sigma+4\lambda_\sigma^\prime\right) \cos^2 \varphi&v_\sigma^2\,\left(\lambda_\sigma - \,12 \lambda_\sigma^\prime\right)\cos \varphi \sin \varphi\\
    \cdot&\cdot&v_\sigma^2\,\left(\lambda_\sigma+4\lambda_\sigma^\prime\right) \sin^2 \varphi
    \end{pmatrix} \; ,
    \label{eq:nondarkscalarmass}
\end{align}
where '$\; \cdot \;$' reflects the symmetric nature of the matrix. The corresponding mixing is,
\begin{equation}
\begin{pmatrix}
\phi^0_\text{R}\\
\sigma_\text{R}\\
\sigma_\text{I}
\end{pmatrix}= \mathbf{R} \begin{pmatrix}
h\\
H_1\\
H_2
\end{pmatrix},
\label{eq:nondarkmixing}
\end{equation}
where $\mathbf{R}$ is a $3\times3$ orthogonal matrix. Note that $h$ and $H_i$ ($i=1,2$) are the mass-eigenstates where we assume the mass ordering $m_h < m_{H_1} < m_{H_2}$, thus $h$ is identified as the SM Higgs boson.

The mass matrix of the dark-neutral scalars, in the $\left(\chi_\text{R}, \xi_\text{R}, \chi_\text{I}, \xi_\text{I} \right)$ basis, reads
\begin{align}
    \bm{\mathcal{M}}^2_{\chi\xi}=\begin{pmatrix}
    \bm{\mathcal{M}}^2_{\chi_\text{R}\xi_\text{R}}&\bm{\mathcal{M}}^2_{\chi_\text{R,I}\xi_\text{R,I}}\\
    \bm{\mathcal{M}}^{2 \; T}_{\chi_\text{R,I}\xi_\text{R,I}}& \bm{\mathcal{M}}^2_{\chi_\text{I}\xi_\text{I}}
    \end{pmatrix},
    \label{eq:darkscalarmass}
\end{align}
where each $2 \times 2$ matrix block is,
\begin{align}
    &\bm{\mathcal{M}}^2_{\chi_\text{R}\xi_\text{R}} = \\
    &\begin{pmatrix}
    \frac{1}{2}\left(v_\sigma^2 \lambda_{\sigma\chi} + v^2 \lambda_{\Phi \chi} \right) + m_\chi^2 + \sqrt{2} \mu_{\chi} v_\sigma \cos \varphi & \frac{v_\sigma}{2}\left[\sqrt{2} \mu_{\chi \xi} \cos \varphi +v_\sigma \left(\lambda_{\sigma \chi \xi} + \lambda_{\sigma \chi \xi}^\prime \right) \cos (2 \varphi)  \right]\\
    \cdot&\frac{1}{2}\left(v_\sigma^2 \lambda_{\sigma\xi} + v^2 \lambda_{\Phi \xi} \right) + m_\xi^2 + \sqrt{2} \mu_{\xi} v_\sigma \cos \varphi
    \end{pmatrix}, \nonumber \\
    &\bm{\mathcal{M}}^2_{\chi_\text{I}\xi_\text{I}} = \nonumber \\
    &\begin{pmatrix}
    \frac{1}{2}\left(v_\sigma^2 \lambda_{\sigma\chi} + v^2 \lambda_{\Phi \chi} \right) + m_\chi^2 - \sqrt{2} \mu_{\chi} v_\sigma \cos \varphi & \frac{v_\sigma}{2}\left[\sqrt{2} \mu_{\chi \xi} \cos \varphi - v_\sigma \left(\lambda_{\sigma \chi \xi} + \lambda_{\sigma \chi \xi}^\prime \right) \cos (2 \varphi)  \right]\\
    \cdot&\frac{1}{2}\left(v_\sigma^2 \lambda_{\sigma\xi} + v^2 \lambda_{\Phi \xi} \right) + m_\xi^2 - \sqrt{2} \mu_{\xi} v_\sigma \cos \varphi
    \end{pmatrix}, \nonumber \\
     &\bm{\mathcal{M}}^2_{\chi_\text{R,I}\xi_\text{R,I}} = \nonumber \\
     &\begin{pmatrix}
   - \sqrt{2} \mu_{\chi} v_\sigma \sin \varphi & \frac{-v_\sigma \sin \varphi}{2}\left[\sqrt{2} \mu_{\chi \xi}  + 2 v_\sigma \left(\lambda_{\sigma \chi \xi} - \lambda_{\sigma \chi \xi}^\prime \right) \cos \varphi  \right]\\
    \frac{v_\sigma \sin \varphi}{2}\left[\sqrt{2} \mu_{\chi \xi}  - 2 v_\sigma \left(\lambda_{\sigma \chi \xi} - \lambda_{\sigma \chi \xi}^\prime \right) \cos \varphi \right]  & \sqrt{2} \mu_{\xi} v_\sigma \sin \varphi
    \end{pmatrix} . \nonumber
\end{align}
The components of the inert singlets are related to mass-eigenstates $\zeta_{i}$ through the $4 \times 4$ orthogonal matrix $\mathbf{V}$:
\begin{equation}
\begin{pmatrix}
\chi_{\text{R}} \\
\xi_{\text{R}} \\
\chi_{\text{I}} \\
\xi_{\text{I}} \\
\end{pmatrix}  = \mathbf{V} \begin{pmatrix}
\zeta_1 \\
\zeta_2 \\
\zeta_3 \\
\zeta_4
\end{pmatrix} \; ,
\label{eq:mixneutralDM}
\end{equation}
where we assume the mass ordering $m_{\zeta_1} < m_{\zeta_2} < m_{\zeta_3} < m_{\zeta_4}$. Hence, the weak-basis fields~$\chi$ and $\xi$ can be written as,
\begin{equation}
\sqrt{2} \chi = \sum_{k=1}^{4} \left(\mathbf{V}_{1 k} + i \mathbf{V}_{3 k} \right) \zeta_k \; , \; \sqrt{2} \xi = \sum_{k=1}^{4} \left(\mathbf{V}_{2 k} + i \mathbf{V}_{4 k} \right) \zeta_k \; .
\label{eq:darkmix}
\end{equation}

After SSB of the EW and $\mathcal{Z}_8$ symmetries, which occurs through nonzero VEVs of $\phi_0$ and $\sigma$ [see Eq.~\eqref{eq:vacuum}],
the tree-level quark mass Lagrangian reads, 
\begin{equation}
- \mathcal{L}_{\text{mass}} = \overline{u_L} \mathbf{M}_u u_R + \left(\overline{d_L} , \overline{B_L} \right) \bm{\mathcal{M}}_{d}^{(0)}\begin{pmatrix} d_R \\ B_R \end{pmatrix} + \text{H.c.} \; , \; \bm{\mathcal{M}}_{d}^{(0)} = \begin{pmatrix} \mathbf{M}_d  & 0 \\ 0 & m_B^{(0)} \end{pmatrix} \; ,
\end{equation}
where $\bm{\mathcal{M}}_{d}^{(0)}$ is the full (tree-level) $4 \times 4$ down-type quark mass matrix, while $\mathbf{M}_{u,d} =  v \Y_{u, d}/\sqrt{2}$ are the $3 \times 3$ SM quark matrices.  
As usual, the SM up and down-type quark mass matrices are bidiagonalized through the unitary transformations $ u_L,d_L  \to \mathbf{V}_{L}^{u,d}\, u_L,d_L$ and $u_R,d_R \to \mathbf{V}_{R}^{u,d}\, u_R,d_R$ [see Eq.~\eqref{eq:massdiag}]:
\begin{equation}
\begin{aligned}
 \mathbf{V}_{L}^{u \dagger} \mathbf{M}_{u} \mathbf{V}_{R}^u = \mathbf{D}_{u} = \text{diag}(m_u, m_c, m_t) \; , \; \mathbf{V}_{L}^{d \dagger} \mathbf{M}_{d} \mathbf{V}_{R}^d = \mathbf{D}_{d} = \text{diag}(\tilde{m}_d, \tilde{m}_s, \tilde{m}_b) \; .
\end{aligned}
\label{eq:diagquakstree}
\end{equation}
Note that, since we have only down-type VLQs in our model, the up-quark sector is SM-like, so that $m_{u,c,t}$ correspond to physical masses. On the other hand, $\tilde{m}_{d,s,b}$ denote the SM down-quark masses at tree level, which will be corrected at higher orders due to the presence of the new down-type VLQs and scalars. These contributions fill the off-diagonal null blocks of the $\bm{\mathcal{M}}_{d}^{(0)}$ matrix. 

Performing the above quark-field rotations, yields the $\mathbf{V}$ quark-mixing matrix is given by Eq.~\eqref{eq:VCKMdef}. However, at this point, the quark mixing matrix matrix is real due to CP invariance. Furthermore, the even VLQ $B$ is a mass eigenstate with mass $m_B$ and does not mix with SM quarks making $\mathbf{V}$ orthogonal. Hence, at tree-level our model is free from FCNC. These arise only at loop level (see Sec.~\ref{sec:flavor}). In summary, at the classical level, the up and down (light) quark sector resembles the SM one, but with a real orthogonal CKM matrix and a VLQ $B$ that is decoupled. Hence CP violation will not be communicated to the quark sector at tree-level. Since the strong CP phase is given by [see Eq.~\eqref{eq:thetaF}]:
\begin{equation}
  \bar{\theta} = \arg[\det(\mathbf{M}_u)] + \arg[\det(\bm{\mathcal{M}}_d)] \; , 
\end{equation}
we obviously have $\bar{\theta} =0$ at tree-level.

%%%%%%%%%%%%%%%%%%%%%%%%%%%%%%%%%%%%%%%%%%%%%%%%%%%%%%%%%%%%%%%%%%%%%%%%%%%%%
\subsection{Complex CKM at one loop with $\bar{\theta}=0$}
\label{sec:oneloopCKM}
%%%%%%%%%%%%%%%%%%%%%%%%%%%%%%%%%%%%%%%%%%%%%%%%%%%%%%%%%%%%%%%%%%%%%%%%%%%%%

%
\begin{table}[t!]
\renewcommand{\arraystretch}{1.7}
\setlength{\tabcolsep}{1pt}
	\centering
	\begin{tabular}{| K{2cm} | K{6cm} |  K{6cm} |}
 \hline
 Dimension & Phase insensitive & Phase sensitive \\
\hline \hline
$5$& $\overline{B_L} B_R \; (\Phi^\dagger \Phi)\; , \; \overline{B_L} B_R \; |\sigma|^2$ & $\overline{B_L} d_R \; \sigma^2 \; , \; \overline{B_L} d_R \; \sigma^{\ast 2}$  \\
		\hline
$6$
& $\overline{q_L} \Phi d_R \; (\Phi^\dagger \Phi)\; , \; \overline{q_L} \Phi d_R \; |\sigma|^2$ &  $\overline{q_L} \Phi B_R \; \sigma^2 \, , \; \overline{q_L} \Phi B_R \; \sigma^{\ast 2}$ \\
		\hline
\multirow{3}{*}{$7$}& $\overline{B_L} B_R \; (\Phi^\dagger \Phi)^2$  &  $\overline{B_L} d_R \; \sigma^2 (\Phi^\dagger \Phi)\; , \; \overline{B_L} d_R \; \sigma^{\ast 2} (\Phi^\dagger \Phi)$ \\
& $\overline{B_L} B_R \; (\Phi^\dagger \Phi) |\sigma|^2$ &  $\overline{B_L} d_R \; \sigma^2 |\sigma|^2 \; , \; \overline{B_L} d_R \; \sigma^{\ast 2} |\sigma|^2$ \\
& $\overline{B_L} B_R \; |\sigma|^4$ & $\overline{B_L} B_R \; \sigma^4 \; , \; \overline{B_L} B_R \; \sigma^{\ast 4}$ \\
		\hline
\multirow{3}{*}{$8$}& \multirow{3}{*}{$\overline{q_L} \Phi d_R \; (\Phi^\dagger \Phi)^2 \; ,\; \overline{q_L} \Phi d_R \; |\sigma|^4$} &  $\overline{q_L} \Phi B_R \; \sigma^2 (\Phi^\dagger \Phi) \; , \; \overline{q_L} \Phi B_R \; \sigma^{\ast 2} (\Phi^\dagger \Phi)$ \\
&  & $\overline{q_L} \Phi B_R \; \sigma^2 |\sigma|^2 \; ,\; \overline{q_L} \Phi B_R \; \sigma^{\ast 2} |\sigma|^2$  \\
&  & $\overline{q_L} \Phi d_R \; \sigma^4 \; ,\; \overline{q_L} \Phi d_R \; \sigma^{\ast 4}$  \\
		\hline
	\end{tabular}
	\caption{Phase-sensitive and insensitive operators up to dim-8 which lead to corrections to the quark-mass terms after SSB.}
\label{tab:operators}
\end{table}
A necessary condition for the higher-order corrections to $\bm{\mathcal{M}}_d ^{(0)}$ to generate a complex effective CKM matrix is that at least one of the correcting mass terms in $\Delta\bm{\mathcal{M}}_d$ is complex -- see Eq.~\eqref{eq:Lmassloopcorrection}. The most intuitive way of investigating how this may happen is to look for higher-order operators which can generate complex mass terms after SCPV. Such operators must be gauge and $\mathcal{Z}_8$ invariant and contain unmatched powers of $\sigma^{(\ast)}$, given that the $\sigma$ VEV phase $\varphi$ is the only source of CP violation in our framework. Then one must check at which loop order those operators arise and compute the corresponding corrections $\Delta\bm{\mathcal{M}}_d$. The relevant effective operators up to dim-8 are given in Table~\ref{tab:operators}. Those depending solely on the bilinears $\Phi^\dagger \Phi$ and/or $|\sigma|^2$ are insensitive to the $\langle \sigma \rangle$ phase $\varphi$ and, thus, only operators with powers of $\sigma$ or $\sigma^*$ may lead to complex quark mass terms. 

\begin{figure}[t!]
    \centering
    \includegraphics[scale=0.9]{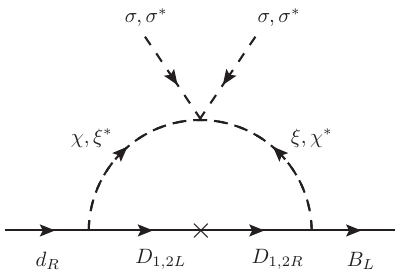} \hspace{+1cm} \includegraphics[scale=0.9]{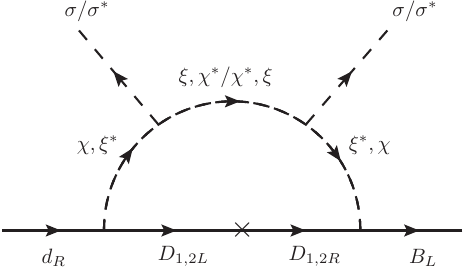} \\
    \vspace{+0.2cm} \includegraphics[scale=0.95]{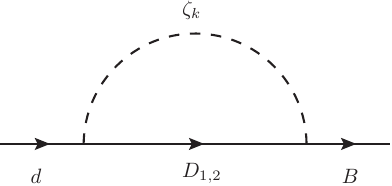}
    \caption{``Dark-mediated'' diagrams for the dim-5 operators $\overline{B_L} d_R\sigma^{(*)2}$ leading to
    $\Delta \mathbf{M}_{Bd}$ after $\mathcal{Z}_8$ symmetry breaking in the weak (top) and mass-basis (bottom), respectively.}
    \label{fig:oneloopdiags}
\end{figure}
The effective operators of Table~\ref{tab:operators} are realized at the quantum level by ``dark-mediated'' diagrams involving odd-sector particles. At one-loop, complex contributions to the quark mass terms stem from the top diagrams (shown in the weak basis) of Fig.~\ref{fig:oneloopdiags}, which correspond to the $\overline{B_L} d_R \; \sigma^2$ and  $\overline{B_L} d_R \; \sigma^{\ast 2}$ phase-sensitive dim-5 operators. In the physical basis, these contributions translate into the bottom diagram of the same figure, where $\zeta_k$ are the dark scalar mass eigenstates -- see Eqs.~\eqref{eq:darkscalarmass}-\eqref{eq:darkmix}. It is worth noticing that there are no one-loop ``dark-mediated'' corrections to the $\overline{d} d$ nor $\overline{d_L} B_R$ terms. This is due to the fact that only the RH down type-quarks $d_R$ couple to the dark-scalars and odd-VLQs at tree level. Notice also that there are one-loop contributions to $\Delta m_B$ but these are real, i.e. phase-sensitive $\overline{B_L} B_R$ operators are not realized at that order. Hence, complex one-loop contributions to $\bm{\mathcal{M}}_d$ are only those stemming from the diagrams of Fig.~\ref{fig:oneloopdiags} which lead to a non-zero $\Delta \mathbf{M}_{Bd}$ term in Eq.~\eqref{eq:Lmassloopcorrection}. 

The relevant interactions for the computation of $\Delta \mathbf{M}_{Bd}$ are the scalar-potential terms~[see Eq.~\eqref{eq:Vpotential}]
\begin{align}
    V &\supset \mu_{\chi} \left(\sigma \chi^2 + \sigma^\ast \chi^{\ast 2} \right) +\mu_{\xi} \left(\sigma \xi^{\ast 2} + \sigma^\ast \xi^2\right) +\mu_{\chi \xi} \left(\sigma \chi^\ast \xi + \sigma^{\ast} \chi \xi^\ast \right) \nonumber \\ 
    & + \lambda_{\sigma \chi \xi} \left(\sigma^2 \chi \xi + \sigma^{\ast 2} \chi^\ast \xi^\ast \right) + \lambda_{\sigma \chi \xi}^\prime \left(\sigma^{2} \chi^\ast \xi^\ast + \sigma^{\ast 2} \chi \xi \right) \; ,
\end{align}
and the Yukawa couplings of Eq.~\eqref{eq:LYuk}. Their importance is apparent from the weak-basis diagrams of Fig.~\ref{fig:oneloopdiags} which realise the lowest-order operator leading to $\Delta \mathbf{M}_{Bd}$ (see Table~\ref{tab:operators}). Performing its computation in the mass basis, and taking into account the coupling definitions in Eq.~\eqref{eq:LYuk}, we obtain
\begin{equation}
\Delta\mathbf{M}_{Bd}^{(1)} = \left[\mathbf{Y}_\chi \; y_{\xi} m_{D_1} \; \mathcal{F}_1\left(m_{D_1},m_{\zeta_k}\right) + \mathbf{Y}_\xi \; y_{\chi} m_{D_2} \; \mathcal{F}_2\left(m_{D_2},m_{\zeta_k}\right)\right]\mathbf{V}_R^d \; ,
\label{eq:MB}
\end{equation}
where $\mathbf{V}_R^d$ is the (real) orthogonal RH rotation of the down-type quarks defined in Eq.~\eqref{eq:diagquakstree}. The loop functions $\mathcal{F}_{1,2}$ read
\begin{align}
\mathcal{F}_{1} \left(m_{D_1} , m_{\zeta_k}\right) = & \; \frac{1}{32 \pi^2} \sum_{k=1}^{4} \left(\mathbf{V}_{1 k} - i \mathbf{V}_{3 k} \right) \left(\mathbf{V}_{2 k} - i \mathbf{V}_{4 k} \right) \frac{m_{\zeta_k}^2}{m_{D_1}^2 - m_{\zeta_k}^2} \ln \left( \frac{m_{D_1}^2}{m_{\zeta_k}^2} \right) \; , \label{eq:F1loop} \\
\mathcal{F}_{2} \left(m_{D_2} , m_{\zeta_k}\right) = & \; \frac{1}{32 \pi^2} \sum_{k=1}^{4} \left(\mathbf{V}_{1 k} + i \mathbf{V}_{3 k} \right) \left(\mathbf{V}_{2 k} + i \mathbf{V}_{4 k} \right) \frac{m_{\zeta_k}^2}{m_{D_2}^2 - m_{\zeta_k}^2} \ln \left( \frac{m_{D_2}^2}{m_{\zeta_k}^2} \right) \; , \label{eq:F2loop}
\end{align}
being $\mathbf{V}$ the $4 \times 4$ orthogonal matrix relating the weak odd-scalar degrees of freedom $(\chi_R,\xi_R,\chi_I,\xi_I)$ to the corresponding mass-eigenstates $\zeta_i$ ($i=1, \cdots, 4$) -- see Eqs.~\eqref{eq:mixneutralDM} and \eqref{eq:darkmix}. Since, in general, the real $\mathbf{Y}_\chi$ and $\mathbf{Y}_\xi$ are not proportional to each other, and the complex loop functions are distinct $\mathcal{F}_1 \neq \mathcal{F}_2$ (implying $\text{arg}[\mathcal{F}_1] \neq \text{arg}[\mathcal{F}_2])$,
the one-loop generated mass matrix $\Delta\mathbf{M}_{Bd}^{(1)}$ is a general $1 \times 3$ complex matrix. This highlights a key feature of our framework: $\sigma$ has no Yukawa couplings to fermions, but only to the dark-scalars and Higgs doublet in the potential $V$ [Eqs.~\eqref{eq:LYuk} and~\eqref{eq:Vpotential}].
As a result, transmitting CP violation to the quark sector can only occur via scalar-potential interactions and interference of loop diagrams that ultimately provide a complex $\mathbf{M}_{Bd}$.

\begin{figure}[t!]
    \centering
    \includegraphics[scale=0.9]{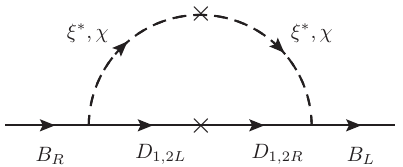} \hspace{+1cm} \includegraphics[scale=0.9]{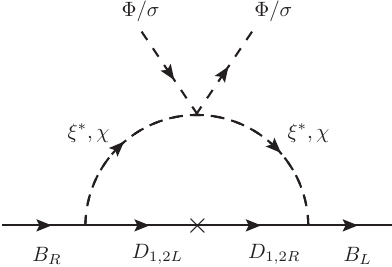} \\
    \vspace{+0.2cm} \includegraphics[scale=0.95]{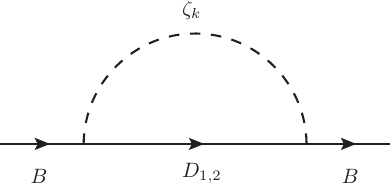}
    \caption{Lowest-order ``dark-mediated'' radiative correction to $\overline{B_L} B_R$ in the weak (top) and mass-basis (bottom).}
    \label{fig:oneloopmB}
\end{figure}
The $\Delta m_B$ one-loop corrections to the bare-mass term $\overline{B_L} B_R$ occur through diagrams like those shown in Fig.~\ref{fig:oneloopmB}. The lowest-order ones involve insertions of the scalar bilinears $m_{\chi,\xi}^2$ [see Eq.~\eqref{eq:Vpotential}] (top leftmost diagram), while $\left(\Phi^\dagger \Phi\right) |\chi|^2,\left(\Phi^\dagger \Phi\right) |\xi|^2$ and $|\sigma|^2|\chi|^2,|\sigma|^2|\xi|^2$ interactions lead to the dim-5 operators shown in Table~\ref{tab:operators} (top rightmost diagrams). The computation in the mass basis (bottom diagram) leads to:
\begin{equation}
\Delta m_B^{(1)} = y_{\xi} y_{\xi}^\prime m_{D_1} \; \mathcal{F}_3\left(m_{D_1},m_{\zeta_k}\right) + y_{\chi} y_{\chi}^\prime m_{D_2} \; \mathcal{F}_4\left(m_{D_2},m_{\zeta_k}\right) \, ,
\label{eq:deltamB}
\end{equation}
with
\begin{align}
\mathcal{F}_{3} \left(m_{D_1} , m_{\zeta_k}\right) = & \; \frac{1}{32 \pi^2} \sum_{k=1}^{4} |\mathbf{V}_{2 k} - i \mathbf{V}_{4 k}|^2 \frac{m_{\zeta_k}^2}{m_{D_1}^2 - m_{\zeta_k}^2} \ln \left( \frac{m_{D_1}^2}{m_{\zeta_k}^2} \right) \; , \label{eq:F3loop} \\
\mathcal{F}_{4} \left(m_{D_2} , m_{\zeta_k}\right) = & \; \frac{1}{32 \pi^2} \sum_{k=1}^{4} |\mathbf{V}_{1 k} - i \mathbf{V}_{3 k} |^2 \frac{m_{\zeta_k}^2}{m_{D_2}^2 - m_{\zeta_k}^2} \ln \left( \frac{m_{D_2}^2}{m_{\zeta_k}^2} \right)  \; . \label{eq:F4loop}
\end{align}
It is straightforward to see from Eqs.~\eqref{eq:deltamB}-\eqref{eq:F4loop} that $\Delta m_B$ is real. At one-loop, the corrected down-quark mass matrix reads, 
\begin{align}
- \mathcal{L}_{\text{mass}} & = \overline{u_L} \mathbf{M}_u u_R + \left(\overline{d_L} , \overline{B}_L \right) \bm{\mathcal{M}}_{d}^{(1)} \begin{pmatrix} d_R \\ B_R \end{pmatrix} + \text{H.c.} \; , \;
\bm{\mathcal{M}}_d^{(1)} = \begin{pmatrix} \mathbf{M}_d  & 0 \\  \Delta\mathbf{M}_{Bd}^{(1)} & m_B^{(0)}+\Delta m_B^{(1)} \end{pmatrix} \, ,
\label{eq:Lmassoneloop}
\end{align}
where $\Delta \mathbf{M}_{Bd}^{(1)}$ ($\Delta m_B^{(1)}$) is complex (real). Notice that the above $\bm{\mathcal{M}}_d^{(1)}$ mass matrix contains an off-diagonal $3 \times 1$ zero block. Thus, and since both $\mathbf{M}_{d}$ and $m_B^{(0)}+\Delta m_B^{(1)}$ are real, we have 
\begin{equation}
\theta_F^{(1)} = \arg[\det(\mathbf{M}_u)\det(\bm{\mathcal{M}}_d)]=0.
\end{equation}
In conclusion, thanks to the $\mathcal{Z}_8$ symmetry, $\overline{\theta}$ vanishes up to one-loop level in our framework, in spite of $\Delta\mathbf{M}_{Bd}^{(1)}$ being complex. As we will see in Sec.~\ref{sec:CKMpheno}, this term provides the seed for transmitting vacuum CP violation to the quark sector and generate a complex CKM quark-mixing matrix. 

%%%%%%%%%%%%%%%%%%%%%%%%%%%%%%%%%%%%%%%%%%%%%%%%%%%%%%%%%%%%%%%%%%%%%%%%%%%%%
\subsection{Corrections beyond one loop}
\label{sec:beyondoneloop}
%%%%%%%%%%%%%%%%%%%%%%%%%%%%%%%%%%%%%%%%%%%%%%%%%%%%%%%%%%%%%%%%%%%%%%%%%%%%%

%
\begin{figure}[t!]
    \centering
    \includegraphics[scale=1]{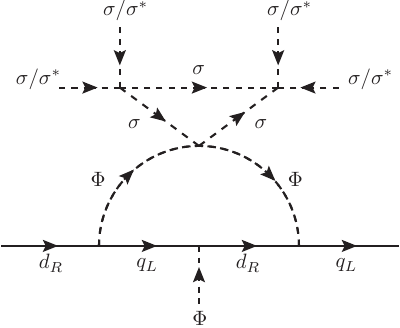}
    \hspace{+0.3cm} \includegraphics[scale=1]{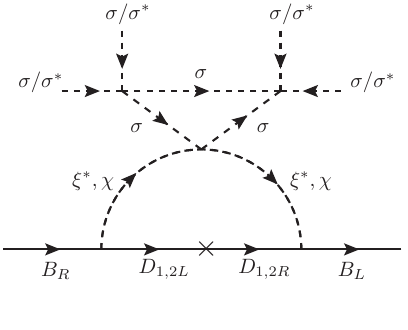} 
    \caption{Two-loop threshold corrections to $\overline{\theta}$, the left (right) diagram contributes to $\Delta \overline{\theta}|_{\Delta \mathbf{M}_{d}}$ ($\Delta \overline{\theta}|_{\Delta m_{B}}$).}
    \label{fig:threshold2}
\end{figure}
At the two-loop level, complex corrections to $ \mathbf{M}_d$ and $m_B$, shown in Fig.~\ref{fig:threshold2}, induce contributions to $\bar{\theta}$ which can be estimated as
    \begin{align}
     \Delta \overline{\theta}|_{\Delta \mathbf{M}_{d}}&\sim \frac{1}{(16 \pi^2)^2} \; \lambda_{\Phi \sigma}\, y_d^2 \,\frac{v_\sigma^2}{v^2}   \,, \\
    \Delta \overline{\theta}|_{\Delta m_B} & \sim \frac{1}{(16 \pi^2)^2} \, \lambda_{\sigma \zeta} \ y_\zeta \,y_\zeta^\prime \ \frac{m_D}{m_B} \frac{v_\sigma^2}{m_\zeta^2}  \label{eq:2loopdmb}\,,
    \end{align}
    where $m_\zeta$ is a typical dark scalar mass, and $y_\zeta^{(\prime)}$ are generic $y_{\xi,\chi}^{(\prime)}$ couplings. Here $\lambda_{\Phi \sigma}$ is the $\left(\Phi^\dagger \Phi\right) |\sigma|^2$ quartic scalar coupling and $\lambda_{\sigma \zeta}$ stands for generic $\lambda_{\sigma \chi} |\sigma|^2 |\chi|^2$ and $\lambda_{\sigma \xi} |\sigma|^2 |\xi|^2$ couplings. For typical values for the SM quark Yukawa couplings $y_d \sim \mathcal{O}(10^{-2})$, the first correction above is under control if
    $\lambda_{\Phi \sigma} \lesssim v^2/v_\sigma^2$. This is reasonable, as the physics accounting for the Higgs hierarchy is likely to also provide a small $\lambda_{\Phi \sigma}$. On the other hand, if all mass scales in Eq.~\eqref{eq:2loopdmb} are of the same order, $\Delta \overline{\theta}|_{\Delta m_B} \lesssim 10^{-10}$
    requires $|\lambda_{\sigma \zeta} \ y_\zeta \,y_\zeta^\prime| \lesssim 10^{-6}$, which can be easily accommodated. In fact, in our framework, the U(1)-sensitive couplings with the dark sector can be naturally small in the 't Hooft sense~\cite{tHooft:1979rat} since the Lagrangian symmetry is enlarged in their absence. Note that, the above contributions come from operators $\overline{q_L}\Phi d_R \sigma^{(\ast) 4}$ and $\overline{B_L} B_R \sigma^{(\ast) 4}$.

    \begin{figure}[t!]
    \centering
    \includegraphics[scale=0.75]{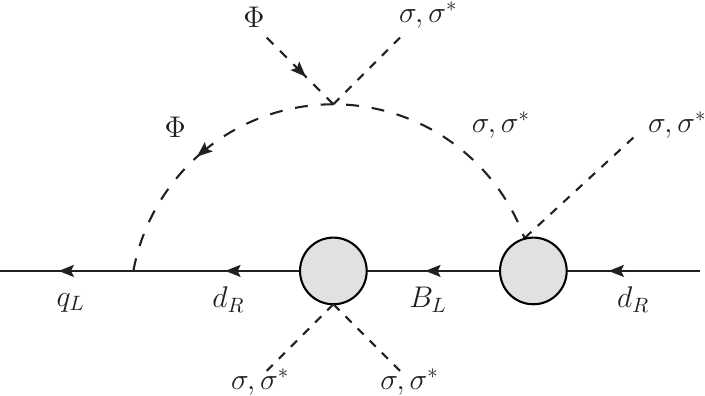} \\
    \hspace{+0.3cm}\includegraphics[scale=0.75]{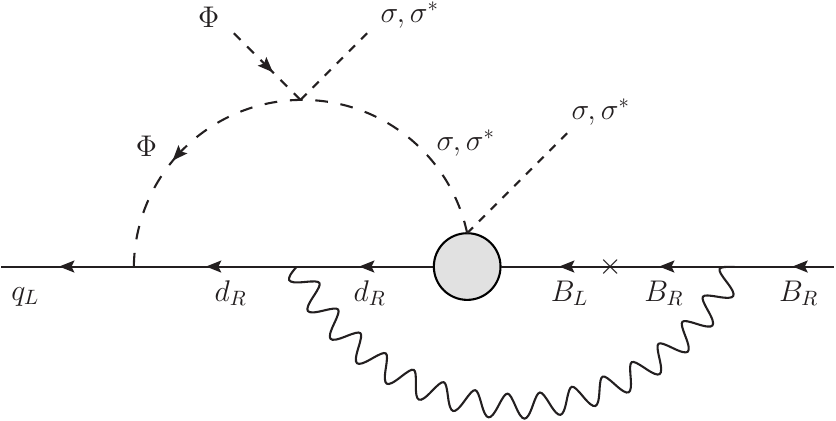}
    \hspace{+0.3cm}\includegraphics[scale=0.75]{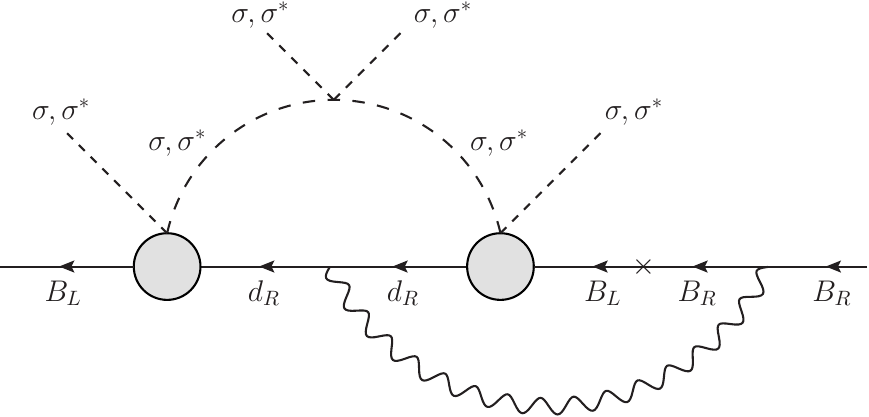} 
    \caption{Threshold corrections to $\overline{\theta}$, involving the blobs that represent the ``dark-mediated'' loops of Fig.~\ref{fig:oneloopdiags}. The top (bottom left) [bottom right] diagram contributes to $\Delta \overline{\theta}|_{\Delta \mathbf{M}_{d }}$ ($\Delta \overline{\theta}|_{\Delta \mathbf{Y}_{d B}}$) [$\Delta \overline{\theta}|_{\Delta m_{B}}$] (see text for details).}
    \label{fig:threshold}
    \end{figure}
    Concerning higher-loop corrections, we have checked that the contributions to $\overline{\theta}$ arise from the three (four) loop diagrams, shown in Fig.~\ref{fig:threshold}, via $\Delta\mathbf{M}_{d,dB}$ ($m_B$), which can be estimated as
    \begin{align}
     \Delta \overline{\theta}|_{\Delta \mathbf{M}_{d B}} &\sim \frac{g^2}{16\pi^2} \Delta \overline{\theta}|_{\Delta \mathbf{M}_{d}}\sim \frac{1}{(16 \pi^2)^2} \; \lambda_{\Phi \sigma} \frac{|\Delta\mathbf{M}_{Bd}|^2}{v_\sigma^2}   \; , \\
    \Delta \overline{\theta}|_{\Delta m_B} & \sim \frac{g^2}{(16 \pi^2)^2} \; \frac{|\Delta\mathbf{M}_{Bd}|^2}{v^2_\sigma} \; ,
    \end{align}
where $g \sim \mathcal{O}(1)$ is a weak coupling and we have considered a $\mathcal{O}(1)$ coupling for the $|\sigma|^4$ term. It is straightforward to see that $|\Delta\mathbf{M}_{Bd}| \lesssim 10^{-3} v_\sigma$ is required to keep these corrections under control
(as long as $\lambda_{\Phi \sigma}$ is made small in a framework where the Higgs mass is stabilized). One may now ask how natural is it to verify this condition in our scenario. In the above estimates, $\Delta\mathbf{M}_{Bd}$ is the one-loop correction in Eq.~\eqref{eq:MB}, with the contributions in Fig.~\ref{fig:oneloopdiags} being roughly estimated as:
\begin{align}
|\Delta\mathbf{M}_{Bd}| &\sim \frac{1}{16 \pi^2}  \lambda_{\sigma \zeta\zeta} |\mathbf{Y}_\zeta| \,y_\zeta \frac{v_\sigma^2}{m_\zeta^2}\,m_D \; ,\label{eq:MBestimation1}\\
|\Delta\mathbf{M}_{Bd}| &\sim \frac{1}{16 \pi^2}   |\mathbf{Y}_\zeta| \,y_\zeta \,\frac{\mu_\zeta^2}{m_\zeta^2} \frac{v_\sigma^2}{m_{\zeta}^2} \,m_D\,,
    \label{eq:MBestimation2}
\end{align}
for the left and right diagram, respectively. Here $\mathbf{Y}_\zeta$ and $y_\zeta$ represent generic $\mathbf{Y}_{\chi,\xi}$ and $y_{\chi,\xi}$ couplings of Eq.~\eqref{eq:LYukchi}, while $\lambda_{\sigma \zeta\zeta}$ and $\mu_{\zeta}$ are quartic and trilinear terms of the scalar potential. From those estimates one sees that, to ensure $|\Delta\mathbf{M}_{Bd}| \lesssim 10^{-3} v_\sigma$ one roughly needs
$ |\mathbf{Y}_\zeta| \,y_\zeta \lesssim  m_\zeta^2/(m_D v_\sigma)$ for $\lambda_{\sigma\zeta\zeta} \lesssim 1$ and $ \mu_\zeta\sim m_\zeta$. This condition is attainable for reasonable values of dark sector couplings and wide mass ranges. In contrast, models where $\mathbf{M}_{Bd}$ is generated at tree-level via a $y_B\sigma^{(\ast)} \overline{B_L}d_R$ have been argued to suffer from a ``quality problem'', requiring a small $y_B \lesssim 10^{-3}$~\cite{Perez:2020dbw,Valenti:2021rdu}. 

As seen before, in the original BBP scenario,  $\Delta \mathbf{M}_{d B}$ and $\Delta m_B$ receive contributions from dim-5 operators of the type $\overline{q_L}\Phi B_R \sigma^{(\ast)}$ and $\overline{B_L} B_R \sigma^{(\ast)2}$, respectively. These affect $\overline{\theta}$ in a way that $\Delta \overline{\theta} \lesssim 10^{-10}$ sets an upper bound on the SCPV scale $v_\sigma \lesssim 10^{3} - 10^{8}$~GeV, for a cutoff $\Lambda$ at the Planck scale~\cite{Choi:1992xp,Dine:2015jga}. This hierarchy between $v_\sigma$ and $\Lambda$ is the essence of the NB ``quality problem''. As recently noted in Ref.~\cite{Asadi:2022vys}, such low SCPV scale may have a drastic impact in cosmology. In our case, the lowest dimension operators that would induce corrections to $\overline{\theta}$ are the dim-6 $y_\Lambda\overline{q_L} \Phi B_R \sigma^{(*)2}$, for which we estimate
\begin{equation}
    \Delta \overline{\theta}|_{\Delta \mathbf{M}_{d B}} \sim \frac{y_\Lambda}{y_d} \frac{|\Delta\mathbf{M}_{Bd}|}{m_B}  \left( \frac{v_\sigma}{\Lambda} \right)^2  \; .
\end{equation}
Taking $|\Delta\mathbf{M}_{Bd}|/m_B \gtrsim \mathcal{O}(1)$ to generate a viable complex CKM matrix, and $y_\Lambda \sim \mathcal{O}(1)$ with $y_d \sim 10^{-5}-1$,
we get that $\Delta \overline{\theta}|_{\Delta \mathbf{M}_{d B}} \lesssim 10^{-10}$ only requires $v_\sigma \lesssim 10^{8} - 10^{13}$~GeV, a milder hierarchy between those scales. Thus, we argue that in contrast to minimal VLQ NB solutions, our framework somehow alleviates the potential "NB quality problem".

In summary, to control beyond one-loop corrections to the strong CP phase we need i) $\lambda_{\Phi \sigma} \lsim v^2/v_\sigma^2$, ii) $|\lambda_{\sigma \zeta} \ y_\zeta \,y_\zeta^\prime| \lesssim 10^{-6}$ and iii) $|\mathbf{M}_{Bd}| \lsim 10^{-3} v_\sigma$. Taking these requirements into account, we will show how at one-loop our model successfully generates a complex CKM matrix in Sec.~\ref{sec:CKMpheno}, as well as understand how the dark-sector can be experimentally probed in Sec.~\ref{sec:DMDSNB}.

%%%%%%%%%%%%%%%%%%%%%%%%%%%%%%%%%%%%%%%%%%%%%%%%%%%%%%%%%%%%%%%%%%%%%%%%%%%%%
\subsection{Quark mixing phenomenology}
\label{sec:CKMpheno}
%%%%%%%%%%%%%%%%%%%%%%%%%%%%%%%%%%%%%%%%%%%%%%%%%%%%%%%%%%%%%%%%%%%%%%%%%%%%%

At one-loop level, the quark mass Lagrangian is given by Eq.~\eqref{eq:Lmassoneloop}, for simplicity in what follows, we adopt the following notation:
\begin{align}
\bm{\mathcal{M}}_d = \begin{pmatrix} \mathbf{M}_d  & 0 \\  \mathbf{M}_{Bd} & m_B \end{pmatrix} = \begin{pmatrix} \mathbf{M}_d  & 0 \\  \Delta\mathbf{M}_{Bd}^{(1)} & m_B^{(0)}+\Delta m_B^{(1)} \end{pmatrix} \, .
\label{eq:Lmassoneloopnotation}
\end{align}
$\bm{\mathcal{M}}_d$ can be bidiagonalized via unitary transformations $(d_{L,R}, B_{L,R}) \to \bm{\mathcal{V}}_{L,R}\, (d_{L,R}, B_{L,R})$ as
\begin{align}
\bm{\mathcal{V}}_L^{d \dagger} \bm{\mathcal{M}}_d \ \bm{\mathcal{V}}_R^d = \bm{\mathcal{D}}_{d} = \text{diag}\,(m_d, m_s, m_B , \overline{m}_B) \; ,
\label{eq:diagdownfullNB}
\end{align}
where $m_{d,s,b}$ are the physical light down-type quark masses, whose values are given in Table~\ref{tab:quarkdata}, while $\overline{m}_B$ denotes the heavy down-type quark mass. The latter is directly constrained by the ATLAS and CMS searches for pair-produced heavy down-type VLQ~\cite{CMS:2018zkf,ATLAS:2018ziw,CMS:2020ttz}, namely at $95\%$ CL~\cite{CMS:2020ttz} we have~\footnote{This bound is obtained by assuming that $B$ couples only/primarily to the third generation of SM quarks. In generic VLQ scenarios the heavy quark(s) can mix with all SM quarks and hence this direct constraint on its mass must be reevaluated~\cite{Panizzi:2014dwa}. In Sec.~\ref{sec:flavor} we show that this bound holds in our framework.},
\begin{equation}
    \overline{m}_B > 1.4 \; \text{TeV} \; .
    \label{eq:collider}
\end{equation}
For a given $ \bm{\mathcal{M}}_d$, $\bm{\mathcal{V}}_L^d$ and $\bm{\mathcal{V}}_R^d$ are determined by diagonalizing the Hermitian matrices $\bm{\mathcal{H}}_{d} = \bm{\mathcal{M}}_{d} \bm{\mathcal{M}}_{d}^{\dagger}$
and $\bm{\mathcal{H}}_{d}^\prime = \bm{\mathcal{M}}_{d}^{\dagger} \bm{\mathcal{M}}_{d}$. 
We remark that the quark mixing structure is totally analogous to that of neutrino mixing in seesaw schemes described in Sec.~\ref{sec:TypeI}. Indeed, quark mixing here is described by a rectangular $3 \times 4$ matrix $\mathbf{W}_{\alpha j}\equiv (\bm{\mathcal{V}}_L)_{\alpha j}$ ($\alpha=d,s,b$, $j=d,s,b, B$):
\begin{equation}
\mathbf{W}_{d} \equiv \left(\mathbf{W}_{\text{light}}^d, \mathbf{W}_B \right) \; ,
\label{eq:3x4}
\end{equation}
where $(\mathbf{W}_{\text{light}}^d)_{\alpha \beta} \equiv (\bm{\mathcal{V}}_L)_{\alpha \beta} $ and $(\mathbf{W}_{B})_\alpha \equiv (\bm{\mathcal{V}}_L)_{\alpha B}$ are $3\times 3$ and $3 \times 1$ matrices, respectively. As we will see shortly, for adequately large VLQ masses the form of this mixing matrix can be determined perturbatively using the method in Ref.~\cite{Schechter:1981cv}. From the above, it is clear that $\mathbf{V}_{L}^{u \dagger} \mathbf{W}_{B}$ defines the mixing between the SM down quarks and the heavy state in the physical up-type quark basis [see Eq.~\eqref{eq:diagquakstree}]. Also, the light-quark mixing matrix $\mathbf{V}$ is no longer unitary and can be written as,
\begin{equation}
\mathbf{V}^\prime = \mathbf{V}_{L}^{u \dagger} \mathbf{W}_{\text{light}}^d \equiv \left(\mathbb{1} - \boldsymbol{\eta} \right) \mathbf{V} \; ,
\label{eq:nonunitCKM}
\end{equation}
with the Hermitian matrix $\boldsymbol{\eta}$ encoding deviations from unitarity of $\mathbf{V}^\prime$ and the unitary CKM matrix $\mathbf{V}$ is parameterized as in Eq.~\eqref{eq:VCKMparam}. In our analysis we use the current best-fit values for the CKM observables given in Table~\ref{tab:quarkdata}.

The use of the quark-seesaw approximation is not only justified but also provides useful insight on important features of NB-type models. We focus on $\bm{\mathcal{H}}_{d}$,
\begin{equation}
\bm{\mathcal{H}}_{d} = \bm{\mathcal{M}}_{d} \bm{\mathcal{M}}_{d}^\dagger = \begin{pmatrix} \mathbf{M}_d \mathbf{M}_d^T  & \mathbf{M}_d \mathbf{M}_{Bd}^\dagger \\ \mathbf{M}_{Bd}\mathbf{M}_d^T & \mathbf{M}_{Bd} \mathbf{M}_{Bd}^\dagger + m_B^2 \end{pmatrix} \; ,
\label{eq:Hd}
\end{equation}
which in the limit $\mathbf{M}_{Bd} \mathbf{M}_{Bd}^\dagger + m_B^2 \gg \mathbf{M}_d \mathbf{M}_d^T$ can be block-diagonalized through $\bm{\mathcal{V}}_{L}^d$ given as follows~\cite{Grimus:2000vj}
\begin{equation}
\bm{\mathcal{V}}_L^d = \begin{pmatrix} 
 \sqrt{\mathbb{1}-\mathbf{\Theta}_d \mathbf{\Theta}_d^{\dagger}} &  \mathbf{\Theta}_d \\
 - \mathbf{\Theta}_d^{\dagger} & \sqrt{\mathbb{1}-\mathbf{\Theta}_d^{\dagger} \mathbf{\Theta}_d}
\end{pmatrix} 
\begin{pmatrix} 
\mathbf{V}_{L}^d &  0 \\
 0 & 1
\end{pmatrix} \, \Rightarrow \bm{\mathcal{V}}_L^{d \dagger} \bm{\mathcal{H}}_d \ \bm{\mathcal{V}}_L^d =
\begin{pmatrix} 
\mathbf{V}_{L}^{d \dagger} \mathbf{M}_{\text{light}}^2 \mathbf{V}_{L}^d & 0 \\
 0 &  \widehat{m}_B^2
\end{pmatrix} \; ,
\label{eq:paramVL}
\end{equation}
where $\mathbf{M}_{\text{light}}$ is the effective light down-type quark mass matrix and $\widehat{m}_B$ the heavy down-type quark mass in the seesaw approximation. 
The former can be diagonalized through a unitary transformation $d_{L}  \to \mathbf{V}_{L}^d \, d_{L}$, satisfying
\begin{equation}
 \mathbf{V}_{L}^{d \dagger}\, \mathbf{M}_{\rm light}^2\,  \mathbf{V}_{L}^d  = \mathbf{\widehat{D}}_{d} = \text{diag}\left(\widehat{m}_d^2, \widehat{m}_s^2, \widehat{m}_b^2\right),
\end{equation}
where $\widehat{m}_d^2, \widehat{m}_s^2, \widehat{m}_b^2$ are the light quark masses. 
The unitary quark mixing matrix in the SM is given as $\mathbf{V} = \mathbf{V}_{L}^{u \dagger} \mathbf{V}_{L}^d$. 
Assuming $\mathbf{M}_{Bd} \mathbf{M}_{Bd}^\dagger + m_B^2 \gg \mathbf{M}_d \mathbf{M}_d^T$ the $3 \times 1$ matrix $\mathbf{\Theta}_d$ of Eq.~\eqref{eq:paramVL} is approximately given by
$\mathbf{\Theta}_d \simeq \mathbf{M}_d \mathbf{M}_{Bd}^\dagger/\widehat{m}_B^2$, so that
\begin{align}
\mathbf{M}_{\text{light}}^2 & \simeq \mathbf{M}_d \mathbf{M}_d^T - \mathbf{\Theta}_d \; \widehat{m}_B^2 \mathbf{\Theta}_d^\dagger \simeq \mathbf{M}_d \mathbf{M}_d^T - \frac{\mathbf{M}_d \mathbf{M}_{Bd}^\dagger \mathbf{M}_{Bd} \mathbf{M}_d^T}{\widehat{m}_B^2} \; , \label{eq:Mlight} \\
\widehat{m}_B^2 &\simeq \mathbf{M}_{Bd} \mathbf{M}_{Bd}^\dagger + m_B^2 \; . \label{eq:heavymass}
\end{align} 
From the seesaw approximation, we can derive the heavy-light quark mixing,
\begin{equation}
\mathbf{V}_L^{u \dagger}\mathbf{W}_{B} = \mathbf{V}_L^{u \dagger} \mathbf{\Theta}_d \simeq \mathbf{V} \frac{\mathbf{V}_L^\dagger \mathbf{M}_d \mathbf{M}_{Bd}^\dagger}{\widehat{m}_B^2} \; ,
\label{eq:heavylightmixing}
\end{equation}
and by expanding Eq.~\eqref{eq:paramVL} up to second order in $\mathbf{\Theta}_d$, we obtain the deviations from unitarity,
\begin{equation}
\boldsymbol{\eta} \simeq \dfrac{1}{2} \mathbf{V}_L^{u \dagger} \mathbf{\Theta}_d \mathbf{\Theta}_d^{\dagger} \mathbf{V}_L^u \simeq \dfrac{1}{2} \mathbf{V}_L^{u \dagger} \frac{\mathbf{M}_d \mathbf{M}_{Bd}^\dagger \mathbf{M}_{Bd} \mathbf{M}_d^T}{\widehat{m}_B^4} \mathbf{V}_L^u \; ,
\label{eq:eta}
\end{equation}
appearing in the light quark mixing matrix $\mathbf{V}^\prime=\left(\mathbb{1} - \boldsymbol{\eta} \right) \mathbf{V}$ of Eq.~\eqref{eq:nonunitCKM}. The above expressions for the effective light and heavy down-type quark masses will be useful in the following.

From the seesaw approximation and the expression for the dark-loop generated $\mathbf{M}_{Bd}$ in Eq.~\eqref{eq:MB}, we can gain some insight on how a complex CKM matrix is generated in our NB-type model. Namely: 
\begin{itemize}

\item Since the only mass matrix containing complex parameters is $\mathbf{M}_{Bd}$, it will be the portal linking SCPV to a complex CKM quark mixing matrix. In order for the second term in $\mathbf{M}_{\text{light}}^2$ of Eq.~\eqref{eq:Mlight} not to be suppressed compared to the first we need 
    \begin{equation}
        \frac{|\mathbf{M}_{Bd}|}{m_B} \gsim 1 \; ,
        \label{eq:MBvsmB}
    \end{equation}
    which, if verified, can lead to a large weak CP-violating phase. Otherwise, $\delta^q$ will be suppressed by~$\sim |\mathbf{M}_{Bd}|^2/m_B^2$. 

    \item From Eq.~\eqref{eq:MB}, we can define the following $1 \times 3$ complex vectors,
    \begin{align}
        \mathbf{Y}_{D_1} \equiv \mathbf{Y}_\chi \; y_{\xi} \; \mathcal{F}_1\left(m_{D_1},m_{\zeta_k}\right) \; ,\;  \mathbf{Y}_{D_2} \equiv \frac{m_{D_2}}{m_{D_1}} \mathbf{Y}_\xi \; y_{\chi} \; \mathcal{F}_2\left(m_{D_2},m_{\zeta_k}\right) \; .
    \end{align}
    In order to successfully transmit SCPV to the quark sector, one needs to ensure that $\mathbf{Y}_{D_{1,2}}$ are not aligned with each other and that their entries are comparable in magnitude. This leads to the following requirement~\cite{Perez:2020dbw}:
    \begin{align}
       r_{12} \equiv \frac{ \left|\vec{\mathbf{Y}}_{D_1} \times \vec{\mathbf{Y}}_{D_2}\right|}{\left|\vec{\mathbf{Y}}_{D_1}\right|^2 + \left|\vec{\mathbf{Y}}_{D_2}\right|^2} \sim \mathcal{O}(1) \; .
       \label{eq:r12}
    \end{align}
    Notice that, $0\leq r_{12} \leq0.5$ since if $\mathbf{Y}_{D_{1}}$ and $\mathbf{Y}_{D_{2}}$ are aligned we obtain $r_{12}=0$, while if $|\mathbf{Y}_{D_{1}}|=|\mathbf{Y}_{D_{2}}|$ and the vectors are
    orthogonal to each other we reach $r_{12}=0.5$.
    
  \item Following the estimate for $\mathbf{M}_{Bd}$ given in Eq.~\eqref{eq:MB}, and taking $|\mathbf{Y}|, y \sim \mathcal{O}(1)$ and $\lambda_{\sigma \zeta} v_\sigma \sim m_{\zeta}$, one sees that $m_D \sim \mathcal{O}(10^3)$~TeV, so that $\widehat{m}_B \gsim 1$ TeV [see Eq.~\eqref{eq:heavymass}], consistently with the LHC bound of Eq.~\eqref{eq:collider}. Moreover, from Eqs.~\eqref{eq:F1loop} and \eqref{eq:F2loop}, it is apparent that to successfully generate a complex $\mathbf{M}_{Bd}$ we need that at least three dark-scalar mass eigenstates mix sizeably among themselves [see Eq.~\eqref{eq:mixneutralDM}]. The latter is due to the fact the $\mathbf{M}_{Bd}$ loop involves both $\chi$ and $\xi$ (see Fig.~\ref{fig:oneloopdiags}). Thus, in order to generate $\widehat{m}_B \gsim 1$ TeV, as well as a complex $\mathbf{V}$, one can have a spectrum where $\zeta_{2,3,4}$ mix sizeably among themselves and $m_{\zeta_1} \ll m_{\zeta_2} \sim m_{\zeta_3} \sim m_{\zeta_4} \sim m_{D_{1,2}} \sim 10^3$ TeV. In what follows we will explore this scenario since it provides a light dark-scalar leading to interesting phenomenology regarding DM (see Sec.~\ref{sec:DMDSNB}).

\end{itemize}
\begin{table}[!t]
\renewcommand{\arraystretch}{1.5}
\centering
\begin{tabular}{|K{2cm}|K{4cm}|K{3cm}|}   
\hline
Sector & Parameters & Scan range \\
\hline
\multirow{3}{*}{Yukawa} & $m_{B}$ & $[10^{-1} , 10^3]$ (TeV)  \\
& $m_{D_{1,2}}$ & $[10^{-1} , 10^8]$ (TeV)  \\
& $|\mathbf{Y}_\chi^{d,s,b}|, |\mathbf{Y}_\xi^{d,s,b}|, \left|y_\chi\right|, \left|y_\xi\right|$ & $\left[10^{-2} , 1\right]$ \\
\hline
\multirow{6}{*}{Scalar} & $\varphi$ &  $[0 , 2 \pi]$ \\
& $m_{\chi}^2 \simeq m_{\xi}^2 $ &  $[10^6 , 10^{12}]$ (TeV$^2$)  \\
& $|\mu_{\chi \xi}| \simeq |\mu_\xi| \simeq m_{\chi}^2/v_\sigma$ & $[10^5 , 10^{11}]$ (TeV) \\
& $|\mu_\chi| \lsim |\mu_\xi| $ & $[10^{-3}, 10^6]$ (TeV) \\
& $|\lambda_{\Phi \chi}|,|\lambda_{\Phi \xi}|$ & $[10^{-10} , 1]$ \\
& $|\lambda_{\sigma \chi}|,|\lambda_{\sigma \xi}|,|\lambda_{\sigma \chi \xi}|  , \; |\lambda_{\sigma \chi \xi}^\prime|$ & $[10^{-5} , 1]$ \\
\hline
\end{tabular}
\caption{Input parameters of our model and corresponding ranges used in our numerical benchmark scan (see text for details). We set $v=246$ GeV and $v_\sigma=10^3$ TeV.}
\label{tab:CKMscan}
\end{table}
  With the above comments in mind, our numerical analysis proceeds as follows:  
\begin{itemize}

\item Without loss of generality, we work in the basis where the up-type quark mass matrix is diagonal and focus on the down-type quark sector. We vary the relevant Yukawa couplings and mass parameters as shown in Table~\ref{tab:CKMscan}. These include the bare mass terms $m_B$ and $m_{D_{1,2}}$ and the Yukawa couplings  $\mathbf{Y}_{\chi,\xi}$ and $y_{\chi,\xi}$. The Higgs doublet and complex scalar singlet VEVs are set to $v=246$ GeV and $v_\sigma=10^3$ TeV, varying the CP phase $\varphi$ in the $[0, 2 \pi]$ range. Furthermore, we scan the quadratic, cubic and quartic couplings entering the dark-scalar mass matrix of Eq.~\eqref{eq:darkscalarmass}. The latter contains a total of $14$ scalar potential parameters determining any dark-scalar mass and mixing spectrum. As already mentioned, we focus on the most interesting phenomenological benchmark scenario with $m_{\zeta_1} \ll m_{D_{1,2}}, m_{\zeta_{2,3,4}}$ and, thus, $\zeta_1$ will be our WIMP DM candidate (see Sec.~\ref{sec:DMDSNB}). To guarantee such mass spectrum, we take $m_{\chi}^2 \simeq m_{\xi}^2 $, $|\mu_{\chi \xi}| \simeq |\mu_\xi| \simeq m_{\chi}^2/v_\sigma$, $|\mu_\chi| \lsim |\mu_\xi| $ and vary these parameters within the ranges indicated in Table~\ref{tab:CKMscan}. By diagonalizing the dark-scalar mass matrix, we obtain $m_{\zeta_{1,2,3,4}}$ and the mixing matrix $\mathbf{V}$ and compute $\mathbf{M}_{Bd}$ using Eq.~\eqref{eq:MB}. The remaining scalar potential parameters, i.e., the even-sector ones and the dark-scalar quartic self-couplings, will be specified in Sec.~\ref{sec:DMDSNB} since they are not relevant at this point.

\item We test the compatibility of our setup by requiring the three down-quark masses $m_{d,s,b}$ and four parameters in the CKM matrix of Eq.~\eqref{eq:VCKMparam} to be compatible with experiment -- see Table~\ref{tab:quarkdata}. To achieve this, we employ a standard $\chi^2$-analysis, by minimizing the function
    \begin{align}
    \chi^2[\left(\mathbf{M}_d\right)_{\alpha \beta}]=\sum_i\dfrac{\{\mathcal{P}_i[\left(\mathbf{Y}_d\right)_{\alpha \beta}]-\mathcal{O}_i\}^2}{\sigma_i^2},
    \label{eq:chisq}
    \end{align}
with respect to the observables associated with each numerically generated $(\mathbf{M}_{Bd},m_B)$ pair (see the first bullet point). In the above, $\mathcal{O}_i$ and $\sigma_i$ denote the corresponding best-fit value and $1\sigma$ experimental uncertainty for the masses and CKM parameters. Moreover, $\mathcal{P}_i$ corresponds to the output value for the observable $i$ obtained by varying the nine real input parameters $\left(\mathbf{M}_d\right)_{\alpha \beta}$ ($\alpha,\beta=d,s,b$) namely, the entries of the SM down-quark mass matrix. Note that here we do not make use of the quark-seesaw approximation but, instead, we diagonalize numerically the full $4 \times 4$ mass matrix $\bm{\mathcal{M}}_d$. Thus, we extract the light quark mixing matrix~$\mathbf{V}^\prime$ and the heavy-light mixing (see Sec.~\ref{sec:flavor}) from $\bm{\mathcal{V}}_L^d$, i.e. via Eqs.~\eqref{eq:diagdownfullNB}, \eqref{eq:3x4} and~\eqref{eq:nonunitCKM}.

\end{itemize}

\begin{figure}[t!]
    \centering 
    \hspace{-1cm} \includegraphics[scale=0.45]{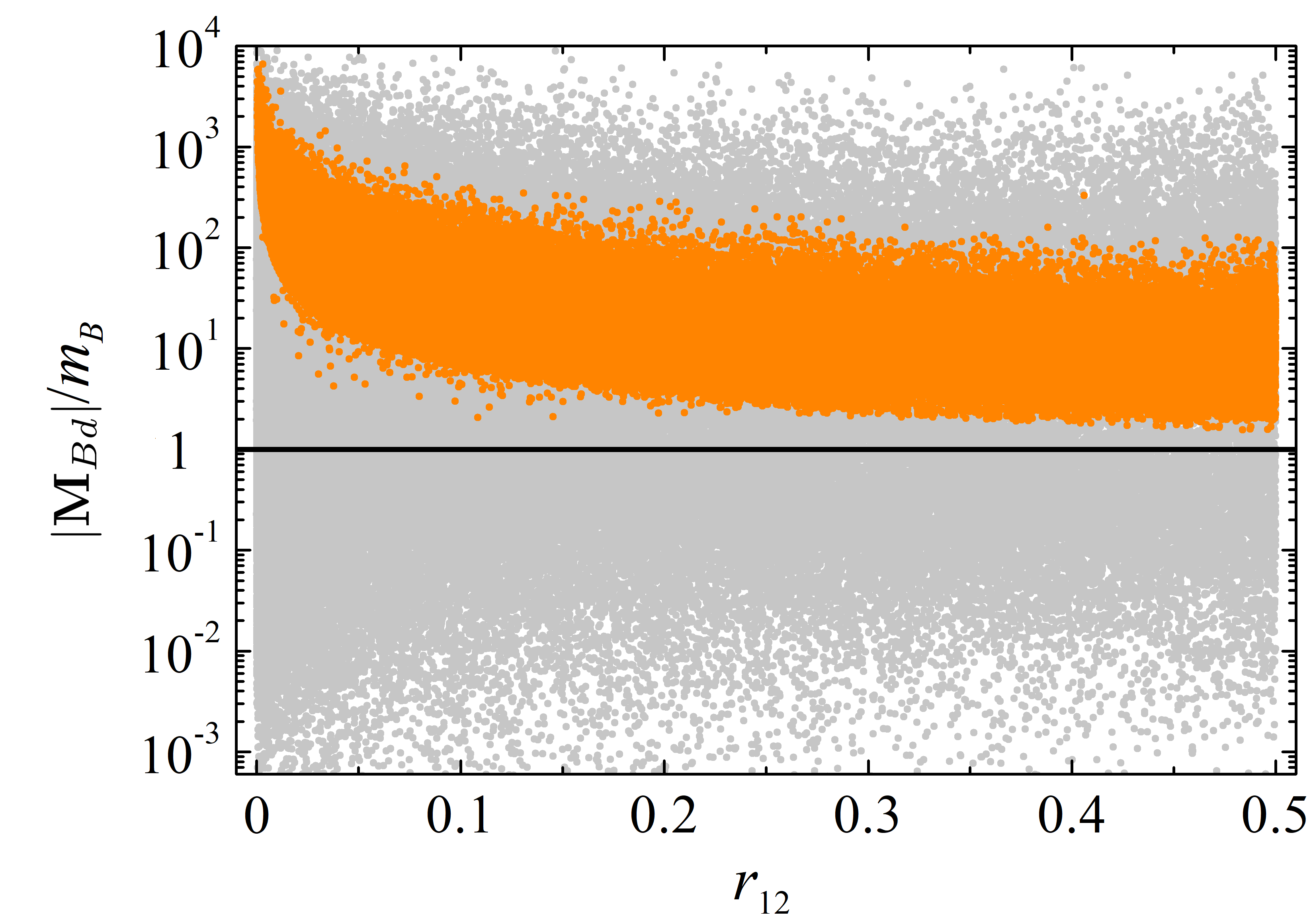} \\
    \includegraphics[scale=0.45]{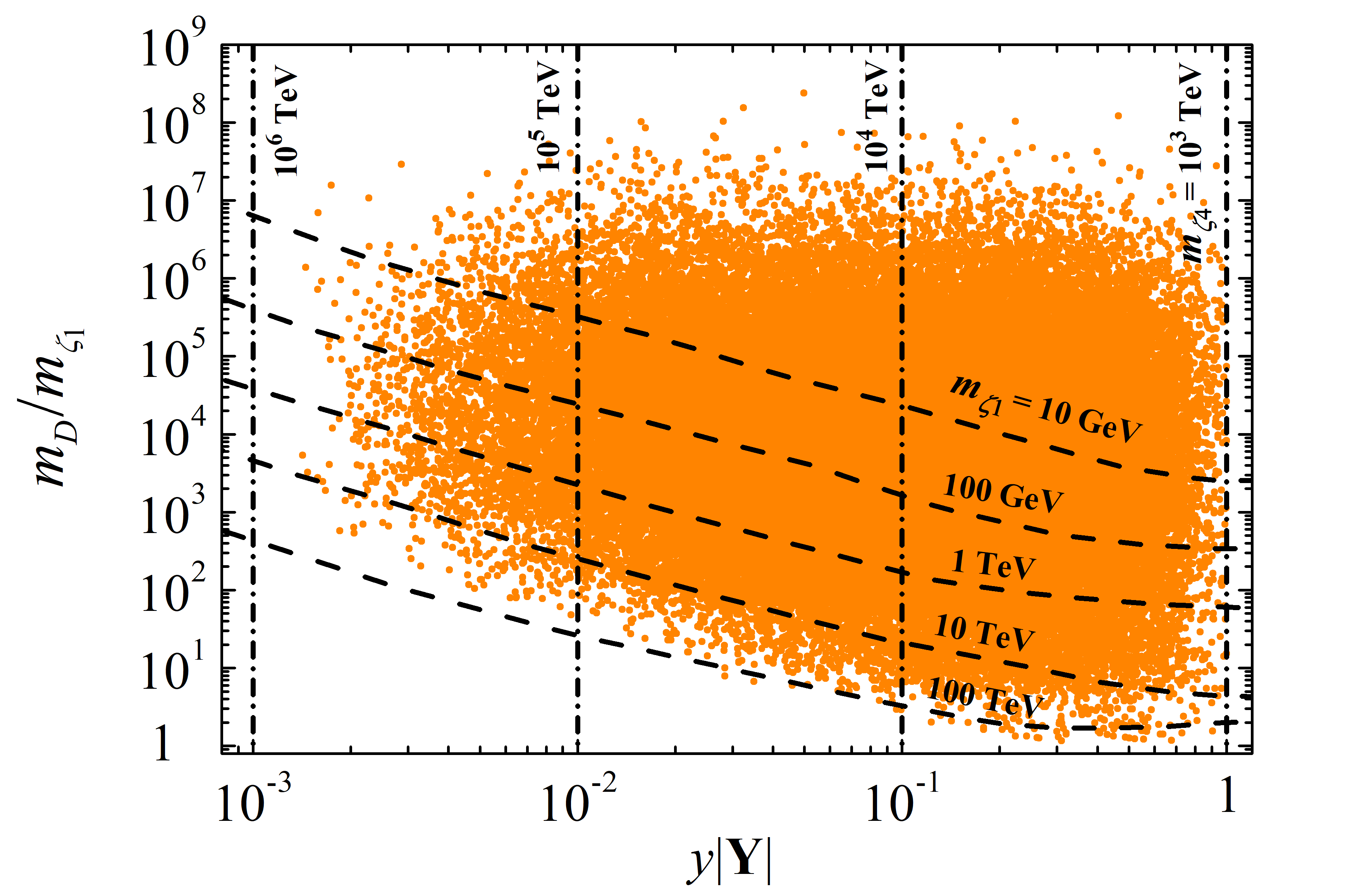}
    \caption{In the top we show $|\mathbf{M}_{Bd}|/m_B$ as a function of $r_{1 2}$ [see Eqs.~\eqref{eq:MBvsmB} and~\eqref{eq:r12}].
    In the bottom panel we plot $m_D/m_{\zeta_1}$ in terms of $y |\mathbf{Y}|$ (see text for details). The orange points are consistent with the quark sector observables at the $1 \sigma$ level and obey the LHC bound of Eq.~\eqref{eq:collider},
    whereas the gray fail either of these restrictions. In the bottom panel, the points above the dashed contours have $m_{\zeta_1}$ below the indicated values, while the same holds for those with $m_{\zeta_4}$ values to the right of the dash-dotted vertical lines.}  
    \label{fig:NBCKM}
\end{figure}
The result of the numerical procedure described above is displayed in Fig.~\ref{fig:NBCKM} where the top panel illustrates generic features of VLQ-NB models, while the bottom one highlights specific features of our framework. In the former, $|\mathbf{M}_{Bd}|/m_B$ is shown against $r_{1 2}$ being the orange points consistent with the quark mass and CKM parameters at the $1 \sigma$ level and with the collider bound in Eq.~\eqref{eq:collider}. The gray points do not fulfill one or both the requirements. We conclude that, consistently with the requirement obtained in the seesaw approximation of Eq.~\eqref{eq:MBvsmB} and with Ref.~\cite{Valenti:2021rdu}, to successfully generate a complex CKM matrix one needs $|\mathbf{M}_{Bd}|/m_B \gsim 2$. Also notice that for $r_{1 2} \gsim 0.3$ we have $2 \lsim |\mathbf{M}_{Bd}|/m_B \lsim 100$. Hence, the requirement in Eq.~\eqref{eq:r12} holds for $|\mathbf{M}_{Bd}| \gsim m_B$. However, as $r_{1 2} \to 0$ we need $|\mathbf{M}_{Bd}|/m_B \gg 1$ in order to obtain $\delta \sim \mathcal{O}(1)$. In the lower panel of the same figure, we show the ratio $m_D/m_{\zeta_1}$ as a function of $y |\mathbf{Y}|$, where we use as reference value for the dark VLQ mass scale the average $m_D = (m_{D_1} + m_{D_2})/2$ and $y |\mathbf{Y}| = (y_\chi |\mathbf{Y}_\xi| + y_\xi |\mathbf{Y}_\chi|)/2$. Above the dashed contours the lightest dark-scalar mass $m_{\zeta_1}$ lies below the indicated values.  The same holds to the right of the dash-dotted vertical lines for the heaviest dark-scalar mass $m_{\zeta_4}$. One can see that viable points lead to a spectrum $m_{\zeta_1} \ll m_{D_{1,2}}, m_{\zeta_{2,3,4}}$ with $m_{\zeta_1} < 100$~TeV, $10^3 \ \text{TeV} \lesssim m_{\zeta_4} \lesssim 10^6 \ \text{TeV}$ and $1<m_D/m_{\zeta_1}<10^8$. This yields a light scalar $\zeta_1$ which is a WIMP DM candidate (see Sec.~\ref{sec:DMDSNB}).

%%%%%%%%%%%%%%%%%%%%%%%%%%%%%%%%%%%%%%%%%%%%%%%%%%%%%%%%%%%%%%%%%%%%%%%%%%%%%
\subsection{Flavor constraints}
\label{sec:flavor}
%%%%%%%%%%%%%%%%%%%%%%%%%%%%%%%%%%%%%%%%%%%%%%%%%%%%%%%%%%%%%%%%%%%%%%%%%%%%%

%
\begin{figure}[t!]
   \centering
   \includegraphics[scale=0.45]{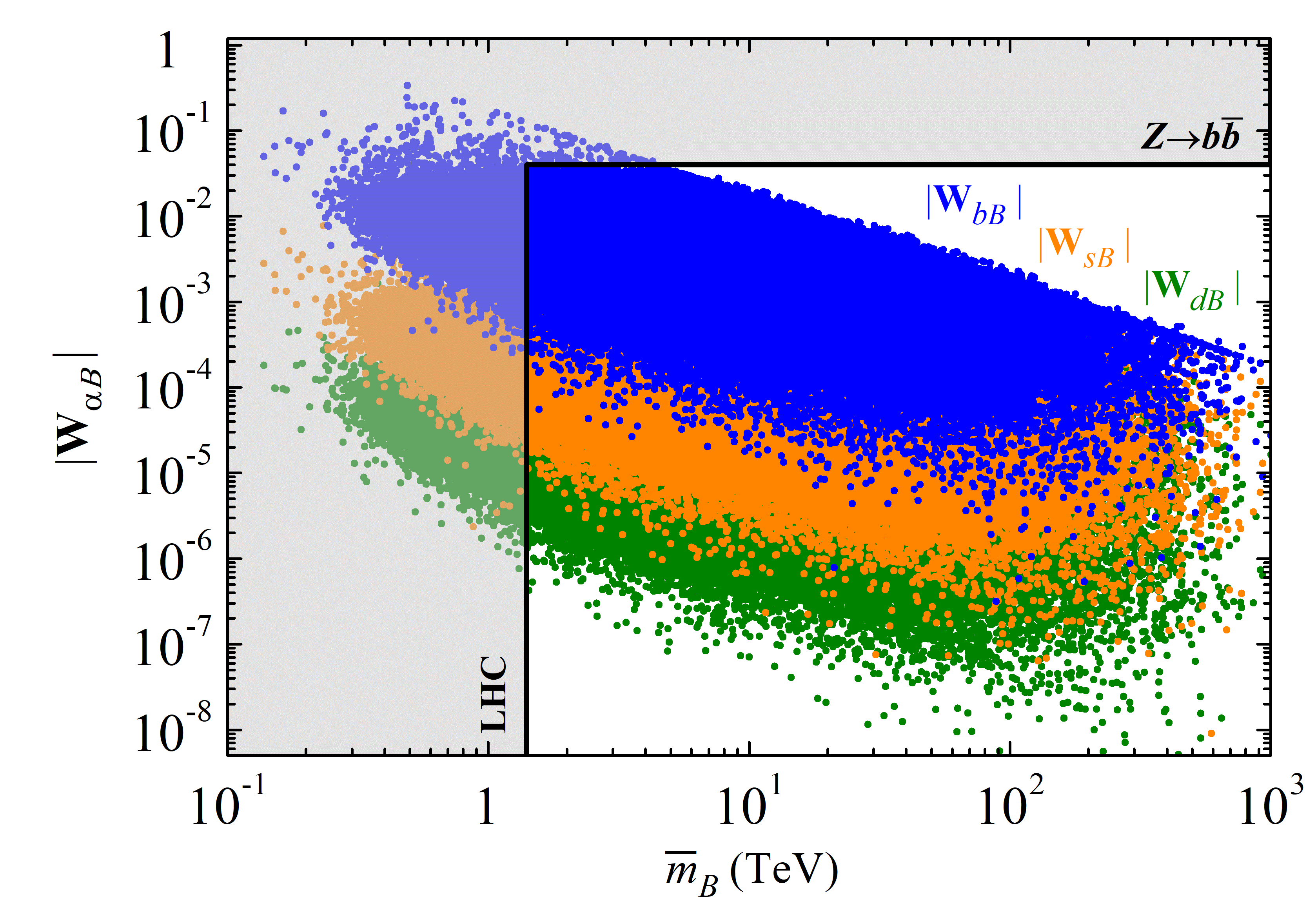}
    \caption{Heavy-light quark mixing $|\mathbf{W}_{\alpha B}|$ ($\alpha = d,s,b$) as a function of the VLQ mass $\overline{m}_B$. The blue (orange) [green] points refer to $|\mathbf{W}_{b B}|$ ($|\mathbf{W}_{s B}|$) [$|\mathbf{W}_{d B}|$]. All points are compatible with the quark sector observables at $1 \sigma$. The points in the gray shaded area are either excluded by the LHC bound of Eq.~\eqref{eq:collider} or $Z\to b \overline{b}$ constraint of Eq.~\eqref{eq:Zbb}.}
    \label{fig:flavor}
\end{figure}
The interactions between the down-type SM quarks and the $W/Z$ gauge bosons are modified due to the presence of the (non-dark) VLQ $B$ in our framework. Namely, the CC and NC interactions become,
\begin{align}
\mathcal{L}_{W^{\pm}} & = - \frac{g}{\sqrt{2}} W_{\mu}^{+} \overline{u} \left[\mathbf{V}^\prime \; \gamma^{\mu} P_{L}  d + \mathbf{V}_L^{u \dagger}\mathbf{W}_{B} \; \gamma^{\mu} P_{L}  B \right] + \text{H.c.} \; ,
\label{eq:Wint} \\
\mathcal{L}_{Z} & \supset \frac{g}{2 c_{W}}  Z_{\mu} \Big[ \overline{d} \mathbf{W}_{\text{light}}^{d \dagger} \mathbf{W}_{\text{light}}^d \gamma^{\mu} P_{R}  d  +  \overline{d} \mathbf{W}_{\text{light}}^{d \dagger} \mathbf{W}_B \gamma^{\mu} P_{R} B \nonumber \\
&+ \overline{B} \mathbf{W}_B^\dagger \mathbf{W}_{\text{light}}^d \gamma^{\mu} P_{R} d + \overline{B} \mathbf{W}_B^\dagger \mathbf{W}_B \gamma^{\mu} P_{R} B \Big] \; .
\label{eq:Zint}
\end{align}
Note that the above effects, are absent at tree-level in our model being only induced at one-loop level. The first term of Eq.~\eqref{eq:Wint} contains the non-unitary quark mixing matrix $\mathbf{V}^\prime$ of Eq.~\eqref{eq:nonunitCKM}, being deviations from unitarity encoded in $\boldsymbol{\eta}$ given in Eq.~\eqref{eq:eta}. The second term of $\mathcal{L}_{W^{\pm}}$ describes heavy-light quark mixing which, in the seesaw approximation, is given by Eq.~\eqref{eq:heavylightmixing}. FCNC in the SM quark sector are controlled by $\mathbf{W}_{\text{light}}^{d \dagger} \mathbf{W}_{\text{light}}^d$ in the first term of Eq.~\eqref{eq:Zint}. These are related to the deviations from unitarity as $\mathbf{W}_{\text{light}}^{d \dagger} \mathbf{W}_{\text{light}}^d \simeq \mathbb{1} - 2 \mathbf{V}^\dagger \boldsymbol{\eta} \mathbf{V}$. From the seesaw formulas [see Eqs.~\eqref{eq:heavylightmixing} and~\eqref{eq:eta}], we notice that the heavy-light mixing is hierarchical. Thus, constraints involving the heavier bottom quark will be the most relevant. Namely, we can restrict $\mathbf{V}_L^{u \dagger}\mathbf{W}_{B}$ (or equivalently $\mathbf{W}_{\text{light}}^{d \dagger} \mathbf{W}_{\text{light}}^d$ or $\boldsymbol{\eta}$) via the ratio $R_b= \Gamma(Z \to b \overline{b})/\Gamma(Z \to \text{hadrons})$~\cite{Aguilar-Saavedra:2013qpa}, by imposing~\cite{Cherchiglia:2020kut,Cherchiglia:2021vhe,Valenti:2021rdu},
\begin{equation}
|\mathbf{W}_{b B}| \lsim 0.04 \; .
\label{eq:Zbb}
\end{equation}
Using the results of the numerical analysis previously 
performed, we plot in Fig.~\ref{fig:flavor} the heavy-light quark mixing parameters $|\mathbf{W}_{\alpha B}|$ as a function of the VLQ mass $\overline{m}_B$. The blue, orange and green points correspond to $\alpha = b,s \ \text{and} \ d$, respectively, and are compatible with the quark mass and CKM parameter values at $1 \sigma$ [see Eq.~\eqref{eq:chisq}]. The points covered by the gray shaded area are either excluded by the LHC bound  \eqref{eq:collider} or by the $R_b$ constraint \eqref{eq:Zbb}. We notice that in our scenario the heavy-light quark mixing is hierarchical with $|\mathbf{W}_{d B}| \ll |\mathbf{W}_{s B}| \ll |\mathbf{W}_{b B}|$. Furthermore, if $\overline{m}_B$ is increased by one order of magnitude each $|\mathbf{W}_{\alpha B}|$ decreases by about two orders of magnitude, as expected from the seesaw formula of Eq.~\eqref{eq:heavylightmixing}. This stems from the $\mathcal{Z}_8$-symmetry of our model that imposes a specific form for the one-loop mass matrix $\bm{\mathcal{M}}_d$ of Eq.~\eqref{eq:Lmassoneloop}. Namely, the symmetry forbids $\overline{q_L} \Phi B_R$ terms which, in generic VLQ scenarios, induce FCNC that strongly constrain the model parameter space. However, since this term is forbidden, heavy-light mixing is hierarchical being proportional to CKM matrix entries and SM Yukawa couplings, being suppressed by the heaviness of $B$. For this reason, deviations from unitarity in our model are not sufficient to explain the CKM non-unitarity hint~\cite{ParticleDataGroup:2024cfk}, which has been addressed in generic VLQ-extensions~\cite{Branco:2021vhs,Botella:2021uxz}. This hierarchical mixing also justifies the direct application of the LHC bound of Eq.~\eqref{eq:collider}, in our framework. From the figure we remark that if $\overline{m}_B>1.4$ TeV is satisfied, $Z \to b \overline{b}$ has little constraining power, since only a very reduced portion of parameter space is excluded, up to $\overline{m}_B \lsim 7$ TeV.

Other observables stemming from precise measurements of rare kaon decays $K_L \to \mu^+ \mu^-$ and $K^+ \to \pi^+ \nu \overline{\nu}$; CP violation in the kaon system; neutral $B$-meson oscillations and rare $B$-meson decays like $B_s \to X_s \ell \overline{\ell}$, $B_{d,s} \to \mu^+ \mu^-$, $B_s \to X_s \gamma$; are usually able to constrain $\Delta F=1,2$ flavor transition operators~\cite{Aguilar-Saavedra:2002phh,Perez:2020dbw,Cherchiglia:2020kut,Valenti:2021rdu,Cherchiglia:2021vhe}. The latter are induced, at tree or loop level, by the presence of down-type VLQs, and depend on the matrix elements of $\mathbf{W}^\dagger_d \mathbf{W}_d$ that controls FCNC [see Eq.~\eqref{eq:Zint}]. However, since in our framework quark-mixing is hierarchical by construction, flavor-violating observables are far less relevant in constraining our parameter space. As shown in Refs.~\cite{Cherchiglia:2019gll,Perez:2020dbw,Cherchiglia:2020kut,Valenti:2021rdu,Cherchiglia:2021vhe}, this is a general feature of models with NB-type VLQs.

%%%%%%%%%%%%%%%%%%%%%%%%%%%%%%%%%%%%%%%%%%%%%%%%%%%%%%%%%%%%%%%%%%%%%%%%%%%%%
\subsection{Dark matter}
\label{sec:DMDSNB}
%%%%%%%%%%%%%%%%%%%%%%%%%%%%%%%%%%%%%%%%%%%%%%%%%%%%%%%%%%%%%%%%%%%%%%%%%%%%%

%
\begin{figure}[t!]
    \centering
    \includegraphics[scale=0.95]{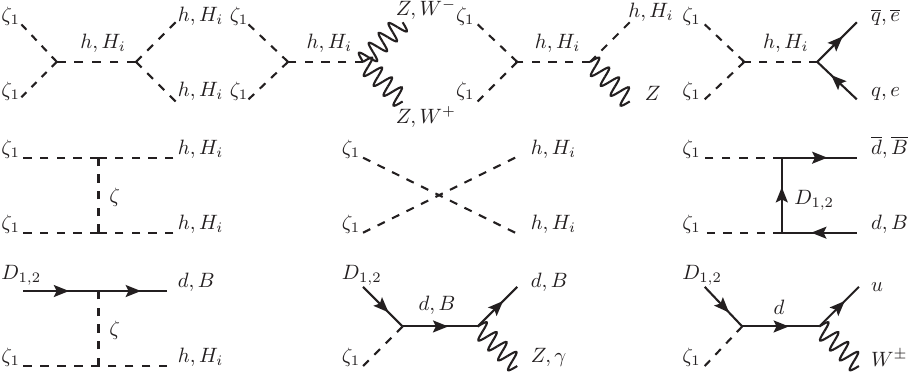}
    \caption{Tree-level DM annihilation and co-annihilation diagrams, where $i=1,2$. Co-annihilation diagrams between $\zeta_1$ and $\zeta_{j}$ ($j=2,3,4$) are obtained by replacing one initial $\zeta_1$ state by $\zeta_j$ in the~$\zeta_1-\zeta_1$ annihilation channels.}
\label{fig:diagsDM}
\end{figure}
Due to the~$\mathcal{Z}_8$ charge assignments given in Table~\ref{tab:modelDSNB}, the scalars $\chi,\xi$ and VLQs $D_{1,2}$ remain odd under a residual $\mathcal{Z}_2$ symmetry after SSB. The lightest among these states will be stable and, since DM must be non-baryonic and electrically neutral, the WIMP DM candidate will be the lightest dark-scalar $\zeta_1$ [see Eq.~\eqref{eq:mixneutralDM}]. In Fig.~\ref{fig:diagsDM} are depicted the DM (co)annihilation channels into even/SM particles that contribute to the thermally average cross section and consequently WIMP DM relic density (see discussion in Sec.~\ref{sec:DM}). In our relic density analysis we use the current observed value obtained by Planck~\cite{Planck:2018vyg} at the $3 \sigma$ level shown in Eq.~\eqref{eq:Oh2Planck}.

\begin{figure}[t!]
    \centering
    \includegraphics[scale=0.55]{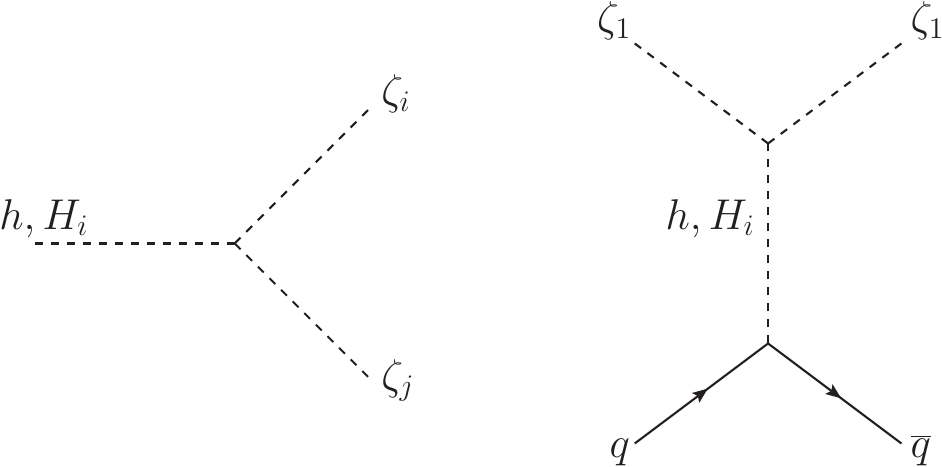}
    \caption{The left diagram contributes to the Higgs invisible decay [see Eq.~\eqref{eq:BRhinv}], while the diagram on the right corresponds to the $h$ and  $H_i$ ($i=1,2$) tree-level contributions to $\sigma^{\text{SI}}$ [see Eq.~\eqref{eq:sigmaSI} and discussion therein].}
    \label{fig:diagsinvDD}
\end{figure}
As discussed in Sec.~\ref{sec:beyondoneloop}, the $\lambda_{\Phi \sigma}$ quartic coupling must be small ($\lambda_{\Phi \sigma} \lsim v^2/v_\sigma^2 \lsim 10^{-8}$), suppressing $\Phi - \sigma$ mixing. Thus, we can safely work in the limit $\lambda_{\Phi \sigma}=0$, for which the Higgs boson $h$ coupling to $\zeta_i$ ($i=1, \cdots, 4$), is given by 
 \begin{equation}
g_{h  i j} = \frac{\lambda_{\Phi \chi}}{2} \left(\mathbf{V}_{1 i} \mathbf{V}_{1 j} + \mathbf{V}_{3 i} \mathbf{V}_{3 j} \right) + \frac{\lambda_{\Phi \xi}}{2} \left(\mathbf{V}_{2 i} \mathbf{V}_{2 j} + \mathbf{V}_{4 i} \mathbf{V}_{4 j}\right) \; , \;
    \end{equation}
where $\mathbf{V}$ is the dark-scalar mixing matrix of Eq.~\eqref{eq:mixneutralDM}, $\lambda_{\Phi \chi}$ and $\lambda_{\Phi \xi}$ are defined in Eq.~\eqref{eq:Vpotential} and $g_{h  1 1}$ is the Higgs-DM coupling. The $g_{h i j}$ play a crucial role in constraining the WIMP DM parameter space, since observables like Higgs invisible decay and WIMP-nucleon SI elastic cross-section $\sigma^{\text{SI}}$, depend strongly on it (see discussion in Sec.~\ref{sec:DM}). Namely, for $m_{\zeta_i} < m_h/2\sim 62.6$~GeV, the dark-scalars contribute to the Higgs invisible decay as shown in the left diagram of Fig.~\ref{fig:diagsinvDD}. The corresponding decay width is,
    \begin{align}
    \Gamma( h \to \zeta_i \zeta_j) = (1+ \delta_{i j})^2 \frac{v^2  g_{h  i j}^2 }{32 \pi m_h} \sqrt{ \frac{(m_h^2 + m_{\zeta_i}^2 - m_{\zeta_j}^2)^2}{m_h^4} - \frac{4 m_{\zeta_i}^2}{m_h^2}} \; ,
    \label{eq:BRhinv}
    \end{align}
    with the BR ratio computed via Eq.~\eqref{eq:BoundBRh1inv_1} where $\Gamma_{h}^{\text{SM}} = 3.2^{+2.8}_{-2.2} \; \text{MeV}$ is the SM Higgs decay width~\cite{ParticleDataGroup:2024cfk}. Current LHC Higgs data imposes the bound of Eq.~\eqref{eq:BoundBRh1inv}. Furthermore, in Fig.~\ref{fig:diagsinvDD} we also show the tree-level diagrams (mediated by even-scalars) that contribute to $\sigma^{\text{SI}}$. Note that, since $\Phi$ and $\sigma$ do not mix, only the SM Higgs $h$ will contribute to $\sigma^{\text{SI}}$, for which we have
\begin{equation}
\sigma^{\text{SI}} = \frac{v^2 g_{h  1 1}^2  f_N^2}{\pi m_h^4} \frac{m_N^2}{(m_N + m_{\zeta_1})^2} \; , 
\label{eq:sigmaSI}
\end{equation}
being $m_N$ and $f_N$ the average nucleon mass and the nucleon form factor, respectively. The typical value used by the \texttt{micrOMEGAs} package~\cite{Belanger:2008sj} is $f_N\simeq 1.8\times10^{-3}$. Current limits on~$\sigma^{\text{SI}}$ come from DD experiments as discussed in Sec.~\ref{sec:DM}.

\begin{figure}[t!]
    \centering
    \hspace{-1cm} \includegraphics[scale=0.45]{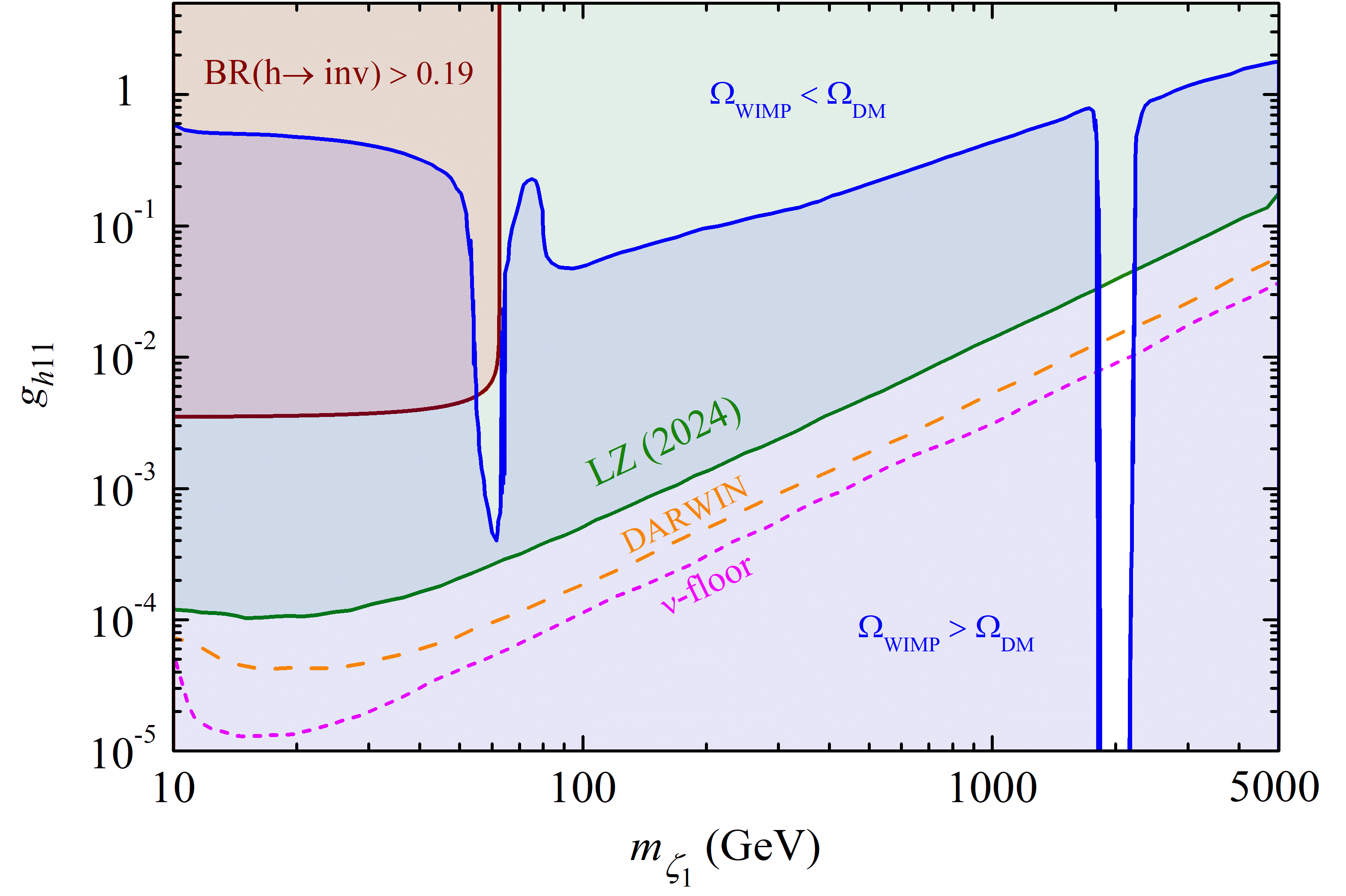}
    \caption{Higgs-DM coupling $g_{h 1 1}$ as a function of the DM mass $m_{\zeta_1}$. Along the blue contour the DM relic density lies in the Planck $3\sigma$ range [see Eq.~\eqref{eq:Oh2Planck}]. The blue shaded region below that contour is ruled out since it leads to overabundant DM. The green shaded region is excluded by the WIMP DD experiment LZ~\cite{LZ:2022lsv,LZ:2024zvo}. The orange-dashed contour indicates the projected sensitivity for DARWIN~\cite{DARWIN:2016hyl}. The pink-dashed line is the "neutrino floor"~\cite{Billard:2013qya}. The bordeau-shaded region is excluded by the LHC bound on the Higgs invisible decay [see Eq.~\eqref{eq:BoundBRh1inv}].}
    \label{fig:DMplot}
\end{figure}
To carry out the DM analysis in the present framework, once again as was done in Sec.~\ref{sec:darkLSS}, we make use of state-of-the art high-energy-physics numerical tools. Namely, we implement our model in \texttt{SARAH}~\cite{Staub:2013tta} and use \texttt{SPheno}~\cite{Porod:2003um} and \texttt{FlavorKit}~\cite{Porod:2014xia} to compute particle masses, mixing, vertices, tadpole equations and BRs. The relic density $\Omega h^2$ and WIMP-nucleon SI elastic scattering cross-section $\sigma^{\text{SI}}$ are computed at tree-level by the \texttt{micrOMEGAs}~\cite{Belanger:2014vza} package. We use \texttt{SSP 1.2.5}~\cite{Staub:2011dp} to establish the link between the aforementioned numerical tools. The parameters of our model vary within the ranges shown in Table~\ref{tab:CKMscan}. For the dark-scalar self-couplings we choose $\lambda_\chi = \lambda_\xi = \lambda_{\chi \xi} = 0.5$. Besides the SM Higgs boson $h$, we have two additional non-dark scalars $H_{1,2}$ for which we set the masses to be $(m_{H_1},m_{H_2}) = (1,4)$~TeV. The latter leads to $\lambda_{\sigma}^{(\prime)} \sim \mathcal{O}(10^{-6})$. As motivated in Sec.~\ref{sec:CKMpheno}, we consider a benchmark hierarchical dark-scalar mass spectrum with $m_{\zeta_1} \sim 1 \,\text{TeV} \ll m_{\zeta_{2,3,4}} \sim (10^3-10^6)$~TeV, and $10\lsim m_{D_{1,2}}\lsim10^8$~TeV for the odd VLQs. In such case, DM coannihilation channels are not relevant. 

Our results are presented in Fig.~\ref{fig:DMplot} where we show the DM constraints in the $(g_{h 1 1},m_{\zeta_1})$ plain, i.e. Higgs-DM coupling vs. DM mass. The LHC constraint on $\BR(h \rightarrow \text{inv})$ of Eq.~\eqref{eq:BRhinv} excludes the bordeau-shaded region with $m_{\zeta_1} \lsim m_h/2$ and $g_{h h 1 1} \gtrsim 5 \times 10^{-3}$. The most stringent constraint comes from the current LZ DD exclusion region shown in green. The correct DM relic density [see Eq.~\eqref{eq:Oh2Planck}] is obtained along the blue solid line which features two dips. The first occurs around $m_{\zeta_1} \sim m_h/2 \simeq 62.6$~GeV due to $s$-channel DM annihilation mediated by the SM Higgs, i.e. $\zeta_1 \zeta_1 \to h^\ast \to \text{nonDM} \ \text{nonDM}$ (top diagrams in Fig.~\ref{fig:diagsDM}). This is typical from the simplest DM model which extends the SM with a singlet dark-scalar~\cite{McDonald:1993ex,Guo:2010hq,Cline:2013gha,Feng:2014vea,Wu:2016mbe,GAMBIT:2017gge,Casas:2017jjg}. It is clear that the most recent, 2024, DD bound from LZ (green shaded region), excludes the minimal scalar-singlet WIMP DM model. However, our framework differs from this simple setup in the sense that the complex singlet $\sigma$ acquires a non-zero VEV and, thus, there are new even scalars $H_{1,2}$ interacting with the DM particle $\zeta_1$. The second dip in the relic density curve occurs via $s$-channel DM annihilation mediated by $H_2$ (see upper leftmost diagram in Fig.~\ref{fig:diagsDM}) at $m_{\zeta_1} \sim m_{H_2}/2 \simeq 2$~TeV. Our dark-sector can be probed by future DD experiments such as DARWIN~\cite{DARWIN:2016hyl} (orange-dashed contour) up to the so-called "neutrino floor"~\cite{Billard:2013qya}.

%%%%%%%%%%%%%%%%%%%%%%%%%%%%%%%%%%%%%%%%%%%%%%%%%%%%%%%%%%%%%%%%%%%%%%%%%%%%%
\subsection{Key ideas and outlook}
%%%%%%%%%%%%%%%%%%%%%%%%%%%%%%%%%%%%%%%%%%%%%%%%%%%%%%%%%%%%%%%%%%%%%%%%%%%%%

In this section, we propose a new solution to the strong CP problem based on the existence of a dark sector containing a viable (scalar) WIMP DM candidate, as seen in Fig.~\ref{fig:DMplot}. In our NB-inspired mechanism, a $\mathcal{Z}_8$ symmetry allows for SCPV while leaving a residual $\mathcal{Z}_2$ to stabilize DM. A complex CKM matrix arises from one-loop corrections to the quark mass matrix mediated by the dark sector -- see Fig.~\ref{fig:oneloopdiags}. In contrast with other proposals, such as the minimal BBP model, here the strong CP phase receives non-zero contributions only at two loops, enhancing naturalness.

Our setup can be embedded in a more general framework aiming at addressing other drawbacks of the SM, besides the strong CP problem and DM (see Sec.~\ref{sec:BSMavenues}). For instance, the VEV of the complex scalar singlet $\sigma$ could be responsible for generating neutrino masses, inducing simultaneously low-energy CP violation in the lepton mixing matrix~\cite{Barreiros:2020gxu}. Additionally, NB constructions naturally incorporate key ingredients for a viable model of EW baryogenesis (see Sec.~\ref{sec:BAU}) -- a new source of CP violation and modifications to the Higgs potential -- induced by the complex scalar singlet can enable a strong first-order EW phase transition~\cite{Cohen:1990py,Cohen:1990it,Nelson:1991ab,Cohen:1993nk,Branco:2001gg}. Alternatively, the scalar singlet besides generating massive neutrinos could play a key role in creating the lepton asymmetry required for leptogenesis as studied in Sec.~\ref{sec:lepto} -- see Refs.~\cite{Barreiros:2022fpi,Murgui:2025scx}. Lastly, this scalar singlet can potentially also address inflation~\cite{Boucenna:2014uma}.

Overall, the work developed in this section opens a window for interesting studies where a dark sector provides a unique solution to several open questions in (astro)particle physics and cosmology.

%------------ 
% CHAPTER 04   
%------------ 

%%%%%%%%%%%%%%%%%%%%%%%%%%%%%%%%%%%%%%%%%%%%%%%%%%%%%%%%%%%%%%%%%%%%%%%%%%%%%
\chapter{Axions} 
\label{chpt:axions}
%%%%%%%%%%%%%%%%%%%%%%%%%%%%%%%%%%%%%%%%%%%%%%%%%%%%%%%%%%%%%%%%%%%%%%%%%%%%%

The idea proposed by Peccei and Quinn~(PQ)~\cite{Peccei:1977hh,Peccei:1977ur}, to solve the strong CP problem -- see Sec.~\ref{sec:strongCPnedmsol} -- consists of adding a $\text{U}(1)_{\text{PQ}}$ global symmetry, under which additional scalars and SM fermions have a charge, such that this new symmetry is classically conserved but anomalous. The explicit breaking of $\text{U}(1)_{\text{PQ}}$ at the quantum level will lead to a contribution of the same form as the $\overline{\theta}$-term -- see Eq.~\eqref{eq:topological} and discussion in Sec.~\ref{sec:strongCP}. Therefore, the transformations of the fields under $\text{U}(1)_{\text{PQ}}$ would allow to reabsorb the $\overline{\theta}$ parameter rendering it unphysical. After this initial proposal, Weinberg and Wilczek~(WW)~\cite{Weinberg:1977ma,Wilczek:1977pj}, pointed out that if the $\text{U}(1)_{\text{PQ}}$ is spontaneously broken, say at a scale $f_a$, it will give rise to a GB. Since this symmetry is also anomalously broken, the GB will gain a non-zero mass at loop level making it a pseudo-GB. The axion particle $a$, corresponds to the pseudo-GB of the $\text{U}(1)_{\text{PQ}}$ symmetry. By preforming an adequate chiral rotation, the $\overline{\theta}$ parameter can be promoted into a dynamical field $\overline{\theta} \rightarrow \overline{\theta} + a/f_a$. Consequently $f_a$ is known as the axion scale or decay constant. In this scenario, the effective minimum energy QCD-vacuum configuration will be:
    \begin{equation}
        \overline{\theta}_{\text{eff}} = \overline{\theta} + \left< \frac{a}{f_a} \right> = 0 \; .
    \end{equation}
    Axions are the excitations around the $\overline{\theta}_{\text{eff}} = 0$ vacuum. This can be verified by explicitly computing, through chiral Lagrangian techniques, the axion potential $V(a)$ and showing that the absolute minimum is located at $\left<a\right> = 0$. Hence, the idea of the PQ solution to the strong CP problem relies on introducing a PQ symmetry preserved by all terms in the Lagrangian except by the axion coupling to the QCD topological term $a G \tilde{G}$. By appropriately shifting the axion field, the $\overline{\theta}$ term is removed from the Lagrangian. 
    
    This chapter provides a comprehensive review of the axion solution to the strong CP problem focusing on the theoretical properties of axions and their phenomenological implications. Numerous excellent reviews, textbooks, and lecture notes exist on this topic, see e.g. Refs.~\cite{Peskin:1995ev,Schwartz:2014sze,Coleman:1985rnk,Srednicki:2007qs,DiLuzio:2020wdo,QuilezLasanta:2019wgl,Irastorza:2018dyq,Adams:2022pbo,link1,link2}, and the present discussion is intended to complement and build upon them. The structure of the chapter is as follows. Sec.~\ref{sec:properties} reviews the properties of QCD axion by analyzing the effective axion Lagrangian, understanding its potential and couplings to gluons, photons and fermions. Next, in Sec.~\ref{sec:models}, the paradigmatic QCD axion models are presented. Sec.~\ref{sec:axionDMcosmo} shows how axions are an excellent DM candidate which can be produced via the so-called misalignment mechanism and discusses the cosmology of pre- and post-inflationary axion DM. Sec.~\ref{sec:experimental} reviews the experimental landscape, including current searches for axions and ALPs, and outlines future directions. Lastly, Sec.~\ref{sec:flavouredaxions} briefly discusses flavor-violating axions which have a rich phenomenology.

%%%%%%%%%%%%%%%%%%%%%%%%%%%%%%%%%%%%%%%%%%%%%%%%%%%%%%%%%%%%%%%%%%%%%%%%%%%%%
\section{Properties of the QCD axion}
\label{sec:properties}
%%%%%%%%%%%%%%%%%%%%%%%%%%%%%%%%%%%%%%%%%%%%%%%%%%%%%%%%%%%%%%%%%%%%%%%%%%%%%

The low-energy effective axion Lagrangian $\mathcal{L}_a^{\text{eff}}$ can be obtained through EFT and/or the PQ current. Starting with the EFT approach, one needs to write down non-renormalizable operators that respect the symmetries of the theory, which in this case besides the SM gauge group $\mathcal{L}_a^{\text{eff}}$ must also be invariant under a PQ type symmetry. Due to the PQ shift property of the axion, its coupling to the SM will either anomalous or derivative. Below the EW scale one has,
\begin{equation}
\mathcal{L}_a^{\text{eff}} = \frac{1}{2} \partial_\mu a \partial^\mu a + \frac{1}{4} g_{a \gamma \gamma}^0 a F \widetilde{F} + \frac{1}{4} g_{a g g}^0 a G \widetilde{G} + \frac{\partial_\mu a}{f_a} \sum_f \frac{c_f^0}{2} \left( \overline{\psi_f} \gamma^\mu \gamma_5 \psi_f\right) \; ,
\label{eq:EFT}
\end{equation}
where $\psi_f$ denotes fermion fields and the axion coupling to photons, gluons and fermions are given by $g_{a \gamma \gamma}^0$, $g_{a g g}^0$ and $c_f^0$, respectively.

Alternatively $\mathcal{L}_a^{\text{eff}}$ can be derived from the $J^\mu_{\text{PQ}}$ current associated to the spontaneously broken $\text{U}(1)_{\text{PQ}}$ symmetry,
\begin{equation}
    J^\mu_{\text{PQ}} \supset f_{\text{PQ}} \partial^\mu a + \sum_f \chi_f \left( \overline{\psi_f} \gamma^\mu \psi_f\right) \, , 
    \label{eq:genPQcurrent}
\end{equation}
where $f_{\text{PQ}}$ is the scale at which the PQ symmetry is spontaneously broken. The above current is conserved apart from anomalies as shown in Sec.~\ref{sec:strongCP}, namely,
\begin{equation}
    \partial_\mu J^\mu_{\text{PQ}} = - \frac{\alpha_e}{8 \pi}   \frac{E}{f_{\text{PQ}}} F \widetilde{F} -  \frac{\alpha_s}{8 \pi} \frac{N}{f_{\text{PQ}}} G \widetilde{G} \; ,
    \label{eq:PQcurrent}
\end{equation}
with $E$ and $N$ being the EM and color anomaly factor, respectively. These depend on the PQ charges associated to fermion fields that run in the loop generating the anomaly and therefore are model-dependent quantities. Namely,
\begin{equation}
    E = 2 \sum_f \left(\chi_L^f - \chi_R^f\right) q^2_f \; , \; N = 2 \sum_f \left(\chi_L^f - \chi_R^f\right) T(R_f) \; , 
\label{eq:EN}
\end{equation}
where for a fermion $\psi_f$ its PQ and electric charge are respectively given by $\chi_f$ and $q_f$. Additionally, $\psi_f$ transforms with representation $R_f$ under SU(3)$_c$ with associated Dynkin index $T(R_f)$. In order to have a viable axion solution to the strong CP problem, the color anomaly factor $N$ must be non-vanishing, so that the axion couples to gluons via $G \tilde{G}$. By inserting the expression for $J^\mu_{\text{PQ}}$ in Eq.~\eqref{eq:PQcurrent} one obtains the axion EOM which are analogously extracted from the following effective Lagrangian,
\begin{equation}
\mathcal{L}_a^{\text{eff}} = \frac{1}{2} \partial_\mu a \partial^\mu a - \frac{\alpha_e}{8 \pi}  \frac{E}{N} \frac{a}{f_a} F \widetilde{F} - \frac{\alpha_s}{8 \pi} \frac{a}{f_a} G \widetilde{G} + \frac{\partial_\mu a}{f_a} \sum_f \frac{\chi_f}{N} \left( \overline{\psi_f} \gamma^\mu \psi_f\right) \; ,
\label{eq:Lacurrent}
\end{equation}
where the axion decay constant $f_a$ is related to $f_{\text{PQ}}$ as follows,
\begin{equation}
    f_a = \frac{f_{\text{PQ}}}{N} \; .
    \label{eq:axionfa}
\end{equation}
The couplings in Eq.~\eqref{eq:Lacurrent} above are related to the ones of Eq.~\eqref{eq:EFT}:
\begin{align}
    g_{a \gamma \gamma}^0 = - \frac{\alpha_e}{2 \pi} \frac{E}{N} \frac{1}{f_a} \; , \; g_{a g g}^0 = - \frac{\alpha_s}{2 \pi} \frac{1}{f_a} \; , \; \frac{c_f^0}{2} = \frac{1}{N} \left(\chi^f_L - \chi^f_R\right)\; .
    \label{eq:couplingsmd}
\end{align}
Note that, the above expressions are model-dependent contributions to the couplings, for a given PQ symmetry the charges of the fields are specified allowing, through Eqs.~\eqref{eq:EN} and~\eqref{eq:couplingsmd}, to determine $E$, $N$, $g_{a \gamma \gamma}^0$, $g_{a g g}^0$ and $c_f^0$.

The QCD axion is a pseudo-GB associated to the spontaneous breaking of a U($1)_{\text{PQ}}$ symmetry which is approximate since it is broken by the anomalous term $a G \tilde{G}$. The latter occurs in the confining QCD group due to its non-perturbative dynamics. As seen previously in Sec.~\ref{sec:strongCP}, the QCD vacuum structure allows for instanton solutions that generate the topological term $G \tilde{G}$. The axion is not massless since through the $a G \tilde{G}$ coupling will be generated the axion potential and mass, being a robust prediction for the QCD axion. In order to derive these model-independent properties of $a$ one needs to resort to chiral perturbation theory~($\chi$PT) to obtain the axion chiral Lagrangian. The term $a G \tilde{G}$ from $\mathcal{L}_a^{\text{eff}}$ in Eq.~\eqref{eq:EFT} can be eliminated by means of the following anomalous axial transformation of the quark fields,
\begin{equation}
    q \rightarrow \exp \left( i \gamma_5 \frac{a}{2 f_a} Q_a \right) q \Rightarrow \mathcal{L}_a^{\text{eff}} \supset \text{Tr}\left(Q_a\right) \frac{\alpha_s}{8 \pi} \frac{a}{f_a} G \widetilde{G} \; ,
\end{equation}
where $Q_a$ is a generic matrix acting on quarks generating an anomalous term which by setting $\text{Tr}\left(Q_a\right)=1$ leads to,
\begin{equation}
    \mathcal{L}_a^{\text{eff}} \supset \frac{1}{2} \partial_\mu a \partial^\mu a + \frac{1}{4} g_{a \gamma \gamma} a F \widetilde{F} + \frac{\partial_\mu a}{2 f_a}  \left( \overline{q} c_q \gamma^\mu \gamma_5 q\right) - \overline{q_L} M_a q_R + \text{H.c.}\; .
\end{equation}
Note that, the $a G \tilde{G}$ was exactly canceled by the quark field transformation above. However, this will affect the remaining terms in the Lagrangian making the couplings dependent on the axion-field and/or $Q_a$. Namely,
\begin{align}
    g_{a \gamma \gamma} &= g_{a \gamma \gamma}^0 - 2 N_c \frac{\alpha_e}{2 \pi f_a} \text{Tr}\left(Q_a Q^2\right) \; , \; c_q = c_q^0 - Q_a  \; , \nonumber \\  M_a &= \exp \left( i \gamma_5 \frac{a}{2 f_a} Q_a \right) M \exp \left( i \gamma_5 \frac{a}{2 f_a} Q_a \right) \; ,
    \label{eq:rotations}
\end{align}
where $N_c = 3$ denotes the number of quark colors. For simplicity, it is enough to consider only the up and down quarks, $q^T = (u,d)$, with 
$Q = \mathrm{diag}(2/3,-1/3)$ and $M = \mathrm{diag}(m_u,m_d)$ denoting the charge and mass matrices, respectively. The non-derivative part of the chiral Lagrangian~$\mathcal{L}_a^{\chi\text{PT}}$, that includes axions and pions, is given at leading order (LO) by~\cite{Srednicki:1985xd,Kaplan:1985dv},
\begin{equation}
    \mathcal{L}_a^{\chi\text{PT}} \supset B_0 \frac{f_\pi^2}{2} \text{Tr}\left(U M_a^\dagger + M_a U^\dagger\right) \; , \; U = e^{i \Pi/f_\pi} \; , \; \Pi = \begin{pmatrix} \pi^0 & \sqrt{2} \pi^+ \\ \sqrt{2} \pi^- &  \pi^0 \end{pmatrix} \; ,
\end{equation}
where $f_\pi = 92.21(14)\,\text{MeV}$~\cite{Zyla:2020zbs} is the pion decay constant. In addition, $B_0$ can be expressed in terms of the QCD quark condensate as
$\left<\overline{q} q\right>$ as $B_0 f_\pi^2 = -2 \left<\overline{q} q\right> \sim \Lambda_{\text{QCD}}^3$,
with $\Lambda_{\text{QCD}}$ denoting the QCD confinement scale. Setting $Q_a = \tfrac{1}{2} \mathrm{diag}(1,1)$, $U$ is expanded in a Taylor series using the expression for $M_a$ in Eq.~\eqref{eq:rotations}, and the resulting expansion is then substituted back into $\mathcal{L}_a^{\chi\text{PT}}$. Hence, at LO considering only the neutral pion $\pi^0$ one obtains,
\begin{align}
V(a,\pi^0) &= - m_\pi^2 f_\pi^2 \sqrt{1- \frac{4 m_u m_d}{(m_u+m_d)^2} \sin^2\left(\frac{a}{2 f_a} \right)} \cos \left(\frac{\pi^0}{f_\pi} - \phi_a \right) \; , \nonumber \\
\tan \phi_a &= \frac{m_u -m_d}{m_u+m_d} \tan \left(\frac{a}{2 f_a} \right) \; .
\end{align}
The above is the axion-pion potential at LO in $\chi$PT
whose absolute minimum is $(a, \pi^0) =(0,0)$.

In the pion ground state $\pi^0 = \phi_a f_\pi$ the axion effective potential is,
\begin{figure}[t!]
    \centering
    \includegraphics[scale=0.75]{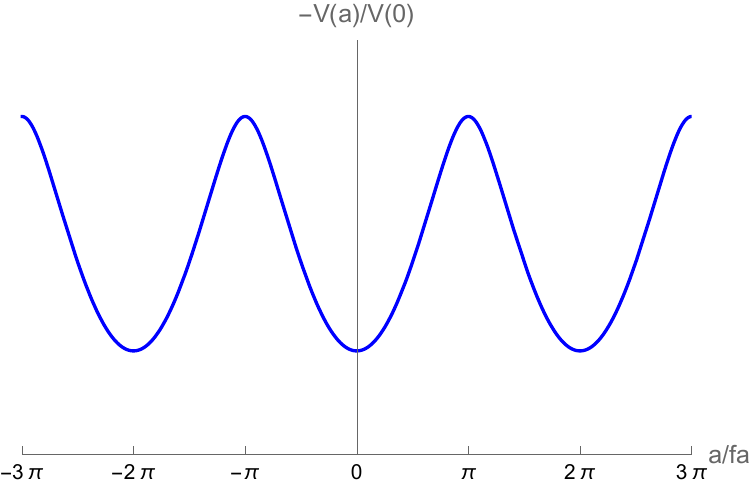}
    \caption{Schematic plot of the axion effective potential $V(a)$ whose expression is given in Eq.~\eqref{eq:Va}.}
    \label{fig:Va}
\end{figure}
\begin{equation}
V(a) = - m_\pi^2 f_\pi^2 \sqrt{1- \frac{4 m_u m_d}{(m_u+m_d)^2} \sin^2\left(\frac{a}{2 f_a} \right)} \; .
\label{eq:Va}
\end{equation}
From Fig.~\ref{fig:Va} one can see that $\left<a\right> = 0$ is indeed the global minimum of the potential. Expanding $V(a)$ around this minimum vacuum configuration leads to the QCD axion mass:
\begin{equation}
    m_a^2 = \frac{m_\pi^2 f_\pi^2}{f_a^2} \frac{m_u m_d}{(m_u+m_d)^2} \; , \; m_a = 5.70(7) \left(\frac{10^{12} \text{GeV}}{f_a}\right) \mu \text{eV} \; ,
    \label{eq:axionmass}
\end{equation}
where the right-hand side expression is the up-to-date next-to-LO~(NLO) calculation of $m_a$~\cite{GrillidiCortona:2015jxo}. The above relation between the axion mass and scale is a model-independent prediction of QCD axion models where the PQ symmetry is only broken by the QCD instantons generating the $G \tilde{G}$ term. By making the choice $Q_a = \frac{1}{2} \text{diag}(1, 1)$ there is axion-pion mass mixing resulting in the pion ground state $\pi^0 = \phi_a f_\pi$. For our purposes, this was enough to derive the axion potential and mass shown above. Note however, that in a more complete picture the axion $a$, neutral $\pi^0$ and the $\eta^0$ scalar mix among themselves. This remark is crucial since it clarifies the above dependence of $m_a$ with the quark masses $m_{u,d}$. Naively one may think that the axion mass should be around the QCD confinement scale $\Lambda_{\text{QCD}}$. However, the breaking of the PQ symmetry through non-perturbative instantons will induce a mixing with the U($1$)$_{\text{A}}$ axial current. In fact, there are two pseudoscalars having anomalous couplings to QCD $a$ and $\eta^\prime$ resulting from the $a-\eta^0$ mixing below the confinement scale. Therefore the axion mass dependence on the product of the quark masses is expected since in the chiral limit~$m_{u,d} \rightarrow 0$ leads to $a$ being a GB of the U($1$)$_{\text{A}}$ axial symmetry of QCD -- see Sec.~\ref{sec:strongCP}. The understanding of the fact that the U($1$)$_{\text{A}}$ symmetry is actually explicitly broken by the QCD anomaly solved what is known as the missing meson problem~\cite{Weinberg:1975ui}.

If one instead chooses $Q_a=M_q^{-1}/\text{Tr}M_q^{-1}$, thereby avoiding axion--pion mass mixing, the physical axion-to-photon coupling can be directly computed from Eq.~\eqref{eq:rotations}, which at LO in \(\chi\)PT leads to
\begin{equation}
    g_{a \gamma \gamma} = \frac{\alpha_e}{2 \pi f_a} \left(\frac{E}{N} - \frac{2}{3} \frac{m_u + 4 m_d}{m_u + m_d}  \right) \; ,
    \label{eq:axioncouplingsgagg}
\end{equation}
where it is clear that in QCD axion models there is always a model-dependent and independent contribution to the axion coupling to SM fields. In order to compute the axion-nucleon (proton and neutron) couplings one needs to make use of non-relativistic theory of nucleons using information on couplings extracted from numerical lattice simulations~\cite{GrillidiCortona:2015jxo}. As for the axion-electron coupling one needs to take into consideration the renormalization group equation effect from the PQ scale to the infrared one providing radiative corrections to the tree-level coupling~\cite{Chang:1993gm,Srednicki:1985xd}.

In summary, the physical axion interaction Lagrangian with photons, protons, neutrons and electrons is written as,
\begin{align}
    \mathcal{L}_a^{\text{eff}} \supset \frac{\alpha_e}{8 \pi}  c_{\gamma} \frac{a}{f_a} F \widetilde{F}
    + c_p \frac{\partial_\mu a}{2 f_a} \overline{p} \gamma_\mu \gamma_5 p +  c_n \frac{\partial_\mu a}{2 f_a} \overline{n} \gamma_\mu \gamma_5 n +  c_e \frac{\partial_\mu a}{2 f_a} \overline{e} \gamma_\mu \gamma_5 e \; ,
    \label{eq:axioneffcoupling}
\end{align}
where the couplings are given by~\cite{GrillidiCortona:2015jxo,DiLuzio:2020wdo},
\begin{align}
    c_{\gamma} &= \frac{E}{N} - 1.92(4) \; , \nonumber \\
    c_{p} &=  -0.47(3) + 0.88(3) c_u^0 - 0.39(2) c_d^0 - c_{\text{sea}} \; , \nonumber \\
    c_{n} &=  -0.02(3) + 0.88(3) c_d^0 - 0.39(2) c_u^0 - c_{\text{sea}} \; , \nonumber \\
    c_{\text{sea}} & = 0.038(5) c_s^0 + 0.012(5) c_c^0 + 0.009(2) c_b^0 + 0.0035(4) c_t^0 \; , \nonumber \\
     c_{e} & = c_e^0 + \frac{3 \alpha_e^2}{4 \pi^2} \left[\frac{E}{N} \log\left(\frac{f_a}{m_e}\right) - 1.92(4) \log\left(\frac{\text{GeV}}{m_e}\right)\right] \; ,
    \label{eq:axioncouplings}
\end{align}
with their model-dependent parts being encoded in $E/N$ and the $c_f^0$ couplings between the axion and fermions $\psi_f$. In the next section we will review QCD axion models and compute these model-dependent couplings for each case. As a last note, the couplings above can be redefined as,
\begin{equation}
    g_{a \gamma \gamma} = \frac{ \alpha_e}{2 \pi f_a} c_{\gamma} \; , \; g_{a e e} = \frac{m_e}{f_a} c_e \; , \; g_{a n} = \frac{m_n}{f_a} c_n \; , \; g_{a p} = \frac{m_p}{f_a} c_p \; .
    \label{eq:expcouplings}
\end{equation}
The above definition for the couplings is the one employed in the literature to plot the experimental constraints, bounds and future sensitivities, in terms of the axion mass and/or its decay constant, as we will see in Sec.~\ref{sec:experimental}.

%%%%%%%%%%%%%%%%%%%%%%%%%%%%%%%%%%%%%%%%%%%%%%%%%%%%%%%%%%%%%%%%%%%%%%%%%%%%%
\section{Paradigmatic QCD Axion Models}
\label{sec:models}
%%%%%%%%%%%%%%%%%%%%%%%%%%%%%%%%%%%%%%%%%%%%%%%%%%%%%%%%%%%%%%%%%%%%%%%%%%%%%

This section presents some paradigmatic axion models. These are UV complete frameworks which at low energy lead to the couplings and properties discussed earlier via the axion effective Lagrangian. It must be noted that vanilla axion models rely on a global PQ symmetry which is susceptible to Planck-scale violations leading to the well-known ``PQ quality problem"~\cite{Georgi:1981pu,Dine:1986bg,Barr:1992qq,Kamionkowski:1992mf,Holman:1992us,Ghigna:1992iv}. Addressing this theoretical challenge lies beyond the scope of this thesis. Nevertheless, it has been shown that high-quality PQ symmetries can emerge as accidental remnants of discrete gauge symmetries~\cite{Chun:1992bn,BasteroGil:1997vn,Babu:2002ic,Dias:2002hz, Harigaya:2013vja}, Abelian gauge symmetries~\cite{Fukuda:2017ylt,Duerr:2017amf,Bonnefoy:2018ibr}, or non-Abelian gauge symmetries~\cite{Randall:1992ut,DiLuzio:2017tjx,Lillard:2018fdt, Lee:2018yak}.

%%%%%%%%%%%%%%%%%%%%%%%%%%%%%%%%%%%%%%%%%%%%%%%%%%%%%%%%%%%%%%%%%%%%%%%%%%%%%
\subsection{Peccei-Quinn-Weinberg-Wilczeck (PQWW)}
\label{sec:PQWW}
%%%%%%%%%%%%%%%%%%%%%%%%%%%%%%%%%%%%%%%%%%%%%%%%%%%%%%%%%%%%%%%%%%%%%%%%%%%%%

%
\begin{table}[t!]
\renewcommand*{\arraystretch}{1.5}
	\centering
	\begin{tabular}{| K{1.5cm} | K{1cm} | K{4.5cm} |  K{1cm} |}
		\hline
 &Fields&\SM&  U($1$)$_{\text{PQ}}$ \\
		\hline
		\multirow{3}{*}{Fermions}
&$u_R$&($\mathbf{3},\mathbf{1}, {2/3}$)& $\chi_1$  \\
&$d_{R}$&($\mathbf{3},\mathbf{1}, {-1/3}$)& $\chi_2$   \\
&$e_{R}$&($\mathbf{1},\mathbf{1}, -1$)& $\chi_2$  \\
		\hline
		\multirow{2}{*}{Scalars}
&$\Phi_1$&($\mathbf{1},\mathbf{2}, {1/2}$)&$\chi_1$ \\	
&$\Phi_2$&($\mathbf{1},\mathbf{2}, {1/2}$)&$-\chi_2$ \\
\hline
	\end{tabular}
	\caption{Original PQWW model model matter content and transformation properties of the fields under the $G_{\text{SM}}$ and $\text{U}(1)_{\text{PQ}}$ symmetries. The PQ charges satisfy $\chi_2/\chi_1 = \tan^2 \beta$ (see text for details).}
	\label{tab:PQWW} 
\end{table}
This is the original model proposed by Peccei and Quinn to solve the strong CP problem~\cite{Peccei:1977hh,Peccei:1977ur}. Afterwards, Weinberg and Wilczeck showed the existence of the pseudo-GB associated to the breaking of the PQ symmetry named the axion~\cite{Weinberg:1977ma,Wilczek:1977pj}. In the PQWW model the SM particle content is extended with a second Higgs doublet and supplemented with a PQ symmetry, as shown in Table~\ref{tab:PQWW}. Note that, we choose the charge assignment leading to the Type-II 2HDM. Namely, the Yukawa Lagrangian is given by,
\begin{equation}
- \mathcal{L}_{\text{Yuk.}} \supset \overline{q_L} \Y_u \tilde{\Phi}_1 u_R  + \overline{q_L} \Y_d \Phi_2 d_R + \overline{\ell_L} \Y_e \Phi_2 e_R  + \text{H.c.} \; .
\label{eq:YukPQWW}
\end{equation}
Note that, the flipped 2HDM version is obtained by charging the RH charged-lepton fields as $e_R \rightarrow \exp(i \chi_1) e_R$. The scalar potential is the same as the U(1)-symmetric 2HDM~\cite{Branco:2011iw}:
\begin{align}
    V(\Phi_1,\Phi_2) &= m_{11}^2 \Phi_1^\dagger \Phi_1 + m_{22}^2 \Phi_2^\dagger \Phi_2 + \frac{\lambda_1}{2} \left(\Phi_1^\dagger \Phi_1\right)^2 + \frac{\lambda_2}{2} \left(\Phi_2^\dagger \Phi_2\right)^2 \nonumber
    \\ & + \lambda_3 \left(\Phi_1^\dagger \Phi_1\right)\left(\Phi_2^\dagger \Phi_2\right) + \lambda_4 \left(\Phi_1^\dagger \Phi_2\right)\left(\Phi_2^\dagger \Phi_1\right) \; ,
    \label{eq:VpotentialPQWW}
\end{align}
where we parametrize the scalar doublets as,
\begin{equation}
     \Phi_{1,2} =
             \begin{pmatrix} \phi_{1,2}^+ \\ \phi_{1,2}^0 = \left(v_{1,2} + \rho_{1,2}\right) e^{i a_{1,2}/v_{1,2}}/\sqrt{2} \end{pmatrix}
    \; ,
\label{eq:parametrisationPQWW}
\end{equation} 
whith the charged components of the Higgs doublets being $\phi_{1,2}^+$, while $a_{1,2}$ denote the neutral-scalar phases, and $\rho_{1,2}$ the radial modes. Upon SSB the Higgs doublets develop non-zero VEVs $\langle \phi_{1,2}^0 \rangle  = v_{1,2}/\sqrt{2}$, breaking both the EW and PQ symmetries, and are related via the $\beta$ angle as shown in Eq.~\eqref{eq:tanbeta}. Note that, the CP-odd components of the Higgs doublets namely the scalars $a_{1,2}$ shown above will lead to two distinct combinations one corresponding to the GB $G^0$ which will be "eaten" by the $Z$-boson and another one being the physical axion $a$. By deriving the orthogonality condition between $a$ and $G^0$ the PQ charges can be specified since $a$ should not mix with $G^0$. This can be done by first identifying the axion $a$ and computing the associated PQ current [see Eq.~\eqref{eq:genPQcurrent}],
\begin{equation}
    J^\mu_{\text{PQ}} = - \chi_1 v_1 \partial^\mu a_1 + \chi_2 v_2 \partial^\mu a_2 + \sum_f \chi_f \left( \overline{\psi_f} \gamma^\mu \psi_f\right) \Rightarrow a = \frac{1}{f_{\text{PQ}}} \left(- \chi_1 v_1 \partial^\mu a_1 + \chi_2 v_2 \partial^\mu a_2\right) \; ,
\end{equation}
where the $\chi_f$ charges for the fermions are the ones shown in Table~\ref{tab:PQWW}, and the PQ scale is:
\begin{equation}
f_{\text{PQ}} = \sqrt{\chi^2_1 v_1^2 + \chi^2_2 v_2^2} \; .
\label{eq:PQWWscale}
\end{equation}
Next, in order for the physical axion $a$ above not to mix with $G^0$ it needs to be invariant under an U($1)_{\text{Y}}$ hypercharge transformation given by,
\begin{equation}
a \rightarrow \frac{1}{f_{\text{PQ}}} \left[- \chi_1 v_1 \partial^\mu \left(a_1 + \frac{\alpha_{\text{Y}}}{2} v_1\right) + \chi_2 v_2 \partial^\mu \left(a_2 + \frac{\alpha_{\text{Y}}}{2} v_2\right) \right] = a + \frac{\alpha_{\text{Y}}}{2 f_{\text{PQ}}} \left(- \chi_1 v_1^2 + \chi_2 v_2^2 \right)\; ,
\end{equation}
with $\alpha_{\text{Y}}$ being an arbitrary multiplicative constant. The invariance requirement leads to the following relation among the PQ charges,
\begin{equation}
    \frac{\chi_2}{\chi_1} = \tan^2\beta \; , \;  \chi_2 = \tan\beta \; , \;  \chi_1 = \frac{1}{\tan\beta} \; , 
\end{equation}
where the second and third equalities were chosen without loss of generality. Hence, from the above and making use of Eq.~\eqref{eq:EN} leads to $N = 3 (1 + \tan \beta)$ showing that this model provides a viable solution to the strong CP problem.

The PQWW framework is also dubbed visible axion model since $f_{\text{PQ}} \sim v$ leads to $m_a \sim 100$ keV and sizable couplings between the axion and SM particles. At the time, soon after this initial proposal was made, the model was ruled out by different experiments looking namely for rare kaon meson decays $K^+ \rightarrow \pi^+ + \text{invisible}$~\cite{Wilczek:1977zn,Donnelly:1978ty,Hall:1981bc,Frere:1981cc,Asano:1981nh,Zehnder:1981qn,Bardeen:1986yb}. The problem lies in the fact that the PQ and EW scale are of the same order. Thus, to accommodate a phenomenologically viable axion model there is the need to decouple the PQ scale from the EW one i.e., $f_{\text{PQ}} \gg v$.

%%%%%%%%%%%%%%%%%%%%%%%%%%%%%%%%%%%%%%%%%%%%%%%%%%%%%%%%%%%%%%%%%%%%%%%%%%%%%
\subsection{Dine-Fischler-Srednicki-Zhitnitsky (DFSZ)}
\label{sec:DFSZ}
%%%%%%%%%%%%%%%%%%%%%%%%%%%%%%%%%%%%%%%%%%%%%%%%%%%%%%%%%%%%%%%%%%%%%%%%%%%%%

%
\begin{table}[t!]
\renewcommand*{\arraystretch}{1.5}
\centering
\begin{tabular}{| K{1.5cm} | K{1cm} | K{4.5cm} |  K{2.5cm} |}
		\hline
 &Fields&\SM&  U($1$)$_{\text{PQ}}$ \\
		\hline
		\multirow{3}{*}{Fermions}
&$u_R$&($\mathbf{3},\mathbf{1}, {2/3}$)& $\chi_1$  \\
&$d_{R}$&($\mathbf{3},\mathbf{1}, {-1/3}$)& $\chi_2$ \\
&$e_{R}$&($\mathbf{1},\mathbf{1}, -1$)& I: $\chi_2$ ; II: $\chi_1$  \\
		\hline
		\multirow{3}{*}{Scalars}
&$\Phi_1$&($\mathbf{1},\mathbf{2}, {1/2}$)&$\chi_1$  \\	
&$\Phi_2$&($\mathbf{1},\mathbf{2}, {1/2}$)&$-\chi_2$  \\
&$\sigma$&($\mathbf{1},\mathbf{1}, 0$)&$(\chi_1 + \chi_2) /2$ \\
\hline
	\end{tabular}
	\caption{DFSZ I and II model model matter content and transformation properties of the fields under the $G_{\text{SM}}$ and $\text{U}(1)_{\text{PQ}}$ symmetries.}
	\label{tab:DFSZ} 
\end{table}
The DFSZ~\cite{Zhitnitsky:1980tq,Dine:1981rt} proposal builds upon the PQWW model by extending it with a complex scalar singlet $\sigma$ with the idea to decouple the PQ scale from the EW one. The DFSZ I and II model particle content and charges of the fields under the PQ symmetry are indicated in Table~\ref{tab:DFSZ}. The DFSZ I model has a Type-II 2HDM Yukawa Lagrangian shown in Eq.~\eqref{eq:YukPQWW}, while the DFSZ II model is the flipped 2HDM version with Yukawa Lagrangian:
\begin{equation}
- \mathcal{L}_{\text{Yuk.}} \supset \overline{q_L} \Y_u \tilde{\Phi}_1 u_R  + \overline{q_L} \Y_d \Phi_2 d_R + \overline{\ell_L} \Y_e \Phi_1 e_R  + \text{H.c.} \; .
\label{eq:YukDFSZII}
\end{equation}
Both DFSZ models feature the following scalar potential:
\begin{align}
    V(\Phi_1,\Phi_2,\sigma) &= m_{11}^2 \Phi_1^\dagger \Phi_1 + m_{22}^2 \Phi_2^\dagger \Phi_2 + \frac{\lambda_1}{2} \left(\Phi_1^\dagger \Phi_1\right)^2 + \frac{\lambda_2}{2} \left(\Phi_2^\dagger \Phi_2\right)^2 \nonumber
    \\
    & + \lambda_3 \left(\Phi_1^\dagger \Phi_1\right)\left(\Phi_2^\dagger \Phi_2\right) + \lambda_4 \left(\Phi_1^\dagger \Phi_2\right)\left(\Phi_2^\dagger \Phi_1\right) \nonumber
    \\
    & + m_{\sigma}^2 |\sigma|^2 + \frac{\lambda_\sigma}{2} |\sigma|^4 + \lambda_{1 \sigma} \Phi_1^\dagger \Phi_1 |\sigma|^2 + \lambda_{2 \sigma} \Phi_2^\dagger \Phi_2 |\sigma|^2 + \left[\lambda \Phi_1^\dagger \Phi_2 \sigma^2 +\text{H.c.} \right] 
    \; .
    \label{eq:VpotentialDFSZOG}
\end{align}
The Higgs doublets are parameterized as in Eq.~\eqref{eq:parametrisationPQWW}, while the new scalar singlet can be parameterized as follows,
\begin{equation}
    \sigma = \frac{v_\sigma + \rho_\sigma}{\sqrt{2}} e^{i a_\sigma /v_\sigma} \; ,
    \label{eq:sigmaPQ}
\end{equation}
with $\rho_\sigma$ being the radial component. The physical axion fields $a$ will correspond to a combination of the CP-odd components~$a_{1,2,\sigma}$, which can be derived through the associated PQ current. Namely,
\begin{align}
    a & = \frac{1}{f_{\text{PQ}}} \left(- \chi_1 v_1 \partial^\mu a_1 + \chi_2 v_2 \partial^\mu a_2 + \frac{\chi_1+\chi_2}{2} v_\sigma \partial^\mu a_\sigma \right) \; , \nonumber \\
    f_{\text{PQ}} & = \sqrt{\chi^2_1 v_1^2 + \chi^2_2 v_2^2 + \frac{(\chi_1+\chi_2)^2}{4} v_\sigma^2} \; .
\end{align}
Considering that $v_\sigma \gg v_{u,d}$ the axion field and PQ scale are approximately given by $a \simeq a_\sigma$ and $f_{\text{PQ}} \simeq v_{\sigma}/\sqrt{2}$. By making use of Eqs.~\eqref{eq:EN} and~\eqref{eq:couplingsmd} one obtains for the DFSZ I,
\begin{align}
    N = 6 \; , \; E=16 \; , \; \frac{E}{N} = \frac{8}{3} \; , \; c_u^0 = \frac{1}{3} \cos^2 \beta \; , \;  c_d^0 = \frac{1}{3} \sin^2 \beta \; , \;  c_e^0 = \frac{1}{3} \sin^2 \beta \; .
\end{align}
For the DFSZ II one has,
\begin{align}
    N = 6 \; , \; E=4 \; , \; \frac{E}{N} = \frac{2}{3} \; , \; c_u^0 = \frac{1}{3} \cos^2 \beta \; , \;  c_d^0 = \frac{1}{3} \sin^2 \beta \; , \;  c_e^0 = -\frac{1}{3} \cos^2 \beta  \; .
\end{align}
Compared to the PQWW model, in the DFSZ case the axion couplings will be suppressed by $\sim v/v_\sigma$ in the expression for the effective Lagrangian of Eq.~\eqref{eq:axioneffcoupling}. Hence, the DFSZ proposal constitutes a viable QCD invisible axion model evading the experimental constraints that ruled out the PQWW setup. Note that in Chapter~\ref{chpt:flavoraxion} we will study a general DFSZ framework in connection to neutrino mass generation and the flavor puzzle leading to interesting axion-fermion coupling phenomenology.

%%%%%%%%%%%%%%%%%%%%%%%%%%%%%%%%%%%%%%%%%%%%%%%%%%%%%%%%%%%%%%%%%%%%%%%%%%%%%
\subsection{Kim-Shifman-Vainshtein-Zakharov (KSVZ)}
\label{sec:KSVZ}
%%%%%%%%%%%%%%%%%%%%%%%%%%%%%%%%%%%%%%%%%%%%%%%%%%%%%%%%%%%%%%%%%%%%%%%%%%%%%

%
\begin{table}[t!]
\renewcommand*{\arraystretch}{1.5}
	\centering
	\begin{tabular}{| K{1.5cm} | K{1cm} | K{4.5cm} | K{1cm} |}
		\hline
 &Fields&\SM& U($1$)$_{\text{PQ}}$ \\
		\hline
		\multirow{2}{*}{Fermions}
&$\Psi_L$&($\mathbf{3},\mathbf{1}, 0$)& $1/2$  \\
&$\Psi_R$&($\mathbf{3},\mathbf{1}, 0$)& $-1/2$ \\
		\hline
		\multirow{1}{*}{Scalars}
&$\sigma$&($\mathbf{1},\mathbf{1}, 0$)& $1$ \\
\hline
	\end{tabular}
	\caption{Original KSVZ model matter content and transformation properties of the fields under the $G_{\text{SM}}$ and $\text{U}(1)_{\text{PQ}}$ symmetries.}
	\label{tab:KSVZ} 
\end{table}
The KSVZ~\cite{Kim:1979if,Shifman:1979if} invisible axion model extends the SM with a vector-like colored fermion $\Psi$ in the fundamental representation of $\text{SU}(3)_{c}$ and singlet under $\text{SU}(2)_{L}$ with zero hypercharge. Additionally, a complex scalar singlet $\sigma$ provides mass to this exotic fermion. The particle content of this original KSVZ and charges of the fields under the PQ symmetry is presented in Table~\ref{tab:KSVZ}. The Lagrangian will contain the following additional terms,
\begin{equation}
- \mathcal{L} \supset \overline{\Psi} i \slashed{D} \Psi + Y_\psi \overline{\Psi_L} \Psi_R \sigma + |\partial_\mu \sigma|^2 + m_\sigma^2 |\sigma|^2 - \frac{\lambda_\sigma}{4} |\sigma|^4 - \lambda_{\Phi \sigma} \Phi^\dagger \Phi |\sigma|^2  + \text{H.c.} \; ,
\end{equation}
where the bare mass term for $\Psi$ is forbidden by the U($1$)$_{\text{PQ}}$ symmetry. Note that, this setup generates a viable color anomaly since $N=1$, and thus solves the strong CP problem -- see Eq.~\eqref{eq:EN}. 

The minimum of the scalar potential, shown above, will break the PQ symmetry when the complex scalar acquires a non-zero VEV $\left<\sigma\right>=v_{\sigma}/\sqrt{2}$. Hence, the axion will be identified as the axial component of the new scalar singlet which is parameterized as in Eq.~\eqref{eq:sigmaPQ}. Thus, it is clear that the axion scale is the scalar VEV $f_{\text{PQ}} = \langle \sigma \rangle$, which can be arbitrarily large $v_\sigma \gg v$. This feature makes this model phenomenologically viable since the axion can be made very light -- invisible -- and the exotic fermion $\Psi$ with mass $M_{\Psi}= v_{\sigma} Y_\psi/\sqrt{2}$ is very heavy, evading experimental constraints. Note that, the particularity of the KSVZ setup is that the SM fermions are not charged under the $\text{U}(1)_{\text{PQ}}$ symmetry. Consequently, the axion has no couplings to the SM fermions and since $\Psi$ has no electric charge $E=0$ [see Eq.~\eqref{eq:EN}]. Nonetheless experiments will be able to probe $a$ solely through the model-independent contributions, to the axion-photon, nucleon and electron couplings [see Eq.~\eqref{eq:axioncouplings}].

In the original KSVZ model the new colored exotic fermions are stable, and can be thermally produced after inflation, which poses cosmological problems~\cite{Nardi:1990ku,Perl:2001xi,Perl:2009zz}. In fact, searches in terrestrial, lunar, and meteoritic materials yield strong limits~\cite{Perl:2001xi,Perl:2009zz,Burdin:2014xma,Hertzberg:2016jie,Mack:2007xj}, practically ruling out such stable baryonic and/or charged relics, unless some mechanism effectively suppresses their density or allows them to decay to ordinary matter~\cite{Nardi:1990ku,DiLuzio:2016sbl,DiLuzio:2017pfr}. This motivates a plethora of different KSVZ setups featuring single/multiple vector-like colored fermions possessing non-trivial representations under $\text{SU}(2)_{L}$ and/or hypercharge, generating $E \neq 0$ and axion-fermion couplings~\cite{DiLuzio:2017pfr,DiLuzio:2016sbl,DiLuzio:2017ogq}.  Namely, in Chapter~\ref{chpt:axionneutrino}, we will explore two novel KSVZ axion frameworks connected to the origin of neutrino masses. Interestingly, in these scenarios the exotic fermions $\Psi$ with $Y \neq 0$ can possibly mix with ordinary quarks allowing for models free of stable colored or charged relics, leading to viable post-inflationary axion DM -- see the upcoming discussion in Sec.~\ref{sec:axionDMcosmo}.

%%%%%%%%%%%%%%%%%%%%%%%%%%%%%%%%%%%%%%%%%%%%%%%%%%%%%%%%%%%%%%%%%%%%%%%%%%%%%
\subsection{Axion-like particles}
\label{sec:ALPs}
%%%%%%%%%%%%%%%%%%%%%%%%%%%%%%%%%%%%%%%%%%%%%%%%%%%%%%%%%%%%%%%%%%%%%%%%%%%%%

As reviewed the QCD axion is a pseudo-GB associated to the spontaneous breaking of a PQ symmetry classically conserved but anomalously broken by the QCD topological term. Besides this specific type of pseudo-GB, one can consider what is know as ALPs which refers to the general pseudo-GB arising from spontaneously broken U(1) symmetries in BSM scenarios. In fact, ALPs share numerous properties with the QCD axion having a similar effective Lagrangian as the one in Eq.~\eqref{eq:EFT} namely derivative couplings to the SM particles inversely proportional to the decay constant. However, ALPs do not solve the strong CP problem and do not acquire their mass through QCD non-perturbative effects as does the QCD axion. Hence, for ALPs their mass and decay constant are independent parameters i.e., the relation in Eq.~\eqref{eq:axionmass} no longer holds, opening up their parameter space. Nonetheless, ALPs appear in BSM scenarios solving other SM puzzles/problems such as providing dynamical generations of neutrino masses based on global U$(1)_L$/U$(1)_{B-L}$ symmetries, which give rise to the Majoron -- see e.g. Refs.~\cite{Chikashige:1980ui,Berezinsky:1993fm}, flavor patterns and mixing~\cite{Froggatt:1978nt}, as well as arising in supersymmetric and string theories~\cite{Witten:1984dg,Conlon:2006tq,Svrcek:2006yi,Choi:2009jt,Arvanitaki:2009fg,Halverson:2017deq}.

%%%%%%%%%%%%%%%%%%%%%%%%%%%%%%%%%%%%%%%%%%%%%%%%%%%%%%%%%%%%%%%%%%%%%%%%%%%%%
\section{Axion dark matter and cosmology}
\label{sec:axionDMcosmo}
%%%%%%%%%%%%%%%%%%%%%%%%%%%%%%%%%%%%%%%%%%%%%%%%%%%%%%%%%%%%%%%%%%%%%%%%%%%%%

The previous sections discussed a variety of UV-complete axion models proposed as elegant solutions to the strong CP problem. The attractiveness of this solution is further increased since the axion turns out to be an excellent DM candidate solving yet another unresolved puzzle of the SM. In fact, axions are naturally light, stable, having weak couplings with ordinary matter and can be non-thermally produced (cold axions) in the early Universe, reproducing the observed value of the DM relic abundance obtained by the Planck satellite data given in Eq.~\eqref{eq:Oh2PlanckOG}~\cite{Planck:2018vyg} -- see Sec.~\ref{sec:DM}. There are two scenarios for axion DM production the pre and post-inflationary cases.

In the pre-inflationary scenario, axion DM production is realized only through the misalignment mechanism~\cite{Preskill:1982cy,Abbott:1982af,Dine:1982ah}. In this case, the PQ symmetry is broken before inflation and never restored during the reheating period of the Universe. At this stage multiple patches of the Universe take different values for the axion field. Inflation will then stretch one of these small patches into what is know as our Universe. This implies that the axion field is constant over the observable Universe $a(\vec{x},t) = a(t) = \theta_0 f_a$, with $\theta_0$ being the initial misalignment angle. When the QCD phase transition takes place for temperature $T \sim \Lambda_{\text{QCD}}$ the axion potential is generated driving the axion field from its initial misalignment towards the vacuum minimum which preserves CP solving the strong CP problem. At high-temperatures near $\Lambda_{\text{QCD}}$, non-perturbative QCD methods fail and therefore one cannot use $\chi$PT to derive $V(a)$, as was done previously. In fact, thermal corrections to the axion mass need to be considered leading to a temperature dependent axion potential $V(a,T)$. The latter can be computed using the dilute instanton gas approximation~(DIGA) leading to~\cite{Callan:1977gz,Gross:1980br,Dimopoulos:2005ac},
\begin{equation}
    V(a,T) = m_a^2(T) f_a^2 \left[1-\cos\left(\frac{a}{f_a}\right) \right] \; ,
    \label{eq:VAT}
\end{equation}
with the axion thermal mass being,
\begin{equation}
    m_a(T) \simeq \alpha \; m_a \left(\frac{T_C}{T} \right)^\beta \; ,
\end{equation}
where with DIGA the coefficients are  approximately $\alpha \sim 0.01$ and $\beta \simeq 4$. Additionally, $T_C$ is the critical temperature where QCD deconfines, the above approximation holds well for $T \gg T_C$. Thus, the temperature dependence of the potential goes as $V(a,T) \sim T^{-8}$.

The axion can be treated as a classical field with its EOM in the expanding Friedman-Robertson-Walker~(FRW) Universe given by,
\begin{figure}[t!]
    \centering
    \includegraphics[scale=0.6]{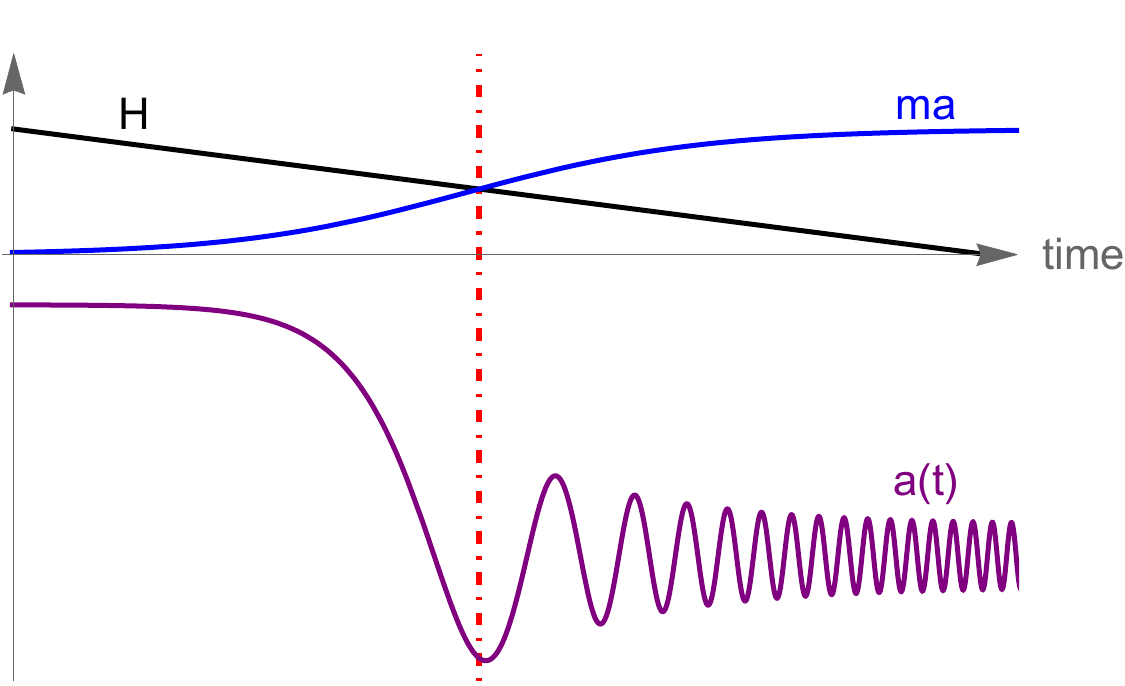}
    \caption{Schematic plot of the axion evolution from misalignment dynamics [see Eq.~\eqref{eq:EOMa} and text for details].}
    \label{fig:misalignment}
\end{figure}
\begin{align}
\Ddot{a} + 3 H \dot{a} + V^\prime(a,T) = 0 &\Rightarrow \Ddot{a} + 3 H \dot{a} + m_a^2(T) f_a \sin\left(\frac{a}{f_a}\right) = 0 \, , \nonumber \\
&\Leftrightarrow \Ddot{\theta} + 3 H \dot{\theta} + m_a^2(T) \sin \theta=0 \; ,
\label{eq:EOMa}
\end{align}
where $H \equiv H(T)$ is the Hubble parameter and we used the cosine type $V(a,T)$ potential shown in Eq.~\eqref{eq:VAT} taking its derivative with respect to the axion field. As usual conditions $\theta(0) \equiv \theta_0$ we  take $0<|\theta_0|<\pi$, with $\theta_0$ being the initial misalignment angle. The evolution of the axion field is illustrated in Fig.~\ref{fig:misalignment}. Namely, in early times of the Universe $H \gg m_a$ and the axion field is constant due to the Hubble friction. When the Hubble parameter becomes comparable to the axion mass $H \sim m_a$ the field becomes dynamical, starting to oscillate with some damping. At later times when $m_a \gg H$ the axion evolves adiabatically with rapid oscillations behaving as CDM at longer time scales that the period of oscillation~\cite{Sikivie:2009qn}.

\begin{figure}[t!]
   \centering
    \includegraphics[scale=0.33]{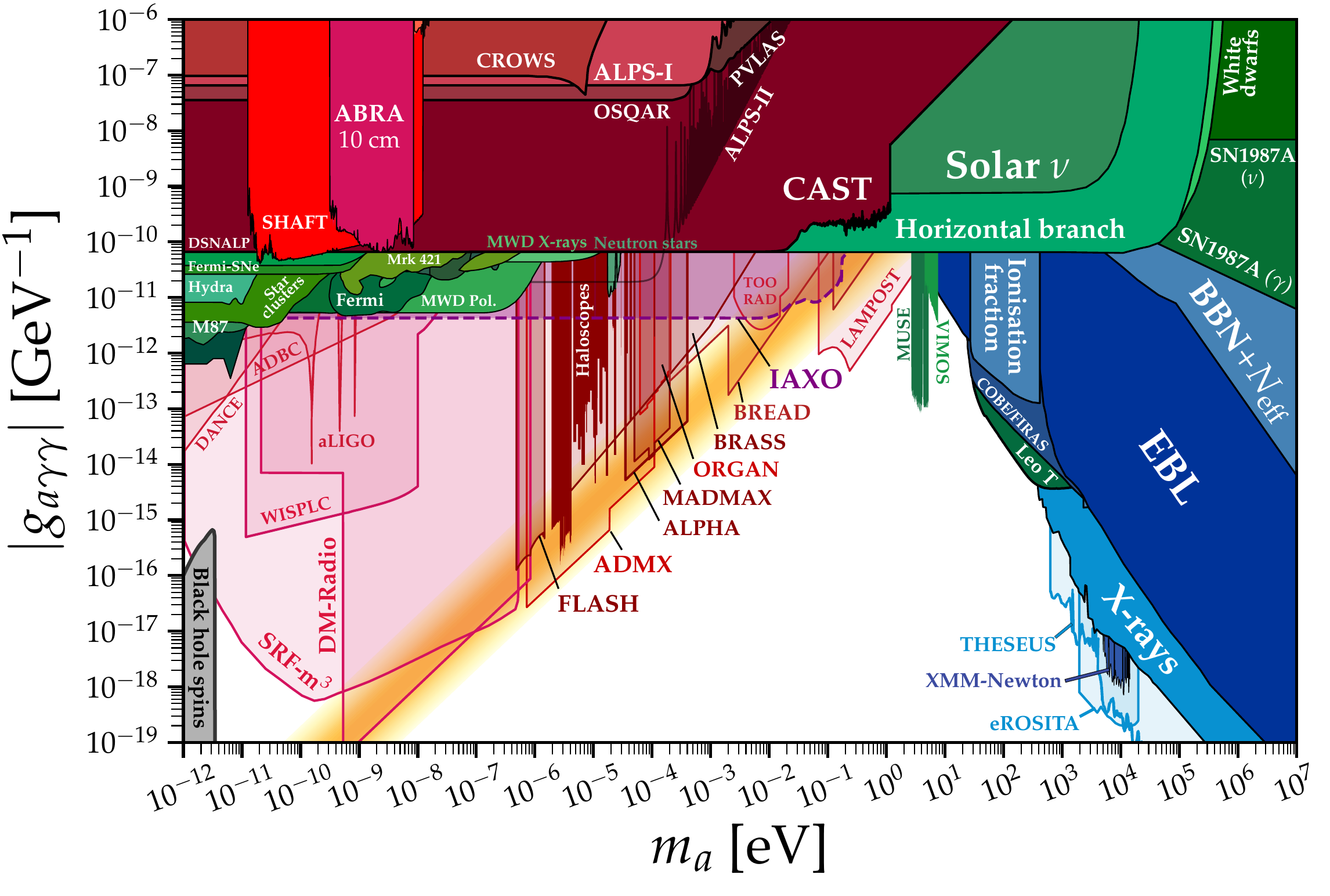}
    \caption{Current (opaque) and projected (transparent) constraints on the axion-photon coupling in terms of the axion mass. Are included a plethora of different searches and bounds coming form astrophysics, solar axions, helioscopes, DM haloscope searches, laboratory experiments, which are described in the main text. The yellow/orange band corresponds to the QCD axion parameter space, where the combination of numerous future experiments are projected to almost fully probe. The figure was taken from Ref.~\cite{Adams:2022pbo}.}
    \label{fig:exp1}
\end{figure}

The relic axion abundance $ \Omega_a h^2$ generated through the misalignment mechanism can be written as a function of its decay constant $f_a$ and initial misalignment angle $\theta_0$ and the average observed value for DM relic density $\Omega_\text{CDM} h^2 \simeq 0.12$~[see Eq.~\eqref{eq:Oh2PlanckOG}]. For $\theta_0 \leq 1$ the following estimation can be obtained~\cite{DiLuzio:2020wdo},
    \begin{equation}
    \Omega_a h^2 \simeq  \Omega_\text{CDM} h^2 \frac{\theta_0^2}{2.15^2} \left(\frac{f_a}{2 \times 10^{11} \ \text{GeV}} \right)^{\frac{7}{6}} \; ,
    \label{eq:relica}
    \end{equation}
Hence, for initial misalignment angles between $0.1<\theta_0<\pi$ the axion decay constant need to be $2 \times 10^{10} \ \text{GeV} <f_a<5\times 10^{12} \ \text{GeV}$ in order to account for all the observed DM relic abundance. As discussed in Sec.~\ref{sec:experimental} and shown in Fig.~\ref{fig:exp1}, this parameter space region is currently being probed by haloscope experiments. 

In pre-inflationary scenarios, the axion leaves an imprint in primordial fluctuations, reflected in the CMB anisotropies and large-scale structure. The resulting isocurvature fluctuations are constrained by cosmic microwave background data~\cite{Beltran:2006sq}, leading to an upper bound on the inflationary scale $H_I$~\cite{DiLuzio:2016sbl}:
\begin{equation}
   H_I \lsim  \frac{0.9\times 10^7}{\Omega_a h^2/\Omega_\text{CDM} h^2} \left(\frac{\theta_0}{\pi} \frac{f_a}{ 10^{11} \ \text{GeV}} \right) \; \text{GeV} \; .
   \label{eq:Iso}
\end{equation}
Taking $\theta_0 \sim \mathcal{O}(1)$ and $\Omega_a h^2=0.12$, leads to a low scale for inflation $H_I \lsim 10^7$ GeV (Planck currently probes $H_I \lsim 10^{13}$ GeV~\cite{Planck:2018vyg}). Nonetheless, the isocurvature bound is dependent on UV physics and the concrete inflationary mechanism at play -- which will not be the object of discussion in this thesis -- and thus can be more relaxed~\cite{Graham:2025iwx}.
    
In the post-inflationary scenario the PQ symmetry is broken after inflation, leading to an observable Universe divided into patches with different values of the axion field. The value of the initial misalignment angle $\theta_0$ is no longer free and, through statistical average, the up-to-date value $\left<\theta_0^2\right> \simeq 2.15^2$~\cite{DiLuzio:2020wdo} has been obtained. Hence, compared to the pre-inflationary scenario, this case is, in principle, more predictive. The requirement  $\Omega_a h^2 = \Omega_{\text{CDM}} h^2 = 0.12$, leads to a prediction for the axion scale $f_a$ (or equivalently mass $m_a$) [see Eq.~\eqref{eq:relica}]. In fact, if only the misalignment mechanism is at the origin of axion abundance, the upper bound of $f_a \lesssim 2 \times 10^{11}$ GeV guarantees that DM is not overproduced. However, the picture is much more complicated since topological defects, strings and DWs, will also contribute to the total axion relic abundance $\Omega_a h^2$~\cite{Bennett:1987vf,Levkov:2018kau,Gorghetto:2018myk,Buschmann:2019icd}. Namely, the complex non-linear dynamics of the resulting network of axion strings and DWs must be carefully resolved through numerical simulations, with their precise contribution remaining an open question and an active area of ongoing research. A detailed analysis of this lies beyond the scope of the present thesis. The cosmological DW problem is absent if $N_{\text{DW}} = 1$, where $N_{\text{DW}}$ corresponds to the vacuum degeneracy stemming from the residual $\mathcal{Z}_{N}$ symmetry, left unbroken from non-perturbative QCD-instanton effects that anomalously break U(1)$_{\text{PQ}}$. Although there are no DWs for $N_{\text{DW}}=1$, axionic string networks can still form. Numerical simulations predict the axion scale $f_a$ to be in the range $[5 \times 10^9,3 \times 10^{11}]$~GeV, in order for the axion to account for the observed CDM abundance, i.e. $\Omega_a h^2 = \Omega_{\text{CDM}} h^2$~\cite{Kawasaki:2014sqa,Klaer:2017ond,Gorghetto:2020qws,Buschmann:2021sdq,Benabou:2024msj}.

%

%%%%%%%%%%%%%%%%%%%%%%%%%%%%%%%%%%%%%%%%%%%%%%%%%%%%%%%%%%%%%%%%%%%%%%%%%%%%%
\section{Experimental searches}
\label{sec:experimental}
%%%%%%%%%%%%%%%%%%%%%%%%%%%%%%%%%%%%%%%%%%%%%%%%%%%%%%%%%%%%%%%%%%%%%%%%%%%%%

This section reviews the current and future experimental axion and ALP searches, which were driven thanks to the invisible axion model proposals. Since the QCD axion presents a tight relation between its mass $m_a$ and scale, setting bounds on the sole parameter $f_a$ and vice-versa, allows to probe and exclude parts of the axion parameter space. These constraints come from the axion couplings to photons, nucleons and electrons stemming from indirect astrophysical and cosmological observations, helioscopes, haloscopes, as well as laboratory searches. For detailed reviews see Refs.~\cite{DiLuzio:2020wdo,Adams:2022pbo} and the references therein. We also refer the reader to the repository~\cite{AxionLimits} which collects up-to-date searches for axions from a broad range of experiments. Figs.~\ref{fig:exp1},~\ref{fig:exp2} and~\ref{fig:exp3} summarize the different searches constraining respectively the axion-photon, electron and nucleon~(neutron and proton) coupling in terms of the axion mass $m_a$ [the couplings are the ones defined in Eq.~\eqref{eq:expcouplings}]. Note that, the yellow band represents the QCD axion model parameter space namely where the KSVZ and DFSZ invisible axion, discussed previously, can be probed. However, the whole parameter space is in principle available for ALPs. 

The current landscape of axion experiments and constraints, as well as future projected sensitivities, can be categorized as follows:
\dblack
\begin{itemize}

\item \textbf{Astrophysical Bounds:} Astrophysical evidence stemming from stellar evolution is a very important tool to probe the axion couplings to photons, electrons and nucleons since these can be thermally produced inside stars. In fact, Fig.~\ref{fig:astrodiag} shows the different processes that can occur in astrophysical objects that involve axion interactions.

\textit{Axion-photon coupling:} Diagram (a) represents the Primakoff process which consists in the conversion of thermally produced photons in the presence of an external electric field of nuclei and electrons $\gamma + Z e \rightarrow a + Z e$ allowing to probe the axion-photon coupling $g_{a \gamma \gamma}$. The Sun is potentially a source of axions, however, the low temperature of its core limits the efficiency of the Primakoff emission rate. Nonetheless, stars lying in Horizontal Branch~(HB) following the Red Giant Branch~(RGB) phase have low core density and high temperature satisfying prime conditions to produce axions via the Primakoff process. To understand the impact of axions on the evolution of HB stars the ratio between the number of stars in the HB and in the upper portion of the RGB is defined~$\mathcal{R}=N_{\text{HB}}/N_{\text{RGB}}$. This parameter in the presence of axions will be modified by the $g_{a \gamma \gamma}$ and axion-electron $g_{a e e}$ couplings namely lowering the $\mathcal{R}$ value. By comparing it to observational evidence and neglecting the axion-electron coupling the following upper bound is obtained~\cite{Ayala:2014pea,Straniero:2015nvc},
    \begin{equation}
      g_{a \gamma \gamma} < 0.65 \times 10^{-10} \ \text{GeV}^{-1} \ (95\% \ \text{CL}) \; ,
    \end{equation}
    known as the HB bound, which is shown in Fig.~\ref{fig:exp1}. Furthermore, the theoretical predictions lead to a larger $\mathcal{R}$ value than the one observed which can hint at the presence of axions. If these are considered to solve this discrepancy one obtains the HB hint~\cite{Ayala:2014pea,Straniero:2015nvc},
    \begin{equation}
      g_{a \gamma \gamma} = \left(0.29 \pm 0.18 \right) \times 10^{-10} \ \text{GeV}^{-1} \ (68\% \ \text{CL}) \; .
    \end{equation}
\begin{figure}[t!]
   \centering
    \includegraphics[scale=0.33]{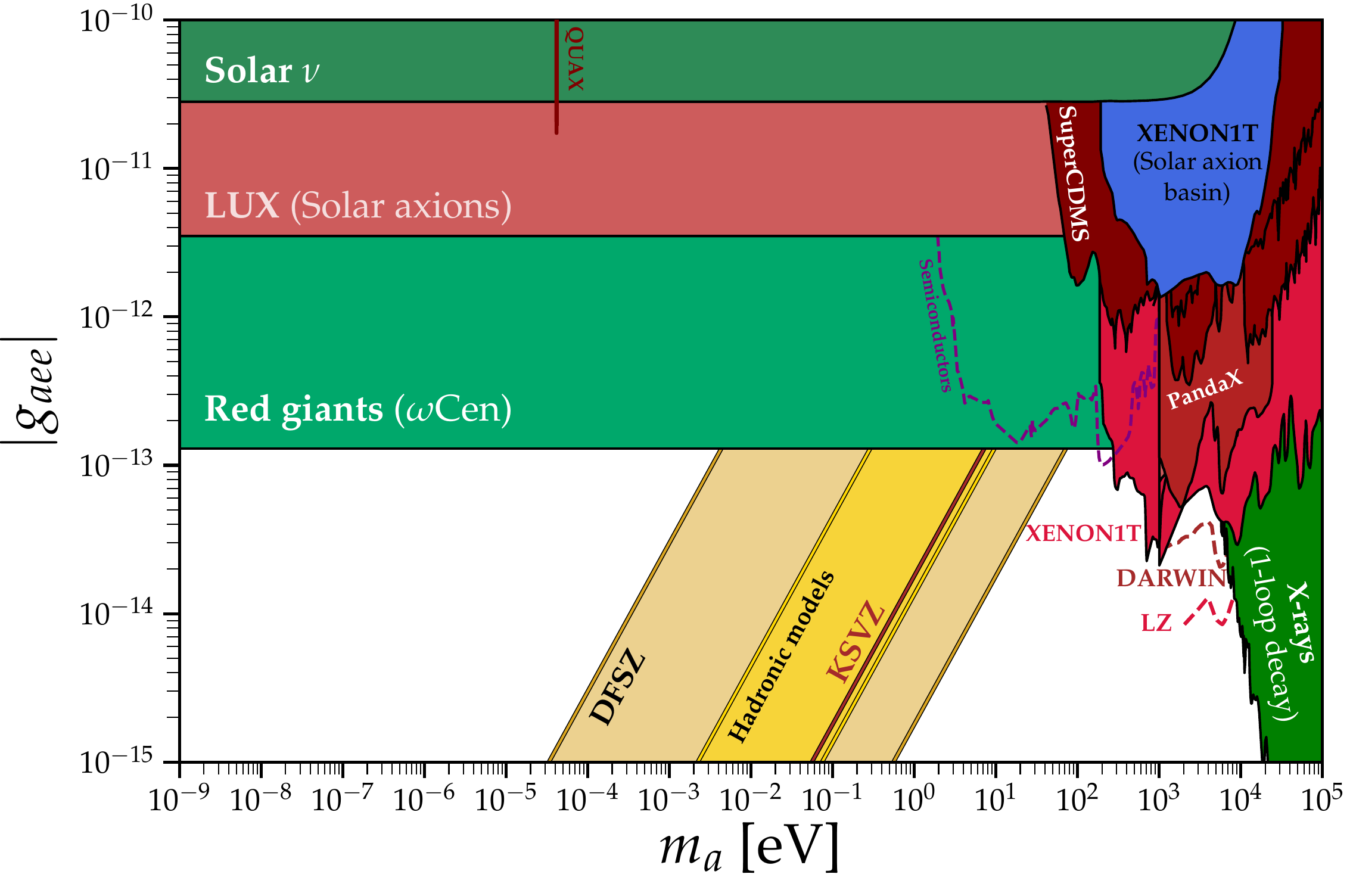}
    \caption{Current and projected constraints on the axion-electron coupling in terms of the axion mass. The astrophysical bounds dominate the experimental constraints along the the majority of the mass range. Are also shown current and future WIMP DD DM experiments. Details are provided in the main text. The yellow band corresponds to the QCD axion parameter space where are indicated the KSVZ and DFSZ invisible axion models discussed in Sec.~\ref{sec:models}. Figure taken from Ref.~\cite{Adams:2022pbo}.}
    \label{fig:exp2}
\end{figure}

\textit{Axion-electron coupling:} Diagram (b) and (c) represents axion-electron/ion bremsstrahlung $e + Z e \rightarrow e + Z e + a$ and Compton scattering $\gamma + e \rightarrow a + e$, respectively. These processes allow to constrain the axion-electron coupling $g_{a e e}$. The former is the dominant production channel of axions in high density environments ruled by electron degeneracy such as white dwarfs~(WD), as well as RGB stars.

The evolution of a WD is determined by its cooling, which is usually governed by photon and neutrino emission, but may also proceed through new channels, such as axion emission. Two possible ways are employed to study WD cooling. The first one relies on the analysis of WDs distribution in terms of their luminosity, this is know as the WD Luminosity function~(WDLF). By cooling a WD luminosity decreases and therefore the efficiency of this process will alter the WDLF. Secondly, one can study WD oscillation period drift which is approximately proportional to the cooling rate. However, the WDLF results on the axion-electron coupling suffer from large theoretical and observational uncertainties~\cite{MillerBertolami:2014rka}. As for the period change since it varies very slowly, large datasets taken over decades are required to obtain accurate results. WD constraints on $g_{a \gamma \gamma}$ are shown in the upper right corner of Fig.~\ref{fig:exp1}.

The study of the luminosity of RGB stars allows to obtain bounds on $g_{a e e}$. Namely, once the hydrogen H in main sequence stars cores is exhausted, their core composed of helium He contracts and they enter the RGB phase of their evolution. During this phase the outer layer of the star expands, its surface cools down and consequently the luminosity increases. Simultaneously the core continues to contract, heats up by burning the remaining H into He, until reaching the He ignition point. When the He-flash occurs the star attains its highest luminosity known as the RGB tip. Probing the luminosity of RGB tip is excellent to understand star cooling which can be altered by axions. From the combined studies of RGB tip from the M5 and M3 galactic clusters the following bound is obtained~\cite{Viaux:2013lha,Straniero:2018fbv},
\begin{equation}
g_{a e e} \leq 3.1 \times 10^{-13} \ (95\% \ \text{CL}) \; ,
\end{equation}
which is indicated in Fig.~\ref{fig:exp2}, being the dominant experimental constraint on $g_{a e e}$ along the majority of the axion mass range.

Combining the hints provided by WD cooling, WDLF and RGB stars give the following $1 \sigma$ preferred interval for the axion-electron coupling~\cite{Giannotti:2017hny},
    \begin{equation}
      g_{a e e} = 1.6^{0.29}_{-0.34}  \times 10^{-13} \; .
    \end{equation}
\begin{figure}[t!]
    \centering
    \begin{subfigure}[b]{0.45\textwidth}
         \centering
         \includegraphics[scale=0.07,trim={0cm 0cm 70cm 0cm},clip]{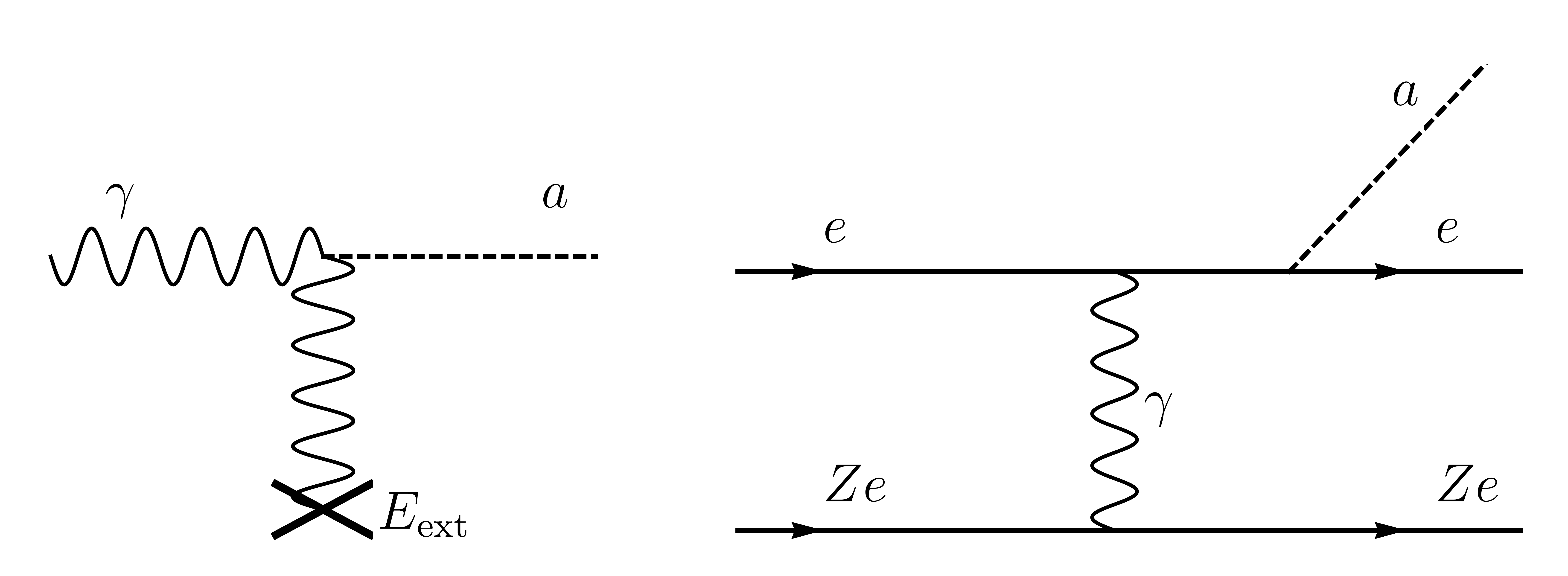}
         \caption{Axion Primakoff process in an external field.}
         \label{fig:aprimakoff}
     \end{subfigure}
     \hspace{+0.5cm}
     \begin{subfigure}[b]{0.45\textwidth}
         \centering
         \includegraphics[scale=0.07,trim={55cm 0cm 0cm 0cm},clip]{Figures/Axions/astro1.png}
         \caption{Axion bremsstrahlung.}
         \label{fig:abrem}
     \end{subfigure} \\
    \begin{subfigure}[b]{0.45\textwidth}
         \centering
         \includegraphics[scale=0.07,trim={0cm 3cm 0cm 0cm},clip]{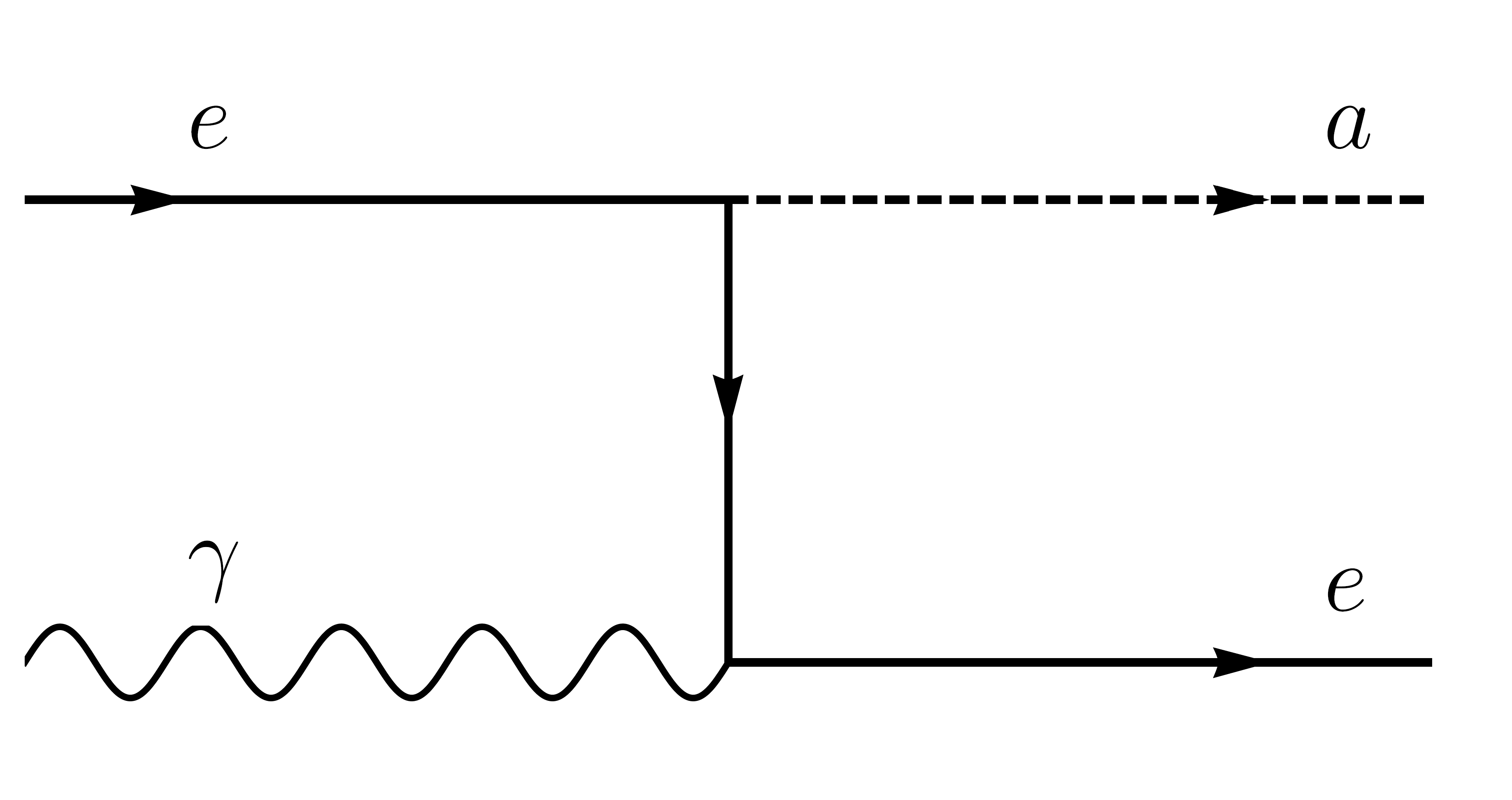}
         \caption{Axion Compton scattering.}
         \label{fig:acompton}
     \end{subfigure}
     \begin{subfigure}[b]{0.45\textwidth}
         \centering
         \includegraphics[scale=0.07]{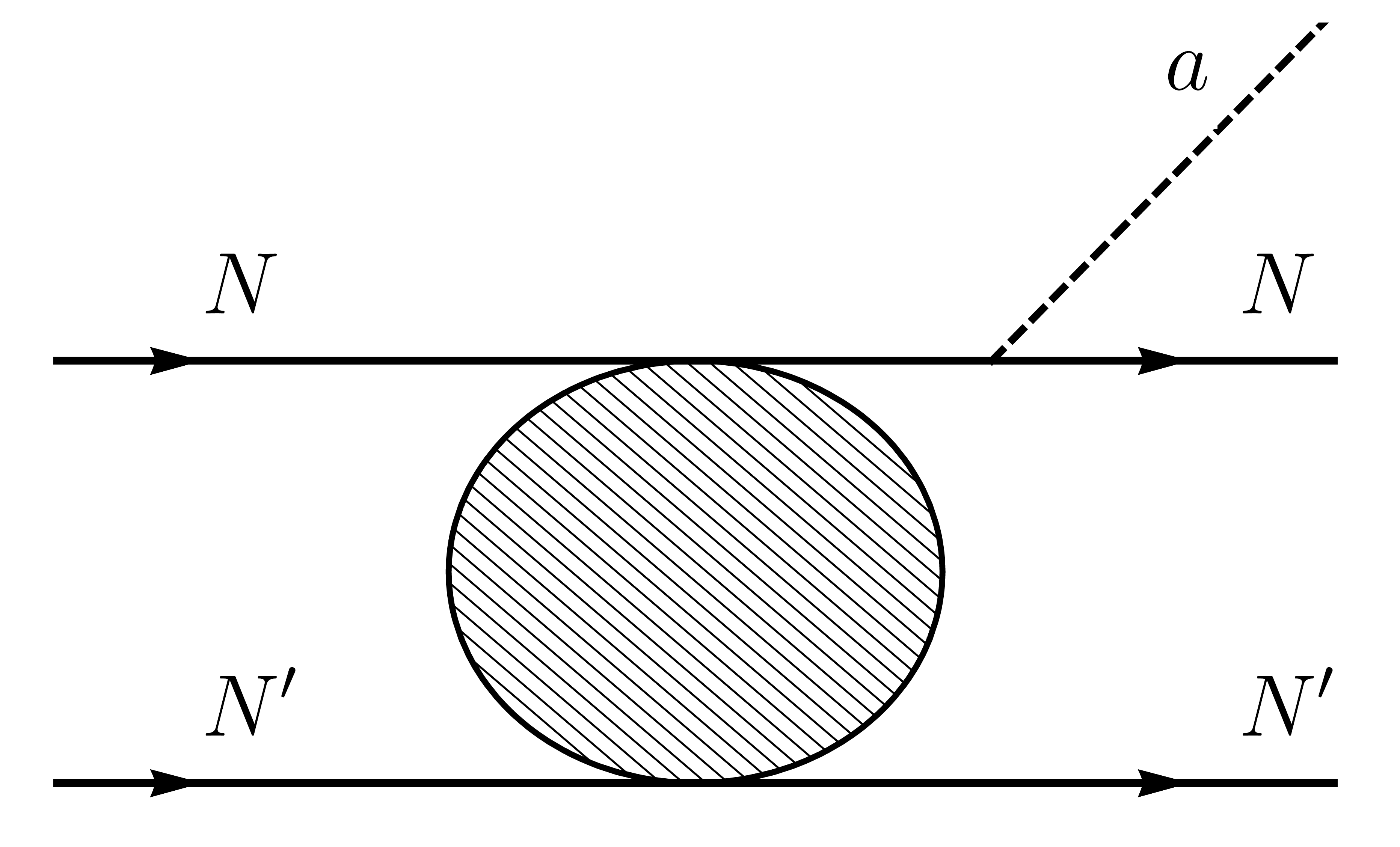}
         \caption{Axion nuclear bremsstrahlung with $N,N^\prime$ being a nucleon i.e., proton or neutron.}
         \label{fig:anucleon}
     \end{subfigure}
    \caption{Axion production mechanisms that can occur in astrophysical objects/sources (see text for details). The diagrams were obtained from Ref.~\cite{DiLuzio:2020wdo}.}
    \label{fig:astrodiag}
\end{figure}

  \textit{Axion-nucleon coupling:} Diagram (d) is the nucleon bremsstrahlung process $N + N^\prime \rightarrow N+N^\prime+a$, with $N$ and $N^\prime$ being either a proton $p$ or neutron $n$, this process provides information on the axion-proton and nucleon couplings respectively labelled~$g_{a n}$ and $g_{a p}$.
  
  Observations of neutron star~(NS) cooling and more importantly on SN signal allows to constrain the axion-nucleon coupling. The more established constraints comes from the observed neutrino signal of SN1987a. The duration of this signal is dependent on the cooling efficiency and is compatible with the fact the SN neutrinos carry $99 \%$ of the total energy released in the SN explosion. A bound can be established for axions since they cannot contribute more than neutrinos to the cooling of SN1987a. The latest results is~~\cite{Keil:1996ju,Raffelt:2006cw}
    \begin{equation}
      g_{a n}^2 + 0.29 g_{a p}^2 + 0.27 g_{a n} g_{a p} \lesssim 3.25 \times 10^{-18} \; ,
      \label{eq:SN1987aaxionbound}
    \end{equation}
    which is taken into account in Figs.~\ref{fig:exp1} and  \ref{fig:exp3}.
    
    Astrophysical evidence can provide model-independent constrains directly on the axion decay constant $f_a$ or equivalently on its mass $m_a$. In fact, insights from axion coupling to gravity can be inferred from \textit{Black Hole~(BH) superradiance}. Whenever the axion's Compton length is of the order of BH radii axion gravitational bound states will form around the BH. The supperadiance phenomena leads to the axion occupation numbers to grow exponentially and efficiently extracts mass and angular momentum from the BH. Thus one can determine the presence or lack off axions by observing BH masses and spins. At present the exclusion region is given by~\cite{Arvanitaki:2014wva,Dafni:2018tvj},
    \begin{equation}
      6 \times 10^{17} \ \text{GeV} \leq f_a \leq 10^{19} \ \text{GeV} \Leftrightarrow 6 \times 10^{-13} \ \text{eV} \leq m_a \leq 10^{-11} \ \text{eV} \; ,
      \label{eq:BHaxionbound}
    \end{equation}
    which is show in Figs.~\ref{fig:exp1} and \ref{fig:exp3}. This analysis does not require axions to be initially present i.e., there is no need to consider DM axions, since superradiance can start from quantum mechanical fluctuations.
    
    \item \textbf{Solar axions and helioscopes:} As mentioned earlier, the Sun constitutes a potential source of axions, which could be produced via the Primakoff process [see diagram (a) of Fig.~\ref{fig:astrodiag}]. Additional processes may also contribute to the conversion of thermally produced photons into axions, namely axio-recombination and de-excitation, bremsstrahlung, and Compton processes (collectively referred to as ABC; see Fig.~\ref{fig:astrodiag}). The total number of axions that could be emitted by the Sun per second is given by~\cite{DiLuzio:2020wdo},
    \begin{equation}
        \frac{d N_a}{d t} = 1.1 \times 10^{39} \left[ \left(\frac{g_{a \gamma \gamma}}{10^{-10} \ \text{GeV}^{-1}} \right)^2 + 0.7 \left( \frac{g_{a e e}}{10^{-12}}\right)^2 \right] \ s^{-1} \; ,
    \end{equation}
    where the Primakoff ($g_{a \gamma \gamma}$) and ABC ($g_{a e e}$) processes provide a similar contributions to the solar axion flux. However, the ABC flux will peak at slightly lower energies than the Primakoff one, allowing to distinguish both contributions. This information can be a discriminant between QCD axion models  (see Sec.~\ref{sec:models}) e.g., for the original KSVZ model $E=0$ there is no axion-electron tree-level coupling and therefore $g_{a e e}/g_{a \gamma \gamma} \approx 0$ while for the DFSZ I $g_{a e e}/g_{a \gamma \gamma} \approx 20 \sin^2 \beta$ and for the  DFSZ II $g_{a e e}/g_{a \gamma \gamma} \approx 12 \cos^2 \beta$.
    
    Helioscopes experiments aim at detecting the solar axion flux. For example, the Sikivie helioscope uses a strong laboratory transverse magnetic field $B$ to coherently convert solar axions into X-ray photons which can be detected experimentally. The conversion probability is given by~\cite{Graham:2015ouw},
    \begin{equation}
        P_{a \rightarrow \gamma} = \left(\frac{g_{a \gamma \gamma} B L}{2}\right)^2 \frac{\sin^2(q L/2)}{(qL/2)^2} \; ,
        \label{eq:Pag}
    \end{equation}
    where $L$ and $q=q_\gamma-q_a$ are the magnet length and momentum transfer between the axion and photon induced by the magnet, respectively. In Fig.~\ref{fig:exp1}, numerous exclusion regions and projected sensitivities of helioscopes experiments are shown. Notably, the CERN Solar Axion Experiment~(CAST)~\cite{Zioutas:1998cc,CAST:2013bqn} excludes $g_{a \gamma \gamma} \geq 0.66 \times 10^{-10} \ \text{GeV}^{-1}$ for masses $m_a \leq 20$ meV. Additionally, the future International Axion Observatory~(IAXO)~\cite{Irastorza:2011gs,Armengaud:2014gea} (and BabyIAXO~\cite{IAXO:2019mpb}) is expected to probe $g_{a \gamma \gamma}$ down to $(10^{-12}-10^{-11}) \ \text{GeV}^{-1}$ reaching QCD axion models parameter space (yellow band).

    Although DD experiments are designed to probe the WIMP DM parameter space -- see Sec.~\ref{sec:DM} -- they provide complementary probes on axions. Experiments such as LZ~\cite{LUX:2016ggv}, PandaX~\cite{PandaX-II:2016vec} and XENON1T~\cite{XENON:2018voc}, are able to detect solar axions through the so-called axioelectric effect. These experiments set constraints on the axion-electron coupling which as seen in Fig.~\ref{fig:exp2} are just able to cover a small portion of parameter space not already constrained by astrophysical bounds. One can infer that future sensitivities of LZ~\cite{LUX-ZEPLIN:2018poe}, DARWIN~\cite{DARWIN:2016hyl}, PandaX~\cite{PandaX:2018wtu} and XENONnT~\cite{XENON:2020kmp} will not bring a clear cut competition to the astrophysical constraints. 

\begin{figure}[t!]
   \centering
    \includegraphics[scale=0.08]{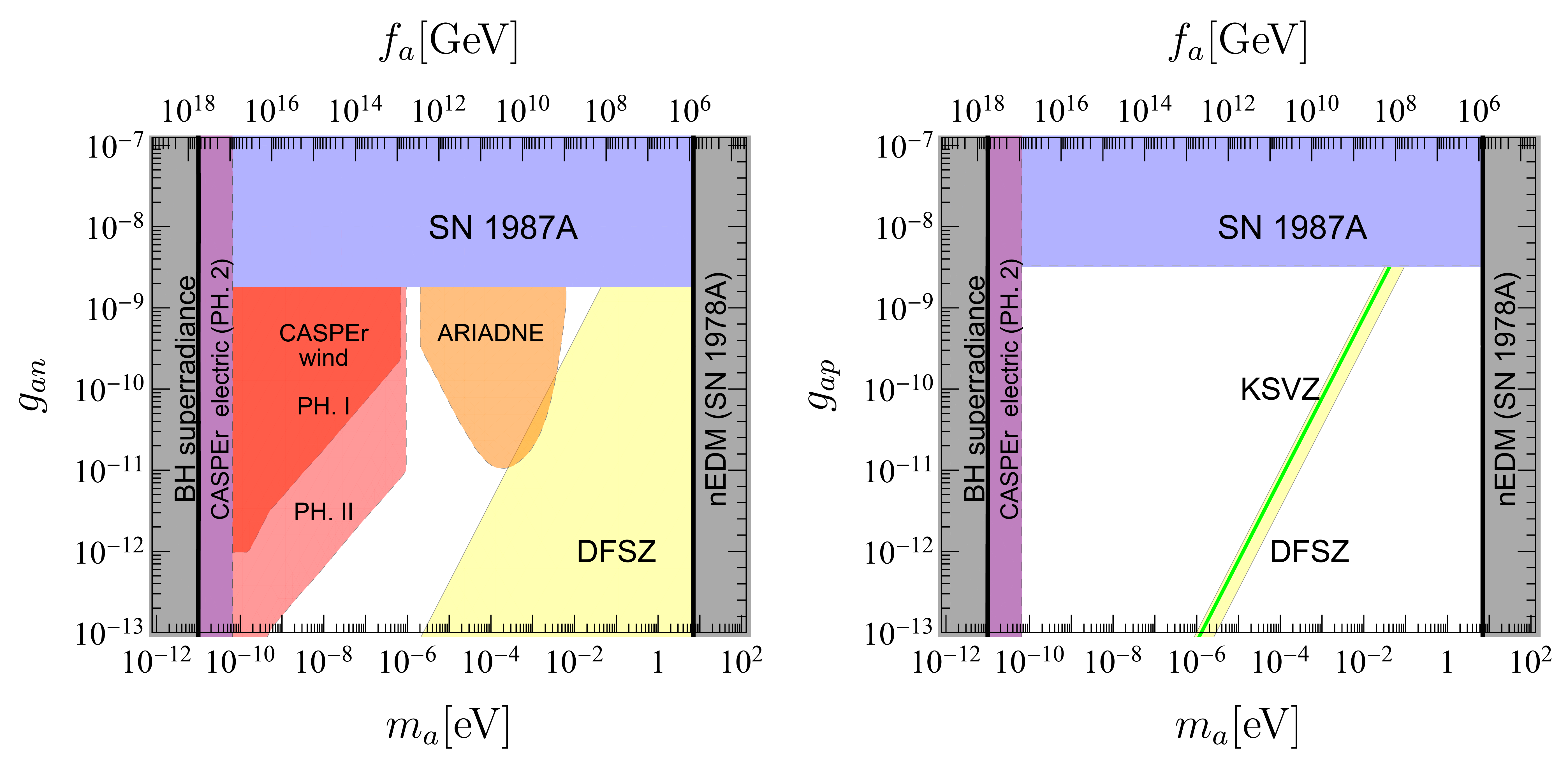}
    \caption{Constraints on axion-nucleon interactions for neutron (left) and proton (right) in terms of the axion mass (bottom axis) and decay constant (upper axis). Are included astrophysical constraints stemming from SN1987a, BH superradiance, as well as future experiments sensitive to $g_{a n}$ CASPERr and ADRIANE (see text for details). The yellow band corresponds to the DFSZ axion parameter space (left and right plots) while the green one to the KSVZ axion (right plot). Notably, the axion-proton coupling is still largely unprobed. Figures can be found in Ref.~\cite{DiLuzio:2020wdo}.}
    \label{fig:exp3}
\end{figure}

    \item \textbf{Haloscopes and DM axions:} The haloscope experiments aim at detecting non-relativistic axions with the assumption that they make up the totality of the observed DM in the Universe. Since the Milky Way has a DM halo with a local DM density estimated to be around $0.45 \ \text{GeV}/\text{cm}^3$ if $\Omega_a h^2 = \Omega_{DM} h^2$ one should expect locally the cold axion density to be about $4.5 \times 10^{14} (\mu\text{eV}/m_a) \ \text{cm}^{-3}$. Experimental techniques have been developed in order to detect these DM axions the most conventional one uses a resonant cavity haloscope~\cite{Sikivie:1983ip}. The idea is to exploit the axion conversion through the Primakoff process in a strong magnetic field which embeds the resonant microwave cavity. The resonant cavity frequency is tuned with the axion one in such a way that a resonance peak could be observed in the DM halo spectrum. Current constraints from haloscopes like ADMX~\cite{ADMX:2001dbg,Stern:2016bbw,Lewis:2017gno,ADMX:2018gho,ADMX:2019uok,ADMX:2020ote,ADMX:2021nhd}, RBF~\cite{DePanfilis:1987dk}, CAPP~\cite{CAPP:2020utb} and HAYSTAC~\cite{HAYSTAC:2020kwv}, rule out the magenta shaded region in Fig.~\ref{fig:exp1} for the axion-to-photon coupling $g_{a \gamma \gamma}$. From all theses experiments the most impressive is the Axion Dark Matter eXperiment~(ADMX)~\cite{ADMX:2001dbg,Stern:2016bbw,Lewis:2017gno,ADMX:2018gho,ADMX:2019uok,ADMX:2020ote,ADMX:2021nhd}. In fact, from Fig.~\ref{fig:exp1} it is clear that ADMX has already reached the KSVZ and DFSZ QCD axion region (yellow band) for masses $m_a \sim 3 \ \mu$eV. The future ADMX generations are expected to probe the entire QCD axion parameter space for masses $1 \ \mu \text{eV} \lesssim m_a \lesssim 100 \ \mu$eV.

    Some future experiments are projected to probe higher mass regions relying on an ultra low temperature cavity. Namely, ORGAN~\cite{McAllister:2017lkb} will probe the QCD axion region for masses $60 \ \mu \text{eV} \lesssim m_a \lesssim 210 \ \mu$eV. Additionally, MADMAX~\cite{MADMAX:2019pub} takes a different approach, employing movable dielectric discs to enhance the photon signal and is projected to swipe the axion mass region $50 \ \mu \text{eV} \lesssim m_a \lesssim 100 \ \mu$eV. Moreover, the Axion Longitudinal Plasma HAloscope (ALPHA)~\cite{Lawson:2019brd,Wooten:2022vpj,ALPHA:2022rxj} is projected to probe the QCD axion band mass region $[20,200] \ \mu \text{eV}$, being complementary to MADMAX. 
    
    For lower mass ranges $ 0.2 \ \mu \text{eV} \lesssim m_a \lesssim 40\ \mu$eV the proposal to test the axion DM premise is through the detection of radio signals stemming from axion conversion to photons in NS magnetosphere~(DM-Radio)~\cite{Hook:2018iia,Edwards:2019tzf,Leroy:2019ghm}. Also, ``A Broadband/Resonant Approach to Cosmic Axion Detection with an Amplifying B-field Ring Apparatus" (ABRACADABRA)~\cite{Ouellet:2018beu}, is expected to probe the mass region below $\sim 1 \mu \text{eV}$ and reach the yellow QCD axion band, being an important complementary probe of axion DM compared to ``traditional" haloscope setups like ADMX.

    In the left plot of Fig.~\ref{fig:exp3} the projected sensitivity (in 5-7 years) on $g_{a n}$ of the Cosmic Axion Spin Precession Experiment (CASPEr)~\cite{JacksonKimball:2017elr} is shown. This experiment exploit the nEDM coupling to axions by trying to measure through nuclear magnetic resonance techniques techniques the oscillating axion field which if present alters the nEDM value.
    
    \item \textbf{Laboratory searches:} Searches for axions produced in laboratory provide a very important setup to probe the axion/ALPs parameter space. In fact, these types of experiments avoid uncertainties related to the indirect astrophysical measurements namely of the axion flux. In Fig.~\ref{fig:exp1} are show the current/future constraints/sensitivity of laboratory experiments namely, Any Light Particle Search~(ALPS) I~\cite{Ehret:2010mh} and II~\cite{Bahre:2013ywa}, OSQAR~\cite{OSQAR:2015qdv}, PVLAS~\cite{DellaValle:2015xxa} and in the left plot of Fig.~\ref{fig:exp3} is shown the projected sensitivity of ADRIANE~\cite{ARIADNE:2017tdd}.
    
    Focusing on Fig.~\ref{fig:exp1}. The ALPS-I searches use the light shinning through a wall method. The idea is to use a laser beam to produce axions which pass through a wall which is opaque for light. Axions are then reconverted thanks to a transverse magnetic field into photons [see Eq.~\eqref{eq:Pag}]. Currently, ALPS-I has already probed the $m_a \lesssim 100 \ \mu$eV mass region for $g_{a \gamma \gamma} \gtrsim 10^{-7} \ \text{GeV}^{-1}$, not being competitive with the CAST experiment (see previous discussion and figure). The next generation of this experiment ALPS-II is set to surpass CAST reaching $g_{a \gamma \gamma}$ down to $\sim 5 \times 10^{-10} \ \text{GeV}^{-1}$ for masses below $100 \ \mu$eV. The current strongest limit on $g_{a \gamma \gamma}$ for these type of experiments was set by OSQAR namely, for $m_a \leq 300 \mu$eV it excludes $g_{a \gamma \gamma} > 3.5 \times 10^{-8} \ \text{GeV}^{-1}$ at $95 \%$ CL. The PVLAS experiment reaches comparable constraints as OSQAR. Overall, the sensitivities of these experiments are way above the ones required to probe the QCD axion region, nonetheless they probe ALPs territory.
    
    Looking now at the left plot of Fig.~\ref{fig:exp3}. Using nuclear magnetic resonance techniques~\cite{Arvanitaki:2014dfa} ADRIANE can potentially detect axions being sensitive only to the $g_{a n}$ coupling. Compared to other laboratory searches ADRIANE is the only one expected to reach the DFSZ axion region (yellow band).
    
\end{itemize}
Note that, as shown in the right plot of Fig.~\ref{fig:exp3} the axion-proton coupling $g_{a p}$ will remain in the foreseeable future completely unprobed by the experimental searches mentioned above being only constrained by astrophysical bounds. Overall the experimental prospects for the search of axions and ALPs are very promising. Within the next decades the QCD axion parameter space will be in principle severely constrained.

%%%%%%%%%%%%%%%%%%%%%%%%%%%%%%%%%%%%%%%%%%%%%%%%%%%%%%%%%%%%%%%%%%%%%%%%%%%%%
\section{Flavor-violating axions}
\label{sec:flavouredaxions}
%%%%%%%%%%%%%%%%%%%%%%%%%%%%%%%%%%%%%%%%%%%%%%%%%%%%%%%%%%%%%%%%%%%%%%%%%%%%%

%
\begin{table}[t!]
    \renewcommand*{\arraystretch}{1.5}
    \centering
    \begin{tabular}{|c|c|c|c|c|}
    \hline 
     Most Restrictive Processes & $\; \alpha, \beta \;$   & $\mathbf{F}^V_{\alpha \beta}$ (GeV)  & $\mathbf{F}^A_{\alpha \beta}$ (GeV)  & Experiment \& Ref. \\
    \hline
    $K^+\rightarrow \pi^+ a$ & $\; s,d \;$ & \renewcommand{\arraystretch}{1.0}\begin{tabular}[c]{@{}c@{}} $6.8\times10^{11}$ \\ $(2\times10^{12})$\end{tabular} & -  &  \renewcommand{\arraystretch}{1.0}\begin{tabular}[c]{@{}c@{}} E949 \& E787~\cite{E949:2007xyy} \\ (NA62 \& KOTO~\cite{MartinCamalich:2020dfe})\end{tabular}
     \\
    \hline
    $\Lambda\rightarrow$ $n$ $a$ (Supernova)  & $\; s,d \;$ & $7.4\times10^9$ & $5.4\times 10^9$ & PDG~\cite{ParticleDataGroup:2024cfk}
     \\
    \hline
    $D^+\rightarrow \pi^+ a$   & $\; c,u \;$ & \renewcommand{\arraystretch}{1.0}\begin{tabular}[c]{@{}c@{}} $9.7\times10^{7}$ \\ $(5\times10^{8})$\end{tabular} & - &     \renewcommand{\arraystretch}{1.0}\begin{tabular}[c]{@{}c@{}} CLEO~\cite{CLEO:2008ffk}  \\ (BES III~\cite{MartinCamalich:2020dfe}) \end{tabular}
     \\
    \hline
    $\Lambda_c \rightarrow$ $p$ $a$ & $\; c,u \;$ & \renewcommand{\arraystretch}{1.0}\begin{tabular}[c]{@{}c@{}} $1.4\times10^{5}$ \\ $(2\times10^{7})$\end{tabular} & \renewcommand{\arraystretch}{1.0}\begin{tabular}[c]{@{}c@{}} $1.2\times10^{5}$ \\ $(2\times10^{7})$\end{tabular} &  
    \renewcommand{\arraystretch}{1.0}\begin{tabular}[c]{@{}c@{}} PDG~\cite{ParticleDataGroup:2024cfk} \\ (BES III~\cite{MartinCamalich:2020dfe}) \end{tabular}
    \\
    \hline
    $B^{+,0}\rightarrow K^{ +,0}a$  & $\; b,s \;$ & \renewcommand{\arraystretch}{1.0}\begin{tabular}[c]{@{}c@{}} $3.3\times10^{8}$ \\ $(3\times10^{9})$\end{tabular} & - &     \renewcommand{\arraystretch}{1.0}\begin{tabular}[c]{@{}c@{}} BABAR~\cite{BaBar:2013npw}   \\ (BELLE II~\cite{MartinCamalich:2020dfe}) \end{tabular}
    \\
    \hline
    $B^{+,0}\rightarrow K^{\ast +,0}a$ & $\; b,s \;$ & - & \renewcommand{\arraystretch}{1.0}\begin{tabular}[c]{@{}c@{}} $1.3\times10^{8}$ \\ $(1\times10^{9})$\end{tabular} &   \renewcommand{\arraystretch}{1.0}\begin{tabular}[c]{@{}c@{}} BABAR~\cite{BaBar:2013npw}   \\ (BELLE II~\cite{MartinCamalich:2020dfe}) \end{tabular}
    \\
    \hline
    $B^{+}\rightarrow \pi^{+}a$ & $\; b,d \;$ & \renewcommand{\arraystretch}{1.0}\begin{tabular}[c]{@{}c@{}} $1.1\times10^{8}$ \\ $(3\times10^{9})$\end{tabular} & - &  \renewcommand{\arraystretch}{1.0}\begin{tabular}[c]{@{}c@{}} BABAR~\cite{BaBar:2004xlo}   \\ (BELLE II~\cite{MartinCamalich:2020dfe}) \end{tabular} 
    \\
    \hline
    $B^{+,0}\rightarrow \rho^{ +,0}a$ & $\; b,d \;$ & - &  $(1\times10^{9})$ & (BELLE II~\cite{MartinCamalich:2020dfe})
    \\
    \hline
    $K^{+}\rightarrow \pi^{ +}a$ (loop) & $\; t,u \;$ & $3\times10^8$ &  $3\times10^8$ &  E949 \& E787~\cite{E949:2007xyy}
    \\
    \hline
    $K^{+}\rightarrow \pi^{ +}a$ (loop) & $\; t,c \;$ & $7\times10^8$ &  $7\times10^8$ &  E949 \& E787~\cite{E949:2007xyy}
    \\
    \hline
\end{tabular}
\caption{The most relevant $90\%$ CL lower bounds on the scales of flavor-violating axion-quark couplings $\mathbf{F}_{\alpha \beta}^{V,A}$ defined in Eq.~\eqref{eq:Fdef}, with future projections in parentheses (obtained from Ref.~\cite{MartinCamalich:2020dfe}).}
\label{tab:QuarkConstraints}
\end{table}
\begin{table}[t!]
    \renewcommand*{\arraystretch}{1.5}
    \centering
    \begin{tabular}{|c|c|c|c|c|}
    \hline 
     Most Restrictive Processes & $\; \alpha,\beta \;$   & $\mathbf{F}_{\alpha\beta}$ (GeV)    & Experiment \& Ref. \\
    \hline
    Star Cooling & $\; e,e \;$ & $4.6\times10^9$  & WDs~\cite{MillerBertolami:2014rka} 
\\
    \hline
    Red Giants & $\; e,e \;$ & $6.4\times10^9$  & TRGB~\cite{Capozzi:2020cbu,Bottaro:2023gep}
     \\
    \hline
   Star Cooling  & $\; \mu,\mu \;$  & $3.2\times 10^7$ & SN$1987$a$_{\mu \mu}$~\cite{Bollig:2020xdr,Croon:2020lrf,Caputo:2021rux}
     \\
    \hline
    $\mu \rightarrow$ $e$  $a$ $\gamma$   & $\; \mu,e \;$ & $(5.1 -  8.3)\times10^8$  & Crystal Box~\cite{Bolton:1988af}
     \\
    \hline
    $\tau \rightarrow$ $e$  $a$  & $\; \tau,e \;$ & \renewcommand{\arraystretch}{1.0}\begin{tabular}[c]{@{}c@{}} $4.3\times10^{6}$ \\ $(7.7\times10^{7})$\end{tabular}  &  \renewcommand{\arraystretch}{1.0}\begin{tabular}[c]{@{}c@{}} ARGUS~\cite{ARGUS:1995bjh} \\ (BELLE II~\cite{Calibbi:2020jvd}) \end{tabular}
    
    \\
    \hline
    $\tau \rightarrow$ $\mu$  $a$   & $\; \tau,\mu \;$ & \renewcommand{\arraystretch}{1.0}\begin{tabular}[c]{@{}c@{}} $3.3\times10^{6}$ \\ $(4.9\times10^{7})$\end{tabular}  & \renewcommand{\arraystretch}{1.0}\begin{tabular}[c]{@{}c@{}} ARGUS~\cite{ARGUS:1995bjh} \\ (BELLE II~\cite{Calibbi:2020jvd})  \end{tabular}
    \\
    \hline
\end{tabular}
\caption{The most relevant $95\%$ CL, and $90\%$ CL for the Crystal Box experiment, lower bounds on the scales of flavor-violating axion-lepton couplings $\mathbf{F}_{\alpha \beta}$ defined in Eq.~\eqref{eq:Fdef}, with future projections in parentheses (obtained from Ref.~\cite{Calibbi:2020jvd}).}
\label{tab:LeptonConstraints}
\end{table}
The PQ symmetry, originally proposed to solve the strong CP problem, does not need to act universally across fermion generations. Unlike gauge symmetries, global anomalous symmetries are not constrained to be flavor-blind, a possibility already noted in the seminal works~\cite{Bardeen:1977bd,Davidson:1981zd,Wilczek:1982rv}. This has motivated models in which the PQ symmetry also acts as a horizontal U(1) flavor symmetry, potentially accounting for the observed hierarchies in fermion masses and mixings. Namely, in Chapter~\ref{chpt:flavoraxion} we will study minimal flavored-PQ symmetries in the context of the DFSZ model with type-I seesaw neutrino masses. In such scenarios the axion couples to fermions of different generations leading to a rich flavor-violating axion-fermion couplings phenomenology. Furthermore, we will present in Chapter~\ref{chpt:axionneutrino}, Sec.~\ref{sec:colormediatedDirac}, a unified KSVZ framework where Dirac neutrino masses are generated radiatively via the exotic quarks added to solve the strong CP problem. By performing a systematic analysis of all VLQ representations that mix with the SM quarks we show that these also induce flavor-violating axion-quark interactions. In this section are gathered the constraints on axion flavor violating couplings to quarks and charged leptons~\cite{MartinCamalich:2020dfe,Alonso-Alvarez:2023wig}, useful for our studied in the upcoming chapters of this thesis. We will use the following notation to describe the effective flavor-violating axion-fermion interactions~\cite{MartinCamalich:2020dfe,Calibbi:2020jvd}:
\begin{align}
    \mathcal{L}_{aff} = \frac{\partial_\mu a}{2 f_a} \overline{f_\alpha} \gamma^\mu \left(\mathbf{C}^{V, f}_{\alpha \beta} + \mathbf{C}^{A, f}_{\alpha \beta} \gamma_5 \right) f_\beta \; 
    ,
\label{eq:axionFermionLagrangian}
\end{align}
where $f=u,d,e$ the $3\times3$ matrices $\mathbf{C}^V$ and $\mathbf{C}^A$ denote the vector and axial flavor-violating quark interactions. In Tables~\ref{tab:QuarkConstraints} and~\ref{tab:LeptonConstraints}, are provided the most restrictive constraints on
\begin{align}
\mathbf{F}_{\alpha \beta}^{V,A} \equiv 2f_a/|\mathbf{C}_{\alpha \beta}^{V,A}| \;, \; \mathbf{F}_{\alpha \beta} \equiv 2f_a/\sqrt{|\mathbf{C}_{\alpha \beta}^{V}|^2 + |\mathbf{C}_{\alpha \beta}^{A}|^2} \; ,
\label{eq:Fdef}
\end{align}
for quarks~\cite{MartinCamalich:2020dfe,Alonso-Alvarez:2023wig} and leptons~\cite{Calibbi:2020jvd}, respectively (for a recent review on flavor-violating constraints see Ref.~\cite{MartinCamalich:2025srw}). The experiments shown in the tables provide a sensitive probe of BSM axion models, complementary to other new physics searches at colliders or rare processes in the flavor sector, as well as to the axion experiments described in Sec.~\ref{sec:experimental}.

%------------
% CHAPTER 05    
%------------

%%%%%%%%%%%%%%%%%%%%%%%%%%%%%%%%%%%%%%%%%%%%%%%%%%%%%%%%%%%%%%%%%%%%%%%%%%%%%
\chapter{Axion-frameworks with color-mediated neutrino masses} 
\label{chpt:axionneutrino}
%%%%%%%%%%%%%%%%%%%%%%%%%%%%%%%%%%%%%%%%%%%%%%%%%%%%%%%%%%%%%%%%%%%%%%%%%%%%%

In this chapter, we explore a novel idea in which neutrino masses are generated at the quantum level via colored mediators that also provide a solution to the strong CP problem based on our works of Refs.~\cite{Batra:2023erw,Batra:2025gzy}. Although reconciling the axion paradigm with an appealing mechanism of neutrino mass generation has been explored in the literature recently, see e.g. Refs.~\cite{Salvio:2015cja,Ballesteros:2016euj,Ballesteros:2016xej,Clarke:2015bea,Sopov:2022bog,Gu:2016hxh,Peinado:2019mrn,Dias:2020kbj,delaVega:2020jcp,Berbig:2022pye}, these frameworks do not unify the origin of neutrino masses with the axion solution. In contrast, our new classes of KSVZ-type axion models (see Chapter~\ref{chpt:axions} on axions and Sec.~\ref{sec:KSVZ}) provide a unified explanation for three otherwise unrelated issues: small neutrino masses, the strong CP problem and DM (see Sec.~\ref{sec:BSMavenues} for discussion of BSM avenues). This unifying approach allows for a direct connection between axion physics and the neutrino sector. We study two distinct classes of such frameworks, distinguished by the nature of the light neutrinos:
\begin{itemize}

    \item Sec.~\ref{sec:colormediatedMajorana} closely follows our work of Ref.~\cite{Batra:2023erw}, where we propose a \textbf{Majorana framework} with neutrino masses arising at the two-loop level through diagrams involving colored fermions and scalars. The PQ symmetry, under which these exotic fermions are chiral, ensures a viable axion solution while allowing for LNV. The resulting effective neutrino mass operator is consistent with the Majorana nature of neutrinos and motivates the search for $0\nu\beta\beta$ decay -- see discussion in Sec.~\ref{sec:neutrinoobservables}. According to the black-box theorem, a positive observation of $0\nu\beta\beta$ would confirm that at least one neutrino is a Majorana particle~\cite{Schechter:1981bd}. However, so far, $0\nu\beta\beta$ decay has not been observed~\cite{Jones:2021cga,Cirigliano:2022oqy,Dolinski:2019nrj}. This could very well indicate that neutrinos, like the remaining charged fermions of the SM, are after all Dirac fermions.
    
    \item  The Dirac approach to neutrino mass generation has also been widely explored in recent literature, and typically requires extra symmetries to forbid operators that lead to Majorana neutrino masses~\cite{Aranda:2013gga}. 
This is the case for the type-I~\cite{Ma:2014qra,Addazi:2016xuh,Bonilla:2017ekt} or type-II~\cite{Valle:2016kyz,Reig:2016ewy,Bonilla:2016zef}
tree-level Dirac seesaw mechanism.
Radiative Dirac neutrino masses has also been discussed, e.g. within the context of the scotogenic framework ~\cite{Bonilla:2016diq,Bonilla:2018ynb,Leite:2020bnb,Leite:2020wjl} -- for a comprehensive analysis see Ref.~\cite{CentellesChulia:2024iom}, see also Sec.~\ref{sec:scotogenic}. Thus, it is natural to consider a Dirac version of the color-mediated neutrino mass idea. In fact, Sec.~\ref{sec:colormediatedDirac} closely follows our work of Ref.~\cite{Batra:2025gzy}, where we propose a \textbf{Dirac framework} with the light neutrinos acquiring mass via a one-loop diagram involving VLQs and scalar leptoquarks. Here, the global PQ symmetry not only addresses the strong CP problem but also enforces lepton number conservation, thereby ensuring the Dirac character of neutrinos.

\end{itemize}
On cosmological grounds, the structure of the models motivates different axion DM scenarios (see discussion in Sec.~\ref{sec:axionDMcosmo}). In the Majorana framework, we consider the pre-inflationary regime, while the Dirac framework favors a post-inflationary scenario. In both cases, the colored fermions introduce rich phenomenology. Depending on their PQ charge assignments and representations under the SM gauge group, the models predict distinct axion-photon couplings that can be tested in haloscope and helioscope experiments previously discussed in Sec.~\ref{sec:experimental}. Moreover, in the Dirac case, mixing between the exotic and SM quarks induces axion-quark flavor-violating interactions as the ones presented in Sec.~\ref{sec:flavouredaxions}, with potential implications for meson mixing, rare decays, and top-quark phenomenology.

%%%%%%%%%%%%%%%%%%%%%%%%%%%%%%%%%%%%%%%%%%%%%%%%%%%%%%%%%%%%%%%%%%%%%%%%%%%%%
\section{Framework for Majorana neutrino masses}
\label{sec:colormediatedMajorana}
%%%%%%%%%%%%%%%%%%%%%%%%%%%%%%%%%%%%%%%%%%%%%%%%%%%%%%%%%%%%%%%%%%%%%%%%%%%%%

%
\begin{table}[t!]
\renewcommand*{\arraystretch}{1.5}
	\centering
	\begin{tabular}{| K{1.5cm} | K{1.5cm} | K{5cm} | K{1.5cm} | K{2.5cm} |}
		\hline
&Fields&\SM&    U($1$)$_{\text{PQ}}$ & Multiplicity\\
		\hline
\multirow{2}{*}{Fermions}&$\Psi_L$&($(p,q),2 n \pm 1, 0$)& $\omega$  & $n_\Psi$\\
&$\Psi_R$&($(p,q),2 n \pm 1, 0$)&  $0$ & $n_\Psi$ \\
		\hline
\multirow{3}{*}{Scalars}&$\sigma$&($\mathbf{1},\mathbf{1}, 0$)& $\omega$ & 1\\
&$\eta$&($(p,q),2 n, 1/2$)& $0$ & $n_\eta$\\
&$\chi_i$&($(p,q)_i,2 n \pm 1, 0$)& $0$ & $n_\chi$ \\
\hline
	\end{tabular}
	\caption{
          Matter content and quantum numbers for our new KSVZ models with two-loop neutrino masses.
          Here, $\omega$ is the PQ charge, and $n = 1,2, \cdots$, $p>q = 0,1,2, \cdots$.}
	\label{tab:general} 
\end{table}
The original KSVZ model~\cite{Kim:1979if,Shifman:1979if} -- presented in Sec.~\ref{sec:KSVZ} -- extends the SM with vector like fermions $\Psi_{L,R}$ in the fundamental representation of SU(3)$_c$, singlets under SU(2)$_L$, and with $Y=0$. A complex scalar singlet $\sigma$ breaks a U($1$)$_{\text{PQ}}$ symmetry spontaneously, providing mass to those exotic fermions. The phase of $\sigma$ corresponds to the axion field $a$. The fact that LH and RH exotic fermions carry different PQ charges ensures the anomalous axion-gluon coupling, required to solve the strong CP problem. 

In this section, following closely our work of Ref.~\cite{Batra:2023erw}, we show that generic $\Psi_{L,R}$ fields in the SU($3)_c$ complex representation $(p,q)$ with $p>q = 0,1,2, \cdots$ can act as Majorana neutrino-mass mediators at the two-loop level. Two scalars $\eta,\chi$ with complex SU($3)_c$ transformation properties are also required to put our mechanism at work. Both $\Psi_{L,R}$ and $\chi$ are hyperchargeless and transform as the same odd SU($2)_L$ representation denoted by $2n \pm 1$. In contrast, $\eta$ has $Y=1/2$ and transforms as an even SU($2)_L$ representation denoted by 
$2n$. As in the original KSVZ prescription, the complex scalar singlet $\sigma$ with nonzero PQ charge $\omega$ is responsible for U($1$)$_{\text{PQ}}$ breaking, giving rise to the axion and to $\Psi_{L,R}$ masses (note that only $\Psi_L$ carries PQ charge $\omega$,
$\Psi_R$ is neutral). Table~\ref{tab:general} lists all the new fields and their transformation properties under the SM and PQ symmetries.

The relevant new Yukawa terms are given by
\begin{align}
- \mathcal{L}_{\text{Yuk.}} &\supset \mathbf{Y}_{\Psi} \overline{\Psi_L} \Psi_R \sigma
                           + \frac{1}{2} \mathbf{Y}_{\chi_j} {\Psi^T_R}~C~\chi_j \Psi_R + \mathbf{Y}_i \ \overline{\ell_L} \; \eta^\ast_i \Psi_R + \text{H.c.} \; ,
\label{eq:LYukgen}
\end{align}
where for simplicity, from now on we omit color and SU(2)$_L$ indices. The multiplicities of $\Psi$, $\eta$ and $\chi$ are $n_\Psi$, $n_\chi$ and $n_\eta$, respectively. Thus $\mathbf{Y}_{\Psi}$, $\mathbf{Y}_{\chi_j}$ and $\mathbf{Y}_i$ are $n_\Psi \times n_\Psi$, $n_\Psi \times n_\Psi$ and $3\times n_\Psi$ complex Yukawa matrices, respectively,
with $i=1, \cdots, n_\eta$ and $j=1, \cdots, n_\chi$.

The key scalar-potential terms responsible for neutrino mass generation are
\begin{align}
    V & \supset \mu_{ijk} \chi_i\chi_j\chi_k + \kappa_{ij} \eta_i^\dagger \Phi  \chi_j + \lambda_{ijk} \Phi^\dagger \eta_i \chi_j \chi_k + \text{H.c.} \; ,
    \label{eq:Vneutrino}
\end{align}
where $\Phi=(\phi^+ , \phi^0)^T$ is the usual SM Higgs doublet. To preserve the SU($3)_c$ symmetry the colored scalars $\eta$ and $\chi$ must not acquire a VEV, so that the only VEVs are $\vev{\sigma} = v_\sigma/\sqrt{2}$ breaking U($1$)$_{\text{PQ}}$, and, as usual $\vev\phi^0 = v/\sqrt{2}$, triggering EWSB.

The PQ field $\sigma = (v_\sigma + \rho) \exp(i a /v_\sigma)/\sqrt{2}$ contains the axion $a$ and the radial mode $\rho$. Once $\sigma$ develops a non-zero $v_\sigma$, the PQ symmetry is spontaneously broken at a scale $f_{\text{PQ}} = \langle \sigma \rangle= v_\sigma/\sqrt{2}$, leading to the axion decay constant [see Eq.~\eqref{eq:PQWWscale}]
\begin{equation}
    f_a = \frac{f_{\text{PQ}}}{N} = \frac{v_{\sigma}}{\sqrt{2} N} \; ,
    \label{eq:axionfacolor}
\end{equation}
where $N$ is the color anomaly factor. To be viable, the axion solution to the strong CP problem requires a non-vanishing anomaly factor $N$ to ensure an axion-gluon coupling. For the models in Table~\ref{tab:general} we get [see Eq.~\eqref{eq:EN}]
\begin{align}
N &= 2 \, n_\Psi \, \omega \, (2 n \pm 1) \; T(p,q) \; ,
\label{eq:Nmodel}
\end{align}
with $T(p,q)$ the Dynkin index of the SU($3)_c$ representation $(p,q)$. As expected, $N$ depends on the multiplicity of the colored fermions $n_\Psi$ and on the $\Psi_L$ PQ charge $\omega$.

%%%%%%%%%%%%%%%%%%%%%%%%%%%%%%%%%%%%%%%%%%%%%%%%%%%%%%%%%%%%%%%%%%%%%%%%%%%%%
\subsection{Two-loop Majorana neutrino masses}
%%%%%%%%%%%%%%%%%%%%%%%%%%%%%%%%%%%%%%%%%%%%%%%%%%%%%%%%%%%%%%%%%%%%%%%%%%%%%

%
    \begin{figure*}[t!]
        \centering
        \includegraphics[scale=0.8]{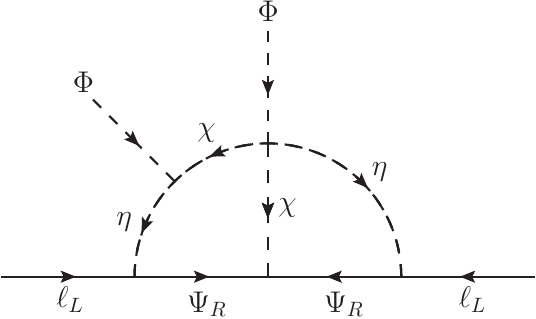} \hspace{+0.2cm}
        \includegraphics[scale=0.8]{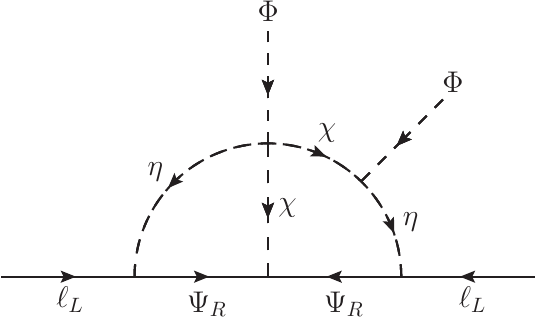} \\ \vspace{+0.5cm}
        \includegraphics[scale=0.8]{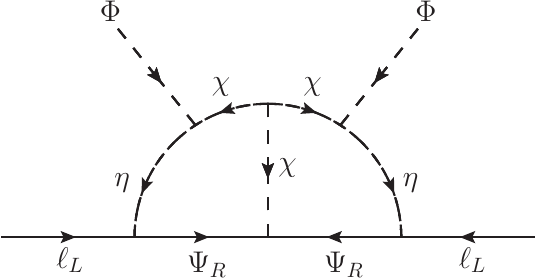}
        \caption{Two-loop neutrino-mass generation diagrams mediated by the colored particles of Table~\ref{tab:general}.}
    \label{fig:neutrino}
    \end{figure*}
With the Yukawa and scalar interactions of Eqs.~\eqref{eq:LYukgen} and~\eqref{eq:Vneutrino}, two-loop Majorana neutrino masses arise from the diagrams in Fig.~\ref{fig:neutrino}~\footnote{Two-loop neutrino mass diagrams have been systematically classified in Refs.~\cite{AristizabalSierra:2014wal,Cai:2017jrq,Avila:2025qsc}. Our topologies also arise in many specific schemes, such as those in Refs.~\cite{Cheng:1980qt,Zee:1985id,Babu:1988ki,Babu:2002uu,Ma:2007gq,Bonilla:2016diq,Aoki:2014cja,Okada:2014qsa,Ho:2016aye,Ho:2017fte,Baek:2017qos}.}. Neutrino masses are mediated by colored particles, $\Psi$, $\eta$ and $\chi$, transforming under complex SU($3)_c$ representations. Moreover, since $\ell_L$ and $\Phi$ are SU($2)_L$ doublets, $\eta$ must lie in an even SU($2)_L$ representation, whereas $\Psi$ and $\chi$ must be in an odd representation. The coupling between $\ell_L$ and $\Psi$ requires $\eta$ to have $Y=1/2$. In our scenario $\Psi$ and $\chi$ carry no hypercharge, ensuring the Majorana nature of light neutrinos.

Among all generic scenarios in Table~\ref{tab:general}, the simplest consistent realization of our idea is for $\Psi_{L,R}$ and $\eta$ to transform as triplets of SU(3)$_{c}$.
Since the SU(3) invariant coming from $(p,q) \otimes (p,q) \otimes (p,q)$ for $(p,q) \equiv \mathbf{3}$ is antisymmetric, we need at least two $\chi$ scalars one being a triplet of color while the other is an anti-sextet. The minimal required multiplicity for the remaining particles is $n_\Psi= 2$ and ~$n_\eta=1$. Concerning SU(2)$_{L}$, $\eta$ is a doublet while $\Psi$ and $\chi$ are singlets. 
For $\omega=1/2$, this setup predicts $N=1$, just as in the original KSVZ model~\cite{Kim:1979if,Shifman:1979if} (see Sec.~\ref{sec:KSVZ}). This minimal scenario simply extends the original KSVZ proposal with extra colored scalars $\eta$ and $\chi$, which mediate neutrino-mass generation. The resulting light-neutrino mass matrix is generically written as~\cite{AristizabalSierra:2014wal,Aoki:2014cja,Hati:2024ppg}
\begin{align}
    (\M_\nu)_{\alpha \beta} &= \frac{N_\text{c}}{(16 \pi^2)^2} \ \tilde{Y}_{a\alpha}^j \ (\tilde{Y}_{\chi})_{a b}^k \ \tilde{Y}_{b \beta}^l \ \tilde{\mu}_{j k l} \ \mathcal{I}_{a b}^{j k l} \; ,
     \label{eq:mnu}
    \end{align}
where $j,k,l=1, \cdots, 6$ and $a,b=1,2$. $N_c=6$ is the color factor, $\tilde{Y}$ and $\tilde{Y}_\chi$ are Yukawa couplings, while $\tilde{\mu}$ denotes an effective cubic scalar coupling, all written now in the mass basis. The loop function $\mathcal{I}_{a b}^{j k l}$ can be found in Refs.~\cite{AristizabalSierra:2014wal,Aoki:2014cja,Hati:2024ppg}. The above result can be estimated by
 \begin{align}
    (M_\nu)_{\alpha \beta} & \sim \ 0.1 \ \text{eV} \left(\frac{\tilde{Y}_{a\alpha}^j \ (\tilde{Y}_{\chi})_{a b}^k \ \tilde{Y}_{b \beta}^l}{10^{-3}} \right) \ \left(\frac{\tilde{\mu}_{j k l}}{10^8 \ \text{GeV}}\right) \left(\frac{v}{246 \ \text{GeV}}\right)^2 \ \left(\frac{10^8 \ \text{GeV}}{m_\zeta}\right)^2 \; ,
      \label{eq:mnuapprox}
\end{align}
where $m_\zeta = \sqrt{\lambda_{\text{eff}}} f_{\text{PQ}}$ is an effective colored scalar mass scale running in the loop with $\lambda_{\text{eff}}$ being some quartic coupling parameter. A typical value for the PQ breaking scale is $f_{\text{PQ}} \sim 10^{12}$ GeV, so that axions account for the observed DM relic abundance (see Sec.~\ref{sec:axionDMcosmo}). Hence, the scalars are expected to be heavy. The smallness of $\tilde{Y}_\chi$ and $\tilde{\mu}_{j k l}$ in Eq.~(\ref{eq:mnu}) is symmetry-protected in
t'Hooft's sense~\cite{tHooft:1979rat}, as the Lagrangian acquires an additional U(1) symmetry in their absence. Note also that, with only two copies of $\Psi$ ($n_\Psi=2$), one of the three light neutrinos is predicted to be massless due to the missing partner nature~\cite{Schechter:1980gr} of the underlying radiative seesaw mechanism. Furthermore, cLFV processes would be mediated at one-loop by the charged colored scalars and exotic fermions, but with very small rates~\footnote{Up to color factors, expressions for such rates resemble those of similar scenarios, e.g. scotogenic models~\cite{Toma:2013zsa,Vicente:2014wga,Ahriche:2016cio,Reig:2018mdk,Mandal:2019oth,Barreiros:2022aqu,Chun:2023vbh} -- see discussions in Secs.~\ref{sec:scotogenic} and~\ref{sec:cLFV}.}. In its minimal version, the above scenario implies that there is no cancellation in the $0_\nu \beta \beta$ amplitude, even for NO neutrino masses~\cite{Reig:2018ztc,Barreiros:2018bju,Avila:2019hhv}, and one finds that, for IO, rates fall inside the expected sensitivities of the next round of experiments~\cite{KamLAND-Zen:2022tow,GERDA:2019ivs,GERDA:2020xhi,Adams:2022jwx} -- see discussion in Sec.~\ref{sec:neutrinoobservables}.

%%%%%%%%%%%%%%%%%%%%%%%%%%%%%%%%%%%%%%%%%%%%%%%%%%%%%%%%%%%%%%%%%%%%%%%%%%%%%
\subsection{Axion-to-photon coupling, dark matter and cosmology}
%%%%%%%%%%%%%%%%%%%%%%%%%%%%%%%%%%%%%%%%%%%%%%%%%%%%%%%%%%%%%%%%%%%%%%%%%%%%%

Indirect astrophysical and cosmological observations, as well as laboratory searches constrain the axion parameter space due to its couplings to photons, nucleons and electrons -- for an extended discussion see Sec.~\ref{sec:experimental}. We now examine how to probe the various scenarios of Table~\ref{tab:general} through their corresponding axion-to-photon coupling $g_{a \gamma \gamma}$.

In the KSVZ setup, the only chiral fermions charged under U($1$)$_{\text{PQ}}$ are the new exotic fermions. Therefore, there are no model-dependent contributions to the axion coupling to nucleons and electrons. The model dependent contributions to the axion-to-photon coupling are given by the ratio $E/N$ between the EM $E$ and color anomaly factors $N$ as shown in Eqs.~\eqref{eq:EN},~\eqref{eq:axioncouplingsgagg} and~\eqref{eq:axioncouplings} of Sec.~\ref{sec:properties}. For our class of models in Table~\ref{tab:general}, we have
\begin{equation}
    \frac{E}{N} = \frac{d(p,q)}{(2 n \pm 1) T(p,q)} \sum_{j=0}^{2n \pm 1 - 1} \left(\frac{2n \pm 1 - 1}{2} - j \right)^2 \; ,
    \label{eq:EKSVZ}
\end{equation}
with $d(q,p)$ being the dimension of SU(3)$_c$ representation. One sees that $E/N = 0$, as long as the hyperchargeless $\Psi_{L,R}$ are SU($2)_L$ singlets, as in the original KSVZ model -- see Sec.~\ref{sec:KSVZ}. For higher weak multiplet representations $E/N \neq 0$ -- see Table~\ref{tab:EN}.
\begin{table}[t!]
\renewcommand*{\arraystretch}{1.5}
	\centering
\begin{tabular}{| K{1.5cm} | K{1.5cm} | K{1.5cm} | K{1.5cm} |  K{1.5cm} | K{1.5cm} |  K{1.5cm} |}
        \hline
\multicolumn{2}{|c|}{\multirow{2}{*}{$E/N$}} & \multicolumn{5}{c|}{SU($2)_L$}\\
		\cline{3-7}
\multicolumn{2}{|c|}{} & $\mathbf{3}$ & $\mathbf{5}$ & $\mathbf{7}$ & $\mathbf{9}$ & $\mathbf{11}$ \\
		\hline 
        \multirow{5}{*}{\rotatebox[origin=c]{90}{SU($3)_c$}}
&$\mathbf{3}$ & 4 & 12 & 24 & 40 & 60 \\
&$\mathbf{6}$ & 8/5 & 24/5 & 48/5 & 16 & 24 \\
&$\mathbf{10}$ & 8/9 & 8/3 & 16/3 & 80/9 & 40/3 \\
&$\mathbf{15}$ & 1 & 3 & 6 & 10 & 15 \\
&$\mathbf{15'}$ & 4/7 & 12/7 & 24/7 & 40/7 & 60/7 \\
        \hline
\end{tabular}
\caption{$E/N$ values for various $\mathrm{SU(3)}_c \otimes \mathrm{SU(2)}_L$ representation choices for $\Psi$ [see Table~\ref{tab:general} and Eq.~\eqref{eq:EKSVZ}].}
\label{tab:EN}
\end{table}
\begin{figure}[!t]
    \centering
      \includegraphics[scale=0.55]{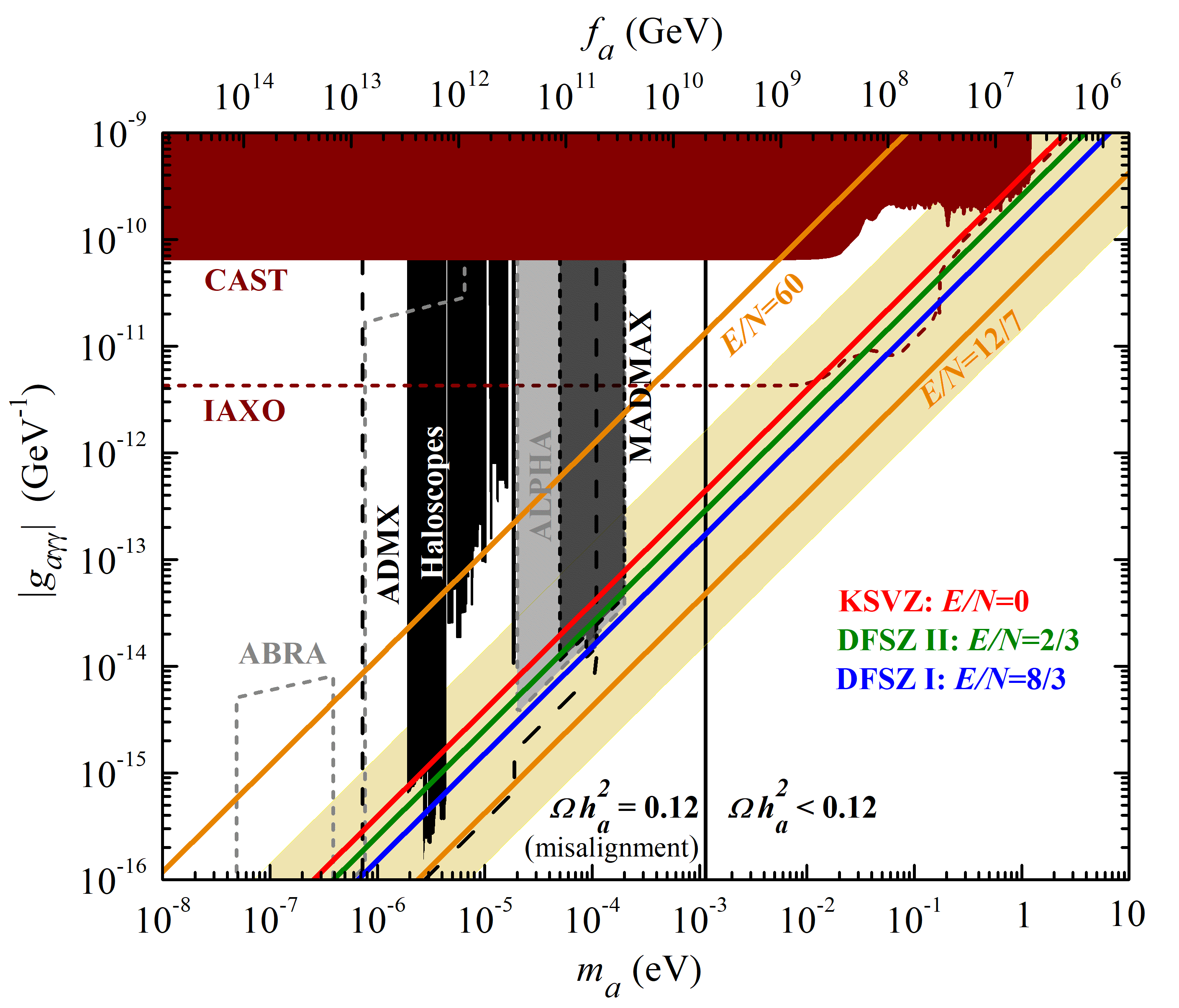}
      \caption{Axion-to-photon coupling $|g_{a \gamma \gamma}|$ versus axion mass $m_a$ (bottom axis) and decay constant $f_a$ (top axis). The orange lines correspond to $E/N$ values leading to maximum and minimum $|g_{a \gamma \gamma}|$ for the representations shown in Table~\ref{tab:EN}. The KSVZ and DFSZ I and II predictions are indicated by the red, blue, and green lines, respectively. The yellow shaded region refers to the usual QCD axion window~\cite{DiLuzio:2016sbl}. Current constraints from helioscopes like CAST~\cite{CAST:2017uph} exclude the bordeau shaded region, while haloscopes like ADMX~\cite{ADMX:2018gho,ADMX:2019uok,ADMX:2021nhd}, RBF~\cite{DePanfilis:1987dk}, CAPP~\cite{CAPP:2020utb} and HAYSTAC~\cite{HAYSTAC:2020kwv}, rule out the black region. Projected sensitivities of IAXO~\cite{Shilon_2013}, ADMX~\cite{Stern:2016bbw}, MADMAX~\cite{Beurthey:2020yuq}, ALPHA~\cite{ALPHA:2022rxj} and ABRACADABRA~\cite{Ouellet:2018beu} are indicated by the dashed bordeau, dashed black, short-dash black (dark-gray shaded region), dotted (light-gray shaded region) and gray-dashed contours, respectively. To the right of the black vertical line, axion DM is under-abundant. In the left region $\Omega h^2_a = 0.12$, for the pre-inflationary case featuring the misalignment mechanism.}
    \label{fig:gaggMajorana}
\end{figure}
In Fig.~\ref{fig:gaggMajorana}, we display by oblique solid lines the axion-photon coupling $|g_{a\gamma \gamma}|$ -- see Eqs.~\eqref{eq:EN},~\eqref{eq:axioncouplingsgagg} and~\eqref{eq:axioncouplings} -- in terms of~$m_a$ (bottom axis) and~$f_a$ (top axis) -- see Eq.~\eqref{eq:axionmass}. The orange lines delimit the band of $E/N$ values leading to the maximum and minimum $|g_{a\gamma \gamma}|$, corresponding to $E/N=60$ for~$\Psi \sim (\mathbf{3},\mathbf{11},0)$ and $E/N=12/7$ for~$\Psi \sim (\mathbf{15}^\prime,\mathbf{5},0)$, respectively (see Table~\ref{tab:EN}). The $|g_{a\gamma \gamma}|$ corresponding to the popular KSVZ and DFSZ-I and II schemes are shown by the solid red, blue, and green lines, respectively -- see Sec.~\ref{sec:models}. The minimal KSVZ model featuring two-loop neutrino masses predicts $E/N=0$ (solid red line). In the same plot we show the current bounds and future sensitivities from helioscopes and haloscopes -- see Sec.~\ref{sec:experimental}. The CAST helioscope experiment~\cite{CAST:2017uph} excludes the bordeau-shaded region, while haloscopes ADMX~\cite{ADMX:2018gho,ADMX:2019uok,ADMX:2021nhd}, RBF~\cite{DePanfilis:1987dk}, CAPP~\cite{CAPP:2020utb}, and HAYSTAC~\cite{HAYSTAC:2020kwv} exclude the black region. One sees that the future IAXO experiment is expected to probe $g_{a \gamma \gamma}$ down to $(10^{-12}-10^{-11}) \ \text{GeV}^{-1}$ reaching the popular QCD axion model predictions for $m_a \sim 0.1$ eV (bordeau-dashed contour). Out of all haloscope experiments, ADMX has already reached the KSVZ and DFSZ QCD axion lines for masses $m_a \sim 3 \ \mu$eV. Upcoming ADMX (black-dashed contour) should probe the full landscape of QCD axion models for masses $1 \ \mu \text{eV} \lesssim m_a \lesssim 100 \ \mu$eV. Moreover, ALPHA~\cite{Lawson:2019brd,Wooten:2022vpj,ALPHA:2022rxj} (light-gray shaded region) and MADMAX~\cite{Beurthey:2020yuq} (dark-gray shaded region) are projected to cover the region $20 \ \mu \text{eV} \lesssim m_a \lesssim 120 \ \mu$eV and $50 \ \mu \text{eV} \lesssim m_a \lesssim 120 \ \mu$eV, respectively. Lastly, ABRACADABRA~\cite{Ouellet:2018beu} (dashed-gray contour), is expected to probe the mass region below $\sim 1 \mu \text{eV}$ and reach the yellow QCD axion band.

Axions are paradigmatic DM candidates as reviewed in Sec.~\ref{sec:axionDMcosmo}. Note that, in analogy with the original KSVZ model studied in Sec.~\ref{sec:KSVZ}, our framework features cosmologically stable baryonic and charged relics: the lightest state stemming from the colored fields $\Psi$, $\eta$, or $\chi$. Namely, after symmetry breaking, due to color and EM symmetries, the Lagrangian exhibits an unbroken accidental $\mathcal{Z}_3$ symmetry where $(\Psi,\eta,\chi) \rightarrow e^{i 2\pi/3} (\Psi,\eta,\chi)$, stabilizing the lightest colored state [see Eqs.~\eqref{eq:LYukgen} and~\eqref{eq:Vneutrino}]. As mentioned before, this stable baryonic and charged relics are practically ruled out~\footnote{If electrically neutral, these relics might form viable bound-state DM~\cite{Reig:2018mdk,Reig:2018ztc}.}. Thus, for our proposed scenario it is judicious to assume a pre-inflationary axion DM scenario, where the abundance of stable baryonic or charged relics will be washed out during inflation, as well as topological defects. Nonetheless, in Sec.~\ref{sec:possibilityHLmixing} we will identify models where the exotic fermions $\Psi$ with $Y \neq 0$ can mix with ordinary quarks allowing scenarios free of stable colored or charged relics, leading to viable post-inflationary axion DM.

\begin{figure}[!t]
    \centering
      \includegraphics[scale=0.5]{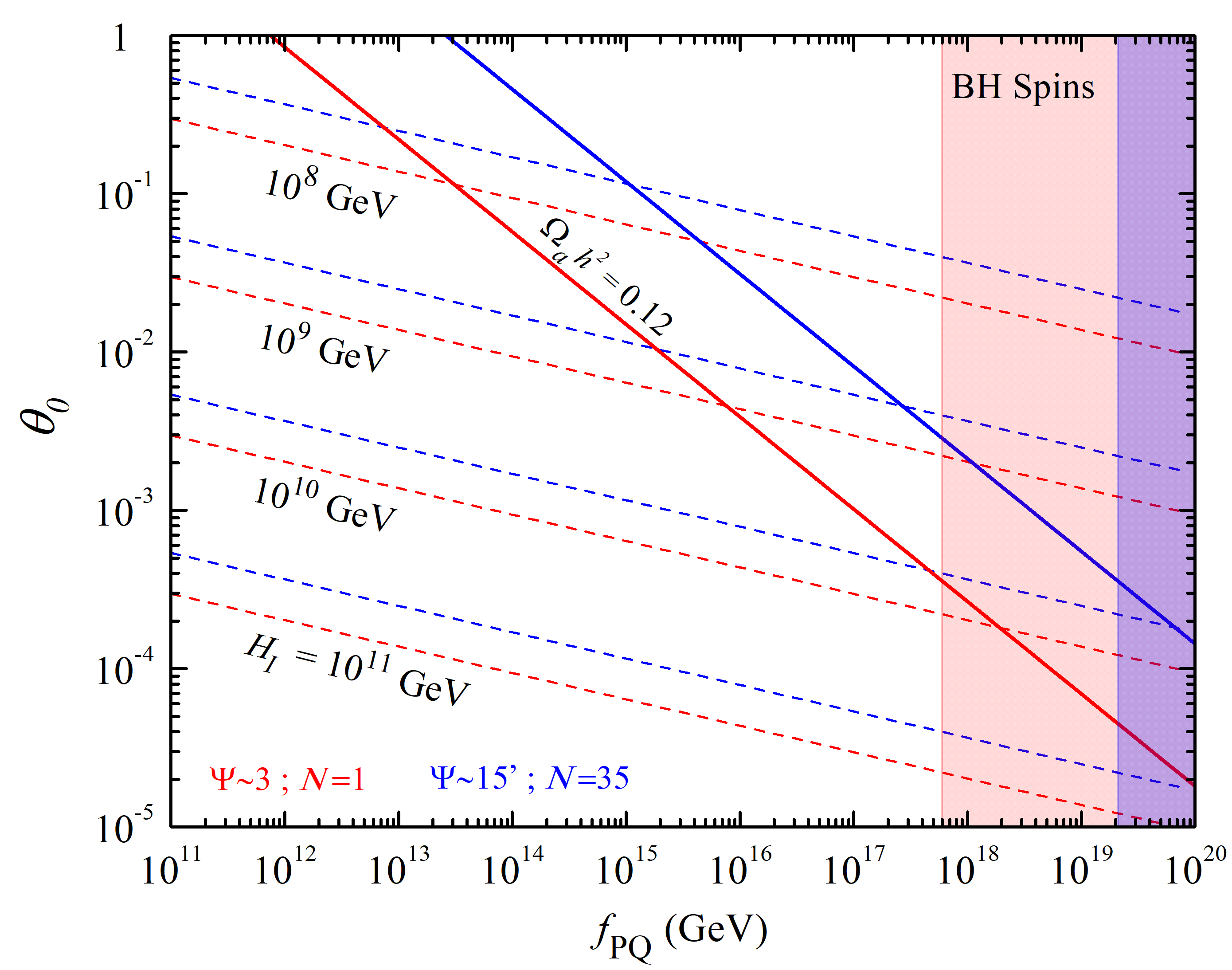}
      \caption{
        Misalignment angle $\theta_0$ as a function of $f_{\text{PQ}}$.
        In red (blue) we show the scenario for $\Psi \sim 3$ ($\Psi \sim 15^\prime$) under SU($3)_c$, singlet under SU($2)_L$, with $\omega=1/2$ and $n_\Psi=2$.
        Along the solid lines $\Omega_a h^2=0.12$, above (below) them we have DM over- or underabundance.
        Above the dashed lines the value of the inflationary scale $H_I$ lies below the indicated value [see Eq.~\eqref{eq:Iso}].
        Vertical bands are excluded by BH superradiance bound of Eq.~\eqref{eq:BHaxionbound}.
      }
\label{fig:DM}
\end{figure}
In Fig.~\ref{fig:DM}, we display the initial misalignment angle $\theta_0$ -- see Eq.~\eqref{eq:relica} -- as a function of $f_{\text{PQ}}$ [see Eq.~\eqref{eq:axionfacolor}]. We highlight two cases, in red and blue, where the fermions, singlets under SU($2)_L$, transform as $\Psi \sim 3$ and $\Psi \sim 15^\prime$ under SU($3)_c$, respectively. We take $\omega=1/2$ and $n_\Psi=2$. Along the solid lines we have $\Omega_a h^2=0.12$ [see Eq.~\eqref{eq:relica}]. The region above these lines is excluded since it implies DM overabundance. BH superradiance leads to the bounds of Eq.~\eqref{eq:BHaxionbound} excluding the shaded bands in the figure. Taking $\theta_0 \sim \mathcal{O}(1)$ leads to $f_a \gsim 5 \times 10^{11}$ GeV, a region currently being probed by haloscope experiments -- see Fig.~\ref{fig:gaggMajorana}. The dashed lines indicate different values of the inflationary scale $H_I$. Above these lines $H_I$ is below the indicated value, in agreement with the isocurvature bound of Eq.~\eqref{eq:Iso}. The allowed region for a given $H_I$ lies above the dashed and below the solid contours. Taking $\theta_0 \sim \mathcal{O}(1)$ and $\Omega_a h^2=0.12$, we get a low scale for inflation $H_I \lsim 10^7$ GeV (Planck currently probes $H_I \lsim 10^{13}$ GeV~\cite{Planck:2018vyg}).

%%%%%%%%%%%%%%%%%%%%%%%%%%%%%%%%%%%%%%%%%%%%%%%%%%%%%%%%%%%%%%%%%%%%%%%%%%
\subsection{The possibility for heavy-light quark mixing}
\label{sec:possibilityHLmixing}
%%%%%%%%%%%%%%%%%%%%%%%%%%%%%%%%%%%%%%%%%%%%%%%%%%%%%%%%%%%%%%%%%%%%%%%%%%
	
Here we present the possibility for the exotic fermions $\Psi$ having non-zero hypercharge, so that they can mix with SM quarks, and discuss its consequences. The potential advantage of such a scenario is that the mixing with quarks can open decay channels for $\Psi$ suppressing the relic density in the current Universe. In order for $\Psi$ to mix with SM quarks, these must transform as triplets of SU(3)$_c$ and either singlets, doublets or triplets of SU(2)$_L$~\cite{delAguila:2000aa,delAguila:2000rc,Aguilar-Saavedra:2013qpa,DiLuzio:2016sbl}. Moreover, $\Psi$ need to be in odd SU(2)$_L$ representations for neutrino masses to be successfully generated at the two-loop level as in Fig.~\ref{fig:neutrino}. Thus, there are only two cases to consider regarding SU(2)$_L$, iso-singlets or iso-triplets. Note that since the fermions $\Psi_{L,R}$ now carry hypercharge, so the scalars $\eta,\chi_i$ also need to carry appropriate hypercharges as we will discuss shortly.

In order for the $\Psi$ to mix with quarks, their hypercharge needs to be either $Y=-1/3$ or $Y=2/3$ leading to exotic-ordinary quark mixing. However, such a mixing in general will lead to two-body proton decays mediated by the scalar leptoquarks present in the model. For example the $p \to \pi^0 + e^+$ decay can happen through $\chi_i$ mediation. The non-observation of this decay channel sets a very stringent bound on the proton decay lifetime $\tau_p > 1.6 \times 10^{34}$ years, which requires the exotic scalar mediator mass $m_{\chi_2} \gsim 10^{15}$ GeV, to be very heavy near the typical GUT scale~\cite{Super-Kamiokande:2016exg}. Since the typical colored scalar masses are given by $m_{\zeta}=\sqrt{\lambda_{\text{eff}}} f_{\text{PQ}}$, the proton decay bounds sets a heavy mediator mass scale at odds with the post-inflationary DM scenario which requires $f_a \lesssim 2 \times 10^{11}$ GeV.

\begin{figure}[t!]
		\centering
		\includegraphics[scale=0.8]{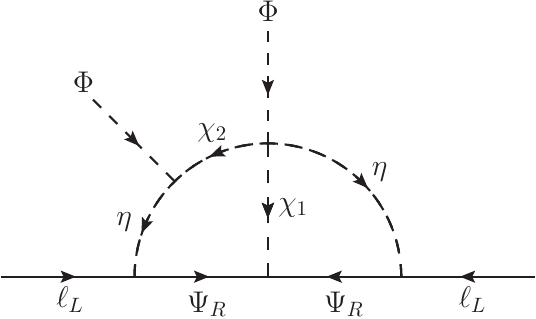} \hspace{+0.2cm}
		\includegraphics[scale=0.8]{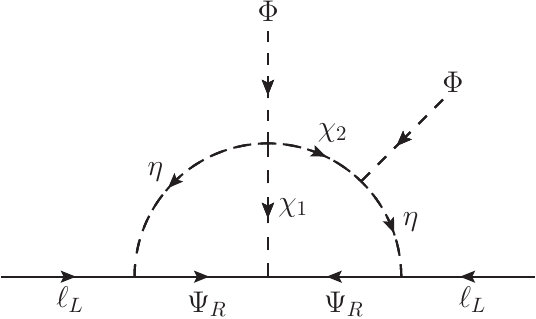} \\ \vspace{+0.5cm}
		\includegraphics[scale=0.8]{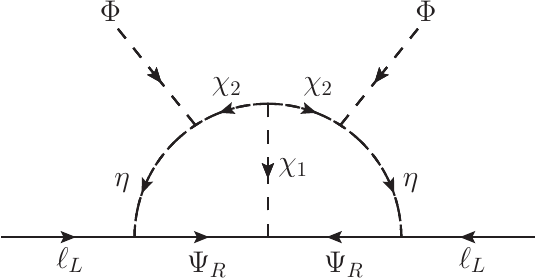}
		\caption{ Two-loop neutrino mass generation diagrams mediated by the colored particles of Table~\ref{tab:modelYneq0}.}
		\label{fig:neutrinoYneq0}
	\end{figure}

The problem of proton decay in the hypercharge non-zero models can be avoided, by modifying the charges of the scalars $\eta, \chi_i$ under the PQ symmetry, so that after the breaking of the PQ symmetry, an appropriate residual symmetry survives unbroken, forbidding the proton decay. We now illustrate this idea by an example. The ${\rm SU(3)}_c \otimes {\rm SU(2)}_L \otimes {\rm U(1)}_Y \otimes {\rm U(1)}_{\rm PQ}$ charges of the particles are given in Table.~\ref{tab:modelYneq0}.   
	\begin{table}[t!]
		\renewcommand*{\arraystretch}{1.5}
		\centering
		\begin{tabular}{| K{1.5cm} | K{1.5cm} | K{5cm} |  K{1.5cm} |K{2.5cm} |}
			\hline
			&Fields&\SM&    U($1$)$_{\text{PQ}}$ & Multiplicity\\
			\hline
			\multirow{2}{*}{Fermions}
			&$\Psi_L$&($\mathbf{3},\mathbf{1}, -1/3 \ (2/3)$)& $0$  & $n_\Psi$\\
			&$\Psi_R$&($\mathbf{3},\mathbf{1}, -1/3 \ (2/3)$)&  $-1/n_\Psi$ & $n_\Psi$ \\
			\hline
			\multirow{4}{*}{Scalars}
			&$\sigma$&($\mathbf{1},\mathbf{1}, 0$)& $1/n_\Psi$ & $1$\\
			&$\eta$&($\mathbf{3},\mathbf{2}, 1/6 \ (7/6)$)& $-1/n_\Psi$ & $n_\eta $\\
			&$\chi_1$&($\overline{\mathbf{6}},\mathbf{1}, 2/3 \ (-4/3)$)& $2/n_\Psi$ & $n_{\chi_1}$ \\
			&$\chi_2$&($\mathbf{3},\mathbf{1}, -1/3 \ (2/3)$)& $-1/n_\Psi$ & $n_{\chi_2}$ \\
			\hline
		\end{tabular}
		\caption{ Matter content and charges for the KSVZ model with two-loop neutrino masses featuring exotic fermion mixing with SM down (up) quarks.}
		\label{tab:modelYneq0} 
	\end{table}
    The U$(1)_Y$ charges outside parentheses will lead to mixing between down-type quarks and $\Psi$, while those inside parentheses will lead to mixing between up-type quarks and $\Psi$. With these modified charges, the neutrino mass can as in the two-loop diagrams given in Fig.~\ref{fig:neutrinoYneq0}.
	The relevant new Yukawa terms are given by:  
	\begin{align}
	- \mathcal{L}_{\text{Yuk.}} \supset \mathbf{Y}_{\Psi} \overline{\Psi_L} \Psi_R \sigma + \frac{1}{2} \mathbf{Y}_{\chi_{1 j}} {\Psi^T_R}~C~\chi_{1 j} \Psi_R + \mathbf{Y}_i \overline{L} \eta^\ast_i \Psi_R + \mathbf{m} \overline{\Psi_L} d_R (+ \mathbf{m} \overline{\Psi_L} u_R) + \text{H.c.} \; ,
	\end{align}
while the scalar-potential terms triggering neutrino mass generation are,
	\begin{align}
		V \supset \mu_{ijk} \chi_{1 i} \chi_{2 j}\chi_{2 k} + \kappa_{ij} \eta_i^\dagger \Phi  \chi_{2 j} + \lambda_{ijk} \Phi^\dagger \eta_i \chi_{1 j} \chi_{2 k} + \text{H.c.} \; .
		\label{eq:VneutrinoYneq0}
	\end{align}
	A few comments on these models are in order:
	\begin{itemize} 
		
		\item The minimal multiplicities required for non-zero neutrino masses is $n_{\Psi}=2$ and $n_{\eta, \chi_1, \chi_2} = 1$. The case $n_\Psi>2$ and/or $n_\eta>1$ leads to three massive light neutrinos.  
		
		\item For $Y=-1/3$ ($Y=2/3$) the term $\overline{\Psi_L} d_R$ ($\overline{\Psi_L}  u_R$) is present in the Lagrangian. This allows for $\Psi$ to decay into ordinary matter. Thus, if $m_\Psi < m_{\eta,\chi_1,\chi_2}$, the lightest colored particle mediating neutrino masses is no longer stable and one can envisage a post-inflationary axion DM scenario.
        
        \item These models feature a DW number $N_{\text{DW}}=1$. Therefore, as discussed in Sec.~\ref{sec:axionDMcosmo} current numerical simulations in post-inflationary scenario for $N_{\text{DW}}=1$ predict the axion scale $f_a$ to be in the range $[5 \times 10^9,3 \times 10^{11}]$~GeV, in order for the axion to account for the observed CDM abundance, i.e. $\Omega_a h^2 = \Omega_{\text{CDM}} h^2$~\cite{Kawasaki:2014sqa,Klaer:2017ond,Gorghetto:2020qws,Buschmann:2021sdq,Benabou:2024msj}.
        
	\item The model with $Y=-1/3$ ($Y=2/3$), leads to $E/N=2/3$ ($E/N=8/3$), the same prediction for $g_{a\gamma\gamma}$ as DFSZ II (DFSZ I) [see Fig.~\ref{fig:gaggMajorana}].  
		
	   \item These models require colored scalars that carry non-zero PQ charges, such that phenomenologically dangerous
                  terms that enable proton decay are absent from the Lagrangian.  
                  In fact, after symmetry breaking, an accidental unbroken $\mathcal{Z}_6$ symmetry remains in the Lagrangian where the fields transform as:
                  $(\ell_L,e_R) \to z^3 (\ell_L,e_R)$, $(q_L, d_R, u_R, \Psi_{L,R}, \chi_1) \to z^2 (q_L, d_R, u_R, \Psi_{L,R}, \chi_1)$ and $(\eta,\chi_2) \rightarrow z^5 (\eta,\chi_2)$,
                  with $z$ being the sixth root of unity. Hence the lowest dimension proton decay operators, $d u u e$, $q q u e$, $q \ell d u$, $q q q \ell$, are forbidden by symmetry. If no PQ charge is assigned to the colored scalars, for the exotic fermion mixing with SM down-type quarks case, the terms $d_R^c \chi_2 u_R$ and $u_R^c \chi_2^\ast e_R$ will be present in the Lagrangian, leading to proton decay at tree-level, e.g. $p \to \pi^0 + e^+$.
		
	\end{itemize}
The iso-triplet version of these models only allows $\Psi_R$ to mix with SM quarks. For these cases there is no way to forbid dangerous proton decay interactions while simultaneously allowing exotic-ordinary quark mixing.
%In the upcoming section, we propose a color-mediated Dirac neutrino mass framework featuring heavy–light quark mixing, which leads to a more predictive axion DM scenario, while potentially dangerous proton decay operators are forbidden by symmetry.

%%%%%%%%%%%%%%%%%%%%%%%%%%%%%%%%%%%%%%%%%%%%%%%%%%%%%%%%%%%%%%%%%%%%%%%%%%%%%
\section{Framework for Dirac neutrino masses}
\label{sec:colormediatedDirac}
%%%%%%%%%%%%%%%%%%%%%%%%%%%%%%%%%%%%%%%%%%%%%%%%%%%%%%%%%%%%%%%%%%%%%%%%%%%%%

%
\begin{table}[t!]
\renewcommand*{\arraystretch}{1.5}
	\centering
	\begin{tabular}{| K{3cm} | K{2.5cm} | K{5cm} | K{3cm} |}
		\hline 
&Fields&\SM&  U$(1)_{\text{PQ}}$   \\
		\hline
		\multirow{3}{*}{Leptons} 
&$\ell_L$&($\mathbf{1},\mathbf{2}, {-1/2}$)& $1/6$   \\
&$e_R$&($\mathbf{1},\mathbf{1}, {-1}$)& {$1/6$}    \\
&$\nu_R$&($\mathbf{1},\mathbf{1}, {0}$)& {$4/6$}  \\ \hline
Vector-like quarks &$\Psi_{1,2 L};\Psi_{1,2 R}$&($\mathbf{3},\mathbf{n}_\Psi, y_\Psi$)& {$\mathcal{Q}_{\text{PQ}}; \mathcal{Q}_{\text{PQ}}-1/2$}  \\ 
\hline
\multirow{2}{*}{Scalars}  &$\Phi$&($\mathbf{1},\mathbf{2}, 1/2$)& {$0$}  \\
&$\sigma$&($\mathbf{1},\mathbf{1}, 0$)& {$1/2$}  \\ \hline
\multirow{2}{*}{Scalar Leptoquarks} &$\eta$&($\mathbf{3},\mathbf{n}_\eta \equiv \mathbf{n}_\Psi \pm \mathbf{1}, y_\Psi+1/2$)& {$\mathcal{Q}_{\text{PQ}}-4/6$}   \\
&$\chi$&($\mathbf{3},\mathbf{n}_\Psi, y_\Psi$)& {$\mathcal{Q}_{\text{PQ}}-4/6$}   \\		
\hline
	\end{tabular}
	\caption{ Field content and transformation properties under \SM and U$(1)_\text{PQ}$. Here, $\mathbf{n}_{\Psi,\eta}$ indicate the SU(2)$_L$ representations of $\Psi$ and $\eta$, while $y_\Psi$ is the hypercharge of $\Psi$ and $\mathcal{Q}_{\text{PQ}}$ is a PQ charge parameter. Specific assignments are given in Table~\ref{tab:postcharges}.}
	\label{tab:generalDirac} 
\end{table}
\begin{table}[t!]
\renewcommand*{\arraystretch}{1.5}
	\centering
	\begin{tabular}{| K{0.5cm} | K{1cm} | K{1cm} | K{1cm} | K{4.5cm} | K{4cm} |}
		\hline 
$\mathbf{n}_\Psi$ & $y_\Psi$ & $\mathcal{Q}_{\text{PQ}}$ &$\mathbf{n}_\eta$ & Mixing terms & Other decay terms  \\
		\hline \hline
		\multirow{4}{*}{$\mathbf{1}$} 
&\multirow{2}{*}{$-1/3$} &$1/2$& \multirow{4}{*}{$\mathbf{2}$} & $\overline{q_L} \Phi \Psi_R, \overline{\Psi_L} \sigma d_R $& $\overline{\ell_L} \tilde{\eta} d_R  $ \\
& &$0$& & $\overline{\Psi_L} d_R$ & $ \overline{q_L} \eta \nu_R $ \\
\cline{2-3} \cline{5-6} 
&\multirow{2}{*}{$2/3$} &$1/2$& & $\overline{q_L} \tilde{\Phi} \Psi_R, \overline{\Psi_L} \sigma u_R$& $\overline{\ell_L} \tilde{\eta} u_R, \overline{q_L} \eta e_R $ \\
& &$0$& & $\overline{\Psi_L} u_R$ & - \\
        \hline \hline
        \multirow{4}{*}{$\mathbf{2}$} 
&\multirow{2}{*}{$1/6$} &$1/2$& \multirow{4}{*}{$\mathbf{1},\mathbf{3}$} & $\overline{q_L} \Psi_R$& $\overline{\ell_L} \chi^\ast d_R$ \\
& &$0$& & $\overline{q_L} \sigma \Psi_R,\overline{\Psi_L} \Phi d_R, \overline{\Psi_L} \tilde{\Phi} u_R$ & - \\
\cline{2-3} \cline{5-6}
&$-5/6$ & \multirow{2}{*}{$0$} & & $\overline{\Psi_L} \tilde{\Phi} d_R$& - \\
\cline{2-2} \cline{5-6}
&$7/6$ & & & $\overline{\Psi_L} \Phi u_R$ & - \\
        \hline \hline
        \multirow{4}{*}{$\mathbf{3}$} 
&\multirow{2}{*}{$-1/3$} & \multirow{4}{*}{$0$} &$\mathbf{2}$ & \multirow{2}{*}{$\overline{q_L} \Phi \Psi_R$}& $\overline{\ell_L} \tilde{\eta} d_R  $ \\
& & & $\mathbf{4}$ & & - \\
\cline{2-2} \cline{4-6}
&\multirow{2}{*}{$2/3$} & & $\mathbf{2}$ &\multirow{2}{*}{$\overline{q_L} \tilde{\Phi} \Psi_R$}& $\overline{\ell_L} \tilde{\eta} u_R, \overline{q_L} \eta e_R $ \\
& & & $\mathbf{4}$ & & - \\
        \hline
	\end{tabular}
	\caption{Charge assignments for the new colored fields given in terms of the $\mathbf{n}_\Psi$, $y_\Psi$, $\mathcal{Q}_{\text{PQ}}$ and $\mathbf{n}_\eta$ parameters (column 1-4) of Table~\ref{tab:generalDirac}. Allowed VLQ and leptoquark Yukawa interactions with SM quark and lepton fields leading to heavy-light quark mixing and exotic particle decays are given in columns 5 and 6, respectively (see text for details).}
\label{tab:postcharges} 
\end{table}
In this section we extend the color-mediated neutrino mass idea to Dirac neutrinos, following closely our work of Ref.~\cite{Batra:2025gzy}. Namely, in Tables~\ref{tab:generalDirac} and~\ref{tab:postcharges}, we present the particle content and transformation properties of the fields under the SM and PQ symmetries in our color-mediated Dirac neutrino mass models. In order to implement massive Dirac neutrinos we add three RH neutrino fields $\nu_R$ to the SM particle content. In addition, to address the strong CP problem, we add chiral VLQs $\Psi_{L,R}$ charged under a global PQ symmetry U(1)$_{\text{PQ}}$, which is spontaneously broken by a complex scalar singlet~$\sigma$. Color-mediated radiative Dirac neutrino masses are generated by adding to the scalar sector two leptoquark multiplets $\eta$ and $\chi$. Before proceeding, a few important comments on our construction are in order. Namely: 
\begin{itemize}
    \item The PQ symmetry ensures the Dirac nature of neutrinos and forbids bare Majorana neutrino mass terms. The PQ-breaking scalar $\sigma$ may lead to effective Majorana neutrino mass generation operators, written generically as,
    \begin{align} 
    \mathcal{L}_{\text{Maj.}} = \frac{\boldsymbol{c}_{\text{Maj.}}}{2 \Lambda^{n + n^\prime + 1}} \; (\bar{\ell_L^c} \tilde{\Phi}^\ast) (\tilde{\Phi}^\dagger \ell_L) \; \sigma^n \sigma^{\ast n^\prime} + \text{H.c.} \; ,
    \label{eq:MajoranaOperators}
    \end{align}
    where $\boldsymbol{c}_{\text{Maj.}}$ is a dimensionless Majorana-type coupling matrix and $\Lambda$ parameterizes some high-energy scale. The case $n=n^\prime=0$ corresponds to the well-known Weinberg operator~\cite{Weinberg:1980wa} -- see Eq.~\eqref{eq:weinbergop} and discussion in Sec.~\ref{sec:Weinberg}. Notice, however, that the lepton PQ assignments in Table~\ref{tab:generalDirac} were specifically chosen such that there is a residual $\mathcal{Z}_3$ symmetry under which $(\ell_L,e_R,\nu_R) \to \omega (\ell_L,e_R,\nu_R)$ and $(\eta,\chi) \to \omega^2 (\eta,\chi)$. This forbids all the above Majorana type operators. In contrast, the effective Dirac neutrino mass operators 
    \begin{equation}
    \mathcal{L}_{\text{Dirac}} = \frac{\boldsymbol{c}_{\text{Dirac}}}{\Lambda^{n+n^\prime}} \; (\bar{\ell_L} \tilde{\Phi} \nu_R) \; \sigma^n \sigma^{\ast n^\prime} + \text{H.c.} \; ,
    \label{eq:DiracOperators}
    \end{equation}
    are allowed by the $\mathcal{Z}_3$ symmetry with the lowest dimension operator being the dimension 5 $(\bar{\ell_L} \tilde{\Phi} \nu_R) \; \sigma^\ast$, i.e. for $n=0,n^\prime=1$. This method of residual $\mathcal{Z}_N$ symmetries forbidding Majorana neutrino masses has been previously employed in the context of a $\text{U}(1)_{B-L}$ symmetry in Ref.~\cite{Bonilla:2018ynb}. The tree-level Yukawa interaction $\bar{\ell_L} \tilde{\Phi} \nu_R$ is forbidden by the PQ symmetry, which is necessary for radiative Dirac neutrino mass generation. In the above, $\boldsymbol{\kappa}_{\text{Dirac}}$ is a dimensionless Dirac-type neutrino coupling matrix. In Sec.~\ref{sec:Diracneutrinos} we will present the explicit one-loop realization of this dimension-5 operator. 
    \item All our models feature mixing between the VLQ $\Psi$ and the ordinary SM quarks. Namely, seven possible VLQ representations mix with the SM quarks at tree-level~\cite{delAguila:2000aa,delAguila:2000rc,Aguilar-Saavedra:2013qpa,DiLuzio:2016sbl,Alves:2023ufm}, as shown in Table~\ref{tab:postcharges} (see fifth column). In the next Sec.~\ref{sec:quarksector}, we discuss heavy-light quark mixing properties in detail for each model and in Sec.~\ref{sec:axionflavorviolatingDirac} we study the associated flavor-violating axion coupling phenomenology -- see discussion in Sec.~\ref{sec:flavouredaxions}. We will focus on these seven color-mediated Dirac neutrino mass models since they allow the new exotic colored particles $\Psi,\eta,\chi$ to decay into ordinary matter. Otherwise one would have stable baryonic and charged relics which may pose cosmological difficulties~\cite{Perl:2001xi,Perl:2009zz,Burdin:2014xma,Hertzberg:2016jie,Mack:2007xj} -- see Sec.~\ref{sec:possibilityHLmixing}. Thus, our framework favors a post-inflationary axion DM cosmology, as discussed in Sec.~\ref{sec:axionDMphotonDirac} -- for details see Sec.~\ref{sec:axionDMcosmo}.

    \item There is no proton decay in our models. The scalar leptoquarks $\eta$ and $\chi$ lead to the lepton-quark interaction terms given in the last column of Table~\ref{tab:postcharges}. 
    However, an accidental baryon number symmetry under which the SM quark fields and the new colored ones are equally charged forbids dangerous proton decay operators such as the dimension 6 $duue$, $qque$, $q\ell d u$, $qqq\ell$, etc. 
    This distinctive feature of our construction ensures that, besides protecting the \emph{Diracness} of neutrinos, the same PQ symmetry necessary to solve the strong CP problem, also ensures the stability of the proton.
    
\end{itemize}
Our framework is a KSVZ-type axion solution to the strong CP problem. To implement the axion solution to the strong CP problem it is required that $N \neq 0$. For the class of models given in Table~\ref{tab:generalDirac}, the PQ charge assignments lead to color-anomaly factor $N=1,2,3$ for the cases where the VLQ $\Psi$ are iso-singlet, doublets, and triplets, respectively -- see Eq.~\eqref{eq:EN} and Sec.~\ref{sec:properties}.

%%%%%%%%%%%%%%%%%%%%%%%%%%%%%%%%%%%%%%%%%%%%%%%%%%%%%%%%%%%%%%%%%%%%%%%%%%%%%
\subsection{Heavy-light quark mixing}
\label{sec:quarksector}
%%%%%%%%%%%%%%%%%%%%%%%%%%%%%%%%%%%%%%%%%%%%%%%%%%%%%%%%%%%%%%%%%%%%%%%%%%%%%

For all the models in Tables~\ref{tab:generalDirac} and~\ref{tab:postcharges} the colored fermion Yukawa Lagrangian contains 
\begin{align}
    -\mathcal{L}_\mathrm{Yuk.}^{\text{c}} \supset -\mathcal{L}_\mathrm{Yuk.}^{\text{c} , 0} = \Y_d \overline{q_L} \Phi d_R + \Y_u \overline{q_L} \Tilde{\Phi} u_R + \mathbf{Y}_\Psi \overline{\Psi_L} \Psi_R \sigma + \mathrm{H.c.} \, ,
\end{align}
the usual SM quark mass terms and that of the VLQs, generated by the PQ breaking scalar. The $3\times 3$ matrix $\Y_{d,u}$ and the $2 \times 2$ matrix $\mathbf{Y}_\Psi$ are in general complex. After SSB, we obtain the following colored fermion mass Lagrangian,
\begin{align}
    -\mathcal{L}_\mathrm{mass}^{\text{c}} \supset -\mathcal{L}_\mathrm{mass}^{\text{c} , 0} & = \overline{d_L} \mathbf{M}_d d_R +  \overline{u_L} \mathbf{M}_u u_R +  \overline{\Psi_L} \mathbf{M}_\Psi \Psi_R + \mathrm{H.c.} \, , \nonumber \\
    \mathbf{M}_d & = \frac{v}{\sqrt{2}} \Y_d \; , \; \mathbf{M}_u = \frac{v}{\sqrt{2}} \Y_u \; ,\; \mathbf{M}_\Psi = \frac{v_\sigma}{\sqrt{2}} \mathbf{Y}_\Psi \; .
\end{align}
In this section, we will discuss the specific models presented in Table~\ref{tab:postcharges}, and compute the heavy-light quark mixing, i.e. the mixing between the heavy PQ scale mass VLQ $\Psi$, introduced to solve the strong CP problem and mediate neutrino mass generation, and the ordinary SM quarks with masses proportional to the EW scale. In what follows, for simplicity, we discuss the PQ charge assignments $\mathcal{Q}_\text{PQ}$ leading to a few heavy-light mixing terms presented in column 5 of Table~\ref{tab:postcharges}. Namely, for all cases we take $\mathcal{Q}_\text{PQ} = 0$, except for $\Psi \sim (\mathbf{3}, \mathbf{2}, 1/6)$ where we take $\mathcal{Q}_\text{PQ} = 1/2$. The heavy-light mixing is crucial for the computation of the flavor-violating quark-axion couplings, whose phenomenological implications are discussed in Sec.~\ref{sec:axionflavorviolatingDirac}.

There are a total of seven models depending on the VLQ representation: 
\begin{itemize}
    \item \underline{$\Psi \sim (\mathbf{3}, \mathbf{1}, -1/3)$:} For this case we have (see Table~\ref{tab:postcharges}),
\begin{align}
    -\mathcal{L}_\mathrm{Yuk.}^{\text{c}} & =  \mathcal{L}_\mathrm{Yuk.}^{c,0} + \mathbf{M}_{\Psi d} \overline{\Psi_L} d_R + \mathrm{H.c.} \, ,
    \label{eq:Lyuk1down}
\end{align}
where $\mathbf{M}_{\Psi d}$ is a $2 \times 3$ bare mass matrix. Defining, $D_{L,R} = (d,\Psi)^T_{L,R}$, after SSB, we can write the mass Lagrangian in the compact form:
\begin{align}
-\mathcal{L}_\mathrm{mass}^{\text{c}} & = \overline{D_L} \boldsymbol{\mathcal{M}}_d D_R 
    + \overline{u_L} \mathbf{M}_u u_R + \mathrm{H.c.} \; , \; \boldsymbol{\mathcal{M}}_d = \begin{pmatrix}
        \mathbf{M}_d & 0 \\
        \mathbf{M}_{\Psi d} & \mathbf{M}_\Psi
    \end{pmatrix} \; .
    \label{eq:Lcompact1down}
\end{align}
This case leads to an analogous full quark mass matrix as in the original BBP model, being the minimal NB-realization, discussed in Sec.~\ref{sec:darkNB}. We will partially repeat the diagonalization procedure in Sec.~\ref{sec:CKMpheno} of the full quark mass matrix above with a new notation that allows to then extend the procedure to the remaining six VLQ representation cases.

The SM up-quark mass matrix is diagonalized in the standard fashion through the unitary transformations $u_{L,R} \to \mathbf{V}_{L,R}^u\, u_{L,R}$, determined by diagonalizing the Hermitian matrices $\mathbf{H}_{u} = \mathbf{M}_{u} \mathbf{M}_{u}^{\dagger}$
and $\mathbf{H}_{u}^\prime = \mathbf{M}_{u}^{\dagger} \mathbf{M}_{u}$ -- see Sec.~\ref{sec:fermionmassmix}. Namely, 
\begin{align}
\mathbf{V}_L^{u \dagger} \mathbf{M}_u \mathbf{V}_R^u = \mathbf{D}_{u} = \text{diag}\,(m_u, m_c, m_t)\,,
\label{eq:diagupSM}
\end{align}
where $m_{u,c,t}$ are the physical light up-type quark masses. The  full quark mass matrix $\boldsymbol{\mathcal{M}}_d$ can be diagonalized through the bi-unitary transformations $(d, \Psi)_{L,R} \to \boldsymbol{\mathcal{V}}_{L,R}^d\, (d, \Psi)_{L,R}$ as 
\begin{align}
\boldsymbol{\mathcal{V}}_L^{d \dagger} \boldsymbol{\mathcal{M}}_d \ \boldsymbol{\mathcal{V}}_R^d = \boldsymbol{\mathcal{D}}_{d} = \text{diag}\,(m_d, m_s, m_b , \tilde{M}_{\Psi_1}, \tilde{M}_{\Psi_{2}})\,,
\label{eq:diagdownfull}
\end{align}
where $m_{d,s,b}$ are the physical light down-type quark masses and $\tilde{M}_{\Psi_{1,2}}$ are the heavy quark masses.
For a given $\boldsymbol{\mathcal{M}}_d$, $\boldsymbol{\mathcal{V}}_L^d$ and $\boldsymbol{\mathcal{V}}_R^d$ are determined by diagonalizing the Hermitian matrices $\boldsymbol{\mathcal{H}}_{d} = \boldsymbol{\mathcal{M}}_{d} \boldsymbol{\mathcal{M}}_{d}^{\dagger}$
and $\boldsymbol{\mathcal{H}}_{d}^\prime = \boldsymbol{\mathcal{M}}_{d}^{\dagger} \boldsymbol{\mathcal{M}}_{d}$. 
 
Since the heavy quark mass scale is set by the PQ breaking scale, in what follows we adopt the seesaw approximation $M_d \ll M_{\Psi d}\,  M_\Psi$.  
Hence, we start by block diagonalizing the Hermitian matrix $\boldsymbol{\mathcal{H}}_{d}$, obtaining for the light and heavy sector mass
\begin{equation}
\mathbf{\Lambda}_{d} \simeq \mathbf{M}_d \left(\mathbb{1}_3-\mathbf{M}_{\Psi d}^{\dagger}\mathbf{\Lambda}_{\Psi}^{-1} \mathbf{M}_{\Psi d}\right) \mathbf{M}_d^\dagger \; , \; \mathbf{\Lambda}_{\Psi}^d \simeq \mathbf{M}_\Psi \mathbf{M}_\Psi^\dagger + \mathbf{M}_{\Psi d} \mathbf{M}_{\Psi d}^\dagger \; .
\end{equation}
Furthermore, for $M_{\Psi d} \ll M_\Psi$, $\mathbf{\Lambda}_d$ becomes,
\begin{equation}
\mathbf{\Lambda}_{d} \simeq \mathbf{M}_d \mathbf{M}_d^\dagger - \mathbf{M}_d
\mathbf{M}_{\Psi d}^{\dagger}(\mathbf{M}_\Psi \mathbf{M}_\Psi^\dagger)^{-1} \mathbf{M}_{\Psi d}\mathbf{M}_d^\dagger
\; .
\end{equation}
In this limit, the leading contributions to light down-quark masses come mainly from the SM Yukawa interactions, with the contributions from the heavy-light quark mixing being sub-leading. The matrix $\mathbf{\Lambda}_d$ is diagonalized through the $3 \times 3$ unitary transformation $d_{L} \to \mathbf{V}_{L}^d\, d_{L}$, leading to the well-known Cabibbo-Kobayashi-Maskawa~(CKM) quark mixing matrix $\mathbf{V}_{\text{CKM}} = \mathbf{V}^{u\,\dagger}_L \mathbf{V}^d_L$ appearing in quark charged-current interactions. Furthermore, the heavy sector mass matrix $\mathbf{\Lambda}_\Psi^d$ is diagonalized through the $2 \times 2$ unitary transformation $\Psi_{L} \to \mathbf{V}_{L}^\Psi\, \Psi_{L}$.

The block diagonalisation procedure above was achieved by performing appropriate unitary transformations on the LH quark fields. Similarly, we can diagonalize the Hermitian matrix $\boldsymbol{\mathcal{H}}_d^\prime$ by rotating the right-handed quark fields. We parameterize the unitary transformations $\boldsymbol{\mathcal{V}}_{L,R}^d$ of Eq.~\eqref{eq:diagdownfull}, that relate the weak and mass basis, as follows:
\begin{align}
    \begin{pmatrix}
    \bm{d} \\
    \bm{\Psi}
    \end{pmatrix}_{L,R}
    =  \begin{pmatrix}
    (\mathbb{1}_{3}- \mathbf{\Theta}^d \mathbf{\Theta}^{d \dagger})^{1/2} & \mathbf{\Theta}^d \\
    -\mathbf{\Theta}^{d \dagger} & (\mathbb{1}_{2}-  \mathbf{\Theta}^{d \dagger} \mathbf{\Theta}^d)^{1/2}
    \end{pmatrix}_{L,R}
    \begin{pmatrix}
    \mathbf{V}^d & 0 \\
    0 & \mathbf{V}^\Psi
    \end{pmatrix}_{L,R}
    \begin{pmatrix}
    d\\
   \Psi
    \end{pmatrix}_{L,R} \; .
    \label{eq:qtrans}
\end{align}
The mixing patterns between light and heavy quarks are given by the $3\times 2$ matrices:
\begin{equation}
    \mathbf{\Theta}_R^d \simeq \mathbf{M}_{\Psi d}^\dagger\, \mathbf{M}_\Psi^{\dagger -1}\; , \;
    \mathbf{\Theta}_L^d \simeq \mathbf{M}_d\,\mathbf{M}_{\Psi d}^\dagger\,(\mathbf{\Lambda}_\Psi^{d})^{-1}\simeq \mathbf{M}_d\,\mathbf{\Theta}_R^d\, \mathbf{M}_\Psi^{-1}\,.
    \label{eq:HLmixingrefdown}
\end{equation}
Note that, since $\Theta_L^d \sim \mathcal{O}(v/f_{\text{PQ}}) \Theta_R^d$, with $v/f_{\text{PQ}} \sim \mathcal{O}(10^{-10}) \ll 1$, hence $\mathbf{\Theta}_L^d\ll \mathbf{\Theta}_R^d \ll 1$ and, therefore, we can safely neglect contributions coming from $\mathbf{\Theta}_L^d$. Since $\mathbf{M}_{\Psi d}$ is an arbitrary bare mass term we can choose its scale $M_{\Psi d}$ close to the $M_\Psi \sim \mathcal{O}(f_{\text{PQ}})$ scale, leading to a sizable $\Theta_R^d \sim M_{\Psi d}/M_\Psi$. This will translate into sizable flavor-violating quark-axion couplings as discussed in Sec.~\ref{sec:axionflavorviolatingDirac}. 

    \item \underline{$\Psi \sim (\mathbf{3}, \mathbf{1}, 2/3)$:} The quark Yukawa Lagrangian is given by (see Table~\ref{tab:postcharges}),
\begin{align}
    -\mathcal{L}_\mathrm{Yuk.}^{\text{c}} & =  \mathcal{L}_\mathrm{Yuk.}^{c,0} + \mathbf{M}_{\Psi u} \overline{\Psi_L} u_R + \mathrm{H.c.} \, ,
\end{align}
where $\mathbf{M}_{\Psi u}$ is a $2 \times 3$ bare mass matrix. Defining, $U_{L,R} = (u,\Psi)^T_{L,R}$, after SSB, we can write the mass Lagrangian in the compact form:
\begin{align}
-\mathcal{L}_\mathrm{mass}^{\text{c}} & = \overline{U_L} \boldsymbol{\mathcal{M}}_u U_R 
    + \overline{d_L} \mathbf{M}_d d_R + \mathrm{H.c.} \; , \; \boldsymbol{\mathcal{M}}_u = \begin{pmatrix}
        \mathbf{M}_u & 0 \\
        \mathbf{M}_{\Psi u} & \mathbf{M}_\Psi
    \end{pmatrix} \; .
    \label{eq:Lcompact1up}
\end{align}
We repeat the procedure outlined for the case $\Psi \sim (\mathbf{3}, \mathbf{1}, -1/3)$, making the replacement $d \rightarrow u$ for the up-type sector. Thus, in total analogy, here we can safely neglect contributions coming from $\mathbf{\Theta}_L^u$. However, $\mathbf{\Theta}_R^u$ can lead to sizable axion flavor-violating couplings (see Sec.~\ref{sec:axionflavorviolatingDirac}).
    
    \item \underline{$\Psi \sim (\mathbf{3}, \mathbf{2}, 1/6)$:} Defining, $\Psi_{L,R} \equiv (T,B)_{L,R}^T$, the colored fermion Yukawa Lagrangian for this case is (see Table~\ref{tab:postcharges}),
\begin{align}
    -\mathcal{L}_\mathrm{Yuk.}^{\text{c}} & =  \mathcal{L}_\mathrm{Yuk.}^{c,0} + \mathbf{M}_{q \Psi} \overline{q_L} \Psi_R + \mathrm{H.c.} \, ,
\end{align}
where $\mathbf{M}_{q \Psi}$ is a $3 \times 2$ bare mass matrix. Defining, $D_{L,R} = (d,B)^T_{L,R}$ and $U_{L,R} = (u,T)^T_{L,R}$, after SSB, we obtain,
\begin{align}
-\mathcal{L}_\mathrm{mass}^{\text{c}} & = \overline{D_L} \boldsymbol{\mathcal{M}}_d D_R 
    + \overline{U_L} \boldsymbol{\mathcal{M}}_u U_R 
    + \mathrm{H.c.} \; , \; \nonumber \\
\boldsymbol{\mathcal{M}}_d & = \begin{pmatrix}
        \mathbf{M}_d & \mathbf{M}_{q \Psi}  \\
         0 & \mathbf{M}_\Psi
    \end{pmatrix} \; , \; \boldsymbol{\mathcal{M}}_u = \begin{pmatrix}
        \mathbf{M}_u & \mathbf{M}_{q \Psi} \\
        0 & \mathbf{M}_\Psi
    \end{pmatrix} \; .
    \label{eq:Lcompact2SM}
\end{align}
Repeating the block-diagonalization procedure outlined above, we obtain the mixing patterns between light and heavy quarks for this specific model which are given by the following $3\times 2$ matrices:
\begin{align}
    \mathbf{\Theta}_L^{d,u} \simeq \mathbf{M}_{q \Psi} \, \mathbf{M}_\Psi^{-1}\; , \;
    \mathbf{\Theta}_R^{d,u} & \simeq \mathbf{M}_{d,u}^{\dagger}\,\mathbf{M}_{q \Psi} \,(\mathbf{\Lambda}_\Psi)^{-1} \simeq \mathbf{M}_{d,u}^\dagger\,\mathbf{\Theta}_L^{d,u}\, (\mathbf{M}_\Psi^{\dagger})^{-1} \, , \nonumber \\
\mathbf{\Lambda}_{\Psi} & \simeq \mathbf{M}_\Psi^\dagger \mathbf{M}_\Psi + \mathbf{M}_{q\Psi}^\dagger \mathbf{M}_{q \Psi} \; .
\label{eq:HLmixing2SM}
\end{align}
Note that, for this case, $\mathbf{\Theta}_R^{d,u}$ can be safely neglected while $\mathbf{\Theta}_L^{d,u}$ can be sizable leading to interesting flavor violating axion phenomenology (see Sec.~\ref{sec:axionflavorviolatingDirac}). 
    
    \item \underline{$\Psi \sim (\mathbf{3}, \mathbf{2}, -5/6)$:} In this model $\Psi_{L,R} \equiv (B,Y)_{L,R}^T$, with $B$ and $Y$ having electric charge $-1/3$ and $-4/3$, respectively (see Table~\ref{tab:postcharges}). The Yukawa Lagrangian is,
\begin{align}
    -\mathcal{L}_\mathrm{Yuk.}^{\text{c}} & =  \mathcal{L}_\mathrm{Yuk.}^{c,0} + \Y_{\Psi d} \overline{\Psi_L} \tilde{\Phi} d_R + \mathrm{H.c.} \, ,
    \label{eq:Lyuk2down}
\end{align}
where $\mathbf{Y}_{\Psi d}$ is a $2 \times 3$ complex Yukawa matrix. Defining, $D_{L,R} = (d,B)^T_{L,R}$, after SSB, we have,
\begin{align}
-\mathcal{L}_\mathrm{mass}^{\text{c}} & = \overline{D_L} \boldsymbol{\mathcal{M}}_d D_R 
    + \overline{u_L} \mathbf{M}_u u_R + \mathbf{M}_\Psi \overline{Y_L} Y_R  
    + \mathrm{H.c.} \; , \nonumber \\
    \boldsymbol{\mathcal{M}}_d & = \begin{pmatrix}
        \mathbf{M}_d & 0  \\
         \mathbf{M}_{\Psi d} & \mathbf{M}_\Psi
    \end{pmatrix} \; , \; \mathbf{M}_{\Psi d} = \frac{v}{\sqrt{2}} \mathbf{Y}_{\Psi d} \; .
    \label{eq:Lcompact2down}
\end{align}
Block-diagonalizing the above mass matrix leads to the analogous heavy-light mixing patterns as for the $\Psi \sim (\mathbf{3},\mathbf{1},-1/3)$ [see Eq.~\eqref{eq:HLmixingrefdown}]. However, here $\mathbf{M}_{\Psi d}$  is no longer an arbitrary bare mass term but is given by a Yukawa interaction with the Higgs doublet as shown in the above equation.
Consequently, for this case we have $\mathbf{\Theta}_R^d \sim \mathcal{O}(v/f_{\text{PQ}}) \sim \mathcal{O}(10^{-10})$ and $\mathbf{\Theta}_L^d \sim \mathcal{O}(v/f_{\text{PQ}} \mathbf{\Theta}_R^d) \sim \mathcal{O}(10^{-20})$, thus both heavy light-mixing parameters are completely negligible.

\item \underline{$\Psi \sim (\mathbf{3}, \mathbf{2}, 7/6)$:} In this model $\Psi_{L,R} \equiv (X,T)_{L,R}^T$, with $T$ and $X$ having electric charge $2/3$ and $5/3$, respectively (see Table~\ref{tab:postcharges}). We have,
\begin{align}
    -\mathcal{L}_\mathrm{Yuk.}^{\text{c}} & =  \mathcal{L}_\mathrm{Yuk.}^{c,0} + \Y_{\Psi u} \overline{\Psi_L} \Phi u_R + \mathrm{H.c.} \, ,
    \label{eq:Lyuk2up}
\end{align}
where $\mathbf{Y}_{\Psi u}$ is a $2 \times 3$ complex Yukawa matrix. Defining $U_{L,R} = (u,T)^T_{L,R}$ leads to
\begin{align}
-\mathcal{L}_\mathrm{mass}^{\text{c}} & = \overline{d_L} \mathbf{M}_d d_R 
    + \overline{U_L} \boldsymbol{\mathcal{M}}_u U_R + \mathbf{M}_\Psi \overline{X_L} X_R  
    + \mathrm{H.c.} \; , \nonumber \\
    \boldsymbol{\mathcal{M}}_u & = \begin{pmatrix}
        \mathbf{M}_u & 0 \\
        \mathbf{M}_{\Psi u}  & \mathbf{M}_\Psi
    \end{pmatrix} \; , \; \mathbf{M}_{\Psi u} = \frac{v}{\sqrt{2}} \mathbf{Y}_{\Psi u} \; .
    \label{eq:Lcompact2up}
\end{align}
The heavy-light mixing patterns are analogous to the ones for $\Psi \sim (\mathbf{3},\mathbf{1}, 2/3)$ with the replacement $d \rightarrow u$ in Eq.~\eqref{eq:HLmixingrefdown}. However, as shown above, here $\mathbf{M}_{\Psi u}$ is no longer an arbitrary bare mass term but is proportional to the EW scale.
Consequently, for this case we have $\mathbf{\Theta}_R^u \sim \mathcal{O}(v/f_{\text{PQ}}) \sim \mathcal{O}(10^{-10})$ and $\mathbf{\Theta}_L^u \sim \mathcal{O}(v/f_{\text{PQ}} \mathbf{\Theta}_R^u) \sim \mathcal{O}(10^{-20})$, being completely negligible.
 
    \item \underline{$\Psi \sim (\mathbf{3}, \mathbf{3}, -1/3)$:} Defining $\Psi_{L,R} = (T,B,Y)_{L,R}^T$, the quark Yukawa Lagrangian is (see Table~\ref{tab:postcharges})
\begin{align}
    -\mathcal{L}_\mathrm{Yuk.}^{\text{c}} & =  \mathcal{L}_\mathrm{Yuk.}^{c,0} + \Y_{q \Psi} \overline{q_L} \Phi \tau^a \Psi_R^a + \mathrm{H.c.} \, ,
\end{align}
where $\mathbf{Y}_{q \Psi}$ is a $3 \times 2$ complex Yukawa matrix and $\tau^a$ ($a=1,2,3$) are the Pauli matrices. With $D_{L,R} = (d,B)^T_{L,R}$ and $U_{L,R} = (u,T)^T_{L,R}$, after SSB, 
\begin{align}
-\mathcal{L}_\mathrm{mass}^{\text{c}} & = \overline{D_L} \boldsymbol{\mathcal{M}}_d D_R 
    + \overline{U_L} \boldsymbol{\mathcal{M}}_u U_R 
    + \mathrm{H.c.} \; , \; \nonumber \\
\boldsymbol{\mathcal{M}}_d &= \begin{pmatrix}
        \mathbf{M}_d & -\mathbf{M}_{q \Psi} \\
         0 & \mathbf{M}_\Psi
    \end{pmatrix} \; , \; \boldsymbol{\mathcal{M}}_u = \begin{pmatrix}
        \mathbf{M}_u & \sqrt{2} \mathbf{M}_{q \Psi} \\
        0  & \mathbf{M}_\Psi
    \end{pmatrix} \; , \; \mathbf{M}_{q \Psi} = \frac{v}{\sqrt{2}} \mathbf{Y}_{q \Psi} \; .
\end{align}
The block-diagonalization procedure leads to:
\begin{align}
\mathbf{\Theta}_L^{d} \simeq - \mathbf{M}_{q \Psi} \, \mathbf{M}_\Psi^{-1}\; , \;
\mathbf{\Theta}_R^{d} & \simeq - \mathbf{M}_{d}^{\dagger}\,\mathbf{M}_{q \Psi} \,(\mathbf{\Lambda}_\Psi^d)^{-1} \simeq - \mathbf{M}_{d}^\dagger\,\mathbf{\Theta}_L^{d}\, (\mathbf{M}_\Psi^{\dagger})^{-1} \, , \nonumber \\
\mathbf{\Lambda}_{\Psi}^d & \simeq \mathbf{M}_\Psi^\dagger \mathbf{M}_\Psi + \mathbf{M}_{q\Psi}^\dagger \mathbf{M}_{q \Psi} \; ; \nonumber \\
\mathbf{\Theta}_L^{u} \simeq \sqrt{2} \mathbf{M}_{q \Psi} \, \mathbf{M}_\Psi^{-1}\; , \;
\mathbf{\Theta}_R^{u} & \simeq \sqrt{2}\mathbf{M}_{u}^{\dagger}\,\mathbf{M}_{q \Psi} \,(\mathbf{\Lambda}_\Psi^u)^{-1} \simeq \sqrt{2} \mathbf{M}_{u}^\dagger\,\mathbf{\Theta}_L^{u}\, (\mathbf{M}_\Psi^{\dagger})^{-1} \, , \nonumber \\
\mathbf{\Lambda}_{\Psi}^u & \simeq \mathbf{M}_\Psi^\dagger \mathbf{M}_\Psi + 2 \mathbf{M}_{q\Psi}^\dagger \mathbf{M}_{q \Psi} \; .
\label{eq:HLmixing3down}
\end{align}
For this case $\mathbf{\Theta}_{R,L}^{d,u}$ are completely negligible.

\item \underline{$\Psi \sim (\mathbf{3}, \mathbf{3}, 2/3)$:} Defining $\Psi_{L,R} = (X,T,B)_{L,R}^T$, we have (see Table~\ref{tab:postcharges}),
\begin{align}
    -\mathcal{L}_\mathrm{Yuk.}^{\text{c}} & =  \mathcal{L}_\mathrm{Yuk.}^{c,0} + \Y_{q \Psi} \overline{q_L} \tilde{\Phi} \tau^a \Psi_R^a + \mathrm{H.c.} \, ,
    \label{eq:Lyuk3up}
\end{align}
where $\mathbf{Y}_{q \Psi}$ is a $3 \times 2$ complex Yukawa matrix and $\tau^a$ ($a=1,2,3$) are the Pauli matrices. Defining, $D_{L,R} = (d,B)^T_{L,R}$ and $U_{L,R} = (u,T)^T_{L,R}$, after SSB, we have,
\begin{align}
-\mathcal{L}_\mathrm{mass}^{\text{c}} & = \overline{D_L} \boldsymbol{\mathcal{M}}_d D_R 
    + \overline{U_L} \boldsymbol{\mathcal{M}}_u U_R 
    + \mathrm{H.c.} \; , \; \nonumber \\
\boldsymbol{\mathcal{M}}_d &= \begin{pmatrix}
        \mathbf{M}_d & - \sqrt{2} \mathbf{M}_{q \Psi}  \\
         0 & \mathbf{M}_\Psi
    \end{pmatrix} \; , \; \boldsymbol{\mathcal{M}}_u = \begin{pmatrix}
        \mathbf{M}_u &  \mathbf{M}_{q \Psi} \\
        0  & \mathbf{M}_\Psi
    \end{pmatrix} \; , \; \mathbf{M}_{q \Psi} = \frac{v}{\sqrt{2}} \mathbf{Y}_{q \Psi} \; .
\end{align}
The block-diagonalization procedure results in,
\begin{align}
\mathbf{\Theta}_L^{d} \simeq - \sqrt{2} \mathbf{M}_{q \Psi} \, \mathbf{M}_\Psi^{-1}\; , \;
\mathbf{\Theta}_R^{d} & \simeq - \sqrt{2} \mathbf{M}_{d}^{\dagger}\,\mathbf{M}_{q \Psi} \,(\mathbf{\Lambda}_\Psi^d)^{-1} \simeq - \sqrt{2} \mathbf{M}_{d}^\dagger\,\mathbf{\Theta}_L^{d}\, (\mathbf{M}_\Psi^{\dagger})^{-1} \, , \nonumber \\
\mathbf{\Lambda}_{\Psi}^d & \simeq \mathbf{M}_\Psi^\dagger \mathbf{M}_\Psi + 2 \mathbf{M}_{q\Psi}^\dagger \mathbf{M}_{q \Psi} \; ; \; \nonumber \\
\mathbf{\Theta}_L^{u} \simeq \mathbf{M}_{q \Psi} \, \mathbf{M}_\Psi^{-1}\; , \;
\mathbf{\Theta}_R^{u} & \simeq \mathbf{M}_{u}^{\dagger}\,\mathbf{M}_{q \Psi} \,(\mathbf{\Lambda}_\Psi^u)^{-1} \simeq \mathbf{M}_{u}^\dagger\,\mathbf{\Theta}_L^{u}\, (\mathbf{M}_\Psi^{\dagger})^{-1} \, , \nonumber \\
\mathbf{\Lambda}_{\Psi}^u & \simeq \mathbf{M}_\Psi^\dagger \mathbf{M}_\Psi + \mathbf{M}_{q\Psi}^\dagger \mathbf{M}_{q \Psi} \; .
\label{eq:HLmixing3up}
\end{align}
Once again, for this case $\mathbf{\Theta}_{R,L}^{d,u}$ are completely negligible.
    
\end{itemize}
In summary, the models with VLQ $\Psi$ representations $(\mathbf{3}, \mathbf{1}, -1/3)$, $(\mathbf{3}, \mathbf{1}, 2/3)$ and $(\mathbf{3}, \mathbf{2}, 1/6)$, are the only ones where heavy-light quark mixing can be sizable. Specifically, in the RH down quark, RH up quark, and LH up and down quark sectors, respectively. 

%In what follows we explain how our unified framework leads to neutrino mass generation and solves the cosmological DM problem. We will then study the associated phenomenology of each scenario with special emphasis on the heavy-light quark mixing and its implications for flavor-violating-axion couplings to quarks.

%%%%%%%%%%%%%%%%%%%%%%%%%%%%%%%%%%%%%%%%%%%%%%%%%%%%%%%%%%%%%%%%%%%%
\subsection{Radiative Dirac neutrino masses}
\label{sec:Diracneutrinos} 
%%%%%%%%%%%%%%%%%%%%%%%%%%%%%%%%%%%%%%%%%%%%%%%%%%%%%%%%%%%%%%%%%%%%
 
\begin{figure}[t!]
        \centering
        \includegraphics[scale=0.9]{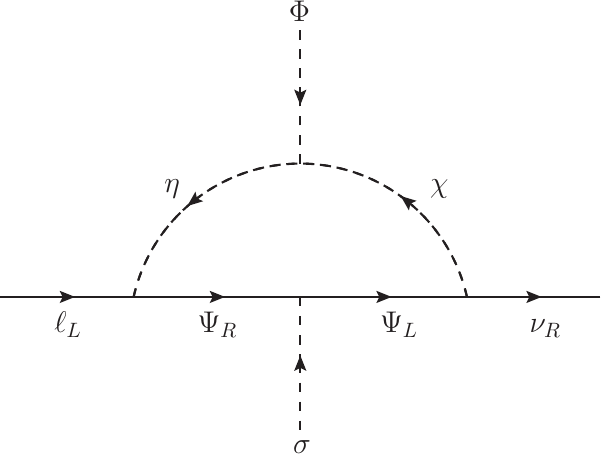}
        \caption{One-loop color-mediated Dirac neutrino masses (see Tables~\ref{tab:generalDirac} and~\ref{tab:postcharges}).}
    \label{fig:neutrinoDirac1loopcolor}
    \end{figure}

The Yukawa interactions responsible for neutrino mass generation are:
\begin{equation}
    - \mathcal{L}_{\text{Yuk.}}^{\nu} = \mathbf{Y}_{\eta} \overline{\ell_L} \; \tilde{\eta} \Psi_R + \mathbf{Y}_{\chi} \overline{\Psi_L} \chi \nu_R + \mathbf{Y}_{\Psi} \overline{\Psi_L} \Psi_R \sigma + \text{H.c.} \; .
    \label{eq:LYukgenYneq0}
\end{equation}
The most general scalar potential allowed by the symmetries of our models is given as
\begin{align}
V & = \sum_S \left[ \mu_{S}^2 S^{\dag}S +\frac{\lambda_S}{4}\left(S^{\dag}S\right)^2\right] + \sum_{S\neq S^\prime}
\frac{\lambda_{SS^\prime}}{2}\left(S^{\dag}S\right)\left({S^\prime}^{\dag}S^\prime\right) + \tilde{\lambda}_{\Phi\eta}\left(\Phi^{\dag}\eta\right)\left(\eta^{\dag}\Phi\right) \nonumber \\
&+ \left(\kappa\, \eta^\dagger \Phi \, \chi + \mathrm{H.c.}\right) \; ,
\label{eq:VpotDirac}
\end{align}
where $S,S^\prime = \Phi, \sigma, \eta, \chi$, with the relevant scalar-potential term for neutrino mass generation being the cubic term. Together with the Yukawa terms above these interactions trigger one-loop color-mediated Dirac neutrino masses via the diagram shown in Fig.~\ref{fig:neutrinoDirac1loopcolor}. 

In order to determine the neutrino mass matrix we must define couplings in the mass-eigenstate basis. In Sec.~\ref{sec:quarksector} we computed the $\boldsymbol{\mathcal{V}}_{L,R}^{q}$ matrices which describe the mixing between $\Psi$ and SM quarks as seen in Eqs.~\eqref{eq:diagdownfull} and~\eqref{eq:qtrans}, with $q$ representing the up/down sector, depending on the VLQ representation. Thus, the Yukawa matrices in Eq.~\eqref{eq:LYukgenYneq0} are in the mass-eigenstate basis:
\begin{align}
(\tilde{\mathbf{Y}}_\eta)_{\alpha j} & = \sum_{k=1}^{2} (\mathbf{Y}_\eta)_{\alpha k} (\boldsymbol{\mathcal{V}}_R^q)_{(3+k) j} \; , \; (\tilde{\mathbf{Y}}_\chi)_{i \beta} = \sum_{k=1}^{2} (\boldsymbol{\mathcal{V}}_L^{q \ast})_{(3+k) i} (\mathbf{Y}_\chi)_{k \beta} \; ,
 \end{align}
where $\alpha, \beta$ are flavor indices, $j,k$ run from 1 to 5, with the first three indices referring to the SM quarks and the remaining two corresponding to the heavy VLQ. Without loss of generality, we assume charged-leptons are in their mass basis, and also that $\mathbf{M}_\Psi \equiv v_\sigma \mathbf{Y}_\Psi/\sqrt{2} = \text{diag}(M_{\Psi_1}, M_{\Psi_2})$. Namely, their masses are given by $\mathbf{\tilde{m}}=(m_{q_1},m_{q_2},m_{q_3},\tilde{M}_{\Psi_1},\tilde{M}_{\Psi_2})$ where $q_{1,2,3}$ denote the SM quarks. 

The $\eta$ and $\chi$ (mass eigenstates) with the same electric charge as the colored fermion $\Psi$, namely $Q_{\Psi}$, will mediate Dirac neutrino masses at the one-loop level together with $\Psi$. Hence, after SSB, the colored scalars with electric charge $Q_{\Psi}$, in the basis $(\eta_{Q_{\Psi}}, \chi_{Q_{\Psi}})$, will mix through,
\begin{align}
    \boldsymbol{\mathcal{M}}_{\eta \chi}^2 = \begin{pmatrix}
        m^2_{\eta} & \frac{\kappa v}{\sqrt{2}} \\
        \frac{\kappa v}{\sqrt{2}} & m^2_{\chi} 
    \end{pmatrix} \; , \;
    m^2_{\eta} &= \mu_{\eta}^2 + \frac{1}{4} \left[(\lambda_{\Phi \eta}+\tilde{\lambda}_{\Phi \eta}) v^2 + \lambda_{\sigma \eta} v_\sigma^2\right] \; , \nonumber \\ 
    m^2_{\chi} &= \mu_{\chi}^2 + \frac{1}{4} (\lambda_{\Phi \chi} v^2 + \lambda_{\sigma \chi} v_\sigma^2) \; ,
\end{align}
where the VEV $\langle \sigma\rangle = v_\sigma /\sqrt{2} \equiv f_{\text{PQ}} = N f_a$ breaks the PQ symmetry -- see Eqs.~\eqref{eq:axionfa} and~\eqref{eq:PQWWscale}. The corresponding mass eigenstates $\zeta_{1,2}$ are expressed through the $2 \times 2$ rotation matrix $\mathbf{R}$,
\begin{align}
\begin{pmatrix}
        \eta_{Q_{\Psi}} \\ \chi_{Q_{\Psi}}
    \end{pmatrix}= \mathbf{R} \begin{pmatrix}
        \zeta_1 \\ \zeta_2
    \end{pmatrix} = \begin{pmatrix}
        \cos \alpha & -\sin \alpha \\
        \sin \alpha & \cos \alpha
    \end{pmatrix} \begin{pmatrix}
        \zeta_1 \\ \zeta_2
    \end{pmatrix}  \; , \; \tan{2 \alpha} = \frac{\sqrt{2}\,\kappa\, v}{ m^2_\eta - m^2_{\chi} } \; ,
    \label{eq:rotationscalars}
\end{align}
with masses given by,
\begin{align}
    m_{\zeta_{1,2}}^2 = \frac{1}{2} \left[ m_\eta^2 + m_\chi^2 \pm \sqrt{(m_\eta^2 - m_\chi^2)^2 + 2 \kappa^2 v^2} \right] \; .
\end{align}
The remaining mass eigenstates stemming from the weak multiplet $\eta$ ($\chi$) with distinct electric charges are degenerate in mass with value $m_\eta$ ($m_\chi$).

From the above, the neutrino mass matrix is given by~\cite{CentellesChulia:2024iom},
\begin{align}
(\mathbf{M}_\nu)_{\alpha \beta} & = \frac{N_c}{16 \pi^2 \sqrt{2}} \sum_{j, k= 1}^{5} (\tilde{\mathbf{Y}}_\eta)_{\alpha j} (\tilde{\mathbf{Y}}_\chi)_{j \beta} \; \mathbf{R}_{1 k} \mathbf{R}_{2 k} \; \tilde{m}_{j} \frac{m_{\zeta_k}^2}{\tilde{m}_{j}^2-m_{\zeta_k}^2} \ln\left(\frac{\tilde{m}_{j}^2}{m_{\zeta_ k}^2}\right) \; ,
\label{eq:neutrinomass}
\end{align}
where the color factor $N_c=3$. From this, we can estimate the neutrino masses as
\begin{align}
(M_\nu)_{\alpha \beta} & \sim 0.1 \text{eV} \; \left(\frac{(\tilde{Y}_\eta)_{\alpha j} (\tilde{Y}_\chi)_{j \beta}}{10^{-2}}\right)  \left(\frac{\kappa}{ 10^2 \; \text{GeV}}\right) \; \left(\frac{\tilde{M}_{\Psi j}}{10^{12} \; \text{GeV}}\right) \; \left(\frac{10^{10} \; \text{GeV}}{m_{\zeta_k}}\right)^2 \; ,
\end{align}
where we took typical values for the VLQ and colored leptoquark masses which are given by the PQ scale $f_{\text{PQ}}=v_\sigma/\sqrt{2}$. As we will see in Sec.~\ref{sec:axionDMphotonDirac}, $f_a \sim \mathcal{O}(10^{12})$ GeV is the typical scale required for the axion particle to account for the observed DM abundance.

Notice that the parameter $\kappa$ can be naturally small in the t' Hooft sense~\cite{tHooft:1979rat}, as in the limit $\kappa \to 0$, the theory exhibits a larger symmetry. It follows that the smallness of Dirac neutrino masses is naturally controlled by the smallness of the $\zeta_{1,2}$ mass splitting. Moreover, the minimal number of $\Psi$ species to successfully generate the two observed neutrino mass splittings and lepton mixing is 2, which we choose here for simplicity. Hence, these minimal color-mediated Dirac neutrino mass models predict a massless light neutrino. However, adding an extra $\Psi$ species would result in three non-zero light neutrino masses. As already mentioned, the underlying PQ symmetry in our construction enforces the {\em Diracness} of neutrinos.

%%%%%%%%%%%%%%%%%%%%%%%%%%%%%%%%%%%%%%%%%%%%%%%%%%%%%%%%%%%%%%%%%%%%
\subsection{Axion-to-photon coupling, dark matter and cosmology}
\label{sec:axionDMphotonDirac}
%%%%%%%%%%%%%%%%%%%%%%%%%%%%%%%%%%%%%%%%%%%%%%%%%%%%%%%%%%%%%%%%%%%%

The color-mediated Majorana neutrino mass scenario, studied in Sec.~\ref{sec:colormediatedMajorana}, in its original version features hypercharless colored fermions which strongly disfavor the more predictive post-inflationary axion DM scenrio. As shown in Sec.~\ref{sec:possibilityHLmixing}, it is required for the exotic fermions to mix with SM quarks and decay into ordinary matter to envisage moving past the pre-inflationary scenario. Notably, our color-mediated Dirac neutrino mass models feature heavy VLQs that mix with SM quarks as analyzed in detail in Sec.~\ref{sec:quarksector}. Thus, the current framework allows for the post-inflationary scenario. As referred in Sec.~\ref{sec:axionDMcosmo}, a way to avoid the cosmological DW problem -- which affects the dynamics of post-inflationary axion DM production -- would be to have a DW number $N_{\text{DW}} \equiv N = 1$, which is related to the residual $\mathcal{Z}_N$ symmetry left unbroken after PQ-symmetry breaking.
This feature occurs for the models of Tables~\ref{tab:generalDirac} and~\ref{tab:postcharges} containing an iso-singlet VLQ [see Eq.~\eqref{eq:Nmodel}]. However, for the cases with an iso-doublet and iso-triplet exotic quark, we have $N_{\text{DW}}=2$ and $N_{\text{DW}}=3$, respectively, unavoidably leading to DWs in the early Universe~\cite{Lazarides:2018aev}.
Note that for $N_{\text{DW}}=1$, although there are no DWs, string networks can still form, with current numerical simulations predicting the axion decay constant $f_a$ to lie around ($5 \times 10^9 - 3 \times 10^{11}$) GeV, such that $\Omega_a h^2 = \Omega_{\text{CDM}} h^2$~\cite{Buschmann:2021sdq,Gorghetto:2020qws,Klaer:2017ond,Kawasaki:2014sqa}. In contrast, in the pre-inflationary scenario, since U(1)$_{\text{PQ}}$ breaking occurs before inflation all topological defects, strings, and DWs, will be washed away. 
It follows that, even in the iso-doublet and triplet models, one may introduce a bias term in the scalar potential which slightly breaks the residual $\mathcal{Z}_{N}$ symmetry. 
This would lift the vacuum degeneracy and render the DW-string network unstable~\cite{Sikivie:1982qv}. The subsequent decay of cosmic strings and DWs can lead to very interesting stochastic gravitational wave signatures~\cite{Roshan:2024qnv,Servant:2023mwt,Morais:2023ciz}.

The models of Tables~\ref{tab:generalDirac}  and~\ref{tab:postcharges} will provide a distinct prediction for the axion-to-photon coupling $g_{a \gamma \gamma}$ -- see Eqs.~\eqref{eq:EN},~\eqref{eq:axioncouplingsgagg} and~\eqref{eq:axioncouplings}.
\begin{figure}[!t]
    \centering
      \includegraphics[scale=0.175]{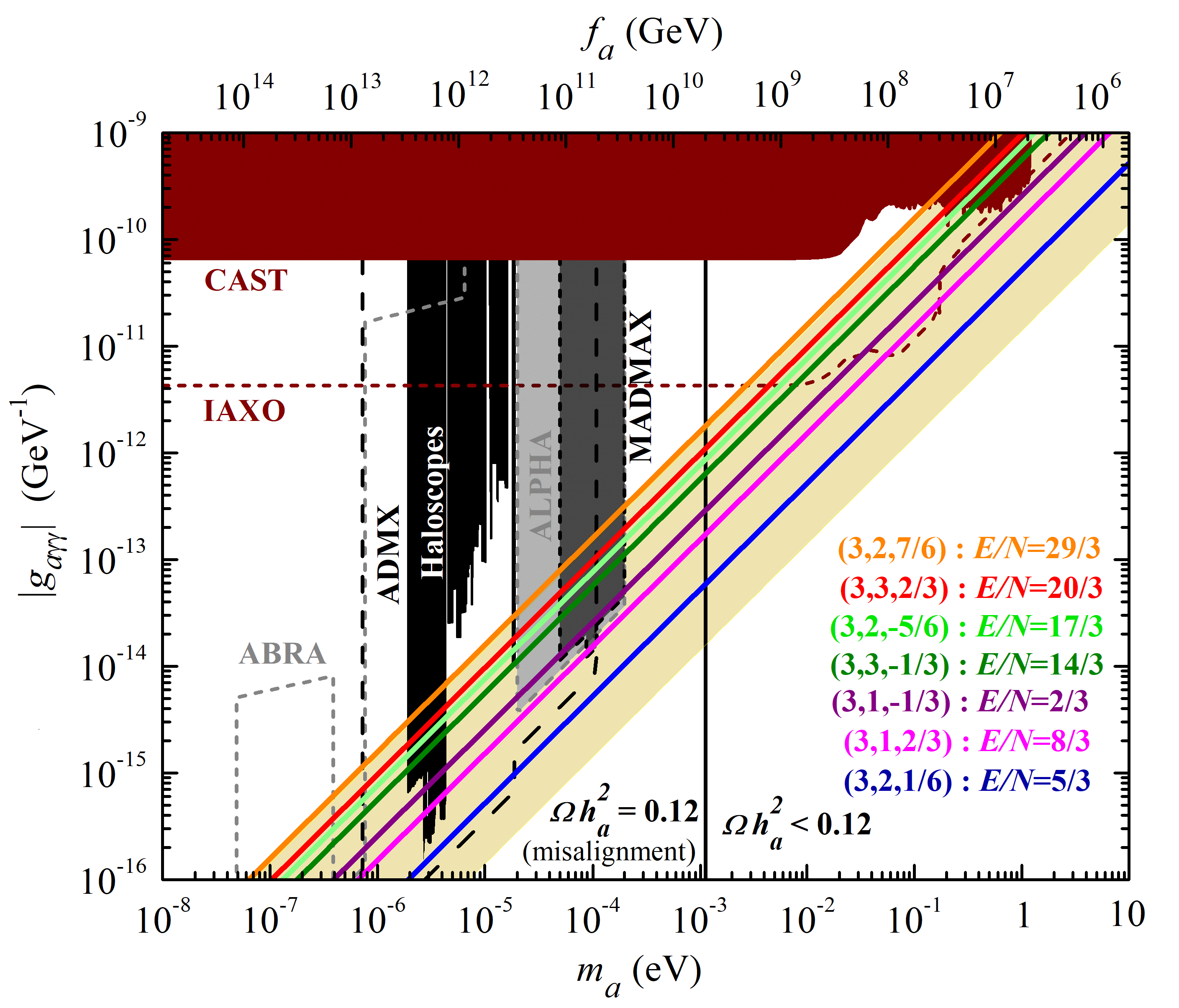}
    \caption{$|g_{a \gamma \gamma}|$ versus $m_a$ (bottom axis) and $f_a$ (top axis). The colored solid lines give the $E/N$ values of various Dirac color-mediated neutrino mass models -- see Tables~\ref{tab:generalDirac} and~\ref{tab:postcharges}, and text for details. The remaining elements in the figure are the same as in Fig.~\ref{fig:gaggMajorana}.}
    \label{fig:gaggDirac}
\end{figure}
In Fig.~\ref{fig:gaggDirac}, we display $|g_{a\gamma \gamma}|$ in terms of the axion mass~$m_a$ (bottom axis) and decay constant~$f_a$ (top axis), showing current bounds and future sensitivities from helioscope and haloscope experiments. The colored oblique lines indicate the $|g_{a\gamma \gamma}|$ predictions for the scenarios presented in Tables~\ref{tab:generalDirac}  and~\ref{tab:postcharges}. Note that, the iso-singlet up- (pink line) and down-type (purple line) VLQ predictions are identical to the popular DFSZ-I and DFSZ-II schemes~\cite{Zhitnitsky:1980tq,Dine:1981rt}, respectively -- see Sec.~\ref{sec:DFSZ}. The upper and lower limits of $|g_{a\gamma \gamma}|$ correspond to the exotic fermion representations $\Psi \sim (\mathbf{3},\mathbf{2},7/6)$ (orange line) and $\Psi \sim (\mathbf{3},\mathbf{2},1/6)$ (dark blue line), respectively. The helioscope CAST~\cite{CAST:2017uph} excludes $|g_{a \gamma \gamma}| \geq 0.66 \times 10^{-10} \ \text{GeV}^{-1}$ for masses $m_a \leq 20$ meV (bordeau-shaded region). The forthcoming IAXO~\cite{Shilon_2013} experiment is expected to probe the axion-to-photon coupling $g_{a \gamma \gamma}$ down to $(10^{-12}-10^{-11}) \ \text{GeV}^{-1}$ for $m_a \sim 0.1$ eV (bordeau-dashed contour), which scrutinizes all our models except for $\Psi \sim (\mathbf{3},\mathbf{2},1/6)$ (dark blue line). The ADMX haloscope has already excluded a considerable part of the parameter space of our scenarios. A notable exception occurs for $\Psi \sim (\mathbf{3},\mathbf{2},1/6)$ (dark blue line), for masses $m_a \sim 3 \ \mu$eV corresponding to $f_a \sim 10^{12}$ GeV. Future ADMX generations (dash-dotted black contour) are expected to probe all our models for masses $1 \ \mu \text{eV} \lesssim m_a \lesssim 20 \ \mu$eV or equivalently for scales $5 \times 10^{11} \ \text{GeV} \lesssim f_a \lesssim 10^{13}$ GeV. ALPHA~\cite{Lawson:2019brd,Wooten:2022vpj,ALPHA:2022rxj} (light-gray shaded region) and MADMAX~\cite{Beurthey:2020yuq} (dark-gray shaded region) are projected to cover all our frameworks except for $\Psi \sim (\mathbf{3},\mathbf{1},2/3)$ (pink line) and $\Psi \sim (\mathbf{3},\mathbf{2},1/6)$ (dark blue line). These experiments provide a sensitive probe of our models, complementary to other new physics searches at colliders or rare processes in the flavor sector.

%%%%%%%%%%%%%%%%%%%%%%%%%%%%%%%%%%%%%%%%%%%%%%%%%%%%%%%%%%%%%%%%%%%%
\subsection{Flavor-violating axion couplings}
\label{sec:axionflavorviolatingDirac}
%%%%%%%%%%%%%%%%%%%%%%%%%%%%%%%%%%%%%%%%%%%%%%%%%%%%%%%%%%%%%%%%%%%%

As in the conventional KSVZ scenarios~\cite{Kim:1979if,Shifman:1979if} -- see Sec.~\ref{sec:KSVZ} -- in our framework the SM quarks are not charged under the PQ-symmetry, resulting in no direct model-dependent couplings with the axion. Nonetheless, all color-mediated Dirac neutrino mass models in Tables~\ref{tab:generalDirac} and~\ref{tab:postcharges} exhibit mixing between the heavy VLQ $\Psi$ and the ordinary SM quarks. This will induce flavor-violating axion-quark couplings. The heavy-light quark mixing properties of the various models are studied in detail in Sec.~\ref{sec:quarksector}. They are described in terms of the $3\times 2$ $\mathbf{\Theta}_{X}^{q}$ mixing matrices of Eqs.~\eqref{eq:HLmixingrefdown},~\eqref{eq:HLmixing2SM},~\eqref{eq:HLmixing3down} and~\eqref{eq:HLmixing3up}, with $X=L,R$ and $q=d,u$ depending on the specific model. Without loss of generality, we can take $\mathcal{Q}_\text{PQ} = 0$ for all models, except for $\Psi \sim (\mathbf{3}, \mathbf{2}, 1/6)$ where we take $\mathcal{Q}_\text{PQ} = 1/2$ (see Table~\ref{tab:postcharges}). These have the least number of mixing terms and hence simpler quark mass matrices. The results, summarized in Table~\ref{tab:axionflavorviolating}, can be easily generalized to the other cases.
\begin{table}[t!]
\renewcommand*{\arraystretch}{1.5}
	\centering
	\begin{tabular}{| K{0.5cm} | K{1cm} | K{1cm} | K{3cm} | K{7.5cm} |}
		\hline 
$\mathbf{n}_\Psi$ & $y_\Psi$ & $\mathcal{Q}_{\text{PQ}}$ & Mixing terms & $\mathbf{\Theta}_{X}^{q}$ mixing parameter \\
		\hline \hline
\multirow{2}{*}{$\mathbf{1}$} & $-1/3$ &\multirow{2}{*}{$0$}& $ \mathbf{M}_{\Psi d} \overline{\Psi_L} d_R$ & $\boxed{\Theta_{R}^{d} \sim M_{\Psi d}/M_\psi} \ , \ \Theta_{L}^{d} \sim (v/M_\Psi) Y_d \Theta_{R}^{d}$ \\
& $2/3$ & & $ \mathbf{M}_{\Psi u} \overline{\Psi_L} u_R$ & $\boxed{\Theta_{R}^{u} \sim M_{\Psi u}/M_\psi} \ , \ \Theta_{L}^{u} \sim (v/M_\Psi) Y_u \Theta_{R}^{u}$ \\
        \hline \hline
        \multirow{3}{*}{$\mathbf{2}$} 
&$1/6$ &$1/2$& $\mathbf{M}_{q \Psi} \overline{q_L} \Psi_R$& $\boxed{\Theta_{L}^{d,u} \sim M_{q \Psi}/M_\psi} \ , \ \Theta_{R}^{d,u} \sim (v/M_\Psi) Y_{d,u} \Theta_{L}^{d,u}$ \\
\cline{2-5}
&$-5/6$ & \multirow{2}{*}{$0$} & $\mathbf{Y}_{\Psi d} \overline{\Psi_L} \tilde{\Phi} d_R$& $\Theta_{R}^{d} \sim (v/M_\psi) Y_{\Psi d} \ , \ \Theta_{L}^{d} \sim (v/M_\Psi) Y_d \Theta_{R}^{d}$ \\
&$7/6$ & & $ \mathbf{Y}_{\Psi u} \overline{\Psi_L} \Phi u_R$ & $\Theta_{R}^{u} \sim (v/M_\psi) Y_{\Psi u} \ , \  \Theta_{L}^{u} \sim (v/M_\Psi) Y_u \Theta_{R}^{u}$ \\
        \hline \hline
        \multirow{2}{*}{$\mathbf{3}$} 
&$-1/3$ & \multirow{2}{*}{$0$} & $ \mathbf{Y}_{q \Psi} \overline{q_L} \Phi \Psi_R$& $\Theta_{R}^{d} \sim (v/M_\psi) Y_{q \Psi} \ , \ \Theta_{L}^{d} \sim (v/M_\Psi) Y_d \Theta_{R}^{d}$ \\
&$2/3$ & &$\mathbf{Y}_{q \Psi} \overline{q_L} \tilde{\Phi} \Psi_R$& $\Theta_{R}^{u} \sim (v/M_\psi) Y_{q \Psi} \ , \ \Theta_{L}^{u} \sim (v/M_\Psi) Y_u \Theta_{R}^{u}$ \\
        \hline
	\end{tabular}
	\caption{ Heavy-light quark mixing terms and $\mathbf{\Theta}_{X}^{q}$ parameter for the various models in Tables~\ref{tab:generalDirac} and~\ref{tab:postcharges} characterized by the VLQ $\Psi$ representation. We highlight via a box the heavy-light quark mixing parameters that can be sizable (see text for details).}
\label{tab:axionflavorviolating} 
\end{table}

Among the seven potential models, three exhibit significant heavy-light quark mixing, resulting in substantial axion-flavor violating couplings. Specifically, VLQ $\Psi$ representations $(\mathbf{3}, \mathbf{1}, -1/3)$, $(\mathbf{3}, \mathbf{1}, 2/3)$ and $(\mathbf{3}, \mathbf{2}, 1/6)$, are the only ones with sizable heavy-light quark mixing, in the RH-down-quark, RH-up-quark and LH-up- and down-quark sectors, respectively. This is highlighted in the table with boxes around the relevant sizable $\mathbf{\Theta}_{X}^{q}$ parameters. For the iso-singlet model with $\Psi \sim (\mathbf{3}, \mathbf{1}, -1/3)$, we have a sizable $\mathbf{\Theta}_R^d$. In fact, since $\mathbf{M}_{\Psi d}$ is an arbitrary bare mass term we can take its scale $M_{\Psi d}$ close to the $M_\Psi \sim \mathcal{O}(f_{\text{PQ}})$ scale, implying a sizable $\Theta_R^d \sim M_{\Psi d}/M_\Psi$. In contrast, $\mathbf{\Theta}_L^d \sim \mathcal{O}(v/f_{\text{PQ}}) \mathbf{\Theta}_R^d \sim \mathcal{O}(10^{-10}) \mathbf{\Theta}_R^d$, is completely negligible. The same holds for $\Psi \sim (\mathbf{3}, \mathbf{1}, 2/3)$, with the substitution $d \rightarrow u$. On the other hand, the iso-doublet case with $\Psi \sim (\mathbf{3}, \mathbf{2}, 1/6)$ will feature sizable heavy-light mixing $\mathbf{\Theta}_L^{d,u}$ with $\mathbf{\Theta}_R^{d,u}$ being completely negligible. As for the rest of the models, the heavy-light quark mixing is not controlled by an arbitrary bare mass parameter, but instead by a Yukawa coupling with the Higgs doublet, resulting in further $v/f_{\text{PQ}}\sim \mathcal{O}(10^{-10})$ suppression making the mixing negligible. Sizable mixing translates into sizable flavor-violating quark-axion couplings, as discussed below.

The heavy-light quark mixing induces FCNCs in the quark sector, involving the SM $Z$ boson and Higgs field $h$, as well as flavor-changing couplings of the axion field $a$. The VLQ $\Psi$ are chirally charged under the PQ symmetry with their LH and RH PQ charge difference matching $\mathcal{Q}_L - \mathcal{Q}_R = \mathcal{Q}_\sigma = 1/2$, where $\mathcal{Q}_\sigma$ is the PQ charge of $\sigma$. Depending on the specific model, either the LH or RH chiral component of $\Psi$ will be charged under PQ (see the $\mathcal{Q}_\text{PQ}$ charge values in Tables~\ref{tab:generalDirac} and~\ref{tab:postcharges}). Generically, $\Psi_{j X}$ transforms under the PQ symmetry as $\Psi_{j X}\to \exp(i \mathcal{Q}_X a /v_\sigma) \Psi_{j X}$. By rotating the axion field, we can remove it from the Yukawa Lagrangian and it will appear in the kinetic term for $\Psi_{j X}$. Since $\Psi_{j X}$ mixes with the SM quarks through $\mathbf{\Theta}_{X}^{q}$, one obtains the flavor-violating axion interactions with quarks shown in Eq.~\eqref{eq:axionFermionLagrangian} presented in Sec.~\ref{sec:flavouredaxions}. In our models the flavor-violating axion couplings with quarks are given by:
\begin{align}
    \mathbf{C}^{V,q}_{\alpha \beta} & = \mathbf{C}^{A,q}_{\alpha \beta} = \mathbf{C}^{q}_{\alpha \beta} =\frac{1}{2} \mathcal{Q}_X (\tilde{\mathbf{\Theta}}_X^q)_{\alpha \beta} \; , \; \tilde{\mathbf{\Theta}}_{X}^{q} = {\mathbf{V}_{X}^{q}}^\dagger \mathbf{\Theta}_{X}^{q} {\mathbf{\Theta}_{X}^{q}}^\dagger \mathbf{V}_{X}^{q} \; ,
    \label{eq:FCNCa}
\end{align}
where $\mathbf{C}^{V,A}_{\alpha \beta}$ represent the vector and axial couplings, which are Hermitian matrices in flavor space.

In Refs.~\cite{MartinCamalich:2020dfe,Alonso-Alvarez:2023wig}, a detailed study of flavor-violating axion couplings to quarks resulted in an exhaustive list of constraints stemming from rare processes, Higgs physics, meson oscillations, among others have been discussed. These constraints were applied to the color-mediated Majorana neutrino mass framework~\cite{Batra:2023erw, Hati:2024ppg}. The goal in \cite{Hati:2024ppg} was to account for an enhanced $B^+\to K^+ + E_{\text{miss}}$ rate at Belle-II with a model featuring heavy-light mixing in the down-quark sector. The heavy-colored scalars mediating neutrino mass generation had masses around the TeV scale in order to explain the aforementioned flavor anomaly. Here, we consider more natural values for the masses of the heavy-colored scalars arising from the leptoquarks $\eta$ and $\chi$, and their mixing.  Since these scalars couple to $\sigma$ via a quartic interaction in the scalar potential of Eq.~\eqref{eq:VpotDirac}, their masses should be close to the PQ breaking scale $f_{\text{PQ}} \sim \mathcal{O}(10^{12})$ GeV, if one assumes $\mathcal{O}(1)$ scalar potential couplings (see discussion in Sec.~\ref{sec:Diracneutrinos}). 

Dirac color-mediated models, with $\Psi \sim (\mathbf{3}, \mathbf{1}, -1/3)$ and $\Psi \sim (\mathbf{3}, \mathbf{2}, 1/6)$, could in principle explain such quark flavor anomalies, thanks to the new colored scalars and sizable interactions in the down-quark sector controlled by $\mathbf{C}^{V,A}_{\alpha \beta}$ (see boxed expressions in Table~\ref{tab:axionflavorviolating}). We do not pursue such a detailed analysis here, and will instead focus on axion flavor-violating couplings. Moreover, since a detailed study of the flavor violating axion to down-quark coupling was performed in Ref.~\cite{Hati:2024ppg}, here we wish to discuss the up-quark sector. 
\begin{figure}[!t]
    \centering
    \includegraphics[scale=0.55]{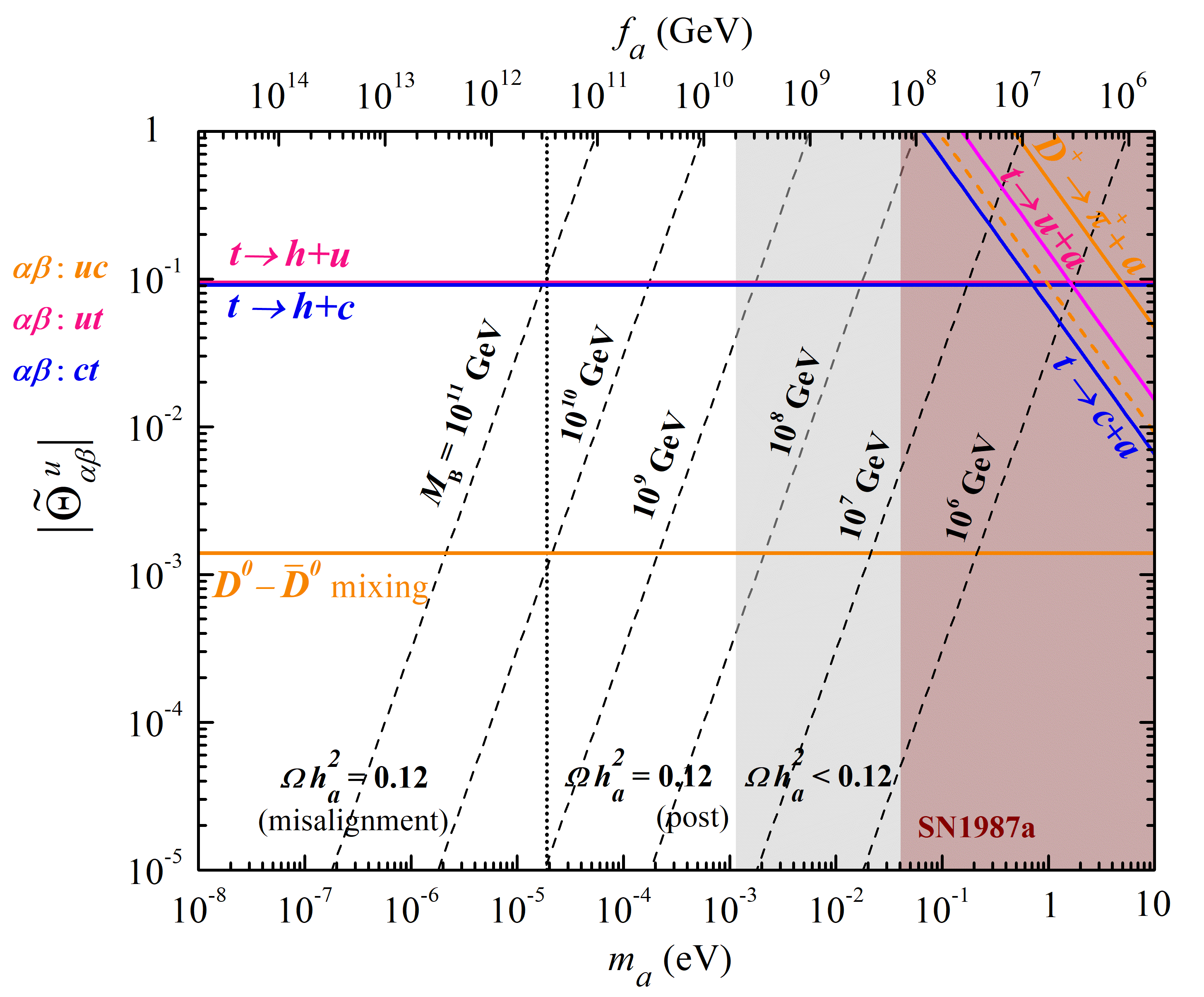}
    \caption{Flavor-violating axion to up-quark coupling, encoded in $|\tilde{\Theta}^{u}_{\alpha\beta}|$, versus $m_a$ (bottom axis) and $f_a$ (top axis). Constraints in orange (pink) [blue], only apply to the case where $\alpha \beta \equiv uc$ ($ut$) [$ct$]~\cite{MartinCamalich:2020dfe,Alonso-Alvarez:2023wig} -- see Table~\ref{tab:QuarkConstraints} and Sec.~\ref{sec:flavouredaxions}. Our benchmarks are indicated by black-dashed lines with $M_{\text{B}} \in [10^{6} - 10^{11}]$ GeV. SN1987a -- see Eq.~\eqref{eq:SN1987aaxionbound} -- excludes the bordeau shaded region~\cite{Carenza:2019pxu}. In the gray shaded region axion DM is underabundant, while in the white region $\Omega h^2_a = 0.12$, for the pre-inflationary case. Considering the post-inflationary scenario for $N_{\text{DW}} =1$ the axion mass is restricted to the white region to the right of the vertical dotted black line where $\Omega h^2_a = 0.12$~\cite{Buschmann:2021sdq,Gorghetto:2020qws,Klaer:2017ond,Kawasaki:2014sqa}.}
    \label{fig:axionfvup}
\end{figure}
Among all the Dirac color-mediated neutrino mass models only two can have sizable axion-to-up-quark flavor violating couplings. Namely, the models with $\Psi \sim (\mathbf{3},\mathbf{1},2/3)$ and $\Psi \sim (\mathbf{3},\mathbf{1},1/6)$, with the mixing encoded in $\mathbf{\Theta}^{u}_R$ and $\mathbf{\Theta}^{u}_L$, respectively (see Table~\ref{tab:axionflavorviolating}). Following Eqs.~\eqref{eq:HLmixingrefdown}, replacing $d$ by $u$,  and ~\eqref{eq:HLmixing2SM}, we express the mixing parameter of Eq.~\eqref{eq:FCNCa} as, 
\begin{equation}
\mathbf{\tilde{\Theta}}^u_{\alpha \beta} =  \sum_{k=1}^2 \frac {(\mathbf{M}_{\text{B}})_{k\alpha} (\mathbf{M}_{\text{B}})_{k\beta} } {\mathbf{M}_{\Psi k}^2} = \sum_{k=1}^2 \frac {(\mathbf{M}_{\text{B}})_{k\alpha} (\mathbf{M}_{\text{B}})_{k\beta} } {(N f_a \mathbf{Y}_{\Psi k})^2} \,, 
\end{equation}
where the bare mass matrices $\mathbf{M}_{\Psi u}$ and $\mathbf{M}_{q \Psi}$ are given by $\mathbf{M}_{\text{B}}$. Note that, without loss of generality, we assume that $\mathbf{M}_{\Psi}$ is in the diagonal basis and, for simplicity, neglect CKM mixing. In Fig.~\ref{fig:axionfvup}, we present the axion-to-up-quark flavor-violating couplings via $|\tilde{\Theta}^{u}_{\alpha\beta}|$ [see Eq.~\eqref{eq:FCNCa}], in terms of the axion mass $m_a$ [or equivalently $f_a$ -- see Eq.~\eqref{eq:axionfa}]. We took six benchmarks for the bare mass parameter $M_{\text{B}}$, setting the scale between $10^{6} - 10^{11}$ GeV, taking $\mathbf{M}_{\Psi} = N f_a$. These cases are indicated by black-dashed lines. Moreover, from the SN1987a bound of Eq.~\eqref{eq:SN1987aaxionbound} we can obtain the lower bound on $f_a > 1.4 \times 10^8$ GeV~\cite{Carenza:2019pxu}, see the orange dashed region in the figure. As seen in Fig.~\ref{fig:gaggDirac} the QCD axion can account for the observed DM relic abundance, in the post-inflationary scenario for $N_{\text{DW}}=1$, the axion decay constant is restricted to the range $f_a \in [5 \times 10^9, 3 \times 10^{11}]$ GeV~\cite{Adams:2022pbo, Buschmann:2021sdq, Gorghetto:2020qws, Klaer:2017ond, Kawasaki:2014sqa} (middle band). 

The constraints in orange (pink) [blue], thoroughly studied in Refs.~\cite{MartinCamalich:2020dfe,Alonso-Alvarez:2023wig} are summarized in Table~\ref{tab:QuarkConstraints} of Sec.~\ref{sec:flavouredaxions}, only apply to the case where $\alpha \beta \equiv uc$ ($ut$) [$ct$]. For the up-type mixing, the constraints imposed by rare quark flavor processes $D^{+}\to \pi^{+} \, + a$ are shown by solid and dashed orange lines to indicate current and future sensitivity. For top decays, $t\to u \, + a$  and $t\to c \, + a$ are presented using the pink and blue lines, respectively, stemming from the $K^+ \rightarrow \pi^+ + a$ (loop) bounds. These constraints lie within the bordeau shaded region already excluded by SN1987a. However, since heavy-light quark mixing also leads to flavor changing neutral currents ($Z$ boson) and couplings (Higgs boson $h$), $D^0-\overline{D}^0$ meson mixing (horizontal orange line) and top decays into the Higgs $t\to h \, + u$ (horizontal pink line) and $t\to h \, + c$ (horizontal blue line), can constrain $|\tilde{\Theta}^{u}_{\alpha\beta}|$ independently of the axion scale~$f_a$. In summary, taking into the account the DM requirement and all the above complementary constraints, the bare mass parameter of our models is restricted to lie within the range $M_B \in [10^8 - 10^{10}]$ GeV, where for $\alpha\beta = uc$ the mixing is bounded as $|\tilde{\Theta}^{u}_{\alpha\beta}|\lsim 10^{-3}$, while for $\alpha\beta = ut,ct$ we have $|\tilde{\Theta}^{u}_{ut,ct}|\lsim 10^{-1}$.

%%%%%%%%%%%%%%%%%%%%%%%%%%%%%%%%%%%%%%%%%%%%%%%%%%%%%%%%%%%%%%%%%%%%
\section{Key ideas and outlook} 
\label{sec:conclcolormediated}
%%%%%%%%%%%%%%%%%%%%%%%%%%%%%%%%%%%%%%%%%%%%%%%%%%%%%%%%%%%%%%%%%%%%

We have presented two complementary frameworks that establish a unified origin for small neutrino masses and the resolution of the strong CP problem within the KSVZ-type axion paradigm. In the Majorana neutrino scenario, the neutrino mass arises radiatively at the two-loop level via the exchange of colored fermions and scalars, which also serve as mediators of PQ symmetry breaking. This setup naturally predicts LNV and can be searched at experiments looking for $0\nu\beta\beta$.

In contrast, the Dirac neutrino scenario realizes neutrino masses at the one-loop level through interactions involving VLQs and scalar leptoquarks. Here, the PQ symmetry ensures the Dirac nature of light neutrinos, simultaneously addressing the strong CP problem and suppressing dangerous baryon- and lepton-number violating operators. This framework allows for mixing between the exotic quarks and SM fermions, which not only supports a predictive post-inflationary axion cosmology but also gives rise to flavor-violating axion couplings constrained by rare meson decays, top-quark processes, and astrophysical bounds.

Both constructions yield models leading to distinct axion-photon couplings offering a pathway to test them at haloscope and helioscope experiments such as ADMX, MADMAX, and IAXO. Furthermore, the colored field content inherent in both frameworks motivates different cosmological scenarios for axion DM: the Majorana case is naturally embedded in a pre-inflationary PQ-breaking regime, whereas the Dirac case is compatible with post-inflationary dynamics. In the latter scenario, topological defects can play an important role, where for example the decay of axion DWs and string network could lead to interesting signals at gravitational wave observatories~\cite{Roshan:2024qnv}.

In summary, the proposed unified models offer structurally minimal yet phenomenologically rich avenues to address three major open questions in fundamental physics: the origin of neutrino masses, the strong CP problem, and the nature of DM. The color-mediated neutrino mass idea can also be tested beyond the phenomenology of the axion couplings. For example, by implementing it within extended gauge groups, or by incorporating flavor symmetries, one could get intriguing insights into the flavor puzzle and/or flavor anomalies~\cite{Hati:2024ppg}.

%------------
% CHAPTER 06   
%------------

%%%%%%%%%%%%%%%%%%%%%%%%%%%%%%%%%%%%%%%%%%%%%%%%%%%%%%%%%%%%%%%%%%%%%%%%%%%%%
\chapter{Flavored Peccei-Quinn symmetries in the DFSZ model} 
\label{chpt:flavoraxion}
%%%%%%%%%%%%%%%%%%%%%%%%%%%%%%%%%%%%%%%%%%%%%%%%%%%%%%%%%%%%%%%%%%%%%%%%%%%%%

The flavor puzzle has motivated a plethora of SM extensions featuring additional fields and new symmetries~\cite{Ishimori:2010au,Morisi:2012fg,King:2017guk,Petcov:2017ggy,Feruglio:2019ybq,Xing:2020ijf}. This chapter focuses on global U(1) Abelian symmetries that give rise to texture zeros in the Yukawa~\cite{Grimus:2004hf,Dighe:2009xj,Adhikary:2009kz,Dev:2011jc,GonzalezFelipe:2014zjk,Samanta:2015oqa,Kobayashi:2018zpq,Rahat:2018sgs,Nath:2018xih,Barreiros:2018ndn,Barreiros:2018bju,Correia:2019vbn,Camara:2020efq,Barreiros:2020gxu,Barreiros:2022aqu,Rocha:2024twm} and mass matrices~\cite{Ludl:2014axa,Ludl:2015lta,GonzalezFelipe:2014zjk,Cebola:2015dwa}. When implemented within the SM, these symmetries generate fermion mass and mixing structures that are ruled out by current experimental data~\cite{Correia:2019vbn,Camara:2020efq,Rocha:2024twm}. The simplest extension of the SM that successfully accommodates U(1) flavor symmetries is the 2HDM~\cite{Branco:2011iw}. Namely, they have been employed in 2HDM frameworks to address the flavor puzzles in the quark~\cite{Ferreira:2010ir} and lepton~\cite{GonzalezFelipe:2014zjk,Correia:2019vbn,Camara:2020efq} sectors separately. A combined analysis of both sectors was performed in our work of Ref.~\cite{Rocha:2024twm}. It was shown that Abelian flavor symmetries provide a natural mechanism to suppress and control FCNCs, which are tightly constrained by quark flavor observables and also induce contributions to cLFV processes, currently being targeted in ongoing experimental searches.

Axion frameworks provide a gateway to address multiple BSM phenomena in a complementary way -- see Chapter~\ref{chpt:axions}. Notably, as shown in Sec.~\ref{sec:axionDMcosmo}, axion relics can be generated non-thermally in the early Universe via the so-called misalignment mechanism, thus accounting for the observed DM relic density~\cite{Preskill:1982cy,Abbott:1982af,Dine:1982ah}. Furthermore, the DFSZ framework, presented in Sec.~\ref{sec:DFSZ}, has been studied in connection with the flavor puzzle in the quark sector, with the PQ symmetry promoted to a flavor symmetry in the same spirit as Abelian flavor symmetries~\cite{Bardeen:1977bd,Davidson:1981zd,Wilczek:1982rv,Celis:2014zaa,Celis:2014jua,Celis:2014iua,Cox:2023squ,delaVega:2021ugs}, leading to flavor-violating axion couplings to fermions as discussed in Sec.~\ref{sec:flavouredaxions}. In Ref.~\cite{Cox:2023squ}, the authors provide a classification of flavored DFSZ models that are free of DWs, i.e. featuring a DW number $N_{\text{DW}} = 1$, focusing exclusively on the quark sector. It would be interesting to systematically study axion flavor models that also incorporate neutrino mass generation -- see Chapter~\ref{chpt:neutrinodarksectors}. Namely, the axion–neutrino connection through the type-I seesaw mechanism (see Sec.~\ref{sec:TypeI}), where the VEV of the scalar singlet provides the Majorana masses for RH neutrinos, has been explored in the context of the DFSZ axion~\cite{Volkas:1988cm,Clarke:2015bea,Sopov:2022bog,RVolkas:2023jiv,Matlis:2023eli}, giving rise to the so-called $\nu$DFSZ model. Similar constructions have also been considered within the KSVZ scenario~\cite{Salvio:2015cja,Ballesteros:2016euj,Ballesteros:2016xej} (see Sec.~\ref{sec:KSVZ}). As studied in detail in the previous Chapter~\ref{chpt:axionneutrino}, within the KSVZ framework, we proposed that exotic colored mediators generate radiatively both Majorana~\cite{Batra:2023erw} (Sec.~\ref{sec:colormediatedMajorana}) and Dirac~\cite{Batra:2025gzy} (Sec.~\ref{sec:colormediatedDirac}) neutrino masses.

In this chapter, we consider flavored PQ symmetries in the minimal $\nu$DFSZ model which features two-Higgs doublets and a complex singlet in the scalar sector and two RH neutrinos generating type-I seesaw neutrino masses. The PQ symmetry necessary to implement the axion solution to the strong CP problem will act as our flavor symmetry. Namely, we are interested in scenarios where the PQ symmetry maximally restricts the quark and lepton mass matrices, such that these contain the least possible number of independent parameters required to reproduce the fermion masses, mixing and CP violation data, while satisfying all relevant phenomenological constraints. In contrast to Ref.~\cite{Cox:2023squ}, here, for the first time, are systematically classified the minimal complete quark and lepton flavor scenarios considering cases that may feature $N_{\text{DW}} \neq 1$ and incorporate neutrino masses via the seesaw mechanism. The lepton sector plays a crucial role in axion coupling phenomenology. The discussion presented in this chapter follows closely our work of Ref.~\cite{Rocha:2025ade} extending the analysis performed in our work of Ref.~\cite{Rocha:2024twm}.

%%%%%%%%%%%%%%%%%%%%%%%%%%%%%%%%%%%%%%%%%%%%%%%%%%%%%%%%%%%%%%%%%%%%%%%%%%%%%
\section{The flavored $\nu$DFSZ model}
%%%%%%%%%%%%%%%%%%%%%%%%%%%%%%%%%%%%%%%%%%%%%%%%%%%%%%%%%%%%%%%%%%%%%%%%%%%%%

In our minimal $\nu$DFSZ, the scalar field content consists of two-Higgs doublets, $\Phi_{1,2} \sim (1,2,1)$, and a complex singlet scalar, $\sigma \sim (1,1,0)$, as in the DFSZ model (see Sec.~\ref{sec:DFSZ}). However, we also extended the SM fermion field content with two RH neutrinos, $\nu_{R_{1,2}}\sim(1,1,0)$, and impose a general global U(1)$_{\text{PQ}}$, under which the various fields transform as:
\begin{align}
&\Phi_k \rightarrow \exp\left({i \zeta \chi_{k} }\right) \Phi_k \;,\;q_{L_\alpha} \rightarrow  \exp\left({i \zeta \chi_{q_\alpha}^L }\right) q_{L \alpha} \;,\;d_{R_\alpha} \rightarrow \exp\left({i \zeta \chi_{d_\alpha}^R }\right) d_{R \alpha} 
\;,\;u_{R \alpha} \rightarrow \exp\left({i \zeta \chi_{u_\alpha}^R }\right) u_{R \alpha}, \nonumber \\
&\sigma \rightarrow \exp\left({i \zeta \chi_\sigma }\right) \sigma \;,\;\ell_{L \alpha} \rightarrow  \exp\left({i \zeta \chi_{\ell_\alpha}^L }\right) \ell_{L \alpha} \;,\;e_{R \alpha} \rightarrow \exp\left({i \zeta \chi_{e_\alpha}^R }\right) e_{R \alpha} \;,\;\nu_{R j} \rightarrow  \exp\left({i \zeta \chi_{\nu_j}^R }\right) \nu_{R j}, 
\label{eq:PQSym}
\end{align}   
where $k = 1,2 $ labels the Higgs doublets, $\alpha = 1,2,3$ the SM fermion generations and $j =1,2$ the number of RH neutrinos. The PQ charges are denoted by $\chi$'s and $\zeta$ is a continuous parameter.

Given the field transformation properties under ${\rm G}_{\rm SM}$ and the PQ symmetries of Eq.~\eqref{eq:PQSym}, the scalar potential is [see Eq.~\eqref{eq:VpotentialDFSZOG}]:
\begin{align}
    V(\Phi_1,\Phi_2,\sigma) &= m_{11}^2 \Phi_1^\dagger \Phi_1 + m_{22}^2 \Phi_2^\dagger \Phi_2 + \frac{\lambda_1}{2} \left(\Phi_1^\dagger \Phi_1\right)^2 + \frac{\lambda_2}{2} \left(\Phi_2^\dagger \Phi_2\right)^2 \nonumber
    \\
    & + \lambda_3 \left(\Phi_1^\dagger \Phi_1\right)\left(\Phi_2^\dagger \Phi_2\right) + \lambda_4 \left(\Phi_1^\dagger \Phi_2\right)\left(\Phi_2^\dagger \Phi_1\right) \nonumber
    \\
    & + m_{\sigma}^2 |\sigma|^2 + \frac{\lambda_\sigma}{2} |\sigma|^4 + \lambda_{1 \sigma} \Phi_1^\dagger \Phi_1 |\sigma|^2
    + \lambda_{2 \sigma} \Phi_2^\dagger \Phi_2 |\sigma|^2
    +
    \left[V_{\text{PQ}}(\Phi_1,\Phi_2,\sigma) +\text{H.c.} \right] 
    \; ,
    \label{eq:VpotentialDFSZ}
\end{align}
where, depending on the scalar field PQ charges, two classes of models can be identified for which $V_{\text{PQ}}$ is given by:
\begin{align}
 {\rm Cubic \,Model:}\,&\;  V_{\text{PQ}}(\Phi_1,\Phi_2,\sigma) =
    \kappa \Phi_2^\dagger \Phi_1 \sigma \; \;,\;\; 
    -\chi_{2}+\chi_{1} + \chi_{\sigma} = 0 \  \text{(mod $2\pi$)}
    \; , \nonumber \\
    {\rm Quartic \,Model:}\,& \;
   V_{\text{PQ}}(\Phi_1,\Phi_2,\sigma) = \lambda \Phi_2^\dagger \Phi_1 \sigma^2
    \; \;,\;\;
    \; \; 
   -\chi_{2}+\chi_{1} + 2\chi_{\sigma} = 0 \  \text{(mod $2\pi$)} \; .
\label{eq:VPQoperator}
\end{align}
Notice that $V_{\text{PQ}}$ is a necessary term in the scalar sector since, if neither of the PQ charge relations shown above are satisfied, the full scalar potential exhibits a U(1)$^3$ symmetry. This results in an additional massless neutral GB, besides the one corresponding to the longitudinal component of the $Z$-boson, and the axion, which gets a mass via non-perturbative QCD effects, making it a pseudo-GB. 

The scalar doublets and singlet are parameterized as in Eqs.~\eqref{eq:parametrisationPQWW} and~\eqref{eq:sigmaPQ}, respectively. With the given parameterization, the physical axion is written as,
\begin{equation}
    a = \frac{1}{v_a} \sum_{i=1,2,\sigma} \chi_i a_i v_i \; , \quad 
    v_a^2 = \sum_{i=1,2,\sigma} \chi_i^2 v_i^2 \; .
    \label{eq:axionandvev}
\end{equation}
Note that, assuming a strong hierarchy among the singlet and doublet VEVs $v_\sigma \gg v_1,v_2$, the physical axion and $v_a$ essentially stem from $\sigma$ and $v_\sigma$, respectively. Imposing that the PQ current is orthogonal to the hypercharge current enforces $ v_1^2\chi_{1} + v_2^2\chi_{2}=0$ and guarantees absence of kinetic mixing between the axion and the $Z$-boson. To satisfy this orthogonality condition, along with PQ invariance of $V$, we set:
\begin{align}
    \chi_{1} = -\sin^2 \beta \equiv - s_\beta^2 \; , \quad 
    \chi_{2} = \cos^2 \beta \equiv c_\beta^2 \; , \quad 
     \chi_{\sigma} = 
    \begin{cases}
        \;\; 1 \quad &\text{(Cubic Model)} \\
        \;\; \dfrac{1}{2} \quad &\text{(Quartic Model)} 
    \end{cases} \; , \; \tan \beta = v_2/v_1\,,
    \label{eq:PQsymmetry}
\end{align}
without loss of generality.

The Yukawa and mass terms invariant under ${\rm G}_{\rm SM}\times {\rm U}(1)_{\rm PQ}$ are:
\begin{align}
    -\mathcal{L}_\mathrm{Yuk.} & = \overline{q_L}\left(\Y^d_1 \Phi_1 + \Y^d_2 \Phi_2\right)d_R 
    + \overline{q_L} \left(\Y^u_1 \Tilde{\Phi}_1 + \Y^u_2 \Tilde{\Phi}_2\right) u_R + \overline{\ell_L}\left(\Y^e_1 \Phi_1 + \Y^e_2 \Phi_2\right)e_R \, \nonumber
    \\
    &+ \overline{\ell_L}\left(\Y^{D \ast}_{1} \Tilde{\Phi}_1 + \Y^{D \ast}_{2} \Tilde{\Phi}_2\right)\nu_R 
    + \frac{1}{2} \overline{\nu_R} \left(
    \mathbf{M}^0_R +  \Y^R_1 \sigma + \Y^R_2 \sigma^\ast
    \right) \nu_R^c
    + \mathrm{H.c.} \; ,
    \label{eq:Lyuk2hdm}
\end{align}
where the Yukawa matrices $\mathbf{Y}^{f (D)}_{1,2}$ ($f = d, u, e$) are $3 \times 3$ (or $3 \times 2$) general complex matrices, while $\mathbf{Y}^R_{1,2}$ and the bare Majorana mass matrix $\mathbf{M}_R^0$ are $2 \times 2$ symmetric matrices. Invariance under the PQ symmetry requires,
\begin{equation}
\begin{aligned}    
    (\mathbf{Y}_{1,2}^{f})_{\alpha \beta} &= \exp\left[{i (\Theta_{1,2}^{f})_{\alpha \beta}} \right](\mathbf{Y}_{1,2}^{f})_{\alpha \beta} \; , \quad &(\mathbf{Y}_{1,2}^{D})_{\alpha j} &= \exp\left[{i (\Theta_{1,2}^{D})_{\alpha j}} \right](\mathbf{Y}_{1,2}^{D})_{\alpha j} \; ,
    \\
    (\mathbf{Y}_{1,2}^{R})_{i j} &= \exp\left[{i (\Theta_{1,2}^{R})_{i j}} \right](\mathbf{Y}_{1,2}^{R})_{i j} \; , \quad &(\mathbf{M}_{R}^0)_{ij} &= \exp\left[{i (\Theta_0^R)_{ij}}\right] (\mathbf{M}^{0}_R)_{ij} \; , 
\label{eq:Invariance}
\end{aligned}    
\end{equation}
where $\alpha,\beta$ run over the three SM fermion generations, $i,j=1,2$, and
\begin{gather}
(\Theta^{d}_{1,2})_{\alpha\beta}  =\chi_{d_\beta}^R-\chi_{q_\alpha}^L+\chi_{1,2} \; , \quad 
(\Theta^u_{1,2})_{\alpha\beta}=\chi_{u_\beta}^R-\chi_{q_\alpha}^L-\chi_{ 1,2} \; , \nonumber
\\
(\Theta^e_{1,2})_{\alpha\beta}=\chi_{e_\beta}^R-\chi_{q_\alpha}^L+\chi_{ 1,2} \; , \quad
(\Theta^D_{1,2})_{\alpha j} =\chi_{\nu_j}^R-\chi_{\ell_\alpha}^L - \chi_{ 1,2} \; , \nonumber
\\
(\Theta^R_0)_{ij}=\chi_{\nu_j}^R +\chi_{\nu_i}^R  \; , \quad 
(\Theta^R_1)_{ij}=\chi_{\nu_j}^R +\chi_{\nu_i}^R - \chi_{\sigma} \; , \quad   
(\Theta^R_2)_{ij}=\chi_{\nu_j}^R +\chi_{\nu_i}^R + \chi_{\sigma} \;  .
\label{eq:canonicalcharges}
\end{gather}
Alternatively, Eqs.~\eqref{eq:Invariance} can be reformulated as follows:
\begin{equation}
\begin{aligned}
(\Theta^{f}_{1,2})_{\alpha \beta} & \neq 0 \  \text{(mod $2\pi$)}  \Leftrightarrow  (\mathbf{Y}^{f}_{1,2})_{\alpha \beta} =  0 \; , \quad  &(\Theta^{D}_{1,2})_{\alpha j} &\neq 0 \  \text{(mod $2\pi$)} \Leftrightarrow (\mathbf{Y}^{D}_{1,2})_{\alpha j} =  0 \,,
\\
(\Theta^{R}_{1,2})_{i j} & \neq 0 \  \text{(mod $2\pi$)}  \Leftrightarrow  (\mathbf{Y}^{R}_{1,2})_{i j} =  0 \; , \quad &(\Theta^R_{0})_{ij} &\neq 0 \  \text{(mod $2\pi$)} \Leftrightarrow (\mathbf{M}_R^0)_{ij} =  0 \;.
\label{eq:YukawaZeros}
\end{aligned}
\end{equation}
Thus, the PQ symmetry may forbid certain Yukawa and bare mass terms in Eq.~\eqref{eq:Lyuk2hdm}, leading to texture-zero patterns for $\mathbf{Y}^{f,D,R}_{1,2}$ and $\mathbf{M}^{0}_R$.

SSB is triggered when the neutral components of the Higgs doublets and the scalar singlet acquire nonzero VEVs, resulting in the fermion-mass Lagrangian:
\begin{align}
    -\mathcal{L}_{\text {mass }}=\overline{d_L} \mathbf{M}_{d} d_R+\overline{u_L} \mathbf{M}_{u} u_R+\overline{e_L} \mathbf{M}_{e} e_R+\overline{\nu_L} \mathbf{M}_D^* \nu_R+\frac{1}{2} \overline{\nu_R} \mathbf{M}_R \nu_R^c+\text {H.c.}  \; ,
        \label{eq:Mass2hdm}
\end{align}
with
\begin{align}
    \mathbf{M}_{f}=\frac{v_1}{\sqrt{2}} \mathbf{Y}^{f}_1 + \frac{v_2 }{\sqrt{2}} \mathbf{Y}^{f}_2 
    \;,\; 
    \quad \mathbf{M}_D=\frac{v_1}{\sqrt{2}} \mathbf{Y}^{D}_1 + \frac{v_2 }{\sqrt{2}} \mathbf{Y}^{D}_2
    \; , \;
    \quad \mathbf{M}_R=\mathbf{M}_R^0+\frac{v_\sigma}{\sqrt{2}}\left(\mathbf{Y}^R_1+\mathbf{Y}^R_2 \right) \; .
\label{Eq:MassMatrices}
\end{align}
The quark mass matrices can be brought to the quark physical basis through the unitary transformations of Eqs.~\eqref{eq:massdiag} and~\eqref{eq:hermitianmassdiag}, yielding the CKM quark mixing matrix $\mathbf{V}$ defined in Eq.~\eqref{eq:VCKMdef} and parameterized as in Eq.~\eqref{eq:VCKMparam}-- see discussion in Sec.~\ref{sec:fermionmassmix}. Furthermore, in the limit $\mathbf{M}_D \ll \mathbf{M}_R$, the effective (light) neutrino mass matrix is given by the well-known type-I seesaw formula of Eq.~\eqref{eq:TypeIMeff} -- see discussion in Sec.~\ref{sec:TypeI}. Charged-lepton fields can be rotated to their physical basis via Eqs.~\eqref{eq:massdiag} and~\eqref{eq:hermitianmassdiag}. The active-neutrino fields are brought to the physical basis via the rotations defined in Eq.~\eqref{eq:leptonmixing}. The diagonalization procedure results in the lepton mixing matrix defined in Eq.~\eqref{eq:CCleptoU} parameterized as in Eq.~\eqref{eq:PMNSparam} -- see in discussion Sec.~\ref{sec:neutrino}. Note that, since there are only two RH neutrinos $\nu_R$ are included in our model, the lightest neutrino remains massless, i.e. $m_1=0$ ($m_3=0$) for NO (IO).

Since the PQ symmetry is anomalous with respect to QCD, the transformation of Eq.~\eqref{eq:Lyuk2hdm} into the mass basis~\eqref{eq:Mass2hdm}, is reflected in the QCD topological term via the color anomaly factor -- see Eq.~\eqref{eq:EN} and Sec.~\ref{sec:properties} -- which in terms of the PQ charges of Eq.~\eqref{eq:PQSym} is given by:
\begin{align} 
    N = \sum_{\alpha=1}^3 (2\chi_{q_\alpha}^L - \chi_{u_\alpha}^R - \chi_{d_\alpha}^R) \; ,
    \label{eq:Ncolor}
\end{align}
To implement the axion solution to the strong CP problem it is required that $N \neq 0$. Furthermore, from Eq.~\eqref{eq:axionandvev}, we identify the axion decay constant as [see Eqs.~\eqref{eq:axionfa} and~\eqref{eq:PQWWscale}],
\begin{equation}
    f_a = \frac{f_{\text{PQ}}}{N} = \frac{v_{a}}{N} \; .
\end{equation}
The relation between $m_a$ and $f_a$ at NLO is given in Eq.~\eqref{eq:axionmass} being a model-independent prediction of the QCD axion.

%%%%%%%%%%%%%%%%%%%%%%%%%%%%%%%%%%%%%%%%%%%%%%%%%%%%%%%%%%%%%%%%%%%%%%%%%%%%%
\section{Minimal flavored Peccei-Quinn symmetries}
\label{sec:symmetries}
%%%%%%%%%%%%%%%%%%%%%%%%%%%%%%%%%%%%%%%%%%%%%%%%%%%%%%%%%%%%%%%%%%%%%%%%%%%%%

In general, the quark and lepton mass matrices are completely arbitrary, containing more independent parameters than those experimentally measured. Our goal is to identify the set of maximally-restrictive mass matrices that are both compatible with fermion data and realizable through the flavored PQ symmetries, taking as reference our previous work of Ref.~\cite{Rocha:2024twm}. In this way, the same symmetry imposed to solve the strong CP problem, also addresses the observed fermion masses and mixing patterns.

A given set of texture-zero mass matrices is realizable through the PQ symmetry if it satisfies the following set of equations [see Eqs.~\eqref{eq:YukawaZeros} and~\eqref{Eq:MassMatrices}]:
\begin{align}
(\mathbf{M}_f)_{\alpha \beta} &= 0 \Leftrightarrow (\Theta^f_1)_{\alpha \beta} \neq 0 \  \text{(mod $2\pi$)} \ \wedge \ (\Theta^f_2)_{\alpha \beta} \neq 0 \ \text{(mod $2\pi$)} \; , \nonumber \\
(\mathbf{M}_D)_{\alpha j} &= 0 \Leftrightarrow (\Theta^D_1)_{\alpha j} \neq 0 \  \text{(mod $2\pi$)} \ \wedge \ (\Theta^D_2)_{\alpha j} \neq 0 \ \text{(mod $2\pi$)} \; , \nonumber \\
(\mathbf{M}_R)_{ij} &= 0 \Leftrightarrow (\Theta^R_{0})_{ij} \neq 0 \  \text{(mod $2\pi$)} \ \wedge \ (\Theta^R_{1})_{ij} \neq 0 \ \text{(mod $2\pi$)} \ \wedge \ (\Theta^R_{2})_{ij} \neq 0 \ \text{(mod $2\pi$)} \; .
\label{eq:canonicalchargesSystem}
\end{align}
To assess compatibility with quark, charged-lepton, and neutrino data within the $\nu$DFSZ framework [see Eq.~\eqref{Eq:MassMatrices}], we employ as usual a standard $\chi^2$-analysis, with the function:
\begin{equation}
\chi^2(x) = \sum_i \frac{\left[\mathcal{P}_i(x) - \mathcal{O}_i\right]^2}{\sigma_i^2} \; ,
\label{eq:chi2}
\end{equation}
where $x$ denotes the input parameters, i.e., the matrix elements of $\mathbf{M}_{d}$, $\mathbf{M}_{u}$, $\mathbf{M}_{e}$, $\mathbf{M}_{D}$ and $\mathbf{M}_{R}$; $\mathcal{P}_i(x)$ is the model output for a given observable with best-fit value $\mathcal{O}_i$, and $\sigma_i$ denotes its $1\sigma$ experimental uncertainty. In our search for viable sets, we use the current quark and lepton data reported in Tables~\ref{tab:quarkdata} and~\ref{tab:leptondata}, respectively, and require that the $\chi^2$-function is minimized with respect to ten observables in the quark sector: the six quark masses $m_{d,s,b}$ and $m_{u,c,t}$, as well as the CKM parameters -- the three mixing angles $\theta_{12,23,13}^q$ and the CP-violating phase $\delta^q$; and nine observables in the lepton sector: three charged-lepton masses $m_{e,\mu,\tau}$, the two neutrino mass-squared differences $\Delta m^2_{21}, \Delta m^2_{31}$, and the lepton mixing matrix parameters -- the three mixing angles $\theta_{12,23,13}^\ell$ and the Dirac CP-violating phase $\delta^\ell$. Note that, for the neutrino oscillation parameters, $\chi^2(x)$ is computed using the one-dimensional profiles $\chi^2(\sin^2 \theta^\ell_{ij})$ and $\chi^2(\Delta m^2_{ij})$, and the two-dimensional~(2D) distribution $\chi^2(\delta^\ell,\sin^2 \theta^\ell_{23})$ for $\delta^\ell$ and $\theta_{23}^\ell$ given in Ref.~\cite{deSalas:2020pgw}. In our analysis, we consider a set of mass matrices to be compatible with data if the observables in Tables~\ref{tab:quarkdata} and~\ref{tab:leptondata} fall within the $1\sigma$ range at the $\chi^2$-function minimum.
\begin{table}[t!]
    \renewcommand*{\arraystretch}{0.35}
        \centering
            \begin{tabular}{|r|cc|}
                \cline{2-3}
                \multicolumn{1}{c|}{} & & \\
                \multicolumn{1}{c|}{} & \multicolumn{2}{c|}{Decomposition} \\[5pt]
                \hline
                & & \\ 
                Texture \; \; \; \; \; \; \; & $\mathbf{Y}_1^{d}$ & $\mathbf{Y}_2^{d}$ \\
                & & \\
                \hline
                & & \\ & & \\  
                \renewcommand{\arraystretch}{1.0} $4_{3}^{d}\sim \begin{pmatrix}
                0 & 0 & \times \\
                0 & \times & \times \\
                \times & \times & 0\\
                \end{pmatrix}$ \; \;    &    
                \renewcommand{\arraystretch}{1.0} $\begin{pmatrix}
                0 & 0 & \times \\
                0 & \times & 0 \\
                \times & 0 & 0 \\
                \end{pmatrix}$ &
                \renewcommand{\arraystretch}{1.0} $\begin{pmatrix}
                0 & 0 & 0 \\
                0 & 0 & \times \\
                0 & \times & 0 \\
                \end{pmatrix}$ 
                \\
                & & \\ & & \\
                \renewcommand{\arraystretch}{1.0} $5_{1}^{d}\sim \begin{pmatrix}
                0 & 0 & \times \\
                0 & \times & 0 \\
                \times & 0 & \times\\
                \end{pmatrix}$ \; \;   
                &   
                \renewcommand{\arraystretch}{1.0} $\begin{pmatrix}
                0 & 0 & \times \\
                0 & 0 & 0 \\
                \times & 0 & 0 \\
                \end{pmatrix}$ &
                \renewcommand{\arraystretch}{1.0} $\begin{pmatrix}
                0 & 0 & 0 \\
                0 & \times & 0 \\
                0 & 0 & \times \\
                \end{pmatrix}$
                \\
                & & \\ & & \\
                \hline
                & & \\ 
                 Texture  \; \; \; \; \; \; \; & $\mathbf{Y}_1^{u}$ & $\mathbf{Y}_2^{u}$ \\
                & & \\ 
                \hline
                & & \\ & & \\
                \renewcommand{\arraystretch}{1.0} $\mathbf{P}_{12}{5}_{1}^u\mathbf{P}_{23}\sim \begin{pmatrix}
                0 & 0 & \times \\
                0 & \bullet & 0 \\
                \times & \times & 0 \\
                \end{pmatrix}$ \; \; 
                &    
                \renewcommand{\arraystretch}{1.0} $\begin{pmatrix}
                0 & 0 & \times \\
                0 & 0 & 0 \\
                0 & \times & 0 \\
                \end{pmatrix}$ &
                \renewcommand{\arraystretch}{1.0} $\begin{pmatrix}
                0 & 0 & 0 \\
                0 & \bullet & 0 \\
                \times & 0 & 0 \\
                \end{pmatrix}$ 
                \\ & & \\ & & \\
                \renewcommand{\arraystretch}{1.0} $\mathbf{P}_{123}5_{1}^u \mathbf{P}_{12}\sim \begin{pmatrix}
                0 & \times & \bullet \\
                0 & 0 & \times \\
                \times & 0 & 0\\
                \end{pmatrix}$ \; \; 
                &    
                \renewcommand{\arraystretch}{1.0} $\begin{pmatrix}
                0 & \times & 0 \\
                0 & 0 & \times \\
                0 & 0 & 0 \\
                \end{pmatrix}$ &
                \renewcommand{\arraystretch}{1.0} $\begin{pmatrix}
                0 & 0 & \bullet \\
                0 & 0 & 0 \\
                \times & 0 & 0 \\
                \end{pmatrix}$ 
                \\ & & \\ & & \\
                \renewcommand{\arraystretch}{1.0} $\mathbf{P}_{12}{4}_{3}^u\sim \begin{pmatrix}
                0 & \bullet & \times \\
                0 & 0 & \times \\
                \times & \times & 0\\
                \end{pmatrix}$ \; \;     
                &    
                \renewcommand{\arraystretch}{1.0} $\begin{pmatrix}
                0 & 0 & \times \\
                0 & 0 & 0 \\
                0 & \times & 0 \\
                \end{pmatrix}$ &
                \renewcommand{\arraystretch}{1.0} $\begin{pmatrix}
                0 & \bullet & 0 \\
                0 & 0 & \times \\
                \times & 0 & 0 \\
                \end{pmatrix}$ 
                \\ & & \\ & & \\
                \renewcommand{\arraystretch}{1.0} $\mathbf{P}_{321}4_{3}^u\mathbf{P}_{23}\sim \begin{pmatrix}
                0 & \bullet & \times \\
                \times & 0 & \times \\
                0 & \times & 0\\
                \end{pmatrix}$ \; \; 
                &           
                \renewcommand{\arraystretch}{1.0} $\begin{pmatrix}
                0 & 0 & \times \\
                \times & 0 & 0 \\
                0 & \times & 0 \\
                \end{pmatrix}$ &
                \renewcommand{\arraystretch}{1.0} $\begin{pmatrix}
                0 & \bullet & 0 \\
                0 & 0 & \times \\
                0 & 0 & 0 \\
                \end{pmatrix}$ 
                \\ & & \\ & & \\
                \hline
                & & \\
                 Texture  \; \; \; \; \; \; \;  & $\mathbf{Y}_1^{e}$ & $\mathbf{Y}_2^{e}$ \\
                & & \\
                \hline
                & & \\ & & \\  
                \renewcommand{\arraystretch}{1.0} $4_{3}^{e}\sim \begin{pmatrix}
                0 & 0 & \times \\
                0 & \times & \times \\
                \times & \times & 0\\
                \end{pmatrix}$ \; \;    &    
                \renewcommand{\arraystretch}{1.0} $\begin{pmatrix}
                0 & 0 & \times \\
                0 & \times & 0 \\
                \times & 0 & 0 \\
                \end{pmatrix}$ &
                \renewcommand{\arraystretch}{1.0} $\begin{pmatrix}
                0 & 0 & 0 \\
                0 & 0 & \times \\
                0 & \times & 0 \\
                \end{pmatrix}$ 
                \\
                & & \\ & & \\
                \renewcommand{\arraystretch}{1.0} $5_{1}^{e}\sim \begin{pmatrix}
                0 & 0 & \times \\
                0 & \times & 0 \\
                \times & 0 & \times\\
                \end{pmatrix}$ \; \;   
                &   
                \renewcommand{\arraystretch}{1.0} $\begin{pmatrix}
                0 & 0 & \times \\
                0 & 0 & 0 \\
                \times & 0 & 0 \\
                \end{pmatrix}$ &
                \renewcommand{\arraystretch}{1.0} $\begin{pmatrix}
                0 & 0 & 0 \\
                0 & \times & 0 \\
                0 & 0 & \times \\
                \end{pmatrix}$
                \\
                & & \\ & & \\
                \hline
            \end{tabular}
    \caption{Realizable decomposition into Yukawa matrices of the quark and charged-lepton mass matrices for the texture pairs of Tables~\ref{tab:quarkcharges} and~\ref{tab:leptoncharges}. A matrix entry ``$0$" denotes a texture zero, ``$\times$" and~``$\bullet$" are a real positive and complex entry, respectively.} 
    \label{tab:Matricescharged}
\end{table}
\begin{table}[t!]
    \renewcommand*{\arraystretch}{0.35}
        \centering
        \begin{tabular}{|r|cccccc|}
        \cline{2-7}
        \multicolumn{1}{c|}{} & & & & & &\\
        \multicolumn{1}{c|}{} & \multicolumn{6}{c|}{Decomposition} \\[5pt]
        \hline
        &  &  & &  &  &  \\ 
         Texture \;\;\;\;\;\; &   \multicolumn{3}{c}{\; \; \; \;  $\Y_1^D$} & \multicolumn{3}{c|}{  $\Y_2^D$ }    \\ 
        &  &  & &  &  &  \\ 
        \hline
        &  &  & &  &  &  \\ &  &  & &  &  &  \\ 
            \renewcommand{\arraystretch}{1.0} $2_{1}^{D}\sim \begin{pmatrix}
            \bullet & 0 \\
            \bullet & \bullet \\
            0 & \bullet \\
            \end{pmatrix}$ \; \; 
            & \multicolumn{3}{c}{\; \; \;  \renewcommand{\arraystretch}{1.0} $ \begin{pmatrix}
            0 & 0  \\
            \bullet & 0  \\
            0 & \bullet  \\
            \end{pmatrix}$ }
            & \multicolumn{3}{c|}{\ \renewcommand{\arraystretch}{1.0} $ \begin{pmatrix}
            \bullet & 0  \\
            0 & \bullet  \\
            0 & 0  \\
            \end{pmatrix}$ \; } \\
        &  &  & &  &  &  \\ &  &  & &  &  &  \\ 
            \renewcommand{\arraystretch}{1.0} $2_{2}^{D} \sim 
            \begin{pmatrix}
            \bullet & \bullet \\
            \bullet & 0 \\
            0 & \bullet \\
            \end{pmatrix}$ \; \; 
            & \multicolumn{3}{c}{\; \; \; \renewcommand{\arraystretch}{1.0} $ \begin{pmatrix}
            \bullet & 0  \\
            0 & 0  \\
            0 & \bullet  \\
            \end{pmatrix}$ }
            & \multicolumn{3}{c|}{\renewcommand{\arraystretch}{1.0} $ \begin{pmatrix}
            0 & \bullet  \\
            \bullet & 0  \\
            0 & 0  \\
            \end{pmatrix}$\;} \\
        &  &  & &  &  &  \\ &  &  & &  &  &  \\ 
            \renewcommand{\arraystretch}{1.0} $3_{1}^{D}\sim \begin{pmatrix}
            \bullet & \bullet \\
            \bullet & 0 \\
            0 & 0 \\
            \end{pmatrix}$ \; \; 
            & \multicolumn{3}{c}{\; \; \; \renewcommand{\arraystretch}{1.0} $ \begin{pmatrix}
            0 & \bullet  \\
            \bullet & 0  \\
            0 & 0  \\
            \end{pmatrix}$ }
            & \multicolumn{3}{c|}{\renewcommand{\arraystretch}{1.0} $ \begin{pmatrix}
            \bullet & 0  \\
            0 & 0  \\
            0 & 0  \\
            \end{pmatrix}$\;} \\
        &  &  & &  &  &  \\ &  &  & &  &  &  \\ 
            \renewcommand{\arraystretch}{1.0}$3_{2}^{D}\sim \begin{pmatrix}
            \bullet & 0 \\
            \bullet & 0 \\
            0 & \bullet \\
            \end{pmatrix}$ \; \; 
            & \multicolumn{3}{c}{\; \; \; \renewcommand{\arraystretch}{1.0} $ \begin{pmatrix}
            0 & 0  \\
            \bullet & 0  \\
            0 & 0  \\
            \end{pmatrix}$ }
            & \multicolumn{3}{c|}{\renewcommand{\arraystretch}{1.0} $ \begin{pmatrix}
            \bullet & 0  \\
            0 & 0  \\
            0 & \bullet  \\
            \end{pmatrix}$ \;} \\
        &  &  & &  &  &  \\ &  &  & &  &  &  \\ 
            \renewcommand{\arraystretch}{1.0} $3_{3}^{D}\sim \begin{pmatrix}
            \bullet & 0 \\
            0 & \bullet \\
            0 & \bullet \\
            \end{pmatrix}$ \; \; 
            & \multicolumn{3}{c}{\; \; \; \renewcommand{\arraystretch}{1.0} $ \begin{pmatrix}
            \bullet & 0  \\
            0 & 0  \\
            0 & \bullet  \\
            \end{pmatrix}$ }
            & \multicolumn{3}{c|}{\renewcommand{\arraystretch}{1.0} $ \begin{pmatrix}
            0 & 0  \\
            0 & \bullet  \\
            0 & 0  \\
            \end{pmatrix}$ \;}\\
        &  &  & &  &  &  \\ &  &  & &  &  &  \\ 
            \renewcommand{\arraystretch}{1.0} $3_{4}^{D}\sim \begin{pmatrix}
            0 & 0 \\
            \bullet & 0 \\
            \bullet & \bullet \\
            \end{pmatrix}$ \; \; 
            & \multicolumn{3}{c}{\; \; \; \renewcommand{\arraystretch}{1.0} $ \begin{pmatrix}
            0 & 0  \\
            0 & 0  \\
            \bullet & 0  \\
            \end{pmatrix}$ }
            & \multicolumn{3}{c|}{\renewcommand{\arraystretch}{1.0} $ \begin{pmatrix}
            0 & 0  \\
            \bullet & 0  \\
            0 & \bullet  \\
            \end{pmatrix}$\;} \\
        &  &  & &  &  &  \\ &  &  & &  &  &  \\ 
        \hline
        &  &  & &  &  &  \\        
        Texture \;\;\;\;\;\; & \multicolumn{2}{c}{$\mathbf{M}^0_R$} & \multicolumn{2}{c}{$\mathbf{Y}^1_R$} & \multicolumn{2}{c|}{$\mathbf{Y}^2_R$} \\
        &  &  & &  &  &  \\
        \hline
        &  &  & &  &  &  \\ &  &  & &  &  &  \\ 
        \renewcommand{\arraystretch}{1.0} $1_{1,1}^{R}\sim \begin{pmatrix}
            0 & \times \\
            \cdot & \times \\
            \end{pmatrix}$ \; \;  
            & \multicolumn{2}{c}{\renewcommand{\arraystretch}{1.0} $\begin{pmatrix}
            0 & 0  \\
            \cdot & \times  \\
            \end{pmatrix}$}
            & \multicolumn{2}{c}{\renewcommand{\arraystretch}{1.0} $\begin{pmatrix}
            0 & \times  \\
            \cdot & 0  \\
            \end{pmatrix}$ }
            & \multicolumn{2}{c|}{\renewcommand{\arraystretch}{1.0} $\begin{pmatrix}
            0 & 0  \\
            \cdot & 0  \\
            \end{pmatrix}$} \\
        &  &  & &  &  &  \\ &  &  & &  &  &  \\ 
        \renewcommand{\arraystretch}{1.0} $1_{1,2}^{R}\sim \begin{pmatrix}
            0 & \times \\
            \cdot & \times \\
            \end{pmatrix}$ \; \;  
            & \multicolumn{2}{c}{\renewcommand{\arraystretch}{1.0} $\begin{pmatrix}
            0 & 0  \\
            \cdot & 0  \\
            \end{pmatrix}$}
            & \multicolumn{2}{c}{\renewcommand{\arraystretch}{1.0} $\begin{pmatrix}
            0 & \times  \\
            \cdot & 0  \\
            \end{pmatrix}$ }
            & \multicolumn{2}{c|}{\renewcommand{\arraystretch}{1.0} $\begin{pmatrix}
            0 & 0  \\
            \cdot & \times  \\
            \end{pmatrix}$} \\
        &  &  & &  &  &  \\ &  &  & &  &  &  \\ 
        \renewcommand{\arraystretch}{1.0} $1_{1,3}^{R}\sim \begin{pmatrix}
            0 & \times \\
            \cdot & \times \\
            \end{pmatrix}$ \; \;  
            &\multicolumn{2}{c}{ \renewcommand{\arraystretch}{1.0} $\begin{pmatrix}
            0 & 0  \\
            \cdot & 0  \\
            \end{pmatrix}$}
            & \multicolumn{2}{c}{\renewcommand{\arraystretch}{1.0} $\begin{pmatrix}
            0 & 0  \\
            \cdot & \times  \\
            \end{pmatrix}$ }
            & \multicolumn{2}{c|}{\renewcommand{\arraystretch}{1.0} $\begin{pmatrix}
            0 & \times  \\
            \cdot & 0  \\
            \end{pmatrix}$} \\
        &  &  & &  &  &  \\ &  &  & &  &  &  \\ 
        \renewcommand{\arraystretch}{1.0} $1_{1,4}^{R}\sim \begin{pmatrix}
            0 & \times \\
            \cdot & \times \\
            \end{pmatrix}$ \; \;  
            & \multicolumn{2}{c}{\renewcommand{\arraystretch}{1.0} $\begin{pmatrix}
            0 & 0  \\
            \cdot & \times  \\
            \end{pmatrix}$}
            & \multicolumn{2}{c}{\renewcommand{\arraystretch}{1.0} $\begin{pmatrix}
            0 & 0  \\
            \cdot & 0  \\
            \end{pmatrix}$ }
            & \multicolumn{2}{c|}{\renewcommand{\arraystretch}{1.0} $\begin{pmatrix}
            0 & \times  \\
            \cdot & 0  \\
            \end{pmatrix}$} \\
        &  &  & &  &  &  \\ &  &  & &  &  &  \\ 
        \renewcommand{\arraystretch}{1.0} $1_{2}^{R}\sim \begin{pmatrix}
            \times & \times \\
            \cdot & 0 \\
            \end{pmatrix}$ \; \;
            & \multicolumn{2}{c}{\renewcommand{\arraystretch}{1.0} $\begin{pmatrix}
            0 & 0  \\
            \cdot & 0  \\
            \end{pmatrix}$ }
            & \multicolumn{2}{c}{\renewcommand{\arraystretch}{1.0} $\begin{pmatrix}
            0 & \times  \\
            \cdot & 0  \\
            \end{pmatrix}$ }
            & \multicolumn{2}{c|}{\renewcommand{\arraystretch}{1.0} $\begin{pmatrix}
            \times & 0  \\
            \cdot & 0  \\
            \end{pmatrix}$} \\
        &  &  & &  &  &  \\ &  &  & &  &  &  \\ 
        \renewcommand{\arraystretch}{1.0} $2_{1}^{R}\sim \begin{pmatrix}
            0 & \times \\
            \cdot & 0 \\
            \end{pmatrix}$ \; \;    
            & \multicolumn{2}{c}{\renewcommand{\arraystretch}{1.0} $\begin{pmatrix}
            0 & \times \\
            \cdot & 0 \\
            \end{pmatrix}$ }
            & \multicolumn{2}{c}{\renewcommand{\arraystretch}{1.0} $\begin{pmatrix}
            0 & 0  \\
            \cdot & 0  \\
            \end{pmatrix}$ }
            & \multicolumn{2}{c|}{\renewcommand{\arraystretch}{1.0} $\begin{pmatrix}
            0 & 0  \\
            \cdot & 0  \\
            \end{pmatrix}$} \\
        &  &  & &  &  &  \\ &  &  & &  &  &  \\ 
        \hline
        \end{tabular}
    \caption{Realizable decomposition into Yukawa matrices of the neutrino mass matrices for the texture pairs of Table~\ref{tab:leptoncharges}. A matrix entry ``$0$" denotes a texture zero, ``$\times$" and~``$\bullet$" are a real positive and complex entry, respectively. The symmetric character of the Majorana matrix is marked by a ``$\cdot$".} 
    \label{tab:Matricesneutrino}
\end{table}
\begin{table}[t!]
    \centering
    \renewcommand*{\arraystretch}{1.5}
    \begin{tabular}{|c|c|ccc|c|}
            \hline
            \; Model \; &
            ($\mathbf{M}_d$,$\mathbf{M}_u)$  &
            $\chi^L_{q_\alpha} + \frac{s_\beta^2}{3} $ &
            $\chi^R_{d_\alpha} - \frac{2 s_\beta^2}{3}$
            & $\chi^R_{u_\alpha} + \frac{4 s_\beta^2}{3}$ & $\;\; N \;\;$
            \\ \hline
            $\text{Q}_\text{1}^\text{I}$ &
            \multirow{2}{*}{$(4_{3}^d,\mathbf{P}_{12}{5}_{1}^u\mathbf{P}_{23})$}              &      $\left(0,1,2 \right)$    &    $\left(2,1,0 \right)$  &  $\left(3,2,0 \right)$   &  $\;\;-2\;\;$ 
            \\
            $\text{Q}_\text{1}^\text{II}$ &
                       &      $\left(0,-1,-2 \right)$    &    $\left(-3,-2,-1 \right)$  &  $\left(-2,-1,1 \right)$   &  $\;\; 2\;\;$ 
            \\
            \hline
            $\text{Q}_\text{2}^\text{I}$ &
            \multirow{2}{*}{$(4_{3}^d,\mathbf{P}_{123}5_{1}^u \mathbf{P}_{12})$}             &      $\left(0,1,2 \right)$    &    $\left(2,1,0 \right)$  &  $\left(3,0,1 \right)$  &  $\;\;-1\;\;$ 
            \\
            $\text{Q}_\text{2}^\text{II}$ &
                          &      $\left(0,-1,-2 \right)$    &    $\left(-3,-2,-1 \right)$  &  $\left(-2,1,0 \right)$  &  $\;\; 1\;\;$ 
            \\
            \hline
            $\text{Q}_\text{3}^\text{I}$ &
            \multirow{2}{*}{$(5_{1}^d,\mathbf{P}_{12}{4}_{3}^u)$}     &      $\left(0,-1,1 \right)$    &    $\left(1,-2,0 \right)$  &  $\left(2,1,0 \right)$ &     $\;\;-2\;\;$
            \\
            $\text{Q}_\text{3}^\text{II}$ &
                 &      $\left(0,1,-1 \right)$    &    $\left(-2,1,-1 \right)$  &  $\left(-1,0,1 \right)$ &     $\;\;2\;\;$
            \\
            \hline
            $\text{Q}_\text{4}^\text{I}$ &
             \multirow{2}{*}{$(5_{1}^d,\mathbf{P}_{321}4_{3}^u\mathbf{P}_{23})$}         &      $\left(0,-1,1\right)$    &    $\left(1,-2,0 \right)$  &  $\left(-1,1,0 \right)$ & $\;\;1\;\;$ \\
            $\text{Q}_\text{4}^\text{II}$ &
                    &      $\left(0,1,-1\right)$    &    $\left(-2,1,-1 \right)$  &  $\left(2,0,1 \right)$ & $\;\;-1\;\;$ \\
            \hline 
        \end{tabular}
    \caption{
        Maximally-restrictive quark models ${\rm Q}^{\text{I,II}}_{1-4}$ made out of matrix pairs ($\mathbf{M}_d$,$\mathbf{M}_u$) compatible with data at~$1 \sigma$ CL (see Table~\ref{tab:quarkdata}). The superscript ``I" stands for the Yukawa decompositions given in Table~\ref{tab:Matricescharged}, while for models with superscript``II" the Yukawa textures are switched $\mathbf{Y}_1^{u,d} \leftrightarrow \mathbf{Y}_2^{u,d}$.
        $N$ is the color anomaly factor [see Eq.~\eqref{eq:Ncolor}]. Quark field charges $\chi_{q_\alpha}^L$, $\chi_{d_\alpha}^R$ and $\chi_{u_\alpha}^R$, follow the notation of Eq.~\eqref{eq:PQSym}.}
    \label{tab:quarkcharges}
\end{table}
\begin{table}[t!]
\centering
\renewcommand{\arraystretch}{1.5}
\begin{tabular}{|c|c|cccc|}
      \hline
      \; Model \; & ($\mathbf{M}_{e}$,$\mathbf{M}_D$,$\mathbf{M}_R$) &
      $\chi_\sigma$ &
      $\chi^L_{\ell_\alpha} - s_\beta^2$ &
      $\chi^R_{e_\alpha} - 2s_\beta^2$ &
      $\chi^R_{\nu_j}$ 
      \\ 
      \hline
      \renewcommand{\arraystretch}{1.0}\begin{tabular}[c]{@{}c@{}} $ $ \\ $\text{L}_\text{1}$\\ $ $\end{tabular} &
      \renewcommand{\arraystretch}{1.0}\begin{tabular}[c]{@{}c@{}} $ $ \\ $(4_{3}^{e},2_{1}^{D},2_{1}^{R})$\\ $ $\end{tabular} &
      \renewcommand{\arraystretch}{1.0}\begin{tabular}[c]{@{}c@{}}$ $\\ $1/2$\\ $ $\end{tabular} &
      \renewcommand{\arraystretch}{1.0}\begin{tabular}[c]{@{}c@{}}$ $\\ $\left(-3/2,-1/2,1/2 \right)$\\ $ $\end{tabular} &
      \renewcommand{\arraystretch}{1.0}\begin{tabular}[c]{@{}c@{}}$ $\\ $\left(1/2,-1/2,-3/2 \right)$\\ $ $\end{tabular} &
      \renewcommand{\arraystretch}{1.0}\begin{tabular}[c]{@{}c@{}}$ $\\ $\left(-1/2,1/2 \right)$\\ $ $\end{tabular} 
      \\
      \hline
      \renewcommand{\arraystretch}{1.4}\begin{tabular}[c]{@{}c@{}}$\text{L}_\text{2}$\\ $\text{L}_\text{3}$\end{tabular} &
      \renewcommand{\arraystretch}{1.4}\begin{tabular}[c]{@{}c@{}}$(4_{3}^{e},3_{1}^{D},1_{1,1}^{R})$\\ $(4_{3}^{e},3_{1}^{D},1_{1,2}^{R}) $\end{tabular} &
      \renewcommand{\arraystretch}{1.4}\begin{tabular}[c]{@{}c@{}} $1$\\ $1/2$\end{tabular} &
      \renewcommand{\arraystretch}{1.4}\begin{tabular}[c]{@{}c@{}} $\left(0,1,2 \right)$\\ $\left(-1/4,3/4,7/4\right)$\end{tabular} &
      \renewcommand{\arraystretch}{1.4}\begin{tabular}[c]{@{}c@{}} $\left(2 ,1,0 \right)$\\ $\left({7}/{4},{3}/{4},-{1}/{4}\right)$\end{tabular} &
      \renewcommand{\arraystretch}{1.4}\begin{tabular}[c]{@{}c@{}} $\left(1,0 \right)$\\ $\left({3}/{4},-{1}/{4} \right)$\end{tabular}
      \\ 
      \hline
      \renewcommand{\arraystretch}{1.0}\begin{tabular}[c]{@{}c@{}}$ $\\ $\text{L}_\text{4}$\\ $ $\end{tabular} &
      \renewcommand{\arraystretch}{1.0}\begin{tabular}[c]{@{}c@{}}$ $\\ $(4_{3}^{e},3_{2}^{D},1_{2}^{R})$\\ $ $\end{tabular} &
      \renewcommand{\arraystretch}{1.0}\begin{tabular}[c]{@{}c@{}}$ $\\ $1$\\ $ $\end{tabular} &
      \renewcommand{\arraystretch}{1.0}\begin{tabular}[c]{@{}c@{}}$ $\\ $\left(-{3}/{2},-{1}/{2},{1}/{2}  \right)$\\ $ $\end{tabular} &
      \renewcommand{\arraystretch}{1.0}\begin{tabular}[c]{@{}c@{}}$ $\\ $\left( {1}/{2},-{1}/{2},-{3}/{2} \right)$\\ $ $\end{tabular} &
      \renewcommand{\arraystretch}{1.0}\begin{tabular}[c]{@{}c@{}}$ $\\ $\left(-{1}/{2},{3}/{2} \right)$\\ $ $\end{tabular}
      \\ 
      \hline
      \renewcommand{\arraystretch}{1.0}\begin{tabular}[c]{@{}c@{}}$ $\\ $\text{L}_\text{5}$\\ $ $\end{tabular} &
      \renewcommand{\arraystretch}{1.0}\begin{tabular}[c]{@{}c@{}}$ $\\ $(4_{3}^{e},3_{2}^{D},1_{1,3}^{R})$\\ $ $\end{tabular} &
      \renewcommand{\arraystretch}{1.0}\begin{tabular}[c]{@{}c@{}}$ $\\ $1$\\ $ $\end{tabular} &
      \renewcommand{\arraystretch}{1.0}\begin{tabular}[c]{@{}c@{}}$ $\\ $\left(-{5}/{2},-{3}/{2},-{1}/{2}   \right)$\\ $ $\end{tabular} &
      \renewcommand{\arraystretch}{1.0}\begin{tabular}[c]{@{}c@{}}$ $\\ $\left(-{1}/{2} ,-{3}/{2},-{5}/{2} \right)$\\ $ $\end{tabular} &
      \renewcommand{\arraystretch}{1.0}\begin{tabular}[c]{@{}c@{}}$ $\\ $\left(-{3}/{2},{1}/{2} \right)$\\ $ $\end{tabular} 
      \\ 
      \hline
      \renewcommand{\arraystretch}{1.0}\begin{tabular}[c]{@{}c@{}}$ $\\ $\text{L}_\text{6}$ \\ $ $ \end{tabular} &
      \renewcommand{\arraystretch}{1.0}\begin{tabular}[c]{@{}c@{}}$ $\\ $(4_{3}^{e},3_{3}^{D},1_{1,3}^{R})$ \\ $ $ \end{tabular} &
      \renewcommand{\arraystretch}{1.0} \begin{tabular}[c]{@{}c@{}}$ $\\ $1$\\ $ $\end{tabular} &
      \renewcommand{\arraystretch}{1.0}\begin{tabular}[c]{@{}c@{}}$ $\\ $\left(-{3}/{2},-{1}/{2} ,{1}/{2} \right)$\\ $ $\end{tabular} &
      \renewcommand{\arraystretch}{1.0}\begin{tabular}[c]{@{}c@{}}$ $\\ $\left({1}/{2},-{1}/{2},-{3}/{2}  \right)$\\ $ $\end{tabular} &
      \renewcommand{\arraystretch}{1.0} \begin{tabular}[c]{@{}c@{}}$ $\\ $\left(-{3}/{2},{1}/{2} \right)$\\ $ $\end{tabular} 
      \\ 
      \hline
      \renewcommand{\arraystretch}{1.0}\begin{tabular}[c]{@{}c@{}}$ $\\ $\text{L}_\text{7}$\\ $ $\end{tabular} &
      \renewcommand{\arraystretch}{1.0}\begin{tabular}[c]{@{}c@{}}$ $\\ $(4_{3}^{e},3_{3}^{D},1_{2}^{R})$\\ $ $\end{tabular} &
      \renewcommand{\arraystretch}{1.0}\begin{tabular}[c]{@{}c@{}}$ $\\ $1$\\ $ $\end{tabular} &
      \renewcommand{\arraystretch}{1.0}\begin{tabular}[c]{@{}c@{}}$ $\\ $\left(-{1}/{2},{1}/{2},{3}/{2} \right)$\\ $ $\end{tabular} &
      \renewcommand{\arraystretch}{1.0}\begin{tabular}[c]{@{}c@{}}$ $\\ $\left({3}/{2},{1}/{2},-{1}/{2}  \right)$\\ $ $\end{tabular} &
      \renewcommand{\arraystretch}{1.0}\begin{tabular}[c]{@{}c@{}}$ $\\ $\left(-{1}/{2},{3}/{2} \right)$\\ $ $\end{tabular} 
      \\
      \hline
      \renewcommand{\arraystretch}{1.4}  \begin{tabular}[c]{@{}c@{}} $\text{L}_\text{8}$\\ $\text{L}_\text{9}$ \end{tabular} &
      \renewcommand{\arraystretch}{1.4}  \begin{tabular}[c]{@{}c@{}} $(4_{3}^{e},3_{4}^{D},1_{1,4}^{R})$\\ $(4_{3}^{e},3_{4}^{D},1_{1,3}^{R})$ \end{tabular} &
      \renewcommand{\arraystretch}{1.4}\begin{tabular}[c]{@{}c@{}} $1$\\ ${1}/{2}$\end{tabular} &
      \renewcommand{\arraystretch}{1.4}\begin{tabular}[c]{@{}c@{}} $\left(-3,-2,-1\right)$\\ $\left(-{11}/{4},-{7}/{4},-{3}/{4} \right)$\end{tabular} &
      \renewcommand{\arraystretch}{1.4}\begin{tabular}[c]{@{}c@{}} $\left(-1,-2,-3 \right)$\\ $\left(-{3}/{4},-{7}/{4},-{11}/{4} \right)$\end{tabular} &
      \renewcommand{\arraystretch}{1.4}\begin{tabular}[c]{@{}c@{}} $\left(-1,0 \right)$\\ $\left(-{3}/{4},{1}/{4} \right)$\end{tabular} 
      \\ 
      \hline
      \renewcommand{\arraystretch}{1.4}\begin{tabular}[c]{@{}c@{}} $\text{L}_\text{10}$ \\ $\text{L}_\text{11}$ \end{tabular} &
      \renewcommand{\arraystretch}{1.4}\begin{tabular}[c]{@{}c@{}} $(5_{1}^{e},2_{2}^{D},1_{1,4}^{R})$\\ $(5_{1}^{e},2_{2}^{D},1_{1,3}^{R})$\end{tabular} &
      \renewcommand{\arraystretch}{1.4}\begin{tabular}[c]{@{}c@{}} $1$\\ ${1}/{2}$\end{tabular} &
      \renewcommand{\arraystretch}{1.4}\begin{tabular}[c]{@{}c@{}} $\left(-1,-2,0 \right)$\\ $\left(-{3}/{4},-{7}/{4},{1}/{4} \right)$\end{tabular} &
      \renewcommand{\arraystretch}{1.4}\begin{tabular}[c]{@{}c@{}} $\left(0,-3, -1 \right)$\\ $\left({1}/{4},-{11}/{4},-{3}/{4} \right)$\end{tabular} &
      \renewcommand{\arraystretch}{1.4}\begin{tabular}[c]{@{}c@{}} $\left(-1,0 \right)$\\ $\left(-{3}/{4},{1}/{4} \right)$\end{tabular} 
      \\ 
      \hline
\end{tabular}
\caption{Maximally-restrictive lepton models ${\rm L}_{1-11}$ made out of the matrix sets ($\mathbf{M}_e$,$\mathbf{M}_D$,$\mathbf{M}_R$) compatible with data at~$1 \sigma$ CL for both NO and IO (see Table~\ref{tab:leptondata}). The lepton field charges $\chi_{\ell_\alpha}^L$, $\chi_{e_\alpha}^R$ and $\chi_{\nu_j}^R$ follow the notation of Eq.~\eqref{eq:PQSym}.} 
\label{tab:leptoncharges}
\end{table}
Our search for maximally restrictive matrix sets is simplified by considering only physically equivalent sets, known as equivalence classes~\cite{Ludl:2014axa,Ludl:2015lta,GonzalezFelipe:2014zjk,Correia:2019vbn,Camara:2020efq,Rocha:2024twm}. In particular, any set of mass matrices belongs to the same equivalence class if they can be transformed into one another through the following weak-basis permutations:
\begin{align}
    \mathbf{M}_d &\rightarrow \mathbf{P}_q^T \mathbf{M}_d \mathbf{P}_d\; , \;  \mathbf{M}_u \rightarrow \mathbf{P}_q^T \mathbf{M}_u \mathbf{P}_u \; , \nonumber \\
    \mathbf{M}_e &\rightarrow \mathbf{P}_\ell^T \mathbf{M}_e \mathbf{P}_e\; , \; \mathbf{M}_D \rightarrow \mathbf{P}_\ell^T \mathbf{M}_D \mathbf{P}_\nu\; , 
    \mathbf{M}_R \rightarrow \mathbf{P}_\nu^T \mathbf{M}_R \mathbf{P}_\nu \; ,
    \;
    \label{eq:Permutations}
\end{align}
where $\mathbf{P}_{\nu}$ are $2 \times 2$ permutation matrices, while the remaining $\mathbf{P}$'s are $3 \times 3$ permutation matrices.  Thus, we restrict our search for quark matrix sets to the equivalence classes derived in Ref.~\cite{Ludl:2014axa}. For the lepton sector, we adopt the charged-lepton equivalence classes from Ref.~\cite{Ludl:2015lta}, where every possible permutation of $\mathbf{M}_e$ via $\mathbf{P}_\ell^T$ and $\mathbf{P}_e$ is considered. Additionally, $\mathbf{M}_D$ matrices related by column permutations belong to the same class. Therefore, we consider only one representative from each class. With this approach, for $\mathbf{M}_R$ we must account for all possible textures.

Our results for the maximally-restrictive sets $\left(\mathbf{M}_{d},\mathbf{M}_{u},\mathbf{M}_{e},\mathbf{M}_{D},\mathbf{M}_{R}\right)$ are summarized in Tables~\ref{tab:Matricescharged} and~\ref{tab:Matricesneutrino}, where we present, respectively, the quark, charged lepton and neutrino mass matrix textures (leftmost column) and their corresponding decompositions (rightmost columns) in terms of the original Yukawa and mass matrices of the Lagrangian -- see Eqs.~\eqref{eq:Mass2hdm} and \eqref{Eq:MassMatrices}. All together, the number of independent parameters in the maximally-restrictive sets of fermion mass matrices matches the number of observables (masses and mixing parameters) in the quark and lepton sectors, meaning that all mass matrix elements are determined by data. Consequently, all non-vanishing Yukawa couplings are known up to a $s_\beta$ or $c_\beta$ factor. 

The PQ charges that realize the decompositions of Tables~\ref{tab:Matricescharged} and~\ref{tab:Matricesneutrino}, are listed in Table~\ref{tab:quarkcharges} and~\ref{tab:leptoncharges}. A few comments are in order: 
\begin{itemize}

\item \textbf{Minimal flavor patterns for quarks:} The maximally-restrictive quark textures realizable by a U$(1)$ Abelian flavor symmetry were found in Ref.~\cite{Rocha:2024twm}, where, without loss of generality, $\chi_1=0$, $\chi_2=1$, and $\chi^L_{q_1}=0$. To promote this framework to a flavored PQ scenario, i.e., to comply with Eq.~\eqref{eq:PQsymmetry}, we shift the flavor charges by $-2Y s_\beta^2$, where $Y$ is the hypercharge of the corresponding field~\footnote{This is possible due to the freedom to redefine the PQ charges through an admixture of the colour-anomaly-free global hypercharge and baryon number.}. Besides the models presented in Ref.~\cite{Rocha:2024twm}, which we label with the superscript ``I'', here we also consider those with $\mathbf{Y}_1^{u,d} \leftrightarrow \mathbf{Y}_2^{u,d}$, marked with ``II'', by redefining the PQ quark charges as $\chi^{L,R} \to -\chi^{L,R} + 2 Y (1+2\chi_1) - 2Y_{q_L}$. This ensures that $\chi^L_{q_1} = -2Y_{q_{L_1}} s_\beta^2$, as in the ``I'' cases. Note that, the I/II Yukawa ordering has physical implications regarding the predictions for the axion-to-photon couplings as shown in Sec.~\ref{sec:axionDMphotonflavor}.

The PQ charge assignments of the maximally-restrictive quark models, along with the corresponding anomaly factor, are presented in Table~\ref{tab:quarkcharges}. The maximally restrictive matrix structures are shown in Table~\ref{tab:Matricescharged}.

\item \textbf{Minimal flavor patterns for leptons:} The maximally-restrictive matrix pairs $(\mathbf{M}_{e}, \mathbf{M}_D, \mathbf{M}_R)$, are identified by examining the equivalence classes with the largest number of texture zeros. We then check whether they can be realized by PQ symmetries solving Eqs.~\eqref{eq:canonicalchargesSystem}, and whether they are compatible with the current lepton data given in Table~\ref{tab:leptondata}. If none of the equivalence classes under consideration passes this test, we add a non-zero entry to the mass matrix pairs and repeat the process until compatibility is achieved. This methodology, which is similar to that of previous works~\cite{GonzalezFelipe:2016tkv,Correia:2019vbn,Camara:2020efq,Rocha:2024twm}, led to eleven different models. The PQ charges~\footnote{Note that performing Yukawa permutations in both the quark and lepton sectors would lead to a physically equivalent scenario as the non-permuted case. Since we have considered permutations in the quark sector, we do not consider them in the lepton sector.} for each model are shown in Table~\ref{tab:leptoncharges} (their respective textures can be found in Tables~\ref{tab:Matricescharged} and~\ref{tab:Matricesneutrino}). 

As already mentioned, the pair ($\mathbf{M}_{e}$, $\mathbf{M}_\nu$), where $\mathbf{M}_\nu$ is determined from $\mathbf{M}_D$ and $\mathbf{M}_R$ using~\eqref{eq:TypeIMeff}, contains the same number of parameters as the number of lepton observables given by those in Table~\ref{tab:leptondata} and the single Majorana phase $\alpha$. By construction, since there are only two RH neutrinos one of the three light neutrinos is predicted to be massless due to the missing partner nature~\cite{Schechter:1980gr} of the underlying type-I seesaw mechanism. Hence, the framework presented here is testable at future experiments looking for $0_\nu \beta \beta$ -- see discussion in Sec.~\ref{sec:neutrinoobservables}.

Although for the sake of compatibility with fermion mass and mixing data, any lepton model can be combined with any quark model, the choice of a specific lepton model is relevant for the axion-to-photon coupling as discussed in Sec.~\ref{sec:axionDMphotonflavor}.

\end{itemize}
We conclude this section by remarking that the minimal flavor models found compatible with data, eight for quarks and eleven for leptons, will result in a total of eighty-eight combined models. In what follows -- Sec.~\ref{sec:axionpheno} -- we will investigate  the properties of these models regarding axion phenomenology.

%%%%%%%%%%%%%%%%%%%%%%%%%%%%%%%%%%%%%%%%%%%%%%%%%%%%%%%%%%%%%%%%%%%%%%%%%%%%%
\section{Axion phenomenology}
\label{sec:axionpheno}
%%%%%%%%%%%%%%%%%%%%%%%%%%%%%%%%%%%%%%%%%%%%%%%%%%%%%%%%%%%%%%%%%%%%%%%%%%%%%

%In this section we present the axion-to-photon coupling predictions for the different models identified in the previous section, discuss axion DM and cosmology, and we investigate possible constraints on axion flavor violating couplings to quarks and charged leptons.

%%%%%%%%%%%%%%%%%%%%%%%%%%%%%%%%%%%%%%%%%%%%%%%%%%%%%%%%%%%%%%%%%%%%
\subsection{Axion-to-photon coupling, dark matter and cosmology}
\label{sec:axionDMphotonflavor}
%%%%%%%%%%%%%%%%%%%%%%%%%%%%%%%%%%%%%%%%%%%%%%%%%%%%%%%%%%%%%%%%%%%%

%
\begin{table}[t!]
\centering
\renewcommand*{\arraystretch}{1.5}
\begin{tabular}{|c|c|}
\hline
Models & $E/N$ \\ \hline
$(\text{Q}_\text{2}^\text{II},\text{L}_\text{1-9})$ & -10/3  \\
$(\text{Q}_\text{2}^\text{II},\text{L}_\text{10,11})$ & -4/3  \\
$(\text{Q}_\text{1}^\text{II},\text{L}_\text{1-9})$ & -1/3  \\
$(\text{Q}_\text{1,2/3,4}^\text{II,I/II,I},\text{L}_\text{10,11/1-9})$ & 2/3  \\
$(\text{Q}_\text{3}^\text{II},\text{L}_\text{10,11})$ & 5/3  \\
$(\text{Q}_\text{1,2/3,4}^\text{I},\text{L}_\text{1-9/10,11})$ & 8/3 \\
$(\text{Q}_\text{3}^\text{I},\text{L}_\text{1-9})$ & 11/3 \\
$(\text{Q}_\text{4}^\text{II},\text{L}_\text{10,11})$ & 14/3 \\
$(\text{Q}_\text{4}^\text{II},\text{L}_\text{1-9})$ & 20/3 \\
\hline
\end{tabular}
\caption{$E/N$ values for the quark and lepton model combinations identified in Sec.~\ref{sec:symmetries} (see Tables~\ref{tab:quarkcharges} and~\ref{tab:leptoncharges}).}
\label{tab:ENDFSZ}
\end{table}
We now examine how to probe the various scenarios of Tables~\ref{tab:quarkcharges} and~\ref{tab:leptoncharges} through their corresponding axion-to-photon couplings $g_{a \gamma \gamma}$, which depends on the $E/N$ ratio -- see Eqs.~\eqref{eq:EN},~\eqref{eq:axioncouplingsgagg} and~\eqref{eq:axioncouplings}. To compute the EM anomaly factor $E$, both quark and lepton PQ charges need to be specified. The eighty eight combinations of Q and L models (see Tables~\ref{tab:quarkcharges} and~\ref{tab:leptoncharges}) lead to nine distinct $E/N$ values, as shown in Table~\ref{tab:ENDFSZ} for the minimal flavored PQ $\nu$DFSZ models identified in Sec.~\ref{sec:symmetries}. The combinations $(\text{Q}_\text{1,2/3,4}^\text{I},\text{L}_\text{1-9/10,11})$ and $(\text{Q}_\text{1,2/3,4}^\text{II,I/II,I},\text{L}_\text{10,11/1-9})$ lead to the same $E/N$ as in the DFSZ-I and DFSZ-II schemes~\cite{Zhitnitsky:1980tq,Dine:1981rt}, namely $E/N=8/3$ and $E/N=2/3$, respectively.

\begin{figure}[!t]
    \centering
      \includegraphics[scale=0.55]{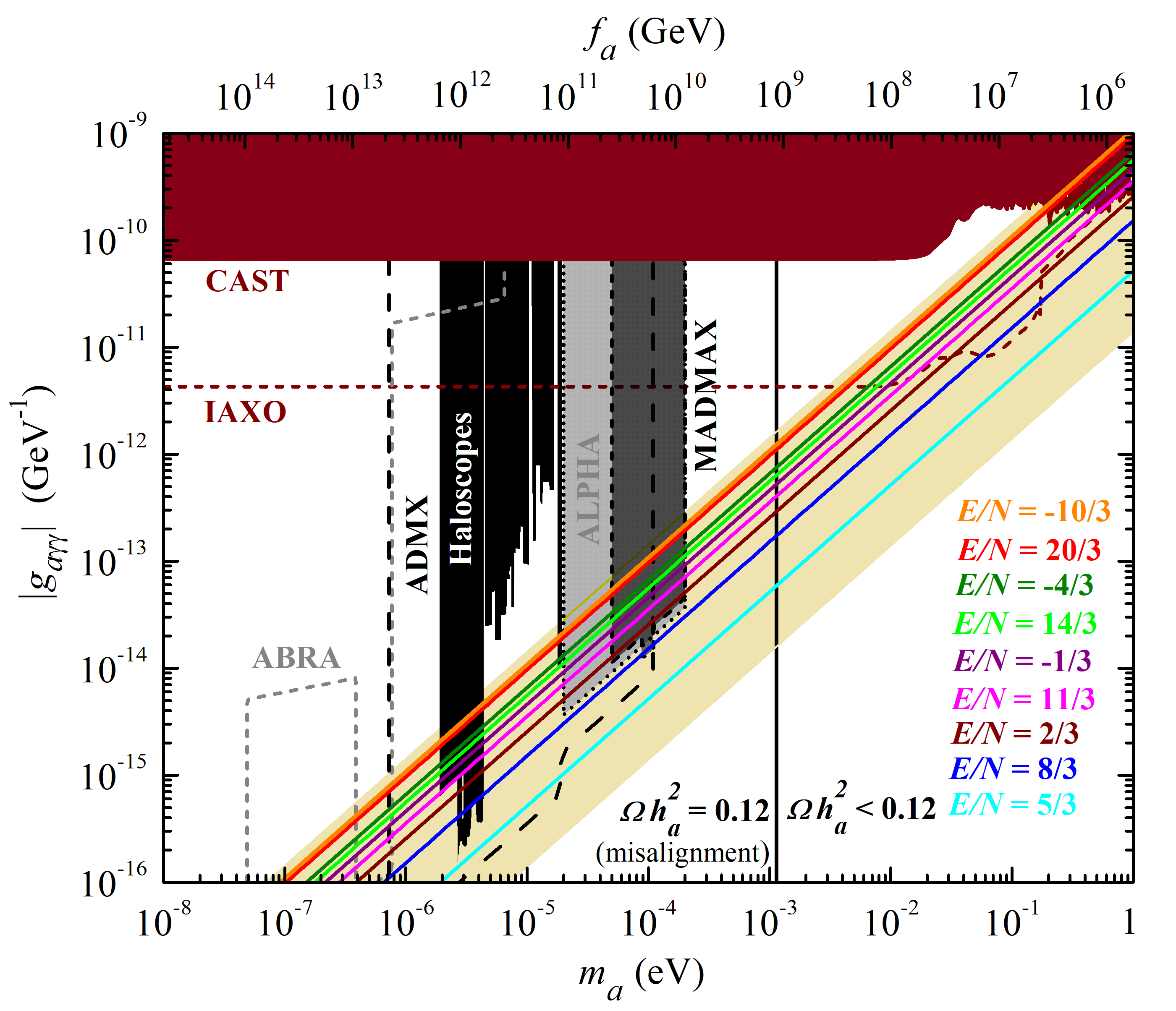}
    \caption{$|g_{a \gamma \gamma}|$ versus $m_a$ (bottom axis) and $f_a$ (top axis). Colored solid lines indicate $E/N$ values for the different models identified in Sec.~\ref{sec:symmetries} -- see Table~\ref{tab:ENDFSZ} and text for details. The remaining elements in the figure are the same as in Fig.~\ref{fig:gaggMajorana}.}
\label{fig:gaggDFSZ}
\end{figure}
In Fig.~\ref{fig:gaggDFSZ}, we display $|g_{a\gamma \gamma}|$ in terms of the axion mass~$m_a$ (bottom axis) and decay constant~$f_a$ (top axis), showing current bounds and future sensitivities from helioscopes and haloscopes. The colored oblique lines indicate the $|g_{a\gamma \gamma}|$ predictions for the model combinations of Table~\ref{tab:ENDFSZ}. A few comments are in order:
\begin{itemize}
    \item The maximum and minimum values for $|g_{a\gamma \gamma}|$ correspond to the $(\text{Q}_\text{2}^\text{II},\text{L}_\text{1-9})$ and $(\text{Q}_\text{3}^\text{II},\text{L}_\text{10,11})$ models, leading to $E/N=-10/3$ (orange line) and $E/N=5/3$ (cyan line), respectively. Naturally, the distinct $|g_{a\gamma \gamma}|$ predictions are within the yellow shaded region corresponding to the usual QCD axion window $|E/N -1.92| \in [0.07,7]$~\cite{DiLuzio:2016sbl}.

    \item For $m_a \gtrsim 0.2$~eV we can roughly say that the only models not excluded by the CAST helioscope (bordeau shaded region) are those with $E/N=2/3,\,8/3$ and 5/3. IAXO~\cite{Shilon_2013} (bordeau-dashed contour) is expected to scrutinize all our models except for $(\text{Q}_\text{3}^\text{II},\text{L}_\text{10,11})$ that leads to the minimal $|g_{a\gamma \gamma}|$ value for $E/N=5/3$ (cyan line).
    
    \item Out of all the haloscope experiments (black shaded region), ADMX currently excludes a considerable part of the parameter space of our scenarios, for axion masses $m_a \sim 3 \ \mu$eV, corresponding to $f_a \sim 10^{12}$ GeV and $g_{a \gamma \gamma}$ down to $(2 \times 10^{-16}) \ \text{GeV}^{-1}$. Again, model $(\text{Q}_\text{3}^\text{II},\text{L}_\text{10,11})$ (cyan line) is an exception. Future ADMX generations (dashed black contour) are expected to probe all our models for masses $1 \ \mu \text{eV} \lesssim m_a \lesssim 20 \ \mu$eV or equivalently scales $5 \times 10^{11} \ \text{GeV} \lesssim f_a \lesssim 10^{13}$ GeV, with $g_{a \gamma \gamma}$ down to $(10^{-16}-10^{-15}) \ \text{GeV}^{-1}$. ALPHA~\cite{Lawson:2019brd,Wooten:2022vpj,ALPHA:2022rxj} (light-gray shaded region) and MADMAX~\cite{Beurthey:2020yuq} (dark-gray shaded region) are projected to cover all our frameworks except for $(\text{Q}_\text{1,2/3,4}^\text{I},\text{L}_\text{1-9/10,11})$ (blue line) and $(\text{Q}_\text{3}^\text{II},\text{L}_\text{10,11})$ (cyan line), that lead to $E/N=8/3$ and $E/N=5/3$, respectively~(see Table~\ref{tab:ENDFSZ}).

    \item As discussed in Sec.~\ref{sec:axionDMcosmo}, axion DM is conceivable in two distinct cosmological scenarios, the pre and post inflationary cases. The pre-inflationary scenario, is possible in the region to the left of the black vertical line [see Eq.~\eqref{eq:relica}]. In contrast, the right region leads to under-abundant axion DM. For the post-inflationary axion DM case, as mentioned before, the cosmological DW problem is absent if $N_{\text{DW}} = 1$. This occurs for the quark models Q$_{2,4}$ of Table~\ref{tab:quarkcharges} for any lepton model in Table~\ref{tab:leptoncharges}, as long as $V_{\rm PQ}$ in Eq.~\eqref{eq:VPQoperator} is the one with $\chi_\sigma = 1$ (cubic model). All remaining cases lead to DWs in the early Universe~\cite{Lazarides:2018aev}. Although there are no DWs for $N_{\text{DW}}=1$, axionic string networks can still form. Current numerical simulations predict the axion scale $f_a$ to be in the range $[5 \times 10^9,3 \times 10^{11}]$~GeV, in order for the axion to account for the observed CDM abundance, i.e. $\Omega_a h^2 = \Omega_{\text{CDM}} h^2$~\cite{Kawasaki:2014sqa,Klaer:2017ond,Gorghetto:2020qws,Buschmann:2021sdq,Benabou:2024msj}. Note that, in Ref.~\cite{Cox:2023squ}, all quark flavor patterns with $N_{\text{DW}}=1$ were identified in the DFSZ model. Our partial results for the quark sector agree with the ones found in that reference.
    
\end{itemize}
Overall, haloscopes and helioscopes provide an important way to probe our models, complementary to other new physics searches in the flavor sector, as it will be explored in the upcoming section.

%%%%%%%%%%%%%%%%%%%%%%%%%%%%%%%%%%%%%%%%%%%%%%%%%%%%%%%%%%%%%%%%%%%%%%%%%%%%%
\subsection{Flavor-violating axion couplings to fermions}
\label{sec:axionflavorviolating}
%%%%%%%%%%%%%%%%%%%%%%%%%%%%%%%%%%%%%%%%%%%%%%%%%%%%%%%%%%%%%%%%%%%%%%%%%%%%%

The strong hierarchy among the singlet and doublet VEVs $v_\sigma \gg v_1,v_2$, leads to a decoupling between the scalar particles of the softly-broken U(1) 2HDM and the singlet~\cite{Sopov:2022bog}. In this framework, the presence of new charged and neutral scalars that couple with fermions introduces new physics contributions in flavor processes. The tree-level FCNCs will be controlled by the matrices,
\begin{align}
    \mathbf{N}_{f} &= 
    \frac{v}{\sqrt{2}} 
    \mathbf{U}_L^{f \dagger}
    \left(s_\beta\Y_1^{f} - c_\beta\Y_2^{f} \right)
    \mathbf{U}_R^{f} \; ,
\label{eq:NMatrices}
\end{align}
with the Yukawa couplings $\Y_{1,2}^{f}$ ($f=u,d,e$) and unitary rotations $\mathbf{U}_{L,R}^{f}$ defined as in Eqs.~\eqref{eq:Lyuk2hdm} and~\eqref{eq:massdiag}, respectively. As shown in Refs.~\cite{Correia:2019vbn,Camara:2020efq,Rocha:2024twm}, the maximally restrictive U(1) flavor symmetries -- promoted to a PQ symmetry here -- can, in some cases, naturally suppress such dangerous FCNCs effects. In fact, the mass matrices labeled ``$5$" in Table~\ref{tab:Matricescharged}, feature an isolated non-zero entry in a specific row and column, corresponding to the mass of a particular fermion—referred to as the ``decoupled state". Therefore, for each model with a type ``$5$" matrix, we can have three decoupled states, which we identify with a superscript on the model tag. For instance, if $u$ is the decoupled state in $\text{Q}_1$, we refer to it as $\text{Q}_1^u$~\footnote{Since permuting the Yukawa couplings results in the same texture for $\mathbf{N}_{f}$, we omit the I/II superscript in this section for simplicity.}. Note that models with decoupled states have zero entries in the $\mathbf{N}_{f}$ matrices of Eq.~\eqref{eq:NMatrices}, which control the strength of FCNCs,
\begin{align}
    \text{Q}^{d}_{1-2},\text{Q}^{u}_{3-4},\text{L}^e_{10-11}&: \mathbf{N}_{d,u,e} \sim \begin{pmatrix}
    \times & 0 & 0 \\
    0 & \times & \times \\
    0 & \times & \times \\
    \end{pmatrix} \; , \; \nonumber
    \\
    \text{Q}^{s}_{1-2},\text{Q}^{c}_{3-4},\text{L}^\mu_{10-11} &: \mathbf{N}_{d,u,e} \sim \begin{pmatrix}
    \times & 0 & \times  \\
    0 & \times & 0 \\
    \times & 0 & \times \\
    \end{pmatrix} \; , \;  
    \text{Q}^{b}_{1-2},\text{Q}^{t}_{3-4},\text{L}^\tau_{10-11} : \mathbf{N}_{d,u,e} \sim \begin{pmatrix}
    \times & \times & 0  \\
    \times & \times & 0 \\
    0 & 0 & \times \\
    \end{pmatrix} \; .
\label{eq:Nudtextures}
\end{align}
Thanks to this symmetry-induced FCNCs suppression mechanism, it was found in our work of Ref.~\cite{Rocha:2024twm}, that some of our models comply with stringent quark sector flavor constraints, with new scalars below the TeV scale, within the reach of current experiments such as the LHC and testable at future facilities. Specifically, for down/strange decoupled models strong constraints stemming from the mass differences of the meson-antimeson systems $K$, $B_d$ and $B_s$, as well as CP violation encoded in the $\epsilon_K$ parameter, are automatically fulfilled, leading to neutral BSM scalar masses as low as $300$ GeV.

In the $\nu$DFSZ model, the axion also has flavor-violating couplings to fermions. Namely, writing  Eq.~\eqref{eq:Lyuk2hdm} in the mass basis~\eqref{Eq:MassMatrices}, modifies the fermion kinetic terms resulting in the effective axion-fermion interactions shown in Eq.~\eqref{eq:axionFermionLagrangian} presented in Sec.~\ref{sec:flavouredaxions}. Note that, diagonal vector couplings are unphysical and are therefore set to zero~\cite{MartinCamalich:2020dfe,Calibbi:2020jvd}. In our models the flavor-violating axion couplings are given by:
\begin{align}
    \mathbf{C}^{V,f} = \frac{1}{N} \left(\mathbf{U}^{f \dagger}_R \boldsymbol{\chi}^{R}_f \mathbf{U}_R^f + \mathbf{U}^{f \dagger}_L \boldsymbol{\chi}^{L}_f \mathbf{U}_L^f \right) \; , \;  \mathbf{C}^{A,f} = \frac{1}{N} \left(\mathbf{U}^{f \dagger}_R \boldsymbol{\chi}^{R}_f \mathbf{U}_R^f -\mathbf{U}^{f \dagger}_L \boldsymbol{\chi}^{L}_f \mathbf{U}_L^f \right) \; ,
\label{eq:axionFermionCoupling}
\end{align}
with $\boldsymbol{\chi}^{L,R}_f = \diag(\chi_1,\chi_2,\chi_3)^{L,R}_f$. In Sec.~\ref{sec:flavouredaxions}, we provide in Tables~\ref{tab:QuarkConstraints} and~\ref{tab:LeptonConstraints}, the most restrictive constraints on the above axion couplings to quarks and leptons, respectively. In complete analogy with Higgs FCNCs, flavored-PQ symmetries will also control axion flavor-violating couplings. In fact, for 5-type mass matrix structure, the structure of $\mathbf{C}^{V,A}$ mirrors that of the $\mathbf{N}_f$ matrices [see Eq.~\eqref{eq:Nudtextures}], i.e.
\begin{align}
    \text{Q}^{d}_{1-2},\text{Q}^{u}_{3-4},\text{L}^e_{10-11}&: \mathbf{C}^{d,u,e} \sim \begin{pmatrix}
    \times & 0 & 0 \\
    0 & \times & \times \\
    0 & \times & \times \\
    \end{pmatrix} \; , \; \nonumber
    \\
    \text{Q}^{s}_{1-2},\text{Q}^{c}_{3-4},\text{L}^\mu_{10-11} &: \mathbf{C}^{d,u,e} \sim \begin{pmatrix}
    \times & 0 & \times  \\
    0 & \times & 0 \\
    \times & 0 & \times \\
    \end{pmatrix} \; , \;  
    \text{Q}^{b}_{1-2},\text{Q}^{t}_{3-4},\text{L}^\tau_{10-11} : \mathbf{C}^{d,u,e} \sim \begin{pmatrix}
    \times & \times & 0  \\
    \times & \times & 0 \\
    0 & 0 & \times \\
    \end{pmatrix} \; ,
\end{align}
where for simplicity we omitted the $V,A$ superscript. Thanks to this mechanism, some flavor-violating constraints are automatically satisfied, i.e. they do not bound the axion mass~\cite{Cox:2023squ}. Specifically, a model with either $\alpha$ or $\beta$ decoupled is consistent with the bounds set on $|\mathbf{C}_{\alpha \neq \beta}^{V,A}|$ (see Tables~\ref{tab:QuarkConstraints} and~\ref{tab:LeptonConstraints}), for any axion mass. However, note that this mechanism does not apply to flavor-conserving constraints, i.e. for $\alpha = \beta$. On the other hand, the mass matrices labeled ``$4$'' lead to $\mathbf{C}^{V,A}$ matrices without zero entries. All axion models listed in Tables~\ref{tab:quarkcharges} and~\ref{tab:leptoncharges}, share some interesting properties, specifically:
\begin{itemize}
    
    \item The off-diagonal entries of $\mathbf{C}^{V,f}$ and $\mathbf{C}^{A,f}$, defined in Eq.~\eqref{eq:PQsymmetry}, are independent of the angle $\beta$. This is evident since the PQ charges of  different $f$-fermion families depend equally on this parameter, implying that only the diagonal couplings exhibit $\beta$ dependence. This results from unitarity of the mixing matrices in Eq.~\eqref{eq:massdiag}.

    \item In our models, it can be shown that all the axion couplings to fermions can be made real. Thus, certain strong constraints do not apply, as those coming from the $D-\bar{D}$ mixing CP-violating phase~\cite{MartinCamalich:2020dfe}. Instead, the weaker CP-conserving constraint applies.

    \item Yukawa permutations in the quark models do not affect the physical axion-fermion couplings. Therefore, in this section, models that differ only by a Yukawa permutation, such as $\text{Q}_1^\text{I}$ and $\text{Q}_1^\text{II}$, can be treated together (we denote them as $\text{Q}_{1}$).

    \item Lepton models allowing for the cubic ($\chi_\sigma=1$) and quartic ($\chi_\sigma=1/2$) $V_{\rm PQ}$ yield the same $\mathbf{C}^{V,A}$ couplings. Consequently, in this section, lepton models with the same textures can be studied simultaneously, independently from the form of $V_{\rm PQ}$.
    
\end{itemize}
\begin{figure}[!t]
    \centering
    \includegraphics[scale=0.55]{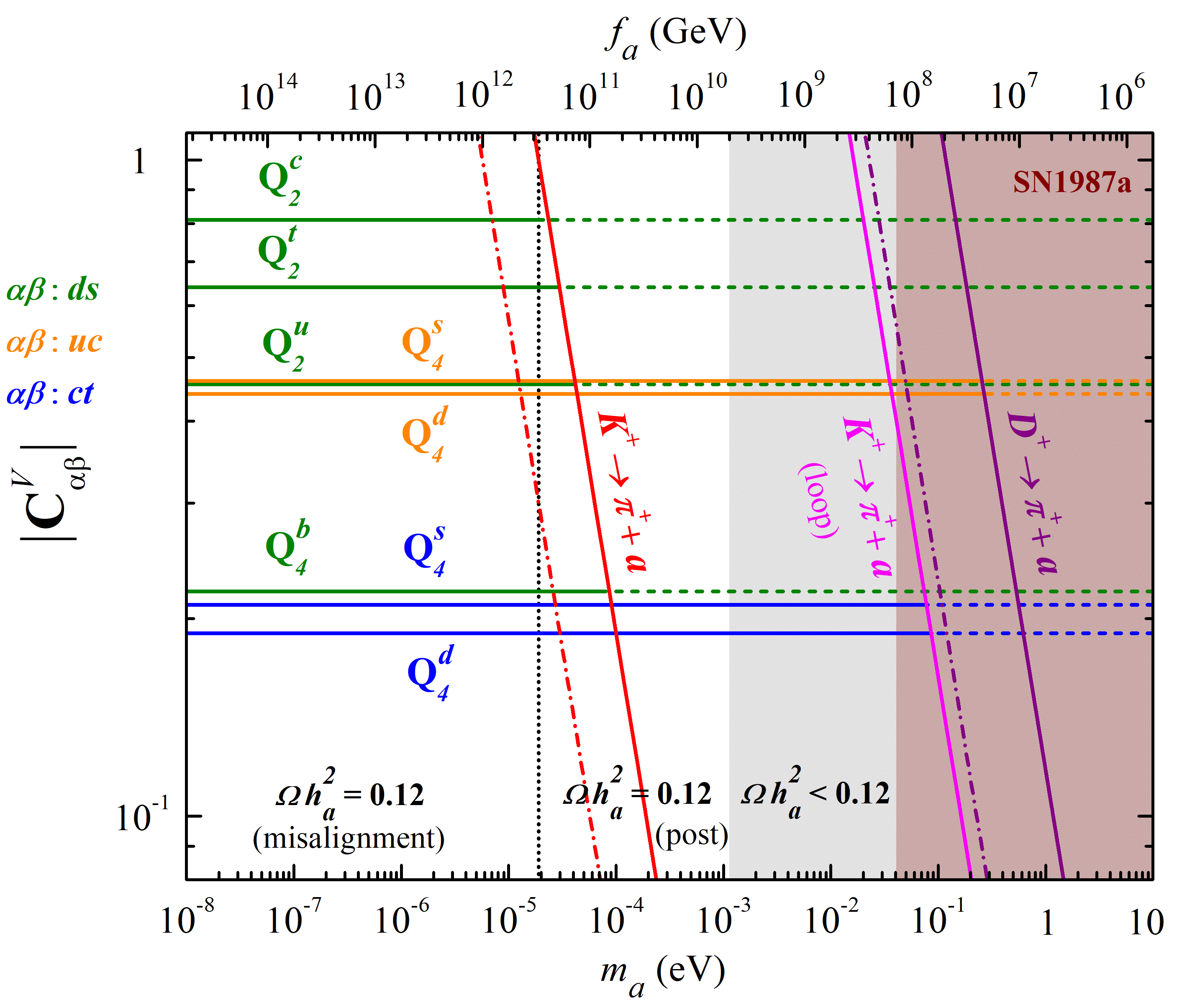}
    \caption{Vector flavor-violating axion-quark couplings $|\mathbf{C}^V_{\alpha\beta}|$, versus $m_a$ (bottom axis) and $f_a$ (top axis). We present the most restricted couplings $\alpha \beta = ds$ ($uc$) [$ct$], indicated by horizontal green (orange) [blue] lines, for models featuring $N_{\text{DW}} =1$ (see Table~\ref{tab:quarkcharges}). The dashed part of the lines are currently excluded by the red (purple) [magenta] oblique bound from $K^+ \rightarrow \pi^+ + a$ ($D^+ \rightarrow \pi^+ + a$) [$K^+ \rightarrow \pi^+ + a$ (loop)]~\cite{MartinCamalich:2020dfe,Alonso-Alvarez:2023wig} (see Table~\ref{tab:QuarkConstraints}). Future experimental sensitivities are indicated by oblique dash-dotted lines. The remaining elements in the figure are the same as in Fig.~\ref{fig:axionfvup}.}
    \label{fig:quarks}
\end{figure}
\begin{figure}[!t]
    \centering
    \includegraphics[scale=0.55]{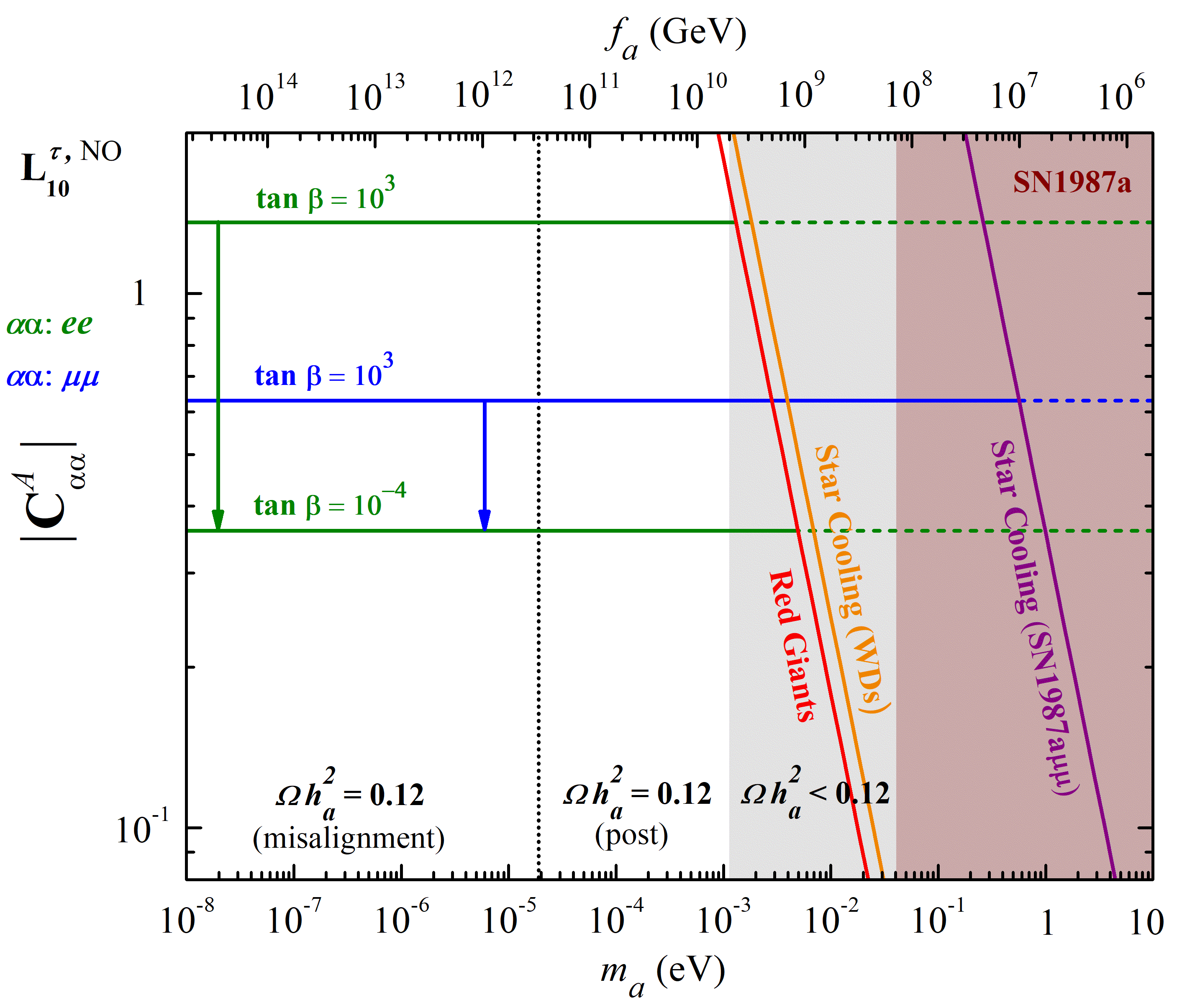}
    \caption{Axial diagonal axion-lepton couplings $|\mathbf{C}^A_{\alpha\alpha}|$, versus $m_a$ (bottom axis) and $f_a$ (top axis). We present the most restricted couplings $\alpha \alpha = ee$ ($\mu\mu$), indicated by horizontal green (blue) lines, for the lepton model $\text{L}_{10}^{\tau,\text{NO}}$ with $N_{\text{DW}} =1$ (see Table~\ref{tab:leptoncharges}). The dashed part of these lines are currently excluded by the red [purple] oblique bound from Red Giants~\cite{Capozzi:2020cbu,Bottaro:2023gep} [Star Cooling (SN1987$_a\mu\mu$)~\cite{MartinCamalich:2020dfe,Alonso-Alvarez:2023wig}]. We also indicate via an orange oblique line the constraint on the $ee$ coupling stemming from Star Cooling White Dwarfs~(WDs)~\cite{MartinCamalich:2020dfe,Alonso-Alvarez:2023wig} (see Table~\ref{tab:LeptonConstraints}). The horizontal lines correspond to the $|\mathbf{C}^A_{\alpha\alpha}|$ values for the indicated $\tan \beta$. The remaining elements follow the color code of Figs.~\ref{fig:axionfvup} and~\ref{fig:quarks}.}
    \label{fig:leptons}
\end{figure}
In the forthcoming numerical analysis, we consider the best fit values of the mass matrix elements for the models listed in Tables~\ref{tab:quarkcharges} and~\ref{tab:leptoncharges}, obtained through the $\chi^2$ analysis described in Sec.~\ref{sec:symmetries}. This enables the computation of the axion-fermion couplings through Eqs.~\eqref{eq:axionFermionCoupling}, which are then confronted with the most constraining present and future bounds, collected in Tables~\ref{tab:QuarkConstraints} and~\ref{tab:LeptonConstraints}. In Figs~\ref{fig:quarks} and~\ref{fig:leptons}, we present the most restricted couplings $\mathbf{C}^{V,A}_{\alpha \beta}$, indicated by horizontal lines, for some interesting quark and lepton minimal models, respectively, featuring $N_{\text{DW}} =1$. These results deserve several comments:
\begin{itemize}
    \item In models that are not $d$ or $s$ decoupled, $m_a$ is constrained by the $K^+ \rightarrow \pi^+ a$ decay as
    \begin{align}
        m_a \leq 
        2  \left(\frac{5.70\times 10^{12}\text{ GeV}}{\mathbf{F}^{V}_{ds}}\right)   |\mathbf{C}^{V}_{ds}  |^{-1}\;  \mu\text{eV} \;.
    \end{align}
    Considering the $\mathbf{F}^{V}_{ds}$ lower bounds in Table~\ref{tab:QuarkConstraints}, this relation exclude the regions on the right of the red oblique lines in Fig.~\ref{fig:quarks}. The values of the couplings $\mathbf{C}^V_{ds}$ for models $\text{Q}_{2}^{u,c,t}$ and $\text{Q}_{4}^{b}$ are shown by green horizontal lines. In these scenarios, the axion mass $m_a$ lies within the range $[10^{-5}, 10^{-4}]$ eV. Interestingly, for these cases the more predictive post-inflationary axion DM scenario (see Sec.~\ref{sec:axionDMcosmo}) is almost excluded (white region to the right of the vertical black-dotted line).

    Considering the axion mass bound, when combining the quark and lepton models permitting $N_{\text{DW}} = 1$, as shown in Table~\ref{tab:LeptonVSQuarkMassBound}, it becomes clear that the lepton constraints do not restrict the $\text{Q}_{2}^{u,c,t}$ and $\text{Q}_{4}^{b}$ models.
    
    \item For $d$ or $s$ decoupled models, the $K^+ \rightarrow \pi^+ a$ limit is automatically satisfied. As shown in Fig.~\ref{fig:quarks}, where the couplings $\mathbf{C}^V_{ds (uc)}$ for the models $\text{Q}_{4}^{d,s}$ are represented by  blue (orange) horizontal lines, other quark constraints impose a bound on $m_a$ comparable to that set by SN1987a, which excludes masses above $4 \times 10^{-2}$ eV~\cite{Carenza:2019pxu} [see Eq.~\eqref{eq:SN1987aaxionbound}].
    
    \item In Fig.~\ref{fig:leptons} we show the $\mathbf{C}^A_{ee,\mu\mu}$ limits for $\text{L}_{10}^\tau$ (the results for the remaining lepton models follow the same pattern). The most stringent bounds come from Red Giants being comparable to one stemming from Star Cooling (WDs), which constrains $\mathbf{C}^A_{ee}$ and imply  
    \begin{align}
        m_a \leq 
        2  \left(\frac{5.70\times 10^{12}\text{ GeV}}{\mathbf{F}_{ee}}\right) \left|\dfrac{s_\beta^2}{N} + \mathbf{C}^A_{ee}(\beta=0)  \right|^{-1} \; \mu\text{eV} \; ,
        \label{Eq:LeptonConservingContraint}
    \end{align}
    where we take the $\mathbf{F}_{ee}$ bounds from Table~\ref{tab:LeptonConstraints}. Note that this bound on $m_a$ depends on the angle $\beta$, as it arises from a flavor-conserving constraint. Therefore, we must account for the limits on $\beta$ imposed by perturbativity of Yukawa couplings. Requiring $|(\mathbf{Y}^f_{1,2})_{\alpha \beta}| \leq \sqrt{4\pi}$ leads to the following lower and upper limits on $t_\beta$:
    \begin{align}
        t^2_\beta \leq \frac{2\pi v^2}{|(\mathbf{M}_1^f)_{\alpha \beta}|^2} - 1,
        \quad
        t^2_\beta \geq {1}/{\left( \frac{2\pi v^2}{|(\mathbf{M}_2^f)_{\alpha \beta}|^2} - 1 \right)} \; ,
        \label{Eq:YukawaPerturbativity}
    \end{align}
    where, without loss of generality, we assume $\beta$ lies in the first quadrant. We have used the notation $\mathbf{M}_1^f$ ($\mathbf{M}_2^f$) for a mass matrix element originated from $\mathbf{Y}^f_{1}$ ($\mathbf{Y}^f_{2}$). At the end, the upper and lower bounds on $t_\beta$ are determined by the maximum values of $|(\mathbf{M}_1^f)|$ and $|(\mathbf{M}_2^f)|$, respectively. Using the above equation and the results of the $\chi^2$ fit for the mass matrix elements, we find that the quark Yukawas impose the most stringent bounds on $t_\beta$, with models $\text{Q}_{4}^{d,s}$ yielding $0.05 \leq t_\beta \leq 3.5$.
    Consequently, for a given set of $d,s$ decoupled mass matrices, the $m_a$ upper bound is within the range obtained from Eq.~\eqref{Eq:LeptonConservingContraint} considering the allowed interval for $\beta$. This contrasts with the cases $\text{Q}_{2}^{u,c,t}$ and $\text{Q}_{4}^{b}$ discussed above.

\end{itemize}

In Table~\ref{tab:LeptonVSQuarkMassBound} we show the $m_a$ bounds obtained when combining quark and lepton models with $N_{\text{DW}} =1$. We conclude that lepton constraints, rather than quark ones, set the axion mass bound in the $d$ or $s$ decoupled models within the range $m_a \in [10^{-3},10^{-2}]$ eV. Therefore, the axion mass can be up to two orders of magnitude larger than in the previously analyzed scenarios. As a result, the whole post-inflationary region remains viable while still accommodating flavor-violating axion couplings. Although we have only covered flavor-conserving lepton constraints, we have checked that the lepton flavor-violating ones are too weak to significantly bound the axion mass.
\begin{table}[t!]
\renewcommand*{\arraystretch}{1.5}
\centering
\resizebox{\textwidth}{!}{\begin{tabular}{|c|c c c c c c c c c|}
\multicolumn{10}{c}{\textbf{Axion mass upper bound (meV)}}
\\
\hline
Model & $\text{L}_{2}$ & $\text{L}_{4}$ & $\text{L}_{5}$ & $\text{L}_{6}$ & $\text{L}_{7}$ & $\text{L}_{8}$ & $\text{L}_{10}^e$ & $\text{L}_{10}^\mu$ & $\text{L}_{10}^\tau$ \\
 \hline
  $\text{Q}_\text{4}^{d,s}$ & $[1.3,4.1]$ & $[0.6,0.9]$ & $[0.7,1.1]$ & $[1.3,3.9]$ & $[0.9,1.6]$ & $[0.7,1.1]$ & $[1.8,23.6]$ & $[1.7,11.4]$ & $[1.4,4.9]$ \\
 \hline
\multicolumn{1}{|c|}{} & \multicolumn{9}{c|}{For all lepton models above} \\
 \hline
$\text{Q}_\text{2}^u$ & \multicolumn{9}{c|}{$3.7\times10^{-2}$} \\
$\text{Q}_\text{2}^c$ & \multicolumn{9}{c|}{$2.1\times10^{-2}$} \\
$\text{Q}_\text{2}^t$ & \multicolumn{9}{c|}{$1.5\times10^{-1}$} \\
$\text{Q}_\text{4}^b$ & \multicolumn{9}{c|}{$7.8\times10^{-2}$} \\
\hline
\end{tabular}}
\caption{Upper bounds for $m_a$ (in meV) extracted from constraints on axion-to-fermion couplings of Tables~\ref{tab:QuarkConstraints} and~\ref{tab:LeptonConstraints}, for NO models of Tables~\ref{tab:quarkcharges} and~\ref{tab:leptoncharges} with $N_{\text{DW}}=1$. As explained in the main text, for $\text{Q}_\text{4}^{d,s}$ the bounds depend on $t_\beta$ which, taking into account Yukawa perturbativity [see Eq.~\eqref{Eq:YukawaPerturbativity}], reflect on the ranges given in the second row of the table for each lepton model.}
\label{tab:LeptonVSQuarkMassBound}
\end{table}
%

%%%%%%%%%%%%%%%%%%%%%%%%%%%%%%%%%%%%%%%%%%%%%%%%%%%%%%%%%%%%%%%%%%%%%%%%%%%%%
\section{Key ideas and outlook}
\label{sec:conclflavoraxion}
%%%%%%%%%%%%%%%%%%%%%%%%%%%%%%%%%%%%%%%%%%%%%%%%%%%%%%%%%%%%%%%%%%%%%%%%%%%%%

We studied minimal $\nu$DFSZ axion models with flavored PQ symmetries and type-I seesaw neutrino masses. We performed a systematic analysis to identify the minimal models, in which the PQ symmetries impose the most restrictive quark and lepton flavor patterns, that are compatible with the observed fermion masses, mixing angles and CP-violating phases. We found eight maximally-restrictive quark flavor textures and eleven for the lepton sector for a combined eighty-eight possible models. In all cases, the number of independent parameters matches the number of observables. By construction, one neutrino is massless and, consequently, our framework can be tested at upcoming experiments looking for $0_\nu \beta \beta$ such as SNO+ II, LEGEND, and nEXO. 

We have also investigated several aspects related with axion phenomenology for each model, namely axion DM production in pre and post-inflationary cosmology. The most predictive scenarios are those which are free from DW problem. Namely, for $N_{\text{DW}}=1$ within the post-inflationary case, current numerical simulations of axion strings restrict the axion decay constant to the interval $5\times 10^9$ GeV to $3 \times 10^{11}$ GeV. In Fig.~\ref{fig:gaggDFSZ}, we show that helioscopes and haloscopes are able to probe our models via their distinct axion-to-photon coupling predictions (see Table~\ref{tab:ENDFSZ}). Future experiments such as IAXO, ADMX and MADMAX, will further scrutinize the models presented here. We also investigated how axion-to-fermion flavor violating couplings are constrained. We show that all our models, thanks to minimal flavored PQ symmetries, provide a natural framework to suppress flavor-violating couplings, as well as FCNCs in the Higgs sector. In fact, these symmetries lead to decoupled fermion states in the mass matrices, which impose restricted patterns in the $\mathbf{N}_f$ ($f=d,u,e$) matrices that control Higgs FCNCs. In particular, zero off-diagonal $\mathbf{N}_f$ entries, that otherwise contribute to the highly constraining flavor processes. This mechanism is also reflected in the vector and axial couplings of the axion to fermions. We find that in the quark sector the most stringent constraint is set by $K^+ \rightarrow \pi^+ + a$ in the $ds$ couplings, being automatically satisfied for models with a down or strange quark decoupled state. Strikingly, as shown in Fig.~\ref{fig:quarks}, for certain models with $N_{\text{DW}}=1$, the post-inflationary possibility for axion DM is excluded by this decay. In the lepton sector, current experimental flavor-violating constraints are not as relevant as those stemming from Red Giants and Star Cooling. These restrict the diagonal $ee$ and $\mu\mu$ axion-lepton couplings which depend on the VEV ratio of the two-Higgs doublets $\tan \beta$, providing a complementary constraint to that coming  from Higgs physics and scalar mediated FCNCs contributing to rare quark and lepton processes. For models with down or strange quark decoupled we found that the lepton Red Giants and Star Cooling constraints set a stronger bound on the axion mass (or scale) than the quark constraints -- see Table~\ref{tab:LeptonVSQuarkMassBound}.

To conclude, we have shown that flavored PQ symmetries in the minimal $\nu$DFSZ provide an appealing framework to address the flavor puzzle in the quark and lepton sector, neutrino mass generation, the strong CP problem and DM. This opens the possibility for further studies using axion frameworks as a gateway to tackle open problems in cosmology, particle and astroparticle physics.

%------------ 
% CHAPTER 07   
%------------ 

%%%%%%%%%%%%%%%%%%%%%%%%%%%%%%%%%%%%%%%%%%%%%%%%%%%%%%%%%%%%%%%%%%%%%%%%%%%%%
\chapter{Conclusions} 
\label{chpt:concl}
%%%%%%%%%%%%%%%%%%%%%%%%%%%%%%%%%%%%%%%%%%%%%%%%%%%%%%%%%%%%%%%%%%%%%%%%%%%%%

It is well established that the SM of particle physics, despite its remarkable empirical success, fails to account for several key observations: neutrino oscillations, DM, and the matter-antimatter asymmetry of the Universe. These constitute unambiguous signals of physics BSM, requiring theoretical extensions that introduce new particle content and/or symmetries. In addition to these phenomenological motivations, the SM leaves open a number of theoretically compelling questions -- such as the flavor puzzle and the strong CP problem -- which further guide the exploration of BSM frameworks.

In this thesis, we investigate BSM avenues guided by two main philosophies. First, we pursue unified frameworks capable of simultaneously addressing multiple outstanding problems in (astro)particle physics and cosmology. Second, we emphasize the testability of our proposals by systematically analyzing their phenomenological implications and experimental signatures. This approach to model building reveals deep connections between seemingly unrelated sectors and problems, enabling a more comprehensive understanding of the underlying physics and ultimately provide a stepping stone into the broader theory of Nature. The main results and conclusions of this thesis are summarized below.

\subsubsection{Dark-sectors at the origin of neutrino masses}

In Chapter~\ref{chpt:neutrinodarksectors}, we review UV complete realizations of the Weinberg operator which generates Majorana neutrino masses. After presenting the paradigmatic tree-level canonical seesaw mechanisms, we discuss their low-scale variants the ISS and LSS. These rely on approximate U(1) symmetries broken by small LNV parameters triggering light neutrino mass generation. The heavy mediators can have masses around the TeV scale with sizeable couplings, making these frameworks testable at experimental facilities, e.g. via observable cLFV signals. Next, making use of the quantum character of the theory, we connect radiative neutrino mass generation to dark-sectors. The latter are stable under an imposed symmetry as simple as $\mathcal{Z}_2$, as is the case in the scotogenic model. These dark-sector particles will trigger Majorana neutrino masses at loop-level, with the lightest of them being a viable DM candidate of WIMP type, that can be probed at DD detection experiments. This interesting idea -- dark-sectors at the origin of Majorana neutrino masses -- lead to our original work of Ref.~\cite{Batra:2023bqj}, which we study at the end of the chapter in Sec.~\ref{sec:darkLSS}. Namely, we propose for the first time in the literature the dark linear seesaw mechanism, where the small LNV parameter required to trigger the LSS is radiatively ``seeded'' by a dark sector. We have considered the minimal setup containing only one $\nu_R-S_R$ heavy neutrino pair, as well as a single dark vector-like fermion $f_{L,R}$. In such a way we are able to explain the two neutrino mass splittings, predicting a massless light neutrino with interesting signals for $0_\nu \beta \beta$. The model can accommodate either fermionic or scalar DM. We analyzed the latter possibility with DM being a mixed state of SU$(2)_L$ singlet and doublet scalars. This scenario can be directly probed at future DD experiments such as LZ, XENONnT and DARWIN. Furthermore, our proposal leads to sizeable cLFV signatures -- via interplay from both the dark and seesaw sectors -- testable at future facilities like MEG II and Belle II.

\subsubsection{The generation of non-trivial CP-violating effects from the vacuum}

In Chapter~\ref{chpt:SCPV} we examine the possibility that the complex VEV of a scalar singlet $\sigma$ serves as the common origin of CP-violating effects, such as LCPV and a realistic CKM matrix. For SCPV originating from $\sigma$ to be successfully transmitted and give rise to these effects, it is necessary to extend the SM with additional fermionic content. These new fermions must couple directly to $\sigma$, and specific conditions must be met to ensure that the CP-violating phase induced by the vacuum cannot be rotated away through field redefinitions. With this setup in mind, we explore two scenarios:
\begin{itemize}
    \item In this Sec.~\ref{sec:lepto}, based on our work of Ref.~\cite{Barreiros:2022fpi}, we study thermal leptogenesis in the type-I seesaw framework extended by complex scalar singlets. CP invariance is imposed at the Lagrangian level, such that the complex VEVs of the scalar singlets constitute the sole source of CP violation. These VEVs simultaneously induce Dirac and Majorana CP-violating phases at the EW scale, as well as high-energy CP violation relevant for successful leptogenesis. Compared to the vanilla type-I seesaw leptogenesis scenario, the scalars induce novel radiative corrections to the CP asymmetry generated when the heavy neutrinos decay into leptons, and provide new tree-level CP-violating three-body decay processes. We computed for the fist time in the literature the CP asymmetry for an arbitrary number of RH neutrinos, complex scalar singlets and Higgs doublets, determined the unflavored BEs by taking into account new decay channels, as well as $\Delta L=1$ and $\Delta N =2$ scatterings. To illustrate how SCPV can simultaneously lead to non-trivial low- and high-energy CP-violation, we studied a simple model featuring in addition to the SM particle content one complex singlet, two RH neutrinos and a new scalar doublet, and we impose a $\mathcal{Z}_8$ flavor symmetry. This corresponds to the minimal model that allows for the possibility of SCPV and compatibility with neutrino oscillation data. Interestingly, the model allows only novel contributions to the CP asymmetry induced by the scalar-singlet couplings to the RH neutrinos. After numerically solving the BEs, we showed that we can successfully account for the observed BAU value in a wide range of parameter space. 

    \item In this Sec.~\ref{sec:darkNB}, following our work of Ref.~\cite{Camara:2023hhn}, we propose a new NB model to address the strong CP problem based on the existence of a dark sector containing viable scalar singlet WIMP DM. Once again, we make use of a $\mathcal{Z}_8$ symmetry that allows for SCPV, with the particularity that after symmetry breaking -- via the complex VEV of $\sigma$ -- is left a residual $\mathcal{Z}_2$ symmetry that stabilizes our DM candidate. In our scenario, a complex CKM matrix arises from one-loop corrections to the quark mass matrix mediated by the dark sector. In contrast to the minimal BBP model, here the strong CP phase receives non-zero contributions only at two loops and we alleviate the so-called “NB-quality problem”. We analyzed the quark sector phenomenology in detail, demonstrating that a realistic CKM phase arises over a broad parameter space. We also examined flavor constraints and DM predictions, the latter of which are testable in ongoing and upcoming DD experiments. In particular, the complex scalar singlet responsible for SCPV opens a new annihilation channel, enhancing the thermally averaged cross section. This allows our setup to evade current stringent LZ DD bounds while successfully reproducing the observed CDM relic abundance.
    
\end{itemize}

\subsubsection{Axion frameworks with color-mediated neutrino masses}

In Chapter~\ref{chpt:axionneutrino}, we have presented two complementary frameworks that unify the origin of neutrino masses with the resolution of the strong CP problem via a KSVZ-type axion. In Ref.~\cite{Batra:2023erw} we introduce the novel idea of color-mediated Majorana neutrino masses. Namely, light neutrino masses arise at two loops mediated by colored states linked to PQ breaking. In the follow-up Dirac scenario proposed in Ref.~\cite{Batra:2025gzy}, neutrino masses are generated at one loop via VLQs and scalar leptoquarks, with the PQ symmetry enforcing \emph{Diracness} and suppressing baryon and lepton number violation. The Dirac case also predicts flavor-violating axion couplings to quarks, subject to stringent constraints from meson decays, top-quark processes, and astrophysical bounds. Both constructions yield distinctive axion-photon couplings, offering discovery prospects at helioscopes and haloscopes such as IAXO, ADMX and MADMAX. Moreover, they motivate complementary cosmologies for axion DM production: pre-inflationary for the Majorana case, post-inflationary for the Dirac. Together, these frameworks provide a unified and testable path toward resolving three fundamental problems in (astro)particle physics and cosmology.

\subsubsection{Minimal flavored Peccei-Quinn symmetries}

In Chapter~\ref{chpt:flavoraxion}, we presented the results of our work of Ref.~\cite{Rocha:2025ade}, where we investigated minimal $\nu$DFSZ axion models with flavored PQ symmetries and type-I seesaw neutrino masses, identifying the most constrained textures consistent with observed fermion masses, mixings, and CP violation. All viable models feature a predictive structure where one neutrino is massless, making them testable in future $0\nu\beta\beta$ searches. The axion sector exhibits rich phenomenology across both pre- and post-inflationary cosmologies for axion DM production. Models free from the DW problem are especially predictive, with axion-photon couplings accessible to upcoming haloscope and helioscope experiments. The flavor structure of the PQ symmetry naturally suppresses axion-mediated flavor violation and Higgs-induced FCNCs. In the quark sector, rare kaon decays provide the leading constraints, while in the lepton sector, stellar cooling bounds dominate, sensitively probing axion couplings to electrons and muons. Altogether, these models provide a unified framework connecting the flavor puzzle, neutrino mass generation, the strong CP problem, and DM.

\vspace{0.3cm}

Overall, this thesis develops minimal and unified BSM frameworks that address multiple fundamental problems in (astro)particle physics and cosmology—establishing deep connections between the origin of neutrino masses, the nature of DM, low- and high-energy CP violation, the matter–antimatter asymmetry, the NB solution to the strong CP problem, axion physics, and the flavor puzzle. Emphasizing experimental testability, the models constructed here yield predictive structures with rich phenomenology across the intensity, energy, and cosmic frontiers. This work offers a coherent step toward a more complete theory BSM. As the experimental landscape advances with next-generation neutrino, DM, axion, and collider searches, the ideas developed here are poised to be tested, refined, and expanded -- shedding light on the fundamental laws of Nature.

\printbibliography[heading=bibintoc]

% Appendices
\appendix

%--------------------------------------------------------------------
%				APPENDIX A    
%--------------------------------------------------------------------

%%%%%%%%%%%%%%%%%%%%%%%%%%%%%%%%%%%%%%%%%%%%%%%%%%%%%%%%%%%%%%%%%%%%%%%%%%%%%
\chapter{General aspects of Boltzmann equations}
\label{chpt:genBEs}
%%%%%%%%%%%%%%%%%%%%%%%%%%%%%%%%%%%%%%%%%%%%%%%%%%%%%%%%%%%%%%%%%%%%%%%%%%%%%

In this appendix are collected the expressions, formulas and notation used in this thesis to write down BEs that encode particle abundance generation for WIMP DM in Sec.~\ref{sec:DM} and leptogenesis Sec.~\ref{sec:BEs}. We work with the relativistic formulation of classical BEs in the FRW metric, and assume quantum coherence effects to be negligible. For recent examples where a quantum treatment of BEs is reviewed see Refs.~\cite{Biondini:2017rpb,Garbrecht:2018mrp,Klaric:2021cpi}.

In the early Universe interactions among particles in the thermal bath and the expansion of the Universe influence the microscopic time evolution of particle number densities and asymmetries which is described by a coupled system of BEs. Considering the number density~$n_\psi$ of a particle species $\psi$, the BEs take the following form~\cite{Kolb:1990vq}
    \begin{align}
    \frac{d n_\psi}{d t} + 3 H n_\psi & = - \sum_{i, j, \cdots} \left[ \gamma\left( \psi \rightarrow i + j + \cdots \right) - \gamma\left( i + j + \cdots \rightarrow \psi \right) \right] \nonumber \\
    & - \sum_{a, i, j, \cdots} \left[\gamma\left(\psi + a \rightarrow i + j + \cdots\right) - \gamma\left(i + j + \cdots \rightarrow \psi + a\right)\right] \; .
    \label{eq:BEsgeneric1}
    \end{align}
    In the above equation, the left-hand side takes into account effects of the expansion of the Universe, while the right-hand side is the collision term involving interactions. The number density is given by,
    \begin{align}
    n_\psi = \frac{g_\psi}{(2 \pi)^3} \int d^3 p_\psi f_\psi \; ,
    \label{eq:ndensity}
    \end{align}
    where $g_\psi$ and $f_\psi$ are the number of internal degrees of freedom and phase space distribution of the particle species $\psi$. For example, $g_{N} = 2$ for Majorana neutrinos, $g_{\ell_L} = 2$ for lepton doublet components, $g_{e_R}=1$ for charged-lepton singlet fields and $g_\Phi=2$ for Higgs doublet components. Furthermore, the Hubble parameter is,
    \begin{align}
    H(T) = \sqrt{\frac{4 \pi^3 g_{\ast}}{45}} \frac{T^2}{M_{\text{Pl}}} \; ,
    \label{eq:Hubble}
    \end{align}
    where $g_{\ast}$ is the effective number of relativistic degrees of freedom which in the SM case is $g_{\ast} = 106.75$ and $M_{\text{Pl}}\simeq 1.22 \times 10^{19}$~GeV is the Planck mass. For a general process $\gamma_{\psi + a + b + \cdots  \rightarrow i + j + \cdots} \equiv \gamma\left(\psi + a + b + \cdots  \rightarrow i + j + \cdots\right)$, involving $\psi$, the collision term is written as:
     \begin{align}
    \gamma_{\psi + a + b + \cdots  \rightarrow i + j + \cdots} & = \int \frac{d^3 p_\psi}{(2 \pi)^3 2 E_\psi} \frac{d^3 p_a}{(2 \pi)^3 2 E_a} \frac{d^3 p_b}{(2 \pi)^3 2 E_b} \cdots \frac{d^3 p_i}{(2 \pi)^3 2 E_i} \frac{d^3 p_j}{(2 \pi)^3 2 E_j} \cdots  \label{eq:collision} \\ 
    & \times (2 \pi)^4 \delta^4\left(p_\psi + p_a + p_b + \cdots  \rightarrow p_i + p_j + \cdots\right)  \nonumber \\ 
    & \times \left|\mathcal{M}\left(\psi + a + b + \cdots  \rightarrow i + j + \cdots\right) \right|^2 f_\psi f_a f_b \cdots (1 \pm f_i) (1 \pm f_j) \; , \nonumber
    \end{align}
    with the phase space integrals containing $p_\psi$ and $E_\psi$, being the momentum and energy of a given particle $\psi$ with mass $m_\psi$. In the above, the Dirac $\delta$-function accounts for momentum conservation and the squared S-matrix element $\left|\mathcal{M}\left(\psi + a + b + \cdots  \rightarrow i + j + \cdots\right) \right|^2$ is summed over the internal degrees of freedom of incoming and outgoing particles taking into consideration appropriate symmetry factors. Furthermore, the upper (lower) sign in $(1 \pm f_i)$ refers to bosons (fermions).

    Working in the dilute gas approximation one may consider $(1 \pm f_i) \simeq 1$. Furthermore, elastic scatterings will only affect the phase space densities of particles while inelastic scatterings change their number densities. Assuming that the elastic scatterings are fast enough to maintain kinetic equilibrium, in comparison to the inelastic ones, and making use of the Maxwell-Boltzmann equilibrium distribution, the phase space and number densities are related through $f_\psi(E_\psi, T) = n_\psi e^{- E_\psi / T}/n_\psi^{\text{eq}}$, where,
    \begin{align}
    n_{\psi}^{\text{eq}} = g_\psi \frac{m_\psi^2}{2 \pi^2} T \; \mathcal{K}_2\left(\frac{m_\psi}{T}\right) \; ,
    \label{eq:equi}
    \end{align}
    with $\mathcal{K}_n(x)$ being the modified Bessel function of order $n$. 

    To automatically take into account effects of the expansion of the Universe, it is convenient to use $Y_\psi$ which normalizes the particle number density to the entropy~$S$ or the particle number $N_\psi$ in the comoving volume $R_\ast(t)^3 = n_\gamma^{\text{eq}}(t)^{-1}$, which are defined as~\footnote{In an isentropically expanding Universe (entropy is conserved), $N_\psi$ and $Y_\psi$ are related by a constant.}:
    \begin{align}
    Y_\psi &= \frac{n_\psi}{S} \; , \; S(T) = \frac{2 \pi^2}{45} g_{\ast S} T^3 \; , \label{eq:ournotationDM} \\
    N_\psi &= \frac{n_\psi}{n_\gamma^{\text{eq}}} \; , \; n_{\gamma}^{\text{eq}}(T) = g_\gamma \frac{T^3}{\pi^2} \; ,
    \label{eq:ournotation}
    \end{align}
    with $g_{\ast S}$ being the effective number of relativistic related to the entropy density, usually $g_\ast \simeq g_{\ast S}$.
    
    Under the stated assumptions and using the change of variable of Eq.~\eqref{eq:ournotation}, used in Sec.~\ref{sec:lepto}, the BEs in Eq.~\eqref{eq:BEsgeneric1} become,
    \begin{align}
    n_\gamma^{\text{eq}} z_\psi H(z_\psi) \frac{d N_\psi}{d z_\psi} & = - \sum_{i, j, \cdots} \left[\frac{N_\psi}{N_\psi^{\text{eq}}} \gamma^{\text{eq}}_{\psi \rightarrow i + j + \cdots } - \frac{N_i N_j \cdots}{N_i^{\text{eq}} N_j^{\text{eq}} \cdots} \gamma^{\text{eq}}_{i + j + \cdots \rightarrow \psi} \right] \nonumber \\
    & - \sum_{a, i, j, \cdots} \left[\frac{N_\psi N_a}{N_\psi^{\text{eq}} N_a^{\text{eq}}} \gamma^{\text{eq}}_{\psi + a \rightarrow i + j + \cdots} - \frac{N_i N_j \cdots}{N_i^{\text{eq}} N_j^{\text{eq}} \cdots} \gamma^{\text{eq}}_{i + j + \cdots \rightarrow \psi + a}\right] \; ,
    \label{eq:BEsgeneric2}
    \end{align}
    where $z_\psi=m_\psi/T$ and $H(z_\psi) \equiv H(T= m_\psi/z)$ [see Eq.~\eqref{eq:Hubble}]. In a dilute gas one only considers decays and two particle scatterings, as well as their back reactions. For the decay one has,
    \begin{align}
         \gamma^{\text{eq}}_{\psi \rightarrow i + j + \cdots} = n_\psi^{\text{eq}} \Gamma_{\psi \rightarrow i + j + \cdots} \frac{\mathcal{K}_1(z_\psi)}{\mathcal{K}_2(z_\psi)} \; ,
         \label{eq:reacdecay}
     \end{align}
     where $\Gamma_{\psi \rightarrow i + j + \cdots}$ is the decay rate of the process $\psi \rightarrow i + j + \cdots$ calculated in the center of mass frame of particle $\psi$. The reaction density for a two-body scattering is given by,
    \begin{align}
         \gamma^{\text{eq}}_{\psi + a \rightarrow i + j + \cdots} = \frac{T}{64 \pi^4} \int_{(m_\psi + m_a)^2}^{\infty} ds \; \hat{\sigma}(s)  \sqrt{s} \mathcal{K}_1\left(\frac{\sqrt{s}}{T}\right) \; ,
         \label{eq:reacscattering}
    \end{align}
    where $s$ is the squared center-of mass energy and $\hat{\sigma}(s)$ is the reduced cross section for the process $\psi + a \rightarrow i + j + \cdots$. The latter is related to the usual total cross section $\sigma(s)$ through,
    \begin{align}
    \hat{\sigma}(s) = \frac{8}{s} \left[ (p_\psi\cdot p_a)^2 - m_\psi^2 m_a^2 \right] \sigma(s) \; .
    \label{eq:reducedsigma}
    \end{align}

    It is useful to define the following decay and scattering variables, which are just a rescaled version of the quantities in eqs.~\eqref{eq:reacdecay} and~\eqref{eq:reacscattering}, respectively given by
    \begin{align}
         D_{\psi \rightarrow i + j + \cdots } &= \frac{\gamma^{\text{eq}}_{\psi \rightarrow i + j + \cdots}}{n_\gamma^{\text{eq}} N_\psi^\text{eq} z_\psi H(z_\psi)} = K_{\psi \rightarrow i + j + \cdots} z_\psi \frac{\mathcal{K}_1(z_\psi)}{\mathcal{K}_2(z_\psi)}\; , \nonumber \\ 
          K_{\psi \rightarrow i + j + \cdots} &= \frac{\Gamma_{\psi \rightarrow i + j + \cdots}}{H(T=m_\psi)} \; ,
         \label{eq:Decaygen}
    \end{align}
   and,
    \begin{align}
        S_{\psi + a \rightarrow i + j + \cdots} = \frac{\gamma^{\text{eq}}_{\psi + a \rightarrow i + j + \cdots}}{\left(n_\gamma^{\text{eq}}\right)^2 N_\psi^\text{eq} N_a^\text{eq} z_\psi H(z_\psi)} \; .
        \label{eq:Scatteringgen}
    \end{align}
    Neglecting the CP-violating effects and under the assumption of CPT symmetry, energy conservation implies that $\gamma^{\text{eq}}_{\psi + a \rightarrow i + j + \cdots} = \gamma^{\text{eq}}_{i + j + \cdots\rightarrow \psi + a }$. Hence, the inverse-decay parameter is related to the decay variable defined above, as follows,
    \begin{align}
         {ID}_{\psi \rightarrow i + j + \cdots } =  \frac{N_\psi^\text{eq}}{N_i^\text{eq} N_j^\text{eq}} D_{\psi \rightarrow i + j + \cdots } \; .
         \label{eq:InvDecaygen}
    \end{align}
    %

%--------------------------------------------------------------------
%				APPENDIX B    
%--------------------------------------------------------------------

%%%%%%%%%%%%%%%%%%%%%%%%%%%%%%%%%%%%%%%%%%%%%%%%%%%%%%%%%%%%%%%%%%%%%%%%%%%%%
\chapter{Scattering cross-sections for scalar-singlet assisted leptogenesis}
\label{chpt:scatterings}
%%%%%%%%%%%%%%%%%%%%%%%%%%%%%%%%%%%%%%%%%%%%%%%%%%%%%%%%%%%%%%%%%%%%%%%%%%%%%

In this appendix we collect the expressions of the reduced cross sections for the $\Delta L = 1$ (Fig.~\ref{fig:DeltaLeq1_scattering}) and~$\Delta N = 2$ (Fig.~\ref{fig:DeltaNeq2_scattering}) two-body scattering processes included in our analysis of the BEs in Sec.~\ref{sec:BEs} and numerical computations of Sec.~\ref{sec:BAUmodel}.

In Fig.~\ref{fig:DeltaLeq1_scattering}, diagrams \subref{fig:DeltaLeq1_usual} are the usual ones occurring in type-I seesaw leptogenesis. The reduced cross-section for the $s$-channel mediated $N_i \ell_{\alpha} \to q_L u_R$ process is given by,
    \begin{align}
    \hat{\sigma}_s(N_i \ell_{\alpha} \rightarrow {q_L}_\beta {u_R}_\gamma)&=  \dfrac{1}{16\pi}\dfrac{\left(s-M_i^2\right)^2}{s^2}\sum_{a,b=1}^{n_H}\Y^{a\ast }_{\alpha i}\Y^b_{\alpha i} (\Y^a_{u})_{\beta \gamma}(\Y^{b\ast}_{u})_{\beta \gamma}\; ,
    \end{align}
    and for the $t$-channel mediated $N_i u_R  \to \ell_{\alpha} q_L$ and $N_i q_L \to \ell_{\alpha} u_R$ processes:
    \begin{align}
    \hat{\sigma}_t(N_i {u_R}_\gamma  \rightarrow \ell_{\alpha} q_L) &= \dfrac{1}{16\pi}\dfrac{s-M_i^2}{s}\sum_{a,b=1}^{n_H}\Y^{a }_{\alpha i}\Y^{b\ast}_{\alpha i} (\Y^a_{u})_{\beta \gamma}(\Y^{b\ast}_{u})_{\beta \gamma} \nonumber\\
    & \times \left[1-\dfrac{m_{\Phi_a}^2(M_i^2-m_{\Phi_a}^2)}{(m_{\Phi_b}^2-m_{\Phi_a}^2)(s-M_i^2)} \log\left(\dfrac{s-M_i^2+m_{\Phi_a}^2}{m_{\Phi_a}^2}\right)\right. \nonumber\\
    &\left.+\dfrac{m_{\Phi_b}^2(M_i^2-m_{\Phi_b}^2)}{(m_{\Phi_b}^2-m_{\Phi_a}^2)(s-M_i^2)}\log\left(\dfrac{s-M_i^2+m_{\Phi_b}^2}{m_{\Phi_b}^2}\right)\right]
    \; , \label{eq:Higgsthermal} \\
    \hat{\sigma}_t(N_i q_L \to \ell_{\alpha} {u_R}_\gamma)&=\hat{\sigma}_t(N_i {u_R}_\gamma  \rightarrow \ell_{\alpha} q_L) \, .
    \end{align}
    The expressions above are written in terms of an arbitrary number of Higgs doublets and generic quark Yukawa matrices $\mathbf{Y}_u$. We assume that all quarks couple diagonally to the first Higgs doublet, which in the alignment limit corresponds to the SM Higgs doublet. Furthermore, we only included the dominant top-quark contribution to the above scattering processes in our numerical analysis. The $t$-channel diagrams are mediated by the Higgs and diverge in the limit where we neglect the Higgs masses $m_{\Phi_{a,b}} = 0$. Hence, as commonly done in the literature~\cite{Plumacher:1996kc,Plumacher:1997ru,Plumacher:1998ex,Luty:1992un}, we introduce a Higgs mass $m_{\Phi_{a,b}}/M_i = 10^{-5}$~\cite{Buchmuller:2004nz,LeDall:2014too,Hahn-Woernle:2009jyb}. As previously noted the numerical results are not affected in a substantial way by the chosen prescription.
    
    In Fig.~\ref{fig:DeltaLeq1_scattering}, the new scattering labelled \subref{fig:DeltaLeq1_Nltophih} corresponds to the $N_i \ell_{\alpha} \to \Phi_a h_k$ process, which has the following reduced cross sections,
    \begin{align}
    &\hat{\sigma}_s(N_i \ell_{\alpha} \rightarrow \Phi_a h_k) = \sum_{b,c=1}^{n_H} \frac{\mathbf{Y}_{\alpha i}^{b}\mathbf{Y}_{\alpha i}^{c*} \tilde{\mu}_{a b, k} \tilde{\mu}_{a c, k}^\ast  }{32 \pi} \frac{(s - m_{h_k}^2) (s-M_i^2)^2}{s^4} \; ,
    \end{align}
    \begin{align}
    \hat{\sigma}_t(N_i \ell_{\alpha} \rightarrow \Phi_a h_k) &=\sum_{j,l=1}^{n_R} \frac{(1+\delta_{ij})(1+\delta_{il})|\mathbf{Y}_{\alpha i}^a|^2}{32\pi s^2(M_j^2-M_l^2)}\left\{sM_i(M_j\mathbf{\Delta}_{ij}^k\mathbf{\Delta}_{il}^k+M_l\mathbf{\Delta}_{ij}^{k*}\mathbf{\Delta}_{il}^{k*})\right. \nonumber \\
    &\left.\times\left(M_j^2\log\left[\dfrac{(s-M_i^2)(s-m_{h_k}^2)+sM_j^2}{sM_j^2}\right]-(j\leftrightarrow l)\right)\right.\nonumber\\
    &\left.+sM_jM_l(s-M_i^2)\mathbf{\Delta}_{ij}^k\mathbf{\Delta}_{il}^{k*}\log\left[\dfrac{(s-M_i^2)(s-m_{h_k}^2)M_j^2+sM_j^2M_l^2}{(s-M_i^2)(s-m_{h_k}^2)M_l^2+sM_j^2M_l^2}\right] \right.\nonumber\\ 
    &\left.  -\mathbf{\Delta}_{ij}^{k*}\mathbf{\Delta}_{il}^{k}(M_j^2-M_l^2)(s-M_i^2)(s-m_{h_k}^2) \right.\nonumber\\ 
    &\left. +\mathbf{\Delta}_{ij}^{k*}\mathbf{\Delta}_{il}^{k}\Big( sM_j^2(s+M_j^2-m_{h_k}^2)\right.\nonumber\\ 
    &\left.\times\log\left[\dfrac{(s-M_i^2)(s-m_{h_k}^2)+sM_j^2}{sM_j^2}\right]-(j\leftrightarrow l)\Big)\right\} ,
    \end{align}
    \begin{align}
    \hat{\sigma}_{s-t}(N_i \ell_{\alpha} \rightarrow \Phi_a h_k) &= \sum_{b=1}^{n_H}\sum_{j=1}^{n_R}\dfrac{(1+\delta_{ij})}{16\pi s^3} \Big\{sM_j  \text{Re}\left[\mathbf{Y}_{\alpha i}^b\mathbf{Y}_{\alpha i}^{a*}\tilde{\mu}_{ab,k}(M_iM_j\mathbf{\Delta}_{ij}^{k}-(s-M_i^2)\mathbf{\Delta}_{ij}^{k*})\right]\nonumber\\
    &\times \log\left[\dfrac{(s-M_i^2)(s-m_{h_k}^2)+sM_j^2}{sM_j^2}\right] \nonumber\\
    &-M_i(s-M_i^2)(s-m_{h_k}^2)\text{Re}\left[\mathbf{Y}_{\alpha i}^b\mathbf{Y}_{\alpha i}^{a*}\tilde{\mu}_{ab,k}\mathbf{\Delta}_{ij}^{k}\right]\Big\},
    \end{align}
    which refer to the $s$ and $t$-channel contributions, as well as their interference. The above expression for the $s$-channel contribution matches the result presented in Ref.~\cite{LeDall:2014too}. In case the scalar potential parameter $\mu$ is present in the Lagrangian, the $s$-channel diagram is dominant when compared to the $t$-channel one, due to the usual logarithmic dependence with the mediator mass obtained in $t$-channel cross sections~\cite{LeDall:2014too,Alanne:2017sip,Alanne:2018brf}.
    
    Furthermore, diagrams \subref{fig:DeltaLeq1_Nhtophil} for $N_i h_k \to \ell_{\alpha} \Phi_a$ lead to,
    \begin{align}
     \hat{\sigma}_s(N_i h_k \rightarrow \ell_{\alpha} \Phi_a) &= \sum_{j,l=1}^{n_R} \frac{(1+\delta_{i j})(1+\delta_{i l})\mathbf{Y}_{\alpha l}^{a}\mathbf{Y}_{\alpha j}^{a*}}{32 \pi} \frac{\sqrt{\rho(s,M_i^2,m_{h_k}^2)}}{s (s-M_j^2)(s-M_l^2)} \nonumber \\
    &  \times \Big[(s+M_i^2-m_{h_k}^2)   \left(s \mathbf{\Delta}^k_{i j} \mathbf{\Delta}^{k \ast}_{i l}  + M_j M_l \mathbf{\Delta}^{k \ast}_{i j} \mathbf{\Delta}^k_{i l}\right) \nonumber \\
    & + 2 s M_i  \left(M_l \mathbf{\Delta}^k_{i j} \mathbf{\Delta}^k_{i l}+M_j \mathbf{\Delta}^{k*}_{i j} \mathbf{\Delta}^{k*}_{i l}\right) \Big], \label{eq:sNihktolaphia} \\
   \hat{\sigma}_t(N_i h_k \rightarrow \ell_{\alpha} \Phi_a) &= \sum_{b,c=1}^{n_H} \frac{\mathbf{Y}_{\alpha i}^{b}\mathbf{Y}_{\alpha i}^{c\ast}\tilde{\mu}_{a b, k}\tilde{\mu}_{a c, k}^\ast}{32 \pi s} \Big\{\log \Big[m_{h_k}^2(m_{h_k}^2-2s) + (M_i^2-s)^2 \nonumber \\
    & + (s-M_i^2+m_{h_k}^2)\sqrt{\rho(s,M_i^2,m_{h_k}^2)} + 2 M_i^2 \Gamma_i^2\Big] \nonumber \\
    &- \log \Big[m_{h_k}^2(m_{h_k}^2-2s) + (M_i^2-s)^2  \nonumber \\
    &- (s-M_i^2+m_{h_k}^2)\sqrt{\rho(s,M_i^2,m_{h_k}^2)} + 2 M_i^2 \Gamma_i^2\Big]\Big\}, \label{eq:tRIS}
    \end{align}
    where $\rho(x,y,z)=(x-y-z)^2-4yz$ and the heavy neutrino total decay widths $\Gamma_i$ are shown in Eq.~\eqref{eq:neutrinototal}. As mentioned in Sec.~\ref{sec:BEs} and shown in diagram \subref{fig:RIS2} of Fig.~\ref{fig:diagsRIS}, for our 2RH neutrino case study of Sec.~\ref{sec:model}, the $N_2$ mediated $s$-channel process $N_1 h_k \rightarrow \ell_{\alpha} \Phi_a$ contains a RIS that must be subtracted. Moreover, the Higgs mediated $t$-channel contribution also contains RIS. In fact, for this case, $N_i h_k \rightarrow \ell_{\alpha} \Phi_a$ can be decomposed as $N_i \rightarrow \ell_{\alpha} (\Phi_b \rightarrow \Phi_b) h_k \rightarrow \Phi_a$, where $\Phi_b$ is produced on-shell (see Fig.~\ref{fig:DeltaLeq1_scattering}). The first part corresponds to the heavy neutrino decay $N_i \rightarrow \ell_{\alpha} \Phi_b$ already accounted for in the BEs. To remove this RIS we follow the procedure outlined in Ref.~\cite{Giudice:2003jh}. One needs to regulate the Higgs propagator via the external heavy neutrino $N_i$ decay width $\Gamma_i$, i.e. $t \rightarrow t + i M_i \Gamma_i$, which was noticed first in other contexts~\cite{Ginzburg:1995js,Melnikov:1996iu}. Upon integration, the result is a linearly divergent term $\propto 1/(M_i \Gamma_i)$, in the limit $M_i \Gamma_i \rightarrow 0$, corresponding to a Dirac delta function $\delta(t)$. This identifies an on-shell mediator corresponding to a RIS which is then removed. The above accounts for this subtraction. The $t$-channel is consistent with Ref.~\cite{LeDall:2014too}. However, our $s$-channel result is distinct. Here, we obtain the correct "$+$" sign before the last term.

    The last of the $\Delta L =1$ diagrams corresponds to $N_i \Phi_a \to \ell_{\alpha} h_k$, labeled \subref{fig:DeltaLeq1_Nphitohl}:
    \begin{align}
   \hat{\sigma}_{t_1}(N_i \Phi_a \to \ell_{\alpha} h_k)  &= \sum_{j,l=1}^{n_R}\dfrac{(1+\delta_{ij})(1+\delta_{il})|\mathbf{Y}_{\alpha i}|^2}{16\pi s(M_j^2-M_l^2)}\left\{\Big[(s-M_i^2)\mathbf{\Delta}_{ij}^{k*}\mathbf{\Delta}_{il}^{k}\right.\\
    &\left.-M_i(M_j\mathbf{\Delta}_{ij}^{k}\mathbf{\Delta}_{il}^{k}+M_l\mathbf{\Delta}_{ij}^{k*}\mathbf{\Delta}_{il}^{k*})\Big] \right.\nonumber\\
    &\left.\times\left(M_j^2\log\left[\dfrac{(s-M_i^2)(s-m_{h_k}^2)+sM_j^2}{sM_j^2}\right]-(j\leftrightarrow l)\right)\right.\nonumber\\
    &\left.-\mathbf{\Delta}_{ij}^{k}\mathbf{\Delta}_{il}^{k*}M_jM_l \right.\nonumber\\
    &\left.\times\left((s+M_j^2-m_{h_k}^2)\log\left[\dfrac{(s-M_i^2)(s-m_{h_k}^2)+sM_j^2}{sM_j^2}\right]-(j\leftrightarrow l)\right)\right\} , \nonumber\\
    \hat{\sigma}_{t_2}(N_i \Phi_a \to \ell_{\alpha} h_k) &= \sum_{b,c=1}^{n_H} \frac{\mathbf{Y}_{\alpha i}^{b}\mathbf{Y}_{\alpha i}^{c*}  \tilde{\mu}_{a b, k}\tilde{\mu}_{a c, k}^*}{32 \pi s}  \log \left[ \frac{s^2 (s-M_i^2-m_{h_k}^2)^2 + s^2 M_i^2 \Gamma_i^2}{ M_i^4 m_{h_k}^4 + s^2 M_i^2 \Gamma_i^2 } \right], 
    \end{align}
    where $t_1$ ($t_2$) refers to the $t$-channel diagram shown on the left (right) mediated by a heavy neutrino (Higgs). Once again the Higgs mediated $t$-channel contains a RIS, being the reason why the external heavy neutrino decay width $\Gamma_i$ appears in the expression above. We follow the same procedure described before for the $N_i h_k \to \ell_{\alpha} \Phi_a$ process to subtract this RIS [see Eq.~\eqref{eq:tRIS}]. The quantity $\hat{\sigma}_{t_2}(N_i \Phi_a \to \ell_{\alpha} h_k)$ is in agreement with Ref.~\cite{LeDall:2014too}. 
    
    In Fig.~\ref{fig:DeltaNeq2_scattering}, the diagrams \subref{fig:DeltaNeq2_NNtohh} correspond to the $N_i N_j \to h_p h_l$ process:
    \begin{align}
    \hat{\sigma}_s(N_i N_j \to h_p h_l) &=  \sum_{k,q=1}^{2 n_S} \frac{\tilde{\mu}_{kpl}\tilde{\mu}_{qpl}^*(1+\delta_{i j})^2}{16 \pi(s-m_{h_k}^2)(s-m_{h_q}^2)}  \frac{\sqrt{\rho(s,M_i^2,M_j^2)}}{s}\frac{\sqrt{\rho(s,m_{h_p}^2,m_{h_l}^2)}}{s} \nonumber \\ 
    &\times \left\{(s-M_i^2-M_j^2)\text{Re}\left[\mathbf{\Delta}^k_{i j}\mathbf{\Delta}^{q*}_{i j}\right] - 2 M_i M_j \text{Re}\left[\mathbf{\Delta}^k_{i j}\mathbf{\Delta}^q_{i j}\right] \right\} \nonumber \\ 
    & \times (1+\delta_{kp}+\delta_{kl}+\delta_{pl}+2\delta_{kp}\delta_{kl}\delta_{pl}) \nonumber \\ 
    & \times  (1+\delta_{qp}+\delta_{ql}+\delta_{pl}+2\delta_{qp}\delta_{ql}\delta_{pl})\;,
    \end{align}
    \begin{align}
    \hat{\sigma}_t(N_i N_j \to h_p h_l) &= \sum_{k,q=1}^{n_R} \frac{(1+\delta_{i k})(1+\delta_{jk})(1+\delta_{i q})(1+\delta_{j q})}{16 \pi s^2 (M_k^2-M_q^2)}\left\{-(M_k^2-M_q^2)\sqrt{\rho(s,M_i^2,M_j^2)}\right.\nonumber\\
    &\left.\times\sqrt{\rho(s,m_{h_p}^2,m_{h_l}^2)}\text{ Re}\left[\mathbf{\Delta}^{p}_{i k}\mathbf{\Delta}^{p*}_{i q}\mathbf{\Delta}^{l*}_{j k}\mathbf{\Delta}^{l}_{j q}\right]\right.\nonumber\\
    &\left.+\left[s\left(\log\left[(m_{h_l}^2-m_{h_p}^2)(M_i^2-M_j^2)\right.\right.\right.\right. \nonumber\\
    &\left.\left.+(m_{h_l}^2+m_{h_p}^2+M_i^2+M_j^2-2M_k^2-s)s-\sqrt{\rho(s,M_i^2,M_j^2)}\sqrt{\rho(s,m_{h_p}^2,m_{h_l}^2)}\right]\right.\nonumber\\
    &\left.\left. -\log\Big[(m_{h_l}^2-m_{h_p}^2)(M_i^2-M_j^2)+(m_{h_l}^2+m_{h_p}^2+M_i^2+M_j^2-2M_k^2-s)s \right.\right.\nonumber\\
    &\left.\left.+\sqrt{\rho(s,M_i^2,M_j^2)}\sqrt{\rho(s,m_{h_p}^2,m_{h_l}^2)}\Big]\right)\right.\nonumber\\
    &\left.\times\left(2M_iM_jM_k\left(M_q\text{ Re}\left[\mathbf{\Delta}^{p}_{i k}\mathbf{\Delta}^{p}_{i q}\mathbf{\Delta}^{l}_{j k}\mathbf{\Delta}^{l}_{j q}\right]+M_k\text{ Re}\left[\mathbf{\Delta}^{p*}_{i k}\mathbf{\Delta}^{p*}_{i q}\mathbf{\Delta}^{l}_{j k}\mathbf{\Delta}^{l}_{j q}\right]\right)\right.\right.\nonumber\\
    &\left.\left.+M_j(M_i^2+M_k^2-m_{h_p}^2)\left(M_k\text{ Re}\left[\mathbf{\Delta}^{p}_{i k}\mathbf{\Delta}^{p*}_{i q}\mathbf{\Delta}^{l}_{j k}\mathbf{\Delta}^{l}_{j q}\right]+M_q\text{ Re}\left[\mathbf{\Delta}^{p*}_{i k}\mathbf{\Delta}^{p}_{i q}\mathbf{\Delta}^{l}_{j k}\mathbf{\Delta}^{l}_{j q}\right]\right)\right.\right.\nonumber\\
    &\left.\left.+M_i(M_j^2+M_k^2-m_{h_l}^2)\left(M_k\text{ Re}\left[\mathbf{\Delta}^{p*}_{i k}\mathbf{\Delta}^{p*}_{i q}\mathbf{\Delta}^{l*}_{j k}\mathbf{\Delta}^{l}_{j q}\right]+M_q\text{ Re}\left[\mathbf{\Delta}^{p}_{i k}\mathbf{\Delta}^{p}_{i q}\mathbf{\Delta}^{l*}_{j k}\mathbf{\Delta}^{l}_{j q}\right]\right)\right.\right.\nonumber\\
    &\left.\left.+M_k M_q(M_i^2+M_j^2-s)\text{ Re}\left[\mathbf{\Delta}^{p*}_{i k}\mathbf{\Delta}^{p}_{i q}\mathbf{\Delta}^{l*}_{j k}\mathbf{\Delta}^{l}_{j q}\right]\right.\right.\nonumber\\
    &\left.\left.\left.+\left(M_i^2M_j^2+M_k^2(M_k^2-s)+m_{h_l}^2m_{h_p}^2-m_{h_l}^2(M_i^2+M_k^2)-m_{h_p}^2(M_j^2+M_k^2)\right) \right.\right.\right.\nonumber\\
    &\left.\left.\left.\times\text{Re}\left[\mathbf{\Delta}^{p}_{i k}\mathbf{\Delta}^{p*}_{i q}\mathbf{\Delta}^{l*}_{j k}\mathbf{\Delta}^{l}_{j q}\right]\right)-(k\leftrightarrow q)\right]\right\}\; , \\
    \hat{\sigma}_u(N_i N_j \to h_p h_p) &= \hat{\sigma}_t(N_i N_j \leftrightarrow h_p h_p)\; , \\
    \hat{\sigma}_{s-t}(N_i N_j \to h_p h_l) &=\sum_{k=1}^{n_R}\sum_{q=1}^{2n_S}\dfrac{(1+\delta_{ik})(1+\delta_{jk})(1+\delta_{ij})(1+\delta_{qp}+\delta_{ql}+\delta_{pl}+2\delta_{qp}\delta_{ql}\delta_{pl})}{8\pi s^2(s-m_{h_q}^2)}\nonumber\\
    &\times\text{ Re}[\tilde{\mu}_{qpl}]\left\{-\sqrt{\rho(s,m_{h_p}^2,m_{h_l}^2)}\sqrt{\rho(s,M_i^2,M_j^2)} 
    \right.\nonumber\\
    &\left.\times\left(M_i\text{ Re}\left[\mathbf{\Delta}^{q}_{i j}\mathbf{\Delta}^{p}_{i k}\mathbf{\Delta}^{l*}_{j k}\right]+M_j\text{ Re}\left[\mathbf{\Delta}^{q}_{i j}\mathbf{\Delta}^{p*}_{i k}\mathbf{\Delta}^{l}_{j k}\right]\right)\right.\nonumber\\
    &\left.+s\left(\log\left[(m_{h_l}^2-m_{h_p}^2)(M_i^2-M_j^2)+(m_{h_l}^2+m_{h_p}^2+M_i^2+M_j^2-2M_k^2-s)s\right.\right.\right.\nonumber\\
    &\left.\left.+\sqrt{\rho(s,M_i^2,M_j^2)}\sqrt{\rho(s,m_{h_p}^2,m_{h_l}^2)}\right]\right.\nonumber\\
    &\left.
    -\log\left[(m_{h_l}^2-m_{h_p}^2)(M_i^2-M_j^2)+(m_{h_l}^2+m_{h_p}^2+M_i^2+M_j^2-2M_k^2-s)s\right.\right.\nonumber\\
    &\left.\left.\left.-\sqrt{\rho(s,M_i^2,M_j^2)}\sqrt{\rho(s,m_{h_p}^2,m_{h_l}^2)}\right]\right)\right.\nonumber\\
    &\left.\times\left(2M_iM_jM_k\text{ Re}\left[\mathbf{\Delta}^{q}_{i j}\mathbf{\Delta}^{p}_{i k}\mathbf{\Delta}^{l}_{j k}\right]+M_k(M_i^2+M_j^2-s)\text{ Re}\left[\mathbf{\Delta}^{q*}_{i j}\mathbf{\Delta}^{p}_{i k}\mathbf{\Delta}^{l}_{j k}\right]\right.\right.\nonumber\\
    &\left.\left.+M_j(M_i^2+M_k^2-m_{h_p}^2)\text{ Re}\left[\mathbf{\Delta}^{q}_{i j}\mathbf{\Delta}^{p*}_{i k}\mathbf{\Delta}^{l}_{j k}\right]\right.\right.\nonumber\\
    &\left.\left.+M_i(M_j^2+M_k^2-m_{h_l}^2)\text{ Re}\left[\mathbf{\Delta}^{q*}_{i j}\mathbf{\Delta}^{p*}_{i k}\mathbf{\Delta}^{l}_{j k}\right]\right)\right\} \; , \\
    \hat{\sigma}_{s-u}(N_i N_j \to h_p h_p) &= \hat{\sigma}_{s-t}(N_i N_j \leftrightarrow h_p h_p)\; , 
    \end{align}
    \begin{align}
    \hat{\sigma}_{t-u}(N_i N_j \to h_p h_p) &= \sum_{k,q=1}^{n_R} \frac{(1+\delta_{i k})(1+\delta_{jk})(1+\delta_{i q})(1+\delta_{j q})}{8 \pi s^2 (2m_{h_p}^2+M_i^2+M_j^2-M_k^2-M_q^2-s)}\left\{-\text{Re}\left[\mathbf{\Delta}^{p}_{i k}\mathbf{\Delta}^{p*}_{i q}\mathbf{\Delta}^{p*}_{j k}\mathbf{\Delta}^{p}_{j q}\right]\right.\nonumber\\
    &\left.
    \times(2m_{h_p}^2+M_i^2+M_j^2-M_k^2-M_q^2-s)\sqrt{\rho(s,M_i^2,M_j^2)}\sqrt{\rho(s,m_{h_p}^2,m_{h_p}^2)}\right.\nonumber\\
    &\left.+\left[s\log\left[\dfrac{(2m_{h_p}^2+M_i^2+M_j^2-2M_k^2-s)s-\sqrt{\rho(s,M_i^2,M_j^2)}\sqrt{\rho(s,m_{h_p}^2,m_{h_p}^2)}}{(2m_{h_p}^2+M_i^2+M_j^2-2M_k^2-s)s+\sqrt{\rho(s,M_i^2,M_j^2)}\sqrt{\rho(s,m_{h_p}^2,m_{h_p}^2)}}\right]\right.\right.\nonumber\\
    &\left.\left.
    \times\Big(M_iM_j\left(2M_kM_q \text{Re}\left[\mathbf{\Delta}^{p}_{i k}\mathbf{\Delta}^{p}_{i q}\mathbf{\Delta}^{p}_{j k}\mathbf{\Delta}^{p}_{j q}\right]+(M_i^2+M_j^2-2m_{h_p}^2) \right.\right.\right.\nonumber\\
    &\left.\left.\left. \times\text{Re}\left[\mathbf{\Delta}^{p*}_{i k}\mathbf{\Delta}^{p*}_{i q}\mathbf{\Delta}^{p}_{j k}\mathbf{\Delta}^{p}_{j q}\right]\right) +M_j\left(M_k(m_{h_p}^2+2M_i^2+M_j^2-M_k^2-s) \right.\right.\right.\nonumber\\
    &\left.\left.\left. \times \text{Re}\left[\mathbf{\Delta}^{p}_{i k}\mathbf{\Delta}^{p*}_{i q}\mathbf{\Delta}^{p}_{j k}\mathbf{\Delta}^{p}_{j q}\right]+M_q(M_i^2+M_k^2-m_{h_p}^2) \text{Re}\left[\mathbf{\Delta}^{p*}_{i k}\mathbf{\Delta}^{p}_{i q}\mathbf{\Delta}^{p}_{j k}\mathbf{\Delta}^{p}_{j q}\right]\right)\right.\right.\nonumber\\
    &\left.\left.+M_i\left(M_k(m_{h_p}^2+M_i^2+2M_j^2-M_k^2-s)\text{ Re}\left[\mathbf{\Delta}^{p*}_{i k}\mathbf{\Delta}^{p*}_{i q}\mathbf{\Delta}^{p*}_{j k}\mathbf{\Delta}^{p}_{j q}\right]
    \right.\right.\right.\nonumber\\
    &\left.\left.\left.
    +M_q(M_j^2+M_k^2-m_{h_p}^2) \text{Re}\left[\mathbf{\Delta}^{p}_{i k}\mathbf{\Delta}^{p}_{i q}\mathbf{\Delta}^{p*}_{j k}\mathbf{\Delta}^{p}_{j q}\right]\right)\right.\right.\nonumber\\
    &\left.\left.
    +M_k M_q(M_i^2+M_j^2-s) \text{Re}\left[\mathbf{\Delta}^{p*}_{i k}\mathbf{\Delta}^{p}_{i q}\mathbf{\Delta}^{p*}_{j k}\mathbf{\Delta}^{p}_{j q}\right]\right.\right.\nonumber\\
    &\left.\left.
    +\left(M_i^2M_j^2-m_{h_p}^4-M_k^2(M_k^2-M_i^2-M_j^2-2m_{h_p}^2+s)\right)
    \right.\right.\nonumber\\
    &\left.\left. \times \text{Re}\left[\mathbf{\Delta}^{p}_{i k}\mathbf{\Delta}^{p*}_{i q}\mathbf{\Delta}^{p*}_{j k}\mathbf{\Delta}^{p}_{j q}\right]\Big)-(k\leftrightarrow q)\right]\right\}\; . 
    \end{align}
     The above scattering process was considered in Ref.~\cite{AristizabalSierra:2014uzi} in the limit of a single heavy neutrino and one Yukawa coupling. Here we provide the complete expressions.
    
   Finally, diagrams \subref{fig:DeltaNeq2_NNtohhphiphi} contribute to $N_i N_j \to \Phi_a \Phi_b$:
    \begin{align}
    \hat{\sigma}_s(N_i N_j \to \Phi_a \Phi_b) &= \sum_{k,l=1}^{2 n_S} \frac{\tilde{\mu}_{a b, k}\tilde{\mu}_{a b, l}^*(1+\delta_{i j})^2}{16 \pi(s-m_{h_k}^2)(s-m_{h_l}^2)}  \frac{\sqrt{\rho(s,M_i^2,M_j^2)}}{s}  \nonumber \\ 
    & \times \left\{-2 M_i M_j \text{Re}\left[\mathbf{\Delta}^k_{i j}\mathbf{\Delta}^l_{i j}\right] +(s-M_i^2-M_j^2)\text{Re}\left[\mathbf{\Delta}^k_{i j}\mathbf{\Delta}^{l*}_{i j}\right] \right\}\; , \\
    \hat{\sigma}_t(N_i N_j \to \Phi_a \Phi_b) &=\sum_{\alpha,\beta=e,\mu,\tau}\dfrac{\mathbf{Y}^{a*}_{\alpha j}\mathbf{Y}^{a}_{\beta j}\mathbf{Y}^b_{\alpha j}\mathbf{Y}^{b*}_{\beta j}}{16\pi s}\Big\{\dfrac{s}{2}\log\left(\dfrac{s-M_i^2-M_j^2+\sqrt{\rho(s,M_i^2,M_j^2)}}{s-M_i^2-M_j^2-\sqrt{\rho(s,M_i^2,M_j^2)}}\right)\nonumber\\
    &-\sqrt{\rho(s,M_i^2,M_j^2)}\Big\} , \\
    \hat{\sigma}_{s-t}(N_i N_j \to \Phi_a \Phi_b) &=\sum_{\alpha=e,\mu,\tau}\sum_{k=1}^{2n_S} \dfrac{1}{16\pi s(s-m_{h_k}^2)}\left\{-\sqrt{\rho(s,M_i^2,M_j^2)} \right.\\
    &\left. \times \left(M_i\text{ Re}\left[\mathbf{Y}_{\alpha j}^b\mathbf{Y}_{\alpha i}^{a*}\tilde{\mu}_{a b, k}^*\mathbf{\Delta}^{k*}_{i j}\right]+M_j\text{ Re}\left[\mathbf{Y}_{\alpha j}^b\mathbf{Y}_{\alpha i}^{a*}\tilde{\mu}_{a b, k}^*\mathbf{\Delta}^{k}_{i j}\right]\right)\right.\nonumber\\
    &\left.+M_iM_j\log\left(\dfrac{s-M_i^2-M_j^2+\sqrt{\rho(s,M_i^2,M_j^2)}}{s-M_i^2-M_j^2-\sqrt{\rho(s,M_i^2,M_j^2)}}\right)\right.\nonumber\\
    &\left. \times\left(M_i\text{ Re}\left[\mathbf{Y}_{\alpha j}^b\mathbf{Y}_{\alpha i}^{a*}\tilde{\mu}_{a b, k}^*\mathbf{\Delta}^k_{i j}\right]+M_j\text{ Re}\left[\mathbf{Y}_{\alpha j}^b\mathbf{Y}_{\alpha i}^{a*}\tilde{\mu}_{a b, k}^*\mathbf{\Delta}^{k*}_{i j}\right]\right)\right\}. \nonumber
    \end{align}
    The $s$-channel contribution was already computed in Ref.~\cite{LeDall:2014too} and our results are consistent. Furthermore, the $t$-channel occurs in vanilla type-I seesaw leptogenesis and was already computed in another context in Ref.~\cite{Plumacher:1998ex} and our results are also in agreement.

\end{document}